\pdfoutput=1
\documentclass[a4paper,11pt,titlepage,oneside]{book}
\usepackage[T1]{fontenc}									
\usepackage[utf8]{inputenc}									
\usepackage{lmodern}										
\usepackage[activate={true,nocompatibility},final,tracking=true,kerning=true,spacing=true,factor=1100,stretch=20,shrink=20]{microtype}
\microtypecontext{spacing=nonfrench}
\usepackage[hmargin=2.5cm,vmargin=2.5cm,bindingoffset=1.5cm,includefoot]{geometry}		
\usepackage{xcolor}										
\usepackage{fancyhdr}										
\usepackage{multicol}										
\usepackage{chngcntr}										
\usepackage{tocloft}										
\usepackage{etoolbox}										
\usepackage{pdfpages}										
\usepackage{emptypage}										
\usepackage[grey]{quotchap}									
\usepackage[framemethod=tikz]{mdframed}								
\usepackage{setspace}
\usetikzlibrary{matrix,patterns,calc,shapes,shapes.misc,arrows}
\usepackage{pdflscape}										
\usepackage{enumerate}										
\usepackage{afterpage}										
\usepackage{calc}
\usepackage{relsize}
\usepackage{cite}										
\usepackage[explicit]{titlesec}
\usepackage{amsmath}										
\usepackage{amssymb}										
\usepackage{amsfonts}										
\usepackage{amsthm}										
\usepackage{stmaryrd}										
\usepackage{dsfont}										
\usepackage{mathtools}										
\usepackage{cancel}										
\usepackage{accents}										
\usepackage{graphicx}										
\usepackage{float}										
\usepackage{tabulary}										
\usepackage{booktabs}										
\usepackage[font=footnotesize]{caption}								
\usepackage{multirow}
\usepackage{hyperref}							
\newcommand{\overbar}[1]{\mkern 1.5mu\overline{\mkern-1.5mu#1\mkern-1.5mu}\mkern 1.5mu} 	
\newcommand{\ubar}[1]{\mkern 1.5mu\underline{\mkern-1.5mu#1\mkern-1.5mu}\mkern 1.5mu} 	
\graphicspath{{./Images/}}
\counterwithout{footnote}{chapter}								
\cftsetindents{chapter}{0.5 cm}{0.5 cm}								
\cftsetindents{section}{1 cm}{1 cm}
\cftsetindents{subsection}{1.5 cm}{1.5 cm}
\cftsetindents{subsubsection}{2 cm}{2 cm}
\makeatletter											
	\renewcommand\part{%
	\if@openright
		\cleardoublepage
	\else
		\clearpage
	\fi
	\thispagestyle{empty}%
	\if@twocolumn
		\onecolumn
		\@tempswatrue
	\else
		\@tempswafalse
	\fi
	\null\vfil
	\secdef\@part\@spart}
\makeatother
\allowdisplaybreaks										
\AtBeginEnvironment{quote}{\singlespace\vspace{-\topsep}\small}
\AtEndEnvironment{quote}{\vspace{-\topsep}\endsinglespace}
\makeatletter
\newcommand*{\gmshow@textheight}{\textheight}
\newdimen\gmshow@@textheight
\g@addto@macro\landscape{%
  \gmshow@@textheight=\hsize
  \renewcommand*{\gmshow@textheight}{\gmshow@@textheight}%
}
\def\Gm@vrule{%
  \vrule width 0.2pt height\gmshow@textheight depth\z@
}%
\makeatother
\newlength\chapnumb
\titleformat{\chapter}[block]
{\normalfont\sffamily}{}{0pt}
{\parbox[b]{\chapnumb}{%
   \fontsize{120}{110}\selectfont\thechapter}%
  \parbox[b]{\dimexpr\textwidth-\chapnumb\relax}{%
    \raggedleft%
    \hfill{\LARGE#1}\\
    \rule{\dimexpr\textwidth-\chapnumb\relax}{0.4pt}}}
\titleformat{name=\chapter,numberless}[block]
{\normalfont\sffamily}{}{0pt}
{\parbox[b]{\chapnumb}{%
   \mbox{}}%
  \parbox[b]{\dimexpr\textwidth-\chapnumb\relax}{%
    \raggedleft%
    \hfill{\LARGE#1}\\
    \rule{\dimexpr\textwidth-\chapnumb\relax}{0.4pt}}}

\makeatletter
\def\@chapter[#1]#2{\ifnum \c@secnumdepth >\m@ne
                       \if@mainmatter
                         \refstepcounter{chapter}%
                         \typeout{\@chapapp\space\thechapter.}%
                         \addcontentsline{toc}{chapter}%
                                   {\protect\numberline{\thechapter}#1}%
                       \else
                         \addcontentsline{toc}{chapter}{#1}%
                       \fi
                    \else
                      \addcontentsline{toc}{chapter}{#1}%
                    \fi
                    \chaptermark{#1}%
                    \if@twocolumn
                      \@topnewpage[\@makechapterhead{#2}]%
                    \else
                      \@makechapterhead{#2}%
                      \@afterheading
                    \fi}
\makeatother

\begin{document}
\frontmatter
	\hypersetup{pageanchor=false}
\begin{titlepage}
	\centering
	\begin{figure}[t]
	\centering
	\includegraphics[width=7cm]{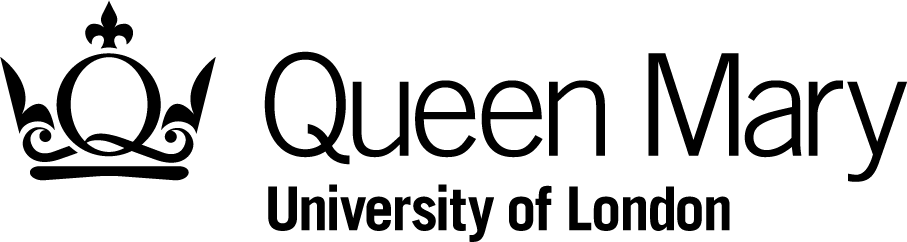}
	\end{figure}
	\vspace{-1cm}
	{\scshape \LARGE Queen Mary University of London\par}
	{\scshape \Large Centre for Research in String Theory\par}
	\vspace{2cm}
	\hrule height 1.1pt
	\vspace{0.5mm}
	\hrule
	\vspace{8pt}
	{\scshape \Huge Exotic Aspects of\\ Extended Field Theories\par}
	\vspace{8pt}
	\hrule
	\vspace{0.5mm}
	\hrule height 1.1pt
	\vspace{4cm}
	{\scshape \Huge Ray Otsuki\par}
	\vspace{2cm}
	{\scshape \Large Supervisor\par}
	{\scshape \LARGE Prof. David S. Berman\par}
	\vspace{1cm}
	{\Large\today\par}
	\vspace{1cm}
    \large
	\begin{gather*}
	\text{Submitted in partial fulfilment of the requirements }\\
	\text{of the Degree of Doctor of Philosophy}.
	\end{gather*}
\thispagestyle{empty}
\end{titlepage}
\hypersetup{pageanchor=true}

\pagenumbering{arabic}
\hspace{0pt}
\vfill
\vfill
\hspace{0pt}
\vfill
\hspace{0pt}
\clearpage
\mainmatter
	\setcounter{tocdepth}{2}
	\setcounter{page}{2}
		\chapter*{Statement of Originality}
\begingroup
\setlength{\parindent}{0cm}
I, Ray Otsuki, confirm that the research included within this thesis is my own work or that where it has been carried out in collaboration with, or supported by others, that this is duly acknowledged below and my contribution indicated. Previously published material is also acknowledged below.
\vspace{0.3cm}\par
I attest that I have exercised reasonable care to ensure that the work is original, and does not to the best of my knowledge break any UK law, infringe any third party’s copyright or other Intellectual Property Right, or contain any confidential material.
\vspace{0.3cm}\par
I accept that the College has the right to use plagiarism detection software to check the electronic version of the thesis.
\vspace{0.3cm}\par
I confirm that this thesis has not been previously submitted for the award of a degree by this or any other university.
\vspace{0.3cm}\par
The copyright of this thesis rests with the author and no quotation from it or information derived from it may be published without the prior written consent of the author.
\vspace{0.3cm}\par
Signature:
\begin{figure}[H]
\includegraphics[width=5cm]{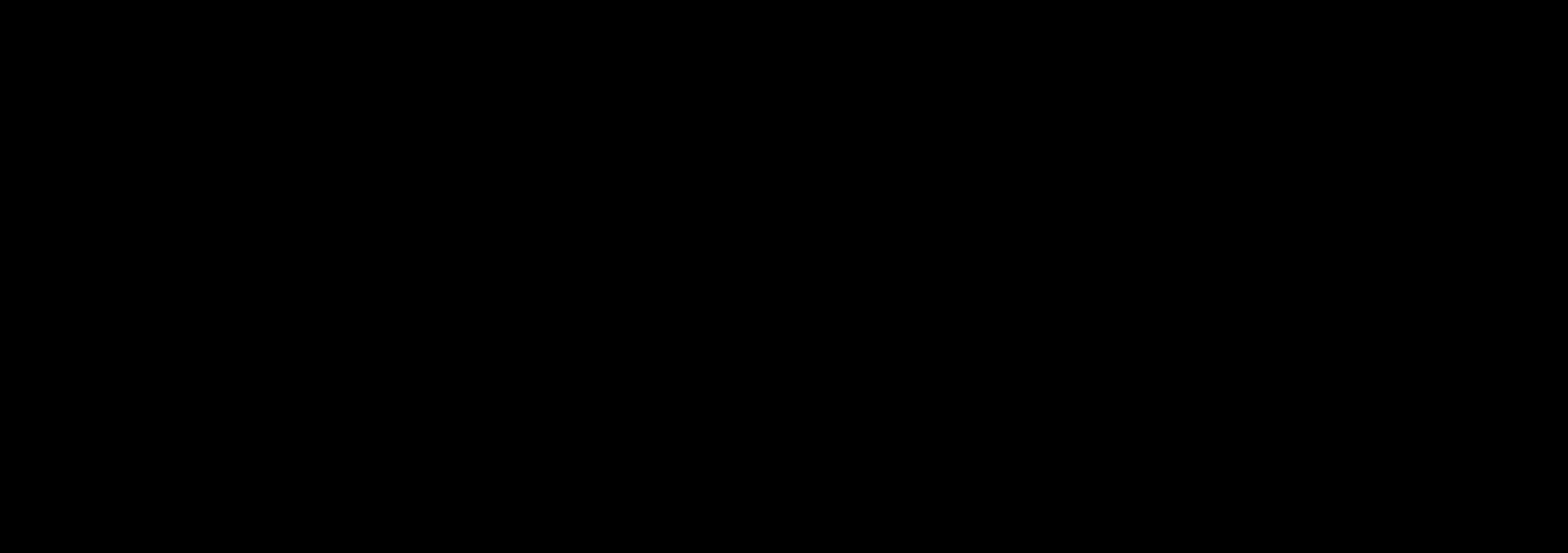}
\end{figure}
Date: \today
\vspace{1cm}\par
\textbf{Details of collaborations and publications:} The work presented in this thesis is based on research carried out in collaboration with my supervisor Professor David S.\ Berman. It is based on the material published in the papers \cite{Berman:2018okd,Berman:2019izh,Otsuki:2019owg,Berman:2019efr}. The work \cite{Berman:2018okd} was further conducted in collaboration with Edvard T.\ Musaev and \cite{Berman:2019izh} was conducted with Chris D.\ A.\ Blair. Additionally, the paper \cite{Bakhmatov:2017les} was published in collaboration with Ilya Bakhmatov, Axel Kleinschmdit and Edvard T.\ Musaev but the material therein has not been included in this thesis.
\endgroup

	\chapter*{Abstract}
Extended field theories (ExFTs) are a relatively young class of theories that lie at the intersection of Kaluza-Klein theory and the remarkable dualities of string- and M-theory. Whereas the original Kaluza-Klein construction unified the local symmetries of an Einstein-Maxwell-dilaton theory into diffeomorphisms in one dimension higher, ExFTs aim for a much more ambitious goal: to unify the local symmetries of supergravity fields into a single symmetry manifest on a higher-dimensional space. Depending on whether we start with Type II or 11-dimensional supergravity, we obtain double and exceptional field theory respectively which we collectively refer to as ExFTs. At the cost of being forced into a generalised notion of diffeomorphisms that fail to close onto an algebra, ExFTs embody a powerful paradigm built on the idea of unification---of symmetries, of fields and of solutions.\par
However, ExFTs are much more than just a rewriting of supergravities. They have been found to contain much more than was originally put into their construction and, in this thesis, we discuss some of the more exotic aspects of these theories. We describe a novel solution in exceptional field theory that unifies a whole family of so-called `exotic branes' into a single solution on the extended space. We follow this with the construction of a maximally non-Riemannian solution whose reduction to the usual spacetime is free of the scalar moduli that typically plague dimensional reductions. In the final part, we consider reductions between exceptional field theories and illustrate, amongst other things, that we can have ExFTs defined on local patches that nevertheless cannot be related by even the duality transformations of the lower-dimensional theory.

	\chapter*{Acknowledgements}
The work presented here could not have been conducted without the continual support and encouragement of my supervisor, Professor David S.\ Berman. His experience and wisdom in the field, coupled to an exceptional enthusiasm and sense of humour, have been a dependable guide throughout my time as a Ph.D.\ student and it is with great pleasure that I extend the warmest of thanks to my mentor.\par
I would also like to extend my thanks to the rest of my collaborators Ilya Bakhmatov, Chris D.\ A.\ Blair, Axel Kleinschmidt and Edvard T.\ Musaev for the many learning experiences that they offered.\par
Within the group at QMUL, I would also like to thank my fellow Ph.D.\ students who made these trying years manageable. Although there are too many to name, I would like to thank my peers in particular: Arnau Robert Koemans Collado, Zolt\'{a}n Laczk\'{o} and Christopher Harry Lewis-Brown. To those who have already graduated, I thank them for welcoming me to the group. To those who have yet to conclude, I wish them the best of luck with their studies. I would also like to thank all of the other academics at CRST for their role in making the group the hub of ideas that it is.\par
Finally, I would like to end by thanking my family and friends without whom I could not have made it this far.\par
\vspace{5cm}
\begin{center}
The work presented here was supported by an STFC studentship.
\end{center}

		\phantomsection \cleardoublepage
		\addcontentsline{toc}{part}{Table of Contents}
		\tableofcontents
		\phantomsection \cleardoublepage
		\addcontentsline{toc}{part}{List of Tables and Figures}
		\listoftables
		\vspace{-0.5cm}
		\listoffigures
\part*{Introduction to\\ Extended Field Theories}
\addcontentsline{toc}{part}{Introduction to Extended Field Theories}
	\chapter{Introduction}
\section{Introduction}
Symmetries of theories have become a staple of modern physics; they form the basis of Yang-Mills gauge theories underlying the Standard Model or the spacetime symmetries underlying special and general relativity. String theory is no different but, in addition to symmetries, it also admits \emph{dualities}---equivalent descriptions of the same physics but in different theories. \emph{T-duality} is one such example. It is an inherently stringy phenomenon which hinges on the fact that strings, unlike point particles, are extended objects which allows them to \emph{wind} around non-contractible cycles. More concretely, the spectrum of a closed string propagating in a background geometry with a compact direction of radius $R$ is given by
\begin{align}\label{eq:StringSpectrum}
M^2 = \frac{n^2}{R^2} + \frac{w^2 R^2}{{\alpha^\prime}^2} + \frac{2}{\alpha^\prime} \left( N + \tilde{N} - 2 \right)\,.
\end{align}
Here, $n$ and $w$ are the quantum numbers of momentum and winding modes respectively and $(N,\tilde{N})$ are left- and right-moving oscillators of the string. Thus, a string receives a contribution of $n/R$ for each of its $n$ units of momenta that it possesses and a contribution of $wR/\alpha^\prime = 2w \pi RT$ for each of its $w$ units of winding it possesses (here, $T = 1/2\pi \alpha^\prime$ is the tension of the string). This is supplemented by the \emph{level-matching condition}
\begin{align}\label{eq:LevelMatching}
N - \tilde{N} = nw\,.
\end{align}
Both of these equations possess a curious invariance under the inversion of the radius $R \leftrightarrow \alpha^\prime /R$ and the exchange of momentum and winding modes $n \leftrightarrow w$---a fact that has come to be known as T-duality. In order to understand why the string spectrum contains such an invariance, it is is constructive to consider the two limits $R \rightarrow \infty$ and $R \rightarrow 0$. The usual intuition of particle physics is recovered in the decompactification limit $R \rightarrow \infty$ in which the circular dimension is expanded to a macroscopic direction. It thus becomes too expensive for the string to wind it and, consequently, the winding modes (whose mass scale as $R / \alpha^\prime$) become too heavy to excite and so decouple from the low-energy dynamics. The only states that are left are the momentum modes (whose mass scale as $1 / R$) which reform the continuum of energy states in the strict limit. Conversely, in the limit $R\rightarrow 0$, the momentum modes become heavy and decouple whilst the winding modes become light. This time it is the momentum modes that start to form a continuum. Whilst this transformation is a symmetry of the bosonic string, it is not a symmetry of the Type II superstring. Instead, it maps the Type IIA and Type IIB string theories into each other---a duality between the two.\par
This is quite a remarkable statement. Another way to state it is that one can exchange winding and momentum modes in the string provided that we also adjust the background geometry in a manner that renders the physics invariant. The transformations of the background fields are dictated by the \emph{Buscher rules} \cite{BUSCHER198759,Alvarez:1994dn} in which one can explicitly see how the components of the metric and 2-form mix. As remarkable as it, one downside of the Buscher rules is the lack of deeper insight that it offers; it gives us a prescription of how to obtain the dual geometry but offers no reasons as to why the transformations should be of that form. The last of these observations leads us to the question of whether there is a neater description of T-duality. In particular, one might wonder if it is possible to reformulate the theory in a manner such that this equivalence is promoted to a manifest symmetry in which dual theories are unified into a single formulation.\par
Double Field Theory (DFT) claims to be the answer. Based on the insight gained from earlier works \cite{Hull:2004in,Hull:2006va,Siegel:1993th,Siegel:1993xq}, DFT was developed by contributions from various authors\cite{Hull:2009mi,Hohm:2010xe,Hohm:2010jy,Hohm:2010pp,Hohm:2011dv,Hohm:2017wtr}. In this thesis, we shall employ the `generalised metric formulation' developed in those papers but it is worth mentioning that there were parallel developments in the same area by a different group that constructed the `doubled-yet-gauged' formulation of DFT \cite{Jeon:2011vx,Jeon:2012kd,Lee:2013hma}. In both descriptions, the action of T-duality is realised on a doubled spacetime in which the $D$ usual coordinates have been supplemented by an equal number of \emph{winding coordinates}. This allows for the momentum and winding modes of the string to be treated on an equal footing and T-duality transformations act linearly on this doubled spacetime\footnote{The situation is actually more subtle than we have suggested; we actually realise a closely related \emph{continuous} hidden symmetry that arises from toroidally compactified string theories rather than the discrete T-duality group. The latter is interpreted on the doubled space as an ambiguity in how the physical spacetime is identified in the presence of isometries. We shall expand on this in more detail in Chapter~\ref{ch:IntroExFT} and continue with this slight abuse of terminology that has become standard in the field.}.\par
In order to construct an action that is manifestly symmetric under this duality, we first need to define fields that transform covariantly under this symmetry. In particular, the metric and Kalb-Ramond 2-form are combined into a \emph{generalised metric} on this doubled space whilst the dilaton is shifted to transform as a scalar under these transformations. The local symmetries of supergravity are then reinterpreted to act on these covariant fields. In particular, just as the Lie derivative of general relativity (GR) generates the infinitesimal diffeomorphisms of the metric, we construct a \emph{generalised Lie derivative} that generates the infinitesimal diffeomorphisms of the metric and 2-form upon acting on the generalised metric. The other local symmetry of DFT is an analogue of the local Lorentz symmetry of GR.\par
With these symmetries in hand, it is then a matter of constructing a lift of the supergravity action to the doubled space that respects all of these symmetries. Of course, having started by doubling the spacetime, we then require some prescription to drop half of the coordinates in a manner that allows us to recover the theory that we started with. This is done through the \emph{section condition}, which are a set of constraints on the coordinate dependences of the fields that specify which of the $2D$ coordinates are to be regarded as the physical spacetime in a particular duality frame. So long as our solutions and equations of motion respect these constraints, the above construction yields a rewriting of supergravity that is duality-symmetric.\par
Of course, we have only given a very schematic presentation of the construction. We have suppressed many of the finer details and will leave those for the more technical introduction to DFT (or rather, ExFTs in general) in Chapter~\ref{ch:IntroExFT}. For now, we note excellent review on DFT can be found in \cite{Geissbuhler:2013uka,Berman:2013eva,Aldazabal:2013sca}. Various extensions of the DFT formalism beyond the universal sector have been constructed; the R-R fields of Type II theories were treated in \cite{Hohm:2011dv,Hohm:2011cp,Hohm:2011zr,Jeon:2012kd}, heterotic extensions were considered in \cite{Hohm:2011ex,Malek:2016vsh} (see also \cite{Hohm:2013nja}) and the supersymmetric completions were considered in \cite{Coimbra:2011nw,Jeon:2011sq,Jeon:2012hp}. Of particular note are the series of works by various groups that extend DFT to higher orders in $\alpha^\prime$, based on the highly restrictive constraints that duality-covariance places on higher-derivative corrections\cite{Godazgar:2013bja,Hohm:2013jaa,Hohm:2014xsa,Hohm:2015mka,Bedoya:2014pma,Coimbra:2014qaa,Marques:2015vua,Lee:2015kba,Baron:2017dvb}.\par
As versatile as DFT is, T-duality is not the only duality of string theory; we also have the \emph{S-duality} of the Type IIB string which inverts the string coupling, rescales the string length and mixes the NS-NS and R-R sectors. Finally, the combination of S- and T-duality gives rise to \emph{U-duality} in 11 dimensions and the symmetry groups governing U-duality transformations are the \emph{Julia-Cremmer duality groups} (see, for example, \cite{Duff:1996aw,Witten:1995zh,Townsend:1995kk,Obers:1998fb} for developments that occurred in this area in the late '90s). Analogous discussions to the one we had above still holds in 11 dimensions but, this time, U-duality exchanges momentum modes with wrapping modes of branes. It is then possible to construct theories that make this U-duality manifest (with the same caveat on terminology that we made for DFT above). Owing to the fact that the Julia-Cremmer groups turn out to be the exceptional Lie groups, the theories thus constructed have been dubbed \emph{exceptional field theory} (EFT). As we wish to emphasise the similarities between DFT and EFT, we shall refer to them collectively as \emph{Extended Field Theories} (ExFTs).\par
Just as in DFT, one defines a generalised metric that places the metric and $p$-forms (this time of 11-dimensional supergravity, compactified on an $n$-torus) on an equal footing. One crucial difference between DFT and EFT is that the latter adopts a Kaluza-Klein type ansatz to couple a $d$-dimensional \emph{external} spacetime to the now extended (rather than doubled) \emph{internal space}. The `glue' between the two spaces comes in the form of generalised vectors and we now require our action to be invariant under diffeomorphisms on the external space, as well as all the symmetries that we have mentioned previously. The technical details are quite involved and it was not until very recently that these theories were constructed in full. As with DFT, we shall reserve a fuller description of their construction to Chapter~\ref{ch:IntroExFT}.\par
Due to it still being relatively young, EFTs currently lack review articles\footnote{After this thesis was written, the review \cite{Berman:2020tqn} appeared.} and so we direct the reader to the original EFT papers \cite{Hohm:2014fxa,Hohm:2013uia,Hohm:2013vpa,Abzalov:2015ega,Musaev:2015ces,Hohm:2015xna,Berman:2015rcc} for the finite cases $2 \leq n \leq 8$ (see also \cite{Berman:2010is,Berman:2011jh,Berman:2011cg,Malek:2012pw,Berman:2012vc,Hillmann:2009ci,Hohm:2013jma} for precursors to EFT and \cite{Bossard:2017aae,Bossard:2018utw} for progress on the $n=9$ EFT) and their supersymmetric extensions \cite{Baguet:2016jph,Godazgar:2014nqa,Musaev:2015pla}.\par
Having spent considerable effort in constructing these theories, one is interested in what they have to offer. Curiously, they have demonstrated themselves to be much more than the rewritings of supergravities that they may appear to be. One of the early successes of DFT was the realisation of that one could relax the section condition to carry out a generalised Scherk-Schwarz reductions to gauged supergravities \cite{Aldazabal:2011nj,Dibitetto:2012rk,Grana:2012rr,Berman:2012uy,Aldazabal:2013mya,Aldazabal:2013via}. In particular it was found that they could give higher-dimensional \emph{geometric} origins to whole orbits of non-geometric fluxes that previously were thought not to follow from the standard supergravity compactifications. One such example is the Romans' massive supergravity \cite{1986PhLB..169..374R} whose DFT and EFT lifts were constructed in \cite{Hohm:2011cp,Ciceri:2016dmd} respectively.\par
ExFTs have also been shown to accommodate highly non-perturbative states called \emph{exotic branes} that are predicted by duality transformations \cite{Obers:1998fb,deBoer:2010ud,deBoer:2012ma,Fernandez-Melgarejo:2018yxq}. One of the most striking features of solutions of ExFTs are their ability to unify multiple solutions of string- and M-theory into single solutions on the extended space. Just as the membrane of M-theory gave a higher-dimensional origin to the string and D2-brane, ExFTs give higher-dimensional origins to multiple objects in string- and M-theory, including exotic branes\cite{Berkeley:2014nza,Berman:2014jsa,Bakhmatov:2016kfn,Bakhmatov:2017les,Kimura:2018hph,Lee:2018gxc,Rudolph:2016sxe,Blair:2013noa,Blair:2016xnn,Blair:2017hhy}. Such solutions shall be the basis of Chapters~\ref{ch:NonGeometricE7} and \ref{ch:Map}.\par
Another interesting development in this area followed from the realisation in \cite{Lee:2013hma} that certain backgrounds that lack a conventional spacetime interpretation could still be studied in the doubled sigma model \cite{Hull:2004in}. This was exploited in \cite{Ko:2015rha} to describe the Gomis-Ooguri scaling limit of closed strings \cite{Gomis:2000bd} (which results in a string that respects a non-relativistic Galilean symmetry rather than a Lorentz symmetry) within DFT. This was followed by \cite{Morand:2017fnv,Cho:2018alk,Cho:2019ofr} that expanded upon this idea to include non-relativistic geometries, ultra-relativistic geometries and even theories with no geometric structure at all. Of particular interest to us is the last of these, which was dubbed the `maximally non-Riemannian' solution of DFT, and it was demonstrated in \cite{Cho:2018alk} that a Kaluza-Klein reduction on such a space is entirely free of the scalar moduli that typically plague such reductions. We shall discuss their work and provide an EFT analogue of the maximally non-Riemannian solution in Chapters~\ref{ch:DFTNonRiemannian} and \ref{ch:E8NonRiemannian}.\par
We finally end with some remarks on closely related formalisms. Firstly, we have generalised geometries and exceptional generalised geometries \cite{Hull:2007zu,Pacheco:2008ps,Coimbra:2011nw,Coimbra:2011ky,Coimbra:2012af,Coimbra:2012yy}, whose developments were sparked by the works of Hitchin and Gualtieri \cite{Hitchin:2004ut,Gualtieri:2003dx}. There, the spacetime itself is not extended but the tangent space is extended. The ExFT formalism also shares many aspects with the $E_{11}$ program pioneered by West \textit{et al.\ }\cite{Tumanov:2016abm,Kleinschmidt:2003jf,West:2004kb,West:2011mm,Riccioni:2009xr,West:2003fc,West:2004iz,Cook:2008bi,Tumanov:2015yjd}. We shall occasionally make reference to these in the main text.
\section{Structure of this Thesis}
This thesis is structured as follows. In Chapter~\ref{ch:IntroExFT}, we begin with a more technical introduction to the basic ideas behind ExFTs that will be assumed for the rest of this thesis. The bulk of the text will then present three studies on exotic aspects of ExFTs.\par
We begin in Chapter~\ref{ch:NonGeometricE7} with an introduction to exotic branes and construct a novel solution in $E_{7(7)}$ EFT that unifies a number of exotic branes in both Type II string- and M-theory into a single solution on the extended space. Each of the individual solutions can be obtained by an appropriate rotation on the internal space and then reduced to supergravity via the section condition. With an appropriate choice of isometries, these rotations implement the U-duality transformations between those objects. In Chapter~\ref{ch:Map} we continue our discussion of exotic branes and describe a procedure that generates all duality-related objects, as characterised by their mass formulae. We also consider an enumeration of such objects at each power of $g_s$ and compare our results with the literature before making some remarks that conclude our study of exotic branes.\par
We then turn our focus towards non-Riemannian backgrounds in ExFTs. Chapter~\ref{ch:DFTNonRiemannian} introduces previous work in the field that recently gave rise to a full classification of admissible backgrounds in DFT, beyond the usual Riemannian parametrisation. Chapter~\ref{ch:E8NonRiemannian} follows this with a construction of an analogue of the `maximally non-Riemannian' solution of that classification within $E_{8(8)}$ EFT. The result is a particular 3-dimensional topological theory that had previously been obtained only as a truncation of the full EFT and so we demonstrate that it can instead be understood as a full solution of EFT, albeit a non-Riemannian solution.\par
Chapter~\ref{ch:EFTReductions} makes some steps towards filling a gap in the literature, regarding reductions of ExFTs. Whilst DFT-to-DFT and EFT-to-DFT have been studied by various authors, little work has been done on EFT-to-EFT reductions. We focus on the cases $E_{8(8)} \rightarrow E_{7(7)}$ and $\operatorname{SL}(5) \rightarrow \operatorname{SL}(3) \times \operatorname{SL}(2)$ and give some examples of the reductions of the generalised coordinates, generalised metric, section conditions and actions. Whilst most of the results are in line with expectation, we raise some possibilities that we believe may be of interest for future works.\par
This concludes the meat of our expositions and we make some closing remarks in Chapter~\ref{ch:Conclusion}. Finally, Appendices~\ref{app:ExoticBackgrounds},\ref{app:ExoticWeb} and \ref{app:MOrigin} collect some supplementary material regarding exotic branes whilst Appendices~\ref{app:ExoticFieldstrength} and \ref{app:CosetProjectorTraces} collect some calculations that would be unconstructive in the main text.
\section{Notation}
We have attempted to standardise the notation as much as possible. Where possible, we have opted to respect the following conventions \emph{in general}
\begin{itemize}
	\item The solution-generating group of an ExFT is denoted $G$ and its maximal compact subgroup is denoted $H$. Any other subgroup, particularly in the context of Chapters~\ref{ch:DFTNonRiemannian} and \ref{ch:E8NonRiemannian}, are denoted $\tilde{H}$.
	\item Representations of $G$ that are relevant to ExFTs are denoted $R_1, R_2, \ldots$, except for the adjoint representation which we denote $\mathbf{adj.}$.
	\item The generalised Lie derivative and projectors onto representations of $G$ are denoted by blackboard bold font: $\mathbb{L}, \mathbb{P}_{R_i}$ respectively.
	\item Objects that are fundamental to ExFTs are denoted by calligraphic capital letters: $\mathcal{M}_{MN}, \mathcal{A}_{\mu}{}^M, \mathcal{F}_{\mu \nu}{}^M, \mathcal{P}_{MN}{}^{KL}$ etc.
	\item The dimension of the supergravity is denoted $D$. For EFT, we have $D=11$ which is split into $d$ external coordinates $x^\mu$ and $n$ internal coordinates $y^m$. The latter are extended to a set of $\operatorname{dim} R_1$ generalised coordinates $Y^M$. We use
	\begin{itemize}
		\item Greek characters $\mu, \nu, \ldots =1,2, \ldots, d$ to index the external space.
		\item Lower case Roman letters $m,n, \ldots = 1, 2, \ldots, n$ to index the M-theory section.
		\item Fraktur font $\mathfrak{m}, \mathfrak{n} = 1, 2, \ldots n-1$ to index the Type IIA coordinates.
		\item Roman font $\mathrm{m}, \mathrm{n}, \ldots = 1, 2, \ldots, n-1$ to index the Type IIB section.
		\item Capital Roman letters $M,N,\ldots = 1, 2, \ldots, \operatorname{dim} R_1$ to index the extended internal space.
		\item `Early' Greek characters $\alpha, \beta, \ldots = 1, 2, \ldots, \operatorname{dim} \textbf{adj.}$ to index adjoint representations of various groups or $\alpha, \beta = 1,2$ to index $\operatorname{SL}(2)$ representations in various contexts.
	\end{itemize}
\end{itemize}
Any deviations from the above have been noted in the main text in such a way as to hopefully avoid any confusion. In any case, particular care has been taken to spell out the meaning of indices or coordinates when they first appear in that context.\par
Finally, the permutation symbol $\varepsilon$ and Levi-Civita tensor $\epsilon$ are defined as follows:
\begin{align*}
\begin{array}{ll}
\varepsilon_{\mu_1 \ldots \mu_n} \coloneqq \begin{cases}
\begin{array}{ll}
+1\,, & \text{Even permutation of indices}\\
-1\,, & \text{Odd permutation of indices}\\
0\,, & \text{Otherwise}\\
\end{array}
\end{cases} & 
\varepsilon^{\mu_1 \ldots \mu_n} \coloneqq {(-1)}^t \varepsilon_{\mu_1 \ldots \mu_n}\,,\\
\epsilon_{\mu_1 \ldots \mu_n} \coloneqq \sqrt{|g|} \varepsilon_{\mu_1 \ldots \mu_n}\,, &
\epsilon^{\mu_1 \ldots \mu_n} \coloneqq \frac{1}{\sqrt{|g|}} \varepsilon^{\mu_1 \ldots \mu_n}\,,\\
\end{array}
\end{align*}
where $t$ is the number of time-like directions in the indices $\mu_i$.

	\chapter{Introduction to Extended Field Theories}\label{ch:IntroExFT}
In this chapter, we give a more technical introduction to the relevant aspects of extended field theories. Whereas much of the literature on ExFTs is split cleanly into dealing with either DFT or EFT, we have instead attempted to give a description that emphasises the similarities of the structures common to both in an attempt to highlight the key concepts that go into constructing these theories. We shall cover the extended spacetime of ExFTs, and the local and global symmetries that act on these spaces. This will lead us to the \emph{generalised Lie derivative} whose closure imposes a crucial consistency condition, called the \emph{section condition}, on ExFTs. We introduce the field content of the theories, including a tensor hierarchy-like structure in EFT, before discussing the action and equations of motion of ExFTs.\par
Whilst there are excellent reviews on DFT available in \cite{Aldazabal:2013sca,Geissbuhler:2013uka,Berman:2013eva}, a general introduction to EFTs is still lacking. Although this is partly due to how young the field is, much of it is also probably due to the idiosyncrasies of each EFT which make treating them all at the same time rather difficult. Where possible, we shall try to speak in general terms that apply to all ExFTs but we shall inevitably have to restrict to individual theories when discussing some of the detail. We have summarised some useful facts about the various ExFTs in Table~\ref{tab:Summary} and the next few sections shall be devoted to understanding what each of the symbols mean.
\begin{landscape}
\begingroup
\captionsetup{width=0.85\linewidth}
\begin{table}
\centering
\begin{tabulary}{\linewidth}{LLLCCCCCCCC}
\toprule
$n$ & $G$ & $H$ & $\operatorname{dim}G/H$ & $\alpha$ & $\omega$ & $\gamma$ & $r$ & $R_1$ & $R_2$ & $\textbf{adj.}$\\
\midrule
$-$ & $\operatorname{GL}(D)$ & $\operatorname{SO}(D)$ & $D(D+1)/2$ & 1 & 0 & 0 & 0 & $\mathbf{D}$ & $-$ & $\mathbf{D^2}$\\
$-$ & $\operatorname{O}(D,D)$ & $\operatorname{O}(D) \times \operatorname{O}(D)$ & $D^2$ & 2 & 0 & 1 & 0 & $\mathbf{2D}$ & $\mathbf{1}$ & $\mathbf{D (2D-1)}$\\
2 & $\operatorname{SL}(2) \times \mathbb{R}^+$ & $\operatorname{SO}(2)$ & 3 & $-$ & $-1/7$ & $4/3$ & 0 & $\mathbf{2}_1 \oplus \mathbf{1}_{-1}$ & $\mathbf{2}_0$ & $\mathbf{3}$\\
3 & $\operatorname{SL}(3) \times \operatorname{SL}(2)$ & $\operatorname{SO}(3) \times \operatorname{SO}(2)$ & 7 & (2,3) & $-1/6$ & 2 & 0 & $\mathbf{(3,2)}$ & $(\overbar{\mathbf{3}},\mathbf{1})$ & $\mathbf{(8,1)\oplus(1,3)}$\\
4 & $\operatorname{SL}(5)$ & $\operatorname{SO}(5)$ & 14 & 3 & $-1/5$ & 3 & 0 & $\mathbf{10}$ & $\overbar{\mathbf{5}}$ & $\mathbf{24}$\\
5 & $\operatorname{SO}(5,5)$ & $\operatorname{SO}(5) \times \operatorname{SO}(5)$ & 25 & 4 & $-1/4$ & 5 & 0 & $\mathbf{16}$ & $\mathbf{10}$ & $\mathbf{45}$\\
6 & $E_{6(6)}$ & $\operatorname{USp}(8)$ & 42 & 6 & $-1/3$ & 10 & 0 & $\mathbf{27}$ & $\mathbf{\overbar{27}}$ & $\mathbf{78}$\\
7 & $E_{7(7)}$ & $\operatorname{SU}(8)$ & 70 & 12 & $-1/2$ & 28 & 0 & $\mathbf{56}$ & $\mathbf{1} \oplus \mathbf{133}$ & $\mathbf{133}$\\
8 & $E_{8(8)}$ & $\operatorname{SO}(16)$ & 128 & 60 & $-1$ & 189 & $2/15$ & $\mathbf{248}$ & $\mathbf{1 \oplus 248 \oplus \mathbf3875}$ & $\mathbf{248}$\\
\bottomrule
\end{tabulary}
\caption[A summary of some pertinent facts regarding ExFTs.]{A summary of some of the important fact that are used in ExFTs. Note that we have included GR, with $G = \operatorname{GL}(D)$, for comparison. The solution-generating group $G$ is the hidden symmetry that appears when supergravity is compactified on $T^n$. Next to it is its maximal compact subgroup $H$. The generalised metric is understood as a representative of the coset $G/H$ and the number of degrees of freedom that it encodes is counted by $\operatorname{dim} G/H$. The numbers $\alpha$ and $\omega$ are $G$-dependent constants that appear in the generalised Lie derivative. In particular, $\alpha$ is the coefficient in front of the adjoint projector and $\omega$ is a universal weight that modifies the weight of an object under the generalised Lie derivative. For $n= 3$, the two values $\alpha = (2,3)$ denotes the coefficients of the two projectors ${\mathbb{P}}_{(8,1)}$ and ${\mathbb{P}}_{(1,3)}$. The last three columns denote particular representations of $G$ that are relevant to ExFTs: $R_1$ is the coordinate representation, $R_2$ is the section representation and $\textbf{adj.}$ is the adjoint representation of $G$. The remaining columns $\gamma \coloneqq Y^{MN}{}_{MN} / \operatorname{dim}R_1$ and $r \coloneqq 1/(2\alpha) \mathcal{M}_{MN} Y^{MN}{}_{KL} \mathcal{M}^{KL}$ denote constants that involve the $Y$-tensor of each theory but are not required until Chapter~\ref{ch:E8NonRiemannian}. Note the modified representations $R_2$ for $n=7,8$, relative to the paper \cite{Berman:2012vc} that introduced them, which did not allow for tensor densities as later papers have.}
\label{tab:Summary}
\end{table}
\endgroup
\end{landscape}
\section{The Coordinate Representation}
The first step is to choose whether we seek to study T-duality or U-duality, for which we choose either $G = \operatorname{O}(D,D; \mathbb{R})$ or $G = E_{n(n)}(\mathbb{R})$ respectively. Note that these are \emph{not} the duality groups themselves, being over $\mathbb{R}$ rather than $\mathbb{Z}$; they are the hidden symmetry groups that emerge from Type II and 11-dimensional supergravity reduced on an $n$-torus. Nevertheless, the duality groups do arise as discrete subgroups of the continuous groups that we consider and we shall highlight the relation between $G$ and the duality groups later on. We shall be careful in our terminology and distinguish between the continuous groups (which we shall henceforth refer to as the \emph{solution-generating groups}) and the true duality groups. We shall also drop the $\mathbb{R}$ qualifier in $G$ and it shall be assumed that we are talking about the solution-generating groups, rather than the duality groups, unless explicitly stated.\par
For DFT, we start off by doubling the $D$-dimensional space to obtain an $2D$-dimensional extended spacetime. This is in the spirit of the generalised geometry discussed, for example, in \cite{Coimbra:2012yy,Coimbra:2011nw,Gualtieri:2003dx,Hitchin:2004ut} in which the transformations of the diffeomorphisms and gauge transformations of the NS-NS sector are combined into generalised diffeomorphisms on an extended tangent bundle $E = TM \oplus T^\ast M$. However, DFT goes one step further and doubles not only the tangent bundle but the entire spacetime. We then define coordinates $Y^M$ (with $M =1,2, \ldots, 2D$) on this doubled spacetime which are composed from the usual coordinates $X^m$, conjugate to momenta, and \emph{winding coordinates} ${\tilde{X}}_m$, conjugate to the winding modes of the string:
\begin{align}
Y^M = (X^m, {\tilde{X}}_m)\,,
\end{align}
Together, these transform in the \emph{coordinate representation} $R_1 = \mathbf{2D}$ of $\operatorname{O}(D,D)$. Transformations in $G$ then act linearly on the coordinates in the expected fashion.\par
For EFT, we instead start off with a nominal splitting of an 11-dimensional space into a $d$-dimensional \emph{external space} $\mathcal{M}^d$ and the remainder $\mathcal{M}^n$ (such that $d+n =11$). The latter is augmented to an \emph{(extended) internal space} of dimension $\operatorname{dim} R_1$, where $R_1$ is a particular representation of $E_{n(n)}$ called the coordinate representation as in DFT. Denoting the coordinates on this internal space as $Y^M$ (again with $M = 1, \ldots, \operatorname{dim}R_1$) these representations are chosen such that, when decomposed under $\operatorname{GL}(n)$, they produce the usual coordinates plus extra coordinates conjugate to the wrapping modes of the branes of M-theory on $T^n$. For example, when $n=4$, the solution-generating group is $G = \operatorname{SL}(5)$ and the coordinate representation is $R_1 = \mathbf{10}$ of $G$. We thus have $d = 11-4 = 7$ external coordinates and 10 extended internal coordinates. Decomposing under $\operatorname{GL}(4)\subset \operatorname{SL}(5)$, the latter split into
\begin{align}
Y^M = (y^m, y_{mn})
\end{align}
where $m = 1,\ldots,4$ and $y_{mn}$ is an antisymmetric set of ${}^4 C_2 = 6$ coordinates that are identified as the duals of the wrapping modes the M2 brane. Note that the 4-dimensional internal space before the enhancement is too small to allow wrappings of the M5 brane and so we do not obtain any coordinates corresponding to M5 wrapping modes in the extended space. These then complement the external coordinates $x^\mu$, with $\mu =1, \ldots, 7$ to define a $(7 + 10)$-dimensional extended spacetime.\par
In this language, GR can be considered as an ExFT with group $G = \operatorname{GL}(D)$, $d=0$ and $n=D$. It is the `trivial' ExFT in the sense that, in addition to possessing no external space (like DFT), the internal space is also not enlarged by brane wrapping modes at all. The coordinate representation in this case is simply the representation $\mathbf{D}$.
\section{The Field Content}\label{sec:FieldsExFT}
The field content of DFT (considering only the NS-NS sector) is given simply by
\begin{align}
\text{DFT: } \{ d, \mathcal{M}_{MN}\}
\end{align}
where $d$ is a shifted dilaton that transforms as an $\operatorname{O}(D,D)$ scalar, related to the supergravity dilaton by
\begin{align}
d = \phi - \frac{1}{4} \ln g\,,
\end{align}
and $\mathcal{M}_{MN}$ is the \emph{generalised metric} that is parametrised by the metric $g$ and Kalb-Ramond 2-form $B_{(2)}$. On the other hand, the EFT field content is given by
\begin{align}
\text{EFT: } \{ g_{\mu \nu}, \mathcal{M}_{MN}, \mathcal{A}_{\mu}{}^M, \ldots \}\,,
\end{align}
where $g_{\mu \nu}$ is a metric on the external space and $\mathcal{M}_{MN}$ is a generalised metric on the internal space. The remaining fields $\mathcal{A}_\mu{}^M,\ldots$ straddle both the external and internal spaces and define a tensor hierarchy of generalised gauge fields whose structure we shall defer to a later section. As mentioned previously, the lack of an external space in DFT removes the need of such gauge fields, resulting in the simpler field content.\par
We see that the generalised metric $\mathcal{M}_{MN}$ is common to both DFT and EFT. In general, for an ExFT with group $G$, the generalised metric is a representative of the coset $G/H$ where $H$ is the maximal compact subgroup of $G$ (these have been collected for the various ExFTs in Table~\ref{tab:Summary}). In the case of DFT, we have $H = \operatorname{O}(D) \times \operatorname{O}(D)$ and the dimension of the coset $\operatorname{dim}G/H = D^2$ enumerates the degrees of freedom of the metric and Kalb-Ramond 2-form:
\begin{align}
D^2 = \underbrace{\frac{D(D+1)}{2}}_{g} + \underbrace{\frac{D(D-1)}{2}}_{B_{(2)}}\,.
\end{align}
Concretely, the DFT generalised metric can be parametrised by the block matrix
\begin{align}\label{eq:DFTGenMetric}
\mathcal{M}_{MN} & = \begin{pmatrix} g_{mn} - B_{mp} g^{pq} B_{qn} & B_{mp} g^{pn}\\
- g^{mp} B_{pn} & g^{mn}
\end{pmatrix}\,.\\
&  = \begin{pmatrix}
\delta_m^p & B_{mp}\\
0 & \delta^m_p
\end{pmatrix}
\begin{pmatrix}
g_{pq} & 0\\
0 & g^{pq}
\end{pmatrix}
\begin{pmatrix}
\delta^q_n & 0\\
- B_{qn} & \delta_q^n
\end{pmatrix}\,.
\end{align}
This matrix also appears in the Hamiltonian formulation of the string worldsheet action when one combines the momenta $P_m$ and velocities $\dot{X}^m$ into a $2D$-dimensional vector.\par
Similarly for EFT, the dimension of the coset is equal to the total number of degrees of freedom of 11-dimensional supergravity compactified on $T^n$. Returning to the $\operatorname{SL}(5)$ EFT example, we have $H = \operatorname{SO}(5)$ and so the dimension of the coset is equal to $24 - 10 = 14$ which are split into the degrees of freedom of an $n$-dimensional metric $g_{mn}$ on the internal space and the M-theory 3-form potential coupling to the M2-brane:
\begin{align}
14 = \underbrace{\frac{4(4+1)}{2}}_{g} + \underbrace{{}^4 C_3}_{A_{(3)}}\,.
\end{align}
As mentioned before, the internal space is too small to contain the wrapping modes of the M5 and its electric potential $A_{6}$ accordingly does not enter into $\mathcal{M}_{MN}$. The generalised metric can then be parametrised by
\begin{align}
\mathcal{M}_{MN} = \begin{pmatrix}
g_{m n} + \frac{1}{2} A_{m pq} g^{pq, st} A_{st n} & \frac{1}{\sqrt{2}} A_{m pq} g^{pq,kl}\\
\frac{1}{\sqrt{2}} g^{mn,pq} C_{pqk} & g^{mn,kl}
\end{pmatrix}\,,
\end{align}
where $g^{mn,kl} \coloneqq g^{m[k|} g^{n|l]}$ is to be thought of as a metric on the antisymmetric representation. If we move to larger $E_{n(n)}$, then the $A_{(6)}$ potential (and, indeed, more exotic objects such as the dual graviton) start entering into the generalised metric and the parametrisation becomes rather involved. By $G = E_{8(8)}$, the generalised metric is a $248 \times 248$ matrix, split into $7 \times 7$ block matrices. Nonetheless, there is a concrete description for obtaining the parametrisation of all generalised metrics, provided one restricts to a particular choice of Borel gauge. This has been described in various places, including \cite{Berman:2011jh,Lee:2016qwn} for $n=4, \ldots, 7$  EFT and \cite{Berman:2015rcc,Malek:2012pw,Godazgar:2013rja} for the remaining $n$.\par
GR again fits into the ExFT description nicely; interpreting it as the dynamics of the coset $\operatorname{GL}(D)/ \operatorname{SO}(D)$, the `generalised metric' of GR consists of only the metric whose $D(D+1)/2$ components equal the dimension of the coset.
\section{The Generalised Lie Derivative}\label{sec:GenLieYTensor}
Thus far, we have only specified only a global symmetry $G$ acting on the internal coordinates. We shall demand two additional local symmetries of our theory. The first is the residual external diffeomorphisms that act on the external space, spanned by $x^\mu$, and the latter are \emph{generalised internal diffeomorphisms} that augment the diffeomorphisms on the internal space by $p$-form gauge transformations. Thus, they encode the local symmetries of the supergravity fields contained in $\mathcal{M}_{MN}$. Of course, the external space is entirely absent from DFT and GR and so they require only the generalised diffeomorphisms on the internal space.\par
In GR, the only local symmetries are the diffeomorphisms of the metric $g$ (the only field in GR) which is generated infinitesimally by the Lie derivative. However its action does not generalise readily to our extended spacetimes which contain multiple representations of $\operatorname{GL}(n)$ (such as the antisymmetric representation we saw above for $\operatorname{SL}(5)$ EFT) and so we seek some prescription that allows an appropriate generalisation. The resolution turns out to be a reinterpretation of the action of the usual Lie derivative on a vector as the sum of a transport term plus an adjoint-valued matrix:
\begin{align}
\mathcal{L}_V W^M = V^N\partial_N W^M - {(\partial \times_{\text{ad.}} V)}^M{}_N W^N\,.
\end{align}
Here, $\times_{\text{ad.}}$ is a projection onto the adjoint representation of $\operatorname{GL}(D)$ such that the second term is to be understood as an arbitrary adjoint-valued matrix. By analogy, we shall define a \emph{generalised Lie derivative} $\mathbb{L}_V$ of this form that likewise encodes all of the local symmetries of the supergravity fields on the internal space:
\begin{align}\label{eq:GenLieAdjoint}
{\mathbb{L}}_{U} V^M \coloneqq U^N \partial_N V^M - \alpha  {\left({\mathbb{P}}_{\text{adj.}} \right)}^M{}_N{}^P{}_Q \partial_P U^Q V^N + \lambda (V) \partial_N U^N V^M\,.
\end{align}
We have additionally allowed for a weight term $\lambda (V)$ for the (generalised) vector $V^M$. Here, ${\mathbb{P}}_{\text{adj.}}$ is a projector onto the adjoint representation\footnote{In the case of $G = \operatorname{SL}(3) \times \operatorname{SL}(2)$ EFT, the projector onto the adjoint is instead replaced with a sum of the projectors onto the adjoint of $\operatorname{SL}(3)$ and the adjoint of $\operatorname{SL}(2)$.} of the solution-generating group $G$ and $\alpha$ is some constant appropriate to the group $G$ being considered. The extension of the action of the generalised Lie derivative to arbitrary $G$-tensors is obtained in the usual fashion by requiring that a $G$-scalar transforms only with the transport term.\par
Since the supergravity fields are encoded in the generalised metric, we demand that $\mathbb{L}$ generates the local supergravity symmetries in the following sense:
\begin{align}\label{eq:LocalSymmetriesCorrespondence}
\delta_V \mathcal{M}_{MN} \coloneqq \mathbb{L}_V \mathcal{M}_{MN} \qquad \Leftrightarrow \qquad \text{Local symmetries of the supergravity}\,.
\end{align}
The simplest $\operatorname{GL}(D)$ case has generators and projector onto the adjoint representation given by
\begin{align}
{(t_Q{}^P)}_M{}^N = \delta^M_Q \delta^P_N\,, \qquad {\left( {\mathbb{P}_{\text{adj.}}} \right)}^M{}_N{}^P{}_Q = \delta^M_Q \delta^P_N\,.
\end{align}
We thus recover the usual Lie derivative (on a vector with weight $\lambda = 0$) if we take $\alpha = 1$:
\begin{align}
{\mathbb{L}}_U V^M = U^N \partial_N V^M - V^N \partial_N U^M = \mathcal{L}_U V^M\,.
\end{align}
Thus, the generalised Lie derivative for GR is really just the conventional Lie derivative and its action on the `generalised' metric---which is really just the spacetime metric $g$ itself for $G = \operatorname{GL}(D)$---recovers the local symmetries of $g$ by definition.\par
The simplest non-trivial example is DFT with $G = \operatorname{O}(D,D)$. The local symmetries of the supergravity fields are given by diffeomorphisms and the gauge transformations of the Kalb-Ramond field which combine into a local $G_{\text{SUGRA}} = \operatorname{Diff} (M) \ltimes \Omega^2_{\text{cl.}}(M)$ symmetry. Explicitly, we have the infinitesimal transformations
\begin{align}\label{eq:NSNSSymmetries}
\delta g_{mn} = \mathcal{L}_u g\,, \qquad \delta B_{(2)} = \mathcal{L}_u B_{(2)} - \textrm{d} \Lambda_{(1)}\,, \qquad \delta \phi = \mathcal{L}_u \phi\,.
\end{align}
The group $G = \operatorname{O}(D,D)$ possesses an invariant
\begin{align}\label{eq:ODDEta}
\eta_{MN} = \begin{pmatrix}
0 & \delta_m^n\\
\delta^m_n & 0
\end{pmatrix}\,,
\end{align}
satisfying
\begin{align}
h^T \eta h = \eta\, \qquad \forall \; h \in \operatorname{O}(D,D)\,,
\end{align}
that is used to raise and lower $\operatorname{O}(D,D)$ indices. We may write the adjoint projector in terms of $\eta$ as
\begin{align}
{\left( \mathbb{P}_{\text{adj.}} \right)}^M{}_N{}^P{}_Q= \frac{1}{2} \left( \delta^P_N \delta^M_Q - \eta^{MP} \eta_{NQ} \right)\,,
\end{align}
from which it follows that the generalised Lie derivative acts on a generalised vector in DFT as
\begin{align}
\mathbb{L}_U V^M = U^N \partial_N V^M - (V^N \partial_N U^M - V_N \partial^M U^N) + \lambda(V)\partial_N V^N V^M\,.
\end{align}
It is then a matter of algebra to show that the choice $\alpha =2$ results in the components of $\mathbb{L}_V \mathcal{M}_{MN}$ reproducing the local supergravity transformations \eqref{eq:NSNSSymmetries}, via \eqref{eq:LocalSymmetriesCorrespondence}, once we have combined the diffeomorphism parameter and 1-form parameter into a single $\operatorname{O}(D,D)$-valued vector on the extended space:
\begin{align}
U^M = \begin{pmatrix}
u^m\\ \Lambda_m
\end{pmatrix}\,.
\end{align}
The transformation of the supergravity dilaton is also recovered from the transformation of the doubled dilaton in an analogous fashion.\par
The story is general and holds for the exceptional groups as well, with the only difference being that the local symmetries of the fields now include the relevant gauge transformations of the M-theory potentials rather than of the Kalb-Ramond 2-form. For $n=4$, we have $\operatorname{G}_{\text{SUGRA}} = \operatorname{Diff}(M) \ltimes \Omega^3_{\text{cl.}}(M)$ and the generalised vector in this case consists of a vector parameter for the conventional Lie derivative and a 2-form parameter for the gauge symmetry of $A_{(3)}$. For $n=6$, we have $G_{\text{SUGRA}} = \operatorname{Diff}(M) \ltimes \left( \Omega^3_{\text{cl}} (M) \times \Omega^6_{\text{cl}} (M) \right)$ such that $V^M$ additionally includes a gauge parameter for the 6-form potential $A_{(6)}$ that couples electrically to the M5-brane.\par
In all cases, it may be verified that any group invariants are preserved by the generalised Lie derivative. For example, we have $\mathbb{L}_V  \eta_{MN} = 0$ in DFT. Thus, $\mathbb{L}_V$ generate a local $G$-action (as might be guessed from the appearance of the projector in its definition). In this sense, it is clear that the gauge transformations of the potentials are instrumental in enhancing the na\"{i}ve residual $\operatorname{GL}(n)$ symmetry of the unextended internal space into a full $\operatorname{O}(D,D)$- or $E_{n(n)}$-symmetry (of course this enhancement does not occur in GR where there are no $p$-form potentials to enhance the local symmetries).
\section{The Section Condition}\label{sec:YTensor}
An equivalent description of the generalised Lie derivative is the one described in \cite{Berman:2012vc} in which the generalised Lie derivative is, instead, considered as a conventional Lie derivative on the extended space but with corrections governed by the so-called \emph{Y-tensor} which is formed from $G$-invariants. Concretely, we take the ansatz
\begin{align}\label{eq:GenLieYTensor}
{\mathbb{L}}_{U} V^M \coloneqq {[U, V]}^M + Y^{MN}{}_{KL} \partial_N U^K V^L + \left(\lambda (V)- \omega \right) \partial_N U^N V^M\,,
\end{align}
where ${[U,V]}^M = U^N \partial_N V^M - V^N \partial_N U^M = \mathcal{L}_U V^M$ and $\omega$ is a \emph{universal weight}, unique to each ExFT, that modifies the weight $\lambda(V)$ to an \emph{effective weight} $\lambda(V) - \omega$. We now move onto the issues of closure. In the forms given here, it is not obvious that the generalised Lie derivative should close onto an appropriate mathematical structure. However, from the work of \cite{Gualtieri:2003dx,Hitchin:2004ut}, it is known that the `geometrisation' of the bosonic symmetries of Type II supergravity leads to a Courant algebroid. In more detail, the \emph{Dorfman derivative} (the equivalent of the generalised Lie derivative in generalised geometry) does not define an antisymmetric bracket like the Lie derivative does in conventional Riemannian geometry\footnote{Indeed, there exists a generalisation of the Lie algebra, called \emph{Leibniz algebra}, for which the bracket is not antisymmetric.}. This in itself is not a problem; we may simply define the \emph{Courant bracket} as the antisymmetric portion of the Dorfman derivative. However, neither the Dorfman derivative (reinterpreted as a bilinear bracket) nor the Courant bracket satisfy the Jacobi identity. In particular, the Courant bracket fails the Jacobi identity by an exact term and so cannot define a Lie algebroid. It instead defines an exact Courant algebroid.\par
This failure translates over to ExFTs as well, though the underlying algebroid structure is no longer a Courant algebroid\footnote{See, for example, \cite{Mori:2019slw,Chatzistavrakidis:2019huz}, building on the earlier work of \cite{Vaisman:2012ke}, for a discussion of the underlying algebroid structure in DFT. The appropriate description for EFT has yet to be explored.}. Following generalised geometry, we define the \emph{C-bracket} (or \emph{E-bracket}, depending on whether we are considering DFT or EFT) to be the antisymmetric portion of the generalised Lie derivative:
\begin{align}
\llbracket U, V \rrbracket \coloneqq \frac{1}{2} \left( \mathbb{L}_U V - \mathbb{L}_V U \right)\,.
\end{align}
The generalised Lie derivative thus differs from this bracket by a term symmetric in its arguments that we denote by $(\!( \cdot, \cdot )\!)$:
\begin{align}
\mathbb{L}_U V^M  = {\llbracket U, V \rrbracket}^M + {(\!( U,V)\!)}^M\,.
\end{align}
The failure of the Jacobi identity is then quantified by the trilinear \emph{Jacobiator}
\begin{align}\label{eq:Jacobiator}
{\operatorname{Jac}(U,V,W)}^M \coloneqq \llbracket \llbracket U,V \rrbracket, W \rrbracket + \text{cycles.}
\end{align}
With this in mind, we may compute the commutator of two generalised Lie derivatives. Assuming no symmetry properties of the $Y$-tensor, one finds
\begin{align}
\begin{array}[b]{l}
\left( \mathbb{L}_U \mathbb{L}_V - \mathbb{L}_V \mathbb{L}_U \right) W^M =  {\mathbb{L}_{\llbracket U, V \rrbracket}}^M\\
- Y^{KL}{}_{PQ} \mathrlap{\biggl[  \left(\partial_K U^M \partial_L V^P - \partial_K V^M \partial_L U^P \right) W^Q + \frac{1}{2} \left( \partial_L U^P V^Q - U^Q \partial_L V^P \right) \partial_K W^M \biggr]}\\
+ \frac{1}{2} \left[ Y^{ML}{}_{PQ} \delta^K_R - Y^{MK}{}_{TR} Y^{TL}{}_{PQ} \right] \left( \partial_{K} \partial_L U^P V^Q W^R - U^Q \partial_K \partial_L V^P\right) W^R \\
+ \partial_K U^P \partial_L V^Q W^R \biggl[
Y^{MK}{}_{PT} Y^{TL}{}_{QR} - Y^{ML}{}_{QT} Y^{TK}{}_{PR}\\
\qquad \qquad + \frac{1}{2} \left( Y^{MK}{}_{TR} Y^{TL}{}_{QP} - Y^{ML}{}_{TR} Y^{TK}{}_{PQ} \right) \\
\qquad \qquad - \left( \frac{1}{2} Y^{MK}{}_{PQ} \delta^L_R  - Y^{ML}{}_{QP} \delta^K_R \right)  - Y^{MK}{}_{QR} \delta^L_P + Y^{ML}{}_{PR} \delta^K_Q \biggr]\,.\\
\end{array}
\end{align}
The first line is a closure onto an algebroid whilst the remaining lines are obstructions to this closure. We thus demand certain constraints on the $Y$-tensor that will allow us to remove these terms and determine its form explicitly. In particular, the terms in the third line and below can be made to vanish by constraints on the form of the $Y$-tensor since they include terms that are quadratic in $Y$ (and $\delta$). However, the second line cannot since it is linear in the $Y$-tensor; it must instead be imposed by hand on the theory. We shall comment on this in a moment.\par
Noting that the final three lines can be split into two groups which can be respectively rewritten in terms of (anti-)symmetrisation of indices as 
\begingroup
\renewcommand{\arraystretch}{1.5}
\begin{align}
\begin{array}[b]{l}
Y^{MK}{}_{PT} Y^{TL}{}_{QR} + \frac{1}{2} Y^{MK}{}_{TR} Y^{TL}{}_{QP} - Y^{ML}{}_{QT} Y^{TK}{}_{PR} - \frac{1}{2} Y^{ML}{}_{TR} Y^{TK}{}_{PQ} = \\
\qquad 2 Y^{M(K|}{}_{[P|T} Y^{T|L)}{}_{|Q]R} + Y^{M(K|}{}_{TR} Y^{T|L)}{}_{[QP]} + 2 Y^{M[K|}{}_{(P|T} Y^{T|L]}{}_{|Q)R}\\
\qquad + Y^{M[K|}{}_{TR} Y^{T|L]}{}_{(QP)}\,,
\end{array}
\end{align}
\endgroup
and
\begingroup
\renewcommand{\arraystretch}{1.5}
\begin{align}
\begin{array}[b]{l}
- \frac{1}{2}Y^{MK}{}_{PQ} \delta^L_R + \frac{1}{2} Y^{ML}{}_{QP} \delta^K_R - Y^{MK}{}_{QR} \delta^L_P + Y^{ML}{}_{PR} \delta^K_Q = \\
\qquad - 2 Y^{M(K}{}_{[Q|R} \delta^{L)}_{|P]} - Y^{M(K}{}_{[PQ]} \delta^{L)}_R - 2 Y^{M[K}{}_{(Q|R} \delta^{L]}_{|P)} - Y^{M[K}{}_{(PQ)} \delta^{L]}_R \,,
\end{array}
\end{align}
\endgroup
we thus demand that the $Y$-tensor is subject to the following set of constraints to remove the obstructions to closure:
\begingroup
\renewcommand{\arraystretch}{1.5}
\begin{align}
Y^{KL}{}_{PQ} \partial_K \otimes \partial_L & = 0\,,\label{eq:SectionCondition}\\
\left[ Y^{MK}{}_{PQ} \delta^{L}_R - Y^{ML}{}_{TR} Y^{TK}{}_{PQ} \right] \partial_{(K} \otimes \partial_{L)} & = 0\,,\label{eq:YTensorConstraint}\\
\begin{array}[b]{l}
\biggl[ 2 Y^{MK}{}_{[P|T} Y^{TL}{}_{|Q]R} + Y^{MK}{}_{TR} Y^{TL}{}_{[QP]}\\
\qquad \qquad \qquad - 2 Y^{MK}{}_{[Q|R} \delta^{L}_{|P]} - Y^{MK}{}_{[PQ]} \delta^{L}_R \biggr]
\end{array}
\partial_{(K} \otimes \partial_{L)} & = 0\,,\\
\begin{array}[b]{l}
\biggl[ 2 Y^{MK}{}_{(P|T} Y^{TL}{}_{|Q)R} + Y^{MK}{}_{TR} Y^{TL}{}_{(QP)}\\
\qquad \qquad \qquad - 2 Y^{MK}{}_{(Q|R} \delta^{L}_{|P)} - Y^{MK}{}_{(PQ)} \delta^{L}_R \biggr]
\end{array}
\partial_{[K} \otimes \partial_{L]} & = 0\,.
\end{align}
\endgroup
The notation that we use is standard in the literature. The symbol $\otimes$ is understood as meaning that the derivative acts either on two distinct objects or the same object. For example, \eqref{eq:SectionCondition} is undestood to mean
\begin{align}
Y^{MN}{}_{KL} \partial_M \bullet \partial_N \bullet = 0 \,,\qquad Y^{MN}{}_{KL} \partial_M \partial_N \bullet = 0\,.
\end{align}
The first is referred to as the \emph{strong constraint} whilst the latter is referred to as the \emph{weak constraint}. It is clear that closure of the gauge algebra requires both to be applied but, in the case of DFT, only the weak constraint has an interpretation in string theory\cite{Berman:2013eva}. It can be understood as a rewriting of the level-matching condition (though not the modified form \eqref{eq:LevelMatching}) in doubled space, given by $P^M P_M = \alpha^\prime (N - \tilde{N})$. For the massless sector $N = \tilde{N} = 1$, this reduces to doubled light-cone condition $P^2 = 0$ which is equivalent to the weak constraint. In particular, due to its importance, \eqref{eq:SectionCondition} is called the \emph{section condition}. Solutions to this constraint deserve more discussion and we shall expand on it in more detail in Section~\ref{sec:SolvingSection}. We can equivalently interpret the section condition as a particular subrepresentation $R_2$, which we call the \emph{section representation}, being projected out from the product of two coordinate representations:
\begin{align}
{(\partial \otimes \partial)}|_{R_2} = 0\,.
\end{align}
For $2 \leq n \leq 6$, it is sufficient to take the $Y$-tensor to be proportional to the projector onto the representation $R_2$ (which we have listed for each ExFT in Table~\ref{tab:Summary}) to satisfy the remaining constraints. At $n=7$, this ansatz fails but may be rectified by the additional of an additional piece formed from the its symplectic invariant. At $n=8$, the construction breaks down and the $Y$-tensor alone is not sufficient for closure of the generalised Lie derivative. We shall defer a discussion of the closure of the $E_{8(8)}$ generalised Lie derivative for Chapter~\ref{ch:E8NonRiemannian}. Nevertheless, the above construction is sufficient to determine the form the $Y$-tensor for all ExFTs \cite{Berman:2015rcc,Berman:2012vc,Cederwall:2015ica}, apart from $E_{8(8)}$ EFT, and these are listed in Table~\ref{tab:YTensors}. Note that the $Y$-tensor is symmetric in the upper and lower indices for all ExFTs except $E_{7(7)}$ and $E_{8(8)}$. However, the $Y$-tensor for these two remain symmetric under the \emph{simultaneous} exchange of upper and lower indices.\par
With these $Y$-tensors in hand, all of the constraints apart from \eqref{eq:SectionCondition} are satisfied and we are left with the only obstruction to closure given by the term proportional to the $Y$-tensor, which we rewrite as $\Delta_{\text{sec.}}$:
\begin{align}
{[ \mathbb{L}_U , \mathbb{L}_V]} =  {\mathbb{L}_{\llbracket U, V \rrbracket}} + \Delta_{\text{sec.}}\,.
\end{align}
We have thus landed on a derivation for which, upon imposing
\begin{align}\label{eq:Section}
Y^{KL}{}_{PQ} \partial_K \otimes \partial_L & = 0
\end{align}
as a separate constraint on our theory, the remaining obstructions $\Delta_{\text{sec.}}$ vanish and the generalised diffeomorphisms are guaranteed to close by construction. Due to this constraint, there exists a class of parameters for which the generalised Lie derivative produces terms that vanish under the section condition and thus generate a trivial action. These are called \emph{trivial transformations} and are generated by vectors of the form $Y^{MN}{}_{KL} \partial_N \chi^{KL}$, for arbitrary $\chi^{KL}$. In fact, one may verify that the transformation induced by such trivial parameters is given by
\begingroup
\renewcommand{\arraystretch}{1.5}
\begin{align}\label{eq:TrivialParameters}
\begin{array}[b]{ll}
\mathbb{L}_{Y \partial X} V^M & = Y^{PN}{}_{KL} (\partial_N \chi^{KL} \partial_P V^M + (\lambda - \omega) \partial_N \partial_P \chi^{KL} V^M)\\
	& \quad +  (Y^{MR}{}_{SQ} Y^{SN}{}_{KL} - Y^{MN}{}_{KL} \delta^R_Q ) V^Q \partial_R \partial_N \chi^{KL}\,,\\
\end{array}
\end{align}
\endgroup
where the first line vanishes under the section condition and the second by the constraint \eqref{eq:YTensorConstraint} on the $Y$-tensor. The Jacobiator \eqref{eq:Jacobiator} of the E-bracket is an example of such a trivial parameter.\par
It may be verified that the $Y^{MN}{}_{PQ}$ can be decomposed into a basis where the indices $N$ and $P$ represent $R_1 \otimes {\overbar{R}}_1$ in which the $Y$-tensor takes on the form
\begin{align}\label{eq:YAdjProjRelation}
Y^{MN}{}_{PQ} = \delta^M_P \delta^N_Q - \alpha {\left( \mathbb{P}_{\text{adj.}} \right)}^M{}_Q{}^N{}_P - \omega \delta^M_Q \delta^N_P
\end{align}
for the theory-dependent constants $(\alpha, \omega)$ listed in Table~\ref{tab:Summary}. Inserting this into \eqref{eq:GenLieYTensor}, we recover the other form of the generalised Lie derivative \eqref{eq:GenLieAdjoint} and so we see that the two descriptions are equivalent.
\begin{table}
\centering
\begin{tabulary}{\textwidth}{LL}
\toprule
ExFT & $Y$-tensor\\
\midrule
$\text{GR}$ & $Y^{MN}{}_{PQq} = 0$\\
$\text{DFT}$ & $Y^{MN}{}_{PQ} = \eta^{MN} \eta_{PQ}$\\
$n = 2$ & $Y^{\alpha z}{}_{\gamma z} = Y^{\alpha z}{}_{z \gamma} = Y^{z \alpha}{}_{\gamma z} = Y^{z \alpha}{}_{z \gamma} = \delta^\alpha_\gamma \text{ (all others 0)}$\\
$n = 3$ & $Y^{m\alpha, n \beta}{}_{p \gamma, q \delta} = 4 \delta^{mn}_{pq} \delta^{\alpha \beta}_{\gamma \delta}$\\
$n = 4$ & $Y^{m_1 m_2, n_1 n_2}{}_{p_1 p_2, q_1 q_2} = 3! \delta^{m_1 m_2 n_1 n_2}_{p_1 p_2 q_1 q_2}$\\
$n = 5$ & $Y^{MN}{}_{PQ} = \frac{1}{2} {\left( \Gamma_a \right)}^{MN} {\left( \Gamma^a \right)}_{PQ}$\\
$n = 6$ & $Y^{MN}{}_{PQ} = 10 d^{MNK} d_{PQK}$\\
$n = 7$ & $Y^{MN}{}_{PQ} = - 12 {\left( t^{\alpha} \right)}^{MN} {\left( t_\alpha \right)}_{PQ} - \frac{1}{2} \Omega^{MN} \Omega_{PQ}$\\
$n = 8$ & $Y^{MN}{}_{PQ} = - f^M{}_{QR} f^{RN}{}_P + 2 \delta^{(M}_P \delta^{N)}_Q$\\
\bottomrule
\end{tabulary}
\caption[The Y-tensors of various ExFTs.]{Y-tensors for the various theories. Each symbol is a $G$-invariant and the indices take on the values $1, 2, \ldots, \operatorname{dim} R_1$. We have already covered the $\operatorname{O}(D,D)$ structure for DFT and the $n=2$ and $n=3$ cases are self-explanatory apart for the indices. For $n=3$, we have $G = \operatorname{SL}(3) \times \operatorname{SL}(2)$ and the coordinate representation is $R_1 = (\mathbf{3,2})$. We thus have $m,n = 1,2,3$ and $\alpha = 1,2$ indexing the fundamental representation of the two factors. The symbol $\delta^{mn}_{pq} =\delta^m_{[p} \delta^n_{q]}$ denotes a generalised Kronecker delta (similarly for $\delta^{\alpha \beta}_{\gamma \delta}$). The coordinate representation for $n=4$ is the antisymmetric representation of $\operatorname{SL}(5)$ and so each pair $m_1 m_2$ is implicitly antisymmetrised (note the different contraction conventions to \cite{Musaev:2015ces} for $n=4$, leading to a different scaling of the Y-tensor). The remaining cases have obvious ranges for the generalised indices $M, N = 1, \ldots, \operatorname{dim} R_1$. At $n=5$, the $Y$-tensor is given in terms of gamma matrices whilst $n=6$ is given in terms of the totally symmetric $d$-symbol. The $n=7$ case is given in terms of the generators $t_\alpha$, valued in the $R_1 = \mathbf{56}$ representation, and the symplectic form $\Omega_{MN}$. Finally, the $n=8$ case is given in terms of its structure constants (note that $R_1 = \mathbf{248}$ is the adjoint representation of $E_{8(8)}$). The case $n=8$ is anomalous and comes with a caveat that shall be discussed separately in Chapter~\ref{ch:E8NonRiemannian}.}
\label{tab:YTensors}
\end{table}
\section{The Gauge structure of EFTs}\label{sec:GaugeStructureExFT}
Before we write down the actions for ExFTs, we return briefly to the field content of EFTs. Thus far, we have ignored the $d$ dimensional external space that EFTs possess that needs to be consistently adjoined to the extended internal structure. If we view the splitting $\mathcal{M}^{11} \rightarrow \mathcal{M}^{d} \times \mathcal{M}^{n}$ that we started off with as a Kaluza-Klein type decomposition, it should hopefully be clear that the first of the generalised gauge fields $\mathcal{A}_{\mu}{}^M$ of EFT is best understood as a generalised Kaluza-Klein vector.\par
What is less obvious is that this necessitates the introduction of higher-form potentials, in the spirit of the tensor hierarchy of gauged supergravities. The details  are dependent on the particular EFT that we choose and so we elect to focus on the gauge structure of $E_{7(7)}$ EFT as an example. The explicit computations are outlined in \cite{Hohm:2013uia} and we shall only summarise the main features of the gauge structure of EFTs. Our starting point is to define a generalised Lie-covariant derivative
\begin{align}
\mathcal{D}_\mu \coloneqq \partial_\mu - \mathbb{L}_{\mathcal{A}_\mu}\,.
\end{align}
The transformation of the gauge field $\mathcal{A}_\mu{}^M$ can be determined by demanding that this covariant derivative transforms covariantly such that $\delta_\Lambda (\mathcal{D}_\mu V) = \mathbb{L}_{\Lambda} (\mathcal{D}_\mu V)$. It is simple to show that this is indeed the case if $\mathcal{A}$ transforms according to
\begin{align}
\delta \mathcal{A}_\mu{}^M = \mathcal{D}_\mu \Lambda^M\,,
\end{align}
which resembles the gauge transformation of a Yang-Mills gauge field. The obvious candidate for a field strength of $\mathcal{A}_\mu{}^M$ is then the non-Abelian Yang-Mills field strength (where the commutator of fields has been replaced by the $E$-bracket):
\begin{align}
F_{\mu \nu}{}^M \coloneqq 2 \partial_{[\mu} \mathcal{A}_{\nu]}{}^M - {\llbracket \mathcal{A}_\mu, \mathcal{A}_\nu \rrbracket}^M\,.
\end{align}
This is also the same fieldstrength that appears in the commutator of two derivatives:
\begin{align}\label{eq:Commutator}
[\mathcal{D}_\mu, \mathcal{D}_\nu] = \mathbb{L}_{-F_{\mu \nu}}\,.
\end{align}
However, computing the general variation of this object, it can be shown that this object is not covariant under the gauge transformations of $\mathcal{A}$. Following the ideas from gauged supergravities, one may wish to try to amend this by the addition of a 2-form potential $\mathcal{B}_{\mu \nu \alpha}$ via a St\"{u}kelberg-type coupling to obtain
\begin{align}
{\mathcal{F}_{\mu \nu}^\circ}^M \coloneqq F_{\mu \nu}{}^M - 12 {(t^\alpha)}^{MN} \partial_N \mathcal{B}_{\mu \nu \alpha}\,.
\end{align}
where $t^\alpha$ are the generators of $E_{7(7)}$, valued in the coordinate representation $R_1 = \mathbf{56}$ of $E_{7(7)}$. As pointed out in \cite{Abzalov:2015ega}, we can view this as the freedom to add any trivial parameters to $F_{\mu \nu}{}^M$ without affecting the gauge structure, in light of its appearance in \eqref{eq:Commutator}. Whilst an analogous modification is sufficient for $n \leq 6$, it fails for $n=7,8$ because of the Hodge duality of forms in low dimensions. Since an external 1-form is dual to a $d-3$ form, in $n = 6$ (equivalently, $d=5$) this gives corrections to the 3-form field strength but this does not enter into the action. However at $d=4$, vectors are dual to vectors and so this introduces corrections to the 2-form fieldstrength that appears in the action. The situation is exacerbated at $d=3$ where vector-scalar duality introduces corrections to the 1-form scalar current and thus appears amongst the fields (as well as the action) as an extra vector field. We emphasise that the hierarchical structure here can be truncated at an appropriate order since the effect of the compensating fields on the fieldstrength of one degree higher will eventually become undetectable at the level of the action. The covariantisation of subsequent fieldstrength can then be neglected, having added only a finite number of compensating fields. The recent paper \cite{Fernandez-Melgarejo:2019pvx} gives a discussion of the parametrisation of these $p$-form gauge fields in more detail.\par
In the case of interest, we are thus required to consider an additional 2-form correction to give the fully covariantised fieldstrength
\begin{align}\label{eq:E7CovariantisedFieldstrength}
\mathcal{F}_{\mu \nu}{}^M \coloneqq F_{\mu \nu}{}^M - 12 {(t^\alpha)}^{MN} \partial_N \mathcal{B}_{\mu \nu \alpha} - \frac{1}{2} \Omega^{MN} \mathcal{B}_{\mu \nu N}\,,
\end{align}
where $\mathcal{B}_{\mu \nu N}$ is \emph{covariantly constrained} on the first index. By this we mean that it is treated like a partial derivative with respect to the section condition such that
\begin{align}
Y^{MN}{}_{PQ} C_M \otimes C_N = 0\,, \qquad C_M \in \{ \partial_M, \mathcal{B}_{\mu \nu M}, \ldots\}\,.
\end{align}
The fieldstrength \eqref{eq:E7CovariantisedFieldstrength} can then be shown to transform covariantly as outlined in \cite{Hohm:2013uia}. The fieldstrengths of the compensating fields then appear in the Bianchi identity for $\mathcal{F}$, given by
\begin{align}
3 \mathcal{D}_{[\mu} \mathcal{F}_{\nu \rho]}{}^M = - 12 {(t^\alpha)}^{MN} \partial_N \mathcal{H}_{\mu \nu \rho \alpha}- \frac{1}{2} \Omega^{MN} \mathcal{H}_{\mu \nu \rho N}\,.
\end{align}
Since the 3-form fieldstrengths $\mathcal{H}_{\mu \nu \rho \alpha}$ and $\mathcal{H}_{\mu \nu \rho N}$ do not enter into the action, they do not need not be covariantised by the addition of further compensating 3-form gauge fields.\par
We close this section by mentioning the work \cite{Hohm:2017wtr} which blurs the line between DFT and EFT by constructing an enhanced DFT for $d=3$ in which vector-scalar duality enhances the symmetry group to $\operatorname{O}(d+1, d+1)$. The construction there more closely aligns with that of EFT, including the gauge structure that we have just described.
\section{Actions for ExFTs}\label{sec:SolvingSection}
We are now in a position to write down the actions for ExFTs. In the case of DFT, we write down all two-derivative gauge invariant terms and fix coefficients such that it reduces correctly to the Type II supergravity action upon applying the section condition. We give only the result of \cite{Hohm:2010pp} which found that the DFT action was given by
\begin{align}
\mathcal{S}_{\text{DFT}} = \int \textrm{d}^D X \textrm{d}^D \tilde{X} e^{-2d} \mathcal{R}\,,
\end{align}
with
\begingroup
\renewcommand{\arraystretch}{1.5}
\begin{align}
\begin{array}[b]{rl}
\mathcal{R} = & \frac{1}{8} \mathcal{M}^{MN} \partial_M \mathcal{M}^{KL} \partial_N \mathcal{M}_{KL}  - \frac{1}{2} \mathcal{M}^{MN} \partial_M \mathcal{M}^{KL} \partial_K \mathcal{M}_{NL}\\
	& + 4 \mathcal{M}^{MN} \partial_M \partial_N d - \partial_M \partial_N \mathcal{M}^{MN} - 4 \mathcal{M}^{MN} \partial_M d \partial_N d + 4 \partial_M \mathcal{M}^{MN} \partial_N d\,.
\end{array}
\end{align}
\endgroup
In the case of DFT, there is only one inequivalent solution to the section condition. Expanding out the $Y$-tensor in terms of the $\operatorname{O}(D,D)$ structure \eqref{eq:ODDEta} gives the form of the DFT section condition more commonly encountered:
\begin{align}\label{eq:DFTSection}
\partial^m \otimes \partial_m = 0\,.
\end{align}
The canonical solution to this constraint is to drop all dependencies on the winding coordinates, $\partial^m = 0$, for which the DFT action reduces (after integrating by parts) to the action of the NS-NS sector of Type II supergravity:
\begin{align}
S = \int \textrm{d}^D x \sqrt{-g} e^{-2\phi} \left[ R + 4 \partial_\mu \phi \partial^\mu \phi - \frac{1}{12} H_{(3)}^2 \right]\,.
\end{align}
Indeed, as mentioned above, the coefficients in the DFT action are chosen such that we obtain this reduction.\par
In the case of ExFTs, the structure is more complicated due to the presence of the external space and generalised gauge fields. In particular, all terms must now be covariant with respect to both internal generalised diffeomorphisms and external (conventional) diffeomorphisms and must reduce to the 11-dimensional supergravity action upon solving the section condition. The details again differ by theory, but every EFT shares a common sector consisting of a covariantised Einstein-Hilbert action, a kinetic term for the scalar sector and a potential term\footnote{The $E_{8(8)}$ potential has an additional term
\begin{align}
-\frac{1}{7200} f^{NQ}{}_P f^{MS}{}_R \mathcal{M}^{PK} \partial_M \mathcal{M}_{QK} \mathcal{M}^{RL} \partial_N \mathcal{M}_{SL}
\end{align}
that is required for gauge invariance of the potential.
}:
\begin{align}
\mathcal{S}_{\text{EFT}} & = \int \textrm{d}^d x\textrm{d}^{\operatorname{dim}} \rho_1 Y \left( \mathcal{L}_{\text{EH}} + \mathcal{L}_{\text{sc.}} + eV + \ldots \right)\\
\mathcal{L}_{\text{EH}} & = e R\\
\mathcal{L}_{\text{sc.}} & = \frac{e}{4\alpha} g^{\mu \nu} \mathcal{D}_\mu \mathcal{M}^{MN} \mathcal{D}_\nu \mathcal{M}_{MN}\\
V & =
\begingroup
\renewcommand{\arraystretch}{1.5}
\begin{array}[t]{l}
\frac{1}{4\alpha} \mathcal{M}^{MN}\partial_M \mathcal{M}^{KL} \partial_N \mathcal{M}_{KL} - \frac{1}{2} \mathcal{M}^{MN} \partial_N\mathcal{M}^{KL} \partial_L \mathcal{M}_{MK}\\
+ \partial_M \ln e \partial_N \mathcal{M}^{MN}+ \mathcal{M}^{MN} \partial_M \ln e\partial_N \ln e + \frac{1}{4} \mathcal{M}^{MN} \partial_M g^{\mu \nu} \partial_N g_{\mu \nu} \,,
\end{array}\label{eq:EFTPotential}
\endgroup
\end{align}
where $e = \sqrt{- \operatorname{det}g}$ is the vielbein of the external space. The ellipsis hides terms that are unique to each ExFT. For example, the $E_{7(7)}$ EFT action contains an additional Yang-Mills piece plus a topological term whilst $E_{8(8)}$ contains a Chern-Simons piece. We shall be concerned with these two theories in particular in Chapter~\ref{ch:EFTReductions} and so shall defer their discussion until then. We note that, in even $d$, the action above is incomplete; it is really a pseudo-action which needs an additional duality constraint on the generalised fieldstrength $\mathcal{F}$. In $E_{7(7)}$ EFT one must manually impose the twisted self-duality constraint
\begin{align}
\mathcal{F}_{\mu \nu}{}^M = - \frac{e}{2} \varepsilon_{\mu \nu \rho \sigma} \Omega^{MN} \mathcal{M}_{NK} \mathcal{F}^{\rho \sigma K}
\end{align}
on the equations of motion to choose a symplectic frame that selects out 28 physical electric vectors from the 56 encoded in $\mathcal{A}_\mu{}^M$.\par
Note that the form of the potential here is a reworking of the form discussed in \cite{Berman:2011jh}. Upon splitting off the trace component of the generalised metric, one may re-substitute its equation of motion back into the action to obtain the form \eqref{eq:EFTPotential}. The advantage of that form is that it more closely resembles the Einstein-Hilbert action, expanded in terms of derivatives of the metric, which allows for an easier comparison with GR.\par
In the case of EFTs, there are only two inequivalent solutions to the section constraint and these generically pick out a $\operatorname{GL}(n)$ or $\operatorname{GL}(n-1) \times \operatorname{SL}(2)$ subgroup of $G$ (necessarily breaking $G$-covariance), corresponding to the \emph{M-theory section} and \emph{Type IIB section} respectively. Upon dropping coordinate dependencies on all but the usual coordinates in the extended space, one recovers the full dynamics of 11-dimensional (resp.\ Type II) supergravity, rearranged into an $n+d$ (resp.\ $n+d-1$) Kaluza-Klein split of the fields. Note that we land on a \emph{rewriting} of the supergravities under a KK split, rather than some truncation, since all of the fields still depend on 11 (resp.\ 10) spacetime coordinates. The Type IIA theory can further be obtained from a circle reduction of the M-theory section and so is not an independent solution of the section condition.\par
We are now ready to discuss the appearance of duality transformations in ExFTs. They arise naturally from the continuous solution-generating group $G$ in the presence of isometries as an ambiguity in the way we may pick out the section from the extended spacetime. More concretely, the fields depend on fewer coordinates than is specified by the section condition and so the physical spacetime may be completed with different choices of coordinates without violating the section condition. It is these choices that are related by duality transformations.\par
For example, the DFT wave described in \cite{Berman:2014jsa} only depends on $D-2$ transverse coordinates which is two less than the number of coordinate dependencies allowed by the DFT section condition \eqref{eq:DFTSection}. It is natural to include the time direction in the section as well but then there are two choices for the final coordinates, which they call $z \in Y^m$ (which lies in the usual coordinates) and $\tilde{z} \in \tilde{Y}_m$ (which lies in the dual coordinates). Picking $z$ as the completion of the section yields the fundamental string whereas picking $\tilde{z}$ yields the wave solution. Both solutions are equally valid solutions to the section constraint and are related by a T-duality transformation $z \xleftrightarrow{T} \tilde{z}$.\par
The power of ExFTs is that this gives a prescription for generating duality-related solutions. Provided that the internal space has a sufficient number of isometries, we can generate whole classes of solutions that descend from a single solution in ExFT and that differ only by the way the section is identified within the extended space. Conversely, it is possible to write down solutions in ExFT that can unify many supergravity solutions into a single solution on the extended space. Just as, for example, the M5-brane in M-theory gives a higher-dimensional origin to both the NS5-brane and D4-branes, solutions in ExFTs give higher-dimensional origins to multiple branes of string- and M-theory, each of which are related by duality transformation. This has been leveraged, for example in \cite{Berkeley:2014nza,Blair:2016xnn,Berman:2014jsa,Bakhmatov:2016kfn,Kimura:2018hph,Blair:2014zba,Berman:2014hna,Bakhmatov:2017les,Berman:2019biz}.
\section{The Projector on the Equations and Motion}\label{sec:ProjectedEOM}
With the action for ExFTs in hand, the next natural step is to obtain their equations of motion. For the external metric and generalised gauge fields, one may proceed as normal and vary the action in the usual fashion. However, there is one subtlety regarding the equations of motion for the generalised metric. Since it is constrained to parametrise the coset $G/H$ in terms of the supergravity fields, the na\"{i}ve equations obtained from varying the action alone are not sufficient. One must impose a projector onto the equations of motion by hand in order to enforce this condition. In particular, if we vary the action with respect to the (inverse) generalised metric, we may denote this as
\begin{align}
\delta \mathcal{S} = \int \textrm{d}^d x \textrm{d}^{\operatorname{dim} \rho_1} Y \mathcal{K}_{MN} \delta \mathcal{M}^{MN}\,, \qquad \mathcal{K}_{MN} \coloneqq \frac{\delta \mathcal{L}}{\delta \mathcal{M}^{MN}}\,.
\end{align}
Here, the variation $K_{MN}$ is with respect to an \emph{arbitrary} matrix $\mathcal{M}^{MN}$. In order to ensure that the variation is consistent with the coset structure that $\mathcal{M}_{MN}$ is required to parametrise, we instead require the projected set of equations
\begin{align}\label{eq:ProjectedEOM}
\mathcal{P}_{MN}{}^{KL} \mathcal{K}_{KL} = 0
\end{align}
for a coset projector $\mathcal{P}_{MN}{}^{KL}$. This was explicitly demonstrated for the DFT and $\operatorname{SL}(5)$ EFT case in \cite{Berkeley:2014nza,Rudolph:2016sxe} where the parametrisation of the generalised metric in terms of the supergravity fields was explicitly inserted in and then re-covariantised with respect to $G$. This was found to be given by
\begin{align}
\mathcal{P}_{MN}{}^{KL} = \frac{1}{\alpha} \left( \delta^{(K}_M \delta^{L)}_N - \omega \mathcal{M}_{MN} \mathcal{M}^{KL} - \mathcal{M}_{MP} Y^{P(K}{}_{NQ} \mathcal{M}^{L)Q} \right)\,.
\end{align}
In fact, we shall show in Chapter~\ref{ch:E8NonRiemannian} that this projector also appears in the transformation of the generalised metric and holds more generally for at least $3 \leq n \leq 8$ (it is also expected to hold for $n=2$ as well, though we shall not pursue this).

\part{Exotic Branes in ExFTs}
	\chapter{The Non-Geometric Solution in \texorpdfstring{$E_{7(7)}$}{E7(7)} EFT}\label{ch:NonGeometricE7}
One of the remarkable aspects of string theory is the presence of non-perturbative branes whose tensions scale as $g_s^{-1}$ (the D-branes) or as $g_s^{-2}$ (the Neveu-Schwarz five branes). The study of these branes in string theory over the last 20 years has revealed much about the connection between quantum field theories and gravity and have been a huge part of the construction of M-theory where there are no perturbative brane states.\par
Following the work of \cite{Obers:1998fb,deBoer:2010ud,deBoer:2012ma}, and others, it was realised that string theory also contains so-called `exotic brane' states whose tensions scale as $g_s^\alpha$ with $\alpha<-2$. These objects typically have low codimension\footnote{By low codimension, we mean branes of codimension-2 (`defect branes'), codimension-1 (`domain wall') and codimension-0 (`space-filling branes')} and so potentially suffer from various pathologies that come with low-codimension objects. Nevertheless there is now a substantial corpus of work in the area including \cite{Bergshoeff:2011se,Bergshoeff:2012ex,Bergshoeff:2012pm} where such branes have been shown to play an important role in duality symmetries. Of course it is also interesting to speculate what such branes would correspond to in a dual holographic theory with masses scaling as $\frac{N^\alpha}{\lambda^\alpha}$ after taking a 't Hooft limit. These states in the dual field theory with higher $N$ dependences should then be related to multiple traces as in the giant graviton story \cite{Corley:2001zk}\footnote{We thank Sanjaye Ramgoolam for discussions on this.}.\par
Apart from being exotic due to their novel scaling, these branes were also curious objects that appeared to lack a well-defined global description as supergravity solution; a key part of their construction is to use elements of the duality group to patch together local solutions such that (globally) these branes end up with monodromies valued in the duality groups. For the case of U-duality, these produce examples of Hull's U-folds\cite{Hull:2006va} which obviously contain S-folds and T-folds amongst their reductions. The obvious question to ask is then `if they are not solutions of supergravity, then what are they solutions of?'\par
For a while Exceptional Field Theory (EFT) appeared to look like a rather nice answer looking for a question. Here, we argue that EFT is the natural setting for studying exotic branes. Below we shall review the relevant aspects of EFT in more detail to set the scene and establish conventions but first let us state some of the ideas behind EFT relevant to us now. One of the key ideas from M-theory is that branes in Type IIA and Type IIB are descendants of a smaller set of branes in the higher-dimensional theory of eleven dimensional supergravity. In addition to branes descending to branes, there are particular instances of branes originating from purely geometric solutions in eleven dimensional supergravity---the D0-brane is a null wave solution in eleven dimensions and the D6-brane is an eleven-dimensional Kaluza-Klein monopole of the sort described by Gross, Sorkin and Perry \cite{Sorkin:1983ns,Gross:1983hb}. Ideally we would like a theory with no central charges and no additional external sources. EFT has a chance of being this theory. As was shown in \cite{Berman:2014hna} and based on work in DFT \cite{Berkeley:2014nza,Berman:2014jsa} the membrane, five-brane and their bound states all come from a single EFT solution, namely the EFT version of the superposition of a wave and monopole. Thus the EFT superalgebra does not contain central charges for these states. Just as the D0 is part of the wave solution in M-theory, and thus its IIA central charge has its origin as an eleven dimensional momentum, so are all the usual M-theory branes in EFT. The next question then is to investigate the role of exotic branes in EFT. This has begun with the works \cite{Sakatani:2014hba,Bergshoeff:2015cba,Bakhmatov:2016kfn,Bergshoeff:2016ncb, Lombardo:2016swq,Bakhmatov:2017les} and others. Ultimately one might wish that all the branes in string and M-theory, including the exotic ones, descend from a single object in EFT. The hope for this is that any object in the same duality orbit must come from a single solution in EFT. So why hasn't this been already achieved?\par
A key problem for EFT is that one picks a particular exceptional group $E_{n(n)}$ and splits spacetime between internal and external spaces. This split respects the $E_{n(n)}$ symmetry (by construction) but does not respect the higher $E_{d+n(d+n)}$ symmetry. Thus, there are objects that are connected through higher $E_{d+n(d+n)}$ symmetries that cannot be related to each other by $E_{n(n)}$ transformations, leading to them being viewed as separate objects in the $E_{n(n)}$-symmetric theory. It is then clear that the above idea of a single unifying object may only be realised within the full $E_{11}$ theory \cite{West:2001as,West:2003fc,West:2004kb,West:2004iz,Cook:2008bi,Cook:2009ri,Tumanov:2015yjd}.\par
In this part, we concentrate on the brane solutions of the $E_{7(7)}$ EFT and construct a single solution that gives rise to the codimension-2 exotic branes in Type IIA, IIB and M-theory. We then look further at what sort of exotic branes may exist beyond those contained in this solution.\par
We begin, in Section~\ref{sec:ExoticBranes}, with a description of these exotic branes in slightly more detail along with an explanation of the notation used to denote these branes. In Section~\ref{sec:NonGeomSoln}, we demonstrate why ExFTs provide an ideal playground in which to probe these exotic branes by explicitly constructing a single solution in $E_{7(7)}$ EFT which unifies many of the exotic branes described to date. This section is perhaps best thought of as complementary to the work described in \cite{Berman:2014hna}, in a manner that we shall describe later (Figure~\ref{fig:Brane}).
\section{Overview of Exotic Branes}\label{sec:ExoticBranes}
It has long been known that low codimension objects possess non-standard features, regardless of the $g_s$ scaling of their tension\footnote{Unless explicitly stated, we shall work in the string frame throughout in which D-branes possess a tension scaling with $g_s^{-1}$.}; the D7-brane in Type IIB (codimension-2 in $D=10$) already modifies the spacetime asymptotics, the D8-brane in Type IIA (codimension-1 in $D=10$) terminates spacetime at a finite distance due to a fast running of the dilaton and the D9-brane is space-filling. However, it so happens that the vast majority of exotic branes are also low-codimension objects. For example, the NS7-brane (the S-dual of the D7-brane and later reclassified as a $7_3$) is a codimension-2 object that is also an exotic brane since its tension scales as $g_s^{-3}$, making it more non-perturbative than the other conventional branes. It has since become customary to organise these exotic states in terms of the $g_s$-scaling of their tensions which we now discuss.\par
The embedding of the T-duality groups within the U-duality groups,
\begin{align}
E_{n(n)} \supset \operatorname{O}(n-1,n-1) \times \mathbb{R}^+\,,
\end{align}
induces a grading of the tension of the branes which may be characterised by a single number $\alpha \leq 0$. The highest values of $\alpha$ correspond to the well-known branes, which are interesting in their own right, but it is the lower powers of $g_s$ that are of interest to us:
\begin{itemize}
	\item \parbox{2cm}{$\alpha=0$:} \parbox{3cm}{Fundamental} $\text{F}1 \equiv 1_0$, P
	\item \parbox{2cm}{$\alpha=-1$:} \parbox{3cm}{Dirichlet} $\text{D}p \equiv p_1$
	\item \parbox{2cm}{$\alpha=-2$:} \parbox{3cm}{Solitonic} $\text{NS5} \equiv 5_2 \xrightarrow{T} \text{KK5} \equiv 5_2^1 \xrightarrow{T} 5_2^2$
	\item \parbox{2cm}{$\alpha\leq - 3$:} \parbox{3cm}{Exotic} e.g. $p_3^{7-p}, 0_4^{(1,6)}$
\end{itemize}
As mentioned previously, these exotic branes are generically low-codimension objects that are additionally non-geometric in a manner that we characterise as follows. Although all exotic branes \emph{require} duality transformations in order to patch correctly (in addition to the conventional diffeomorphisms and gauge transformations), they may or may not admit local descriptions in terms of the supergravity fields. Those that do, we call \emph{globally non-geometric objects} since it is only at the global level that the usual description breaks down. These include the T-folds and U-folds of Hull \cite{Hull:2006va}. If they do not admit a local supergravity description, for example by possessing an explicit dependence on winding or wrapping coordinates (whose interpretation in string theory we shall discuss later), they are called \emph{locally non-geometric objects}. Since their existence and behaviour is so closely tied to the duality transformations of string- and M-theory, ExFTs are an obvious candidate in which to study these objects as they geometrise these pathologies in a manner that we demonstrate later.\par
Whilst these exotic branes are admittedly rather non-standard (although their first appearance in the literature goes back to, at least, \cite{Obers:1998fb}), some of the better-behaved of these exotic branes were explicitly argued to exist in string theory in \cite{deBoer:2010ud,deBoer:2012ma} via the supertube effect acting on conventional branes. Separate to this, there has been a series of papers \cite{Kimura:2013fda,Kimura:2013zva,Kimura:2018hph} on the description of exotic 5-branes within in the GLSM formalism and on the effective world-volume actions of exotic five-branes \cite{Chatzistavrakidis:2013jqa,Kimura:2014upa,Kimura:2016anf}. It thus follows that a better understanding of these exotic branes is warranted. See also \cite{Plauschinn:2018wbo} for a recent review of the description of exotic branes within string theory and \cite{Kimura:2016xzd} for a discussion of their supersymmetry projection rules.\par
Before we proceed, we emphasise that we still lack a concrete criterion for a brane to be exotic. As the D7, D8 and D9 show, possessing low codimension is not necessarily indicative of the sort of non-geometry that we seek to study which is characterised by low $g_s^{\alpha}$ scaling of the tension. On the other hand, there exists an exotic $5_2^2$-brane (we shall cover the notation denoting these branes shortly) which, whilst being codimension-2, possesses the same $g_s$ scaling as the NS5 and KK5 but manifests the sort of non-geometry that we are interested in by virtue of the fact that the metric is not single-valued at $\theta =0, 2\pi$ as one traverses around the brane. In fact the $5_2^2$ has become the standard example of Hull's T-fold in that traversing around the brane only returns the original configuration up to a T-duality transformation. We thus see that low codimension alone is generally insufficient for non-geometry; one appears to require low $g_s$ scaling as well. Yet, as the $5_2^2$ demonstrates, there exists states which are non-geometric but still scale as $g_s^{-2}$.\par
The utility of ExFTs is that many of the exotic states that one can construct can be better understood, or at least more elegantly unified, when the duality transformations are realised linearly. Indeed we shall demonstrate in Chapter~\ref{ch:Map} that, not only are codimension$<2$ objects common, they form the majority of the exotic states and may even \emph{require} an ExFT description to make sense. In particular, ExFTs allow for the construction of non-trivial space-filling branes by allowing for a dependence of the fields on the extended coordinates. This is only possible because of the distinguishing feature of ExFT in that they capture winding mode dependences. We shall give a more detailed argument for this in Chapter~\ref{ch:Map}.\par
For the $E_{7(7)}$ solution presented here, we shall focus on codimension-2 exotic states but then later move on to discussing all of the possible exotic states that one should be able to construct. We first briefly discuss the notation used in \cite{deBoer:2012ma} for branes, in which they are characterised by the mass-dependence when wrapping an internal torus (equivalent to a characterisation by their tensions). For Type II states, the mass of a $b_n^{(\ldots, d, c)}$-brane depends linearly on $b$ radii, quadratically on $c$ radii, cubically on $d$ radii, and so on. Additionally, the subscript denotes the power dependence on the string coupling\footnote{Note that $n =- \alpha$ such that $n \geq 0$. This is not to be confused with the $n$ of the exceptional groups $E_{n(n)}$ and context should make the meaning obvious.}. Finally, the power of $l_s$ on the denominator is such that the total mass has units of ${(\text{Length})}^{-1}$ as required. For M-theory states, the notation is very similar except for the absence of the string coupling number $n$ and the role of $l_s$ being taken over by the Planck length $l_P$ in eleven dimensions:
\begin{align}
\text{Type II}: && \text{M}(b_n^{(\ldots, d,c)}) & = \frac{\ldots {(R_{k_1} \ldots R_{k_d})}^3 {(R_{j_1} \ldots R_{j_c})}^2 {(R_{i_1} \ldots R_{i_b})}}{g_s^nl_s^{1 + b + 2c + 3d + \ldots}}\,,\\
\text{M-Theory}: && \text{M}(b^{(\ldots, d,c)}) & = \frac{\ldots {(R_{k_1} \ldots R_{k_d})}^3 {(R_{j_1} \ldots R_{j_c})}^2 {(R_{i_1} \ldots R_{i_b})}}{l_p^{1 + b + 2c + 3d + \ldots}}\,.
\end{align}
For states that appear in both Type IIA and IIB, we may additionally append an A/B suffix to the brane if the theory being discussed is relevant e.g. $0_4^{(1,6)}\text{B}$ for the version of the object appearing in the Type IIB theory. These exotic branes couple electrically to the mixed symmetry potentials\footnote{Note that we shall label the type of potential by the power of $g_s$, schematically labelled $E^{(n)}$. Thus, for $n=0,1,2,3,4,\ldots$, we shall denote the potentials $B, C, D, E, F, \ldots$ The offset is chosen such that the fundamental string ($n=0$) couples to the NS-NS 2-form $B_{2}$, the D$p$-branes ($n=1$) couple to $C_{p+1}$ etc. We shall reserve $A_{(p)}$ to denote the $p$-form potentials that appear in M-theory.}
\begin{align}
{(E^{(n)})}_{1 + b+c_2 + \ldots + c_s,  c_2 + \ldots + c_s, \cdots, c_{s-1} + c_s, c_s} \leftrightarrow b_n^{(c_s, \ldots, c_2)}\,,
\end{align}
where the subscripts on the potentials denote the number of indices in that set. The notation is such that the each set is implicitly antisymmetrised over and contains all the sets of indices to the right of it. For example, the $n=4$ exotic $0_4^{(1,6)}$ brane couples to
\begin{align}
F_{8,7,1} \sim F_{x y_1 \ldots y_6 z, y_1 \ldots y_6 z, z}\,.
\end{align}
See also \cite{Lombardo:2016swq,Lombardo:2017yme,Bergshoeff:2017gpw,Kleinschmidt:2011vu} for a group-theoretic discussion on classifying these mixed-symmetry potentials. Note that the we do not consider the Hopf fibre (and more generally, distinguished isometric directions) as a worldvolume direction and so the KK-monopole in eleven dimensions shall be denoted KK6 (or KK6M) and the monopole in ten dimensions as KK5 (or KK5A/B).
\section{The Non-Geometric Solution in \texorpdfstring{$E_{7(7)}$}{E7(7)} EFT}\label{sec:NonGeomSoln}
\subsection{Overview of \texorpdfstring{$E_{7(7)} \times \mathbb{R}^+$}{E7(7)R+} Exceptional Field Theory}\label{sec:E7EFT}
In this section we work with the $E_{7(7)}$ theory. Although we gave an introduction to ExFTs in Chapter~\ref{ch:IntroExFT}, we collect the relevant aspects of $E_{7(7)}$ EFT in particular for convenience. The coordinate representation $R_1$ of $E_{7(7)}$ EFT is the 56-dimensional fundamental representation which we index with $M = 1, \ldots, 56$. For every EFT, there are exactly two inequivalent (i.e. not related by $E_{n(n)}$ transformations) solutions to the section constraint; the M-theory section and the Type IIB section. For the M-theory section, we decompose the coordinates under $\operatorname{GL}(7)$ as
\begin{align}\label{eq:MSect}
\mathbf{56} \rightarrow {\mathbf{7}}_{+3} + {\overbar{\mathbf{21}}}_{+1} + {\mathbf{21}}_{-1} + {\overbar{\mathbf{7}}}_{-3}\,,
\end{align}
where the subscript denotes the weight under the $\operatorname{GL}(1)$ subgroup. Letting $m,n = 1, \ldots, 7$ denote the vector representation of $\operatorname{GL}(7)$, we may decompose the 56 coordinates of the internal space $Y^M$ to
\begin{align}\label{eq:E7MCoords}
Y^M = ( y^m, y_{mn}, y^{mn}, y_m)\,,
\end{align}
where $y_{mn}$ and $y^{mn}$ are labelled by a pair of antisymmetric indices. The coordinates $y_{mn}$ are dual to the wrapping modes the M2-brane, $y^{mn} \sim \epsilon^{mn k_1 \ldots k_5} y_{k_1 \ldots k_5}$ are dual to the wrapping modes of the M5-brane and $y_{m} \sim y_{n_1 \ldots n_7,m}$ are dual to the wrapping modes of the KK6M.\par
The other section is the Type IIB which corresponds to decomposition under the subgroup $\operatorname{GL}(6) \times \operatorname{SL}(2)$:
\begin{align}
\mathbf{56} \rightarrow {(\mathbf{6,1})}_{+2} + {({\overbar{\mathbf{6}}},\mathbf{2})}_{+1} + {(\mathbf{20,1})}_0 + {(\mathbf{6,2})}_{-1} + {({\mathbf{\overbar{6}}},\mathbf{1})}_{-2}\,.
\end{align}
Denoting $\mathrm{m} , \mathrm{n}, \mathrm{k} = 1, \ldots, 6$ and $\alpha = 1,2$ for the $\operatorname{SL}(2)$ index, this corresponds to
\begin{align}
Y^M = ({\mathrm{y}}^{\mathrm{m}}, {\mathrm{y}}_{\mathrm{m} \alpha}, {\mathrm{y}}_{\mathrm{m} \mathrm{n} \mathrm{k}}, {\mathrm{y}}^{\mathrm{m} \alpha}, {\mathrm{y}}_{\mathrm{m}})\,.
\end{align}
The field content of $E_{7(7)}$ EFT is given as follows:
\begin{align}
\{ g_{\mu \nu}, \mathcal{M}_{MN}, \mathcal{A}_{\mu}{}^M, {\mathcal{B}}_{\mu \nu, \alpha}, {\mathcal{B}}_{\mu \nu}{}^M \}\,,
\end{align}
where $\alpha = 1, \ldots, 133$ is an $E_{7(7)}$ adjoint index and $\mu, \nu = 1, \ldots 4$ ranges over the external space. The first of these fields should hopefully be self-explanatory---it is the metric on the external space---and so we focus on the remainder. The scalar degrees of freedom (from the perspective of the 4-dimensional external space) are held in the generalised metric $\mathcal{M}_{MN}$ which parametrises the coset $(E_{7(7)} \times \mathbb{R}^+)/\operatorname{SU}(8)$. A simple counting reveals that the 70 independent components of the metric, three-form and six-form potentials combine with the extra $\mathbb{R}^+$ scaling generator to match the dimension of the coset, as required. The generalised metric\footnote{Note that we refer to both the unscaled metric ${\tilde{\mathcal{M}}}_{MN}$ and scaled metric $\mathcal{M}_{MN}=e^{-\Delta} {\tilde{\mathcal{M}}}_{MN}$ (with any choice of $e^{-\Delta}$) as `the generalised metric' although, strictly speaking, only the former is a generalised metric of an $E_{n(n)}$ EFT whilst the latter is of an $E_{n(n)} \times \mathbb{R}^+$ EFT.} of $E_{7(7)}$ EFT (without the $\mathbb{R}^+$ scaling), which we denote ${\tilde{\mathcal{M}}}_{MN}$, is given in \cite{Hillmann:2009ci,Lee:2016qwn} (though note that the latter adopts a slightly modified choice of dualised coordinates to those used here, which follow the conventions of \cite{Hillmann:2009ci,Berman:2011jh}) and is also the one used in \cite{Berman:2014hna}. In the absence of internal potentials, it is given by\footnote{We note in passing that there is an alternative parametrisation of the generalised metric that is sometimes used called the `non-geometric parametrisation' (see, for example,  \cite{Lee:2016qwn,Sakatani:2017nfr,Sakatani:2014hba,Andriot:2014uda,Hassler:2013wsa} amongst others) which is closely linked to the $\beta$-supergravity formulation and that may offer an alternative perspective to exotic branes. We shall not pursue this line here.}
\begin{align}\label{eq:E7GenMetric}
{\tilde{\mathcal{M}}}_{MN} = \operatorname{diag} [g_{(7)}^{\frac{1}{2}} g_{mn}; g_{(7)}^{\frac{1}{2}} g^{mn,pq}; g_{(7)}^{-\frac{1}{2}} g^{mn}; g_{(7)}^{-\frac{1}{2}} g_{mn,pq}]\,,
\end{align}
where $g_{mn}$ is the internal metric, $g_{(7)}$ is its determinant and $g_{mn,pq} \coloneqq \frac{1}{2} ( g_{mp} g_{qn} - g_{mq} g_{pn})$ (similarly for $g^{mn,pq}$). With this choice, the generalised metric is a true $E_{7(7)}$ element (with determinant 1). However since we are considering the full EFT, and thus distinguish internal and external spaces, we allow for a relative scaling factor between the two. We thus consider a generalised metric of the form $\mathcal{M}_{MN} = e^{-\Delta} {\tilde{\mathcal{M}}}_{MN}$. Such a generalised metric was constructed in \cite{Berman:2011jh} (see also \cite{Malek:2012pw,Bakhmatov:2017les}) from a non-linear realisation of $E_{11}$ which yielded $e^{-\Delta}= {g_{(7)}}{}^{-1}$, although we note that the external space there has been truncated. Here, we shall adopt a different choice of scaling, 
\begin{align}\label{eq:Rescaling}
e^{-\Delta} = g_{(4)}^{-\frac{1}{4}}\,.
\end{align}
This overall scaling is identified as an extra scalar determining the relative scaling of the internal and external spaces and is the analogue of the $e^\phi = \operatorname{det} g_{\text{ext}}^{\frac{1}{7}}$ introduced in \cite{Blair:2013gqa,Blair:2014zba,Park:2013gaj} and later used in \cite{Bakhmatov:2017les} for the $\operatorname{SL}(5) \times \mathbb{R}^+$ EFT. With this setup, we are able to induce transformations on (the determinant of) the external metric via transformations of the generalised metric $\mathcal{M}_{MN}$.\par
One way to understand this is as follows. From the perspective of gauged supergravities we consider two distinct symmetries. The first is, of course, the Julia-Cremmer $E_{n(n)}$ duality symmetry which itself contains a natural scaling of the internal torus under the embedding $\operatorname{GL}(1) \subset \operatorname{GL}(n) \subset E_{n(n)}$---essentially a rescaling of the coordinates of the internal torus $y^m$ by $y^m \mapsto \lambda y^m$. The second is the so-called \emph{trombone symmetry} which is a well-known global scaling symmetry acting on the supergravity fields as
\begin{align}
g \mapsto \lambda^2 g\,, \qquad A_{(3)} \mapsto \lambda^3 A_{(3)}\,.
\end{align}
This is an on-shell symmetry for $n\leq 9$, being realised only at the level of the equations of motion (the Lagrangian of 11-dimensional supergravity transforms as $\mathcal{L} \mapsto \lambda^{d-2} \mathcal{L}$ for $\lambda \in \mathbb{R}^+$), but is promoted to an off-shell symmetry for $n=9$ (i.e. a symmetry of both the action and equations of motion\footnote{The special case of $n=9$ is perhaps best understood in terms of the allowed gaugings of the dimensionally reduced theory. In $n \leq 9$, the standard $E_{n(n)} \subset G$ gaugings (excluding the trombone symmetry) organise themselves into the embedding tensor $\theta_M{}^\alpha$, transforming under a particular subrepresentation of $R_{V^\ast} \otimes R_{\text{adj.}}$ that is dictated by group theoretic arguments---the so-called `linear' and `quadratic' algebraic constraints that are related to preserving supersymmetry and requiring $E_{n(n)}$ invariance of $\theta_M{}^\alpha$ respectively. In addition, the gaugings of the trombone symmetry organise themselves into a second object $\theta_M$ that transforms under $R_{V^\ast}$. For $n=9$, one finds that these two objects unify into a single object transforming under a single representation of the affine Kac-Moody algebra $E_{9(9)}$ \cite{LeDiffon:2008sh}. Thus, unlike in higher dimensions, generic gaugings in $d=2$ naturally contain trombone gaugings and so the trombone symmetry is promoted to an off-shell symmetry.}).  The extra $\mathbb{R}^+$ factor that we consider here may thus be considered as a combination of these two symmetries and was first considered in the present context in the closely related exceptional generalised geometry in \cite{Coimbra:2011ky,Coimbra:2012af,Coimbra:2011nw}. Just as the conventional embedding tensors of higher duality groups seed both the conventional gaugings and trombone gaugings of lower-dimensional gauged supergravities, the extra $\mathbb{R}^+$ factor may be understood as arising from the truncation of a higher duality group and is thus indispensable in generating other U-duality orbits.\par
The remaining fields $(\mathcal{A}_{\mu}{}^M, \mathcal{B}_{\mu \nu, \alpha}, \mathcal{B}_{\mu \nu, M})$ are a set of generalised gauge fields. The fully gauge-covariant field strength of $\mathcal{A}_\mu{}^M$ is given by
\begin{align}\label{eq:GenFieldStrength}
\mathcal{F}_{\mu \nu}{}^M = F_{\mu \nu}{}^M - 12 {\left(t^\alpha \right)}^{MN} \partial_N \mathcal{B}_{\mu \nu, \alpha} - \frac{1}{2} \Omega^{MN} \mathcal{B}_{\mu \nu, N}\,,
\end{align}
where $F_{\mu \nu}{}^M \coloneqq 2 \partial_{[\mu} \mathcal{A}_{\nu]}{}^M - {\llbracket \mathcal{A}_\mu \mathcal{A}_\nu \rrbracket}^M$ is the na\"{i}ve non-Abelian field strength that is given in terms of the E-bracket (covered below) and ${(t^\alpha)}^{MN}$ are the generators of $E_{7(7)}$, valued in the fundamental representation. This satisfies the Bianchi identity
\begin{align}\label{eq:Bianchi}
3 \mathcal{D}_{[\mu} \mathcal{F}_{\nu \rho]}{}^M = - 12 {\left(t^\alpha\right)}^{MN} \partial_N \mathcal{H}_{\mu \nu \rho, \alpha} - \frac{1}{2} \Omega^{MN} \mathcal{H}_{\mu \nu \rho, N}\,,
\end{align}
where $\mathcal{D}_\mu \coloneqq \partial_\mu - \mathbb{L}_{\mathcal{A}_\mu}$ denotes the Lie-covariantised derivative and $\mathcal{H}_{\mu \nu \rho, \bullet}$ are the (appropriately covariantised) fieldstrengths of the compensating gauge fields $\mathcal{B}_{\mu \nu, \bullet}$. Not all components of this fieldstrength are independent; half of them are related to the remaining components via the twisted self-duality relation,
\begin{align}\label{eq:TwistedSelfDuality}
\mathcal{F}_{\mu \nu}{}^M = - \frac{1}{2} {|g_{(4)}|}^{\frac{1}{2}} \varepsilon_{\mu \nu \rho \sigma} g^{\rho \lambda} g^{\sigma \kappa} \Omega^{MN} \mathcal{M}_{NK} \mathcal{F}_{\lambda \kappa}{}^K\,,
\end{align}
to leave only 28 propagating degrees of freedom.\par
Additionally, the theory possesses two group invariants: a symplectic form $\Omega_{MN}$ and a totally symmetric four-index object $c_{MNPQ}$, though we shall not be needing the latter. Due to the $\mathbb{R}^+$ factor, the former is a weighted symplectic matrix\cite{Aldazabal:2013mya}, of weight $\lambda(\Omega) = \frac{1}{2}$, and it is related to ${\tilde{\Omega}}_{MN} \in \operatorname{Sp}(56) \supset E_{7(7)}$ by
\begin{align}\label{eq:WeightedOmega}
\Omega_{MN} = e^{-\Delta} {\tilde{\Omega}}_{MN}\,.
\end{align}
We adopt the convention that indices are raised and lowered according to
\begin{align}
V^M = \Omega^{MN} V_N\,, \qquad V_M = V^N \Omega_{NM}\,.
\end{align}
The only non-vanishing components of $\Omega_{MN}$ are
\begin{align}
\Omega_m{}^n & = e^{-\Delta} \delta^m_n = - \Omega^n{}_m\,, \qquad \Omega_{mn}{}^{pq} = e^{-\Delta} \delta^{mn}_{pq} = - \Omega^{pq}{}_{mn}\,,
\end{align}
and similarly for the inverse, defined through $\Omega^{MK} \Omega_{NK} = \delta^M_N$.\par
In addition to the global $E_{n(n)} \times \mathbb{R}^+$ symmetry, the theory possesses a number of local symmetries---the general coordinate transformations of the metric and the $p$-form gauge transformations. Analogous to how the Lie derivative generates the algebra of infinitesimal diffeomorphisms in GR, we define a \emph{generalised Lie derivative} $\mathbb{L}$ which generates these local symmetries. In component form, the generalised Lie derivative of a generalised vector $V$, of weight $\lambda(V)$, along $U$ in $E_{7(7)} \times \mathbb{R}^+$ EFT is given by
\begin{align}\label{eq:GenLie}
\mathbb{L}_U V^M = {[U,V]}^M + Y^{MN}{}_{PQ} \partial_N U^P V^Q + \left( \lambda(V) - \frac{1}{2} \right) \partial_N U^N V^M\,,
\end{align}
where $Y^{MN}{}_{PQ}$ is the \emph{Y-tensor}, given in terms of group invariants as\footnote{
\cite{Berman:2012vc} gives an equivalent form in terms of the quartic invariant $c_{MNPQ}$ of $E_{7(7)}$:
\begin{align}
Y^{MN}{}_{PQ} = 12 c^{MN}{}_{PQ} + \delta^{(M}_P \delta^{N)}_Q + \frac{1}{2} \Omega^{MN} \Omega_{PQ}\,,
\end{align}
where the quartic invariant is defined in terms of the generators ${(t_\alpha)}_{MN} = {(t_\alpha)}_{NM}$ and symplectic form as\cite{LeDiffon:2011wt}
\begin{align}\label{eq:Quartic}
{(t^\alpha)}_{MN} {(t_{\alpha})}_{KL} = \frac{1}{12} \tilde{\Omega}_{M(K} \tilde{\Omega}_{L)N} + c_{MNKL}\,.
\end{align}
Indeed, this last relation is used to show equivalence between \eqref{eq:P1} and \eqref{eq:P2} by permuting the indices on the two generators using the symmetry of $c_{MNKL}$. 
}
\begin{align}
Y^{MN}{}_{PQ} & = - 12 {\left( \mathbb{P}_{\text{adj}} \right) }{}^M{}_Q{}^N{}_P + \frac{1}{2} \delta^M_Q \delta^N_P + \delta^M_P \delta^N_Q\\
	& = - 12 {(t^\alpha)}^{MN} {(t_\alpha)}_{PQ} - \frac{1}{2} \Omega^{MN} \Omega_{PQ}\,.
\end{align}
The second line uses the fact that the projector onto the adjoint representation $\mathbb{P}_{\text{adj}}$ is given by
\begin{align}
{\left( \mathbb{P}_{\text{adj}} \right) }{}^M{}_Q{}^N{}_P & = {(t^\alpha)}_Q{}^M {(t_\alpha)}_P{}^N\label{eq:P1}\\
	& = {(t^\alpha)}^{MN} {(t_\alpha)}_{QP} + \frac{1}{24} \delta^M_Q \delta^N_P +\frac{1}{12} \delta^M_P \delta^N_Q - \frac{1}{24} \Omega^{MN} \Omega_{QP}\label{eq:P2}\,,
\end{align}
subject to the normalisation ${\mathbb{P}_{(\text{adj})}}^M{}_N{}^N{}_M = 133$. One sees that, in the form \eqref{eq:GenLie}, there is a naturally defined \emph{effective} weight in the theory given by the bracketed term which naturally singles out $\lambda = \frac{1}{2}$ (indeed, for consistency, one requires that the gauge transformations of the generalised gauge fields have weight $\lambda = \frac{1}{2}$). One may verify that the generalised Lie derivative can equivalently be cast into the form given in \cite{Hohm:2013uia}:
\begin{align}
\mathbb{L}_U V^M & = U^N \partial_N V^M - 12 {\left(\mathbb{P}_{\text{adj}} \right)}^M{}_Q{}^N{}_P \partial_N U^P V^Q + \lambda(V) \partial_N U^N V^M.
\end{align}
We also list some of the properties of the generators of $E_{7(7)} \times \mathbb{R}^+$ that hold \emph{for this chapter}\footnote{Unfortunately, we shall need to adopt a different set of conventions in Chapter~\ref{ch:E8NonRiemannian} which is dictated by a reduction of the $E_{8(8)}$ algebra. This will be described in more detail there.}. With the conventions that we adopt, raising and lowering with the symplectic form, it is easy to check that ${(t_\alpha)}_{MN} = {(t_\alpha)}_{(MN)}$. These generators further satisfy
\begin{align}
{(t_\alpha)}_K{}^{(P} c^{QRS)K} & = 0\,,\\
{(t_\beta)}_M{}^K {(t^\alpha)}_K{}^N & = \frac{19}{8} \delta_M{}^N\,,\\
{(t^\alpha)}^{MK} {(t_\beta)}_{KL} {(t_\alpha)}^{LN} & = - \frac{7}{8} {(t_\beta)}^{MN}\,.
\end{align}
Finally, the $\mathbb{R}^+$ generator is ${(t_0)}_M{}^N = - \delta_M^N$. We may read off the following section conditions for this theory from the $Y$-tensor:
\begin{align}
{(t_\alpha)}^{MN} \partial_{MN} \bullet = 0\,,, \qquad {(t_\alpha)}^{MN} \partial_M \bullet \partial_N \bullet = 0\,, \qquad \Omega^{MN} \partial_M \bullet \partial_N \bullet = 0\,.
\end{align}
Finally, we mention that the pseudo-action for this theory is given by
\begin{align}
{\mathcal{S}}_{E_{7(7)}} = & \int \textrm{d}^4 x \textrm{d}^{56} Y \left( {\mathcal{L}}_{\text{E-H}} + {\mathcal{L}}_{\text{sc.}} + {\mathcal{L}}_{\text{Y-M}} - e_{(4)} V \right) + S_{\text{top.}}\,, \qquad e_{(4)} = \sqrt{-g_{(4)}}\,,\\
{\mathcal{L}}_{\text{E-H}} = & e_{(4)} \hat{R}\,,\\
{\mathcal{L}}_{\text{sc.}} = & \frac{1}{48} e_{(4)} g^{\mu \nu} {\mathcal{D}}_\mu {\mathcal{M}}_{MN} {\mathcal{D}}_\nu {\mathcal{M}}^{MN}\,,\\
{\mathcal{L}}_{\text{Y-M}} = & - \frac{1}{8} e_{(4)} {\mathcal{M}}_{MN} {\mathcal{F}}^{\mu \nu M} {\mathcal{F}}_{\mu \nu}{}^N\,,\\
\mathllap{V =} &
\begingroup
\renewcommand{\arraystretch}{1.5}
\begin{array}[t]{l}
 - \frac{1}{48} {\mathcal{M}}^{MN} \partial_M \mathcal{M}^{KL} \partial_N \mathcal{M}_{KL} + \frac{1}{2} \mathcal{M}^{MN} \partial_M \mathcal{M}^{KL} \partial_K \mathcal{M}_{NL}\\
- \frac{1}{2}  \partial_M \ln |g_{(4)}| \partial_N \mathcal{M}^{MN} - \frac{1}{4} \mathcal{M}^{MN} \partial_M \ln |g_{(4)}| \partial_N \ln |g_{(4)}|\\
- \frac{1}{4} \mathcal{M}^{MN}\partial_M g^{\mu \nu} \partial_N g_{\mu \nu}\,,\\
\end{array}
\endgroup\\
{\mathcal{S}}_{\text{top.}} = & - \frac{1}{24} \int \textrm{d}^5 x \int \textrm{d}^{56} Y \varepsilon^{\mu \nu \rho \sigma \tau} {\mathcal{F}}_{\mu \nu}{}^M {\mathcal{D}}_{\rho} {\mathcal{F}}_{\sigma \tau M}\,,
\end{align}
which is supplemented by the twisted self-duality constraint \eqref{eq:TwistedSelfDuality}. The only new notation here is the covariantised Einstein-Hilbert action in which the partial derivatives have been promoted to the Lie-covariantised derivatives $\partial_\mu \rightarrow \mathcal{D}_\mu$.\par
For this section, we shall denote a particular choice of frame by a superscript such that e.g.\ $g_{\mu \nu}^{5^3}$ denotes the external metric in the $5^3$ frame (or, more generally, $g_{\mu \nu}^{\text{M}}$ for a generic M-theory solution). Here we add to a growing list of solutions in DFT \cite{Berkeley:2014nza,Blair:2016xnn,Berman:2014jsa,Bakhmatov:2016kfn,Kimura:2018hph} and EFT\cite{Blair:2014zba,Berman:2014hna,Bakhmatov:2017les} in which a single solution in the extended space reduces to a number of distinct known solutions in 10 and 11 dimensions upon applying the appropriate section constraint. Although apparently unrelated in the reduced section, they are nonetheless related by a duality transformation (or at least a \emph{solution-generating} transformation) acting linearly on the extended space. The solution presented here is a very close analogue of the solution presented in \cite{Berman:2014hna} which described all of the conventional branes. We shall henceforth refer to that solution the `geometric solution'. The solution that is presented here covers all the non-geometric branes of de Boer and Shigemori \cite{deBoer:2012ma} and will thus be referred to as the `non-geometric solution'. The branes that are contained within these two solutions are indicated in Figure~\ref{fig:Brane}.\par
The set-up that we choose is as follows: we define the external space by the coordinates
\begin{align}
x^\mu = ( t, r, \theta, z)
\end{align} 
and take the external metric to be the same in both the M-theory and Type IIB sections:
\begin{align}\label{eq:Ext}
g_{\mu \nu} = \operatorname{diag} [ - {(HK^{-1})}^{-\frac{1}{2}}, {(HK)}^{\frac{1}{2}}, r^2 {(HK)}^{\frac{1}{2}}, {(HK^{-1})}^{\frac{1}{2}}]\,, \qquad g_{(4)} \coloneqq \operatorname{det} g_{\mu \nu}\,.
\end{align}
The external metrics of all the branes shall be proportional to this e.g.\ for the $5^3$-brane, we have $g_{\mu \nu}^{5^3} = {(HK^{-1})}^{\frac{1}{6}} g_{\mu\nu} = {\left| g_{(7)}^{5^3} \right|}^{-\frac{1}{2}} g_{\mu \nu}$. Here, $H$---a harmonic function in the $r$-$\theta$ plane---and $K$ are given by
\begin{align}
H(r) & = h_0 + \sigma \ln \frac{\mu}{r}\,,\qquad K = H^2 + \sigma^2 \theta^2 \,.
\end{align}
The generalised metric is chosen to be diagonal and, in any given frame, the 56 components of the generalised metric split into 27 components of $(HK^{-1})^{\frac{1}{2}}$, 27 components of ${(HK^{-1})}^{-\frac{1}{2}}$ and one component each of ${(HK^{-1})}^{\frac{3}{2}}$ and ${(HK^{-1})}^{-\frac{3}{2}}$. The non-zero components of the EFT vector $\mathcal{A}_\mu{}^M$ shall always point in the distinguished directions.\par
For the purposes of this solution, we shall set the compensating gauge fields ${\mathcal{B}}_{\mu \nu, \bullet}$ (which are only required to close the gauge structure of the theory which itself is broken upon applying the section condition) to zero, such that their corresponding field strengths $\mathcal{H}_{\mu \nu \rho, \bullet}$ also vanish. We choose the only non-vanishing components of $\mathcal{A}_\mu{}^M$ to be
\begin{align}\label{eq:AAnsatz}
\mathcal{A}_t{}^M & = -H^{-1} K a^M\,, \qquad \mathcal{A}_z{}^M = -K^{-1} \theta \sigma {\tilde{a}}^M\,,
\end{align}
where $a^M$ and ${\tilde{a}}^M$ determine the direction the vector points in the generalised space. Note that these two vectors are related through the definition of the generalised field strength \eqref{eq:GenFieldStrength} in the twisted self-duality condition \eqref{eq:TwistedSelfDuality} and are thus not independent. We leave a proof that this ansatz does indeed satisfy the twisted self-duality constraint and Bianchi identity to Appendix~\ref{app:ExoticFieldstrength}.
\thispagestyle{empty}
\begin{landscape}
\begingroup
\captionsetup{width=0.85\linewidth}
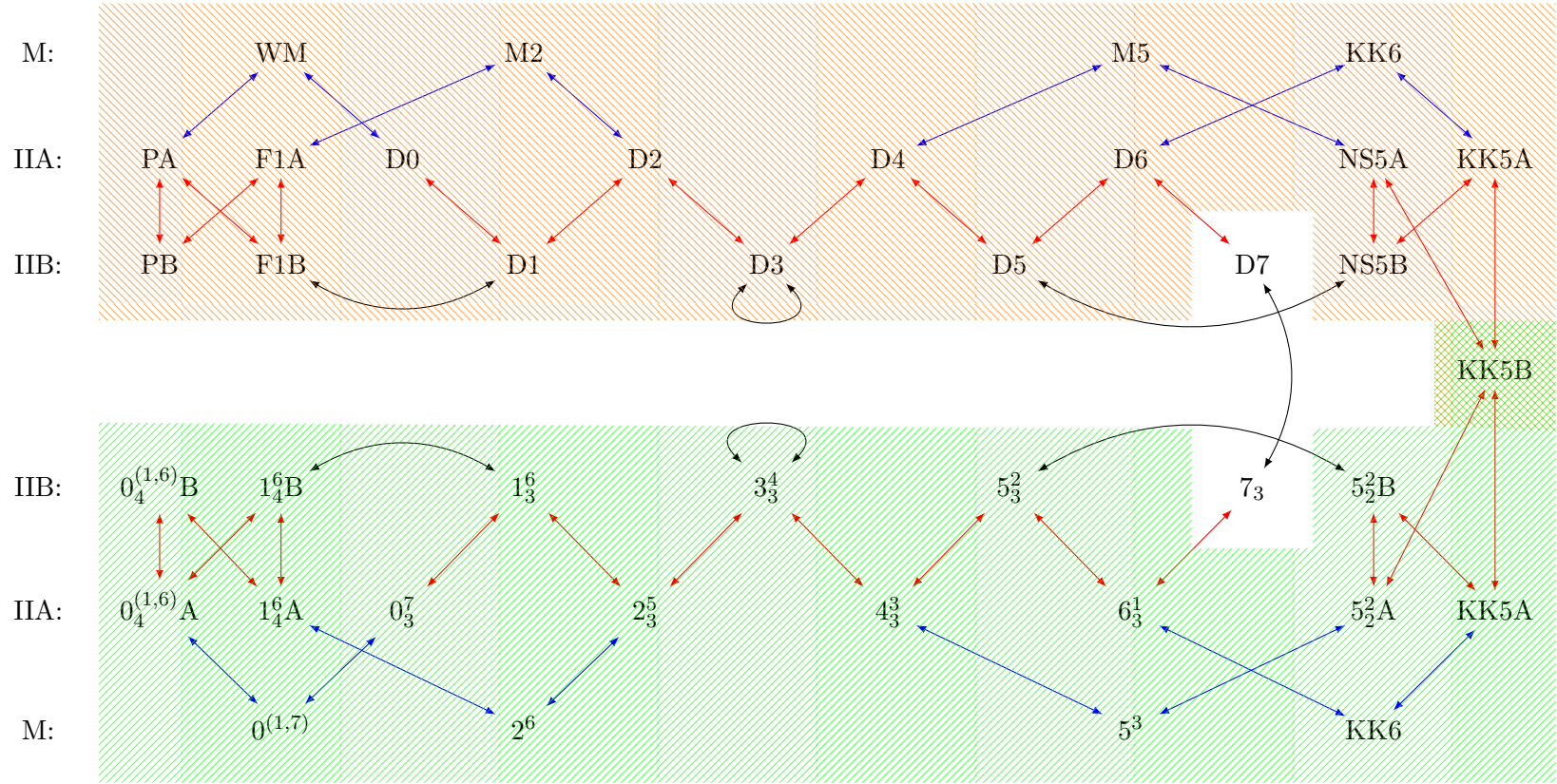
\begin{figure}[t]
\centering
\begin{tikzpicture}
\matrix(M)[matrix of math nodes, row sep=2.45em, column sep=2.4 em, minimum width=2em]{
\clap{\phantom{a}}&&&&&&&&&\clap{\phantom{a}}&\clap{\phantom{a}}&&&\clap{\phantom{a}}\\
\clap{\text{M:}}&\clap{\phantom{a}}&\clap{\text{WM}}&&\clap{\text{M2}}&&&&&\clap{\text{M5}}&&\clap{\text{KK6}}&\clap{\phantom{a}}&\\
\clap{\text{IIA:}}&\clap{\text{PA}}&\clap{\text{F1A}}&\clap{\text{D0}}&&\clap{\text{D2}}&&\clap{\text{D4}}&&\clap{\text{D6}}&&\clap{\text{NS5A}}&\clap{\text{KK5A}}&\\
\clap{\text{IIB:}}&\clap{\text{PB}}&\clap{\text{F1B}}&&\clap{\text{D1}}&&\clap{\text{D3}}&&\clap{\text{D5}}&\clap{\phantom{a}}&\clap{\text{D7}}&\clap{\text{NS5B}}&\clap{\phantom{a}}&\clap{\phantom{a}}\\
\clap{\phantom{a}}&&&&&&&&&\clap{\phantom{a}}&\clap{\phantom{a}}&\clap{\phantom{a}}&\clap{\text{KK5B}}&\clap{\phantom{a}}\\
\clap{\text{IIB:}}&\mathclap{0_4^{(1,6)}\text{B}}&\mathclap{1_4^6\text{B}}&&\mathclap{1_3^6}&&\mathclap{3_3^4}&&\mathclap{5_3^2}&\mathclap{\phantom{a}}&\mathclap{7_3}&\mathclap{5_2^2\text{B}}&\mathclap{\phantom{a}}&\mathclap{\phantom{a}}\\
\clap{\text{IIA:}}&\mathclap{0_4^{(1,6)}\text{A}}&\mathclap{1_4^6\text{A}}&\mathclap{0_3^7}&&\mathclap{2_3^5}&&\mathclap{4_3^3}&&\mathclap{6_3^1}&\mathclap{\phantom{a}}&\mathclap{5_2^2\text{A}}&\clap{\text{KK5A}}&\\
\clap{\text{M:}}&\clap{\phantom{a}}&\mathclap{0^{(1,7)}}&&\mathclap{2^6}&&&&&\mathclap{5^3}&&\clap{\text{KK6}}&\clap{\phantom{\text{a}}}&\clap{\phantom{a}}\\
\clap{\phantom{a}}&&&&&&&&&\clap{\phantom{a}}&\clap{\phantom{a}}&&\clap{\phantom{a}}&\\
	};
\draw[latex-latex, draw=blue, fill=blue] (M-2-3) -- (M-3-2);
\draw[latex-latex, draw=blue, fill=blue] (M-2-3) -- (M-3-4);
\draw[latex-latex, draw=blue, fill=blue] (M-2-5) -- (M-3-3);
\draw[latex-latex, draw=blue, fill=blue] (M-2-5) -- (M-3-6);
\draw[latex-latex, draw=blue, fill=blue] (M-2-10) -- (M-3-8);
\draw[latex-latex, draw=blue, fill=blue] (M-2-10) -- (M-3-12);
\draw[latex-latex, draw=blue, fill=blue] (M-2-12) -- (M-3-10);
\draw[latex-latex, draw=blue, fill=blue] (M-2-12) -- (M-3-13);
\draw[latex-latex, draw=red, fill=red] (M-3-2) -- (M-4-2);
\draw[latex-latex, draw=red, fill=red] (M-3-3) -- (M-4-3);
\draw[latex-latex, draw=red, fill=red] (M-3-2) -- (M-4-3);
\draw[latex-latex, draw=red, fill=red] (M-3-3) -- (M-4-2);
\draw[latex-latex, draw=red, fill=red] (M-3-4) -- (M-4-5);
\draw[latex-latex, draw=red, fill=red] (M-4-5) -- (M-3-6);
\draw[latex-latex, draw=red, fill=red] (M-3-6) -- (M-4-7);
\draw[latex-latex, draw=red, fill=red] (M-4-7) -- (M-3-8);
\draw[latex-latex, draw=red, fill=red] (M-3-8) -- (M-4-9);
\draw[latex-latex, draw=red, fill=red] (M-4-9) -- (M-3-10);
\draw[latex-latex, draw=red, fill=red] (M-3-10) -- (M-4-11);
\draw[latex-latex, draw=red, fill=red] (M-3-12) -- (M-4-12);
\draw[latex-latex, draw=red, fill=red] (M-3-13) -- (M-5-13);
\draw[latex-latex, draw=red, fill=red] (M-3-12) -- (M-5-13);
\draw[latex-latex, draw=red, fill=red] (M-3-13) -- (M-4-12);
\draw[latex-latex] (M-4-3) to[bend right,min distance=5mm] (M-4-5);
\draw[latex-latex] (M-4-9) to[bend right,min distance=5mm] (M-4-12);
\draw[latex-latex] (M-4-7) to[out=-135,in=-45,min distance=10mm](M-4-7);
\draw[latex-latex] (M-4-11) to[bend left] (M-6-11);
\draw[latex-latex] (M-6-7) to[out=45,in=135, min distance=10mm](M-6-7);
\draw[latex-latex] (M-6-9) to[bend left,min distance=5mm] (M-6-12);
\draw[latex-latex] (M-6-3) to[bend left,min distance=5mm] (M-6-5);
\draw[latex-latex, draw=red, fill=red] (M-6-2) -- (M-7-2);
\draw[latex-latex, draw=red, fill=red] (M-6-3) -- (M-7-3);
\draw[latex-latex, draw=red, fill=red] (M-6-2) -- (M-7-3);
\draw[latex-latex, draw=red, fill=red] (M-6-3) -- (M-7-2);
\draw[latex-latex, draw=red, fill=red] (M-7-4) -- (M-6-5);
\draw[latex-latex, draw=red, fill=red] (M-6-5) -- (M-7-6);
\draw[latex-latex, draw=red, fill=red] (M-7-6) -- (M-6-7);
\draw[latex-latex, draw=red, fill=red] (M-6-7) -- (M-7-8);
\draw[latex-latex, draw=red, fill=red] (M-7-8) -- (M-6-9);
\draw[latex-latex, draw=red, fill=red] (M-6-9) -- (M-7-10);
\draw[latex-latex, draw=red, fill=red] (M-7-10) -- (M-6-11);
\draw[latex-latex, draw=red, fill=red] (M-6-12) -- (M-7-12);
\draw[latex-latex, draw=red, fill=red] (M-5-13) -- (M-7-12);
\draw[latex-latex, draw=red, fill=red] (M-5-13) -- (M-7-13);
\draw[latex-latex, draw=red, fill=red] (M-6-12) -- (M-7-13);
\draw[latex-latex, draw=blue, fill=blue] (M-8-3) -- (M-7-2);
\draw[latex-latex, draw=blue, fill=blue] (M-8-3) -- (M-7-4);
\draw[latex-latex, draw=blue, fill=blue] (M-8-5) -- (M-7-3);
\draw[latex-latex, draw=blue, fill=blue] (M-8-5) -- (M-7-6);
\draw[latex-latex, draw=blue, fill=blue] (M-8-10) -- (M-7-8);
\draw[latex-latex, draw=blue, fill=blue] (M-8-10) -- (M-7-12);
\draw[latex-latex, draw=blue, fill=blue] (M-8-12) -- (M-7-10);
\draw[latex-latex, draw=blue, fill=blue] (M-8-12) -- (M-7-13);
\pattern[pattern=north east lines, pattern color=orange, draw opacity=0.6]
	($(M-1-1.south east)!0.5!(M-2-2.north west)$)
	--($(M-5-1.north east)!0.5!(M-4-2.south west)$)
	--($(M-5-10.north east)!0.5!(M-4-11.south west)$)
	--($(M-3-10.south west)!0.5!(M-4-11.north east)$)
	--($(M-4-11.north east)!0.5!(M-3-12.south west)$)
	--($(M-5-11.north east)!0.5!(M-4-12.south west)$)
	--($(M-5-12.north east)!0.5!(M-4-13.south west)$)
	--($(M-5-13.south west)!0.5!(M-6-12.north east)$)
	--($(M-6-14.north west)!0.5!(M-5-13.south east)$)
	--($(M-2-13.north east)!0.5!(M-1-14.south west)$)
	--($(M-1-1.south east)!0.5!(M-2-2.north west)$);
\pattern[pattern=north west lines, pattern color=green, draw opacity=0.6]
	($(M-5-1.south west)!0.5!(M-6-2.north east)$)
	--($(M-9-1.south west)!0.5!(M-8-2.north east)$)
	--($(M-9-13.north east)!0.5!(M-8-14.south west)$)
	--($(M-4-13.south east)!0.5!(M-5-14.north west)$)
	--($(M-5-12.north east)!0.5!(M-4-13.south west)$)
	--($(M-6-12.north east)!0.5!(M-5-13.south west)$)
	--($(M-6-11.north east)!0.5!(M-5-12.south west)$)
	--($(M-7-12.north west)!0.5!(M-6-11.south east)$)
	--($(M-7-10.north east)!0.5!(M-6-11.south west)$)
	--($(M-5-10.south east)!0.5!(M-6-11.north west)$)
	--($(M-5-1.south west)!0.5!(M-6-2.north east)$);
\end{tikzpicture}
\caption[The scope of the geometric and non-geometric solutions in $E_{7(7)}$ EFT.]{The branes that we consider. Red lines denote T-duality, blue lines denote lifts/reductions and black lines denote S-dualities. The hashed green area contains all the branes contained within the non-geometric solution whilst the hashed orange area contains all the branes contained in the geometric solution.}
\label{fig:Brane}
\end{figure}
\endgroup
\end{landscape}
\subsection{M-Theory Section}
For this non-geometric solution, we take a subtly different approach to the solution constructed in \cite{Berman:2014hna}. There, they chose the generalised metric to be a genuine $E_{7(7)}$ element ${\tilde{\mathcal{M}}}_{MN}$, of the form \eqref{eq:E7GenMetric}, and implemented a relative scaling between the internal and external sectors $g_{\mu \nu} \mapsto {\left|g_{(7)}\right|}^{\frac{1}{2}} g_{\mu \nu}$ (reducing to the Einstein frame for the Type II solutions) as a conventional Kaluza-Klein decomposition of the 11-dimensional metric $\textrm{d} s^2 = g_{\mu \nu} \textrm{d} x^\mu \textrm{d} x^\nu + e^{2\alpha} g_{mn}{(\textrm{d} y^m + \mathcal{A}_\mu{}^m \textrm{d} x^\mu)}{(\textrm{d} y^n + \mathcal{A}_\nu{}^n \textrm{d} x^\nu)}$.  In particular, fixing a frame $\text{M}$ in \eqref{eq:E7GenMetric}, there is no preferred scaling of the external metric encoded within ${\tilde{\mathcal{M}}}_{MN}$. Whilst notionally intuitive, the scaling in \cite{Berman:2014hna} requires the section condition to have been imposed and has an \textit{ad hoc} feel in the sense that there appears to be nothing in the EFT framework that compels one to take their particular scaling $g_{\mu \nu} \mapsto {\left| g_{(7)} \right|}^{\frac{1}{2}} g_{\mu \nu}$; it is only imposed to match the known brane metrics. Here we make a small modification. We may use the scaling \eqref{eq:Rescaling} to encode this information directly into the generalised metric such that it forces the required relative scaling of the internal and external sector without imposing the section condition---a perhaps more natural description in terms of EFT. In particular let a superscript/subscript $\text{M}$ denote a given duality frame. Since we wish to impose the scaling of the external metric
\begin{align}\label{eq:ExtScaling}
g_{\mu \nu} = {\left|g_{(7)}^{\text{M}}\right|}^{\frac{1}{2}} g^{\text{M}}_{\mu \nu}\,, \qquad  \Rightarrow \qquad {|g^{\text{M}}_{(4)}|}^{-\frac{1}{4}} = {\left|g_{(7)}^{\text{M}}\right|}^{\frac{1}{2}} {|g_{(4)}|}^{-\frac{1}{4}}\,,
\end{align}
we choose to define the $E_{7(7)} \times \mathbb{R}^+$ element ${\mathcal{M}}_{MN} = {|g_{(4)}|}^{-\frac{1}{4}} {\tilde{\mathcal{M}}}_{MN}$ such that
\begin{align}
{\mathcal{M}}_{MN} & = {\left|g^{\text{M}}_{(4)}\right|}^{-\frac{1}{4}} \operatorname{diag} [ g^{\text{M}}_{mn}; g_{\text{M}}^{mn,pq}; {(g^{\text{M}}_{(7)})}^{-1} g_{\text{M}}^{mn}; {(g^{\text{M}}_{(7)})}^{-1} g^{\text{M}}_{mn,pq}]\,.
\end{align}
In particular all terms, including the scaling of the external metric, are all in the same duality frame. To be explicit, we are constructing a generalised metric $\mathcal{M}_{MN}$ which reproduces the backgrounds of the exotic branes in the following fashion:
\begin{align}\label{eq:NonGeom}
\begin{aligned}
\mathcal{M}_{MN} & = {\left|g^{5^3}_{(4)}\right|}^{-\frac{1}{4}} \operatorname{diag} [ g^{5^3}_{mn}; g_{5^3}^{mn,pq}; {(g^{5^3}_{(7)})}^{-1} g_{5^3}^{mn}; {(g^{5^3}_{(7)})}^{-1} g^{5^3}_{mn,pq}]\\
	& = {\left|g^{2^6}_{(4)}\right|}^{-\frac{1}{4}} \operatorname{diag} [ g^{2^6}_{mn}; g_{2^6}^{mn,pq}; {(g^{2^6}_{(7)})}^{-1} g_{2^6}^{mn}; {(g^{2^6}_{(7)})}^{-1} g^{2^6}_{mn,pq}]\\
	& = {\left|g^{0^{(1,7)}}_{(4)}\right|}^{-\frac{1}{4}} \operatorname{diag} [ g^{0^{(1,7)}}_{mn}; g_{0^{(1,7)}}^{mn,pq}; {(g^{0^{(1,7)}}_{(7)})}^{-1} g_{0^{(1,7)}}^{mn}; {(g^{0^{(1,7)}}_{(7)})}^{-1} g^{0^{(1,7)}}_{mn,pq}]\\
	& = {\left|g^{\text{KK6}}_{(4)}\right|}^{-\frac{1}{4}} \operatorname{diag} [ g^{\text{KK6}}_{mn}; g_{\text{KK6}}^{mn,pq}; {(g^{\text{KK6}}_{(7)})}^{-1} g_{\text{KK6}}^{mn}; {(g^{\text{KK6}}_{(7)})}^{-1} g^{\text{KK6}}_{mn,pq}]\\
	& \qquad \qquad \qquad \qquad \vdots
\end{aligned}
\end{align}
where the vertical dots represent all of the Type IIA and Type IIB branes listed in Tables~\ref{tab:NonGeomIIA} and \ref{tab:NonGeomIIB}.\par
We split the seven internal coordinates into
\begin{align}
y^m = (\xi, \chi, w^a) \equiv (Y^\xi, Y^\chi, Y^a)\,,
\end{align}
where $a = 1, \ldots, 5$. These are promoted to the 56 coordinates of the extended internal space of EFT, indexed by $M=1, \ldots, 56$, according to \eqref{eq:E7MCoords}, which we order by
\begin{align}\label{eq:IntCoords}
Y^M = ( Y^\xi, Y^\chi, Y^a; Y_{\xi \chi}, Y_{\xi a}, Y_{\chi a}, Y_{ab}; Y_{\xi} Y_{\chi}, Y_a; Y^{\xi \chi}, Y^{\xi a} Y^{\chi a}, Y^{ab})\,.
\end{align}
The coordinates with two labels are thus antisymmetric and we have delimited the generalised coordinates by semi-colons for easier identification of the coordinates later. Since the generalised metric that we consider here is diagonal, this unambiguously defines the ordering of the components of the generalised matrix. Under this coordinate splitting, the configurations of the branes that we obtain is summarised in Table~\ref{tab:NonGeomM}.\par
\begin{table}[t]
\centering
\begin{tabulary}{\textwidth}{CLCCCCCCCCC}
\toprule
& & $t$ & $r$ & $\theta$ & $z$ & & $\xi$ & $\chi$ & $w^a$ &\\
\cmidrule{3-6}\cmidrule{8-10}
& $5^3$ & $\ast$ & $\bullet$ & $\bullet$ & $\circ$ & & $\circ$ & $\circ$ & $\ast$\\
& $2^6$ & $\ast$ & $\bullet$ & $\bullet$ & $\circ$ & & $\ast$ & $\ast$ & $\circ$\\
& $0^{(1,7)}$ & $\ast$ & $\bullet$ & $\bullet$ & $\circ$ & & $\odot$ & $\circ$ & $\circ$\\
& KK6 & $\ast$ & $\bullet$ & $\bullet$ & $\odot$ & & $\circ$ & $\ast$ & $\ast$\\
\bottomrule
\end{tabulary}
\caption[The configurations of the M-theory branes that we consider.]{The configurations of the M-theory branes that we consider. Asterisks $\ast$ denote worldvolume coordinates, empty circles $\circ$ denote smeared transverse coordinates and filled circles $\bullet$ denote coordinates that the harmonic function depends on. Finally, $\odot$ denotes an otherwise distinguished direction; the Hopf fibre for the monopole and the cubic direction for the $0^{(1,7)}$.}
\label{tab:NonGeomM}
\end{table}
Upon taking the M-theory section, the components of the EFT vector take on different roles, depending on the direction in which it points in the generalised space:
\begin{align}
\mathcal{A}_\mu{}^M \rightarrow (\mathcal{A}_\mu{}^m, \mathcal{A}_{\mu, mn}, \mathcal{A}_{\mu m}, \mathcal{A}_{\mu}{}^{mn}).
\end{align}
The first of these sources the conventional Kaluza-Klein vector, sourcing a cross-sector coupling of the type seen in the $0^{(1,7)}$, and the third is related to the dual graviton. The remaining two components then source the cross-sector components of the M-theory potentials: $\mathcal{A}_{\mu,mn} \sim A_{\mu mn}$ and $\mathcal{A}_\mu{}^{mn} \sim \epsilon^{mn p_1 \ldots p_5} A_{\mu p_1 \ldots p_5}$.\par
The components of the generalised metric are interchanged by a rotation on the internal space and are thus dependent on the frame chosen but, for the $5^3$ frame, we choose (setting $Q = HK^{-1}$ for convenience)
\begingroup
\renewcommand{\arraystretch}{1.5}
\begin{align}
\mathcal{M}_{MN} & =
\begin{array}[t]{ll}
{|g_{(4)}|}^{-\frac{1}{4}} \operatorname{diag}[
& \!\!\!\! {Q}^{\frac{1}{2}}, {Q}^{\frac{1}{2}}, {Q}^{-\frac{1}{2}} \delta_{(5)}; {Q}^{-\frac{3}{2}}, {Q}^{-\frac{1}{2}} \delta_{(5)}, {Q}^{-\frac{1}{2}} \delta_{(5)}, {Q}^{\frac{1}{2}} \delta_{(10)};\\
& \!\!\!\! {Q}^{-\frac{1}{2}}, {Q}^{-\frac{1}{2}}, {Q}^{\frac{1}{2}} \delta_{(5)}; {Q}^{\frac{3}{2}}, {Q}^{\frac{1}{2}} \delta_{(5)}, {Q}^{\frac{1}{2}} \delta_{(5)}, {Q}^{-\frac{1}{2}} \delta_{(10)}]
\end{array}\label{eq:53GM}\\
& = \begin{array}[t]{ll}
\smash{\left[\smash{ \underbrace{ {Q}^{-\frac{1}{6}}}_{{\left|g_{(7)}^{5^3}\right|}^{\frac{1}{2}}}} {| g_{(4)}|}^{-\frac{1}{4}} \right] \operatorname{diag} [}
& \!\!\!\! {Q}^{\frac{2}{3}}, {Q}^{\frac{2}{3}}, {Q}^{-\frac{1}{3}} \delta_{(5)}; {Q}^{-\frac{4}{3}}, {Q}^{-\frac{1}{3}} \delta_{(5)}, {Q}^{-\frac{1}{3}} \delta_{(5)}, {Q}^{\frac{2}{3}} \delta_{(10)};\\
& \!\!\!\! {Q}^{-\frac{1}{3}}, {Q}^{-\frac{1}{3}}, {Q}^{\frac{2}{3}} \delta_{(5)}; {Q}^{\frac{5}{3}}, {Q}^{\frac{2}{3}} \delta_{(5)}, {Q}^{\frac{2}{3}} \delta_{(5)}, {Q}^{-\frac{1}{3}} \delta_{(10)}]\,,\\
\end{array}\nonumber\label{eq:53GenMetric}
\end{align}
\endgroup
where we stress that the factor in the square brackets is to be identified with the scaling \eqref{eq:ExtScaling}. The notation $\delta_{(p)}$ is used as a shorthand for the identity matrix in $p$ dimensions, with the appropriate index structure. One may verify that this does indeed give the background of the $5^3$ upon applying the section condition if one identifies
\begin{gather}
\begin{gathered}
g_{mn}^{5^3} = \operatorname{diag} [ {Q}^{\frac{2}{3}}, {Q}^{\frac{2}{3}}, {Q}^{-\frac{1}{3}} \delta_{(5)}]\,,\\
{\left| g_{(4)}^{5^3}\right|}^{-\frac{1}{4}} \coloneqq {Q}^{-\frac{1}{6}} {| g_{(4)}|}^{-\frac{1}{4}} \qquad \Rightarrow \qquad  g_{\mu \nu}^{5^3} = {Q}^{\frac{1}{6}} g_{\mu \nu}\,.
\end{gathered}
\end{gather}
For this frame, we choose the EFT vectors to point out of section such that they do not contribute to the metric but rather source the M-theory potentials of the $5^3$:
\begin{align}\label{eq:53EFTVec}
\mathcal{A}_t{}^{\xi \chi} = - H^{-1}K\,, \qquad \mathcal{A}_{z,\xi \chi} = - K^{-1} \theta \sigma\,.
\end{align}
We thus obtain the background of the $5^3$:
\begin{gather}\label{eq:53Metric}
\begingroup
\renewcommand{\arraystretch}{1.5}
\begin{array}[t]{rl}
\textrm{d} s^2_{5^3} & = {(HK^{-1})}^{-\frac{1}{3}} ( -\textrm{d} t^2 + \textrm{d} {\vec{w}}^2_{(5)} ) + {(HK^{-1})}^{\frac{2}{3}} ( \textrm{d} z^2 + \textrm{d} \xi^2 + \textrm{d} \chi^2)\\
	& \qquad  + H^{\frac{2}{3}} K^{\frac{1}{3}} ( \textrm{d} r^2 + r^2 \textrm{d} \theta^2)\,,\\
\end{array}
\endgroup\\
A_{(3)} = -K^{-1} \theta \sigma \textrm{d} z \wedge \textrm{d} \xi \wedge \textrm{d} \chi\,, \qquad A_{(6)} = - H^{-1}K \textrm{d} t \wedge \textrm{d} w^1 \ldots \wedge \textrm{d} w^5\,.
\end{gather}
After applying the coordinate swap
\begin{gather}
Y^M \leftrightarrow Y_M
\end{gather}
and factoring out a new scaling factor, one obtains
\begingroup
\renewcommand{\arraystretch}{1.5}
\begin{align}
\begin{array}[t]{rl}
\mathcal{M}_{MN} = {Q}^{\frac{1}{6}}  {|g_{(4)}|}^{-\frac{1}{4}} \operatorname{diag} [
&\!\!\!\! {Q}^{-\frac{2}{3}}, {Q}^{-\frac{2}{3}}, {Q}^{\frac{1}{3}} \delta_{(5)}; {Q}^{\frac{4}{3}}, {Q}^{\frac{1}{3}} \delta_{(5)}, {Q}^{\frac{1}{3}} \delta_{(5)}, {Q}^{-\frac{2}{3}} \delta_{(10)};\\
&\!\!\!\! {Q}^{\frac{1}{3}}, {Q}^{\frac{1}{3}}, {Q}^{-\frac{2}{3}} \delta_{(5)}; {Q}^{-\frac{5}{3}}, {Q}^{-\frac{2}{3}} \delta_{(5)}, {Q}^{-\frac{2}{3}} \delta_{(5)}, {Q}^{\frac{1}{3}} \delta_{(10)}]\,.\\
\end{array}
\label{eq:26GenMetric}
\end{align}
\endgroup
Identifying
\begin{gather}
\begin{gathered}
g_{mn}^{2^6} = \operatorname{diag}[{Q}^{-\frac{2}{3}}, {Q}^{-\frac{2}{3}}, {Q}^{\frac{1}{3}} \delta_{(5)}]\,,\\
{\left| g_{(4)}^{2^6}\right|}^{-\frac{1}{4}} \coloneqq {Q}^{\frac{1}{6}}  {|g_{(4)}|}^{-\frac{1}{4}} \qquad \Rightarrow \qquad g_{\mu \nu}^{2^6} = {Q}^{-\frac{1}{6}} g_{\mu \nu}\,,
\end{gathered}
\end{gather}
and noting that the EFT vectors get exchanged by the rotation,
\begin{align}
\mathcal{A}_t{}^{\xi \chi} & \mapsto \mathcal{A}_{t, \xi \chi}\,, \qquad \mathcal{A}_{z,\xi \chi} \mapsto \mathcal{A}_z{}^{\xi \chi}\,,
\end{align}
one obtains the background of the $2^6$:
\begingroup
\renewcommand{\arraystretch}{1.5}
\begin{gather}
\begin{array}[t]{rl}
\textrm{d} s^2_{2^6} & = {(HK^{-1})}^{-\frac{2}{3}} ( - \textrm{d} t^2 + \textrm{d} \xi^2 + \textrm{d} \chi^2 ) + {(HK^{-1})}^{\frac{1}{3}} ( \textrm{d} z^2 + \textrm{d} {\vec{w}}^2_{(5)} )\\
	& \qquad  + H^{\frac{1}{3}} K^{\frac{2}{3}} ( \textrm{d} r^2 + r^2 \textrm{d} \theta^2 )\,,
\end{array}\\
A_{(3)} = - H^{-1} K \textrm{d} t \wedge \textrm{d} \xi \wedge \textrm{d} \chi\,,\qquad A_{(6)} = - K^{-1} \sigma \theta \textrm{d} z \wedge \textrm{d} w^1 \wedge \ldots \wedge \textrm{d} w^5\,.
\end{gather}
\endgroup
Further applying the rotations
\begin{gather}
Y^{ab} \leftrightarrow Y_{ab}\,, \qquad Y^{\xi} \leftrightarrow Y_{\xi \chi}, \qquad Y_\xi \leftrightarrow Y^{\xi \chi}\,, \qquad Y^\chi \leftrightarrow Y_\chi\,,\qquad Y^{\chi a} \leftrightarrow Y_{\chi a},
\end{gather}
one obtains the generalised metric
\begingroup
\renewcommand{\arraystretch}{1.5}
\begin{align}
\begin{array}[t]{rl}
\mathcal{M}_{MN} = {Q}^{\frac{1}{2}}  {|g_{(4)}|}^{-\frac{1}{4}} \operatorname{diag} [
& \!\!\!\! Q, 1, \delta_{(5)}; Q^{-1}, Q^{-1} \delta_{(5)}, \delta_{(5)}, \delta_{(10)};\\
& \!\!\!\! Q^{-2}, Q^{-1}, Q^{-1} \delta_{(5)}; 1, \delta_{(5)}, Q^{-1} \delta_{(5)}, Q^{-1} \delta_{(10)};]\,,\\
\end{array}
\label{eq:017GenMetric}
\end{align}
\endgroup
which is consistent with the metric of the $0^{(1,7)}$ if one identifies
\begin{gather}
\begin{gathered}
g_{mn}^{0^{(1,7)}} = \operatorname{diag} [ Q, 1, \delta_{(5)}]\,,\\
{\left| g_{(4)}^{0^{(1,7)}} \right|}^{-\frac{1}{4}} \coloneqq \left[ {Q}^{\frac{1}{2}}  {|g_{(4)}|}^{-\frac{1}{4}} \right] \qquad \Rightarrow \qquad g_{\mu \nu}^{0^{(1,7)}} = {Q}^{-\frac{1}{2}} g_{\mu \nu}\,.
\end{gathered}
\end{gather}
Combining with the generalised vectors that now source the KK-vector and the dual graviton (which does not enter into the background),
\begin{align}
\mathcal{A}_{t,\xi \chi} & \mapsto \mathcal{A}_t{}^\xi\,, \qquad \mathcal{A}_z{}^{\xi \chi} \mapsto \mathcal{A}_{z, \xi}\,,
\end{align}
one obtains the background of the $0^{(1,7)}$:
\begingroup
\renewcommand{\arraystretch}{1.5}
\begin{align}
\begin{array}{rl}
\textrm{d} s^2_{0^{(1,7)}} & = - H^{-1}K \textrm{d} t^2 + {\vec{w}}_{(5)}^2 + \textrm{d} z^2 + \textrm{d} \chi^2 + HK^{-1} {(\textrm{d} \xi - H^{-1}K \textrm{d} t)}^2\\
& \qquad + K (\textrm{d} r^2 + r^2 \textrm{d} \theta^2)\,.
\end{array}
\end{align}
\endgroup
Finally, applying the swap
\begin{align}
Y^{M} \leftrightarrow Y_M\,,
\end{align}
one obtains
\begingroup
\renewcommand{\arraystretch}{1.5}
\begin{align}
\begin{array}[t]{rl}
\mathcal{M}_{MN}= {Q}^{-\frac{1}{2}}  {|g_{(4)}|}^{-\frac{1}{4}} \operatorname{diag} [
& \!\!\!\! {Q}^{-1}, 1, \delta_{(5)}; Q, Q \delta_{(5)}, \delta_{(5)}, \delta_{(10)};\\
& \!\!\!\! {Q}^2, Q, Q \delta_{(5)}; 1, \delta_{(5)}, Q \delta_{(5)}, Q \delta_{(10)}]\,.
\end{array}\label{eq:KK6GenMetric}
\end{align}
\endgroup
One may verify that this is sourced by the background
\begin{gather}
\begin{gathered}
g_{mn}^{\text{KK6}} = \operatorname{diag} [ H^{-1}K, 1, \delta_{(5)}]\,,\\
{\left| g_{(4)}^{\text{KK6}} \right|}^{-\frac{1}{4}} \coloneqq {Q}^{-\frac{1}{2}} {|g_{(4)}|}^{-\frac{1}{4}} \qquad \Rightarrow \qquad g_{\mu \nu}^{\text{KK6}} = {Q}^{\frac{1}{2}} g_{\mu \nu}\,.
\end{gathered}
\end{gather}
The EFT vectors are rotated to
\begin{align}
\mathcal{A}_t{}^\xi & \mapsto \mathcal{A}_{t,\xi}\,, \qquad \mathcal{A}_{z, \xi} \mapsto \mathcal{A}_z{}^\xi\,,
\end{align}
and so, this time, the latter sources a cross-sector coupling. We thus obtain the following background:
\begingroup
\renewcommand{\arraystretch}{1.5}
\begin{align}
\begin{array}[t]{rl}
\textrm{d} s^2 & = - \textrm{d} t^2 + H \left( \textrm{d} r^2 + r^2 \textrm{d} \theta^2 \right) + HK^{-1} \textrm{d} z^2 + H^{-1}K{\left( \textrm{d} \xi - K^{-1} \theta \sigma \textrm{d} z \right)}^2\\
& \qquad + \textrm{d} \chi^2 + \textrm{d} {\vec{w}}^2_{(5)}.
\end{array}
\end{align}
\endgroup
However, focusing on the $\textrm{d}z$ and $\textrm{d} \xi$ terms, one finds that this is a disguised KK6 since
\begin{align}\label{eq:DisguisedKK}
\begin{aligned}
(HK^{-1} + H^{-1} K^{-1} \theta^2 \sigma^2) \textrm{d} z^2 + H^{-1}K \textrm{d} \xi^2 - 2 H^{-1} \theta \sigma \textrm{d} z \textrm{d} \xi = H \textrm{d} \xi^2 + H^{-1} {(\textrm{d} z + \theta \sigma \textrm{d} \xi)}^2.
\end{aligned}
\end{align}
We thus obtain the metric of the KK6:
\begin{align}
\textrm{d} s^2_{\text{KK6}} & = - \textrm{d} t^2 + \textrm{d} \chi^2 + \textrm{d} {\vec{w}}^2_{(5)} + H (\textrm{d} r^2 + r^2 \textrm{d} \theta^2 + \textrm{d} \xi^2 ) + H^{-1} {(\textrm{d} z + \theta \sigma \textrm{d} \xi)}^2.
\end{align}
Note that the harmonic function is smeared in $\xi$ and so this is, more accurately, a \emph{defect} KK monopole.
\subsubsection{Type IIA Reduction}
The reduction to the Type IIA theory is not an independent solution to the section condition but rather a simple re-identification of degrees of freedom in terms of the Type IIA fields. We exploit the fact that the M-theory background, reduced on an isometry $\eta$, is equivalent to a Type IIA background (in the Einstein frame) under the identification
\begin{gather}
\textrm{d} s^2_{\text{M}} = e^{-\frac{\phi}{6}} \textrm{d} s^2_{\text{IIA},\text{E}} + e^{\frac{4\phi}{3}} {(\textrm{d} \eta + A_{(1)})}^2,\\
A_{(3)} = B_{(2)} \wedge \textrm{d} \eta + C_{(3)}\,,\label{eq:MPot}
\end{gather}
where $A_{(p)}$ are the M-theory potentials and $B_{(q)}, C_{(r)}$ are the Type IIA NS-NS and R-R potentials. Splitting the M-theory section into $y^m = (y^\mathfrak{m}, \eta)$, with the Type IIA coordinates $y^{\mathfrak{m}} \equiv Y^{\mathfrak{m}}$ indexed by $\mathfrak{m} = 1, \ldots, 6$, the reduction induces a decomposition of the generalised coordinates according to
\begin{align}\label{eq:IIACoords}
Y^M = (Y^{\mathfrak{m}}, Y^\eta; Y_{\mathfrak{m} \mathfrak{n}}, Y_{\mathfrak{m} \eta}; Y_{\mathfrak{m}}, Y_\eta; Y^{\mathfrak{m} \mathfrak{n}}; Y^{\mathfrak{m} \eta})\,.
\end{align}
As usual, the $g_{\eta \eta}$ component gives the ten-dimensional dilaton. We also see that the external metric is rescaled by $e^{-\frac{\phi}{6}}$, resulting in
\begin{align}\label{eq:IIAExtFrame}
g^{\text{M}}_{\mu \nu} = e^{-\frac{\phi}{6}} {\mathfrak{g}}^{\text{A}}_{\mu \nu} \qquad \Rightarrow \qquad {\left|g^{\text{M}}_{(4)}\right|}^{-\frac{1}{4}} = e^{\frac{\phi}{6}} {\left|{\mathfrak{g}}^{\text{A}}_{(4)} \right|}^{-\frac{1}{4}}\,.
\end{align}
Note that $x^\mu = (t,r,\theta z)$ still indexes the same coordinates on the external space as the M-theory section and the base external metric takes on the same numerical values as in the M-theory frame:
\begin{align}\label{eq:IIAExt}
{\mathfrak{g}}_{\mu \nu} = \operatorname{diag} [ - {(HK^{-1})}^{\frac{1}{2}}, {(HK)}^{\frac{1}{2}}, r^2 {(HK)}^{\frac{1}{2}}, {(HK^{-1})}^{\frac{1}{2}}] \equiv g_{\mu \nu}\,.
\end{align}
We further read off that the determinant of the internal metric decompose according to
\begin{align}
g^{\text{M}}_{(7)} = e^{\frac{\phi}{3}} {\mathfrak{g}}^{\text{A}}_{(6)},
\end{align}
and so the generalised metric decomposes to\footnote{If one wishes to work in the string frame, the analogous decomposition is given by
\begin{align}
\mathllap{\mathcal{M}_{MN}} = {\left| \mathfrak{g}_{(4)}\right|}^{-\frac{1}{4}}  \operatorname{diag} [ {\mathfrak{g}}_{\mathfrak{m} \mathfrak{n}}, e^{2\phi}; e^{2\phi} {\mathfrak{g}}^{\mathfrak{m} \mathfrak{n}, \mathfrak{p} \mathfrak{q}}, {\mathfrak{g}}^{\mathfrak{m} \mathfrak{n}}; e^{4\phi} {\mathfrak{g}}_{(6)}^{-1} {\mathfrak{g}}^{\mathfrak{m} \mathfrak{n}}, e^{2\phi} {\mathfrak{g}}_{(6)}^{-1}; e^{2\phi} {\mathfrak{g}}_{(6)}^{-1} {\mathfrak{g}}_{\mathfrak{m} \mathfrak{n}, \mathfrak{p} \mathfrak{q}}, e^{4\phi} {\mathfrak{g}}_{(6)}^{-1}  {\mathfrak{g}}_{\mathfrak{m} \mathfrak{n}} ]\,,
\end{align}
which follows from taking the decomposition
\begin{align}
\textrm{d} s^2_{\text{M}} = e^{-\frac{2\phi}{3}} \textrm{d} s^2_{\text{IIA,s}} + e^{\frac{4\phi}{3}} {(\textrm{d} \eta + A_{(1})}^2
\end{align}
instead.
}
\begin{align}\label{eq:IIAGenMetric}
\phantom{\mathcal{M}_{MN}} &
\begin{alignedat}{2}
\mathllap{\mathcal{M}_{MN}} = \smash{\underbrace{e^{\frac{\phi}{6}} {\left|{\mathfrak{g}}^{\text{A}}_{(4)} \right|}^{-\frac{1}{4}}}_{\text{from } {| {\mathfrak{g}}_{(4)}^{\text{M}}|}^{-\frac{1}{4}}} \operatorname{diag} [}
& e^{-\frac{\phi}{6}} {\mathfrak{g}}_{\mathfrak{m} \mathfrak{n}}^{\text{A}}, e^{\frac{4\phi}{3}};\\
& e^{\frac{\phi}{3}} g^{\mathfrak{m} \mathfrak{n}, \mathfrak{p} \mathfrak{q}}_{\text{A}}, e^{-\frac{7\phi}{6}} {\mathfrak{g}}^{\mathfrak{m} \mathfrak{n}}_{\text{A}};\\
& e^{-\frac{\phi}{6}} {\mathfrak{g}}^{\text{A}}_{(6)}{}^{-1} {\mathfrak{g}}^{\mathfrak{m} \mathfrak{n}}_{\text{A}}, e^{-\frac{5\phi}{3}} {\mathfrak{g}}^{\text{A}}_{(6)}{}^{-1};\\
& e^{-\frac{2\phi}{3}} {\mathfrak{g}}^{\text{A}}_{(6)}{}^{-1} {\mathfrak{g}}^{\text{A}}_{\mathfrak{m} \mathfrak{n}, \mathfrak{p} \mathfrak{q}}, e^{\frac{5\phi}{6}} {\mathfrak{g}}^{\text{A}}_{(6)}{}^{-1} {\mathfrak{g}}^{\text{A}}_{\mathfrak{m} \mathfrak{n}} ],
\end{alignedat}
\end{align}
where the ordering of components follows that of the coordinates \eqref{eq:IIACoords}. The EFT vector likewise decomposes to 
\begin{align}
\mathcal{A}_\mu{}^M \rightarrow  (\mathcal{A}_{\mu}{}^{\mathfrak{m}}, \mathcal{A}_{\mu}{}^{\eta}; \mathcal{A}_{\mu, \mathfrak{m} \mathfrak{n}}, \mathcal{A}_{\mu, \mathfrak{m} \eta}; \mathcal{A}_{\mu, \mathfrak{m}}, \mathcal{A}_{\mu, \eta}; \mathcal{A}_{\mu}{}^{\mathfrak{m} \mathfrak{n}}, \mathcal{A}_{\mu}{}^{\mathfrak{m} \eta})\,.
\end{align}
As before, the $\mathcal{A}_{\mu}{}^{\mathfrak{m}}$ components sources the KK-vector of the (now) 4+6 split and the $\mathcal{A}_{\mu, \mathfrak{m}}$ is related to the dual graviton. Of the remaining components, the R-R potentials $C_{(1)}, C_{(3)}, C_{(5)}$ and $C_{(7)}$ are encoded in the components $\mathcal{A}_{\mu}{}^{\eta}, \mathcal{A}_{\mu, \mathfrak{m} \mathfrak{n}}, \mathcal{A}_{\mu}{}^{\mathfrak{m} \mathfrak{n}}$ and $\mathcal{A}_{\mu \eta}$ respectively (where the latter two are to be dualised on the internal space) and the NS-NS potentials $B_{(2)}$ and $B_{(6)}$ are held in $\mathcal{A}_{\mu, \mathfrak{m} \eta}$ and $\mathcal{A}_{\mu}{}^{\mathfrak{m} \eta}$ respectively.\par
In order to tabulate the brane configurations that we obtain, it will be convenient to split the five $w^a$ coordinates into $w^a = (u^{\textrm{a}}, v)$ with $\textrm{a} = 1, \ldots, 4$. The results are summarised in Table~\ref{tab:NonGeomIIA}.\par
\begin{table}[t]
\centering
\begin{tabulary}{\textwidth}{CLLCCCCCCCCCC}
\toprule
& & & & & & & & & & \multicolumn{2}{c}{$w^a$} & \\
\cmidrule{11-12}
& Parent & & $t$ & $r$ & $\theta$ & $z$ & & $\xi$ & $\chi$ & $u^{\textrm{a}}$ & $v$ & \\
\cmidrule{2-2}\cmidrule{4-7}\cmidrule{9-12}
& \multirow{2}{*}{$5^3$} & $5_2^2\text{A}$ & $\ast$ & $\bullet$ & $\bullet$ & $\circ$ & & $\times$ & $\circ$ & $\ast$ & $\ast$\\
& & $4_3^3$ & $\ast$ & $\bullet$ & $\bullet$ & $\circ$ & & $\circ$ & $\circ$ & $\ast$ & $\times$\\
& \multirow{2}{*}{$2^6$} & $2_3^5$ & $\ast$ & $\bullet$ & $\bullet$ & $\circ$ & & $\ast$ & $\ast$ & $\circ$ & $\times$\\
& & $1_4^6\text{A}$ & $\ast$ & $\bullet$ & $\bullet$ & $\circ$ & & $\times$ & $\ast$ & $\circ$ & $\circ$\\
& \multirow{2}{*}{$0^{(1,7)}$} & $0_4^{(1,6)}\text{A}$ & $\ast$ & $\bullet$ & $\bullet$ & $\circ$ & & $\odot$ & $\circ$ & $\circ$ & $\times$\\
& & $0_3^7$ & $\ast$ & $\bullet$ & $\bullet$ & $\circ$ & & $\times$ & $\circ$ & $\circ$ & $\circ$\\
& \multirow{2}{*}{KK6} & $6_3^1$ & $\ast$ & $\bullet$ & $\bullet$ & $\circ$ & & $\times$ & $\ast$ & $\ast$ & $\ast$\\
& & KK5A & $\ast$ & $\bullet$ & $\bullet$ & $\odot$ & & $\circ$ & $\ast$ & $\ast$ & $\times$\\
\bottomrule
\end{tabulary}
\caption[The configurations of the Type IIA branes that we consider.]{The configurations of the Type IIA branes that we consider. Note that the coordinates heading the columns are those of the M-theory section. A cross $\times$ denotes the direction that is being reduced on i.e.\ the choice of isometry $\eta$ that leads to the indicated brane.}
\label{tab:NonGeomIIA}
\end{table}
Note that since some of the M-theory solutions are symmetric under certain coordinate transformations, these are not the only reductions that we could have done to obtain the Type IIA branes. For example, the generalised metrics of the $5^3$ and $2^6$ are invariant under the exchange $\xi \leftrightarrow \chi$ and so we could have obtained the $5_2^2\text{A}$ and $1_4^6\text{A}$ by reducing along $\chi$ instead of $\xi$ (although this further requires a re-identification of $\mathcal{A}_{\mu}{}^{\eta \mathfrak{m}} \sim - B_{(6)}$ and $\mathcal{A}_{\mu, \eta \mathfrak{m}} \sim - B_{(2)}$). Likewise, the role of $\chi$ is indistinguishable from any of the $w^a$ in the generalised metrics of the $0^{(1,7)}$ and KK6 and so we may have equally swapped $\chi$ with any one of the $w^a$ coordinates (which we nominally called $v$ in Table~\ref{tab:NonGeomIIA}) and obtained a valid reduction to the KK6A and $0_4^{(1,6)}\text{A}$ by reducing on $\chi$ instead of $v$ (again with a suitable re-identification of the potentials to $\mathcal{A}_{\mu}{}^{\eta} \sim - C_{(1)}$ and $\mathcal{A}_{\mu,\eta} \sim - C_{(7)}$). Nonetheless, the choice given in Table~\ref{tab:NonGeomIIA} is the most symmetric choice of reductions.\par
Since all the reductions given above are along $\xi$ or $v$, we work through two examples in detail to illustrate these two reductions. The first is the reduction of the $5^3$ generalised metric along $\eta = \xi$. We begin by noting that the external metric scales as
\begin{align}
{\left| g_{(4)}^{5^3}\right|}^{-\frac{1}{4}} & = {(HK^{-1})}^{-\frac{1}{6}} {|g_{(4)}|}^{-\frac{1}{4}} = {(HK^{-1})}^{-\frac{1}{6}} {|{\mathfrak{g}}_{(4)}|}^{-\frac{1}{4}} \coloneqq e^{\frac{\phi}{6}} {\left| {\mathfrak{g}}_{(4)}^{5_2^2\text{A}} \right|}^{-\frac{1}{4}}\,,
\end{align}
where we have used \eqref{eq:IIAExtFrame} and \eqref{eq:IIAExt}. The dilaton is obtained from the $\mathcal{M}_{\eta \eta} = \mathcal{M}_{\xi\xi}$ component of the generalised metric in the $5^3$ frame \eqref{eq:53GenMetric}:
\begin{align}
e^{\frac{4\phi}{3}} = {(HK^{-1})}^{\frac{2}{3}}\,.
\end{align}
We thus obtain the scaling of the external metric
\begin{align}
{\left|{\mathfrak{g}}^{5^2_2\text{A}}_{(4)}\right|}^{-\frac{1}{4}} & = {(HK^{-1})}^{-\frac{1}{4}} {| {\mathfrak{g}}_{(4)} |}^{-\frac{1}{4}} \qquad \Rightarrow \qquad {\mathfrak{g}}_{\mu \nu} = {(HK^{-1})}^{\frac{1}{4}} {\mathfrak{g}}_{\mu \nu}\,.
\end{align}
One may verify that the rest of the $5^3$ generalised metric \eqref{eq:53GenMetric} is sourced by
\begin{align}
{\mathfrak{g}}_{\mathfrak{m} \mathfrak{n}}^{5^2_2\text{A}} & = \operatorname{diag} \left[{(HK^{-1})}^{\frac{3}{4}}, {(HK^{-1})}^{-\frac{1}{4}} \delta_{(5)} \right],
\end{align}
when fitted to the form \eqref{eq:IIAGenMetric}. Here, the reduced internal space is spanned by $y^\mathfrak{m} =(\chi, w^a)$. Since the non-vanishing components of the EFT vector in this frame $\mathcal{A}_t{}^{\xi \chi}$ and $\mathcal{A}_{z,\xi \chi}$ lie along the reduction direction $\eta = \xi$, they must source the NS-NS potentials, giving the background of the $5^2_2\text{A}$ in the Einstein frame:
\begingroup
\renewcommand{\arraystretch}{1.5}
\begin{gather}
\begin{gathered}
\begin{array}[t]{rl}
\textrm{d} s^2_{5^2_2\text{A},\text{E}} & = {(HK^{-1})}^{-\frac{1}{4}} \left( -\textrm{d} t^2 + \textrm{d} {\vec{w}}^2_{(5)} \right) + {(HK^{-1})}^{\frac{3}{4}} \left( \textrm{d} z^2 + \textrm{d} \chi^2 \right)\\
& \qquad + H^{\frac{3}{4}} K^{\frac{1}{4}} \left(\textrm{d} r^2 + r^2 \textrm{d} \theta^2 \right)\,,\qquad e^{2(\phi -\phi_0)} = HK^{-1}\,,\\
\end{array}\\
B_{(2)} = - K^{-1} \theta \sigma \textrm{d} z \wedge \textrm{d} \chi\,, \qquad B_{(6)} = - H^{-1} K \textrm{d} t^2 \wedge \textrm{d} w^1 \wedge \ldots \wedge \textrm{d} w^5\,.
\end{gathered}
\end{gather}
\endgroup
The second reduction of the $5^3$ is along $\eta = v \equiv w^5$ and so the coordinates of the Type IIA internal space are $y^{\mathfrak{m}} = (\xi, \chi, u^{\text{a}})$. We begin by rewriting the generalised metric of the $5^3$, given in \eqref{eq:53GenMetric}, adapted to this coordinate splitting: 
\begin{align}\label{eq:SplitGenMetric}
\begin{aligned}
\mathcal{M}_{MN} = {\left|g_{(4)}^{5^3}\right|}^{-\frac{1}{4}} \operatorname{diag}[
& {Q}^{\frac{2}{3}}, {Q}^{\frac{2}{3}}, {Q}^{-\frac{1}{3}} \delta_{(4)}, {Q}^{-\frac{1}{3}};\\
& {Q}^{-\frac{4}{3}}, {Q}^{-\frac{1}{3}} \delta_{(4)}, {Q}^{-\frac{1}{3}}, {Q}^{-\frac{1}{3}} \delta_{(4)}, {Q}^{-\frac{1}{3}},  {Q}^{\frac{2}{3}} \delta_{(6)}, {Q}^{\frac{2}{3}} \delta_{(4)};\\
& {Q}^{-\frac{1}{3}}, {Q}^{-\frac{1}{3}}, {Q}^{\frac{2}{3}} \delta_{(4)}, {Q}^{\frac{2}{3}};\\
& {Q}^{\frac{5}{3}}, {Q}^{\frac{2}{3}} \delta_{(4)}, {Q}^{\frac{2}{3}}, {Q}^{\frac{2}{3}} \delta_{(4)}, {Q}^{\frac{2}{3}},  {Q}^{-\frac{1}{3}} \delta_{(6)}, {Q}^{-\frac{1}{3}} \delta_{(4)}]\,.
\end{aligned}
\end{align}
We now proceed as before and examine the prefactor and $\mathcal{M}_{vv}$ component to obtain the dilaton and relative scaling:
\begin{gather}
\begin{gathered}
{\left| g_{(4)}^{5^3} \right|}^{-\frac{1}{4}} = {(HK^{-1})}^{-\frac{1}{6}} {|{\mathfrak{g}}_{(4)}|}^{-\frac{1}{4}} \coloneqq e^{\frac{\phi}{6}} {\left|{\mathfrak{g}}_{(4)}^{4_3^3} \right|}^{-\frac{1}{4}}\,,\\
{(HK^{-1})}^{-\frac{1}{3}} = e^{\frac{4\phi}{3}}\,.
\end{gathered}
\end{gather}
These can be solved to give
\begin{gather}
\begin{gathered}
e^{2(\phi- \phi_0)} = {(HK^{-1})}^{-\frac{1}{2}}\,,\\
{\left| {\mathfrak{g}}_{(4)}^{4_3^3} \right|}^{-\frac{1}{4}} = {(HK^{-1})}^{-\frac{1}{8}} {|{\mathfrak{g}}_{(4)}|}^{-\frac{1}{4}} \qquad \Rightarrow \qquad {\mathfrak{g}}_{\mu \nu}^{4_3^3} = {(HK^{-1})}^{\frac{1}{8}} {\mathfrak{g}}_{\mu \nu}\,.
\end{gathered}
\end{gather}
One may verify that the rest of the generalised background is sourced, conforming to the Type IIA decomposition \eqref{eq:IIAGenMetric}, by
\begin{align}
{\mathfrak{g}}^{4^3_3}_{\mathfrak{m} \mathfrak{n}} & = \operatorname{diag} [ {(HK^{-1})}^{\frac{5}{8}}, {(HK^{-1})}^{\frac{5}{8}}, {(HK^{-1})}^{-\frac{3}{8}} \delta_{(4)}]\,.
\end{align}
Since the direction being reduced on is not contained in the EFT vector, it reduces trivially to source the 5-form and 3-form R-R potentials. We thus obtain the background of the $4_3^3$ in the Einstein frame:
\begin{gather}
\begin{gathered}
\textrm{d}s^2_{4^3_3,\text{E}} = \begingroup \renewcommand{\arraystretch}{1.5}
\begin{array}[t]{l}
{(HK^{-1})}^{-\frac{3}{8}} \left( -\textrm{d} t^2 + \textrm{d} {\vec{w}}^2_{(4)} \right) + {(HK^{-1})}^{\frac{5}{8}} \left( \textrm{d} z^2 + \textrm{d} \xi^2 + \textrm{d} \chi^2 \right)\\
\qquad + H^{\frac{5}{8}} K^{\frac{3}{8}} ( \textrm{d} r^2 + r^2 \textrm{d} \theta^2)\,, \qquad e^{2(\phi - \phi_0)} = {(HK^{-1})}^{-\frac{1}{2}}\,,\\
\end{array}
\endgroup\\
C_{(3)} = -K^{-1} \theta \sigma \textrm{d} z \wedge \textrm{d} \xi \wedge \textrm{d} \chi\,, \qquad C_{(5)} = - H^{-1} K \textrm{d} t^2 \wedge \textrm{d} w^1 \wedge \ldots \wedge \textrm{d} w^4\,.\\
\end{gathered}
\end{gather}
The remaining reductions of the M-theory solutions are done in exactly the same fashion as described above. The only complication is that the reduction of the KK6 reduces to the KK5A, along $\eta=v$, to give the non-canonical form
\begingroup
\renewcommand{\arraystretch}{1.5}
\begin{align}
\begin{array}[t]{rl}
\textrm{d} s^2_{\text{KK5A}} & = - \textrm{d} t^2 + \textrm{d} \chi^2 + \textrm{d} {\vec{u}}^2_{(4)} + H (\textrm{d} r^2 + r^2 \textrm{d} \theta^2) + HK^{-1} \textrm{d}z^2\\
& \qquad + H^{-1}K {(\textrm{d} \xi - K^{-1} \theta \sigma) \textrm{d} z}^2,
\end{array}
\end{align}
\endgroup
and one needs to apply the same trick \eqref{eq:DisguisedKK} as before to obtain the canonical form of the (defect) KK-monopole.
\subsection{Type IIB Section}
The generalised metric in the Type IIB section, in the absence of internal potentials, is given by
\begin{align}
\mathcal{M}_{MN} = {|{\mathrm{g}}_{(4)}|}^{-\frac{1}{4}} \operatorname{diag} [ {\mathrm{g}}_{\mathrm{m} \mathrm{n}}; {\mathrm{g}}^{\mathrm{m} \mathrm{n}} \gamma^{\alpha \beta}; {\mathrm{g}}_{(6)}^{-1} {\mathrm{g}}_{\mathrm{m} \mathrm{k} \mathrm{p}, \mathrm{n} \mathrm{k} \mathrm{q}}; {\mathrm{g}}^{-1}_{(6)} {\mathrm{g}}_{\mathrm{m} \mathrm{n}} \gamma_{\alpha \beta}; {\mathrm{g}}^{-1}_{(6)} {\mathrm{g}}^{\mathrm{m} \mathrm{n}} ]\,,
\end{align}
where $\mathrm{m}, \mathrm{n} = 1, \ldots, 6$ index the Type IIB section and $\alpha, \beta =1,2$ are $\operatorname{SL}(2)$ indices. Accordingly, ${\mathrm{g}}_{\mathrm{m} \mathrm{n}}, {\mathrm{g}}_{\mu \nu}$ and $\gamma_{\alpha \beta}$ are the metric on the internal space, external space and torus respectively. The external metric is again taken to be of the same form as the M-theory section (though we have modified the notation to conform to the other fields in this section,
\begin{align}
{\mathrm{g}}_{\mu \nu} = \operatorname{diag} [ - {(HK^{-1})}^{-\frac{1}{2}}, {(HK)}^{\frac{1}{2}}, r^2 {(HK)}^{\frac{1}{2}}, {(HK^{-1})}^{\frac{1}{2}} ] \equiv g_{\mu \nu}\,,
\end{align}
and $\gamma_{\alpha \beta}$ is parametrised by the axio-dilaton $\tau = A_{(0)} + i e^{-\phi}$ in the usual fashion
\begin{align}
\gamma_{\alpha \beta} = \frac{1}{\operatorname{Im} \tau} \begin{pmatrix} {|\tau|}^2 & \operatorname{Re} \tau\\ \operatorname{Re} \tau & 1 \end{pmatrix}\,.
\end{align}
Additionally, we have defined ${\mathrm{g}}_{\mathrm{m} \mathrm{k} \mathrm{p}, \mathrm{n} \mathrm{l} \mathrm{q}} \coloneqq {\mathrm{g}}_{\mathrm{m} [ \mathrm{n}|} {\mathrm{g}}_{\mathrm{k}| \mathrm{l}|} {\mathrm{g}}_{\mathrm{p}|\mathrm{q}]}$. The EFT vector splits according to this coordinate decomposition as 
\begin{align}
\mathcal{A}_{\mu}{}^M \rightarrow ( \mathcal{A}_{\mu}{}^{\mathrm{m}}, \mathcal{A}_{\mu, \mathrm{m} \alpha}, \mathcal{A}_{\mu}{}^{\mathrm{m} \mathrm{k} \mathrm{p}}, \mathcal{A}_{\mu}{}^{\mathrm{m} \alpha}, \mathcal{A}_{\mu, \mathrm{m}})\,.
\end{align}
As always, the $\mathcal{A}_{\mu}{}^{\mathrm{m}}$ component is identified as the KK-vector and $\mathcal{A}_{\mu \mathrm{m}}$ is the dual graviton. The $\operatorname{SL}(2)$ index $\alpha$ distinguishes between $C_{(2)}/B_{(2)}$ in $\mathcal{A}_{\mu \mathrm{m} \alpha}$ and $C_{(6)}/ B_{(6)}$ in $\mathcal{A}_{\mu}{}^{\mathrm{m} \alpha}$. In particular, when $\alpha = 1$, the potential is of R-R type and when $\alpha =2$, the potential is of NS-NS type. Finally, the $\mathcal{A}_{\mu}{}^{\mathrm{m} \mathrm{k} \mathrm{p}}$ component, once dualised on the internal space, sources the self-dual 4-form potential $C_{(4)}$.\par
In order to tabulate the brane configurations that we obtain, it will be necessary to split the five $w^a$ coordinates into $w^a = (\omega, {\bar{w}}^{\bar{a}}, {\ubar{w}}_{\ubar{a}})$ with $\bar{a} = 1,2$ and $\ubar{a} = 1,2$. The results are summarised in Table~\ref{tab:NonGeomIIB}. 
\begin{table}[t]
\centering
\begin{tabulary}{\textwidth}{LCCCCCCCCC}
\toprule
& & & & & & & & $w^a$\\
\cmidrule{8-10}
& $t$ & $r$ & $\theta$ & $z$ & & $\zeta$ & $\omega$ & ${\bar{w}}^{\bar{a}}$ & ${\ubar{w}}^{\ubar{a}}$\\
\cmidrule{2-5}\cmidrule{7-10}
$5_2^2\text{B}$ & $\ast$ & $\bullet$ & $\bullet$ & $\circ$ & & $\circ$ & $\ast$ & $\ast$ & $\ast$\\
$5_3^2$ & $\ast$ & $\bullet$ & $\bullet$ & $\circ$ & & $\circ$ & $\ast$ & $\ast$ & $\ast$\\
$3_3^4$ & $\ast$ & $\bullet$ & $\bullet$ & $\circ$ & & $\circ$ & $\ast$ & $\circ$ & $\ast$\\
$1_3^6$ & $\ast$ & $\bullet$ & $\bullet$ & $\circ$ & & $\circ$ & $\ast$ & $\circ$ & $\circ$\\
$1_4^6\text{B}$ & $\ast$ & $\bullet$ & $\bullet$ & $\circ$ & & $\circ$ & $\ast$ & $\circ$ & $\circ$\\
$0_4^{(1,6)}\text{B}$ & $\ast$ & $\bullet$ & $\bullet$ & $\circ$ & & $\circ$ & $\circ$ & $\circ$ & $\circ$\\
KK5B & $\ast$ & $\bullet$ & $\bullet$ & $\odot$ & & $\ast$ & $\circ$ & $\ast$ & $\ast$\\
\bottomrule
\end{tabulary}
\caption{The configurations of the Type IIB branes that we consider.}
\label{tab:NonGeomIIB}
\end{table}
In order to identify the relation between the M-theory section and Type IIB section, we examine the $5^3$ generalised metric. Recall the internal coordinates of the M-theory section were $(\xi, \chi, w^a)$. Of these, the five $w^a$ coordinates enter directly into the Type IIB section but the remaining two coordinates $(\xi, \chi)$ become, loosely speaking, the $\operatorname{SL}(2)$ indices that pick out an M2 wrapping direction. In particular, denoting the six coordinates of the Type IIB section as $Y^{\mathrm{m}} \equiv y^{\mathrm{m}} = ( Y_{\zeta 2}, Y^a)$ and generating the remaining generalised coordinates $Y^M$, the correspondence between the M-theory and Type IIB coordinates are given by
\begin{gather}
Y^{\xi} \equiv Y^{\zeta 1}\,, \quad Y_{\xi} \equiv Y_{\zeta 1}\,, \quad Y^{\chi} \equiv Y_{\zeta}\,, \quad Y^{\chi} \equiv Y^{\zeta}\,, \quad Y^{\xi \chi} \equiv Y^{\zeta 2}\,, \qquad Y_{\xi \chi} \equiv Y_{\zeta 2}\label{eq:MIIB}.\\
Y^{\xi a} \equiv Y_{a 1}\,, \qquad Y_{\xi a} \equiv Y^{a 1}\,, \qquad Y^{\chi a} \equiv Y^{a 2}\,, \qquad Y_{\chi a} \equiv Y_{a 2}\,,\\
Y^a \equiv Y^a\,, \qquad Y^{ab} \equiv Y^{abc}\,, \qquad Y_{ab} \equiv Y^{\zeta a b}\,.
\end{gather}
This gives the generalised metric
\begingroup
\renewcommand{\arraystretch}{1.5}
\begin{align}
\mathcal{M}_{MN} & = {\left| {\mathrm{g}}_{(4)} \right|}^{-\frac{1}{4}} \operatorname{diag} [
\begin{array}[t]{l}
\!\! {Q}^{\frac{1}{2}}, {Q}^{-\frac{1}{2}} \delta_{(5)}; {Q}^{-\frac{1}{2}}, {Q}^{\frac{1}{2}} \delta_{(5)}, {Q}^{-\frac{3}{2}}, {Q}^{-\frac{1}{2}} \delta_{(5)};\\
\!\! {Q}^{\frac{1}{2}} \delta_{(10)}, {Q}^{-\frac{1}{2}} \delta_{(10)};\\
\!\! {Q}^{\frac{1}{2}}, {Q}^{-\frac{1}{2}} \delta_{(5)}, {Q}^{\frac{3}{2}}, {Q}^{\frac{1}{2}} \delta_{(5)}; {Q}^{-\frac{1}{2}}, {Q}^{\frac{1}{2}} \delta_{(5)}]\\
\end{array}\\
& = {Q}^{-\frac{1}{4}}  {|{\mathrm{g}}_{(4)}|}^{-\frac{1}{4}} \operatorname{diag} [
\begin{array}[t]{l}
\!\! {Q}^{\frac{3}{4}}, {Q}^{-\frac{1}{4}} \delta_{(5)}; {Q}^{-\frac{1}{4}}, {Q}^{\frac{3}{4}} \delta_{(5)}, {Q}^{-\frac{5}{4}}, {Q}^{-\frac{1}{4}} \delta_{(5)};\\
\!\! {Q}^{\frac{3}{4}} \delta_{(10)}, {Q}^{-\frac{1}{4}} \delta_{(10)};\\
\!\! {Q}^{\frac{3}{4}}, {Q}^{-\frac{1}{4}} \delta_{(5)}, {Q}^{\frac{7}{4}}, {Q}^{\frac{3}{4}} \delta_{(5)}; {Q}^{-\frac{1}{4}}, {Q}^{\frac{3}{4}} \delta_{(5)}]\,,\\
\end{array}\label{eq:522BGenMetric}
\end{align}
\endgroup
which is consistent with the background of the $5_2^2\text{B}$ if one identifies
\begin{align}
\begin{array}{lcl}
{\mathrm{g}}_{\mathrm{m}\mathrm{n}}^{5_2^2\text{B}} = \operatorname{diag} [ {(HK^{-1})}^{\frac{3}{4}}, {(HK^{-1})}^{-\frac{1}{4}} \delta_{(5)}]\,, & &\\
{\left| {\mathrm{g}}_{(4)}^{5^2_2\text{B}} \right|}^{-\frac{1}{4}} \coloneqq  {(HK^{-1})}^{-\frac{1}{4}}  {|{\mathrm{g}}_{(4)}|}^{-\frac{1}{4}}  & \Rightarrow  & {\mathrm{g}}_{\mu \nu}^{5^2_2\text{B}} = {(HK^{-1})}^{\frac{1}{4}} {\mathrm{g}}_{\mu \nu}\,,\\
\gamma_{\alpha \beta} = \operatorname{diag} [ {(HK^{-1})}^{- \frac{1}{2}}, {(HK^{-1})}^{\frac{1}{2}}] & \Rightarrow & \tau = i {(HK^{-1})}^{-\frac{1}{2}}\,.
\end{array}
\end{align}
Applying \eqref{eq:MIIB} on the potentials \eqref{eq:53EFTVec} to identify the direction that the EFT vector points in the Type IIB frame, we have
\begin{align}
\mathcal{A}_t{}^{\xi \chi} \rightarrow \mathcal{A}_t{}^{\zeta 2} = - H^{-1} K\,,  \qquad \mathcal{A}_{z, \xi \chi} \rightarrow \mathcal{A}_{z, \zeta 2} = - K^{-1} \theta \sigma\,,
\end{align}
of which the first sources $B_{(6)}$ (upon being dualised on the internal space) and the latter sources $B_{(2)}$. We thus obtain the background of the $5^2_2\text{B}$ in the Einstein frame:
\begingroup
\renewcommand{\arraystretch}{1.5}
\begin{gather}
\begin{gathered}
\begin{array}[t]{rl}
\textrm{d} s^2_{5^2_2\text{B},\text{E}} & = {(HK^{-1})}^{-\frac{1}{4}} \left( -\textrm{d} t^2 + \textrm{d} {\vec{w}}^2_{(5)} \right) + {(HK^{-1})}^{\frac{3}{4}} \left( \textrm{d} z^2 + \textrm{d} \zeta^2 \right)\\
& \qquad + H^{\frac{3}{4}} K^{\frac{1}{4}} \left(\textrm{d} r^2 + r^2 \textrm{d} \theta^2 \right), \qquad e^{2(\phi -\phi_0)} = HK^{-1}\,,\\
\end{array}\\
B_{(2)} = - K^{-1} \theta \sigma \textrm{d} z \wedge \textrm{d} \zeta, \qquad B_{(6)} = - H^{-1} K \textrm{d} t^2 \wedge \textrm{d} w^1 \wedge \ldots \wedge \textrm{d} w^5\,.\
\end{gathered}
\end{gather}
\endgroup
Alternatively, we could have reduced the $5^3$ to the $5_3^2$---the S-dual of the $5_2^2\text{B}$. The coordinate identifications that we make between the M-theory and Type IIB coordinates are essentially the same as for the $5_2^2\text{B}$ except for with the $\operatorname{SL}(2)$ index exchanged: $y^{\mathfrak{m}} = (Y_{\zeta 1}, Y^a)$. The generalised metric that one obtains is then
\begingroup
\renewcommand{\arraystretch}{1.5}
\begin{align}
\mathcal{M}_{MN} & = {\left|{\mathrm{g}}_{(4)} \right|}^{-\frac{1}{4}} \operatorname{diag} [
\begin{array}[t]{l}
\!\! {Q}^{\frac{1}{2}}, {Q}^{-\frac{1}{2}} \delta_{(5)}; {Q}^{-\frac{3}{2}}, {Q}^{-\frac{1}{2}} \delta_{(5)}, {Q}^{-\frac{1}{2}}, {Q}^{\frac{1}{2}} \delta_{(5)};\\
\!\! {Q}^{\frac{1}{2}} \delta_{(10)}, {Q}^{-\frac{1}{2}} \delta_{(10)};\\
\!\! {Q}^{\frac{3}{2}}, {Q}^{\frac{1}{2}} \delta_{(5)}, {Q}^{\frac{1}{2}}, {Q}^{-\frac{1}{2}} \delta_{(5)}; {Q}^{-\frac{1}{2}}, {Q}^{\frac{1}{2}} \delta_{(5)}]
\end{array}\\
& = {Q}^{-\frac{1}{4}}  {|{\mathrm{g}}_{(4)}|}^{-\frac{1}{4}} \operatorname{diag} [
\begin{array}[t]{l}
\!\! {Q}^{\frac{3}{4}}, {Q}^{-\frac{1}{4}} \delta_{(5)}; {Q}^{-\frac{5}{4}}, {Q}^{-\frac{1}{4}} \delta_{(5)}, {Q}^{-\frac{1}{4}}, {Q}^{\frac{3}{4}} \delta_{(5)};\\
\!\! {Q}^{\frac{3}{4}} \delta_{(10)}, {Q}^{-\frac{1}{4}} \delta_{(10)};\\
\!\! {Q}^{\frac{7}{4}}, {Q}^{\frac{3}{4}} \delta_{(5)}, {Q}^{\frac{3}{4}}, {Q}^{-\frac{1}{4}} \delta_{(5)}; {Q}^{-\frac{1}{4}}, {Q}^{\frac{3}{4}} \delta_{(5)}]\,,
\end{array}\label{eq:523GenMetric}
\end{align}
\endgroup
which is consistent with the background of the $5_3^2$ if one identifies
\begin{align}
\begin{array}{lcl}
{\mathrm{g}}_{\mathrm{m} \mathrm{n}}^{5_3^2} = \operatorname{diag} [ {(HK^{-1})}^{\frac{3}{4}}, {(HK^{-1})}^{-\frac{1}{4}}\delta_{(5)}]\,, & &\\
{\left| {\mathrm{g}}_{(4)}^{5^2_3} \right|}^{-\frac{1}{4}} \coloneqq {(HK^{-1})}^{-\frac{1}{4}}  {|{\mathrm{g}}_{(4)}|}^{-\frac{1}{4}} & \Rightarrow & {\mathrm{g}}_{\mu \nu}^{5^2_3} = {(HK^{-1})}^{\frac{1}{4}} {\mathrm{g}}_{\mu \nu}\,,\\
\gamma_{\alpha \beta} = \operatorname{diag} [ {(HK^{-1})}^{\frac{1}{2}}, {(HK^{-1})}^{-\frac{1}{2}}] & \Rightarrow & \tau = i {(HK^{-1})}^{\frac{1}{2}}\,.
\end{array}
\end{align}
Additionally, in the identification of coordinates used here, we have
\begin{align}
\mathcal{A}_t{}^{\xi \chi} \rightarrow \mathcal{A}_t{}^{\zeta 1} = - H^{-1} K\,, \qquad \mathcal{A}_{i, \xi \chi} \rightarrow \mathcal{A}_{i, \zeta 1} = - K^{-1} \theta \sigma\,,
\end{align}
and so these source R-R potentials. We thus obtain the background of the $5_3^2$:
\begingroup
\renewcommand{\arraystretch}{1.5}
\begin{gather}
\begin{gathered}
\begin{array}[t]{rl}
\textrm{d} s^2_{5^2_3,\text{E}} & = {(HK^{-1})}^{-\frac{1}{4}} \left( -\textrm{d} t^2 + \textrm{d} {\vec{w}}^2_{(5)} \right) + {(HK^{-1})}^{\frac{3}{4}} \left( \textrm{d} z^2 + \textrm{d} \zeta^2 \right)\\
& \qquad  + H^{\frac{3}{4}} K^{\frac{1}{4}} \left(\textrm{d} r^2 + r^2 \textrm{d} \theta^2 \right)\,, \qquad e^{2(\phi -\phi_0)} = H^{-1}K\,,\\
\end{array}\\
C_{(2)} = - K^{-1} \theta \sigma \textrm{d} z \wedge \textrm{d} \zeta\,, \qquad C_{(6)} = - H^{-1} K \textrm{d} t^2 \wedge \textrm{d} w^1 \wedge \ldots \wedge \textrm{d} w^5\,.\\ 
\end{gathered}
\end{gather}
\endgroup
Note in particular that, being the S-dual of the $5_2^2\text{B}$, the dilaton has been inverted as expected. One may verify that this background could have equivalently been obtained from the $5_2^2\text{B}$ generalised metric by the rotations
\begin{align}
Y^{\mathrm{m}, 1} \leftrightarrow Y^{\mathrm{m},2}\,, \qquad Y_{\mathrm{m}, 1} \leftrightarrow Y_{\mathrm{m},2}\,.
\end{align}
Once we are in a Type IIB frame, we are now free to apply rotations to the generalised metric as before. In order to rotate to the $3_3^4$, we first split the five $w^a$ coordinates to a further 1+2+2 splitting $w^a = (\omega, {\bar{w}}^{\bar{a}},{\ubar{w}}^{\ubar{a}})$, where $\bar{a}$ and $\ubar{a}$ can each take on values 1 or 2. The generalised metric of the $5_2^2\text{B}$ in this coordinate splitting becomes
\begin{align}
\begin{aligned}
\mathcal{M}_{MN} = {\left|{\mathrm{g}}_{(4)} \right|}^{-\frac{1}{4}} \operatorname{diag} [
& {Q}^{\frac{1}{2}}, {Q}^{-\frac{1}{2}}, {Q}^{-\frac{1}{2}} \delta_{(2)}, {Q}^{-\frac{1}{2}} \delta_{(2)};\\
& {Q}^{-\frac{1}{2}}, {Q}^{\frac{1}{2}}, {Q}^{\frac{1}{2}} \delta_{(2)}, {Q}^{\frac{1}{2}} \delta_{(2)},\\
	& \qquad \qquad {Q}^{-\frac{3}{2}}, {Q}^{-\frac{1}{2}}, {Q}^{-\frac{1}{2}} \delta_{(2)}, {Q}^{-\frac{1}{2}} \delta_{(2)};\\
& {Q}^{\frac{1}{2}} \delta_{(2)}, {Q}^{\frac{1}{2}} \delta_{(2)}, {Q}^{\frac{1}{2}}, {Q}^{\frac{1}{2}} \delta_{(4)}, {Q}^{\frac{1}{2}},\\
	& \qquad \qquad {Q}^{-\frac{1}{2}}, {Q}^{-\frac{1}{2}} \delta_{(4)}, {Q}^{-\frac{1}{2}}, {Q}^{-\frac{1}{2}} \delta_{(2)}, {Q}^{-\frac{1}{2}} \delta_{(2)};\\
& {Q}^{\frac{1}{2}}, {Q}^{-\frac{1}{2}}, {Q}^{-\frac{1}{2}} \delta_{(2)}, {Q}^{-\frac{1}{2}} \delta_{(2)},\\
	& \qquad \qquad {Q}^{\frac{3}{2}}, {Q}^{\frac{1}{2}}, {Q}^{\frac{1}{2}} \delta_{(2)}, {Q}^{\frac{1}{2}} \delta_{(2)};\\
& {Q}^{-\frac{1}{2}}, {Q}^{\frac{1}{2}}, {Q}^{\frac{1}{2}} \delta_{(2)}, {Q}^{\frac{1}{2}} \delta_{(2)}]\,.
\end{aligned}
\end{align}
Note, as always, the positions of the semicolons delimit each part of the generalised metric. Then, applying the rotations
\begin{gather}
Y^{\zeta \ubar{a}\ubar{b}} \leftrightarrow Y^{\zeta 2}\,, \qquad Y^{\omega \ubar{a} \ubar{b}} \leftrightarrow Y^{\zeta \bar{a} \bar{b}}\,, \qquad Y^{\omega \bar{a} \bar{b}} \leftrightarrow Y_{\zeta 2}\,, \qquad Y^{\bar{a} \ubar{b}\ubar{c}} \leftrightarrow Y^{\zeta \omega \bar{a}}\,,\\
Y_{\omega 2} \leftrightarrow Y^{\omega 2}\,, \qquad Y_{\bar{a} 2} \leftrightarrow Y^{\bar{a} 2}\,, \qquad Y^{\ubar{a}} \leftrightarrow Y_{\ubar{a}}\label{eq:522B433}\,, \qquad Y^{\ubar{a}1} \leftrightarrow Y^{\ubar{a}2}\,,
\end{gather}
the generalised metric that one obtains is
\begin{align}
\begin{aligned}
\mathcal{M}_{MN} = {\left|{\mathrm{g}}_{(4)} \right|}^{-\frac{1}{4}} \operatorname{diag} [
& {Q}^{\frac{1}{2}}, {Q}^{-\frac{1}{2}}, {Q}^{-\frac{1}{2}} \delta_{(2)}, {Q}^{\frac{1}{2}} \delta_{(2)};\\
& {Q}^{-\frac{1}{2}}, {Q}^{\frac{1}{2}}, {Q}^{\frac{1}{2}} \delta_{(2)}, {Q}^{-\frac{1}{2}} \delta_{(2)},\\
	& \qquad \qquad {Q}^{-\frac{1}{2}}, {Q}^{\frac{1}{2}}, {Q}^{\frac{1}{2}} \delta_{(2)}, {Q}^{-\frac{1}{2}} \delta_{(2)};\\
& {Q}^{-\frac{1}{2}} \delta_{(2)}, {Q}^{\frac{1}{2}} \delta_{(2)}, {Q}^{-\frac{1}{2}}, {Q}^{\frac{1}{2}} \delta_{(4)}, {Q}^{\frac{3}{2}},\\
	& \qquad \qquad {Q}^{-\frac{3}{2}}, {Q}^{-\frac{1}{2}} \delta_{(4)}, {Q}^{\frac{1}{2}}, {Q}^{-\frac{1}{2}} \delta_{(2)}, {Q}^{\frac{1}{2}} \delta_{(2)};\\
& {Q}^{\frac{1}{2}}, {Q}^{-\frac{1}{2}}, {Q}^{-\frac{1}{2}} \delta_{(2)}, {Q}^{\frac{1}{2}} \delta_{(2)},\\
	& \qquad \qquad {Q}^{\frac{1}{2}}, {Q}^{-\frac{1}{2}}, {Q}^{-\frac{1}{2}} \delta_{(2)}, {Q}^{\frac{1}{2}} \delta_{(2)};\\
& {Q}^{-\frac{1}{2}}, {Q}^{\frac{1}{2}}, {Q}^{\frac{1}{2}} \delta_{(2)}, {Q}^{-\frac{1}{2}} \delta_{(2)}]\,,
\end{aligned}
\end{align}
which one may verify is sourced by the background
\begin{align}
\begin{array}{lcl}
\multicolumn{3}{l}{{\mathrm{g}}^{3_3^4}_{\hat{m}\hat{n}} = \operatorname{diag} [ {(HK^{-1})}^{\frac{1}{2}}, {(HK^{-1})}^{-\frac{1}{2}}, {(HK^{-1})}^{-\frac{1}{2}} \delta_{(2)}, {(HK^{-1})}^{\frac{1}{2}} \delta_{(2)}]\,,}\\
{\left| {\mathrm{g}}_{(4)}^{3_3^4} \right|}^{-\frac{1}{4}} = {\left| {\mathrm{g}}_{(4)}\right|}^{-\frac{1}{4}} & \Rightarrow & {\mathrm{g}}_{\mu \nu}^{3_3^4} = {\mathrm{g}}_{\mu \nu}\,,\\
{\gamma}_{\alpha \beta} = \operatorname{diag} [1,1] & \Rightarrow & \tau = i\,.
\end{array}
\end{align}
The EFT vector is rotated to
\begin{align}
\mathcal{A}_t{}^{\zeta 2} \mapsto \mathcal{A}_t{}^{\zeta \ubar{a} \ubar{b}}\,, \qquad  \mathcal{A}_{z, \zeta 2} \mapsto \mathcal{A}_{z}{}^{\omega \bar{a}\bar{b}}\,,
\end{align}
and these both source the self-dual 4-form potential $C_{(4)}$. Thus, the background that one obtains is that of the $3_3^4$:
\begingroup
\renewcommand{\arraystretch}{1.5}
\begin{gather*}
\begin{gathered}
\begin{array}[t]{rl}
\textrm{d} s^2_{3_3^4} & = {(HK^{-1})}^{-\frac{1}{2}} \left( - \textrm{d} t^2 + \textrm{d} \omega^2 + \textrm{d} {\vec{\bar{w}}}^2_{(2)} \right) + {(HK^{-1})}^{\frac{1}{2}} \biggl( \textrm{d} z^2 + \textrm{d} \zeta^2 + \mathrlap{ \textrm{d} {\vec{\ubar{w}}}^2_{(2)} \biggr)}\\
& \qquad + {(HK)}^{\frac{1}{2}} \left( \textrm{d} r^2 + r^2 \textrm{d} \theta^2 \right)\,, \qquad e^{2(\phi - \phi_0)} = 1\,,\\
\end{array}
\end{gathered}\nonumber\\
C_{(4)} = - K^{-1} \theta \sigma \textrm{d} t \wedge \textrm{d} \omega \wedge \textrm{d} {\bar{w}}^{\bar{1}} \wedge \textrm{d} {\bar{w}}^{\bar{2}}\,, \qquad C_{(4)} = - H^{-1} K \textrm{d} z \wedge \textrm{d} \zeta \wedge \textrm{d} {\ubar{w}}^{\ubar{1}} \wedge\textrm{d} {\ubar{w}}^{\ubar{2}}\,.\nonumber
\end{gather*}
\endgroup
Since the dilaton is trivial, there is no distinction between string and Einstein frames and, indeed, the $3_3^4$ is self-dual under S-duality much like the D3-brane. Note that the apparent distinction of the $\alpha=2$ index in, for example, \eqref{eq:522B433} is a consequence of rotating from the $5_2^2\text{B}$; had we rotated from the $5_3^2$ instead, they would instead have been replaced with $\alpha=1$.\par
Using the same coordinate splitting as for the $3_3^4$, we rotate the generalised internal coordinates according to
\begin{gather}
Y^{\bar{a}} \leftrightarrow Y_{\bar{a}}\,, \qquad Y_{\bar{a} 1} \leftrightarrow Y^{\bar{a}1}\,, \qquad Y_{\ubar{a}2} \leftrightarrow Y^{\ubar{a}2}\,, \qquad Y_{\zeta 2} \leftrightarrow Y^{\zeta 2}\,,\\
Y^{\zeta \ubar{a} \ubar{b}} \leftrightarrow Y_{\omega 2}\,, \qquad Y^{\omega \bar{a} \bar{b}} \leftrightarrow Y^{\omega 2}\,, \qquad Y^{\zeta \bar{a} \bar{b}} \leftrightarrow Y^{\omega \ubar{a} \ubar{b}}\,, \qquad Y^{\bar{a} \bar{b} \ubar{a}} \leftrightarrow Y^{\zeta \omega \ubar{a}}\,,
\end{gather}
to obtain the generalised metric of the $1_4^6\text{B}$:
\begin{align}
\phantom{\mathcal{M}_{MN}}&
\begin{alignedat}{2}
\mathllap{\mathcal{M}_{MN}} = {\left|{\mathrm{g}}_{(4)} \right|}^{-\frac{1}{4}} \operatorname{diag} [
& {Q}^{\frac{1}{2}}, {Q}^{-\frac{1}{2}}, {Q}^{\frac{1}{2}} \delta_{(2)}, {Q}^{\frac{1}{2}} \delta_{(2)};\\
& {Q}^{-\frac{1}{2}}, {Q}^{\frac{1}{2}}, {Q}^{-\frac{1}{2}} \delta_{(2)}, {Q}^{-\frac{1}{2}} \delta_{(2)},\\
	& \qquad \qquad {Q}^{\frac{1}{2}}, {Q}^{\frac{3}{2}}, {Q}^{\frac{1}{2}} \delta_{(2)}, {Q}^{\frac{1}{2}} \delta_{(2)};\\
& {Q}^{-\frac{1}{2}} \delta_{(2)}, {Q}^{-\frac{1}{2}} \delta_{(2)}, {Q}^{\frac{1}{2}}, {Q}^{\frac{1}{2}} \delta_{(4)}, {Q}^{\frac{1}{2}},\\
	& \qquad \qquad {Q}^{-\frac{1}{2}}, {Q}^{-\frac{1}{2}} \delta_{(4)}, {Q}^{-\frac{1}{2}}, {Q}^{\frac{1}{2}} \delta_{(2)}, {Q}^{\frac{1}{2}} \delta_{(2)};\\
& {Q}^{\frac{1}{2}}, {Q}^{-\frac{1}{2}}, {Q}^{\frac{1}{2}} \delta_{(2)}, {Q}^{\frac{1}{2}} \delta_{(2)},\\
	& \qquad \qquad {Q}^{-\frac{1}{2}}, {Q}^{-\frac{3}{2}}, {Q}^{-\frac{1}{2}} \delta_{(2)}, {Q}^{-\frac{1}{2}} \delta_{(2)};\\
& {Q}^{-\frac{1}{2}}, {Q}^{\frac{1}{2}}, {Q}^{-\frac{1}{2}} \delta_{(2)}, {Q}^{-\frac{1}{2}} \delta_{(2)}]
\end{alignedat}\\
&
\begin{alignedat}{2}
= {Q}^{\frac{1}{4}} \cdot {\left|{\mathrm{g}}_{(4)} \right|}^{-\frac{1}{4}} \operatorname{diag} [
& {Q}^{\frac{1}{4}}, {Q}^{-\frac{3}{4}}, {Q}^{\frac{1}{4}} \delta_{(2)}, {Q}^{\frac{1}{4}} \delta_{(2)};\\
& {Q}^{-\frac{3}{4}}, {Q}^{\frac{1}{4}}, {Q}^{-\frac{3}{4}} \delta_{(2)}, {Q}^{-\frac{3}{4}} \delta_{(2)},\\
	& \qquad \qquad {Q}^{\frac{1}{4}}, {Q}^{\frac{5}{4}}, {Q}^{\frac{1}{4}} \delta_{(2)}, {Q}^{\frac{1}{4}} \delta_{(2)};\\
& {Q}^{-\frac{3}{4}} \delta_{(2)}, {Q}^{-\frac{3}{4}} \delta_{(2)}, {Q}^{\frac{1}{4}}, {Q}^{\frac{1}{4}} \delta_{(4)}, {Q}^{\frac{1}{4}},\\
	& \qquad \qquad {Q}^{-\frac{3}{4}}, {Q}^{-\frac{3}{4}} \delta_{(4)}, {Q}^{-\frac{3}{4}}, {Q}^{\frac{1}{4}} \delta_{(2)}, {Q}^{\frac{1}{4}} \delta_{(2)};\\
& {Q}^{\frac{1}{4}}, {Q}^{-\frac{3}{4}}, {Q}^{\frac{1}{4}} \delta_{(2)}, {Q}^{\frac{1}{4}} \delta_{(2)},\\
	& \qquad \qquad {Q}^{-\frac{3}{4}}, {Q}^{-\frac{7}{4}}, {Q}^{-\frac{3}{4}} \delta_{(2)}, {Q}^{-\frac{3}{4}} \delta_{(2)};\\
& {Q}^{-\frac{3}{4}}, {Q}^{\frac{1}{4}}, {Q}^{-\frac{3}{4}} \delta_{(2)}, {Q}^{-\frac{3}{4}} \delta_{(2)}]\,.
\end{alignedat}
\end{align}
This is sourced by the background
\begin{align}
\begin{array}{lcl}
\multicolumn{3}{l}{{\mathrm{g}}^{1_4^6\text{B}}_{\hat{m}\hat{n}} = \operatorname{diag} [ {(HK^{-1})}^{\frac{1}{4}}, {(HK^{-1})}^{-\frac{3}{4}},  {(HK^{-1})}^{\frac{1}{4}} \delta_{(2)},  {(HK^{-1})}^{\frac{1}{4}} \delta_{(2)}]\,,}\\
{\left| {\mathrm{g}}_{(4)}^{1_4^6\text{B}} \right|}^{-\frac{1}{4}} = {(HK^{-1})}^{\frac{1}{4}} {\left| {\mathrm{g}}_{(4)}\right|}^{-\frac{1}{4}} & \Rightarrow & {\mathrm{g}}^{1_4^6 \text{B}}_{\mu \nu} = {(HK^{-1})}^{-\frac{1}{4}} {\mathrm{g}}_{\mu \nu}\,,\\
\gamma_{\alpha \beta} = \operatorname{diag} [ {(HK^{-1})}^{\frac{1}{2}}, {(HK^{-1})}^{-\frac{1}{2}}] & \Rightarrow & \tau = i {(HK^{-1})}^{\frac{1}{2}}\,.
\end{array}
\end{align}
The EFT vector is rotated to
\begin{align}
\mathcal{A}_t{}^{\zeta\ubar{a} \ubar{b}} \mapsto \mathcal{A}_{t \omega 2}\,, \qquad \mathcal{A}_z{}^{\omega \bar{a} \bar{b}} \mapsto \mathcal{A}_z{}^{\omega 2}\,,
\end{align}
and so must source the 2-form and 6-form NS-NS potentials. We thus obtain the background of the $1_4^6\text{B}$:
\begin{gather}
\begin{gathered}
\textrm{d} s^2_{1_4^6\text{B}} =
\begingroup \renewcommand{\arraystretch}{1.5}
\begin{array}[t]{l}
{(HK^{-1})}^{-\frac{3}{4}} \left( - \textrm{d} t^2 + \textrm{d} \omega^2 \right) + {(HK^{-1})}^{\frac{1}{4}} \left( \textrm{d} z^2 + \textrm{d} \zeta^2 + \textrm{d} {\vec{\bar{w}}}_{(2)}^2 + \textrm{d} {\vec{\ubar{w}}}_{(2)}^2 \right)\\
\qquad + H^{\frac{1}{4}} K^{\frac{3}{4}} \left( \textrm{d} r^2 + r^2 \textrm{d} \theta^2 \right)\,, \qquad e^{2(\phi - \phi_0)} = {(HK^{-1})}^{-1}\,.\\
\end{array}
\endgroup\\
B_{(2)} = -H^{-1} K \textrm{d} t \wedge \textrm{d} \omega, \qquad B_{(6)} = -K^{-1} \sigma \theta \textrm{d} z \wedge \textrm{d} \zeta \wedge \textrm{d} {\bar{w}}^{\bar{1}} \wedge \textrm{d} {\bar{w}}^{\bar{2}} \wedge \textrm{d} {\ubar{w}}^{\ubar{1}} \wedge \mathrlap{\textrm{d} {\ubar{w}}^{\ubar{2}}},\\
\end{gathered}
\end{gather}
The rotation of the $3_3^4$ to its S-dual $1_3^6$ is very similar, except with the $\operatorname{SL}(2)$ indices exchanged in the rotations above.\par
For the rotation to the $0_4^{(1,6)}\text{B}$, we begin by noting that the horrible form of the generalised metric of the $1_4^6\text{B}$ is a consequence of passing through coordinates adapted to the $3_3^4\text{B}$. We may clean up the generalised metric of the $1_4^6\text{B}$ by choosing more appropriate coordinates. In particular, we define the set of coordinates $v^a = (\zeta, {\bar{w}}^{\bar{a}}, {\ubar{w}}_{\ubar{a}}),$ which are essentially the same grouping as the $w^a$ before, but with $\omega$ exchanged for $\zeta$. Then, the six coordinates of the Type IIB section are $y^{\mathfrak{m}} = (\omega, v^a)$ and the generalised metric becomes
\begin{align}
\begin{aligned}
\mathcal{M}_{MN} = {\left| {\mathrm{g}}_{(4)} \right|}^{-\frac{1}{4}}\operatorname{diag} [
& {Q}^{-\frac{1}{2}}, {Q}^{\frac{1}{2}} \delta_{(5)}; {Q}^{\frac{1}{2}}, {Q}^{-\frac{1}{2}} \delta_{(5)}, {Q}^{\frac{3}{2}}, {Q}^{\frac{1}{2}} \delta_{(5)};\\
& {Q}^{-\frac{1}{2}} \delta_{(10)}, {Q}^{\frac{1}{2}} \delta_{(10)};\\
& {Q}^{-\frac{1}{2}}, {Q}^{\frac{1}{2}} \delta_{(5)}, {Q}^{-\frac{3}{2}}, {Q}^{-\frac{1}{2}} \delta_{(5)}; {Q}^{\frac{1}{2}}, {Q}^{-\frac{1}{2}} \delta_{(5)}]\,.
\end{aligned}
\end{align}
If we apply the rotations
\begin{gather}
Y^{\omega} \leftrightarrow Y_{\omega 2}\,, \qquad Y_{\omega} \leftrightarrow Y^{\omega 2}\,, \qquad Y_{\omega 1} \leftrightarrow Y^{\omega 1}\,, \qquad Y_{a1} \leftrightarrow Y^{a1}\,,\\
Y^{\omega a b} \leftrightarrow Y^{c d e}\,,
\end{gather}
we obtain the generalised metric of the $0_4^{(1,6)}\text{B}$:
\begin{align}
\phantom{\mathcal{M}_{MN}}&
\begin{alignedat}{2}
\mathllap{\mathcal{M}_{MN}} = {\left|{\mathrm{g}}_{(4)} \right|}^{-\frac{1}{4}} \operatorname{diag} [
& {Q}^{\frac{3}{2}},  {Q}^{\frac{1}{2}} \delta_{(5)}; {Q}^{-\frac{1}{2}}, {Q}^{\frac{1}{2}} \delta_{(5)}, {Q}^{-\frac{1}{2}}, {Q}^{\frac{1}{2}}\delta_{(5)};\\
& {Q}^{\frac{1}{2}} \delta_{(10)}, {Q}^{-\frac{1}{2}} \delta_{(10)};\\
& {Q}^{\frac{1}{2}}, {Q}^{-\frac{1}{2}} \delta_{(5)}, {Q}^{\frac{1}{2}}, {Q}^{-\frac{1}{2}}\delta_{(5)}; {Q}^{-\frac{3}{2}}, {Q}^{-\frac{1}{2}} \delta_{(5)}]
\end{alignedat}\\
&
\begin{alignedat}{2}
=  {Q}^{\frac{1}{2}} \cdot {\left|{\mathrm{g}}_{(4)} \right|}^{-\frac{1}{4}} \operatorname{diag} [
& HK^{-1}, \delta_{(5)}; {Q}^{-1}, \delta_{(5)}, {Q}^{-1}, \delta_{(5)};\\
& \delta_{(10)}, {Q}^{-1} \delta_{(10)};\\
& 1, {Q}^{-1}\delta_{(5)}, 1, {Q}^{-1} \delta_{(5)}; {Q}^{-2}, {Q}^{-1} \delta_{(5)}]\,.
\end{alignedat}
\end{align}
One may verify that this is sourced by the background
\begin{align}
\begin{array}{lcl}
\multicolumn{3}{l}{{\mathrm{g}}^{0_4^{(1,6)}\text{B}}_{\mathrm{m} \mathrm{n}} = \operatorname{diag}[HK^{-1}, \delta_{(5)}]\,,}\\
{\left| {\mathrm{g}}_{(4)}^{0_4^{(1,6)}\text{B}}\right|}^{-\frac{1}{4}} = {(HK^{-1})}^{\frac{1}{2}} {\left| {\mathrm{g}}_{(4)} \right|}^{-\frac{1}{4}} & \Rightarrow & {\mathrm{g}}_{\mu \nu}^{0_4^{(1,6)}\text{B}} = {(HK^{-1})}^{-\frac{1}{2}} {\mathrm{g}}_{\mu \nu}\,,\\
\gamma_{\alpha \beta} = \operatorname{diag} [ 1,1] & \Rightarrow & \tau = i\,.
\end{array}
\end{align}
The EFT vector is rotated to source the Kaluza-Klein vector:
\begin{align}
\mathcal{A}_{t,\omega 2} & \mapsto \mathcal{A}_t{}^{\omega}\,, \qquad \mathcal{A}_{z}{}^{\omega 2}  \mapsto \mathcal{A}_{z \omega}\,,
\end{align}
and so one obtains the background of the $0_4^{(1,6)}\text{B}$:
\begingroup
\renewcommand{\arraystretch}{1.5}
\begin{gather}
\begin{gathered}
\begin{array}[t]{rl}
\textrm{d} s^2_{0_4^{(1,6)}\text{B}} & = - {(HK^{-1})}^{-1} \textrm{d} t^2 + HK^{-1} {\left( \textrm{d} \omega - H^{-1}K \textrm{d} t\right)}^2 + \textrm{d} z^2 + \textrm{d} {\vec{v}}^2_{(5)}\\
& \qquad + K \left( \textrm{d} r^2 + r^2 \textrm{d} \theta^2\right)\,, \qquad e^{2(\phi - \phi_0)} = 1\,.
\end{array}
\end{gathered}
\end{gather}
\endgroup
Finally, applying the rotation
\begin{gather}
Y^{\mathrm{m}} \leftrightarrow Y_{\mathrm{m}}\,, \qquad Y^{\mathrm{m} \alpha} \leftrightarrow Y_{\mathrm{m} \alpha}\,, \qquad Y^{\zeta a b} \leftrightarrow Y^{cde}\,,
\end{gather}
one obtains
\begingroup
\renewcommand{\arraystretch}{1.4}
\begin{align}
\mathcal{M}_{MN} & = {\left|{\mathrm{g}}_{(4)} \right|}^{-\frac{1}{4}} \operatorname{diag} [
\begin{array}[t]{l}
\!\! {Q}^{-\frac{3}{2}}, {Q}^{-\frac{1}{2}} \delta_{(5)}; {Q}^{\frac{1}{2}}, {Q}^{-\frac{1}{2}} \delta_{(5)}, {Q}^{\frac{1}{2}}, {Q}^{-\frac{1}{2}}\delta_{(5)};\\
\!\! {Q}^{-\frac{1}{2}} \delta_{(10)}, {Q}^{\frac{1}{2}} \delta_{(10)};\\
\!\! {Q}^{-\frac{1}{2}}, {Q}^{\frac{1}{2}} \delta_{(5)}, {Q}^{-\frac{1}{2}}, {Q}^{\frac{1}{2}}\delta_{(5)}; {Q}^{\frac{3}{2}},  {Q}^{\frac{1}{2}} \delta_{(5)}]\\
\end{array}\\
& = {Q}^{-\frac{1}{2}} \cdot {\left|{\mathrm{g}}_{(4)} \right|}^{-\frac{1}{4}} \operatorname{diag} [
\begin{array}[t]{l}
\!\! Q^{-1}, \delta_{(5)}; Q, \delta_{(5)}, Q, \delta_{(5)};\\
\!\! \delta_{(10)}, Q \delta_{(10)};\\
\!\! 1, Q\delta_{(5)}, 1, Q \delta_{(5)}; {Q}^2, Q\delta_{(5)}]\,.
\end{array}
\end{align}
\endgroup
This is sourced by the background
\begin{align}
\begin{array}{lcl}
\multicolumn{3}{l}{{\mathrm{g}}^{\text{KK5B}}_{\mathrm{m} \mathrm{n}} = \operatorname{diag} [ H^{-1}K, \delta_{(5)}]\,,}\\
{\left| {\mathrm{g}}_{(4)}^{\text{KK5B}}\right|}^{-\frac{1}{4}} = {(HK^{-1})}^{-\frac{1}{2}} {\left| {\mathrm{g}}_{(4)} \right|}^{-\frac{1}{4}} & \Rightarrow & {\mathrm{g}}_{\mu \nu}^{\text{KK5B}} = {(HK^{-1})}^{\frac{1}{2}} {\mathrm{g}}_{\mu \nu}\,,\\
\gamma_{\alpha \beta} = \operatorname{diag} [1,1] & \Rightarrow & \tau = i\,.
\end{array}
\end{align}
The EFT vector is rotated to
\begin{align}
\mathcal{A}_t{}^\omega & \mapsto \mathcal{A}_{t,\omega}\,, \qquad \mathcal{A}_{z \omega} \mapsto \mathcal{A}_z{}^\omega\,,
\end{align}
and so we obtain the background
\begin{gather}
\begin{gathered}
\begin{aligned}
\textrm{d}s^2_{\text{KK5B}} & = -\textrm{d} t^2 + \textrm{d} {\vec{v}}^2_{(5)} + H (\textrm{d} r^2 + r^2 \textrm{d} \theta^2) + HK^{-1} \textrm{d} z + H^{-1}K {(\textrm{d} \omega - K^{-1} \theta \sigma \textrm{d} z)}^2\\
	& = - \textrm{d} t^2 + \textrm{d} {\vec{v}}^2_{(5)} + H( \textrm{d} r^2 + r^2 \textrm{d} \theta^2 + \textrm{d} \omega^2) + H^{-1}{( \textrm{d} z + \theta \sigma \textrm{d} \omega)}^2\,,\\
\end{aligned}\nonumber\\
\end{gathered}\\
e^{2(\phi - \phi_0)} = 1\,.
\end{gather}
Recall that we switched coordinates in the $0_4^{(1,6)}\text{B}$ frame such that it has coordinates $(\omega, v^a)$ with $\zeta$ in $v^a$. Thus, one sees that the transverse 3-space of the KK5B is spanned by $(r,\theta, \omega)$ whereas the 3-transverse space in the KK6A was spanned by $(r,\theta, \zeta)$.
\subsection{Discussion}
The solution presented here shares obvious similarities with the solution in \cite{Berman:2014hna}. More concretely, recall that the geometric solution was constructed with a three-dimensional transverse space. Denoting the coordinates of this transverse space as $(r,\theta, z)$ and the harmonic function as $\tilde{H}$, the potential $\tilde{A}$ sourcing this is obtained by solving $\textrm{d} \tilde{A} = \star_3 \textrm{d} \tilde{H}$. The non-geometric solution, as presented above, is obtained by smearing this solution over $z$ to give the $(H,A)$ used in the solution. However, noting that $HK^{-1}$ is itself harmonic in two dimensions, we may construct the following table 
\begin{table}[H]
\centering
\begin{tabulary}{\textwidth}{LCLCL}
Geometric & & \multicolumn{3}{c}{Non-Geometric}\\
\cmidrule{1-1} \cmidrule{3-5}
3-dimensional on $\mathbb{R}^3$ & & Effective 2-dimensional on $\mathbb{R}^2 \times S^1$ & & 2-dimensional on $\mathbb{R}^2$\\
${\tilde{\mathcal{A}}}_t{}^M = (1 - {\tilde{H}}^{-1}) a^M$ & & $\mathcal{A}_t{}^M = - H^{-1} K a^M$ & & $\mathcal{A}_t{}^M = - \hat{H}^{-1} a^M$\\
${\tilde{\mathcal{A}}}_i{}^M = A_{i} {\tilde{a}}^M$ & & $\mathcal{A}_z{}^M = - K^{-1} \theta \sigma {\tilde{a}}^M$ & & $\mathcal{A}_z{}^M = \hat{A} {\tilde{a}}^M$\\
$\mathcal{M}_{MN} = \{ {\tilde{H}}^{\pm \frac{3}{2}}, {\tilde{H}}^{\pm \frac{1}{2}} \delta_{(27)} \}$ & & $\mathcal{M}_{MN} = \{ {(HK^{-1})}^{\pm \frac{3}{2}}, {(HK^{-1})}^{\pm \frac{1}{2}} \delta_{(27)} \}$ & & $\mathcal{M}_{MN} = \{ {\hat{H}}^{\pm \frac{3}{2}}, {\hat{H}}^{\pm \frac{1}{2}} \delta_{(27)} \}$\\
\end{tabulary}
\end{table}
where
\begin{align}
\textrm{d} {\tilde{A}} & = \star_3 \textrm{d} \tilde{H}, & \textrm{d} A & = \star_3 \textrm{d}H, & \textrm{d} \hat{A} & = \star_2 \textrm{d} \hat{H}\,,\\
\tilde{H} & = 1 + \frac{{\tilde{h}}_0}{\sqrt{r^2 + z^2}}, & H & = h_0 + \sigma \ln \frac{\mu}{r}, & \hat{H} & = HK^{-1}\,,\\
{\tilde{A}} & = \frac{{\tilde{h}}_0 z}{2r^2 \sqrt{r^2 + z^2}} \textrm{d} \theta, & A & = - \sigma \theta \textrm{d} z, & \hat{A} & = - \sigma \theta K^{-1}\,.
\end{align}
The first column shows the variables used in the geometric solution whilst the second column shows the variables used in the non-geometric solution presented here. Finally, the third column shows the variables that would have been used (given in terms of the variables that we did use) had we constructed a genuine 2-dimensional solution from the beginning rather than smearing a 3-dimensional space. We thus see a clear parallel between the geometric solution with transverse space $(r,\theta,z)$ and the non-geometric solution with transverse space $(r,\theta)$, reinterpreted to treat the combination $\hat{H} = HK^{-1}$ as the fundamental object. This shows that we could have constructed the entire non-geometric solution using arbitrary functions $\hat{H}(r,\theta)$, harmonic in 2 dimensions, rather than the smeared codimension-3 harmonic function that we used. Note that the fact that $\mathcal{A}_t{}^M \sim - \hat{H} a^M$ rather than $\mathcal{A}_t{}^M \sim (1-\hat{H}) a^M$ in this interpretation is mostly irrelevant considering the fact that the asymptotics of the harmonic function require some method of regularising the divergence anyway (e.g.\ some anti-brane configuration around the exotic branes to absorb any flux, along the lines of the D8 story).\par 
We end by noting that the non-geometric solution exchanges branes as follows:
\begin{align}
Y^M \leftrightarrow Y_M : \begin{cases}
2^6 \leftrightarrow 5^3\,,\\
0^{(1,7)} \leftrightarrow \text{KK6}\,,
\end{cases}
\end{align}
where $Y^M \leftrightarrow Y_M$ corresponds to $Y^m \leftrightarrow Y_m$ and $Y^{mn} \leftrightarrow Y_{mn}$ in the M-theory section. For the Type IIB solutions, we have
\begin{align}
Y^M \leftrightarrow Y_M: \begin{cases}
5_2^2\text{B} \leftrightarrow 1_4^6\text{B}\,,\\
5_3^2 \leftrightarrow 1_3^6,,\\
3_3^4 \leftrightarrow 3_3^4\,,\\
0_4^{(1,6)} \leftrightarrow \text{KK5B}\,,
\end{cases}
\end{align}
where $Y^M \leftrightarrow Y_M$ for the IIB coordinates corresponds to $Y^{\mathrm{m}} \leftrightarrow Y_{\mathrm{m}}, Y^{\mathrm{m} \alpha} \leftrightarrow Y_{\mathrm{m} \alpha}, Y^{\zeta a b} \leftrightarrow Y^{cde}$. Under this transformation, the $3_3^4$ is self-dual, but with the roles of $(\omega, {\bar{w}}_{(2)}) \leftrightarrow (\zeta, {\ubar{w}}_{(2)})$ exchanged in the resulting metric.

	\chapter{Mapping out the Exotic States}\label{ch:Map}
Whilst we gave an EFT parent to a modest number of exotic branes in the previous chapter, it has become increasingly clear that exotic branes are far more common than were previously thought. Indeed, the branes discussed in the solution above are only a tiny fraction of exotic branes that one may find in ExFT-like theories (specifically, $E_{9(9)}$ and larger). Here, we discuss a very simple algorithm for mapping out all of the branes based on the transformations of the masses under the various dualities. We then compare our results with what has previously appeared in the literature where work has primarily focused on the mixed-symmetry potentials that these branes couple to. Due to the number of branes that we have tabulated, we have only kept the notation, duality transformations and a worked example in the main text, as described in Section~\ref{sec:Mapping}. The resulting inter-connected web of branes that we obtain, ordered by their $g_s$-dependence, have been relegated to Appendix~\ref{app:ExoticWeb}. For now, we shall note that, as $\alpha$ decreases, the size of the orbits generally grows leaving us with a vast taxonomy of branes. The hope is that this taxonomy will allow us to find patterns in the exotic brane structure and point to the existence of unifying solutions in EFT.
\section{Duality Transformations}\label{sec:Mapping}
In the following sections, we map out all the allowed exotic branes down to $\alpha=-7$---the lowest power of $g_s^{\alpha}$ admissible in $E_{7(7)}$ EFT\footnote{Here, we shall work in the string frame throughout.}. The general scheme is to map out all of the allowed S- and T-duality transformations and lifts/reductions of all the branes and to determine the result of a given transformation by the mass of the resulting object. Since T-duality does not change the $g_s$-scaling of the branes' tension/mass, we organise these objects into T-duality orbits ordered by their scaling $\alpha$. Each figure in Appendix~\ref{app:ExoticWeb} corresponds to a single T-duality orbit i.e. every brane in each figure may be reached from any other brane in the same figure by judicious T-dualities alone. A T-duality transformation along the direction $y$ is given by
\begin{align}\label{eq:TTrans}
T_y: R_y \mapsto \frac{l_s^2}{R_y}\,, \qquad g_s \mapsto \frac{l_s}{R_y} g_s\,.
\end{align}
We stress that this process has a natural description in ExFT wherein the duality transformations correspond to different choices of section that generically allow winding mode dependences. It is a well-known fact that the T-duality rules encoded in the Buscher rules or the reduction from M-theory to Type IIA both require an isometry but the ExFT description of this is simply the rotation of coordinates, in and out of section, which does not require an isometry. These extended theories thus afford us a much richer spectrum of branes since one can take duality transformation in directions which classically would not be allowed.\par
For example, in supergravity, whilst a codimension-1 brane may be T-dualised along the transverse direction after smearing the harmonic function in that direction, this removes any dependence of the harmonic function on any of the coordinates and thus become a simple constant which renders it equivalent to the trivial D9-brane. The DFT description of this, however, still allows for a meaningful duality transformation since the dependence of the harmonic function is simply shifted to a dependence on a winding coordinate, rather than being lost entirely. Thus, one may still construct space-filling branes in DFT that remain non-trivial by virtue of this winding mode dependence. A similar story holds for reductions of M-theory branes; a codimension-1 brane in M-theory may be `reduced' along the transverse direction to yield a non-trivial codimension-0 solution in ten dimensions simply because the coordinate dependence is only shifted out of section.\par
The dependence on winding modes pre-dates DFT and has been well-studied in the context of Gauged Linear Sigma Models (GLSM). By comparing their interpretations on either sides of the T-dual pair $\text{NS5} \xleftrightarrow{T} \text{KK5}$, it was shown that such a winding mode dependence may be understood as worldsheet instanton corrections \cite{Tong:2002rq,Harvey:2005ab}. More specifically, the worldsheet instanton corrections of an H-monopole break the isometry in the $S^1$, localising it to an NS5 and this transfers over to the T-dual picture as the breaking of the isometry in the dual circle. Thus, one concludes that the information encoded in a dependence on dual coordinates is equivalent to that of worldsheet instanton corrections. More recently, the GLSM analysis was extended to include the $5_2^2$ \cite{Kimura:2013fda,Lust:2017jox,Kimura:2018ain} and further studied in \cite{Kimura:2018hph} in the context of DFT with similar conclusions that winding mode dependences may be interpreted as worldsheet instanton corrections to the geometry.\par
It is easy to see that the T-duality rules given in \eqref{eq:TTrans} are, taken together, equivalent to the general rule proposed in \cite{Lombardo:2016swq}
\begin{align}
\alpha = -n: \qquad \underbrace{a,a,\ldots, a}_p \xleftrightarrow{T_a}\underbrace{a, a, \ldots, a}_{n-p}\,.
\end{align}
Moving onto S-duality, we first remark that that its action on Type IIB branes alters the $g_s$-scaling of the mass but does not affect the wrapping structure of the brane i.e.\ a $b_{n}^{(\dots, d,c)}$-brane in Type IIB is mapped to some other $b_{n^\prime}^{(\ldots, d,c)}$-brane. S-duality thus maps between the orbits/figures in Appendix~\ref{app:ExoticWeb}. The effect of an S-duality transformation on a brane is encoded in the following:
\begin{align}
S: g_s \mapsto \frac{1}{g_s}\,, \qquad l_s \mapsto g_s^{\frac{1}{2}} l_s\,.
\end{align}
Finally, the lift of each Type IIA brane is determined by using the relations between the ten- and eleven-dimensional constants:
\begin{align}
\begin{rcases}
l_s & = \frac{l_p^{\frac{3}{2}}}{R_\natural^{\frac{1}{2}}}\\
g_s & = {\left( \frac{R_\natural}{l_p} \right)}^{\frac{3}{2}}
\end{rcases} \leftrightarrow
\begin{cases}
R_\natural & = l_s g_s\\
l_p & = g_s^{\frac{1}{3}} l_s\,.
\end{cases}
\end{align}
That this is possible is documented in detail in, for example, \cite{Ortin:2015hya,Obers:1998fb,deBoer:2012ma}.\par
Each brane in M-theory can then be reduced in multiple directions, indicating the existence of other Type IIA branes. The above procedure is repeated iteratively until all possible duality transformations and lifts/reductions have been accounted for.\par
All the figures presented in Appendix~\ref{app:ExoticWeb} were generated by saturating all possible S- and T-duality transformations as well as lifts/reductions. We have chosen to display any even-$\alpha$ branes that appear in both Type II theories as separate nodes such that the number of lines coming out of each brane is always equal to the number of T-dual partners that the brane possesses---this provides a simple verification that all possible T-duality transformations have been accounted for. Note that we shall not include time-like reductions in our analysis. This means that, representing each $b_n^{(\ldots, d, c)}$-brane as a single node, one must always have $l$ lines emanating from each brane where
\begin{align}
l = \begin{cases} (b + c + d + \ldots ) + 1  & \text{ if codimension}\neq 0\,,\\
(b + c + d +  \ldots ) & \text{ if codimension}=0\,.
\end{cases}
\end{align}
Since T-dualising along a transverse direction produces a brane of 1 codimension lower, the special case of codimension-0 branes in eleven dimensions is precisely why our T-duality orbits close. For example, the $0_4^{(2,1,6)}$ (obtained from a double T-duality along the two transverse coordinates of the $0_4^{(1,6)}$) has only three T-dual partners rather than four.\par
The branes presented in Appendix~\ref{app:ExoticWeb} are `complete' to $g_s^{-7}$ in so far as all branes down to that power, whose existence is implied by the above rules, are included. The missing figure references are all for branes of $\alpha \leq -8$ but these are also expected to fall into their own T-duality orbits. For example, at $g_s^{-8}$, the process described above implies the existence of 64 branes in Type IIA and 26 branes in Type IIB. Another 190 further branes are required to organise these into eight complete T-duality orbits. The proliferation of branes is evident and it is not clear whether the process will terminate at finite $g_s^\alpha$ or not (indeed, there is a strong expectation that it should not, leading to an infinite number of branes---we shall comment on this later). Already at $g_s^{-7}$, one finds the implied existence of branes down to $g_s^{-15}$ and at $g_s^{-8}$ there is an implied existence of branes down to $g_s^{-17}$ (the lift of an $0_8^{(6,1,2,0,0)}$ will give rise to a $0_{17}^{(1,0,0,0,0,0,0,6,1,1,0,0)}$ as one of its possible reductions). 
\section{A Partial Example}
To illustrate the procedure of generating the diagrams in Appendix~\ref{app:ExoticWeb}, we give a partial example below. Consider the $0_4^{(1,6)}$-brane in Type IIA. Its mass is given by
\begin{align}
\text{M}(0_4^{(1,6)}) = \frac{R_7^3 {(R_6 \ldots R_1)}^2}{g_s^4 l_s^{16}}\,.
\end{align}
We have three possible distinct T-duality transformations that we may apply (up to renaming of coordinates); a duality transformation along the cubic direction, the quadratic direction or a direction entirely transverse to the brane:
\begingroup
\renewcommand{\arraystretch}{2}
\begin{align}
\begin{array}{llllllll}
& \mathrel{\raisebox{-7pt}{$\nearrow$}} & \xrightarrow{T_8} & \frac{R_7^3 {(R_6 \ldots R_1)}^2}{{\left(\frac{l_s}{R_8} g_s\right)}^4 l_s^{16}} & = & \frac{R_8^4 R_7^3 {(R_6 \ldots R_1)}^2}{g_s^4 l_s^{20}} & = & \text{M} (0^{(1,1,6)}_4\text{B})\,,\\
 \text{M}(0_4^{(1,6)}\text{A})& \rightarrow & \xrightarrow{T_7} & \frac{{\left(\frac{l_s^2}{R_7} \right)}^3 {(R_6 \ldots R_1)}^2}{{\left( \frac{l_s}{R_7} g_s \right)}^4 l_s^{16}} & = & \frac{R_7 {(R_6 \ldots R_1)}^2}{g_s^4 l_s^{14}} & = & \text{M} ( 1_4^6\text{B})\,,\\
& \mathrel{\raisebox{7pt}{$\searrow$}} &\xrightarrow{T_6} & \frac{R_7^3 {\left( \frac{l_s^2} R_6 \right)}^2 {(R_5 \ldots R_1)}^2}{{\left( \frac{l_s}{R_6} g_s \right)}^4 l_s^{16}} & = & \frac{R_7^3 {(R_6 \ldots R_1)}^2}{g_s^4 l_s^{16}} & = & \text{M} ( 0_4^{(1,6)}\text{B})\,.\\
\end{array}
\end{align}
\endgroup
Of these, the first is a novel codimension-1 object that appears only because we are allowing transformations along non-isometric directions (if one is more careful, one should be able to obtain these in the standard supergravity picture by appropriate arraying and smearing of the $0_4^{(1,6)}$). Additionally, note that the appearance of the $0_4^{(1,6)}$ in the Type IIB theory also means that one must have $0_4^{(1,1,6)}\text{A}$- and $1_4^6\text{A}$-branes as well.\par
We now proceed with the example. The respective S-duals of these branes are given by
\begingroup
\renewcommand{\arraystretch}{2}
\begin{align}
\begin{array}{lllllll}
\text{M} (0_4^{(1,1,6)}\text{B}) & \xrightarrow{S} & \frac{R_8^4 R_7^3 {(R_6 \ldots R_1)}^2}{{\left( \frac{1}{g_s} \right)}^4 {\left( g_s^{\frac{1}{2}} l_s\right)}^{20}} & = & \frac{R_8^4 R_7^3 {(R_6 \ldots R_1)}^2}{g_s^{6} l_s^{10}} & = & \text{M}(0_6^{(1,1,6)}\text{B})\,,\\
\text{M} (1_4^6\text{B}) & \xrightarrow{S} & \frac{R_7 {(R_6 \ldots R_1)}^2}{{\left( \frac{1}{g_s} \right)}^4 {\left( g_s^{\frac{1}{2}} l_s\right)}^{14}} & = & \frac{R_7 {(R_6 \ldots R_1)}^2}{g_s^3 l_s^{14}} & = & \text{M}(1_3^6\text{B})\,,\\
\text{M} ( 0_4^{(1,6)}\text{B}) & \xrightarrow{S} & \frac{R_7 {(R_6 \ldots R_1)}^2}{{\left( \frac{1}{g_s} \right)}^4 {\left( g_s^{\frac{1}{2}} l_s\right)}^{16}} & = & \frac{R_7 {(R_6 \ldots R_1)}^2}{{\left( \frac{1}{g_s} \right)}^4 g_s^8 l_s^{16}} & = & \text{M} (0_4^{(1,6)}\text{B})\,.\\
\end{array}
\end{align}
\endgroup
Here, the last two are branes that we have already encountered and it is only the $0_6^{(1,1,6)}\text{B}$ which is novel. Its existence means that there is at least one T-duality orbit at $g_s^{-6}$ which must be fleshed out. One must then map out all allowed S- and T-duals of those objects. Finally, we may lift the $0_4^{(1,6)}\text{A}$ to M-theory by rewriting its mass in terms of M-theory constants:
\begin{align}
\text{M}(0_4^{(1,6)}\text{A}) = \frac{R_7^3 {(R_6 \ldots R_1)}^2}{g_s^4 l_s^{16}} = \frac{R_7^3{(R_6 \ldots R_1)}^2 R_\natural^2}{l_p^{18}} = \text{M} (0^{(1,7)})\,,
\end{align}
where $R_\natural$ is the M-theory circle. Thus, we may deduce that the $0_4^{(1,6)}\text{A}$ is obtained from the $0^{(1,7)}$ by choosing the M-theory circle to correspond to one of the quadratic directions. The existence of the parent brane in M-theory then requires the introduction of other branes in Type IIA. In particular, we have three distinct choices for the reduction of the $0^{(1,7)}$: the M-theory circle may lie along a direction entirely transverse to the brane, along the cubic direction or along one of the quadratic directions. Relabelling coordinates, we have
\begingroup
\renewcommand{\arraystretch}{1.5}
\begin{align}
\text{M}(0^{(1,7)}) & = \frac{R_8^3 {(R_7 \ldots R_1)}^2}{l_p^{18}}\\
	& \rightarrow \begin{cases}
\begin{array}{llllll}
R_\natural = R_9: & \frac{R_8^3 {(R_7^2 \ldots R_1)}^2}{l_p^{18}} & = & \frac{R_8^3{(R_7 \ldots R_1)}^2}{g_s^6 l_s^{18}} & = & \text{M} (0_6^{(1,7)}\text{A})\,,\\
R_\natural = R_8: & \frac{R_\natural^3 {(R_7 \dots R_1)}^2}{l_p^{18}} & = & \frac{{(R_7 \ldots R_1)}^2}{g_s^3 l_s^{18}} & = & \text{M}(0_3^7\text{A})\,,\\
R_\natural = R_7: & \frac{R_8^3 R_\natural^2 {(R_6 \ldots R_1)}^2}{l_p^{18}} & = & \frac{R_8^3 {(R_6 \ldots R_1)}^2}{g_s^4 l_s^{16}} & = & \text{M}(0_4^{(1,6)}\text{A})\,.\\
\end{array}
\end{cases}
\end{align}
\endgroup
The last two are in agreement with the de Boer-Shigemori classification. Just as the $0_3^7$ obtained in this way happens to be in the same $p_3^{7-p}$ T-duality orbit (the only $g_s^{-3}$ orbit) as the $1_3^6\text{B}$ found above, the $0_6^{(1,7)}\text{A}$ obtained here happens to be in the same $g_s^{-6}$ orbit (of which there are multiple) as the $0_6^{(1,1,6)}\text{B}$ found above. We thus see the beginnings of a heavily intertwined, complex structure in these dualities and lifts/reductions. The novel branes only appeared here because we are allowing for dependences on winding and wrapping modes. The number of such branes is seen to quickly proliferate once one starts to apply this procedure iteratively.
\section{Enumerating the Branes}
It quickly becomes clear that generating all the T-duality orbits at each power of $g_s$ quickly becomes unfeasible due to the rapid growth in the numbers of these branes. As a complementary endeavour to the one above, one may instead by more interested in the \emph{number} of branes that appear at each power of $g_s$. We tabulate our results of an exhaustive search of the number of distinct\footnote{We count the same brane appearing in both theories, such as the NS5A/B as different objects.} branes $N_{(\alpha)}$ that one finds at all powers down to $\alpha = -25$, as well as all the branes in M-theory that are required to describe them, in Table~\ref{tab:gsGradingOfExoticBranes}.\par
For the fourth column we have split $N_{(\alpha)}$ into the number of branes that appear only in Type IIA (denoted $A$), the number of branes that appear only in Type IIB (denoted $B$) and the number of branes that appear in both theories (denoted $C$). If $N_{(\alpha)}$ splits cleanly into $A = B = N_{(\alpha)}/2$ (such that $C=0$), this means that every brane appears only in either one of the theories. The D-branes at $\alpha = -1$ follow this pattern since Type IIA/B respectively contain only even/odd D-branes. We consequently designate that power of $g_s$ to be of `R-R' type. Conversely, if $A=B=0$, then all the branes are common to both IIA and IIB and we give that power of $g_s$ the designation `NS-NS'. This is seen, for example, at $\alpha = -2$ with two copies of the 5-brane chain $5_2\text{A/B} \xleftrightarrow{T} 5_2^1\text{B/A} \xleftrightarrow{T} 5_2^2 \text{A/B} \xleftrightarrow{T} 5_2^3\text{B/A} \xleftrightarrow{T} 5_2^4\text{A/B}$. In the language of DFT, the first four are the branes that couple to the components of the generalised flux $F_{AB}{}^C = \{ H_{abc}, f_{ab}{}^c, Q_a{}^{bc}, R^{abc}\}$ but the final object is a less familiar codimension-0 brane called the $5_2^4$ (whose flux necessarily vanishes). This has already been proposed to exist in \cite{Kimura:2018hph} where it was presented as one of the possible solution embedded in the DFT monopole.\par
Looking at Table~\ref{tab:gsGradingOfExoticBranes}, we see a clear pattern; when $n = - \alpha$ is odd, the set of branes is of R-R type. Additionally, when $n=2 \! \mod \, 4$, the branes are of NS-NS type. However, the situation is more complicated when $n = 0\! \mod\, 4$. These powers of $g_s$ are predominantly of NS-NS type but there is a comparatively small set of branes at those powers that break this pattern, and we have denoted this the `NS-NS violation' in the final column. It is expected that these powers of $g_s$ will house predominantly NS-NS orbits, with only a small number of R-R orbits breaking the pattern. The first instance occurring at $\alpha = -4$ contains only 3 orbits, 2 of which are NS-NS and the final one being R-R as seen in the orbits presented in Appendix~\ref{app:ExoticWeb}. In fact the 10 branes that violate full NS-NS correspond to a T-duality chain that mirrors the D-brane chain but headed by the $9_4$-brane (the S-dual of the $\text{D9} = 9_1$) rather than the $9_1$-brane. Particularly striking is how the number of branes grows steadily as the power of $g_s$ decreases to ever more non-perturbative branes powers of $g_s$. Indeed, there is no indication that this will ever terminate and so we expect an infinite number of such exotic states which form obvious candidates for supplying the wrapping coordinates of the infinite-dimensional ExFTs.\par
\afterpage{%
\clearpage
\renewcommand{\arraystretch}{0.85}
\thispagestyle{empty}
\begin{landscape}
\begin{table}[!ht]
\centering
\captionsetup{width=0.9\linewidth}
\begin{tabulary}{1.25\linewidth}{cclcllclclclc}
\toprule
$\alpha$ & Number of Branes $N_{(\alpha)}$ & Type & \multicolumn{9}{c}{Breakdown $N_{(\alpha)} = A + B + 2C$} & NS-NS Violation $A+B$\\
\midrule
0 & 4 & \llap{(}NS-NS) & $4$ & $=$ & $0$ & $+$ & $0$ & $+$ & $2$ & $\times$ & $2$ & $0$ \\
-1 & 10 & R-R & $10$ & $=$ & $5$ & $+$ & $5$ &\\
-2 & 10 & NS-NS & $10$ & $=$ & $0$ & $+$ & $0$ & $+$ & $2$ & $\times$ & $5$ &\\
-3 & 24 & R-R & $24$ & $=$ & $12$ & $+$ & $12$ &\\
-4 & 46 &  & $46$ & $=$ & $5$ & $+$ & $5$ & $+$ & $2$ & $\times$ & $18$ & $10$\\
-5 & 72 & R-R & $72$ & $=$ & $36$ & $+$ & $36$ &\\
-6 & 104 & NS-NS & $104$ & $=$ & $0$ & $+$ & $0$ & $+$ & $2$ & $\times$ & $52$ &\\
-7 & 210 & R-R & $210$ & $=$ & $105$ & $+$ & $105$ &\\
-8 & 280 & & $280$ & $=$ & $12$ & $+$ & $12$ & $+$ & $2$ & $\times$ & $128$ & $24$\\
-9 & 448 & R-R & $448$ & $=$ & $224$ & $+$ & $224$ &\\
-10 & 632 & NS-NS & $632$ & $=$ & $0$ & $+$ & $0$ & $+$ & $2$ & $\times$ & $316$ &\\
-11 & 942 &  R-R & $942$ & $=$ & $471$ & $+$ & $471$ &\\
-12 & 1244 & & $1244$ & $=$ & $36$ & $+$ & $36$ & $+$ & $2$ & $\times$ & $586$ & $72$\\
-13 & 1926 & R-R & $1926$ & $=$ & $963$ & $+$ & $963$ &\\
-14 & 2340 & NS-NS & $2340$ & $=$ & $0$ & $+$ & $0$ & $+$ & $2$ & $\times$ & $1170$ &\\
-15 & 3398 & R-R & $3398$ & $=$ & $1699$ & $+$ & $1699$ &\\
-16 & 4378 & & $4378$ & $=$ & $105$ & $+$ & $105$ & $+$ & $2$ & $\times$ & $2084$ & $210$\\
-17 & 5942 & R-R & $5942$ & $=$ & $2971$ & $+$ & $2971$ &\\
-18 & 7316 & NS-NS & $7316$ & $=$ & $0$ & $+$ & $0$ & $+$ & $2$ & $\times$ & $3658$ &\\
-19 & 10050 & R-R & $10050$ & $=$ & $5025$ & $+$ & $5025$ &\\
-20 & 12252 & & $12252$ & $=$ & $224$ & $+$ & $224$ & $+$ & $2$ & $\times$ & $5902$ & $448$\\
-21 & 16134 & R-R & $16134$ & $=$ & $8067$ & $+$ & $8067$ &\\
-22 & 19388 & NS-NS & $19388$ & $=$ & $0$ & $+$ & $0$ & $+$ & $2$ & $\times$ & $9694$ &\\
-23 & 25320 & R-R & $25320$ & $=$ & $12660$ & $+$ & $12660$ &\\
-24 & 30374 & & $30374$ & $=$& $471$ & $+$ & $471$ & $+$ & $2$ & $\times$ & $14716$ & 942\\
-25 & 38310 & R-R & $38310$ & $=$ & $19155$ & $+$ & $19155$ &\\
\midrule
M & 458124\\
\bottomrule
\end{tabulary}
\caption[Numbers of branes in each theory down to $g_s^{-25}$.]{Number of branes in each theory down to $g_s^{-25}$, as well as the total number of branes required in M-theory to accommodate all of them.}
\label{tab:gsGradingOfExoticBranes}
\end{table}
\end{landscape}
\clearpage
}
We end this section by reiterating that our results were obtained by ignoring any possible reductions or T-dualisations along time-like directions due to the difficulties in interpretations that they give rise to rather than any technical difficulties. However, a recent paper \cite{Fernandez-Melgarejo:2019pvx} has suggested that the pattern that we see above no longer holds if one includes such transformations. Although their analysis is in terms of the mixed symmetry potentials, it is simple to translate their work in terms of the branes that they couple to. They found that the S-dual of the $F_{10,10,7,1}$ potential is an $\alpha = - 10$ potential $L_{10,10,7,1}$ which should couple to a ${(-1)}_{10}^{(1,6,3)}$-brane which we understand as some instantonic object with a zero-dimensional worldvolume, analogous to the $\text{D}(-1)$ instanton. In fact, based on the observation that the D-instanton arises as a solution of a Wick-rotated Euclideanised Type IIB theory \cite{Gibbons:1995vg} that admits a geometric parent via a $(1,1)$-signature reduction of a wave in a $(1,11)$-signature F-theory \cite{Tseytlin:1996ne},  and that $\operatorname{SL}(2) \times \mathbb{R}^+$ EFT realises M-/F-theory duality, we anticipate that it should be relatively straightforward to combine these ideas to describe such objects in EFTs. Since we have argued that EFT allows us to promote the exotic branes to an equal footing to the conventional branes, it should not be too much of a stretch to suggest that EFT should also be able to describe the non-standard objects of the sort described in \cite{Fernandez-Melgarejo:2019pvx}.
\section{M-Theory Origins of Type IIA branes}
In Appendix~\ref{app:MOrigin}, we have collated all of the M-theory lifts of every Type IIA brane that we have introduced up to this point, as well as all of their reductions. Since every parent in M-theory may be reduced in multiple ways, the existence of any one brane in Type IIA indicates the existence of multiple `siblings' obtained in this manner. Every single brane down to $g_s^{-7}$ have been housed in one of the duality orbits and so any gaps in the figure references correspond to lower powers of $\alpha$.\par
The format should be self-explanatory: the left-most brane in each column is an M-theory brane whilst the branes to the right of each brace are all the possible reductions that one can obtain from that brane. We stress that it is only within EFT that one may `reduce' along a non-isometric direction since this corresponds only to a re-identification of section. In particular, one may still `reduce' transverse to a codimension-1 brane in M-theory to give a codimension-0 brane in ten dimensions without worrying about isometries and without obtaining a trivial result; the non-trivial structure is encoded in the dependence on wrapping directions, which distinguishes between the space-filling branes in 10-dimensions.
\section{Comparison with the Literature}
\subsection{Known Exotic Branes in the Literature}
A small subset of the exotic branes presented here have appeared in the literature before. The starting point are, of course, the `standard' branes appearing at $g_s^0$ (P and F1), $g_s^{-1}$ (D$p$-branes) and $g_s^{-2}$ (NS5 and KK5). Additionally, the existence of the $7_3$-brane as the S-dual of the D7 has been known for a long time with much work being conducted in the context of the $(p,q)$ 7-branes of F-theory. This was included amongst the codimension-2 exotic branes of \cite{LozanoTellechea:2000mc,deBoer:2012ma} (note that the former uses an alternative notation to what we use, e.g. the $5_2^2$ here is called an NS$5_2$ there). The latter also gives a detailed exposition of the T-duality chain $\text{NS5}=5_2 \xrightarrow{\text{T}} \text{KK5}=5_2^1 \xrightarrow{\text{T}} 5_2^2$. This prototypical chain was extended to include the $5_2^3$ in \cite{Bakhmatov:2016kfn} and then, more recently, a novel $5_2^4$-brane from DFT considerations in \cite{Kimura:2018hph} (it is worth mentioning that some of their exotic brane junctions in F-theory have been studied in \cite{Kimura:2016yqa}). This whole five-brane chain matches the work presented here, specifically Figure~\ref{fig:52Orbit}.\par
Lower codimension objects are even less well-studied and understood and there is limited literature on the subject. However, it has been known since, at least, \cite{Hull:1998mh,Obers:1998fb} (and references therein) that a massive deformation of 11-dimensional supergravity admits a domain wall solution in M-theory which has since appeared under various names such as the M9 in \cite{Bergshoeff:1998bs} or KK9M in \cite{Ortin:2015hya}. However, as remarked in \cite{Obers:1998fb}, it should more properly be called a KK8 following its mass formula designation $8^{(1,0)}$. It is, perhaps, to be understood as an object that exists only as a lift of the D8-brane of Type IIA. The remaining reductions of the $8^{(1,0)}$ are the $7_3^{(1,0)}$ and $8_4^{(1,0)}$ which were also recorded in \cite{Ortin:2015hya} (though named as the KK8A and KK9A respectively there, with the same caveat as above).\par
Finally, much like the D7-brane, the D9-brane also has an exotic S-dual (previously called an S9 or an NS9) but which we designate as a $9_4\text{B}$, as was done in \cite{Obers:1998fb}.
\subsection{Mixed-Symmetry Potentials in the Literature}\label{sec:MSP}
Recently, much work has been done on the mixed-symmetry potentials that these exotic branes couple to \cite{Bergshoeff:2017gpw,Lombardo:2016swq}. These have focused on trying to classify the T-duality orbits starting from the highest weight representations of the Lie algebra but miss out on the T-duality orbits that require S-dualities and/or lifts to M-theory to obtain. Nonetheless, there is significant overlap between their work and the work presented here and we summarise this in Table~\ref{tab:BranePotentialCorrespondence}.\par
Another piece of work we can compare to is \cite{Kleinschmidt:2011vu} in which a similar set of potentials were derived from $E_{11}$ and the tensor hierarchy associated to it. One may verify that the majority of the potentials that they obtain for Type II coincide with ours. Those that they are missing are, again expected to be those that appear in the $d=2,1,0$ duality groups whilst those that we are missing are expected to turn up at lower $g_s$ scaling (note that they have not organised their results in powers of $g_s$, complicating the comparison of results).
\afterpage{%
\clearpage
\renewcommand{\arraystretch}{0.8}
\thispagestyle{empty}
\begin{landscape}
\begin{table}[!ht]
\centering
\captionsetup{width=0.9\linewidth}
\begin{tabulary}{1.25\linewidth}{CCCLp{7cm}L}
\toprule
\multirow{2}{*}{$\alpha$} & \multicolumn{2}{c}{Potentials} & \multirow{2}{*}{Figure} & \multirow{2}{*}{Conditions} & \multirow{2}{*}{Notes}\\
\cmidrule{2-3}
& IIA & IIB & &\\
\midrule
0 & \multicolumn{2}{c}{$B_2$} & \ref{fig:10Orbit} & & NS-NS\\
-1 & $C_{2n+1}$ & $C_{2n}$ & \ref{fig:11Orbit} & $n \in \{0,1,2,3,4\}$ & R-R\\
-2 &\multicolumn{2}{c}{$D_{6+n,n}$} & \ref{fig:52Orbit} & $n\in \{0,1,2,3,4\}$ & 5-brane chain\\
-3 & $E_{8+n, 2m+1, n}$ & $E_{8+n,2m,n}$ & \ref{fig:532Orbit} & $n \in \{0,1,2\}, m \in \{1,2,3,4\}$ &\\
-4 & \multicolumn{2}{c}{$F_{8+n,6+m,m,n}$} & \ref{fig:146Orbit} & $n \in \{0,1,2\}, m \in \{n, n+1\}$ &\\
& \multicolumn{2}{c}{$F_{9+n,3+m,m,n}$} & \ref{fig:4413Orbit} & $n \in \{0,1\}; m \in \{n, n+1, \ldots, n+5\}$ &\\
& $F_{10,2n+1,2n+1}$ & $F_{10,2n,2n}$ & \ref{fig:7420Orbit} & $n \in \{0, \ldots, 4\}$ &\\
-5 & $G_{9+p,6+n,2m,n,p}$ & $G_{9+p, 6+n, 2m+1, n,p}$ & \ref{fig:2515Orbit} & $p \in \{0,1\}, n \in \{ p, p+1, p+2\}$ \newline $n \leq 2m,2m+1 \leq n+6$ & \\
& $G_{10,4+n,2m+1,n}$ & $G_{10,4+n,2m,n}$ & \ref{fig:5522Orbit} & $n \in \{0, 1, \ldots, 5\}; n \leq 2m, 2m+1 \leq n+4$ &\\
-6 & \multicolumn{2}{c}{$H_{10, 6+n, 2+m, m,n}$} & \ref{fig:3624Orbit} & $n\in\{0,1,2,3\}; m \in \{n, n+1, \ldots, n+4\}$ &\\
& \multicolumn{2}{c}{$H_{9+n,8+n,m+n,m+n-1,n,n}$} & \ref{fig:0617Orbit} & $m\in \{1,2, \ldots, 8\}, n \in \{0,1\}$ & \multicolumn{1}{c}{$\ddagger$}\\
& \multicolumn{2}{c}{$H_{9+p, 7+n, 4+m, m,n,p}$} & \ref{fig:1643Orbit} & $p \in \{0,1\}, n \in \{p,p+1\}$ \newline $m \in \{n, n+1, n+2, n+3\}$ & \multicolumn{1}{c}{$\dagger$}\\
-7 & $I_{10,8+p, n+2, 2m+1, n, p}$ & $I_{10, 8+p, 2+n, 2m, n, p}$ & \ref{fig:1726Orbit} & $p\in\{0,1\}, n \in \{p, p+1, \ldots, p+6\}$ \newline $n \leq 2m, 2m+1 \leq 2+n$ & \multicolumn{1}{c}{$\ddagger$}\\
& $I_{9+p, 8+p, 5+n+p, 2m+1, n+p, p, p}$ & $I_{9+p, 8+p, 5+n+p, 2m, n+p, p, p}$ & \ref{fig:0753Orbit} & $p \in \{0,1\}, n \in \{0,1,2,3\}$ \newline $n+p \leq 2m,2m+1 \leq n+p+5$ & \multicolumn{1}{c}{$\ddagger$}\\
& $I_{9+p,p+n+7,p+n+7,2m,n+p,n+p,p}$ & $I_{9+p,p+n+7,p+n+7,2m+1,n+p,n+p,p}$ & \ref{fig:17160Orbit} & $n \in \{0,1\}, p \in \{0,1\}$, \newline $n+p \leq 2m, 2m+1 \leq n+p+7$ & \multicolumn{1}{c}{$\dagger$}\\
& $I_{10,7+p,4+n+p,2m,n+p,p}$ & $I_{10,7+p,4+n+p,2m+1,n+p,p}$ & \ref{fig:27133Orbit} & $p\in \{0,1,2\}, n \in \{0,1,2,3\}$\newline $n+p \leq 2m, 2m+1, \leq, 4 +n +p$ & \multicolumn{1}{c}{$\dagger$}\\
& $I_{10, 6+n, 6+n, 2m+1, n,n}$ & $I_{10, 6+n, 6+n, 2m, n,n}$ & \ref{fig:3760Orbit} & $n \in \{0,1,2,3\}; n \leq 2m,  2m+1 \leq, 6+n$ &\\
\bottomrule
\end{tabulary}
\caption[The mixed symmetry potentials that the exotic branes couple to.]{The mixed symmetry potentials that the exotic branes couple to. Each of the branes in the referenced T-duality orbit couple to one type of potential, listed in the second and third columns. Those potentials that straddle the two columns are common to both. Most of the lower $g_s^\alpha$ potentials have not been found yet. Note that we use $\dagger$ and $\ddagger$ to indicate that the potential was not included in the analysis of \cite{Bergshoeff:2017gpw} and \cite{Bergshoeff:2017gpw,Fernandez-Melgarejo:2018yxq} respectively. However, a more recent paper \cite{Fernandez-Melgarejo:2019mgd} contains significant overlap with the groups of potentials we have listed here.}
\label{tab:BranePotentialCorrespondence}
\end{table}
\end{landscape}
\clearpage
}
Note that \cite{Fernandez-Melgarejo:2018yxq} obtains many of the branes that we have, through studying U-duality multiplets\footnote{We would like to thank the authors of \cite{Fernandez-Melgarejo:2018yxq} for pointing out an issue with the NS9-brane in an earlier version of this paper which we have now rectified. We have removed three small spurious orbits (one each at $g_s^{-2}, g_s^{-4}$ and $g_s^{-6}$ respectively) following their comments.} and we find good agreement for the portions that overlap, specifically to $g_s^{-7}$. Here, we spell out the correspondence between their potentials (right-hand side) and ours (left-hand side) for the $g_s^{-6}$ and $g_s^{-7}$ potentials only since the other potentials should hopefully be self-evident.
\begin{align}
\begin{array}{ccl}
{\left. \left\{ \begingroup \renewcommand{\arraystretch}{1.25} \begin{array}{c} H_{10,6+n,2+m,m,n} \end{array} \endgroup \right\} \right|}_{m \rightarrow m+n} & \longleftrightarrow & E_{10,6+n,2+m+n,m+n,n}^{(6)}\\
{\left. \left\{ \begingroup \renewcommand{\arraystretch}{1.25} \begin{array}{c} H_{9+p,7+n,4+m,m,n,p} \end{array} \endgroup \right\} \right|}_{p=n=0} & \longleftrightarrow & E_{9,7,4+n,n}^{(6)}\\
{\left. \left\{ \begingroup \renewcommand{\arraystretch}{1.25} \begin{array}{c}  I_{9+p,p+n+7,p+n+7,2m,n+p,n+p,p}\\ I_{9+p,p+n+7,p+n+7,2m+1,n+p,n+p,p}\\ \end{array} \endgroup \right\} \right|}_{p=n=0} & \longleftrightarrow & E_{9,7,7,q}^{(7)}\\
{\left. \left\{ \begingroup \renewcommand{\arraystretch}{1.25} \begin{array}{c} I_{10,7+p,4+n+p,2m,n+p,p}\\  I_{10,7+p,4+n+p,2m+1,n+p,p}\\ \end{array} \endgroup \right\} \right|}_{p=0} & \longleftrightarrow & E_{10,7,4+n,q,n}^{(7)}\\
{\left. \left\{ \begingroup \renewcommand{\arraystretch}{1.25} \begin{array}{c} I_{10,6+n,6+n,2m+1,n,n}\\ I_{10,6+n,6+n,2m,n,n}\\ \end{array} \endgroup \right\}  \right|}_{p=0} & \longleftrightarrow & E_{10,6+n,6+n,q,n,n}^{(7)}\\
\end{array}
\end{align}
\section{Discussion}\label{sec:Discussion}
Throughout the main text, we have focused on two notions of non-geometry. The first signs of non-geometry are the \emph{globally non-geometric objects}, of the type that we started our discussion with, such as the Q-monopole (smeared $5_2^2$). Whilst a globally geometric description of such objects is not possible, since their patching require duality transformations in addition to the conventional diffeomorphisms and gauge transformations, one can still construct local descriptions of these objects through the supergravity fields. These are thus realisations of the T-folds and U-folds proposed by Hull.\par
The second class of non-geometric objects are those that are \emph{locally non-geometric}. These are backgrounds that require a dependence on coordinates outside of the usual spacetime to describe and thus lack even a local description in terms of conventional supergravity. It is hopefully obvious that all the codimension-0 branes that we have argued to exist must necessarily be of this type to admit a non-trivial structure. What is less obvious is that higher codimension objects can also be non-geometric in this sense---the prime example being the $5_2^3$ brane (indeed, this is the context in which this type of non-geometry was first discussed). Explicit construction of the background shows that the structure of the fields necessitates a dependence on at least one winding mode if the solution is to remain non-trivial. We expect this winding mode dependence to then be interpreted as worldsheet instanton corrections in the conventional supergravity lore, as suggested by the GLSM. The vast majority of the branes that we have recorded are expected to be of this type.\par
To these, we add a final speculative type of non-geometry that we shall refer to as, `truly non-geometric backgrounds'---backgrounds that are not related to any geometric background by duality transformations, thus forming entirely disconnected orbits from the ones that we have constructed. The very nature of how we generated our non-geometric backgrounds prevents us from probing such backgrounds and it is not obvious how one might go about constructing such backgrounds. Indeed, it is not clear if such objects even exist and it remains an open question if there are more general objects than the ones we have found.
\subsection{Unification at Larger Duality Groups}
We have touched on it a number of times already but the proliferation of exotic branes is self-evident from the figures in Appendix~\ref{app:ExoticWeb} and Table~\ref{tab:gsGradingOfExoticBranes}. Our procedure does not tell us whether the process will even terminate at all. However, the growing number of DFT and EFT solutions found to date, including the DFT monopole \cite{Berman:2014hna,Bakhmatov:2016kfn,Kimura:2018hph}, the $E_{7(7)}$ geometric solution \cite{Berman:2014hna} and the non-geometric solution presented here, all point to the over-arching theme of unification of branes in higher dimensions. Just as the possible wrappings of the M2 were found to give a unifying description of the F1 and D2 in one dimension higher, multiple branes have lifted to single solutions in DFT and EFT. That they only unify a small fraction of the branes that we have described is not a problem and, indeed, is probably to be expected given the awkward split between internal and external spaces that is inherent in EFT which puts a restriction on which branes can be lifted to the same solution within that EFT. More exciting is the possibility that every single brane that we have presented here should all lift to one unified solution in ExFT at higher duality groups.\par
The rationale behind this claim is as follows. We have already mentioned that many of the novel branes that we have found at codimension-1 and 0 have not been found in the literature simply because previous efforts such as \cite{Bergshoeff:2017gpw,Fernandez-Melgarejo:2018yxq} have always classified them under U-duality representations of the reduced exceptional groups, typically down to $d=3$. It is thus natural to expect that the novel branes presented here will only appear when one considers reductions down to $d=2,1$ or even $d=0$. Put differently, if one were to consider the largest duality groups, one should be able to accommodate more and more of them until every single brane presented here is accounted for. This is also consistent with the observation that the procedure does not appear to have any clear termination point---it is still very much possible that there are an infinite number of exotic branes whose wrapping modes are then used to construct the infinite-dimensional extended spaces of the highest ExFTs.\par
We further note that since every one of the figures are inter-related by S- and T-dualities, the lift to M-theory means that every one of those figures are part of a single U-duality orbit. From the discussion before, it is tempting to call it a single U-duality orbit of $E_{11}$ that fragments to smaller U-duality orbits only when one descends down the $E_n$-series. See \cite{West:2001as,West:2003fc,Tumanov:2015yjd,Tumanov:2016abm,Kleinschmidt:2003jf,Cook:2008bi,West:2004kb,West:2004iz,Cook:2009ri,Cook:2011ir} for a discussion on both standard and exotic branes in the context of the $E_{11}$ program. According to this conjecture one may thus only construct `truly' non-geometric objects (in the sense that we described above such that there are no U-duality transformations that can transform them to geometric solutions) within the smaller duality groups; what appear to be distinct orbits in those groups should successively merge into fewer and fewer orbits of the higher duality groups until one is left with only a single U-duality orbit at $E_{11}$. Thus, whilst we now have two distinct solutions in $E_{7(7)}$ EFT covering different sets of branes and with no apparent way to transition between the two (see Figure~\ref{fig:Brane}), one might expect that these two EFT solutions can be unified into a single solution of a larger EFT (perhaps along with a whole range of other exotic branes).

\part{Non-Riemannian Geometries in ExFTs}
	\chapter{Classification of Non-Riemannian Solutions in DFT}\label{ch:DFTNonRiemannian}
In the previous chapter, we considered exotic branes as a particular class of non-geometric solutions that can be described within ExFTs. Here we consider a rather different class of backgrounds, namely \emph{non-Riemannian backgrounds}. Whereas the solutions in the previous section were characterised by either a lack of a global geometric description, owing to requiring duality transformations to patch correctly, or a lack of a local geometric description, due to a dependence of coordinates outside of the physical spacetime, the solutions we consider here are exotic in that they do not admit even local descriptions in terms of an invertible Riemannian metric. The definition is rather broad and includes various singular limits of the metric that obstruct its inversion.\par
The key to describing such backgrounds is realising that fact that the generalised metric can remain regular in such backgrounds, even if the spacetime metric becomes singular, due to the presence of the off-diagonal terms in the generalised metric that can compensate for it. This fact was already appreciated in \cite{Lee:2013hma} where it appeared in the context of the doubled sigma model. Their work was then extended in \cite{Morand:2017fnv,Cho:2019ofr} to a full characterisation of the possible backgrounds that one can obtain in DFT by solving the $\operatorname{O}(D,D)$ constraints on the generalised metric in generality. The resulting classification is given in terms of two non-negative integers $(n, \bar{n})$ with $0 \leq n + \bar{n} \leq D$ that determine which coset the generalised metric is chosen to parametrise. This allows one to describe various non-Riemannian limits including non-relativistic geometries, such as Newton-Cartan geometries or Gomis-Ooguri-type limits, or ultra-relativistic limits, such as Carrollian geometries, all within a framework that was originally developed to describe regular supergravities.\par
Of particular interest to us is the $(n, \overbar{n}) = (D,0)$ case which gives rise to a `maximally non-Riemannian' solution. Quite remarkably, and in contrast to conventional reductions which lead to scalar moduli in the resulting spectrum, it was shown in \cite{Cho:2018alk} that employing such a background as the internal space of a Kaluza-Klein reduction of DFT led to a background that is moduli-free.\par
The purpose of this chapter and the next is to extend theses ideas to EFT, with a particular focus on $E_{8(8)}$ EFT. In particular, Whilst we shall not be able to give a full classification of the possible parametrisations of the generalised metric as was done in the DFT case, we show that one can construct an analogue of the maximally non-Riemannian solution of DFT that naturally reduces to the `topological phase' of $E_{8(8)}$ EFT that was described in \cite{Hohm:2018ybo}.
\section{The \texorpdfstring{$\operatorname{O}(D,D)$}{O(D,D)} Constraints Revisited}
The material covered in this section was first described in \cite{Morand:2017fnv} and we cover it briefly to illustrate the role played by the choice of coset that the generalised metric is chosen to parametrise. The constraints on the  generalised metric $\mathcal{M}_{MN}$ in DFT read
\begin{align}\label{eq:DFTCompatibilityConditions}
\mathcal{M}_{MN} = \mathcal{M}_{NM}\,, \qquad \mathcal{M}_{MP} \eta^{PQ} \mathcal{M}_{QN} = \eta_{MN}\,.
\end{align}
The general solution to these is worked out in detail in \cite{Morand:2017fnv} where it was found that generic solutions to these constraints could be parametrised by two non-negative integers $(n,\bar{n})$ according to
\begingroup
\renewcommand*{\arraystretch}{2.3}
\begin{align}\label{eq:DFTGenMetricParam}
\mathcal{M}_{MN} & = \begin{pmatrix}
\begingroup
\renewcommand*{\arraystretch}{1.0}
\begin{array}{cc}K_{mn} - B_{mp} H^{pq} B_{qn}\\
+ 2 X^a_{(m} B_{n)q} Y^q_a - 2 {\overbar{X}}^{\overbar{a}}_{(m} B_{n)q} {\overbar{Y}}^q_{\overbar{a}}\\ \end{array}
\endgroup
& B_{mq} H^{qn} + X^a_m Y^n_a - {\overbar{X}}_m^{\overbar{a}} {\overbar{Y}}^n_{\overbar{a}}\\
- H^{m q} B_{qn} + Y^m_a X^a_n - {\overbar{Y}}_{\overbar{a}}^m {\overbar{X}}^{\overbar{a}}_n & H^{mn}
\end{pmatrix}\,,
\end{align}
\endgroup
where $H^{mn}$ and $K_{mn}$ are symmetric tensors and $B_{mn}$ is skew-symmetric. The remaining objects $\{ X_m^a, {\overbar{X}}_m^{\overbar{a}}\}$ and $\{ Y^m_a, {\overbar{Y}}^m_{\bar{a}}\}$ span the kernels of $H^{mn}$ and $K_{mn}$ respectively (with indies running over $a = 1, \ldots, n$ and $\overbar{a} = 1, \ldots , \overbar{n}$) according to
\begin{align}
H^{mn} X_n^a = 0\,, \qquad H^{mn} {\overbar{X}}_n^{\overbar{a}} = 0\,, \qquad K_{mn} Y^n_a = 0\,, \qquad K_{mn} {\overbar{Y}}^n_{\overbar{a}} = 0\,.
\end{align}
In particular, the kernels of both $H^{mn}$ and $K_{mn}$ are both of dimension $n + \overbar{n}$ and so we restrict our considerations to
\begin{align}\label{eq:KernelBounds}
0 \leq n + \overbar{n} \leq D\,, \qquad n , \overbar{n} \in \mathbb{Z}^+\,.
\end{align}
We stress that we do not assume any other properties of the fields, including any invertibility of the fields. Indeed, whilst the form above may suggest that $K_{mn}$ and $H^{mn}$ could be interpreted as a spacetime metric and its inverse respectively, if $K_{mn}$ and $H^{mn}$ are not full rank they will be neither invertible nor inverses of each other. There is a further completeness relation that the fields satisfy, given by
\begin{align}
H^{mq} K_{qn} + Y^m_a X_n^a + {\overbar{Y}}^m_{\overbar{a}} {\overbar{X}}_n^{\overbar{a}} = \delta^m_n\,,
\end{align}
that can be used to show the following compatibility conditions:
\begingroup
\renewcommand{\arraystretch}{1.5}
\begin{align}
\begin{array}{c}
Y^m_a X^b_m = \delta_a^b\,, \qquad {\overbar{Y}}^m_{\overbar{a}} {\overbar{X}}_m^{\overbar{b}} = \delta_{\overbar{a}}^{\overbar{b}}\,, \qquad Y^m_a {\overbar{X}}_m^{\overbar{b}} = {\overbar{Y}}_{\overbar{a}}^m X_m^b = 0\,,\\
H^{mp} K_{pq} H^{qn} = H^{mn}\,, \qquad K_{mp} H^{pq} K_{qn} = K_{mn}\,.\\
\end{array}
\end{align}
\endgroup
Using these relations, one may verify that the trace of the generalised metric is given by $\mathcal{M}^{M}{}_M = 2 (n - \overbar{n})$. As in the usual DFT parametrisation, the action of the $B$-field can be factored out into a conjugation,
\begin{align}
\mathcal{M}_{MN} = \begin{pmatrix}
\delta_m^p & B_{mp}\\
0 & \delta^m_p
\end{pmatrix}
\begin{pmatrix}
K_{pq} & X^a_p Y_a^q - {\overbar{X}}_p^{\overbar{a}} {\overbar{Y}}_{\overbar{a}}^q \\
Y^p_a X^a_q - {\overbar{Y}}^p_{\overbar{a}} {\overbar{X}}^{\overbar{a}}_q & H^{pq}
\end{pmatrix}
\begin{pmatrix}
\delta^q_n& 0\\
- B_{qn} & \delta^n_q
\end{pmatrix}\,,
\end{align}
which demonstrates that its existence is independent of the values of $(n,\bar{n})$.\par
The action of the generalised Lie derivative on the generalised metric, with parameter $X^M = (x^m, \xi_m)$, dictates transformations of the fields:
\begingroup
\renewcommand{\arraystretch}{1.5}
\begin{align}
\begin{array}{c}
\delta H^{mn} = \mathcal{L}_\xi H^{mn}\,, \qquad \delta K_{mn} = \mathcal{L}_\xi K_{mn}\,, \qquad \delta B_{mn} = \mathcal{L}_x B_{mn} + 2 \partial_{[m} {\tilde{\xi}}_{n]}\\
\delta X_m^a = \mathcal{L}_\xi X_m^a\,, \qquad \delta {\overbar{X}}_m^{\overbar{a}} = \mathcal{L}_\xi {\overbar{X}}_m^{\overbar{a}}\,, \qquad \delta Y_a^m = Y_a^m\,, \qquad \delta {\overbar{Y}}_{\overbar{a}}^m = \mathcal{L}_\xi {\overbar{Y}}_{\overbar{a}}^m\,.
\end{array}
\end{align}
\endgroup
The generalised metric, as parametrised in \eqref{eq:DFTGenMetricParam}, possesses the following symmetries. The first is a $\operatorname{GL}(n) \times \operatorname{GL}(\overbar{n})$ symmetry that acts on the indices $a, b,\ldots$ and $\overbar{a}, \overbar{b}, \ldots$ under which the fields transform in the obvious way:
\begingroup
\renewcommand{\arraystretch}{1.5}
\begin{align}
\begin{array}{lcl}
X_m^a \mapsto X_m^a R_a{}^B\,, && Y_a^m \mapsto {\left( R^{-1} \right)}_a{}^b Y_b^m\,,\\
{\overbar{X}}_m^{\overbar{a}} \mapsto {\overbar{X}}_m^{\overbar{b}} {\overbar{R}}_{\overbar{b}}{}^{\overbar{a}}\,, && {\overbar{Y}}_{\overbar{a}}^m \mapsto {\left( {\overbar{R}}^{-1} \right)}_{\overbar{a}}{}^{\overbar{b}} {\overbar{Y}}_{\overbar{b}}^m\,,\\
\end{array}
\end{align}
\endgroup
whilst leaving the remaining fields inert. The second is the less obvious transformations
\begingroup
\renewcommand{\arraystretch}{1.5}
\begin{align}
Y_a^m & \mapsto Y_a^m + H^{mn} V_{na}\\
{\overbar{Y}}_{\overbar{a}}^m & \mapsto {\overbar{Y}}_{\overbar{a}}^m  + H^{mn} {\overbar{V}}_{n \overbar{a}}\\
K_{mn} & \mapsto \begin{array}[t]{ll} K_{mn} - 2 X^a_{(m} K_{n)p} H^{pq} V_{q a} - 2 {\overbar{X}}^{\overbar{a}}_{(m} K_{n) p} H^{pq} {\overbar{V}}_{q \overbar{a}}\\ + (X^a_m V_{pa} + {\overbar{X}}_m^{\overbar{a}} {\overbar{V}}_{p \overbar{a}} ) H^{pq} (X^b_n V_{qb} + {\overbar{X}}_n^{\overbar{b}} {\overbar{V}}_{q \overbar{b}}\end{array}\\
B_{mn} & \mapsto \begin{array}[t]{ll} B_{mn} - 2 X^a_{[m} V_{n]a} + 2 {\overbar{X}}^{\overbar{a}}_{[m} {\overbar{V}}_{n]\overbar{a}}\\ + 2 X^a_{[m} {\overbar{X}}^{\overbar{a}}_{n]} ( Y_a^q {\overbar{V}}_{q \overbar{a}} + {\overbar{Y}}_{\overbar{a}}^q V_{qa} + V_{qa} H^{qp}{\overbar{V}}_{p \overbar{a}})\end{array}
\end{align}
\endgroup
in terms of arbitrary local parameters $V_{ma}$ and ${\overbar{V}}_{m\overbar{a}}$. It can be understood as a generalisation of the Galilean boosts in the Newtonian gravity literature, called Milne boosts.\par
Following the formulation of \cite{Jeon:2011vx,Jeon:2012kd,Lee:2013hma}, we define the symmetric projection matrices
\begin{align}
\mathcal{P}_M{}^N = \frac{1}{2} \left(\delta_M^N + \mathcal{M}_{MP} \eta^{PN}\right)\,, \qquad {\overbar{\mathcal{P}}}_M{}^N = \frac{1}{2} \left(\delta_M^N - \mathcal{M}_{MP} \eta^{PN} \right)\,.
\end{align}
It is easy to verify that these these are complete $\mathcal{P}_M{}^N + {\overbar{\mathcal{P}}}_M{}^N = \delta_M^N$ and thus orthogonal $\mathcal{P}_M{}^N {\overbar{\mathcal{P}}}_N{}^Q = 0$. Additionally, their traces in the general parametrsiation \eqref{eq:DFTGenMetricParam} are given by
\begin{align}\label{eq:DFTProjectorTraces}
{\mathcal{P}}_M{}^M = D + n - \overbar{n}\,, \qquad {\overbar{\mathcal{P}}}_M{}^M = D - n + \overbar{n}
\end{align}
which follow from the orthogonality of the pairs $(X,Y)$ and $(\overbar{X}, \overbar{Y})$ (it is convenient to note that we may disregard the $B$-field to simplify this calculation by the cyclicity of the trace). For convenience, we also note that the inverse relations are given by
\begin{align}\label{eq:ProjectorMEta}
\begin{rcases}
\begin{array}{l}
\mathcal{P}_{MN} = \frac{1}{2}( \eta_{MN} + \mathcal{M}_{MN} )\\
{\overbar{\mathcal{P}}}_{MN} = \frac{1}{2}( \eta_{MN} - \mathcal{M}_{MN} )
\end{array}
\end{rcases}
\qquad \Leftrightarrow \qquad
\begin{cases}
\begin{array}{r}
\eta_{MN} = \mathcal{P}_{MN} + {\overbar{\mathcal{P}}}_{MN}\,,\\
\mathcal{M}_{MN} = \mathcal{P}_{MN} - {\overbar{\mathcal{P}}}_{MN}\,.\\
\end{array}
\end{cases}
\end{align}
The projector on the ExFT equations of motion, introduced in Section~\ref{sec:ProjectedEOM}, can then be written in terms of these projectors as\footnote{Indeed, defining an analogue of the Christoffel connection in doubled geometry gives rise to the particular form of the equations of motion (see, for example \cite{Jeon:2011cn,Aldazabal:2013sca})
\begin{align}
\mathcal{P}_M{}^{(K} {\overbar{\mathcal{P}}}_N{}^{K)} \mathcal{R}_{KL} = 0\,,
\end{align}
where $\mathcal{R}_{KL}$ is an appropriately defined generalised Ricci scalar. Such a form does not seem to be possible in EFT since only the two-index projector $\mathcal{P}_{MN}{}^{KL}$ has an appropriate generalisation in EFT.
}
\begin{align}\label{eq:FactorisedProjectors}
\mathcal{P}_{MN}{}^{KL} & = 2 \mathcal{P}_M{}^{(K} {\overbar{\mathcal{P}}}_N{}^{L)}\,.
\end{align}
 We note in passing that the universal weight $\omega$ of ExFTs generically introduces a term that acts as an obstruction to expressing the projector $\mathcal{P}_{MN}{}^{KL}$ in terms of a pair of projectors $(\mathcal{P}_M{}^N, {\overbar{\mathcal{P}}}_M{}^N)$ in this form. More concretely, since all EFTs possess a non-trivial universal weight, the fact that DFT admits such a factorisation is to be seen more as a quirk of the theory rather than a statement that holds more generally.\par
For later, we also rewrite the $\operatorname{O}(D,D)$ constraint \eqref{eq:DFTCompatibilityConditions} in terms of these projectors. We begin with
\begin{align}
\delta {\mathcal{M}}_{MN} & = \eta_{MP} \delta \mathcal{M}^{PQ} \eta_{QN} = ( \mathcal{P}_M{}^P + {\overbar{\mathcal{P}}}_M{}^P )  \delta \mathcal{M}_{PQ} ( {\mathcal{P}}^Q{}_N + {\overbar{\mathcal{P}}}^Q{}_N)\,.
\end{align}
Of the four terms that arise, only the cross-terms remain since two of them vanish by
\begin{align}
\mathcal{P}_M{}^P \delta \mathcal{M}_{PQ} \mathcal{P}^Q{}_N & = 2 \mathcal{P}_M{}^P \delta \mathcal{P}_{PQ} \mathcal{P}^Q{}_N\\
	& = 2 \left( \delta \mathcal{P}_{MQ} - \delta \mathcal{P}_M{}^P \mathcal{P}_{PQ} \right) \mathcal{P}^Q{}_N\\
	& = 2 \left( \delta \mathcal{P}_{MQ} \mathcal{P}^{Q}{}_N - \delta \mathcal{P}_M{}^Q \mathcal{P}_{QN} \right) = 0\,.
\end{align}
We are thus left with
\begin{align}\label{eq:ODDConstraintProjectors}
\delta \mathcal{M}_{MN} & = \mathcal{P}_M{}^P \delta \mathcal{M}_{PQ} {\overbar{\mathcal{P}}}^Q{}_N + {\overbar{\mathcal{P}}}_M{}^P \delta {\mathcal{M}}_{PQ} \mathcal{P}^Q{}_N\,.
\end{align}
\section{The Park-Morand Classification}
In this section, we study some examples of the backgrounds that are contained within the Park-Morand classification. The list is far from exhaustive and, in addition to the ones covered below, DFT is known to accommodate at least Newton-Cartan geometries $(n, \overbar{n}) = (0,1)$ and Carroll geometries $(n,\overbar{n}) = (D-1,0)$. However, they are not particularly relevant for our discussions and so the reader is directed to the original papers for details of those embeddings.
\subsection{The Usual Parametrisation}
The usual parametrisation of the generalised metric corresponds to $(n, \overbar{n}) = (0,0)$. In this case, both $H^{mn}$ and $K_{mn}$ have full rank (all the null vectors $\{ X, Y, \overbar{X}, \overbar{Y}\}$ vanish) and the completeness relation reduces to the condition that the two are inverses of each other. Upon the obvious identification $g_{mn} = K_{mn}$, we are left with the usual parametrisation of the generalised metric:
\begin{align}
\mathcal{M}_{MN} = \begin{pmatrix}
g_{mn} - B_{mp} g^{pq} B_{qn} & B_{mq} g^{qn}\\
- g^{mq} B_{qn} & g^{mn}
\end{pmatrix}\,.
\end{align}
Given the explicit parametrisation (or using \eqref{eq:DFTProjectorTraces}), it is easy to check that $\mathcal{P}_M{}^M = {\overbar{\mathcal{P}}}_M{}^M = D$ and that, consequently, the dimension of the coset is given by
\begin{align}
{\mathcal{P}}_{MN}{}^{MN} = D^2 = \underbrace{\frac{D(D+1)}{2}}_{\text{in } g_{mn}} + \underbrace{\frac{D(D-2)}{2}}_{\text{in } B_{mn}}\,.
\end{align}
We have mentioned previously that this form of the DFT generalised metric can be constructed from a non-linear realisation of the coset $G/H$, where $H = \operatorname{O}(D) \times \operatorname{O}(D)$ is the maximal compact subgroup of $G = \operatorname{O}(D,D)$. It should come as no surprise that the permissible parametrisations of the generalised metric change the denominator of the coset that $\mathcal{M}_{MN}$ parametrises, since the image of the projectors changes with the parametrisation, and we shall return to this point later.\par
\subsection{The Maximally Non-Riemannian Background}
If we instead choose $(n, \overbar{n}) = (D,0)$, then the barred null vectors $\{{\overbar{X}}_n^{\overbar{a}}, {\overbar{Y}}^m_{\overbar{a}}\}$ vanish. In this case, the solution is maximally non-Riemannian in the sense that the bound \eqref{eq:KernelBounds} is saturated in such a way that the fields $H^{m n}$ and $K_{m n}$ (which were previously identified with the inverse metric and metric respectively) both vanish to leave not even a residual notion of a Riemannian metric. The orthogonality relation between $X$ and $Y$ gives
\begin{align}
Y_a^m X^b_m = \delta_a^b \qquad \Rightarrow \qquad (X^a_n Y_a^m) X_m^b = X_n^b
\end{align}
and so we conclude that $X^a_n Y_a^m = \delta^m_n$, in addition to $X^a_m Y_b^m = \delta^a_b$. Comparing to the completeness relation, which is reduced to
\begin{align}
H^{mq} K_{qn} + Y^m_a X_n^a = \delta^m_n\,,
\end{align}
we see that at least either one of $H^{mn}$ or $K_{mn}$ vanishes. However, the compatibility condition between the two necessitates that the other also vanishes. Thus, the maximally non-Riemannian solution does not admit even a vestigial metric, let alone a non-singular one. We are thus left with only the $B$-field to consider. It is easy to check that the only non-trivial contribution containing it is $2 X^a_{(m} B_{n)q} Y^q_a$ and that it vanishes by the antisymmetry of the $B$-field. We thus end up with
\begin{align}
\mathcal{M}_{MN} & = \begin{pmatrix}
0 & \delta_m^n\\
\delta^m_n & 0
\end{pmatrix}
= \eta_{MN}\,.
\end{align}
Consequently, the various projectors that we have introduced thus far reduce to
\begin{align}\label{eq:VanishingProjectors}
\mathcal{P}_M{}^N = \delta_M^N\,, \qquad {\overbar{\mathcal{P}}}_M{}^N = 0\,, \qquad {\mathcal{P}}_{MN}{}^{KL} = 0\,.
\end{align}
In particular, the last of these is manifestly a solution of the equations of motion of DFT \eqref{eq:ProjectedEOM}, albeit not an `obvious' one. Returning to the $\operatorname{O}(D,D)$ constraint \eqref{eq:ODDConstraintProjectors}, we see that $\mathcal{M}_{MN} = \eta_{MN}$ gives $\delta \mathcal{M}_{MN} = 0$; \emph{the background admits no fluctuations}. This was put to use in \cite{Cho:2018alk} where they considered a Kaluza-Klein reduction of DFT for which the internal space was taken to be this maximally non-Riemannian background. The result was a rigid internal space with no scalar moduli. Put another way, the maximally non-Riemannian solution allows for reductions that are \emph{moduli-free}. It was thus conjectured that the coset structure associated to this parametrisation was $\operatorname{O}(D,D) / \operatorname{O}(D,D) = G/G$, owing to the fact that it does not admit any propagating degrees of freedom. In the context of the doubled sigma model, one finds that this yields Siegel's chiral string\cite{Siegel:2015axg,Casali:2016atr, Casali:2017mss,Lee:2017utr,Lee:2017crr}.
\subsection{The Gomis--Ooguri string}
In addition to the two extremal cases outlined above, there are further non-Riemannian solutions that satisfy the $\operatorname{O}(D,D)$ constraints on the generalised metric. For example, $(n, \overbar{n})= (1,1)$ gives the Gomis--Ooguri string \cite{Gomis:2000bd}. We begin by considering a closed string in flat space $g_{\mu \nu} = \eta_{\mu \nu}$ winding around a circle of radius $R$ in the $x^1 \equiv z$ direction. Splitting the coordinates according to $x^\mu = (x^m, x^a)$, with $x^m = (t, z)$, we rescale the metric of the $x^m$ sector and consider the background
\begin{align}\label{eq:GomisOoguri}
\textrm{d}s^2 = c^2 ( - \textrm{d}t^2 + \textrm{d} z^2 ) + \textrm{d} \vec{x}^2_8\,, \qquad B_{(2)} = (c^2 - \mu) \textrm{d} t \wedge \textrm{d} z\,,
\end{align}
where $c$ is the speed of light and $\mu$ is a constant. Denoting $B_{tz} \equiv B$, the dispersion relation of a string in this background is a modification of \eqref{eq:StringSpectrum} (decorated with factors of $c^2$) given by
\begin{align}\label{eq:GomisOoguriDispersion1}
\frac{1}{c^2} {\left( E + \frac{wRB}{\alpha} \right)}^2 = k^2 + c^2 {\left( \frac{wR}{\alpha^\prime} \right)}^2 + \frac{1}{c^2} {\left( \frac{n}{R} \right)}^2 + \frac{2}{\alpha^\prime} (N + \tilde{N} -2)\,,
\end{align}
where we have split the momentum into energy $E$ and momenta $k^a$ in the $x^a$ directions. The remaining symbols should be self-explanatory; $n$ is the quantum number of momentum in the $z$ direction, $w$ is the winding number around $z$ and $(N, \tilde{N})$ are the excitation numbers of the left- and right-moving oscillators. This is, as always, supplemented by the level-matching condition $N - \tilde{N} = nw$.\par
Taking the non-relativistic limit $c \rightarrow \infty$ and employing a Taylor expansion of the left-hand side, we obtain
\begin{align}
E = \frac{\mu w R}{\alpha^\prime} + \frac{\alpha^\prime k^2}{2 w R} + \frac{N + \tilde{N} -2}{wR}\,.
\end{align}
In particular, one finds both the momentum mode mass and winding mode mass (terms proportional to $n^2$ and $w^2$ respectively) have dropped out since the former vanishes in this limit,
\begin{align}
\lim_{c \rightarrow \infty} \frac{\alpha^\prime n^2}{2 w R c^2} \rightarrow 0,
\end{align}
whilst the latter is cancelled exactly by the divergence in the winding mode charge (the choice of $B_{(2)}$ here is thus crucial in order for this limit to be well-defined). The dispersion relation above is of a Galilean particle with mass and charge $wR/\alpha^\prime$ and chemical potential $\mu$, modified by the intrinsically stringy oscillator contributions \cite{Ko:2015rha}. Remarkably, this limit is singular when inserted into the Polyakov action but remains non-singular if we double the directions $(t,z)$ and lift the fields \eqref{eq:GomisOoguri} to a generalised metric on the doubled space spanned by $X^M = (t,z,\tilde{t}, \tilde{z})$:
\begin{align}
\mathcal{M}_{MN} = \begin{pmatrix}
-2 \mu + \mu^2 G^{-1} & 0 & 0 & 1 - \mu G^{-1} \\
0 & 2 \mu - \mu^2 G^{-1} & 1 - \mu G^{-1} & 0\\
0 & 1 - \mu G^{-1} & - G^{-1} & 0\\
1 - \mu G^{-1} & 0 & 0 & G^{-1}
\end{pmatrix}\,,
\end{align}
where we have defined $G \coloneqq c^2$ to return to the notation of \cite{Berman:2019izh}. This is part of a more general feature of DFT; whilst the backgrounds contained within \eqref{eq:DFTGenMetricParam} are generically singular (with the notable exception of the Riemannian parametrisation) they are nevertheless guaranteed to lift to non-singular, well-defined generalised metrics on the doubled space through its compatibility with the $\operatorname{O}(D,D)$ structure.\par
One may verify that, in the terms of the variables defined in \eqref{eq:DFTGenMetricParam}, the background above corresponds to
\begin{gather}
H^{mn} = 0\,, \qquad K_{mn} = 0\,, \qquad B_{ij} = -\mu \begin{pmatrix} 0 & 1\\ -1 & 0\end{pmatrix}\,,\\
X_m = \frac{1}{\sqrt{2}} \begin{pmatrix}1\\ 1 \end{pmatrix}\,, \qquad Y^m = \frac{1}{\sqrt{2}} \begin{pmatrix}1\\ 1 \end{pmatrix}\,, \qquad {\overbar{X}}_{\overbar{m}} = \frac{1}{\sqrt{2}} \begin{pmatrix}1\\ - 1 \end{pmatrix}\,, \qquad {\overbar{Y}}^{\overbar{m}} = \frac{1}{\sqrt{2}} \begin{pmatrix}1\\ - 1 \end{pmatrix}\,,\nonumber
\end{gather}
in the limit $G \rightarrow \infty$ and so corresponds to a non-Riemannian background with $(n, \overbar{n}) = (1,1)$. The background above can also be understood as a timelike dual to the F1 solution (we work in the string frame)
\begin{gather}
\textrm{d}s^2 = H^{-1} \left( - \textrm{d} t^2 + \textrm{d} z^2  \right) + \textrm{d} \vec{x}_8^2\,, \quad B_{(2)} = - {(H^{-1} -1)} \textrm{d}t \wedge \textrm{d} z\,, \quad e^{-2\phi} = H\,,
\end{gather}
where $H$ is a harmonic function of the transverse coordinates:
\begin{align}
H = 1 + \frac{h}{r^6}\,, \qquad r \equiv | \vec{x}_8|\,.
\end{align}
Doubling the coordinates $(t,z)$ as before, we obtain
\begin{align}
\mathcal{M}_{MN} = 
\begin{pmatrix}
H-2 & 0 & 0 & H -1\\
0 & - (H-2) & H-1 & 0\\
0 & H-1 & - H & 0\\
H-1 & 0 & 0 & H
\end{pmatrix}\,,
\end{align}
together with the doubled dilaton $e^{-2d} = 1$. Dualising in both $t$ and $z$, we obtain the dual background
\begin{align}\label{eq:NegF1GenMetric}
{\tilde{\mathcal{M}}}_{MN} = 
\begin{pmatrix}
\tilde{H} -2 & 0 & 0 & 1-\tilde{H}\\
0 & - (\tilde{H} -2)  & 1-\tilde{H} & 0\\
0 & 1 - \tilde{H} & - \tilde{H} & 0\\
1 - \tilde{H} & 0 & 0 & \tilde{H}
\end{pmatrix}\,,
\end{align}
where we have defined
\begin{align}
\tilde{H} = 2 - H = 1 - \frac{h}{r^6}\,,
\end{align}
which is equivalent to the background
\begin{gather}
\textrm{d}s^2 = {\tilde{H}}^{-1} {(- \textrm{d} t^2 + \textrm{d} \tilde{z}^2)} + \textrm{d}  \vec{x}^2_8\,, \quad B_{(2)} = ( {\tilde{H}}^{-1} - 1)\textrm{d} \tilde{t} \wedge \textrm{d} \tilde{z}\,, \quad e^{-2 \phi} = | \tilde{H}|\,.
\end{gather}
Owing to the difference in sign in the harmonic function, this object has an ADM mass equal to minus that of the F1 and has been dubbed the \emph{negative F1} in the literature (see \cite{Blair:2016xnn} for a discussion of the negative F1 in the context of DFT). It is part of a wider class of objects, called negative branes, that arise when one considers duality transformations along closed time-like curves \cite{Dijkgraaf:2016lym}. Generically quite pathological, they also include Euclidean branes that appear elsewhere in the literature\footnote{Such objects can appear, for example, after a Wick rotation of a standard $(1,9)$ Lorentzian signature string theory.} that possess kinetic terms with the `wrong' sign. Moreover, unlike the standard branes, they possess a singularity when $\tilde{H} = 0$ which has been previously suggested to bound a `bubble' near the brane worldvolume in which the spacetime signature flips for $\tilde{H} < 0$ to one of variants of string- or M-theory with unusual signatures, of the sort studied by Hull in \cite{Hull:1998vg,Hull:1998ym}. Then, the negative tension branes in a $(1,9)$-signature theory (whose pair creation would \emph{release} energy) can be understood as positive tension branes in those exotic-signature theories.\par
The relevance to us is that the DFT generalised metric \eqref{eq:NegF1GenMetric} remains non-singular, even at the transition $\tilde{H} = 0$ for which we have
\begin{align}
{\left. {\tilde{\mathcal{M}}}_{MN} \right|}_{\tilde{H} = 0} = \begin{pmatrix}
-2 & 0 & 0 & 1\\
0 & 2 & 1 & 0\\
0 & 1 & 0 & 0\\
1 & 0 & 0 & 0
\end{pmatrix}\,.
\end{align}
This is precisely of the form of the Gomis-Ooguri limit studied above, with $\mu = 1$. It is then interesting to speculate that perhaps the spacetime turns non-Riemannian at the transition point.
\section{Identifying the Cosets}
 As hinted above, the various parametrisations of the generalised metric are equivalently classified by the coset that it parametrises; fixing a choice of $(n,\overbar{n})$ is equivalent to changing the denominator, which we generically denote as $\tilde{H}$, of the coset $G/\tilde{H}$. For example, the usual parametrisation corresponds to $\tilde{H} = H$ and the maximally non-Riemannian solution corresponds to $\tilde{H} = G$. We shall now discuss the cosets of generic $(n, \overbar{n})$ backgrounds.\par
Given the traces of the projectors \eqref{eq:DFTProjectorTraces} (which determine the dimension of the vector spaces upon which they act), we may introduce a pair of DFT vielbeins $({\ubar{\mathcal{E}}}_{M\ubar{m}}, {\overbar{\mathcal{E}}}_{M\overbar{m}})$ for the projectors according to
\begin{align}
\mathcal{P}_{MN} = {\ubar{\mathcal{E}}}_{M\ubar{m}} {\ubar{\mathcal{E}}}_{N\ubar{n}} {\ubar{\eta}}^{\ubar{m}\ubar{n}} \,, \qquad \overbar{\mathcal{P}}_{MN} =  {\overbar{\mathcal{E}}}_{M\overbar{m}} {\overbar{\mathcal{E}}}_{N\overbar{n}} {\overbar{\eta}}^{\overbar{m}\overbar{n}}\,,
\end{align}
where ${\ubar{\eta}}^{\ubar{m}\ubar{n}}$ and ${\overbar{\eta}}^{\overbar{m} \overbar{n}}$ are $(D + n - \overbar{n}) \times (D + n - \overbar{n})$ and $(D - n + \overbar{n}) \times (D - n + \overbar{n})$ matrices whose signatures we denote $(\ubar{p},\ubar{q})$ and $(\overbar{p}, \overbar{q})$ respectively. These also act as simultaneous vielbeins for $\mathcal{M}_{MN}$ and $\eta_{MN}$ through \eqref{eq:ProjectorMEta}:
\begin{align}
\mathcal{M}_{MN} = {\ubar{\mathcal{E}}}_{M\ubar{m}} {\ubar{\mathcal{E}}}_{N\ubar{n}} \eta^{\ubar{m}\ubar{n}} - {\overbar{\mathcal{E}}}_{M \overbar{m}}  {\overbar{\mathcal{E}}}_{M \overbar{m}} {\overbar{\eta}}^{\overbar{m} \overbar{n}}\,, \quad \eta_{MN} = {\ubar{\mathcal{E}}}_{M\ubar{m}} {\ubar{\mathcal{E}}}_{N\ubar{n}} \eta^{\ubar{m}\ubar{n}} + {\overbar{\mathcal{E}}}_{M \overbar{m}}  {\overbar{\mathcal{E}}}_{M \overbar{m}} {\overbar{\eta}}^{\overbar{m} \overbar{n}}\,.
\end{align}
\par
From $D$, we subtract the $n+ \overbar{n}$ non-Riemannian coordinates that span the kernels of $H^{m n}$ and $K_{m n}$ and categorise the remaining coordinates into signature $(t,s)$ such that $D = n + \overbar{n} + t + s$. It then follows that
\begin{align}
\ubar{p} + \ubar{q} & = D + n - \overbar{n} = s + t + 2n\,,\\
\overbar{p} + \overbar{q} & = D - n + \overbar{n} = s + t - 2 \overbar{n}\,,
\end{align}
and so we have that the flat metrics are given by
\begin{align}
{\ubar{\eta}}_{\ubar{m}\ubar{n}} & = \begin{pmatrix}
{\ubar{\eta}}_{\ubar{a}\ubar{b}} & 0 & 0\\
0 & - \delta_{(n)} & 0\\
0 & 0 & + \delta_{(n)}
\end{pmatrix}\,, \qquad {\ubar{\eta}}_{\ubar{a}\ubar{b}} = \operatorname{diag} ( \underbrace{-1, \ldots, -1}_t, \underbrace{+1, \ldots + 1}_s )\,,\\
{\overbar{\eta}}_{\overbar{m} \overbar{n}} & = \begin{pmatrix}
{\overbar{\eta}}_{\overbar{a} \overbar{b}} & 0 & 0\\
0 & + \delta_{(\overbar{n})} & 0\\
0 & 0 & - \delta_{(\overbar{n})}
\end{pmatrix}\,, \qquad {\overbar{\eta}}_{\overbar{a} \overbar{b}} = \operatorname{diag} ( \underbrace{+1, \ldots, +1}_t, \underbrace{-1, \ldots - 1}_s )\,.
\end{align}
If we define the composite index $\overbar{M} = (\ubar{m}, \overbar{m})$ and combine the pair of vielbein into a generalised vielbein ${\mathcal{E}}_M{}^{\overbar{M}} = ({\ubar{\mathcal{E}}}_M{}^{\ubar{m}}, {\overbar{\mathcal{E}}}_M{}^{\overbar{m}})$, we may define flat metrics through $\mathcal{M}_{MN} = {\mathcal{E}}_M{}^{\overbar{M}} {\mathcal{E}}_N{}^{\overbar{N}} {\overbar{\mathcal{M}}}_{\overbar{M} \overbar{N}}$ and $\eta_{MN} = {\mathcal{E}}_M{}^{\overbar{M}} {\mathcal{E}}_N{}^{\overbar{N}} {\overbar{\eta}}_{\overbar{M} \overbar{N}}$ to obtain
\begin{align}
{\overbar{\mathcal{M}}}_{\overbar{M} \overbar{N}} = \begin{pmatrix}
{\ubar{\eta}}_{\ubar{m}\ubar{n}} & 0\\
0 & - {\overbar{\eta}}_{\overbar{m} \overbar{n}}
\end{pmatrix}\,,\qquad
{\overbar{\eta}}_{\overbar{M} \overbar{N}} = \begin{pmatrix}
{\ubar{\eta}}_{\ubar{m}\ubar{n}} & 0\\
0 & + {\overbar{\eta}}_{\overbar{m} \overbar{n}}
\end{pmatrix}\,.
\end{align}
In this form, it is easy to see that $\mathcal{M}_{MN} \eta^{MN} = {\overbar{\mathcal{M}}}_{\overbar{M} \overbar{N}} {\overbar{\eta}}^{\overbar{M} \overbar{N}} = 2(n- \overbar{n})$ as was previously found. We see that ${\overbar{\eta}}_{\overbar{M} \overbar{N}}$ is preserved by local $\operatorname{O}(t+n, s+n) \times \operatorname{O}(s+\overbar{n}, t+ \overbar{n})$ transformations and so we find that the appropriate coset structure is
\begin{align}\label{eq:DFTCosets}
\frac{\operatorname{O}(D,D)}{\operatorname{O}(t +n , s + n) \times \operatorname{O}(t + \overbar{n} , s + \overbar{n})}\,.
\end{align}
Note in particular that the dimension of the coset is $D^2 - {(n- \overbar{n})}^2$ and so, generically, any choice of coset $\tilde{H}$ with $(n, \overbar{n}) \neq (0,0)$ will have fewer components than the usual parametrisation (see \cite{Cho:2019ofr} for an alternative derivation of this fact).

	\chapter{The Maximally Non-Riemannian Solution in \texorpdfstring{$E_{8(8)}$}{E8(8)} EFT}\label{ch:E8NonRiemannian}
The maximally non-Riemannian parametrisation of the generalised metric in DFT is quite remarkable; it allows for a dimensional reduction of a theory that is free from the unstabilised scalar moduli that have conventionally plagued phenomenological models. Since it was found to be equivalent to setting the generalised metric equal to the $\operatorname{O}(D,D)$ structure $\eta_{MN}$, it is intrinsically tied to the group structure of DFT. A natural question to ask is whether EFT also admits such a solution.\par
The most natural starting point is $E_{8(8)}$ EFT since it is the only finite exceptional group that admits an analogous invariant tensor in the symmetric representation $\operatorname{Sym}(R_1 \otimes R_1)$ that can be set equal to the generalised metric. In $E_{8(8)}$ EFT, the coordinate representation is $R_1 = \mathbf{248}$, which is both the fundamental representation and the adjoint representation of $E_{8(8)}$ and is thus equipped with an invariant Cartan-Killing form $\kappa_{MN}$. We thus consider the choice $\mathcal{M}_{MN} \propto \kappa_{MN}$ and consider its implications for the rest of the theory.\par
We begin with an introduction to the relevant aspects of $E_{8(8)}$ EFT. Note that, for this chapter, we have adopted the conventions of \cite{Hohm:2014fxa} but we shall consider a different set of conventions in Chapter \ref{ch:EFTReductions} that will we be outlined there.
\section{Introduction to \texorpdfstring{$E_{8(8)}$}{E8(8)} EFT}
The coordinate representation of $E_{8(8)}$ EFT is the fundamental/adjoint representation $\mathbf{248}$, which we index by $M, N, \ldots = 1, \ldots, 248$. Denoting its generators $\{T^M\}$, we define the structure constants through the commutator $[T^M, T^N] = - f^{MN}{}_K T^K$. We choose the normalisation of the structure constants
\begin{align}
f^{MPQ} f_{NPQ} = - 60 \delta^M_N
\end{align}
and adopt a non-canonical normalisation for the Killing form,
\begin{align}
\kappa^{MN} \coloneqq \frac{1}{60} \operatorname{Tr} \left( T^M, T^N \right) = \frac{1}{60} f^{MP}{}_Q f^{NQ}{}_P\,,
\end{align}
with which we raise and lower $E_{8(8)}$ indies. Unlike other ExFTs the generalised Lie derivative acting on a vector $V^M$, of weight $\lambda(V)$, has two parts:
\begin{align}\label{eq:E8TotalGenLie}
\mathbb{L}_{(\Lambda, \Sigma)} = \mathbb{L}_\Lambda + \delta_\Sigma\,.
\end{align}
The first is the usual generalised Lie derivative with parameter $\Lambda^M$, given in terms of the projector onto the adjoint representation, as
\begin{align}\label{eq:E8Lie1}
\mathbb{L}_{\Lambda} V^M & = \Lambda^N \partial_N V^M - 60 {\left( \mathbb{P}_{\mathbf{248}} \right)}^M{}_N{}^K{}_L \partial_K \Lambda^L V^N + \lambda(V) \partial_N \Lambda^N V^M\,.
\end{align}
The projector onto the adjoint representation is, in turn, given in terms of the structure constants as
\begin{align}
{\left( \mathbb{P}_{\mathbf{248}} \right)}^M{}_N{}^K{}_L & = \frac{1}{60} f^M{}_{NQ} f^{QK}{}_L \label{eq:248Projector}\,.
\end{align}
One could also rewrite this in terms of the projector onto the $\mathbf{3875}$ representation of $E_{8(8)}$, given by
\begin{align}\label{eq:3875Projector}
{\left( \mathbb{P}_{\mathbf{3875}} \right)}^{MK}{}_{NL} = \frac{1}{7} \delta^M_{(N} \delta^K_{L)} - \frac{1}{56} \kappa^{MK} \kappa_{NL} - \frac{1}{14} f^P{}_N{}^{(M} f_{PL}{}^{K)}\,.
\end{align}
For completeness, we also give the projector onto the singlet representation:
\begin{align}\label{eq:1Projector}
{\left( \mathbb{P}_{\mathbf{1}} \right)}^{MN}{}_{KL} = \frac{1}{248} \kappa_{KL} \kappa^{MN}\,.
\end{align}
The second term in \eqref{eq:E8TotalGenLie} is a novel feature appearing at $n\geq 8$. It is an extra gauge transformation that is parametrised by a generalised parameter,
\begin{align}\label{eq:E8ExtraGaugeTransformation}
\delta_\Sigma V^M = - \Sigma_K f^{KM}{}_L V^L\,,
\end{align}
where $\Sigma_M$ is covariantly constrained in the sense that it is treated equivalently to a derivative under the section condition. Unlike the cases $n \leq 7$, \eqref{eq:E8Lie1} on its own does not close properly under the section condition and the extra gauge transformation is required to ensure proper closure. It further appears as a generic feature of 3-dimensional ExFTs\cite{Hohm:2017wtr,Hohm:2013jma} and its appearance is related to the dual graviton\cite{Hohm:2018qhd}. Taking the combination \eqref{eq:E8TotalGenLie}, the generalised Lie derivative closes if one imposes that the tensor product of two derivatives (or, more generally, any covariantly constrained objects) vanishes if one projects onto the subrepresentation $\mathbf{1 \oplus 248 \oplus3875} \subset \mathbf{248 \otimes 248}$. Using the explicit forms of the generators given above in \eqref{eq:248Projector}, \eqref{eq:3875Projector} and \eqref{eq:1Projector}, this is cast into the more practical form
\begin{align}
\kappa^{MN} C_M \otimes C_N & = 0 \,,\\
f^{KMN} C_M \otimes C_N & = 0\,,\\
{\left( \delta^{(M}_K \delta^{N)}_L - \frac{1}{2} f^{P(M}{}_K f_P{}^{N)}{}_L \right)} C_M \otimes C_N & = 0\,.
\end{align}
where $C_M \in \{ \partial_M, \Sigma_M , \ldots \}$ are covariantly constrained objects in the sense described above. Whilst one might expect this extra gauge transformation in the generalised Lie derivative to spoil the structure discussed in Section~\ref{sec:GenLieYTensor}, we may still assign a $Y$-tensor to $E_{8(8)}$ EFT if we instead consider the composite gauge parameter,
\begin{align}
R^M (\Lambda, \Sigma) = f^{MK}{}_L \partial_K \Lambda^L + \Sigma^M = f^{MK}{}_L \left( \partial_K \Lambda^L + \frac{1}{60} \Sigma_Q f^{QL}{}_K \right)\,,
\end{align}
introduced in \cite{Hohm:2017wtr}. Then, we may write the total generalised Lie derivative \eqref{eq:E8TotalGenLie} in the form
\begin{align}
\mathbb{L}_{(\Lambda, \Sigma)} V^M & = \Lambda^N \partial_N V^M + Y^{MN}{}_{KL} \partial_N R^K V^L + \lambda(V) \partial_N \Lambda^N V^M\,,
\end{align}
with the $Y$-tensor given by
\begin{align}
Y^{MN}{}_{KL} & = 2 \delta^{(M}_K \delta^{N)}_L - f^M{}_{LP} f^{PN}{}_K\,.
\end{align}
Indeed, this is the form that we gave in Table~\eqref{tab:YTensors} and agrees with the construction given in \cite{Cederwall:2015ica,Rosabal:2014rga} where the authors gave an unusual interpretation that the extra gauge parameter $\Sigma^M$, or at least a particular subset of the possible gauge parameters, can be interpreted as a term proportional to a generalised Weitzenb\"{o}ck connection that could give a geometric understanding for this extra gauge transformation.\par
One may verify that this $Y$-tensor encodes the same section conditions by expanding it out in terms of projectors onto irreducible representations of $E_{8(8)}$, for which one finds
\begin{align}\label{eq:CovariantlyConstrainedSectionCondition}
{\left( 62 \mathbb{P}_{\mathbf{1}} + 30 \mathbb{P}_{\mathbf{248}} + 14 \mathbb{P}_{\mathbf{3875}} \right)}_{LK}{}^{MN} C_M \otimes C_N = 0\,.
\end{align}
The trivial parameters (which we recall are parameters whose action on fields under the generalised Lie derivative vanish under the section conditions), in this case, can be split into two types. Firstly, we have parameters $\Lambda$ (in combination with $\Sigma = 0$) of the form\begin{align}
\Lambda^M & = \kappa^{MN} \Omega_N \qquad (\Omega_N \text{ covariantly constrained})\,,\\
\Lambda^M & = {\left( {\mathbb{P}}_{\mathbf{2875}} \right)}^{MN}{}_{KL} \partial_{K} \chi^{KL}\,,
\end{align}
 which produce a trivial action under $\mathbb{L}_{\Lambda}$ where $\chi^{KL}$ is unconstrained. These are the analogues of the trivial parameters found in $n \leq 7$ EFT and DFT. However, due to the extra gauge transformation present in the $E_{8(8)}$ generalised Lie derivative, there are further non-trivial \emph{combinations} of parameters $(\Lambda, \Sigma)$ that generate a trivial action under the full generalised Lie derivative $\mathbb{L}_{(\Lambda, \Sigma)}$. These are of the form
\begingroup
\renewcommand{\arraystretch}{1.5}
\begin{align}
\begin{cases}
\begin{array}{l}
\Lambda^M = f^{MN}{}_K \Omega_{N}{}^K\,,\\
\Sigma_M = \partial_M \Omega_N{}^N + \partial_N \Omega_M{}^N\,,\\
\end{array}
\end{cases}
\end{align}
\endgroup
with $\Omega_{N}{}^K$ covariantly constrained on the first index. The fields of $E_{8(8)}$ EFT were discussed in Section~\ref{sec:FieldsExFT} and Section~\ref{sec:GaugeStructureExFT} but we summarise them here for convenience:
\begin{align}
\{g_{\mu \nu}, \mathcal{M}_{MN}, \mathcal{A}_{\mu}{}^M, \mathcal{B}_{\mu M}\}\,.
\end{align}
These are the external metric, internal generalised metric and two generalised gauge fields. Defining the Lie-covariantised derivative $\mathcal{D}_\mu \coloneqq \partial_\mu - \mathbb{L}_{(\mathcal{A}_\mu, \mathcal{B}_\mu)}$, the field strengths of $\mathcal{A}_\mu{}^M$ and $\mathcal{B}_{\mu M}$ are defined up to trivial gauge parameters by 
\begin{align}
[\mathcal{D}_\mu, \mathcal{D}_\nu] V^M \coloneqq - \mathbb{L}_{(F_{\mu \nu}, G_{\mu \nu})} V^M\,,
\end{align}
where the uncovariantised fieldstrengths are given by
\begingroup
\renewcommand{\arraystretch}{1.5}
\begin{align}
F_{\mu \nu}{}^M & = \begin{array}[t]{l}
2 \partial_{[\mu} {\mathcal{A}}_{\nu]}{}^M - 2 {\mathcal{A}}_{[\mu}{}^N \partial_N {\mathcal{A}}_{\nu]}{}^M + 14 {\left( {\mathbb{P}}_{\mathbf{3875}} \right)}^{MN}{}_{KL} {\mathcal{A}}_{[\mu}{}^K \partial_N {\mathcal{A}}_{\nu]}{}^L\\
\qquad + \frac{1}{4} {\mathcal{A}}_{[\mu}{}^N \partial^M {\mathcal{A}}_{\nu]N} - \frac{1}{2} f^{MN}{}_P f^P{}_{KL} {\mathcal{A}}_{[\mu}{}^K \partial_N {\mathcal{A}}_{\nu]}{}^L\,,
\end{array}\\
G_{\mu \nu M} & = 2 \mathcal{D}_{[\mu} {\mathcal{B}}_{\nu] M} - f^N{}_{KL} {\mathcal{A}}_{[\mu}{}^K \partial_M \partial_N {\mathcal{A}}_{\nu]}{}^L\,.
\end{align}
\endgroup
In the forms above, the failure of the fieldstrengths to transform covaraiantly are all of the form of trivial parameters and so we may add two-form couplings to recovariantise them:
\begin{align}
\mathcal{F}_{\mu \nu}{}^M & = F_{\mu \nu}{}^M 14 {\left( {\mathbb{P}}_{\mathbf{3875}} \right)}^{mn}{}_{kl} \partial_N C_{\mu \nu}{}^{KL}_{(\mathbf{3875})} + \frac{1}{4} \partial^M C_{\mu \nu} + 2 f^{MN}{}_K C_{\mu \nu N}{}^K\,,\\
\mathcal{G}_{\mu \nu M} & = G_{\mu  \nu M} + 2 \partial_N C_{\mu \nu M}{}^N + 2 \partial_M C_{\mu \nu N}{}^N\,.
\end{align}
Whilst we have added compensating two-form fields $C_{\mu \nu}{}^{KL}_{\mathbf{3875}}, C_{\mu \nu}$ and $C_{\mu \nu M}{}^N$ (the last of which is required to be covariantly constrained on the first internal index), all of them eventually drop out of the action and transformation rules and so we do not need to worry about recovariantising their fieldstrengths; the tensor hierarchy structure can be terminated here.\par
The full action for the $E_{8(8)}$ ExFT was constructed in \cite{Hohm:2014fxa} and is given by
\begin{align}\label{eq:E8Action}
S & = \int \textrm{d}^{3}x \textrm{d}^{248} Y e \left( \hat R[g] +  \frac{1}{240}  g^{\mu \nu} \mathcal{D}_\mu \mathcal{M}_{MN} \mathcal{D}_\nu \mathcal{M}^{MN} - V( \mathcal{M},g ) + \frac{1}{e} \mathcal{L}_{CS} \right)\,,
\end{align}
where $e \coloneqq \sqrt{-g}$ is the determinant of the external vielbein. The first term $\hat R[g]$ is a covariantisation of the usual Ricci scalar for the extrnal metric, in which the partial derivative has been replaced by the Lie-covariantised derivative $\partial_\mu \rightarrow \mathcal{D}_\mu$. The second term is a kinetic piece for the scalar sector encoded in the generalised metric. The third piece is a potential $V$ is given by
\begingroup
\renewcommand{\arraystretch}{1.5}
\begin{align}\label{eq:E8Potential}
V & = \begin{array}[t]{l}
- \frac{1}{240} \mathcal{M}^{MN} \partial_M \mathcal{M}^{KL} \partial_N \mathcal{M}_{KL} + \frac{1}{2} \mathcal{M}^{MN} \partial_M \mathcal{M}^{KL} \partial_L \mathcal{M}_{NK}\\
+ \frac{1}{7200} f^{NQ}{}_P f^{MS}{}_R \mathcal{M}^{PK} \partial_M \mathcal{M}_{QK} \mathcal{M}^{RL} \partial_N \mathcal{M}_{SL}\\
 - \frac{1}{2} g^{-1} \partial_M g \partial_N \mathcal{M}^{MN} - \frac{1}{4} \mathcal{M}^{MN} g^{-1}  \partial_M g \, g^{-1} \partial_N g - \frac{1}{4} \mathcal{M}^{MN} \partial_M g^{\mu \nu} \partial_N g_{\mu \nu}\,,\\
\end{array}
\end{align}
\endgroup
whilst the last term is a Chern-Simons term, most conveniently described as a boundary term on an auxiliary space $\Sigma^4$ whose boundary $\partial \Sigma^4$ is the three-dimensional external space that we consider:
\begin{align}
S_{\text{CS}} \sim \int_{\Sigma^4} \textrm{d}^4x \int \textrm{d}^{248} Y \left( \mathcal{F}^M \wedge \mathcal{G}_M - \frac{1}{2} f_{MN}{}^K \mathcal{F}^M \wedge \partial_K \mathcal{G}^{\mathcal{N}} \right)\,.
\end{align}
\section{Generalised Metric and Projector}\label{sec:CosetProjectorsForAllExFTs}
In Section~\ref{sec:ProjectedEOM}, we discussed how the equations of motions for ExFTs are obtained by a \emph{projection} of the equations obtained from varying the action with respect to the generalised metric in order to take into account the coset structure that $\mathcal{M}_{MN}$ must parametrise. In this section, we discuss this in more detail, first commenting in generality before restricting to the $E_{8(8)}$ case later. Our starting point is the action of the generalised Lie derivative on the generalised metric. In all cases for $4 \leq n \leq 7$, one may verify that this may be written as
\begin{align}
\delta_\Lambda \mathcal{M}_{MN} = \Lambda^P \partial_P \mathcal{M}_{MN} + 2 \alpha \mathcal{P}_{MN}{}^{KL} \partial_K \Lambda^P \mathcal{M}_{LP}\,,
\end{align}
where $\mathcal{P}$ (note the suggestive notation) is given in terms of the numbers listed in Table~\ref{tab:Summary} as
\begin{align}\label{eq:CosetProjectorExpanded}
\mathcal{P}_{MN}{}^{KL} = \frac{1}{\alpha} \left( \delta^{(K}_M \delta^{L)}_N - \omega \mathcal{M}_{MN} \mathcal{M}^{KL} - \mathcal{M}_{MQ} Y^{Q(K}{}_{RN} \mathcal{M}^{L)R} \right)\,.
\end{align}
We may equivalently write this in terms of the projector onto the adjoint representation, using \eqref{eq:YAdjProjRelation}, as
\begin{align}\label{eq:CosetProjector}
\mathcal{P}_{MN}{}^{KL} = \mathcal{M}_{MQ} {\left( \mathbb{P}_{\text{adj.}} \right)}^Q{}_{N}{}^{(K}{}_R \mathcal{M}^{L)R}\,.
\end{align}
For $E_{8(8)}$, this must be modified to take the extra gauge transformation into account:
\begin{align}
\delta_{(\Lambda, \Sigma)} \mathcal{M}_{MN} = \Lambda^P \partial_P \mathcal{M}_{MN} + 2 \cdot 60 \mathcal{P}_{MN}{}^{KL} \left( \partial_K \Lambda^P + \frac{1}{60} f^{QP}{}_K \Sigma_Q \right) \mathcal{M}_{PL}
\end{align}
but the associated $\mathcal{P}$ is still of the same form as \eqref{eq:CosetProjector}, with the projector onto the adjoint representation of $E_{8(8)}$ given in \eqref{eq:248Projector}. The final case that we consider is the $n=3$ case which is modified to account for the two groups in $G = \operatorname{SL}(3) \times \operatorname{SL}(2)$:
\begin{align}
\delta_\Lambda \mathcal{M}_{MN} = \Lambda^P\partial_P \mathcal{M}_{MN} + 2\mathcal{M}_{MQ} \left( 2 \mathbb{P}_{(\mathbf{8,1})}{}^Q{}_N{}^{(K}{}_R + 3 \mathbb{P}_{(\mathbf{1,3})}{}^Q{}_N{}^{(K}{}_R \right) \mathcal{M}^{L)R} \partial_K \Lambda^P \mathcal{M}_{LP}\,,
\end{align}
where $\mathbb{P}_{\mathbf{(8,1)}}$ and $\mathbb{P}_{\mathbf{(1,3)}}$ are projectors onto the adjoint representations of the two groups. In all cases $3 \leq n \leq 8$ one can see that it is manifestly symmetric in its upper indices but one may use the fact that it is a group invariant (such that the simultaneous action of $\mathcal{M}^{-1}$ and $\mathcal{M}$ on indices leaves it invariant) to show that it is further symmetric in its lower indices. It is then simple to show that it squares to itself according to $\mathcal{P}_{MN}{}^{KL} \mathcal{P}_{KL}{}^{PQ} = \mathcal{P}_{MN}{}^{PQ}$, thus defining a projector. The dimension of its image is given by the trace
\begin{align}
\mathcal{P}_{MN}{}^{MN} & = \mathcal{M}_{MQ} {\left( \mathbb{P}_{\text{adj.}} \right)}^Q{}_N{}^{(M}{}_R \mathcal{M}^{N)R}\\
	& = \frac{1}{2} \left( {\left(\mathbb{P}_{\text{adj.}} \right)}^M{}_N{}^N{}_M - \frac{1}{\alpha} \mathcal{M}_{MQ} \left[ Y^{QM}{}_{RN} - \delta^Q_R \delta^M_N + \omega \delta^Q_N \delta^M_R \right] \mathcal{M}^{NR} \right)\\
	& = \frac{1}{2\alpha} \left( \operatorname{dim} R_1 (1 - \omega) + \alpha \operatorname{dim} \operatorname{adj.} - \mathcal{M}_{MN} Y^{MN}{}_{QP}\mathcal{M}^{PQ} \right)\,.\label{eq:TraceCosetProjector}
\end{align}
In each case, one may verify numerically that this gives
\begin{align}\label{eq:r}
\mathcal{P}_{MN}{}^{MN} = \operatorname{dim} G/H - r\,, \qquad r \coloneqq \frac{1}{2\alpha} \mathcal{M}_{MN} Y^{MN}{}_{KL} \mathcal{M}^{KL}
\end{align}
except for $E_{8(8)}$ where it instead gives
\begin{align}
\mathcal{P}_{MN}{}^{MN} = \operatorname{dim} G/H + \frac{2}{15} - r\,.
\end{align}
Then, it is a matter of treating each ExFT individually to find (in the usual supergravity parametrisation) that
\begin{align}
r = \begin{cases}
\begin{array}{ll}
0\,, & 3 \leq n \leq 7\\
\frac{2}{15}\,, & n =8\\
\end{array}
\end{cases}
\end{align}
such that $\mathcal{P}_{MN}{}^{MN}$ does indeed give the dimension of the coset $G/H$ for $3 \leq n \leq 8$. Putting it all together, we can thus interpret $\mathcal{P}_{MN}{}^{KL}$ as a projector from $\operatorname{Sym}(R_1 \otimes R_1)$ into the space that $\mathcal{M}_{MN}$ lives. We have thus shown that this is the same object that projects out the equations of motion in \eqref{eq:ProjectedEOM} and that also appeared in the previous chapter. The cases for DFT and $n=4,5$ EFT were already considered in the earlier work \cite{Berkeley:2014nza,Rudolph:2016sxe}, in which it was already conjectured that the form \eqref{eq:CosetProjectorExpanded} holds to at least $n \leq 7$, and so we are left with verifying the cases for EFT with $n=3,6,7,8$. We have collected explicit computations for these cases in Appendix~\ref{app:CosetProjectorTraces}.
\section{Maximally Non-Riemannian Solution in \texorpdfstring{$E_{8(8)}$}{E8(8)} EFT}
In the conventional parametrisation, the generalised metric is a representative of the coset $E_{8(8)}/\operatorname{SO}(16)$ and is typically parametrised in terms of the internal metric and $p$-form fields in a choice of Borel gauge. The $E_{8(8)}$ EFT analogues of the compatibility conditions \eqref{eq:DFTCompatibilityConditions} on the generalised metric that are required to ensure invariance of the action \eqref{eq:E8Action} are
\begin{align}\label{eq:E8CompatibilityConditions}
\mathcal{M}_{MN} \mathcal{M}_{KL} \mathcal{M}_{PQ} f^{NLQ} = - f_{MKP}\,,\qquad \mathcal{M}_{MP} \kappa^{PQ} \mathcal{M}_{QN} = \kappa_{MN}\,,
\end{align}
in addition to the obvious symmetry $\mathcal{M}_{MN} = \mathcal{M}_{NM}$. Since these are evidently much more involved than the DFT case, we shall not give a full classification (although such a classification of permissible EFT generalised metrics would obviously be ideal) and rather focus on constructing an analogue of the maximally non-Riemannian solution. Recalling that the maximally non-Riemannian solution in DFT is equivalently characterised as the choice of generalised metric for which the coset projector vanishes \eqref{eq:VanishingProjectors}, we choose $\mathcal{M}_{MN}$ in $E_{8(8)}$ EFT to satisfy the same condition. In $E_{8(8)}$ EFT, the coset projector is more involved due to the presence of a universal weight $\omega$ and the complicated $Y$-tensor (which we recall has no definite symmetries any more). It is given by
\begin{align}
\mathcal{P}_{MN}{}^{KL} & = \frac{1}{60} \left( \delta^{(K}_M \delta^{L)}_N + \mathcal{M}_{MN} \mathcal{M}^{KL} - \mathcal{M}_{MQ} Y^{Q(K}{}_{RN} \mathcal{M}^{L) R} \right)\\
	& = \frac{1}{60} \mathcal{M}_{MQ} f^Q{}_{NP} f^{P (K }{}_R \mathcal{M}^{L) R}\,.
\end{align}
However, one may verify quite readily that choosing $\mathcal{M}_{MN} = - \kappa_{MN}$ satisfies all the constraints on the generalised metric \eqref{eq:E8CompatibilityConditions}\footnote{Note, in particular, that we require the generalised metric to be equal to \emph{minus} the Killing form in order to satisfy $\mathcal{M}_{MN} \mathcal{M}_{KL} \mathcal{M}_{PQ} f^{NLQ} = - f_{MKP}$ whilst remaining compatible with the simple raising and lowering of indices with the Killing form $\kappa_{MN} \kappa_{KL} \kappa_{PQ} f^{NLQ} = + f_{MKP}$.} and, further, causes the projector to vanish $\left. \mathcal{P}_{MN}{}^{KL} \right|_{\mathcal{M} = - \kappa} = 0$ since $f^{P(KL)} = 0$. Thus, the equations of motion of the generalised metric $\mathcal{P}_{MN}{}^{KL} \mathcal{K}_{KL} = 0$ are satisfied by construction.\par
In addition to projecting out the equations of motion, $\mathcal{P}_{MN}{}^{KL}$ must also project out the fluctuations of the generalised metric and so we have $\delta \mathcal{M}_{MN} = \mathcal{P}_{MN}{}^{KL} \delta \mathcal{M}_{KL}$---this is equivalent to the DFT statement \eqref{eq:ODDConstraintProjectors}. We thus see that, under our choice of generalised metric, $\delta \mathcal{M}_{MN} = 0$ and so fluctuations about this background vanish just as they did in DFT. In particular, all the scalar moduli that would arise under a dimensional reduction of this ExFT end up fixed and the resulting three-dimensional theory is moduli-free.\par
In the Park-Morand classification of the DFT generalised metric, the different non-Riemannian backgrounds arose as parametrisations of different choices of the coset $G/\tilde{H}$. Since all fluctuation must be projected out for the maximally non-Riemannian choices in both DFT and $E_{8(8)}$ EFT, we interpret them to parametrise the coset $G/G$, in line with the fact that they encode no degrees of freedom. The rest of the story is expected to follow that of DFT; from the form of the trace \eqref{eq:TraceCosetProjector}, we see that the only information about the choice of coset enters through the trace of the $Y$-tensor $r$ via the generalised metric. In particular, the choice of $\tilde{H}$ is encoded into the generalised metric by introducing a generalised vielbein ${\mathcal{E}}_M{}^{\overbar{M}}$ such that
\begin{align}
\mathcal{M}_{MN} = {\mathcal{E}}_M{}^{\overbar{M}} {\overbar{\mathcal{M}}}_{\overbar{M} \overbar{N}} {\mathcal{E}}^{\overbar{N}}{}_N\,,
\end{align}
where ${\overbar{\mathcal{M}}}_{\overbar{M} \overbar{N}}$ is the flat metric that is left invariant under local $\tilde{H}$-transformations for a particular choice of $\tilde{H} \subseteq G$. We thus see that, generically, any non-Riemannian parametrisations (where possible) will end up with $r$ degrees of freedom fewer than the usual parametrisation (or, rather, $r + 2/15 \in \mathbb{Z}^+$ for $E_{8(8)}$ EFT to cancel the extra contribution). 
\section{Reduction to the Leibniz-Chern-Simons Theory}
The obvious next step is then to consider the theory resulting from setting $\mathcal{M}_{MN} = - \kappa_{MN}$. Inserting this into the action \eqref{eq:E8Action}, one finds
\begin{align}\label{eq:ReducedE8Action}
\begin{aligned}
\mathcal{S}_{E_{8(8)}} \rightarrow & \int \textrm{d}^3x \int \textrm{d}^{248} Y e \left[ \hat{R}[g] + \frac{1}{4} \kappa^{MN} \left( g^{-1} \partial_M g g^{-1} \partial_N g + \partial_M g^{\mu \nu} \partial_N g_{\mu \nu}\right) \right]\\
	& \qquad + \int_{\Sigma^4} \textrm{d}^4x \int \textrm{d}^{248} Y \left( \mathcal{F}^M \wedge \mathcal{G}_M - \frac{1}{2} f_{MN}{}^K \mathcal{F}^M \wedge \partial_K \mathcal{G}^N \right)\,,
\end{aligned}
\end{align}
where all other terms vanish since $\partial_M \kappa_{KL} = \mathcal{D}_\mu \kappa_{KL} = 0$ (since the Killing form is an invariant tensor and so vanishes under the action of the generalised Lie derivative). Additionally, the two terms on the first line of the action that are proportional to $\kappa^{MN}$ further vanish under the section condition. The presence of such terms serves as a reminder that we are still in the full $E_{8(8)}$ EFT; the remaining fields can still depend on the extended coordinates $Y^M$, subject to the section condition.\par
Before we identify the resulting theory, we first make some remarks on the consequences of this choice of generalised metric. In particular, one might be worried if such a reduction introduces any pathologies in the equations on motion and so we start by making a few remarks. As mentioned previously, the equations of motion for the generalised metric are automatically satisfied by construction and so we focus on the remaining fields. In particular the equations of motion for the external metric is rather involved, mixing all of the fields together. It is given by
\begingroup
\renewcommand{\arraystretch}{1.3}
\begin{align*}
\begin{array}{ll}
0 & = 
{\hat{\mathcal{R}}}_{\mu \nu} - \frac{1}{2} g_{\mu\nu} \left( \hat{\mathcal{R}}[g] + \frac{1}{240}  g^{\rho\sigma} \mathcal{D}_\rho \mathcal{M}_{MN} \mathcal{D}_\sigma \mathcal{M}^{MN} - V( \mathcal{M},g )\right)\\
& + \frac{1}{240} \mathcal{D}_\mu \mathcal{M}_{MN} \mathcal{D}_\nu \mathcal{M}^{MN} + \frac{1}{2}\sqrt{|g|}^{-1} g_{\mu\nu} \partial_M \left( \sqrt{|g|} ( \partial_N \mathcal{M}^{MN} + \mathcal{M}^{MN} \partial_N \ln |g| )\right)\\
& - \frac{1}{2} \sqrt{|g|}^{-1} \partial_M ( \sqrt{|g|} \mathcal{M}^{MN} ) \partial_N g_{\mu\nu} - \frac{1}{2} \mathcal{M}^{MN} g_{\mu \rho} \partial_M g^{\rho \sigma} \partial_N g_{\sigma \nu} - \frac{1}{2} \mathcal{M}^{MN} \partial_M \partial_N g_{\mu\nu}\,,
\end{array}
\end{align*} 
\endgroup
where $\hat{\mathcal{R}}_{\mu \nu}$ is defined as the variation of $\hat{\mathcal{R}}[g]$ with respect to $g^{\mu \nu}$ and is, in particular, independent of the generalised metric. Since the appearance of the generalised metric is always under derivatives (either $\partial_M$ or $\mathcal{D}_\mu$) or contracted with two internal derivatives, they all drop out under the ansatz $\mathcal{M}_{MN} = - \kappa_{MN}$ and we are left with just the Lie-covariantised vacuum Einstein field equations on section. Similarly, the generalised metric only appears in the equations of motion for the generalised gauge fields under the combinations outlined above. For example, the $\mathcal{B}$-field has an equation of motion
\begin{align}
\epsilon^{\rho \mu \nu} \mathcal{F}_{\mu \nu}{}^M = 2g^{\mu \nu} j_{\nu}{}^M\,,
\end{align}
where $j_\mu{}^M f_{ML}{}^K \coloneqq \mathcal{M}^{KQ} \mathcal{D}_\mu \mathcal{M}_{QL}$ is the scalar current, which obviously vanishes in the maximally non-Riemannian solution. In all cases, setting $\mathcal{M}_{MN} = - \kappa_{MN}$ produces no pathologies and so the equations of motion for the fields $(g_{\mu\nu}, \mathcal{A}_{\mu}{}^M, \mathcal{B}_{\mu M})$ are given by varying the action \eqref{eq:ReducedE8Action}.\par
This theory has appeared previously in the literature, though not under this interpretation, as the Leibniz-Chern-Simons theory of \cite{Hohm:2018ybo}. The construction of a Chern-Simons action requires an inner product that satisfies certain invariance conditions and, in \cite{Hohm:2018ybo}, it was argued that the generalised diffeomorphisms of $E_{8(8)}$ EFT can be reinterpreted as a Leibniz algebra equipped with an invariant quadratic form---the necessary ingredients required to construct a gauge-invariant Chern-Simons action. In their construction, they identified the resulting theory as a `topological phase' of the full $E_{8(8)}$ EFT, obtained from the truncation $\mathcal{M}_{MN} = 0$. Here, we have shown that the same theory can be obtained without a truncation and instead arises as a non-singular, maximally non-Riemannian solution $\mathcal{M}_{MN} = - \kappa_{MN}$ of the full EFT. In particular, this solution retains a full unbroken $E_{8(8)}$ symmetry. In this respect, it is fundamentally different from the more natural vacuum $\mathcal{M}_{MN} = \delta_{MN}$ which would break $E_{8(8)}$ to $\operatorname{SO}(16)$. As such, this theory is not just a topological theory with a Chern-Simons term since all of the remaining fields can have non-trivial dependences on the internal space.\par
It is tempting to speculate that such a mechanism may realise the old idea that one can generate a geometry from the spontaneous breaking of an underlying topological phase of gravity, along the lines of what was considered in \cite{Tseytlin:1981ks, Witten:1988xj, Horowitz:1989ng}. However, one major difference between those ideas and the proposals made in ExFTs is that the topological phase may be realised with a maximally non-Riemannian metric rather than a vanishing metric, as originally envisaged.

\part{Reductions of ExFTs}
	\chapter{Relations Amongst Exceptional Field Theories}\label{ch:EFTReductions}
Thus far, the literature on the relationship between ExFTs has remained rather sparse. Previous work in this area has generally focused on DFT-to-DFT reductions such as in gauged DFT (GDFT)\cite{Aldazabal:2011nj,Dibitetto:2012rk,Grana:2012rr,Berman:2012uy,Aldazabal:2013mya,Aldazabal:2013via}, which realised that one could produce DFTs with gauge deformations by a Scherk-Schwarz reduction of an ungauged DFT, or the Kaluza-Klein reduction of the DFT generalised metric. A key exception is \cite{Thompson:2011uw} which considered the reduction from the $\operatorname{SL}(5)$ extended field theory of \cite{Berman:2010is} to $\operatorname{O}(3,3)$ DFT\footnote{See also \cite{Sakatani:2017nfr} which studied the relation between the M-theory and Type IIB solutions of the \emph{same} EFT.}. This was not for the full exceptional field theory but just for the extended space and did not consider alternative reductions. In this paper, we extend these works by examining a host of EFT-to-EFT reductions. We mention that similar ideas have been leveraged in the infinite-dimensional EFTs. In particular, the embedding of $E_{8(8)}$ within $E_{9(9)}$ was used in \cite{Bossard:2017aae,Bossard:2018utw} and reductions within the context of the $E_{11}$ programme has been considered in \cite{Berman:2011jh,West:2011mm,Riccioni:2009xr,Riccioni:2007au,West:2012qz}. The work presented here is complimentary to, but differs slightly from, previous work on the tensor hierarchy in that we work at the level of the action rather than the representations. In addition to this there was early seminal work in \cite{Obers:1999um} predating ExFT where the so-called particle and string multiplets related to various U-duality groups were studied in detail. The coordinates and section constraints for the different EFTs then become related to these particle and string multiplets. Finally, the reduction considered here is similar to the decompactification limit of curvature corrections that were first studied in \cite{Bossard:2016hgy} and later in the context of EFT in \cite{Bossard:2015foa}.\par
In this chapter, we aim to extend these works by introducing some aspects of EFT-to-EFT reductions. Along the way, we shall highlight some unexpected features of these reductions that we believe could embody some aspects of ExFTs that have not been considered before.\par
Whilst each EFT is constructed in the same way each theory nevertheless ends up with rather distinct features, necessitating that EFT-to-EFT reductions be treated on a case-by-case basis. This is in contrast to DFT where one simply needs to know the dimension, $D$ for the relevant $\operatorname{O}(D,D)$ group and everything else (the action, section condition, generalised Lie derivative etc.) is identical. For example, $E_{7(7)}$ EFT comes with a generalised Yang-Mills term in the action, a self-duality constraint on the generalised field strength and a symplectic structure $\Omega_{MN}$, none of which have an obvious origin in $E_{8(8)}$ EFT. Additionally, the $Y$-tensor in $E_{8(8)}$ EFT is not sufficient to close the generalised diffeomorphisms, requiring an extra gauge transformation to ensure closure and it is not obvious what happens to this extra gauge transformation if we reduce to $E_{7(7)}$ EFT.  Other examples of quirks of EFTs include the reducible coordinate representation of $\operatorname{SL}(2) \times \mathbb{R}^+$ EFT or the product group of $n=3$ EFT.\par
Throughout this chapter, an EFT-to-EFT reduction should be understood as a spontaneous symmetry breaking from an  $E_{n(n)}$ EFT to an $E_{n-1(n-1)}$ EFT when there is a generalised isometry present. Note that the further we compactify down from eleven dimensions, the larger the exceptional group becomes and so the usual Kaluza-Klein reduction in supergravity yields an \emph{increase} in the dimension of the exceptional group which is made manifest in the reduced theory. Here, we are going in the opposite direction and spontaneously breaking the exceptional group to a subgroup. In the supergravity literature, this would be considered as an oxidation of the supergravity theory to one dimension higher (note the conflicting terminology; a reduction of the exceptional group corresponds to an oxidation of the spacetime dimension). A useful paper covering aspects of oxidation in supergravity before the ExFT programme is \cite{Keurentjes:2002xc}.\par
In this paper, wherever we need to differentiate between two ExFTs, we shall refer to the larger ExFT (the theory with the larger group $\hat{G}$, though \emph{smaller} external space) as the `parent' theory and adorn all objects/indices in that ExFT with hats $(\widehat{\hphantom{G}})$ to distinguish them from the analogous structure in the `child' theory (whose associated group we denote as $G \subset \hat{G}$). However, when we speak in generality (as we shall for the next section) we shall drop any hats to prevent cluttering the formulae. Hopefully, there should be no ambiguity in doing so.\par
\section{Reduction of the Generalised Metric}\label{sec:ReductionGenMetric}
The generalised metric of any ExFT is a representative of the coset $G/H$ and encodes the scalar degrees of freedom of the theory. The form of the generalised metric depends on the theory but the generalised metric for each EFT (in a Borel gauge) has been known for a while and can be found in \cite{Lee:2016qwn,Berman:2011jh}. Note that one could also work with the generalised vielbein instead (from which one could, of course, work out the generalised metric reduction). This approach has been considered for $E_{7(7)}$ and $E_{6(6)}$ EFT in \cite{Bossard:2014lra}.\par
Here, we shall consider two cases; a reduction of the DFT generalised metric and the reduction of the $\operatorname{SL}(5)$ generalised metric. The first will be useful as an illustrative example of what a reduction may look like in ExFT and agrees with previous results in the area. We will then explicitly reduce the $\operatorname{SL}(5)$ generalised metric and show how both the $\operatorname{O}(3,3)$ generalised metric and the $\operatorname{SL}(3) \times \operatorname{SL}(2)$ generalised metric can both be obtained from the same circle reduction, just with different reduction ansatzes. The reduction to the $\operatorname{O}(3,3)$ DFT matches that described in \cite{Thompson:2011uw,Blair:2018lbh,Berman:2019izh}.
\subsection{\texorpdfstring{$\operatorname{O}(D,D)$}{O(D,D)} DFT to \texorpdfstring{$\operatorname{O}(d,d) \times \operatorname{O}(n,n)$}{O(d,d)xO(n,n)} DFT}
Our starting point is the $\hat{G} = \operatorname{O}(D,D)$ DFT generalised metric
\begin{align}
{\hat{\mathcal{M}}}_{\hat{M} \hat{N}} & = \begin{pmatrix}
{\hat{g}}_{\hat{m} \hat{n}} - {\hat{B}}_{\hat{m} \hat{p}} {\hat{g}}^{\hat{p} \hat{q}} {\hat{B}}_{\hat{q} \hat{n}} & {\hat{B}}_{\hat{m} \hat{p}} {\hat{g}}^{\hat{p} \hat{n}}\\
- {\hat{g}}^{\hat{m} \hat{p}} {\hat{B}}_{\hat{p} \hat{n}} & {\hat{g}}^{\hat{m} \hat{n}}
\end{pmatrix}\,,
\end{align}
where $\hat{M},\hat{N} = ({}^{\hat{m}}, {}_{\hat{m}}) = 1, \ldots,2D$ and $\hat{m}, \hat{n} = 1, \ldots, D$. We now consider a Kaluza-Klein (KK) decomposition of the underlying fields according to
\begin{align}
{\hat{g}}_{\hat{m} \hat{n}} & = \begin{pmatrix}
e^{2\alpha \phi} g_{\mu \nu} + e^{2\beta \phi} A_\mu{}^m g_{mn} A^n{}_\nu & e^{2\beta \phi} A_\mu{}^m g_{mn}\\
e^{2 \beta \phi} g_{mn} A^n{}_\nu & e^{2\beta \phi} g_{mn}\\
\end{pmatrix}\label{eq:KKMetric}\,,\\
{\hat{g}}^{\hat{m} \hat{n}} & = \begin{pmatrix}
e^{-2\alpha \phi} g^{\mu \nu} & - e^{-2 \alpha \phi} g^{\mu \nu} A_\nu{}^n\\
- e^{-2 \alpha \phi} A^m{}_\mu g^{\mu \nu} & e^{-2\alpha \phi} A^m{}_\mu g^{\mu \nu} A_\nu{}^n + e^{-2\beta \phi} g^{mn}
\end{pmatrix}\label{eq:KKInverse}\,,\\
{\hat{B}}_{\hat{m} \hat{n}} & = \begin{pmatrix}
B_{\mu \nu} + B_{\mu n} A^n{}_\nu + A_\mu{}^m B_{m \nu} + A_\mu{}^m B_{mn} A^n{}_\nu & B_{\mu n} + A_\mu{}^m B_{mn}\\
B_{m \nu} + B_{mn} A^n{}_\nu & B_{mn}
\end{pmatrix}\,,
\end{align}
where we have now decomposed the index $\hat{m}$ to $\hat{m} = (\mu, m)$ with ranges $\mu = 1, \ldots, d$ and $m = 1, \ldots, n$ such that $d+n = D$. One finds that this can be reorganised into a generalised Kaluza-Klein ansatz for ${\hat{\mathcal{M}}}_{\hat{M} \hat{N}}$ according to
\begin{align}\label{eq:GeneralisedKK}
{\hat{\mathcal{M}}}_{\hat{M} \hat{N}} & = \begin{pmatrix}
\mathcal{M}_{MN} + \mathcal{A}_M{}^A \mathcal{G}_{AB} \mathcal{A}^B{}_N &  {\mathcal{A}}_M{}^A \mathcal{G}_{AB} + \mathcal{M}_{MN} \mathcal{B}^N{}_B\\
\mathcal{G}_{AB} \mathcal{A}^B{}_N + \mathcal{B}_A{}^M \mathcal{M}_{MN} & \mathcal{G}_{AB} + \mathcal{B}_A{}^M \mathcal{M}_{MN} \mathcal{B}^N{}_B
\end{pmatrix}\,,
\end{align}
where
\begin{align}
\mathcal{M}_{MN} & = \begin{pmatrix}
\begingroup
\setlength\arraycolsep{0pt}
\renewcommand{\arraystretch}{1}
\begin{array}[t]{c}
e^{2\alpha \phi} g_{\mu \nu}\\ - e^{-2\alpha \phi} (B_{\mu \sigma} + A_\mu{}^m B_{m \sigma}) g^{\sigma \rho} ( B_{\rho \nu} + B_{\rho n} A^n{}_\nu)\\
\end{array}
\endgroup & e^{-2\alpha \phi} (B_{\mu \sigma} + A_\mu{}^m B_{m \sigma}) g^{\sigma \nu}\\
&\\
- e^{-2\alpha \phi} g^{\mu \sigma} ( B_{\sigma \nu} + B_{\sigma n} A^n{}_\nu) & e^{-2\alpha \phi} g^{\mu \nu}\\
\end{pmatrix}\\
\mathcal{G}_{AB} & = \begin{pmatrix}
e^{2 \beta \phi} g_{mn} - e^{-2 \beta \phi} B_{mp} g^{pq} B_{qn} & e^{-2 \beta \phi} B_{mp} g^{pn}\\
- e^{-2 \beta \phi} g^{mp} B_{pn} & e^{-2\beta \phi} g^{mn}
\end{pmatrix}\\
{\mathcal{A}}_M{}^A & = \begin{pmatrix} A_\mu{}^m & B_{\mu m} \\ 0 & 0 \end{pmatrix}, \qquad
{\mathcal{A}}^A{}_M = \begin{pmatrix} A^m{}_\mu & 0 \\ - B_{m \mu} & 0 \end{pmatrix} = {\left( {\mathcal{A}}_M{}^A \right)}^T\\
{\mathcal{B}}^M{}_A & = \begin{pmatrix} 0 & 0 \\ - B_{\mu m} & - A_\mu{}^m \end{pmatrix}, \qquad
{\mathcal{B}}_A{}^M = \begin{pmatrix} 0 & B_{m \mu} \\ 0 & - A^m{}_\mu \end{pmatrix} = {\left( {\mathcal{B}}^M{}_A \right)}^T\,.
\end{align}
The new indices are DFT-type doubled indices given by ${}_M = ({}_\mu, {}^\mu)$, ${}_A = ({}_m, {}^m)$. In effect, we have split the parent DFT coordinates into two sets of doubled coordinates ${\hat{Y}}^{\hat{M}} = (Y^M, Y^A)$ for which we recognise $\mathcal{M}_{MN}$ as a generalised metric on the $Y^M$ space (constructed from a metric $e^{2\alpha \phi} g_{\mu \nu}$ and 2-form $B_{\mu \nu} + A_\mu{}^m B_{m \nu}$) and $\mathcal{G}_{AB}$ as a generalised metric on the $Y^A$ space (itself constructed from a metric $e^{2\beta \phi} g_{mn}$ and 2-form $B_{mn}$). Finally, $\mathcal{A}$ and $\mathcal{B}$ take on the roles of generalised Kaluza-Klein vectors, though not in the standard form \eqref{eq:KKMetric} but rather in a manner that treats the internal DFT and external DFT on an equal footing \eqref{eq:GeneralisedKK} by introducing extra $\mathcal{B}$-twisted terms. As such, we shall refer to this type of ansatz as a \emph{generalised Kaluza-Klein ansatz}. We shall comment on this again when we consider the reduction of the $\operatorname{SL}(5)$ generalised metric.\par
Note that if we split the indices of the $\operatorname{O}(D,D)$ structure in the same way, we obtain
\begin{align}
{\hat{\eta}}_{\hat{M} \hat{N}} & = \begin{pmatrix}
\eta_{MN} & 0\\
0 & \eta_{AB}	
\end{pmatrix}
\end{align}
with
\begin{align}
\eta_{MN} & = \begin{pmatrix} 0 & \delta_\mu^\nu\\ \delta^\mu_\nu & 0\end{pmatrix}\,, \qquad \eta_{AB} = \begin{pmatrix} 0 & \delta_m^n\\ \delta^m_n & 0 \end{pmatrix}\,.
\end{align}
Then the generalised Kaluza-Klein vectors can be related to each other as
\begin{align}
\eta_{AB} \mathcal{A}^B{}_N \eta^{NM} = - \mathcal{B}_A{}^M\,,
\end{align}
and it is then tempting to write this in term of the Y-tensor as follows:
\begin{align}
{\hat{Y}}^{MN}{}_{BA} {\mathcal{A}}^B{}_N = - {\mathcal{B}}_A{}^M\,.
\end{align}
Then this gives us a generalised KK ansatz for ExFTs and the usual KK ansatz can then be understood as a particular case since the $Y$-tensor vanishes for $\hat{G} = \operatorname{GL}(d)$ in GR (when viewed as an ExFT with no external space at all), causing all of the extra $\mathcal{B}$-twisted terms to drop out. The fact that $\mathcal{A}$ and $\mathcal{B}$ are related is also consistent with the counting of degrees of freedom: ${\hat{\mathcal{M}}}_{\hat{M} \hat{N}}$ parametrises the coset $\operatorname{O}(D,D) / (\operatorname{O}(D) \times \operatorname{O}(D))$ and thus has $D^2 = {(n+ d)}^2 = n^2 + d^2 + 2 nd$ components. The $d^2$ and $n^2$ components enter into $\mathcal{M}_{MN} \in \operatorname{O}(d,d) / (\operatorname{O}(d) \times \operatorname{O}(d) )$ and $\mathcal{G}_{AB} \in \operatorname{O}(n,n) / (\operatorname{O}(n) \times \operatorname{O}(n))$ respectively and the $2nd$ components entered into $\mathcal{A}_M{}^A \sim \mathcal{B}^M{}_A \sim (A_\mu{}^m, B_{\mu m})$.\par
In order to invert this generalised KK ansatz, we note that we may diagonalise the matrix ${\hat{\mathcal{M}}}_{\hat{M} \hat{N}}$ as follows:
\begin{align}
{\hat{\mathcal{M}}}_{\hat{P} \hat{Q}} = {\hat{\mathcal{E}}}_{\hat{P}}{}^{\hat{M}} {\tilde{\mathcal{M}}}_{\hat{M} \hat{N}} {\hat{\mathcal{E}}}^{\hat{N}}{}_{\hat{Q}}
\end{align}
where
\begin{align*}
{\tilde{\mathcal{M}}}_{\hat{M} \hat{N}} = \begin{pmatrix}
\mathcal{M}_{MN} & 0 \\
0 & \mathcal{G}_{AB}
\end{pmatrix}\,, \qquad {\hat{\mathcal{E}}}_{\hat{P}}{}^{\hat{M}} = \begin{pmatrix}
\delta^M_P & \mathcal{A}_P{}^A\\
\mathcal{B}_C{}^M & \delta^A_C
\end{pmatrix}, \qquad 
{\hat{\mathcal{E}}}^{\hat{N}}{}_{\hat{Q}} = \begin{pmatrix}
\delta^N_Q & \mathcal{B}^N{}_D\\
\mathcal{A}^B{}_Q & \delta^B_D
\end{pmatrix}\,.
\end{align*}
We then proceed by noting that the combinations $\mathcal{A}^A{}_M \mathcal{B}^M{}_B$ and $\mathcal{B}_A{}^M \mathcal{A}_M{}^B$ both vanish such that the inverses of ${\hat{\mathcal{A}}}_{\hat{M}}{}^{\hat{P}}$ and ${\hat{\mathcal{A}}}^{\hat{N}}{}_{\hat{Q}}$ are given by
\begin{align*}
{\left({\hat{\mathcal{E}}}^{-1} \right)}_{\hat{P}}{}^{\hat{N}} & = \begin{pmatrix}
\delta_P^N + \mathcal{A}_P{}^B \mathcal{B}_B{}^N & - \mathcal{A}_P{}^B\\
- \mathcal{B}_C{}^N & \delta^B_C \end{pmatrix}\,, \quad
{\left( {\hat{\mathcal{E}}}^{-1} \right)}^{\hat{Q}}{}_{\hat{R}} = \begin{pmatrix}
\delta^Q_R + \mathcal{B}^Q{}_C \mathcal{A}^C{}_R & - \mathcal{B}^Q{}_E\\
- \mathcal{A}^D{}_R & \delta^D_E
\end{pmatrix}\,.
\end{align*}
The inverse of the generalised KK ansatz is then given by
\begingroup
\renewcommand*{\arraystretch}{1.5}
\begin{align}
{\hat{\mathcal{M}}}^{\hat{M} \hat{N}} & \coloneqq {\left( {\hat{\mathcal{E}}}^{-1} \right)}^{\hat{M}}{}_{\hat{P}} {\left( {\tilde{\mathcal{M}}}^{-1} \right)}^{\hat{P} \hat{Q}}  {\left( {\hat{\mathcal{E}}}^{-1} \right)}_{\hat{Q}}{}^{\hat{N}}\\
& = \begin{pmatrix} \label{eq:InverseGeneralisedKK}
\begingroup
\renewcommand{\arraystretch}{1}
\begin{array}[t]{c}
\mathcal{M}^{MN} +  \mathcal{B}^M{}_E \mathcal{A}^E{}_P {\mathcal{M}}^{PQ} \mathcal{A}_Q{}^F \mathcal{B}_F{}^N\\
\mathcal{B}^M{}_E \mathcal{A}^E{}_P \mathcal{M}^{PN} + {\mathcal{M}}^{MQ} \mathcal{A}_Q{}^F \mathcal{B}_F{}^N\\
+ \mathcal{B}^M{}_C \mathcal{G}^{CD} \mathcal{B}_D{}^N
\end{array}
\endgroup
& 
\begingroup
\renewcommand{\arraystretch}{1}
\begin{array}[t]{c}
-  \left( \delta^M_P + \mathcal{B}^M{}_E \mathcal{A}^E{}_P \right) \mathcal{M}^{PQ} \mathcal{A}_Q{}^B\\
- \mathcal{B}^M{}_C \mathcal{G}^{CB}
\end{array}
\endgroup
\\
\begingroup
\renewcommand*{\arraystretch}{1}
\begin{array}[t]{c}
- \mathcal{A}^A{}_P \mathcal{M}^{PQ} \left( \delta_Q^N + \mathcal{A}_Q{}^F \mathcal{B}_F{}^N \right)\\
- \mathcal{G}^{AD} \mathcal{B}_D{}^N
\end{array}
\endgroup
&
\mathcal{A}^A{}_P \mathcal{M}^{PQ} \mathcal{A}_Q{}^B + \mathcal{G}^{AB}
\end{pmatrix}\,.
\end{align}
\endgroup
The parametrisation that we obtain here matches the one obtained in \cite{Hohm:2013nja}, in which they constructed a tensor hierarchy for DFT (essentially enhancing it to a full EFT with group $\operatorname{O}(D,D)$) which, in turn, was found to be consistent with results from the heterotic theory. We have also constructed the second generalised metric explicitly. So as to compare our results with those in the literature we write their generalised vector as, $A_\mu{}^A[\text{HS}]$ in terms of the fields here as
\begin{align}
A_\mu{}^A[\text{HS}] = \begin{pmatrix}
A_\mu{}^m [\text{here}]\\
- B_{\mu m} [\text{here}]
\end{pmatrix}
\end{align}
and their generalised metric $\mathcal{M}_{MN}[\text{HS}]$ is should be thought of as our $\mathcal{G}_{AB}$, parametrised in terms of the internal components $g_{mn}[\text{here}]$ and $B_{mn} [\text{here}]$. However, the 2-form that they introduce as part of the tensor hierarchy is related to ours by field redefinitions:
\begin{align}
C_{\mu \nu}[\text{HS}] = B_{\mu \nu}[\text{HS}] + \frac{1}{2} A_\mu{}^A [\text{HS}] A_{\nu A}[\text{HS}] = B_{\mu \nu} [\text{here}] +  A_\mu{}^m [\text{here}] B_{m \nu}[\text{here}]
\end{align}
and so $B_{\mu \nu}[\text{HS}] = B_{\mu \nu}[\text{here}] + A_{[\mu|}{}^m [\text{here}]B_{m| \nu]} [\text{here}]$. The crucial difference between these two constructions is the location of the isometries. In the canonical Kaluza-Klein set-up, one must take the internal space (with coordinates $Y^m$) to be an isometry such that all the fields transform covariantly under the symmetries of the reduced theory. When lifting this ansatz to the doubled spacetime, one must presumably require the doubled space $Y^A = (Y^m, Y_m)$ to be an isometry such that any DFT frame will have the same number of isometries. However, this differs from the coordinate dependence of \cite{Hohm:2013nja} where the fields are allowed to depend on (in our notation) $Y^m$ and $Y_m$. 
\subsection{Reduction of the \texorpdfstring{$\operatorname{SL}(5)$}{SL(5)} Generalised Metric}
We now turn to the reduction of the $\hat{G} = \operatorname{SL}(5)$ generalised metric which is given by
\begin{align}
{\hat{\mathcal{M}}}_{\hat{M}\hat{N}} = \begin{pmatrix}
{\hat{g}}_{\hat{m} \hat{n}} + \frac{1}{2} {\hat{C}}_{\hat{m} \hat{k} \hat{l}} {\hat{g}}^{\hat{k} \hat{l}, \hat{p} \hat{q}} {\hat{C}}_{\hat{p} \hat{q} \hat{n}} & \frac{1}{\sqrt{2}} {\hat{C}}_{\hat{m} \hat{k} \hat{l}} {\hat{g}}^{\hat{k} \hat{l}, {\hat{n}_1 {\hat{n}}_2}}\\
\frac{1}{\sqrt{2}} {\hat{g}}^{{\hat{m}}_1 {\hat{m}}_2, \hat{p} \hat{q}} {\hat{C}}_{\hat{p} \hat{q} \hat{n}} & {\hat{g}}^{{\hat{m}}_1 {\hat{m}}_2, {\hat{n}}_1 {\hat{n}}_2}
\end{pmatrix}\,,
\end{align}
where ${\hat{g}}^{\hat{m}\hat{n}, \hat{p} \hat{q}} = \frac{1}{2} ( {\hat{g}}^{\hat{m} \hat{p}} {\hat{g}}^{\hat{q} \hat{n}} - {\hat{g}}^{\hat{m} \hat{q}} {\hat{g}}^{\hat{p} \hat{n}})$ and $\hat{m}, \hat{n}= 1, \ldots, 4$ such that $\hat{M} = 1, \ldots, 10$. We consider the reduction of this metric to the generalised metric of both $\operatorname{O}(3,3)$ DFT and $\operatorname{SL}(3) \times \operatorname{SL}(2)$ EFT. The first of these was explored in \cite{Thompson:2011uw}. Here we give an equivalent description that facilitates the comparison with the reduction to the $\operatorname{SL}(3) \times \operatorname{SL}(2)$ generalised metric.\par
We split the index $\hat{m} = (m, z)$ with $m = 1,2,3$ for the KK ansatz
\begin{align}
{\hat{g}}_{\hat{m} \hat{n}} & = \begin{pmatrix}
e^{2\alpha \phi} g_{mn} + e^{2\beta \phi} A_m A_n & e^{2\beta \phi} A_m\\
e^{2 \beta \phi} A_n & e^{2\beta \phi}
\end{pmatrix}\label{eq:CircleReduction}\,,\\
{\hat{g}}^{\hat{m} \hat{n}} & = \begin{pmatrix}
e^{-2\alpha \phi} g^{mn} & -  e^{-2 \alpha \phi} A^m\\
- e^{- 2 \alpha \phi} A^n & e^{-2 \alpha \phi} A^m A_m + e^{- 2 \beta \phi}
\end{pmatrix}\label{eq:InverseCircleReduction}\,,\\
{\hat{C}}_{\hat{m} \hat{n} \hat{p}} & = \begin{pmatrix}
C_{m n p} + 3 B_{[m n} A_{p]}\\
3B_{mn}
\end{pmatrix}\,,
\end{align}
which induces the reduction of the $\operatorname{SL}(5)$ generalised metric
\begin{align}
{\hat{\mathcal{M}}}_{\hat{M}\hat{N}} = \begin{pmatrix}
{\hat{\mathcal{M}}}_{m n} & {\hat{\mathcal{M}}}_{m z} & {\hat{\mathcal{M}}}_{m}{}^{n_1 n_2} & {\hat{\mathcal{M}}}_{m}{}^{n_1 z}\\
{\hat{\mathcal{M}}}_{z n} & {\hat{\mathcal{M}}}_{z z} & {\hat{\mathcal{M}}}_{z}{}^{n_1 n_2} & {\hat{\mathcal{M}}}_{z}{}^{n_1 z}\\
{\hat{\mathcal{M}}}^{m_1 m_2}{}_n & {\hat{\mathcal{M}}}^{m_1 m_2}{}_z & {\hat{\mathcal{M}}}^{m_1 m_2, n_1 n_2} & {\hat{\mathcal{M}}}^{m_1 m_2, n_1 z}\\
{\hat{\mathcal{M}}}^{m_1 z}{}_n & {\hat{\mathcal{M}}}^{m_1 z}{}_z & {\hat{\mathcal{M}}}^{m_1 z, n_1 n_2} &{\hat{\mathcal{M}}}^{m_1 z, n_1 z}\\
\end{pmatrix}\,,
\end{align}
where
\begin{align}
{\hat{\mathcal{M}}}_{mn} & =\begingroup \renewcommand{\arraystretch}{1.5} \begin{array}[t]{l}
e^{2\alpha \phi} g_{mn} + e^{2\beta \phi} A_m A_n - 9 e^{-2(\alpha + \beta) \phi} B_{mp} g^{pq} B_{qn}\\
\qquad + \frac{1}{2} e^{-4 \alpha \phi} \left( C_{m k l} - 3 B_{[m k} A_{l]} \right) g^{k l, p q} \left( C_{p q n} - 3 A_{[p} B_{q n]} \right)
\end{array}\endgroup\\
{\hat{\mathcal{M}}}_{mz} & = \frac{3}{2} e^{-4 \alpha \phi} (C_{m k l} - 3 B_{[m k} A_{l]} )g^{k l, p q} B_{p q} + e^{2 \beta \phi} A_m\\
{\hat{\mathcal{M}}}_m{}^{n_1 n_2} & = \frac{1}{\sqrt{2}} e^{-4 \alpha \phi} (C_{m k l} - 3 B_{[m k} A_{l]} ) g^{k l, n_1 n_2}\\
{\hat{\mathcal{M}}}_m{}^{n_1 z} & = \frac{1}{\sqrt{2}} e^{-4 \alpha \phi} (C_{m k l} - 3 B_{[m k} A_{l]}) g^{k l, p n_1} A_{p} + \frac{3}{\sqrt{2}} e^{-2(\alpha + \beta)} B_{m p} g^{p n_1}\\
{\hat{\mathcal{M}}}_{zn} & = \frac{3}{2} e^{-4 \alpha \phi} B_{k l} g^{k l, p q} (C_{p q n} - 3 B_{[p q} A_{n]}) + e^{2\beta \phi} A_n\\
{\hat{\mathcal{M}}}_{zz} & = e^{2\beta \phi} + \frac{9}{2} e^{-4\alpha \phi} B_{k l} g^{k l, p q} B_{p q}\\
{\hat{\mathcal{M}}}_z{}^{n_1 n_2} & = \frac{3}{\sqrt{2}} e^{-4 \alpha \phi} B_{p q} g^{p q, n_1 n_2}\\
{\hat{\mathcal{M}}}_z{}^{n_1 z} & = \frac{3}{\sqrt{2}} e^{-4\alpha \phi} B_{k l} A_p g^{k l, p n_1}\\
{\hat{\mathcal{M}}}^{m_1 m_2}{}_n & = \frac{1}{\sqrt{2}} e^{-4\alpha \phi}  g^{m_1 m_2, p q} (C_{p q n} - 3 A_{[p} B_{q n]})\\
{\hat{\mathcal{M}}}^{m_1 m_2}{}_z & = \frac{3}{\sqrt{2}} e^{-4 \alpha \phi} g^{m_1 m_2, k l} B_{k l}\\
{\hat{\mathcal{M}}}^{m_1 m_2, n_1 n_2} & = e^{-4\alpha \phi} g^{m_1 m_2, n_1 n_2}\\
{\hat{\mathcal{M}}}^{m_1 m_2, n_1 z} & = e^{-4 \alpha \phi} g^{m_1 m_2, p n_1} A_p\\
{\hat{\mathcal{M}}}^{m_1 z}{}_n & = - \frac{1}{\sqrt{2}} e^{-4 \alpha \phi} g^{m_1 k, p q} A_k( C_{p q n} - 3B_{[p q} A_{n]}) -\frac{3}{\sqrt{2}} e^{-2 (\alpha + \beta) \phi} g^{m_1 p} B_{p n}\\
{\hat{\mathcal{M}}}^{m_1 z}{}_z & = - \frac{3}{\sqrt{2}} e^{-4\alpha} g^{m_1 k, p q} A_k B_{p q}\\
{\hat{\mathcal{M}}}^{m_1 z, n_1 n_2} & =  - e^{-4 \alpha \phi} A_k g^{m_1 k, n_1 n_2}\\
{\hat{\mathcal{M}}}^{m_1 z, n_1 z} & = \frac{1}{2} e^{-2(\alpha  + \beta) \phi} g^{m_1 n_1} - \frac{1}{2} e^{-4\alpha} A_k g^{m_1 k, p n_1} A_p
\end{align}
We now have two possible reductions, depending on which components we choose to form the reduced generalised metric from\footnote{In principle, one should also be able to do the same for the DFT case considered above.}. In the following, the components that enter into ${\hat{\mathcal{M}}}_{MN}$ and ${\hat{\mathcal{M}}}_{AB}$ in the reduced theory are boxed in red and green respectively. For $\operatorname{SL}(5) \rightarrow \operatorname{SL}(3) \times \operatorname{SL}(2)$ we choose
\begin{equation}\label{eq:SL3SL2Reduction}
{\hat{\mathcal{M}}}_{\hat{M} \hat{N}} =
\begin{tikzpicture}[baseline=(current bounding box.center)]
  \matrix (m)[
    matrix of math nodes,
    nodes in empty cells,
    left delimiter={(},
    right delimiter={)},
    minimum width=2.3cm,
    minimum height=0.6cm,
  ] {
{\hat{\mathcal{M}}}_{m n} &
{\hat{\mathcal{M}}}_{m z} &
{\hat{\mathcal{M}}}_{m}{}^{n_1 n_2} &
{\hat{\mathcal{M}}}_{m}{}^{n_1 z}\\
{\hat{\mathcal{M}}}_{z n} &
{\hat{\mathcal{M}}}_{z z} &
{\hat{\mathcal{M}}}_{z}{}^{n_1 n_2} &
{\hat{\mathcal{M}}}_{z}{}^{n_1 z}\\
{\hat{\mathcal{M}}}^{m_1 m_2}{}_n &
{\hat{\mathcal{M}}}^{m_1 m_2}{}_z &
{\hat{\mathcal{M}}}^{m_1 m_2, n_1 n_2} &
{\hat{\mathcal{M}}}^{m_1 m_2, n_1 z}\\
{\hat{\mathcal{M}}}^{m_1 z}{}_n &
{\hat{\mathcal{M}}}^{m_1 z}{}_z &
{\hat{\mathcal{M}}}^{m_1 z, n_1 n_2} &
{\hat{\mathcal{M}}}^{m_1 z, n_1 z}\\
  } ;
  \draw[draw=red] (m-1-1.south west) rectangle (m-1-1.north east);
  \draw[draw=red] (m-3-1.south west) rectangle (m-3-1.north east);
  \draw[draw=red] (m-1-3.south west) rectangle (m-1-3.north east);
  \draw[draw=red] (m-3-3.south west) rectangle (m-3-3.north east);
  \draw[draw=green] (m-2-2.south west) rectangle (m-2-2.north east);
  \draw[draw=green] (m-4-2.south west) rectangle (m-4-2.north east);
  \draw[draw=green] (m-2-4.south west) rectangle (m-2-4.north east);
  \draw[draw=green] (m-4-4.south west) rectangle (m-4-4.north east);
\end{tikzpicture}\,,
\end{equation}
whilst for $\operatorname{SL}(5) \rightarrow \operatorname{O}(3,3)$, we choose
\begin{equation}\label{eq:O33Reduction}
{\hat{\mathcal{M}}}_{\hat{M} \hat{N}} =
\begin{tikzpicture}[baseline=(current bounding box.center)]
  \matrix (m)[
    matrix of math nodes,
    nodes in empty cells,
    left delimiter={(},
    right delimiter={)},
    minimum width=2.3cm
  ] {
{\hat{\mathcal{M}}}_{m n} &
{\hat{\mathcal{M}}}_{m z} &
{\hat{\mathcal{M}}}_{m}{}^{n_1 n_2} &
{\hat{\mathcal{M}}}_{m}{}^{n_1 z}\\
{\hat{\mathcal{M}}}_{z n} &
{\hat{\mathcal{M}}}_{z z} &
{\hat{\mathcal{M}}}_{z}{}^{n_1 n_2} &
{\hat{\mathcal{M}}}_{z}{}^{n_1 z}\\
{\hat{\mathcal{M}}}^{m_1 m_2}{}_n &
{\hat{\mathcal{M}}}^{m_1 m_2}{}_z &
{\hat{\mathcal{M}}}^{m_1 m_2, n_1 n_2} &
{\hat{\mathcal{M}}}^{m_1 m_2, n_1 z}\\
{\hat{\mathcal{M}}}^{m_1 z}{}_n &
{\hat{\mathcal{M}}}^{m_1 z}{}_z &
{\hat{\mathcal{M}}}^{m_1 z, n_1 n_2} &
{\hat{\mathcal{M}}}^{m_1 z, n_1 z}\\
  } ;
  \draw[draw=red] (m-1-1.north west) rectangle (m-1-1.south east);
  \draw[draw=red] (m-2-3.north east) rectangle (m-1-4.north east);
  \draw[draw=red] (m-4-1.south west) rectangle (m-4-1.north east);
  \draw[draw=red] (m-3-3.south east) rectangle (m-4-4.south east);
  \draw[draw=green] (m-1-1.south east) rectangle (m-3-3.south east);
  \draw[draw=green] (m-1-3.south west) -- (m-3-3.south west);
  \draw[draw=green] (m-2-2.south west) -- (m-2-3.south east);
\end{tikzpicture}\,.
\end{equation}
In both cases we have suppressed the components that enter into the off-diagonal pieces but they should hopefully be clear from the above: each piece takes one component from each quadrant of ${\hat{\mathcal{M}}}_{\hat{M} \hat{N}}$. We shall use the indices $M, N$ to index the coordinate representation of the reduced theory and $A,B$ to index the remaining coordinates. For the $\operatorname{SL}(3) \times \operatorname{SL}(2)$ reduction, we have $Y^M = (Y^m, Y_{m_1 m_2})$ whilst for the $\operatorname{O}(3,3)$ reduction we have $Y^M = (Y^m ,Y_{m_1 z})$.\par
Note that we have introduced a slight abuse of terminology; the direction $z$ plays different roles in the two reductions since it corresponds to the M-theory circle in the $\operatorname{O}(3,3)$ reduction but to the decompactification of one of the directions of the M-theory 4-torus in oxidising $d=7$ to $d=8$. It is thus perhaps better to think of $z$ as just some `distinguished' direction rather than a compactification circle. The shift in perspective is indicated by the change in identification of the dilaton below since the fixing of the dilaton determines the particular embedding of the $\operatorname{SL}(3) \subset \operatorname{SL}(5)$.
\subsubsection{\texorpdfstring{$\operatorname{SL}(5)$}{SL(5)} EFT to \texorpdfstring{$\operatorname{SL}(3) \times \operatorname{SL}(2)$}{SL(3)xSL(2)} EFT Reduction}\label{sec:SL3xSL2GenMetric}
We begin by rescaling the whole $\operatorname{SL}(5)$ generalised metric ${\hat{\mathcal{M}}}_{\hat{M} \hat{N}}$ by ${\hat{g}}^{\frac{1}{5}} = g^{\frac{1}{5}} e^{\frac{6\alpha + 2 \beta}{5}\phi}$ to obtain a determinant 1 generalised metric. From \eqref{eq:SL3SL2Reduction}, we rearrange the components of the $\operatorname{SL}(5)$ generalised metric into blocks that expose the underlying structure that will become apparent in a moment:
\begin{align}
\begin{split}
{\hat{\mathcal{M}}}_{MN} & = g^{\frac{1}{5}} e^{\frac{-14\alpha + 2\beta}{5} \phi}
\begin{pmatrix}
e^{6\alpha \phi} g_{m n} + \frac{1}{2} {\tilde{C}}_{m k l} g^{k l, p q}  {\tilde{C}}_{p q n} & \frac{1}{\sqrt{2}} {\tilde{C}}_{m k l}  g^{k l, n_1 n_2}\\
\frac{1}{\sqrt{2}}g^{m_1 m_2, p q} {\tilde{C}}_{p q n} & g^{m_1 m_2, n_1 n_2}
\end{pmatrix}\\
& \qquad + g^{\frac{1}{5}} e^{\frac{-4 \alpha -8 \beta}{5} \phi}
\begin{pmatrix}
e^{(2 \alpha + 4 \beta) \phi} A_m A_n - \frac{1}{2} {\tilde{B}}_{m p} g^{p q} {\tilde{B}}_{q n} & 0\\
0 & 0\\
\end{pmatrix}\,,
\end{split}\\
\begin{split}
{\hat{\mathcal{M}}}_{MB} & = g^{\frac{1}{5}}e^{\frac{-14\alpha + 2 \beta}{5} \phi}
\begin{pmatrix}
\frac{1}{2\sqrt{2}} {\tilde{C}}_{m k l} g^{k l, p q} {\tilde{B}}_{p q} &  \frac{1}{\sqrt{2}} {\tilde{C}}_{m k l} g^{k l, p n_1} A_p\\
\frac{1}{2} g^{m_1 m_2, p q} {\tilde{B}}_{p q} & g^{m_1 m_2, p n_1} A_p
\end{pmatrix}\\
& \qquad +g^{\frac{1}{5}} e^{\frac{-4\alpha - 8 \beta}{5}}
\begin{pmatrix}
e^{(2 \alpha + 4\beta) \phi} A_m & \frac{1}{2} {\tilde{B}}_{m p} g^{p n_1}\\
0 & 0
\end{pmatrix}\,,
\end{split}\\
\begin{split}
{\hat{\mathcal{M}}}_{AN} & = g^{\frac{1}{5}} e^{\frac{-14\alpha +2 \beta}{5}\phi}
\begin{pmatrix}
\frac{1}{2 \sqrt{2}} {\tilde{B}}_{k l} g^{k l, p q} {\tilde{C}}_{p q n} & \frac{1}{2} {\tilde{B}}_{k l} g^{k l, n_1 n_2}\\
- \frac{1}{\sqrt{2}} g^{m_1 l, p q} A_l {\tilde{C}}_{p q n} & - A_l g^{m_1 l, n_1 n_2}
\end{pmatrix}\\
& \qquad + g^{\frac{1}{5}} e^{\frac{-4\alpha - 8 \beta}{5} \phi}
\begin{pmatrix}
e^{(2 \alpha + 4 \beta) \phi} A_n & 0\\
- \frac{1}{2} g^{m k} {\tilde{B}}_{k n} & 0\\
\end{pmatrix}\,,
\end{split}\\
\begin{split}
{\hat{\mathcal{M}}}_{AB} & =g^{\frac{1}{5}} e^{\frac{-4\alpha - 8\beta}{5}}
\begin{pmatrix}
e^{(2\alpha + 4 \beta) \phi} & 0\\
0 & \frac{1}{2}  g^{m_1 n_1}
\end{pmatrix}\\
& \qquad + g^{\frac{1}{5}}e^{\frac{-14\alpha + 2 \beta}{5} \phi}
\begin{pmatrix}
\frac{1}{4} {\tilde{B}}_{k l} g^{k l, p q} {\tilde{B}}_{p q} & \frac{1}{2} {\tilde{B}}_{k l} A_p g^{k l, p n_1}\\
- \frac{1}{2} g^{m_1 l, p q} A_l {\tilde{B}}_{p q} & - A_k g^{m_1 k, p n_1} A_p
\end{pmatrix}\,,
\end{split}
\end{align}
where we have defined
\begin{align}
{\tilde{C}}_{m n p} & \coloneqq C_{m n p} - 3 B_{[m n} A_{p]}\,,\qquad {\tilde{B}}_{m n} \coloneqq 3 \sqrt{2} B_{m n}\,.
\end{align}
We can make the group structure more explicit by defining dual coordinates
\begin{align}
Y^{\overbar{m}} \coloneqq \varepsilon^{m n_1 n_2} Y_{n_1 n_2}\,,
\end{align}
where $\overbar{m} =1, 2,3$ indexes a distinct $\mathbf{3}$ of $\operatorname{SL}(3)$ to the first one that we indexed by $m = 1, 2,3$ (and we have thus adorned with an overbar to distinguish the two) but that is raised and lowered with the \emph{same} 3-dimensional metric such that $g_{\overbar{m} \overbar{n}} = g_{m n} = g_{m \overbar{n}} = g_{\overbar{m} n}$, in the same way that $g^{m_1 m_2, n_1 n_2}$ contains the same metric degrees of freedom as $g_{mn}$. In terms of these coordinates, we have
\begin{align}
\begin{split}
{\hat{\mathcal{M}}}_{MN} & = g^{\frac{1}{5}} e^{\frac{-14\alpha + 2\beta}{5} \phi}
\begin{pmatrix}
e^{6\alpha \phi} g_{m n} + {\tilde{C}}^2 g_{m n} & \sqrt{\frac{2}{g}} \tilde{C} g_{m \overbar{n}}\\
\sqrt{\frac{2}{g}} \tilde{C} g_{\overbar{m}n} & \frac{2}{g} g_{\overbar{m} \overbar{n}} 
\end{pmatrix}\\
& \qquad + g^{\frac{1}{5}} e^{\frac{-4 \alpha -8 \beta}{5} \phi}
\begin{pmatrix}
e^{(2 \alpha + 4 \beta) \phi} A_m A_n - \frac{1}{2} {\tilde{B}}_{m k} g^{l q} {\tilde{B}}_{q n} & 0\\
0 & 0\\
\end{pmatrix}\,,
\end{split}\\
\begin{split}
{\hat{\mathcal{M}}}_{MB} & = g^{\frac{1}{5}}e^{\frac{-14\alpha + 2 \beta}{5} \phi}
\begin{pmatrix}
\frac{1}{2\sqrt{2}} {\tilde{C}}_{m k l} g^{k l, p q} {\tilde{B}}_{p q} &  \frac{1}{\sqrt{2}} {\tilde{C}}_{m k l} g^{k l, q n_1} A_q\\
\frac{1}{2\sqrt{2}} g^{-\frac{1}{2}} g_{m k} \epsilon^{k p q} {\tilde{B}}_{p q} & g^{-\frac{1}{2}} g_{m k} \epsilon^{k q n_1} A_q
\end{pmatrix}\\
& \qquad +g^{\frac{1}{5}} e^{\frac{-4\alpha - 8 \beta}{5}}
\begin{pmatrix}
e^{(2 \alpha + 4\beta) \phi} A_m & \frac{1}{2} {\tilde{B}}_{m k} g^{k n_1}\\
0 & 0
\end{pmatrix}\,,
\end{split}\\
\begin{split}
{\hat{\mathcal{M}}}_{AN} & = g^{\frac{1}{5}} e^{\frac{-14\alpha +2 \beta}{5}\phi}
\begin{pmatrix}
\frac{1}{2\sqrt{2}} {\tilde{B}}_{k l} g^{k l, p q} {\tilde{C}}_{p q n} & \frac{1}{2} g^{-\frac{1}{2}} {\tilde{B}}_{k l} \epsilon^{k l q} g_{q n}\\
- \frac{1}{\sqrt{2}} g^{m_1 l, p q} A_l {\tilde{C}}_{p q n} & - A_k \epsilon^{m_1 k l} g_{l n}
\end{pmatrix}\\
& \qquad + g^{\frac{1}{5}} e^{\frac{-4\alpha - 8 \beta}{5} \phi}
\begin{pmatrix}
e^{(2 \alpha + 4 \beta) \phi} A_n & 0\\
- \frac{1}{2} g^{m q} {\tilde{B}}_{q n} & 0\\
\end{pmatrix}\,,
\end{split}\\
\begin{split}
{\hat{\mathcal{M}}}_{AB} & =g^{\frac{1}{5}} e^{\frac{-4\alpha - 8\beta}{5}}
\begin{pmatrix}
e^{(2\alpha + 4 \beta) \phi} & 0\\
0 & \frac{1}{2}  g^{m_1 n_1}
\end{pmatrix}\\
& \qquad + g^{\frac{1}{5}}e^{\frac{-14\alpha + 2 \beta}{5} \phi}
\begin{pmatrix}
\frac{1}{4} {\tilde{B}}_{k l} g^{k l, p q} {\tilde{B}}_{p q} & \frac{1}{2} {\tilde{B}}_{k l} A_q g^{k l, q n_1}\\
- \frac{1}{2} g^{m_1 k, p q} A_\lambda {\tilde{B}}_{p q} & - A_k g^{m_1 k, q n_1} A_q
\end{pmatrix}\,.
\end{split}
\end{align}
Note that we have used the fact that ${\tilde{C}}_{m n p} \coloneqq C_{m n p} - 3 B_{[m n} A_{p]}$ is a top form such that ${\tilde{C}}_{m n p} \propto \epsilon_{m n p}$. In particular, we find
\begin{align}
\epsilon^{m k l} {\tilde{C}}_{k l n} = 2{\tilde{C}} \delta^m_n\,, \qquad \tilde{C} \coloneqq \frac{1}{3!} \epsilon^{m n p} C_{m n p}
\end{align}
by taking the trace, giving ${\tilde{C}}_{m n p} = {\tilde{C}} \epsilon_{m n p}$. Under the above splitting, we can rewrite the $\operatorname{SL}(5)$ generalised metric in the same generalised KK ansatz as the DFT case \eqref{eq:GeneralisedKK}
\begin{align}\label{eq:GeneralisedKKAnsatz}
{\hat{\mathcal{M}}}_{\hat{M} \hat{N}} & = \begin{pmatrix}
e^{2A\phi} \mathcal{M}_{MN} + e^{2B\phi} {\mathcal{A}}_M{}^A {\mathcal{G}}_{AB} {\mathcal{A}}^B{}_N & e^{2B\phi} {\mathcal{A}}_M{}^A {\mathcal{G}}_{AB} + e^{2A\phi} {\mathcal{M}}_{MN} {\mathcal{B}}^N{}_B\\
e^{2B\phi} \mathcal{G}_{AB} {\mathcal{A}}^B{}_N + e^{2A\phi} {\mathcal{B}}_A{}^M {\mathcal{M}}_{MN} & e^{2B\phi} \mathcal{G}_{AB} +  e^{2A\phi} {\mathcal{B}}_A{}^M {\mathcal{M}}_{MN} {\mathcal{B}}^N{}_B
\end{pmatrix}\,,
\end{align}
where
\begin{align}
{\mathcal{M}}_{MN} & =  \begin{pmatrix}
e^{6\alpha \phi} g_{m n} + {\tilde{C}}^2 g_{m n} & \sqrt{\frac{2}{g}} \tilde{C} g_{m \overbar{n}}\\
\sqrt{\frac{2}{g}} \tilde{C} g_{\overbar{m} n} & \frac{2}{g} g_{\overbar{m} \overbar{n}} 
\end{pmatrix}\\
{\mathcal{A}}_M{}^A & = \begin{pmatrix}
A_m & {\tilde{B}}_{m p_1}\\
0 & 0
\end{pmatrix}, \quad {\mathcal{A}}^B{}_N = \begin{pmatrix}
A_n & 0\\
- {\tilde{B}}_{q_1 n} & 0\\
\end{pmatrix} = {\left( {\mathcal{A}}^T \right)}^B{}_N\\
\mathcal{B}^N{}_B & = \begin{pmatrix}
0 & 0\\
\frac{1}{4} \varepsilon^{\overbar{m} k l} {\tilde{B}}_{k l} &  \frac{1}{2} \varepsilon^{\overbar{n} p n_1} A_p
\end{pmatrix}\,, \quad \mathcal{B}_A{}^M = \begin{pmatrix}
0 & \frac{1}{4} \varepsilon^{\overbar{m} p q} {\tilde{B}}_{p q}\\
0 &  - \frac{1}{2} A_q \varepsilon^{m_1 q \overbar{m}} 
\end{pmatrix} = {\left( {\mathcal{B}}^T \right)}_A{}^M \label{eq:SL5B}\\
{\mathcal{G}}_{AB} & = \begin{pmatrix}
e^{(2\alpha + 4 \beta) \phi} & 0\\
0 & \frac{1}{2}  g^{m_1 n_1}
\end{pmatrix} \\
e^{2A\phi} & = g^{\frac{1}{5}}e^{\frac{-14\alpha + 2 \beta}{5} \phi}\,, \qquad e^{2B\phi} = g^{\frac{1}{5}}e^{\frac{-4\alpha - 8 \beta}{5} \phi}\,.
\end{align}
Note that both $\mathcal{A}$ and $\mathcal{B}$ do not contain metric degrees of freedom as required. In particular $\mathcal{B}$ is defined with the alternating \emph{symbol} $\varepsilon$ without reference to the metric determinant. As in the DFT case, we have that ${\mathcal{A}}^A{}_M {\mathcal{B}}^M{}_B = {\mathcal{B}}_A{}^M {\mathcal{A}}_M{}^B = \mathbf{0}$ and so the inverse reduction ansatz is given by \eqref{eq:InverseGeneralisedKK} except with $\mathcal{M}_{MN} \rightarrow e^{2A\phi} \mathcal{M}_{MN}$ and $\mathcal{G}_{AB} \rightarrow e^{2B\phi} \mathcal{G}_{AB}$.\par
To demonstrate that $\mathcal{M}_{MN}$ is indeed the $\operatorname{SL}(3) \times \operatorname{SL}(2)$ generalised metric, we define objects upon which the $\operatorname{SL}(2)$ action is manifest:
\begin{align}
C_{(0)} & = \sqrt{\frac{g}{2}} \tilde{C} \,, \qquad e^{\Phi} = e^{-3\alpha \phi} \sqrt{\frac{2}{g}}\,.
\end{align}
Note that, like $\mathcal{B}$ (though unlike $\tilde{C}$), the scalar $C_{(0)} = \frac{1}{\sqrt{2} \cdot 3!} \varepsilon^{m n p} C_{m n p}$ is also defined with the alternating symbol and so does not include the metric degree of freedom that $\tilde{C}$ included; it is an independent degree of freedom from the metric determinant, as required.
Then,
\begin{align}
\mathcal{M}_{MN} = \sqrt{\frac{2}{g}} e^{3\alpha \phi} g_{m n} \otimes \frac{1}{e^{-\Phi}} \begin{pmatrix}
e^{-2\Phi} + C_{(0)}^2 &  C_{(0)}\\
C_{(0)} & 1
\end{pmatrix}\,.
\end{align}
In this form, it is clear that the generalised metric can be factorised into an $\operatorname{SL}(3)$ component and an $\operatorname{SL}(2)$ component as $\mathcal{M}_{MN} = \mathcal{M}_{m n} \otimes \mathcal{M}_{\alpha \beta}$. As in the usual KK ansatz, we have the freedom to fix $\alpha$ to a convenient value. One way to fix it would be to require that we reduce to a generalised metric that also has determinant 1. For the $\operatorname{SL}(3)$ `generalised metric' on the coordinate representation $R_1 = \mathbf{3}$ (which is really just the usual 3-dimensional metric), this occurs for $\mathcal{M}_{m n} = g^{-\frac{1}{3}} g_{m n}$ from which
\begin{align}\label{eq:AlphaFix}
\sqrt{\frac{2}{g}} e^{3\alpha \phi} = g^{-\frac{1}{3}}\,,\qquad \Rightarrow \qquad e^{2\alpha \phi} = {\left( \frac{g}{2^3} \right)}^{\frac{1}{9}}\,.
\end{align}
The other constant $\beta$ can be fixed in the same way as the conventional Kaluza-Klein theory, namely by a choice of frame (by which we mean choice of Weyl scaling to give the string or Einstein frame). Although we shall not conduct the full reduction of the potential, we illustrate what we mean by singling out one of the terms that appears in the reduction. To simplify the analysis, we take $C_{m n p} = B_{m n} = 0$ which implies
\begingroup
\setlength\arraycolsep{0pt}
\begin{align}
{\mathcal{A}}_M{}^A & =\! \begin{pmatrix}
A_m & 0\\
0 & 0\\
\end{pmatrix}\,, \qquad {\mathcal{B}}^M{}_A = \begin{pmatrix}
0 & 0\\
0 & \frac{1}{2} \varepsilon^{\overbar{m} q m_1} A_q
\end{pmatrix}\,,\\
{\hat{\mathcal{M}}}^{\hat{M} \hat{N}} & = \!\begin{pmatrix}
e^{-2A\phi} {\mathcal{M}}^{MN} + e^{-2B\phi} {\mathcal{B}}^M{}_A {\mathcal{G}}^{AB} {\mathcal{B}}_B{}^N & - e^{-2A\phi} {\mathcal{M}}^{MQ} {\mathcal{A}}_Q{}^B - e^{-2B\phi} {\mathcal{B}}^M{}_A {\mathcal{G}}^{AB}\\
- e^{-2A\phi} {\mathcal{A}}^A{}_P {\mathcal{M}}^{PN} - e^{-2B\phi} {\mathcal{G}}^{AB} {\mathcal{B}}_B{}^N & e^{-2A\phi} {\mathcal{A}}^A{}_P {\mathcal{M}}^{PQ} {\mathcal{A}}_Q{}^B + e^{-2B\phi} {\mathcal{G}}^{AB}
\end{pmatrix}\!.\nonumber
\end{align}
\endgroup
Note, in particular, that ${\mathcal{B}}^M{}_A {\mathcal{A}}^A{}_N = \mathbf{0}$. One of the terms in the $\operatorname{SL}(5)$ potential is
\begin{align}
\frac{1}{2} {\hat{\mathcal{M}}}^{\hat{M} \hat{N}} \partial_{\hat{M}} {\hat{\mathcal{M}}}^{\hat{K} \hat{L}} \partial_{\hat{K}} {\hat{\mathcal{M}}}_{\hat{N} \hat{L}} = - \frac{1}{4} g e^{-2B\phi} \partial_{\overbar{m}} A_q \partial^{\overbar{m}} A^q  + \ldots
\end{align}
In particular, the first term contributes to an additional Maxwell term that appears in addition to the $\operatorname{SL}(3) \times \operatorname{SL}(2)$ potential. Taking into account the fact that the $\operatorname{SL}(3) \times \operatorname{SL}(2)$ potential comes with the scaling
\begin{align}
-\frac{1}{12} {\hat{\mathcal{M}}}^{\hat{M} \hat{N}} \partial_{\hat{M}} {\hat{\mathcal{M}}}^{\hat{K} \hat{L}} \partial_{\hat{N}} {\hat{\mathcal{M}}}_{\hat{K} \hat{L}} = - \frac{1}{12} e^{-2A\phi} {\mathcal{M}}^{MN}\partial_M {\mathcal{M}}^{KL} \partial_N {\mathcal{M}}_{KL} + \ldots\,,
\end{align}
one sees that the potential and new Maxwell terms have a relative scaling (up to constant factors) of $g e^{2(A-B)\phi} = g e^{-2(\alpha - \beta) \phi}$ which can be fixed to land on any frame that one may wish by an appropriate choice of $\beta$.

\subsubsection{\texorpdfstring{$\operatorname{SL}(5)$}{SL(5)} EFT to \texorpdfstring{$\operatorname{O}(3,3)$}{O(3,3)} DFT Reduction}
The reduction of the $\operatorname{O}(3,3)$ generalised metric, in the form \eqref{eq:O33Reduction} agrees with \cite{Thompson:2011uw}:
\begin{align}
\begin{split}
{\hat{\mathcal{M}}}_{MN} & = e^{\frac{16 \alpha + 2\beta}{5} \phi} g^{\frac{1}{5}}
\begin{pmatrix}
g_{m n} - {\tilde{B}}_{m p} g^{p q} {\tilde{B}}_{q n} & {\tilde{B}}_{m p} g^{p n_1}\\
- g^{m q} {\tilde{B}}_{q n_1} & g^{m_1 n_1}
\end{pmatrix}\\
& \qquad + g^{\frac{1}{5}} e^{\frac{6 \alpha + 12 \beta}{5}\phi}
\begin{pmatrix}
{\tilde{C}}_{m k l} g^{k l, p q} {\tilde{C}}_{p q n} + A_m A_n & \sqrt{2} {\tilde{C}}_{m k l} g^{k l, p n_1} A_p\\
-\sqrt{2} A_k g^{m_1 k, p q} {\tilde{C}}_{p q n} & -2 A_k g^{m_1 k, q n_1} A_q
\end{pmatrix}\,,
\end{split}\\
{\hat{\mathcal{M}}}_{MB} & = e^{\frac{6\alpha + 12 \beta}{5} \phi} g^{\frac{1}{5}}
\begin{pmatrix}
A_m + \frac{1}{\sqrt{2}} {\tilde{C}}_{m k l} g^{k l, p q} {\tilde{B}}_{p q} & \sqrt{2} {\tilde{C}}_{m k l} g^{k l, n_1 n_2}\\
- g^{m k, p q} A_k {\tilde{B}}_{p q} & - 2 A_k g^{m_1 k, n_1 n_2} 
\end{pmatrix}\,,\\
{\hat{\mathcal{M}}}_{AN} & = e^{\frac{6\alpha + 12 \beta}{5}\phi} g^{\frac{1}{5}}
\begin{pmatrix}
A_\nu + \frac{1}{\sqrt{2}} {\tilde{B}}_{k l} g^{k l, p q} {\tilde{C}}_{p q n} & {\tilde{B}}_{k l} A_{p} g^{k l, p n_1}\\
\sqrt{2} g^{m_1 m_2, p q} {\tilde{C}}_{p q n} & 2 g^{m_1 m_2, k n_1} A_k
\end{pmatrix}\,,\\
{\hat{\mathcal{M}}}_{AB} & = e^{\frac{6\alpha + 12 \beta}{5} \phi} g^{\frac{1}{5}}
\begin{pmatrix}
1 + \frac{1}{2} {\tilde{B}}_{k l} g^{k l, p q} {\tilde{B}}_{p q} & {\tilde{B}}_{k l} g^{k l, n_1 n_2}\\
g^{m_1 m_2, p q} {\tilde{B}}_{p q} & 2 g^{m_1 m_2, n_1 n_2}
\end{pmatrix}\,,
\end{align}
where we have chosen different values of $\alpha$ and $\beta$ from the $\operatorname{SL}(3) \times \operatorname{SL}(2)$ reduction, instead taking
\begin{align}
e^{-(4\alpha + 2 \beta) \phi} = 2\,,
\end{align}
that will enable us to land on the canonical form of the DFT metric. Then, subject to the following identifications
\begin{align}
{\mathcal{M}}_{MN} & = \begin{pmatrix}
g_{mn} - {\tilde{B}}_{mp} g^{pq} {\tilde{B}}_{qn} & {\tilde{B}}_{mp} g^{pn_1}\\
- g^{mp} {\tilde{B}}_{pn_1} & g^{m_1 n_1}
\end{pmatrix}\,,\\
\mathcal{A}_M{}^A & = \begin{pmatrix}
A_m & \frac{1}{\sqrt{2}} {\tilde{C}}_{m p_1 p_2} - \frac{1}{2} A_{[m} {\tilde{B}}_{p_1 p_2]}\\
0 & - \delta^m_{[p_1} A_{p_2]}
\end{pmatrix}\,,\\
\mathcal{A}^B{}_N & = \begin{pmatrix}
A_n & 0\\
\frac{1}{\sqrt{2}}  {\tilde{C}}_{q_1 q_2 n} - \frac{1}{2} {\tilde{B}}_{[q_1 q_2} A_{n]} & A_{[q_1} \delta_{q_2]}^n
\end{pmatrix}\,,\\
{\mathcal{G}}_{AB} & = \begin{pmatrix}
1 + \frac{1}{2} {\tilde{B}}_{kl} g^{kl, pq} {\tilde{B}}_{pq} & {\tilde{B}}_{pq} g^{pq, n_1 n_2}\\
g^{m_1 m_2, pq} {\tilde{B}}_{pq} & 2 g^{m_1 m_2, n_1 n_2}
\end{pmatrix}\,,\\
e^{2A\phi} & = g^{\frac{1}{5}} e^{\frac{16\alpha + 2\beta}{5}}\,, \qquad e^{2B\phi} = g^{\frac{1}{5}} e^{\frac{6\alpha + 12\beta}{5}}\,,
\end{align}
we can rewrite the $\operatorname{SL}(5)$ generalised metric as
\begin{align}\label{eq:KK}
{\hat{\mathcal{M}}}_{MN} & = \begin{pmatrix}
e^{2A \phi} \mathcal{M}_{MN} + e^{2B\phi} {\mathcal{A}}_M{}^A {\mathcal{G}}_{AB} {\mathcal{A}}^B{}_N & e^{2B\phi} {\mathcal{A}}_M{}^A {\mathcal{G}}_{AB}\\
e^{2B\phi} {\mathcal{G}}_{AB} {\mathcal{A}}^B{}_N & e^{2B\phi} {\mathcal{G}}_{AB}
\end{pmatrix}
\end{align}
which is the \emph{doubled KK ansatz} \cite{Thompson:2011uw}. Like the conventional KK ansatz, this can be understood as a particular case of the generalised KK ansatz given by \eqref{eq:GeneralisedKKAnsatz}. One may verify that the pieces of the $\operatorname{SL}(5)$ $Y$-tensor that would have entered into the $\mathcal{B}$-twisted terms under this reduction happen to vanish and so the reduction of the generalised KK ansatz to this doubled KK anstz in this case is non-trivial.\par
We note that the appearance of the $Y$-tensor in the generalised KK ansatz may be justified as follows: in the Kaluza-Klein reduction ansatz, the reduced fields are required to transform under the symmetries of the lower dimensional theory. In ExFTs, these must include the lower-dimensional (generalised) diffeomorphisms and so any appearance of the $Y$-tensor in the reduction ansatz could come about as a compensatory term to ensure the fields transform correctly.\par
\subsection{Some notes on the Reduction of Larger Generalised Metrics}\label{sec:ReductionLargerGenMetric}
Larger generalised metrics are much more difficult to reduce in full; at $E_{6(6)}$ and upwards it also contains the 6-form that couples electrically to the M5 whilst for $E_{8(8)}$, it further contains the dual graviton as propagating degrees of freedom. It is evident that reductions from $E_{6(6)}$ to $\operatorname{SO}(5,5)$ and $E_{8(8)}$ to $E_{7(7)}$ must somehow exclude the 6-form and dual graviton respectively from the reduced generalised metric.\par
Additionally, the generalised metric grows with the coordinate representation. In particular this means that the number of blocks appearing in the generalised metric also grows; for $E_{7(7)}$, the $R_1 = \mathbf{56}$ decomposed under $\operatorname{GL}(7)$ to $\mathbf{7} \oplus \mathbf{21} \oplus \overbar{\mathbf{21}} \oplus \overbar{\mathbf{7}}$ produces $4 \times 4$ block matrices and so it is not clear whether even the generalised KK ansatz \eqref{eq:GeneralisedKK} is sufficient. The case for $E_{8(8)}$ is even worse with $R_1 = \mathbf{248}$. Setting all internal potentials to zero for simplicity, the generalised metric for $E_{8(8)}$ (which we have rescaled to give determinant 1) when decomposed under $\operatorname{GL}(8)$ takes the form
\begin{align}\label{eq:E8GenMetric}
\begin{split}
{\hat{\mathcal{M}}}_{\hat{M} \hat{N}} = \operatorname{diag} \biggl[ & \hat{g} {\hat{g}}_{\hat{m} \hat{n}}, \hat{g} {\hat{g}}^{\hat{m}_1 \hat{m}_2, \hat{n}_1 \hat{n}_2}, {\hat{g}}_{\hat{m}_1 \hat{m}_2 \hat{m}_3, \hat{n}_1 \hat{n}_2 \hat{n}_3}, {\hat{g}}^{\hat{m}_1 \hat{n}_1} {\hat{g}}_{\hat{m}_2 \hat{n}_2} - \frac{1}{8} \delta^{\hat{m}_1}_{\hat{m}_2} \delta^{\hat{n}_1}_{\hat{n}_2}, 1,\\
& {\hat{g}}^{\hat{m}_1 \hat{m}_2 \hat{m}_3, \hat{n}_1 \hat{n}_2, \hat{n}_3}, {\hat{g}}^{-1} {\hat{g}}_{\hat{m}_1 \hat{m}_2, \hat{n}_1 \hat{n}_2}, {\hat{g}}^{-1} {\hat{g}}^{\hat{m} \hat{n}} \biggr]\,,
\end{split}
\end{align}
where ${\hat{m}}_i, {\hat{n}}_i = 1, \ldots, 8$ and we have chosen the conventions
\begin{align}
{\hat{g}}_{{\hat{m}}_1 {\hat{m}}_2, {\hat{n}}_1 {\hat{n}}_2} & \coloneqq {\hat{g}}_{\hat{{m}}_1[{\hat{n}}_1|} {\hat{g}}_{{\hat{m}}_2|{\hat{n}}_2]}\label{eq:G2}\,,\\
{\hat{g}}^{{\hat{m}}_1 {\hat{m}}_2, {\hat{n}}_1 {\hat{n}}_2} & \coloneqq {\hat{g}}^{\hat{{m}}_1[{\hat{n}}_1|} {\hat{g}}^{{\hat{m}}_2|{\hat{n}}_2]} \label{eq:InverseG2}\,,\\
{\hat{g}}_{{\hat{m}}_1 {\hat{m}}_2 {\hat{m}}_3, {\hat{n}}_1 {\hat{n}}_2 {\hat{n}}_3} & \coloneqq {\hat{g}}_{{\hat{m}}_1[{\hat{n}}_1|} {\hat{g}}_{{\hat{m}}_2 |{\hat{n}}_2|} {\hat{g}}_{{\hat{m}}_3 |{\hat{n}}_3]} \label{eq:G3}\,,\\
{\hat{g}}^{{\hat{m}}_1 {\hat{m}}_2 {\hat{m}}_3, {\hat{n}}_1 {\hat{n}}_2 {\hat{n}}_3} & \coloneqq {\hat{g}}^{{\hat{m}}_1[{\hat{n}}_1|} {\hat{g}}^{{\hat{m}}_2 |{\hat{n}}_2|} {\hat{g}}^{{\hat{m}}_3 |{\hat{n}}_3]}\label{eq:InverseG3}\,,
\end{align}
for the metrics on the antisymmetric representations. In principle, one could try the brute-force approach from the previous section and reduce the 8-dimensional internal metric under the standard circle reduction ansatz \eqref{eq:CircleReduction}. The antisymmetrised metrics in this case are given by
\begingroup
\renewcommand*{\arraystretch}{2}
\setlength\arraycolsep{7pt}
\begin{align}
{\hat{g}}_{{\hat{m}}_1 {\hat{m}}_2, {\hat{n}}_1 {\hat{n}}_2} & = \begin{pmatrix}
\begingroup
\renewcommand{\arraystretch}{1}
\begin{array}[t]{c}
e^{4\alpha \phi} g_{m_1 m_2, n_1 n_2}\\ - 2 e^{2(\alpha + \beta)\phi}A_{[m_1} g_{m_2],[n_1} A_{n_2]}
\end{array}
\endgroup & - e^{2(\alpha + \beta)\phi} A_{[m_1} g_{m_2]n_1}\\
e^{2(\alpha + \beta) \phi} g_{m_1[n_1} A_{n_2]} & \frac{1}{2} e^{2(\alpha + \beta)\phi} g_{m_1 n_1}
\end{pmatrix}\,,\\
{\hat{g}}^{{\hat{m}}_1 {\hat{m}}_2, {\hat{n}}_1 {\hat{n}}_2} & = \begin{pmatrix}
e^{-4\alpha \phi} g^{m_1 m_2, n_1 n_2} & e^{-4 \alpha \phi} A^{[m_1} g^{m_2]n_1}\\
- e^{-4\alpha \phi} g^{m_1[n_1} A^{n_2]} & \begingroup \renewcommand{\arraystretch}{1} \begin{array}[t]{c} \frac{1}{2} e^{-2(\alpha + \beta)} g^{m_1 n_1}\\ + \frac{1}{2} e^{-4\alpha \phi} ( g^{m_1 n_1} A \cdot A - A^{m_1} A^{n_1}) \end{array}\endgroup
\end{pmatrix}\,,\\
{\hat{g}}_{{\hat{m}}_1 {\hat{m}}_2 {\hat{m}}_3, {\hat{n}}_1 {\hat{n}}_2 {\hat{n}}_3} & = \begin{pmatrix}
\begingroup
\renewcommand{\arraystretch}{1}
\begin{array}{c}
e^{6\alpha \phi} g_{m_1 m_2 m_3, n_1 n_2 n_3}\\
+ 3 e^{2(2\alpha + \beta)\phi}A_{[m_1} g_{m_2 m_3],[n_1 n_2} A_{n_3]}
\end{array}
\endgroup
& e^{2(2\alpha + \beta)\phi} A_{[m_1} g_{m_2 m_3],n_1 n_2}\\
e^{2(2\alpha + \beta) \phi} g_{m_1 m_2, [n_1 n_2} A_{n_3]} & \frac{1}{3} e^{2(2\alpha + \beta)\phi} g_{m_1 m_2, n_1 n_2}
\end{pmatrix}\,,\\
{\hat{g}}^{{\hat{m}}_1 {\hat{m}}_2 {\hat{m}}_3, {\hat{n}}_1 {\hat{n}}_2 {\hat{n}}_3} & = \begin{pmatrix}
e^{-6\alpha \phi} g^{m_1 m_2 m_3, n_1 n_2 n_3} & - e^{-6\alpha \phi} A^{[m_1} g^{m_2 m_3],n_1 n_2}\\
- e^{-6\alpha \phi} g^{m_1 m_2,[n_1 n_2} A^{n_3]} &
\begingroup
\renewcommand{\arraystretch}{1}
\begin{array}{c}
\frac{1}{3} e^{-2(2\alpha + \beta)\phi} g^{m_1 m_2, n_1 n_2}\\
+ \frac{1}{3} e^{-6 \alpha \phi} g^{m_1 m_2, n_1 n_2} A \cdot A\\
+ \frac{2}{3} e^{-6 \alpha \phi} A^{[m_1} g^{m_2][n_1} A^{n_2]}\\
\end{array}
\endgroup\\
\end{pmatrix}\,.
\end{align}
\endgroup
It is simple to check that ${\hat{g}}^{{\hat{m}}_1 {\hat{m}}_2, {\hat{n}}_1 {\hat{n}}_2}$ and ${\hat{g}}^{{\hat{m}}_1 {\hat{m}}_2 {\hat{m}}_3, {\hat{n}}_1 {\hat{n}}_2 {\hat{n}}_3}$ given above are indeed the inverses of ${\hat{g}}_{{\hat{m}}_1 {\hat{m}}_2, {\hat{n}}_1 {\hat{n}}_2}$ and ${\hat{g}}_{{\hat{m}}_1 {\hat{m}}_2 {\hat{m}}_3, {\hat{n}}_1 {\hat{n}}_2 {\hat{n}}_3}$ respectively if we account for the fact that the contraction of the decomposed indices requires the contraction conventions
\begingroup
\renewcommand*{\arraystretch}{1.5}
\begin{align}
\begin{array}[t]{l}
{\hat{g}}_{{\hat{m}}_1 {\hat{m}}_2, {\hat{p}}_1 {\hat{p}}_2} {\hat{g}}^{{\hat{p}}_1 {\hat{p}}_2, {\hat{n}}_1 {\hat{n}}_2}\\
\qquad = {\hat{g}}_{{\hat{m}}_1 {\hat{m}}_2, p_1 p_2} {\hat{g}}^{p_1 p_2, {\hat{n}}_1 {\hat{n}}_2}  + {\hat{g}}_{{\hat{m}}_1 {\hat{m}}_2, p_1 z} {\hat{g}}^{p_1 z, {\hat{n}}_1 {\hat{n}}_2} + {\hat{g}}_{{\hat{m}}_1 {\hat{m}}_2, z p_2} {\hat{g}}^{z p_2, {\hat{n}}_1 {\hat{n}}_2}\\ 
\qquad = {\hat{g}}_{{\hat{m}}_1 {\hat{m}}_2, p_1 p_2} {\hat{g}}^{p_1 p_2, {\hat{n}}_1 {\hat{n}}_2} + 2 {\hat{g}}_{{\hat{m}}_1 {\hat{m}}_2, p_1 z} {\hat{g}}^{p_1 z, {\hat{n}}_1 {\hat{n}}_2}
\end{array}\\
\begin{array}[t]{l}
{\hat{g}}_{{\hat{m}}_1 {\hat{m}}_2 {\hat{m}}_3, {\hat{p}}_1 {\hat{p}}_2 {\hat{p}}_3} {\hat{g}}^{{\hat{p}}_1 {\hat{p}}_2 {\hat{p}}_3, {\hat{n}}_1 {\hat{n}}_2 {\hat{n}}_3}\\
\qquad = {\hat{g}}_{{\hat{m}}_1 {\hat{m}}_2 {\hat{m}}_3, p_1 p_2 p_3} {\hat{g}}^{p_1 p_2 p_3, {\hat{n}}_1 {\hat{n}}_2 {\hat{n}}_3} + {\hat{g}}_{{\hat{m}}_1 {\hat{m}}_2 {\hat{m}}_3, p_1 p_2 z} {\hat{g}}^{p_1 p_2 z, {\hat{n}}_1 {\hat{n}}_2 {\hat{n}}_3}\\
\qquad \qquad + {\hat{g}}_{{\hat{m}}_1 {\hat{m}}_2 {\hat{m}}_3, p_1 z p_3} {\hat{g}}^{p_1 z p_3, {\hat{n}}_1 {\hat{n}}_2 {\hat{n}}_3} + {\hat{g}}_{{\hat{m}}_1 {\hat{m}}_2 {\hat{m}}_3, z p_2 p_3} {\hat{g}}^{z p_2 p_3, {\hat{n}}_1 {\hat{n}}_2 {\hat{n}}_3}\\
\qquad = {\hat{g}}_{{\hat{m}}_1 {\hat{m}}_2 {\hat{m}}_3, p_1 p_2 p_3} {\hat{g}}^{p_1 p_2 p_3, {\hat{n}}_1 {\hat{n}}_2 {\hat{n}}_3} + 3 {\hat{g}}_{{\hat{m}}_1 {\hat{m}}_2 {\hat{m}}_3, p_1 p_2 z} {\hat{g}}^{p_1 p_2 z, {\hat{n}}_1 {\hat{n}}_2 {\hat{n}}_3}\,.
\end{array}
\end{align}
\endgroup
Then, under these conventions, one may verify that
\begin{align}
{\hat{g}}_{{\hat{m}}_1 {\hat{m}}_2, {\hat{p}}_1 {\hat{p}}_2} {\hat{g}}^{{\hat{p}}_1 {\hat{p}}_2, {\hat{n}}_1 {\hat{n}}_2} & = \delta^{{\hat{n}}_1 {\hat{n}}_2}_{{\hat{m}}_1 {\hat{m}}_2} = \begin{pmatrix}
\delta^{n_1 n_2}_{m_1 m_2} & 0\\ 0 & \frac{1}{2} \delta^{n_1}_{m_1}
\end{pmatrix}\,,\\
{\hat{g}}_{{\hat{m}}_1 {\hat{m}}_2 {\hat{m}}_3, {\hat{p}}_1 {\hat{p}}_2 {\hat{p}}_3} {\hat{g}}^{{\hat{p}}_1 {\hat{p}}_2 {\hat{p}}_3, {\hat{n}}_1 {\hat{n}}_2 {\hat{n}}_3} & = \delta^{{\hat{n}}_1 {\hat{n}}_2 {\hat{n}}_3}_{{\hat{m}}_1 {\hat{m}}_2 {\hat{m}_3}} = \begin{pmatrix}
\delta^{n_1 n_2 n_3}_{m_1 m_2 m_3} & 0\\
0 & \frac{1}{3} \delta^{n_1 n_2}_{m_1 m_2}
\end{pmatrix}\,.
\end{align}
However, the adjoint block $\operatorname{diag}[ {\hat{g}}^{\hat{m}_1 \hat{n}_1} {\hat{g}}_{\hat{m}_2 \hat{n}_2} - \frac{1}{8} \delta^{\hat{m}_1}_{\hat{m}_2} \delta^{\hat{n}_1}_{\hat{n}_2}, 1]$ becomes troublesome as it contributes more off-diagonal terms than the $\operatorname{SL}(5)$ case and it is not clear what the appropriate ansatz should be in this case. As the generalised metric becomes more cumbersome it may be more efficient to consider the reduction of the generators or generalised coordinates for a qualitative picture of the reduction instead. Returning to the $E_{8(8)}$ coordinate representation $R_1 = \mathbf{248}$, we decompose it under $\operatorname{SL}(9)$ to give 
\begin{align}\label{eq:SL9Decomp}
\mathbf{248} \rightarrow \mathbf{80} \oplus \mathbf{84} \oplus \overbar{\mathbf{84}}\,.
\end{align}
In terms of generators, we have
\begin{align}
\{ {\hat{T}}^{\hat{M}} \} & \xrightarrow{\operatorname{SL}(9)} \{ E^{\hat{m}}{}_{\hat{n}}, Z^{\hat{m}_1 \hat{m_2} \hat{m}_3}, Z_{\hat{m}_1 \hat{m}_2 \hat{m}_3} \}\,,
\end{align}
where $\hat{m}, \hat{n} = 1, \ldots, 9$. These satisfy the algebra\cite{Rosabal:2014rga}
\begin{subequations}
\begin{align}
[E^{\hat{m}_1}{}_{\hat{m}_2}, E^{\hat{n}_1}{}_{\hat{n}_2}] & = \delta^{\hat{n}_1}_{\hat{m}_2} E^{\hat{m}_1}{}_{\hat{n}_2} - \delta^{\hat{m}_1}_{\hat{n}_2} E^{\hat{n}_1}{}_{\hat{m}_2}\,,\\
[E^{\hat{m}_{1}}{}_{\hat{m}_2}, Z^{\hat{n}_1 \hat{n}_2 \hat{n}_3}] & = + \left( 3 \delta_{\hat{m}_2}^{[\hat{n}_1} Z^{\hat{n}_2 \hat{n}_3] \hat{m}_1} - \frac{1}{3} \delta^{\hat{m}_1}_{\hat{m}_2} Z^{\hat{n}_1 \hat{n}_2 \hat{n}_3} \right) \,,\\
[E^{\hat{m}_{1}}{}_{\hat{m}_2}, Z_{\hat{n}_1 \hat{n}_2 \hat{n}_3}] & = - \left( 3 \delta_{[\hat{n}_1}^{\hat{m}_1} Z_{\hat{n}_2 \hat{n}_3] \hat{m}_2} - \frac{1}{3} \delta^{\hat{m}_1}_{\hat{m}_2} Z_{\hat{n}_1 \hat{n}_2 \hat{n}_3} \right)\,,\\
[Z^{\hat{m}_1 \hat{m}_2 \hat{m}_3}, Z^{\hat{n}_1 \hat{n}_2 \hat{n}_3}] & = - \frac{1}{3!} \epsilon^{\hat{m}_1 \hat{m}_2 \hat{m}_3 \hat{n}_1 \hat{n}_2 \hat{n}_3 \hat{p}_1 \hat{p}_2 \hat{p}_3} Z_{\hat{p}_1 \hat{p}_2 \hat{p}_3}\,,\\
[Z^{\hat{m}_1 \hat{m}_2 \hat{m}_3}, Z_{\hat{n}_1 \hat{n}_2 \hat{n}_3}] & = 18 \delta^{[\hat{m}_1 \hat{m}_2}_{[\hat{n}_1 \hat{n}_2} E^{\hat{m}_3]}{}_{\hat{n}_3]}\,,\\
[Z_{\hat{m}_1 \hat{m}_2 \hat{m}_3}, Z_{\hat{n}_1 \hat{n}_2 \hat{n}_3}] & = + \frac{1}{3!} \epsilon_{\hat{m}_1 \hat{m}_2 \hat{m}_3 \hat{n}_1 \hat{n}_2 \hat{n}_3 \hat{p}_1 \hat{p}_2 \hat{p}_3} Z^{\hat{p}_1 \hat{p}_2 \hat{p}_3}\,.
\end{align}
\end{subequations}
Under $\operatorname{GL}(8)$, each of the representations in \eqref{eq:SL9Decomp} decompose as
\begin{subequations}
\begin{align}
\mathbf{80} & \xrightarrow{\operatorname{GL}(8)} \mathbf{63}_0 \oplus \mathbf{8}_{+9} \oplus \overbar{\mathbf{8}}_{-9} \oplus \mathbf{1}_0\,,\\
\mathbf{84} & \xrightarrow{\operatorname{GL}(8)} \mathbf{56}_{+3} \oplus \mathbf{28}_{-6}\,,\\
\overbar{\mathbf{84}} & \xrightarrow{\operatorname{GL}(8)} \overbar{\mathbf{56}}_{-3} \oplus \overbar{\mathbf{28}}_{+6}\,,
\end{align}
\end{subequations}
whilst the generators break down according to
\begin{subequations}
\begin{align}
\{ E^{\hat{m}}{}_{\hat{n}} \} & \xrightarrow{\operatorname{GL}(8)} \{ E^m{}_n, E^m{}_9, E^9{}_m, E^9{}_9 \}\,,\\
\{ Z^{\hat{m}_1 \hat{m}_2 \hat{m}_3} \} & \xrightarrow{\operatorname{GL}(8)} \{ Z^{m_1 m_2 m_3}, Z^{m_1 m_2 9} \}\,,\\
\{ Z_{\hat{m}_1 \hat{m}_2 \hat{m}_3} \} & \xrightarrow{\operatorname{GL}(8)} \{Z_{m_1 m_2 m_3}, Z_{m_1 m_2 9} \}\,.
\end{align}
\end{subequations}
Associating the index structures above to each representation , we see that the $E_{8(8)}$ coordinates in this notation are
\begin{align}\label{eq:E8SL9Coords}
{\hat{Y}}^{\hat{M}} = (Y^m{}_9, Y_{m_1 m_2 9}, Y^{m_1 m_2 m_3}, Y^m{}_n, Y^9{}_9, Y_{m_1 m_2 m_3}, Y^{m_1 m_2 9}, Y^9{}_m)\,,
\end{align}
which is the familiar decomposition of $E_{8(8)}$ where each set of coordinates correspond to the usual coordinates and the wrappings modes\footnote{Actually, there is an additional subtlety; the $\mathbf{63} \oplus \mathbf{1}$ contains an additional 8 coordinates over the KK6 wrapping modes which are thought to corresponds to the wrapping modes of non-supersymmetric branes. The string theory interpretation of this is given in \cite{deBoer:2012ma} whilst the $E_{11}$ picture was given in \cite{Kleinschmidt:2011vu}.} of the M2, M5, KK6, $5^3$, $2^6$ and $0^{(1,7)}$ branes. 
We shall determine which components of the $E_{8(8)}$ generalised metric enter into the $E_{7(7)}$ generalised metric by determining how these coordinates fall into into $E_{7(7)}$ representations. In order to do so, we consider the decomposition of the $\operatorname{SL}(9)$ generators under another maximal subgroup $\operatorname{GL}(7)\times \operatorname{SL}(2)$ as a stepping stone to reconstructing full $E_{7(7)}$ representations. The relevant decompositions are
\begin{subequations}\label{eq:SL9SL7}
\begin{align}
\mathbf{80} & \xrightarrow{\operatorname{GL}(7) \times \operatorname{SL}(2)} {\mathbf{(\overbar{7},2)}}_{-9} \oplus {\mathbf{(1,3)}}_{0} \oplus {\mathbf{(48,1)}}_0 \oplus {\mathbf{(1,1)}}_0 \oplus {\mathbf{(7,2)}}_{+9}\,,\\
\mathbf{84} & \xrightarrow{\operatorname{GL}(7) \times \operatorname{SL}(2)} {\mathbf{(7,1)}}_{-12} \oplus {\mathbf{(21,2)}}_{-3} \oplus {\mathbf{(35,1)}}_{+6}\,,\\
\mathbf{\overbar{84}} & \xrightarrow{\operatorname{GL}(7) \times \operatorname{SL}(2)} {\mathbf{(\overbar{35},1)}}_{-6} \oplus {\mathbf{(\overbar{21},2)}}_{+3} \oplus {\mathbf{(\overbar{7},1)}}_{+12}\,,
\end{align}
\end{subequations}
whilst the generators break according to ($\check{m}, \check{n} = 1, \ldots, 7$)
\begin{subequations}
\begin{align}
\{E^{\hat{m}}{}_{\hat{n}} \}& \xrightarrow{\operatorname{GL}(7)\times \operatorname{SL}(2)} \{E^{\check{m}}{}_{\check{n}}, E^{\check{m}}{}_8, E^8{}_{\check{n}}, E^8{}_8, E^{\check{m}}{}_9, E^8{}_9, E^9{}_{\check{n}}, E^9{}_8, E^9{}_9\}\\
\{ Z^{\hat{m}_1 \hat{m}_2 \hat{m}_3} \} & \xrightarrow{\operatorname{GL}(7) \times \operatorname{SL}(2)} \{ Z^{\check{m}_1 \check{m}_2 \check{m}_3}, Z^{\check{m}_1 \check{m}_2 8}, Z^{\check{m}_1 \check{m}_2 9}, Z^{\check{m}_1 89} \}\\
\{ Z_{\hat{m}_1 \hat{m}_2 \hat{m}_3} \} & \xrightarrow{\operatorname{GL}(7) \times \operatorname{SL}(2)} \{ Z_{\check{m}_1 \check{m}_2 \check{m}_3}, Z_{\check{m}_1 \check{m}_2 8}, Z_{\check{m}_1 \check{m}_2 9}, Z_{\check{m}_1 89} \}
\end{align}
\end{subequations}
Being explicit, the exact identification of the $\operatorname{GL}(7)\times \operatorname{SL}(2)$ generators with the representations in \eqref{eq:SL9SL7} are
\begin{subequations}\label{eq:SL7Gen}
\begin{align}
{\mathbf{(\overbar{7},2)}}_{-9}: & \{E^8{}_{\check{n}}, E^9{}_{\check{n}}\}\\
{\mathbf{(48,1)}}_0 \oplus {\mathbf{(1,1)}}_0 \oplus {\mathbf{(1,3)}}_0: & \{E^{\check{m}}{}_{\check{n}}, E^8{}_8, E^8{}_9, E^9{}_8, E^9{}_9\}\\
{\mathbf{(7,2)}}_{+9}: & \{ E^{\check{m}}{}_8, E^{\check{m}}{}_9\}\\
{\mathbf{(7,1)}}_{-12}: & \{ Z^{\check{m}_1 89}\}\\
{\mathbf{(21,2)}}_{-3}: & \{ Z^{\check{m}_1 \check{m}_2 8}, Z^{\check{m}_1 \check{m}_2 9}\}\\
{\mathbf{(35,1)}}_{+6}: & \{ Z^{\check{m}_1 \check{m}_2 \check{m}_3}\}\\
{\mathbf{(\overbar{35},1)}}_{-6}: & \{ Z_{\check{m}_1 \check{m}_2 \check{m}_3}\}\\
{\mathbf{(\overbar{21},2)}}_{+3}: & \{ Z_{\check{m}_1 \check{m}_2 8}, Z_{\check{m}_1 \check{m}_2 9}\}\\
{\mathbf{(\overbar{7},1)}}_{+12}: & \{ Z_{\check{m}_1 89}\}\,.
\end{align}
\end{subequations}
Note that the $\operatorname{SL}(2)$ factor acts on the $T^2$, spanned by the directions $y^8$ and $y^9$, by exchanging $8 \leftrightarrow 9$ as expected.
Comparing to the decomposition of $\mathbf{248}$ under $E_{7(7)}\times \operatorname{SL}(2)$
\begin{align}
\mathbf{248} = & \mathbf{(133,1)}\oplus \mathbf{(56,2)} \oplus \mathbf{(1,3)},
\end{align}
we may reconstruct full $E_{7(7)}$ representations from those appearing in \eqref{eq:SL9SL7} by the $\operatorname{SL}(2)$ representations that appear here:
\begin{subequations}
\begin{align}
{(\mathbf{133,1})} & \xrightarrow{\operatorname{GL}(7) \times \operatorname{SL}(2)} {(\mathbf{7,1})}_{-12} \oplus {(\overbar{\mathbf{35}}, \mathbf{1})}_{-6} \oplus \mathbf{(48,1)}_0 \oplus \mathbf{(1,1)}_0 \oplus \mathbf{(35,1)}_{+6} \oplus {(\overbar{\mathbf{7}}, \mathbf{1})}_{+12}\,,\\
\mathbf{(1,3)} & \xrightarrow{\operatorname{GL}(7) \times \operatorname{SL}(2)} \mathbf{(1,3)}_0\,,\\
{(\mathbf{56,2})} & \xrightarrow{\operatorname{GL}(7) \times \operatorname{SL}(2)} {(\overbar{\mathbf{7}},\mathbf{2})}_{-9} \oplus {(\mathbf{21,2})}_{-3} \oplus {(\overbar{\mathbf{21}}, \mathbf{2})}_{+3} \oplus {(\mathbf{7,2})}_{+9}\,.
\end{align}
\end{subequations}
Note, in particular, the doublet of $\mathbf{56}$ representations; $E_{8(8)}$ is large enough to contain two copies of the fundamental representation of $E_{7(7)}$. If we denote the generators of $E_{7(7)}$, decomposed under $\operatorname{GL}(7) \times \operatorname{SL}(2)$, as
\begin{align}
\{{\hat{T}}^{\hat{M}}\} \xrightarrow{E_{7(7)} \times \operatorname{SL}(2)} \{t^{\alpha}, t^\sharp, t^\natural, t^\flat, t^M, t^{\overbar{M}}\},
\end{align}
we are now ready to identify how the generators of $E_{8(8)}$ descend to $E_{7(7)} \times \operatorname{SL}(2)$. The only non-trivial identification is for $(\mathbf{1,1}) \oplus (\mathbf{1,3})$. Noting that the Cartan generators of $E_{8(8)}$ are $\{E^1{}_2, \ldots E^7{}_8, R^{678} \}$, we decompose this under $E_{7(7)} \times \operatorname{SL}(2)$ by deleting the node corresponding to $E^1{}_2$ in the extended Dynkin diagram leaving the Cartan generators of $E_{7(7)}$ and $\operatorname{SL}(2)$ to be $\{E^2{}_3, \ldots, E^7{}_8, R^{678}\}$ and $\{E^8{}_9\}$ respectively. Here, the generator $E^8{}_9$ corresponds to the extra node in the extended Dynkin diagram or, equivalently, the final node of the gravity line under $E_{8(8)} \rightarrow \operatorname{SL}(9)$. Thus, the $\operatorname{SL}(2)$ triplet must then be formed from the generators $\{E^8{}_8, E^8{}_9, E^9{}_9\} \equiv \{ t^\sharp, t^\natural, t^\flat\}$ and the $(\mathbf{1,1})$ factor must be given by the remaining $\{E^9{}_8\}$ generator. Thus, we end up with the identification of the generators
\begin{subequations}
\begin{align}
t^{\alpha} : & \{Z^{\check{m}_1 89}, Z_{\check{m}_1 \check{m}_2 \check{m}_3}, E^{\check{m}}{}_{\check{n}}, E^9{}_8, Z^{\check{m}_1 \check{m}_2 \check{m}_3}, Z_{\check{m}_1 89}\}\,,\\
(t^\sharp, t^\natural, t^\flat): & \{ E^8{}_8, E^8{}_9, E^9{}_9\}\,,\\
(t^M, t^{\overbar{M}}) : & \{ E^8{}_{\check{n}}, E^9{}_{\check{n}}, Z^{\check{m}_1 \check{m}_2 8}, Z^{\check{m}_1 \check{m}_2 9}, Z_{\check{m}_1 \check{m}_2 8}, Z_{\check{m}_1 \check{m}_2 9}, E^{\check{m}}{}_8, E^{\check{m}}{}_9\}\,.
\end{align}
\end{subequations}
The ranges of the indices should hopefully be self-explanatory: $\alpha = 1,\ldots, 133$ indexes the adjoint representation of $E_{7(7)}$, $M$ and $\overbar{M}$ index distinct 56-dimensional representations and $(\sharp, \natural, \flat)$ denote an $\operatorname{SL}(2)$ triplet of $E_{7(7)}$ singlets. To each of these generators, we assign coordinates with the same index structure in the usual fashion e.g. $Y^{\check{m}_1}$ is associated to $Z^{\check{m}_1 89}$ etc.\par
The above data is now sufficient to reconstruct all of the $E_{7(7)}$ coordinates from the $E_{8(8)}$ coordinates. Since we can trace the origin of the $E_{7(7)}$ generators back to those of $E_{8(8)}$ e.g. $E^{\check{m}}{}_9$ (associated to $Y^{\check{m}}{}_9$) descends from $E^m{}_9$ (associated to $Y^m{}_9$, or the usual coordinates), we may disentangle the two sets of $E_{7(7)}$ generalised coordinates $Y^M$ and $Y^{\overbar{M}}$ by demanding that the geometric wrapping modes of $E_{7(7)}$ descend from the geometric wrapping modes of $E_{8(8)}$. This gives (note the mixing of 8 and 9 indices)\footnote{For the exotic branes, we may identify the branes by dualising in 8 dimensions and/or adding full sets of antisymmetric indices $[ \check{m}_1 \ldots \check{m}_7 8]$ (note that the index 9 may be dropped as it is just a relic of the decomposition we took):
\begin{subequations}
\begin{align}
Y^{\check{m}_1}{}_8 & \equiv Y_{\check{m}_2 \ldots \check{m}_7 8,8} \rightarrow 6^1 \text{=KK6}\\
Y_{\check{m}_1 \check{m}_2 8} & \equiv Y_{\check{m}_1 \ldots \check{m}_7 8, \check{m}_1 \check{m}_2 8} \rightarrow 5^3\\
Y^{\check{m}_1 \check{m}_2 9} & \equiv Y_{\check{m}_1 \ldots \check{m}_7 8, \check{m}_3 \ldots \check{m}_7 8} \rightarrow 2^6\\
Y^9{}_{\check{m}_1} & \equiv Y_{\check{m}_1 \ldots \check{m}_7 8 , \check{m}_1 \ldots \check{m}_7 8, \check{m}_1} \rightarrow 0^{(1,7)}.
\end{align}
\end{subequations}
The identifications of the geometric coordinates should hopefully be self-explanatory.}:
\begin{align}
Y^{M} &: \begin{cases}
Y^{\check{m}}{}_9 & \text{from usual coordinates}\\
Y_{\check{m}_1 \check{m}_29} & \text{from M2}\\
Y^{\check{m}_1 \check{m}_28} & \text{from M5}\\
Y^8{}_{\check{m}} & \text{from KK6}
\end{cases}\\
Y^{\overbar{M}} &: \begin{cases}
Y^{\check{m}}{}_8 & \text{from KK6}\\
Y_{\check{m}_1 \check{m}_28} & \text{from } 5^3\\
Y^{\check{m}_1 \check{m}_29} & \text{from } 2^6\\
Y^9{}_{\check{m}} & \text{from } 0^{(1,7)}
\end{cases}
\end{align}
The remaining $E_{7(7)}$ coordinates are identified with the following on the $E_{8(8)}$ side:
\begin{align}
Y^{\alpha} &: \begin{cases}
Y^{\check{m}_1 8 9} & \text{from } 2^6\\
Y_{\check{m}_1 \check{m}_2 \check{m}_3} & \text{from } 5^3\\
Y^{\check{m}_1}{}_{\check{m}_2} & \text{from KK6}\\
Y^9{}_8 & \text{from } 0^{(1,7)}\\
Y^{\check{m}_1 \check{m}_2 \check{m}_3} & \text{from M5}\\
Y_{\check{m}_1 89} & \text{from M2}
\end{cases}\\
(Y^\sharp, Y^\natural, Y^\flat) &: \begin{cases}
Y^8{}_8 & \text{from KK6}\\
Y^8{}_9 & \text{from usual coordinates}\\
Y^9{}_9 & \text{from KK6}
\end{cases}
\end{align}
Actually, from an earlier footnote on the $E_{8(8)}$ coordinates, $Y^{\check{m}}{}_8$ and $Y^8{}_8$ may need to be identified with the duals of the wrapping modes of non-supersymmetric branes. With this, we see that one of the copies of the $\mathbf{56}$ coordinates descends from the geometric sector of $E_{8(8)}$ whilst the other descends from the non-geometric sector of $E_{8(8)}$. From here, it is simple to reconstruct the decomposition of the simplified generalised metric \eqref{eq:E8GenMetric}, if one so wished. However, this alone will not allow us to reduce the full generalised metric (with non-vanishing internal potentials); the brute force method remains the most direct, though troublesome, method to reduce it.\par
We end with a remark on how such a full reduction may still house new ideas. In the $\operatorname{SL}(5)$ case we saw that different ways of identifying the components that enter into the generalised metric of the reduced theory gave rise to reductions to distinct theories. For each $E_{n(n)}$ EFT, one should be able to reduce to at least the $E_{n-1(n-1)}$ EFT as well as the $\operatorname{O}(n-1, n-1)$ DFT. However, as the size of the generalised metric (as well as the complexity of the reduction ansatz) increases, there is more freedom in how we may pick out the components that enter into the reduced generalised metric. It may then be possible that there exists more choices than the two we have highlighted that lead to reductions to theories that have not yet been studied in the literature, particularly if we do not restrict ourselves to circle reductions as we have done here.
\section{Reduction of the section condition}\label{sec:ReductionSection}
We now consider how the section condition for a given ExFT reduces. We shall consider the reduction of the $E_{8(8)}$ EFT section condition to the $E_{7(7)}$ section condition in detail by an explicit reduction of the $Y$-tensor. We shall be more schematic in the reduction of the $\operatorname{SL}(5)$ section condition but shall reduce it to both the $\operatorname{SL}(3) \times \operatorname{SL}(2)$ and $\operatorname{O}(3,3)$ section conditions.
\subsection{\texorpdfstring{$E_{8(8)}$}{E8(8)} EFT to \texorpdfstring{$E_{7(7)}$}{E8(8)} EFT}
In reducing between EFTs, it quickly becomes clear that we need a consistent set of conventions for both theories that will allow us to reduce one to the other. However, the conventions presented in \cite{Hohm:2014fxa} (whilst, of course, internally consistent) are found to be incompatible with those of \cite{Hohm:2013uia} and so we shall have to modify the conventions of both to conform to a consistent set of rules. In particular, we shall adopt the conventions of \cite{Marrani:2010de,deWit:2002vt} which give compatible reductions of the exceptional structure. We first set up some notation speaking first in generality and then restricting to the cases of interest later. Let the structure constants of an algebra $\mathfrak{g}$ be defined through the commutation relations of the generators in the representation $R$:
\begin{align}
{[t^\alpha, t^\beta]}_M{}^N = f^{\alpha \beta}{}_{\gamma} {(t^\gamma)}_M{}^N
\end{align}
where $\alpha, \beta, \gamma = 1, \ldots, \operatorname{dim} \mathfrak{g}$ are adjoint indices and $M, N = 1, \ldots, \operatorname{dim}R$ denote the indices of some representation  $R$ of $\mathfrak{g}$, which may or may not also be the adjoint representation. The Killing form, with the canonical scaling, is defined as
\begin{align}
{\tilde{\kappa}}^{\alpha \beta} \coloneqq \frac{1}{C_{\mathbf{adj.}}} f^{\alpha \gamma}{}_\delta f^{\beta \delta}{}_\gamma\,,
\end{align}
where $C_{\mathbf{adj.}}$ is the quadratic Casimir in the adjoint representation. The quadratic Casimir of a representation $R$ is defined through the inverse Killing form and generators $t^\alpha$ in $R$ as
\begin{align}\label{eq:CasimirR}
C_{\mathbf{R}} \delta^N_M \coloneqq {\tilde{\kappa}}_{\alpha \beta} {(t^\alpha)}_M{}^P {(t^\beta)}_P{}^N\,.
\end{align} 
However, rather than work with the Killing form itself, we shall define the rescaled bilinear invariant
\begin{align}
\kappa^{\alpha \beta} \coloneqq \operatorname{Tr} (t^\alpha t^\beta) = {(t^\alpha)}_M{}^N {(t^\beta)}_N{}^M\,.
\end{align}
Taking the trace of \eqref{eq:CasimirR}, we obtain
\begin{align}
C_{R} \cdot \operatorname{dim} R = {\tilde{\kappa}}_{\alpha \beta} {(t^\alpha)}_M{}^N {(t^\beta)}_N{}^M = {\tilde{\kappa}}_{\alpha \beta} \kappa^{\alpha \beta}
\end{align}
and so we see that the rescaled and canonical Killing forms are related by
\begin{align}
\kappa^{\alpha \beta} = \frac{C_{R} \cdot \operatorname{dim} R}{\operatorname{dim} \mathfrak{g}} {\tilde{\kappa}}^{\alpha \beta}\,.
\end{align}
It is then easy to check that the inverse rescaled Killing form satisfies
\begin{align}\label{eq:Norm}
\kappa_{\alpha \beta} {(t^\alpha)}_M{}^P {(t^\beta)}_P{}^N = \frac{\operatorname{dim}\mathfrak{g}}{\operatorname{dim}R} \delta^N_M\,.
\end{align}
We thus end up with
\begin{align}
f_{\alpha \gamma \delta} f_{\beta}{}^{\gamma \delta} & = - \frac{\operatorname{dim}\mathfrak{g}}{\operatorname{dim} R} \frac{C_{\mathbf{adj.}}}{C_{R}} \kappa_{\alpha \beta}
\end{align}
and we shall use this rescaled Killing-form (henceforth referred to as just `the Killing form') to raise and lower adjoint indices. Finally, we introduce the Dynkin index of a representation $R$ as
\begin{align}
I_{R} \coloneqq \frac{\operatorname{dim} R}{\operatorname{dim} \mathfrak{g}} C_{R}\,.
\end{align}
In the case that $R$ is the adjoint representation, the Dynkin index of the adjoint representation $I_{\mathbf{adj.}}$ coincides with the dual Coxeter number $g^\vee$ and so we obtain
\begin{align}
\kappa_{\alpha \beta} = - \frac{I_{R}}{g^\vee} f_{\alpha \gamma \delta} f_{\beta}{}^{\gamma \delta}\,.
\end{align}
As a consequence of this normalisation, we end up with the orthogonality of the structure constants
\begin{align}\label{eq:NormalisationGenerators}
f^{\alpha \gamma \delta} f_{\beta \gamma \delta} = - \frac{g^\vee}{I_{R}} \delta^\alpha_\beta\,.
\end{align}
For our purposes, we are interested in $\mathbf{adj.} = R = \mathbf{248}$ for $E_{8(8)}$ and $\mathbf{adj.} = \mathbf{133}$, $R = \mathbf{56}$ for $E_{7(7)}$ which have
\begin{align}
\begin{array}{ll}
g^\vee (E_{8(8)}) = 30, & I_{\mathbf{248}} = 30\\
g^\vee (E_{7(7)}) = 18, & I_{\mathbf{56}} = 6\,.\\
\end{array}
\end{align}
For this section, we shall use $\hat{M}, \hat{N} = 1, \ldots, 248$ and $\alpha, \beta = 1, \ldots, 133$ to index the adjoint representations of $E_{8(8)}$ and $E_{7(7)}$ respectively and $M, N = 1, \ldots, 56$ to index the coordinate representation of $E_{7(7)}$. We define the following Killing forms for $E_{8(8)}$ and $E_{7(7)}$ respectively (adorning hats on $E_{8(8)}$ objects).\par
\begin{align}
{\hat{\kappa}}_{\hat{M} \hat{N}} & = f_{\hat{M} \hat{P}}{}^{\hat{Q}} f_{\hat{N} \hat{Q}}{}^{\hat{P}} \,, \qquad \kappa_{\alpha \beta} = \frac{1}{3} f_{\alpha \gamma}{}^\delta f_{\beta \delta}{}^\gamma\,.
\end{align}
Before we continue, we make a note on differing conventions in the literature. In the original $E_{8(8)}$ EFT paper \cite{Hohm:2014fxa}, the authors define
\begin{align}
{\left( {\hat{\mathbb{P}}}_{\mathbf{248}} \right)}^{\hat{M}}{}_{\hat{N}}{}^{\hat{K}}{}_{\hat{L}} = +\frac{1}{60} {\hat{f}}^{\hat{M}}{}_{\hat{N} \hat{P}} {\hat{f}}^{\hat{P} \hat{K}}{}_{\hat{L}}
\end{align}
with normalisation ${\hat{f}}^{\hat{M} \hat{K} \hat{L}} {\hat{f}}_{\hat{N} \hat{K} \hat{L}} = - 60 \delta^{\hat{M}}_{\hat{N}}$. By contrast, our normalisation is dictated by \eqref{eq:NormalisationGenerators} as
\begin{align}
{\hat{f}}^{\hat{M} \hat{P} \hat{Q}} {\hat{f}}_{\hat{N} \hat{P} \hat{Q}} = - \delta^{\hat{M}}_{\hat{N}}\,,
\end{align}
which is the same convention as that used in \cite{Marrani:2010de}. Since both $E_{8(8)}$ and $E_{7(7)}$ possess invariants with which to identify $R_1$ and ${\overbar{R}}_1$, we shall define our projectors to act on $R_1 \otimes R_1$ with the following conventions:
\begin{align}
{{\mathbb{P}}}_{MN}{}^{KL} {{\mathbb{P}}}_{KL}{}^{PQ} & = { {\mathbb{P}} }_{MN}{}^{PQ}\,, \qquad {{\mathbb{P}}}_{MN}{}^{MN} = \operatorname{rnk} \mathbb{P}\,.
\end{align}
Then, we require 
\begin{align}
{\left({\hat{\mathbb{P}}}_{\mathbf{248}} \right)}_{\hat{K} \hat{L}}{}^{\hat{M} \hat{N}} & = - {\hat{f}}_{\hat{K} \hat{L} \hat{P}} {\hat{f}}^{\hat{P} \hat{M} \hat{N}}\,.
\end{align}
Since the normalisation of the generators have the same sign but the projector differs from \cite{Hohm:2014fxa} by a sign (when we identify their ${\left( {\hat{\mathbb{P}}}_{\mathbf{248}} \right)}^{\hat{M}}{}_{\hat{K}}{}^{\hat{N}}{}_{\hat{L}}$ with our ${\hat{\kappa}}^{\hat{M} \hat{S}} {\left({\hat{\mathbb{P}}}_{\mathbf{248}} \right)}_{\hat{S} \hat{K}}{}^{\hat{N} \hat{T}} {\hat{\kappa}}_{\hat{T} \hat{L}}$) we must compensate by introducing a minus sign for every instance of ${\hat{\mathbb{P}}}_{\mathbf{248}}$ such as in the generalised Lie derivative \eqref{eq:OurGenLie}. Note, however, that we do not need to introduce a minus sign for ${\hat{\mathbb{P}}}_{\mathbf{3875}}$. We also note the conventions of \cite{Cederwall:2015ica} which is closer to ours and differs only by the scaling of the generators.\par
In our conventions, the projectors onto various irreducible representations of $E_{8(8)}$ within $\mathbf{248} \otimes \mathbf{248}$ are given by
\begin{align}
{\left( {\hat{\mathbb{P}}}_{\mathbf{1}} \right)}_{\hat{K} \hat{L}}{}^{\hat{M} \hat{N}}  & = \frac{1}{248} {\hat{\kappa}}_{\hat{K} \hat{L}} {\hat{\kappa}}^{\hat{M} \hat{N}}\,\\
{\left({\hat{\mathbb{P}}}_{\mathbf{248}} \right)}_{\hat{K} \hat{L}}{}^{\hat{M} \hat{N}} & = - {\hat{f}}_{\hat{K} \hat{L} \hat{P}} {\hat{f}}^{\hat{P} \hat{M} \hat{N}}\,\\
{\left( {\hat{\mathbb{P}}}_{\mathbf{3875}} \right)}_{\hat{K} \hat{L}}{}^{\hat{M} \hat{N}} & = \frac{1}{7} \delta^{(\hat{M}}_{\hat{K}} \delta^{\hat{N})}_{\hat{L}} - \frac{1}{56} {\hat{\kappa}}_{\hat{K} \hat{L}} {\hat{\kappa}}^{\hat{M} \hat{N}} - \frac{30}{7} {\hat{f}}^{\hat{P} (\hat{M}}{}_{\hat{K}} {\hat{f}}_{\hat{P}}{}^{\hat{N})}{}_{\hat{L}}\label{eq:3875}\,\\
{\left( {\hat{\mathbb{P}}}_{\mathbf{27000}} \right)}_{\hat{K} \hat{L}}{}^{\hat{M} \hat{N}} & = \frac{6}{7} \delta^{(\hat{M}}_{\hat{K}} \delta^{\hat{N})}_{\hat{L}} + \frac{3}{217} {\hat{\kappa}}_{\hat{K} \hat{L}} {\hat{\kappa}}^{\hat{M} \hat{N}} + \frac{30}{7} {\hat{f}}^{\hat{P} (\hat{M}}{}_{\hat{K}} {\hat{f}}_{\hat{P}}{}^{\hat{N})}{}_{\hat{L}}\,\\
{\left({\hat{\mathbb{P}}}_{\mathbf{30380}} \right)}_{\hat{K} \hat{L}}{}^{\hat{M} \hat{N}} & = \delta^{[\hat{M}}_{\hat{K}} \delta^{\hat{N}]}_{\hat{L}} + {\hat{f}}_{\hat{K} \hat{L} \hat{P}} {\hat{f}}^{\hat{P} \hat{M} \hat{N}}\,.
\end{align}
Simple computations will verify that each of these square to themselves and project onto spaces of the correct dimensions. However, showing that ${\hat{\mathbb{P}}}_{\mathbf{3875}}$ and ${\hat{\mathbb{P}}}_{\mathbf{27000}}$ square to themselves will require the use of the identity
\begin{align}
\begin{aligned}
{\hat{f}}_{\hat{K} \hat{R} \hat{M}} {\hat{f}}_{\hat{L}}{}^{\hat{R} \hat{N}} {\hat{f}}_{\hat{P}}{}^{\hat{M} \hat{S}} {\hat{f}}_{\hat{Q} \hat{N} \hat{S}} & = \frac{1}{300} ( {\hat{\kappa}}_{\hat{K} \hat{Q}} {\hat{\kappa}}_{\hat{L} \hat{P}} + 2 {\hat{\kappa}}_{\hat{K}(\hat{L}} {\hat{\kappa}}_{\hat{P)} \hat{Q}})\\
& \qquad  - \frac{1}{6} ( 2 {\hat{f}}^{\hat{M}}{}_{\hat{K} \hat{P}} {\hat{f}}_{\hat{M} \hat{L} \hat{Q}} - {\hat{f}}^{\hat{M}}{}_{\hat{K} \hat{Q}} {\hat{f}}_{\hat{M} \hat{L} \hat{P}} )\,.\\
\end{aligned}
\end{align}
Noting that
\begin{align}
\operatorname{Sym} \left( \mathbf{248} \otimes \mathbf{248} \right) & = \mathbf{1} \oplus \mathbf{3875} \oplus \mathbf{27000}\,,\\
\operatorname{Asym} \left( \mathbf{248} \otimes \mathbf{248} \right) & = \mathbf{248} \oplus \mathbf{30380}\,,
\end{align}
we see that the projectors further obey the completeness relations
\begin{align}
{\left( {\hat{\mathbb{P}}}_{\mathbf{1}} \right)}_{\hat{K} \hat{L}}{}^{\hat{M} \hat{N}} + {\left( {\hat{\mathbb{P}}}_{\mathbf{3875}} \right)}_{\hat{K} \hat{L}}{}^{\hat{M} \hat{N}} + {\left( {\hat{\mathbb{P}}}_{\mathbf{27000}} \right)}_{\hat{K} \hat{L}}{}^{\hat{M} \hat{N}} & = \delta^{(\hat{M}}_{\hat{K}} \delta^{\hat{N})}_{\hat{L}}\,,\\
{\left( {\hat{\mathbb{P}}}_{\mathbf{248}} \right)}_{\hat{K} \hat{L}}{}^{\hat{M} \hat{N}} + {\left( {\hat{\mathbb{P}}}_{\mathbf{30380}} \right)}_{\hat{K} \hat{L}}{}^{\hat{M} \hat{N}} & = \delta^{[\hat{M}}_{\hat{K}} \delta^{\hat{N}]}_{\hat{L}}\,,
\end{align}
and that the sum of all these projectors gives the identity on $\mathbf{248} \otimes \mathbf{248}$, namely $\delta^{\hat{M}}_{\hat{K}} \delta^{\hat{N}}_{\hat{L}}$. Finally, the relation between ${\hat{\mathbb{P}}}_{\mathbf{248}}$ and ${\hat{\mathbb{P}}}_{\mathbf{3875}}$ is modified to
\begin{align}
{\left( {\hat{\mathbb{P}}}_{\mathbf{3875}} \right) }_{\hat{K} \hat{L}}{}^{\hat{M} \hat{N}} & = \frac{1}{7} \delta^{(\hat{M}}_{\hat{K}} \delta^{\hat{N})}_{\hat{L}} - \frac{1}{56} {\hat{\kappa}}_{\hat{K} \hat{L}} {\hat{\kappa}}^{\hat{M} \hat{N}} - \frac{30}{7} {\hat{f}}^{\hat{P} \hat{M}}{}_{\hat{K}} {\hat{f}}_{\hat{P}}{}^{\hat{N}}{}_{\hat{L}} - \frac{15}{7} {\left( {\hat{\mathbb{P}}}_{\mathbf{248}} \right)}_{\hat{K} \hat{L}}{}^{\hat{M} \hat{N}}\,,
\end{align}
as may be verified by using the Jacobi identity
\begin{align}\label{eq:Jacobi}
{\hat{f}}^{\hat{P} \hat{M} \hat{N}} {\hat{f}}_{\hat{P} \hat{K} \hat{L}} = 2 {\hat{f}}^{\hat{P} [ \hat{M}}{}_{\hat{K}} {\hat{f}}_{\hat{P}}{}^{\hat{N}]}{}_{\hat{L}}\,.
\end{align}
For $E_{7(7)}$, our normalisation agrees with \cite{Hohm:2013uia} but our choice of adjoint projector again differs by a sign:
\begin{align}
{\left( {\mathbb{P}}_{\mathbf{133}} \right)}_{KL}{}^{MN} & = - {(t_\alpha)}_{KL} {(t^\alpha)}^{MN}\,,
\end{align}
where ${\left( t^{\alpha} \right)}_{MN} = {\left( t^\alpha \right)}_M{}^P \Omega_{PN} = {\left(t^\alpha \right)}_{(MN)}$ are the generators of $E_{7(7)}$ and $\Omega_{MN}$ is the invariant symplectic form\footnote{Our conventions for contractions with the symplectic form are the standard in the literature:
\begin{align}
V^M = \Omega^{MN} V_N\,, \qquad V_M = V^N \Omega_{NM}\,,
\end{align}
with normalisation $\Omega^{MK} \Omega_{NK} = \delta^M_N$
}. Thus, as in the $E_{8(8)}$ case, we need to introduce a sign into every instance of ${\mathbb{P}}_{\mathbf{133}}$. Note, however, that the normalisation of the generators is still the same and follows from \eqref{eq:Norm}:
\begin{align}
{(t_\alpha)}_{MP} {(t^\alpha)}^{PN} = - {(t_\alpha)}_M{}^P {(t^\alpha)}_P{}^N =  - \frac{19}{8} \delta^N_M\,.
\end{align}
In our conventions, the projectors of $E_{7(7)}$ onto irreps in $\mathbf{56} \otimes \mathbf{56}$ are
\begin{align}
{\left( {\mathbb{P}}_{\mathbf{1}} \right)}_{KL}{}^{MN} & = \frac{1}{56} \Omega_{KL} \Omega^{MN}\\
{\left( {\mathbb{P}}_{\mathbf{133}} \right)}_{KL}{}^{MN} & = - {(t_\alpha)}_{KL} {(t^\alpha)}^{MN}\\
{\left( {\mathbb{P}}_{\mathbf{1463}} \right)}_{KL}{}^{MN} & = \delta^{(M}_K \delta^{N)}_L + {(t_\alpha)}_{KL} {(t^\alpha)}^{MN}\\
{\left( {\mathbb{P}}_{\mathbf{1539}} \right)}_{KL}{}^{MN} & = \delta^{[M}_K \delta^{N]}_L - \frac{1}{56} \Omega_{KL} \Omega^{MN}\,.
\end{align}
As before, these may be verified to square to themselves and project onto spaces of the correct dimension. Analogous to the $E_{8(8)}$ relations, we have
\begin{align}
\operatorname{Sym} \left( \mathbf{56} \otimes \mathbf{56} \right) & = \mathbf{133} \oplus \mathbf{1463}\,,\\
\operatorname{Asym} \left( \mathbf{56} \otimes \mathbf{56} \right) & = \mathbf{1} \oplus \mathbf{1539}\,,
\end{align}
which requires that the projectors satisfy the completeness relations
\begin{align}
{\left( {\mathbb{P}}_{\mathbf{133}} \right)}_{\hat{K} \hat{L}}{}^{\hat{M} \hat{N}} + {\left( {\mathbb{P}}_{\mathbf{1463}} \right)}_{\hat{K} \hat{L}}{}^{\hat{M} \hat{N}} & = \delta^{(M}_K \delta^{N)}_L\,,\\
{\left( {\mathbb{P}}_{\mathbf{1}} \right)}_{\hat{K} \hat{L}}{}^{\hat{M} \hat{N}} + {\left( {\mathbb{P}}_{\mathbf{1539}} \right)}_{\hat{K} \hat{L}}{}^{\hat{M} \hat{N}} & = \delta^{[M}_K \delta^{N]}_L\,,
\end{align}
with their sum giving the identity $\delta^M_K \delta^N_L$ on $\mathbf{56} \otimes \mathbf{56}$. Finally, since our normalisation of the generators is the same as \cite{Hohm:2013uia}, we still have the relation
\begin{align}\label{eq:E7Gen}
{(t_\alpha)}_M{}^K {(t^\alpha)}_N{}^L & = \frac{1}{24} \delta^K_M \delta^L_N + \frac{1}{12} \delta^K_N \delta^L_M + {(t_\alpha)}_{MN} {(t^\alpha)}^{KL} - \frac{1}{24} \Omega_{MN} \Omega^{KL} \,.
\end{align}
In the conventions that we employ, the generalised Lie derivative of the $E_{8(8)}$ and $E_{7(7)}$ EFT are
\begin{align}\label{eq:OurGenLie}
{\hat{\mathbb{L}}}_{(\hat{\Lambda}, \hat{\Sigma})} {\hat{V}}^{\hat{M}} & =
\begingroup \renewcommand{\arraystretch}{1.5} \begin{array}[t]{l} 
 {\hat{\Lambda}}^{\hat{N}} {\hat{\partial}}_{\hat{N}} {\hat{V}}^{\hat{M}} + 60 {\left({\mathbb{P}}_{\mathbf{248}} \right)}^{\hat{M}}{}_{\hat{K}}{}^{\hat{N}}{}_{\hat{L}} \left( {\hat{\partial}}_{\hat{N}} {\hat{\Lambda}}^{\hat{L}} + \frac{1}{60} {\hat{\Sigma}}_{\hat{R}} {\hat{f}}^{\hat{R} \hat{L}}{}_{\hat{N}} \right) {\hat{V}}^{\hat{K}}\\
\qquad + \hat{\lambda} {\hat{\partial}}_{\hat{N}} {\hat{\Lambda}}^{\hat{N}} {\hat{V}}^{\hat{M}}\,,
\end{array}
\endgroup\\
{\mathbb{L}}_{\Lambda} V^M & = \Lambda^N \partial_N V^M + 12 {\left( {\mathbb{P}}_{\mathbf{133}} \right)}^M{}_K{}^N{}_L \partial_N \Lambda^K V^L + \lambda \partial_N \Lambda^N V^M\,,
\end{align}
where we have further modified the definition of $\hat{\Sigma}$ to take into account the difference in normalisation of the structure constants in $E_{8(8)}$. Here, ${\hat{\Sigma}}_M$ is a parameter for an extra gauge transformation, not present in other ExFTs, that is required for the generalised Lie derivative to close appropriately. It is a constrained parameter (in the sense that it is treated in the same way as a derivative with respect to the section condition) and appears to be a common feature of 3-dimensional ExFTs, appearing in \cite{Hohm:2017wtr,Hohm:2014fxa,Hohm:2013jma}. In this form, we may read off the $Y$-tensor in these conventions as
\begin{align}
{\hat{Y}}^{\hat{M} \hat{N}}{}_{\hat{K} \hat{L}} & = 60 {\left({\hat{\mathbb{P}}}_{\mathbf{248}} \right)}^{\hat{M}}{}_{\hat{L}}{}^{\hat{N}}{}_{\hat{K}} + 2 \delta^{(\hat{M}}_{\hat{K}} \delta^{\hat{N})}_{\hat{L}}\,,\\
Y^{MN}{}_{KL} & = 12 {\left( {\mathbb{P}}_{\mathbf{133}} \right)}^M{}_L{}^N{}_K + \frac{1}{2} \delta^M_L \delta^N_K + \delta^M_K \delta^N_L\label{eq:E7Y}\\
	& = - 12 {(t_\alpha)}_{KL} {(t^\alpha)}^{MN} - \frac{1}{2} \Omega_{KL} \Omega^{MN}\,.
\end{align}
We choose to break the $E_{8(8)}$ generalised coordinates under $E_{7(7)} \times \operatorname{SL}(2)$ according to\footnote{We have condensed the notation from Section \eqref{sec:ReductionLargerGenMetric} for convenience. Here, $Y^{Ma} = (Y^{M1}, Y^{M2}) \equiv (Y^M, Y^{\overbar{M}})$ and $Y^i \equiv (Y^\sharp, Y^\natural, Y^\flat)$ of that section.}
\begin{align}\label{eq:E8E7Decomp}
{\hat{Y}}^{\hat{M}} = (Y^\alpha, Y^{Ma}, Y^i)\,.
\end{align} 
Then, with the normalisation that we employ, the $E_{8(8)}$ Killing form breaks under $E_{7(7)}$ according to
\begin{align}
{\hat{\kappa}}_{\hat{M} \hat{N}} & = \operatorname{diag} \left( \kappa_{\alpha \beta}, \Omega_{MN} \varepsilon_{ab}, g_{ij} \right)\\
{\hat{\kappa}}^{\hat{M} \hat{N}} & = \operatorname{diag} \left( \kappa^{\alpha \beta}, \Omega^{MN} \varepsilon^{ab}, g^{ij} \right)\label{eq:KillingDecomp}
\end{align}
where $g_{ij} = -\delta_{ij}$ is the $\operatorname{SL}(2)$ Killing form (taken to be negative-definite\footnote{Note that the Levi-Civita symbol consequently picks up an extra sign in its contractions:
\begin{align}
\varepsilon_{ikl} \varepsilon_j{}^{kl} = -2 g_{ij}
\end{align}}) that raises and lowers $\operatorname{SL}(2)$ adjoint indices and $\varepsilon_{ab}$ is the $\operatorname{SL}(2)$ invariant (i.e. $i,j = 1, 2,3$ and $a,b, = 1,2$). Here, $\kappa^{\alpha \beta}$ is the Killing form on $E_{7(7)}$ with the scaling defined above. The $E_{8(8)}$ structure constants consistent with this Killing form are
\begin{align}\label{eq:E8StructureConstants}
{\hat{f}}_{\hat{M} \hat{N} \hat{K}} & = \begin{cases}
{\hat{f}}_{\alpha \beta \gamma} & = \frac{1}{\sqrt{5}} f_{\alpha \beta \gamma}\,,\\
{\hat{f}}_{Ma Nb i} & = -\frac{1}{\sqrt{30}} \Omega_{MN} {\left( D_i \right)}_{ab}\,,\\
{\hat{f}}_{Ma Nb \alpha} & = -\frac{1}{\sqrt{5}} {(t_\alpha)}_{MN} \varepsilon_{ab}\,,\\
{\hat{f}}_{ijk} & = \frac{1}{\sqrt{30}} \varepsilon_{ijk}\,.
\end{cases}
\end{align}
Here, the $D_i$ are the representation matrices of the $\mathbf{3}$ of $\operatorname{SL}(2)$ which we take to be anti-Hermitian Pauli matrices
\begin{align}
{(D_i)}_a{}^b = \frac{i}{2} {(\sigma_i)}_a{}^b\,.
\end{align}
We may also identify the adjoint representation as the symmetric representation via
\begin{align}
{(D_i)}^{ab} {(D^i)}^{cd} = - \frac{1}{4} \left( \varepsilon^{ac} \varepsilon^{bd} + \varepsilon^{ad} \varepsilon^{bc} \right)\,.
\end{align}
We are now ready to reduce the generalised Lie derivative \eqref{eq:OurGenLie}. Our ansatz for the reduction is to break the $\operatorname{SL}(2)$ covariance by selecting ${\hat{Y}}^{M1} \equiv Y^M$ to be our $E_{7(7)}$ extended coordinates, and applying the following:
\begin{align}\label{eq:E8E7Ansatz}
{\hat{\partial}}_{M1} & = \partial_{M} \qquad {\hat{\partial}}_{M2} = {\hat{\partial}}_{\alpha} = {\hat{\partial}}_i = 0\,,\\
{\hat{V}}^{\hat{M}} & = ({\hat{V}}^{M1} = V^M , 0, 0, 0)\,,\\
{\hat{\Lambda}}^{\hat{M}} & = ({\hat{\Lambda}}^{M1} = \Lambda^M , 0, 0, 0)\,,\\
{\hat{\Sigma}}^{\hat{M}} & = ({\hat{\Sigma}}^{M1} = \Sigma^M , 0, 0, 0)
\end{align}
i.e. drop all coordinate dependence on the extra extended coordinate $\{ {\hat{Y}}^{M2}, {\hat{Y}}^\alpha, {\hat{Y}}^i \}$ and set all components of the fields to zero, apart from those in the $E_{7(7)}$ section. As such, we are only interested in the $M1$ components of all our objects. Consider the restriction of the $E_{8(8)}$ adjoint projector onto $(\mathbf{56},\mathbf{2})$ indices $Ma$:
\begin{align}\label{eq:248Restriction}
{\left( {\hat{\mathbb{P}}}_{\mathbf{248}} \right)}^{Ma}{}_{Kc}{}^{Nb}{}_{Ld} & = - \frac{1}{5} {(t_{\alpha})}_K{}^M {(t^\alpha)}_L{}^N \delta^a_c \delta^b_d - \frac{1}{120} \delta^M_K \delta^N_L ( \delta^a_d \delta^b_c - \varepsilon^{ab} \varepsilon_{cd})\\ 
& = \frac{1}{60} {\left( 12 {\left( {\mathbb{P}}_{\mathbf{133}}\right)}^M{}_K{}^N{}_L \delta^a_c \delta^b_d  - \frac{1}{2} \delta^M_K \delta^N_L \delta^a_d \delta^b_c \right)} + \frac{1}{120} \delta^M_K \delta^N_L \varepsilon^{ab} \varepsilon_{cd}\,.
\end{align}
Upon setting $a=b=c=d=1$ (i.e.\ restricting entirely to the $E_{7(7)}$ section), the last term drops out and we see that the reduction of the adjoint projector of $E_{8(8)}$ yields the adjoint projector in $E_{7(7)}$ as well as an extra $\delta \delta$ term:
\begin{align}\label{eq:ProjectorReduction}
{\left( {\hat{\mathbb{P}}}_{\mathbf{248}} \right)}^{M1}{}_{K1}{}^{N1}{}_{L1} & = \frac{1}{60} {\left( 12 {\left( {\mathbb{P}}_{\mathbf{133}}\right)}^M{}_K{}^N{}_L - \frac{1}{2} \delta^M_K \delta^N_L \right)}\,.
\end{align}
The generalised Lie derivative then reduces as
\begin{align}
{\hat{\mathbb{L}}}_{(\hat{\Sigma}, \hat{\Lambda})} {\hat{V}}^{\hat{M}} & \longrightarrow\begingroup \renewcommand{\arraystretch}{1.5} \begin{array}[t]{l}  \Lambda^{N} \partial_{N} V^M  + \hat{\lambda} \partial_N \Lambda^N V^M\\
+ \left( 12 {\left( {\mathbb{P}}_{\mathbf{133}} \right)}^M{}_K{}^N{}_L - \frac{1}{2} \delta^M_K \delta_L^N \right) \left( \partial_N \Lambda^L  + \frac{1}{60} \Sigma_R {\hat{f}}^{R L}{}_N \right) V^K \,,\\
\end{array}
\endgroup\\
& \longrightarrow \Lambda^{N} \partial_{N} V^M + 12 {\left( {\mathbb{P}}_{\mathbf{133}} \right)}^M{}_K{}^N{}_L \partial_N \Lambda^L + \left(\hat{\lambda} -\frac{1}{2}\right) \partial_N \Lambda^N V^M \,.
\end{align}
In particular, all terms involving the extra gauge parameter $\hat{\Sigma}$ automatically drop out under our ansatz. We thus recover the $E_{7(7)}$ generalised Lie derivative, where the weight under the $E_{7(7)}$ generalised Lie derivative is identified with a shift of the weight under the $E_{8(8)}$ generalised diffeomorphisms:
\begin{align}
\hat{\lambda} \rightarrow \hat{\lambda} - \frac{1}{2} \coloneqq \lambda\,.
\end{align}
There is a more natural interpretation of this if we employ the $Y$-tensor. The reduction of the $Y$-tensor induced by \eqref{eq:ProjectorReduction} is
\begin{align}\label{eq:E8ToE7YTensor}
{\hat{Y}}^{M1 N1}{}_{K1 L1} & = Y^{MN}{}_{KL}\,.
\end{align}
The effective weight ($\lambda_{\text{eff.}} = \lambda_V + \omega$) term reduces exactly and we are left with
\begin{align}
{\hat{\mathbb{L}}}_{(\hat{\Sigma}, \hat{\Lambda})} {\hat{V}}^{\hat{M}} & = \begingroup \renewcommand{\arraystretch}{1.5} \begin{array}[t]{l} {\hat{\Lambda}}^{\hat{N}} {\hat{\partial}}_{\hat{N}} {\hat{V}}^{\hat{M}}  - {\hat{V}}^{\hat{N}} {\hat{\partial}}_{\hat{N}} {\hat{\Lambda}}^{\hat{M}} + {\hat{Y}}^{\hat{M} \hat{N}}{}_{\hat{K} \hat{L}} {\hat{\partial}}_{\hat{N}} {\hat{\Lambda}}^{\hat{K}}  {\hat{V}}^{\hat{L}} + \lambda_{\text{eff.}} {\hat{\partial}}_{\hat{N}} {\hat{\Lambda}}^{\hat{N}} {\hat{V}}^{\hat{M}}\\ \qquad - {\hat{\Sigma}}_{\hat{K}} {\hat{f}}^{\hat{K} \hat{M}}{}_{\hat{N}} {\hat{V}}^{\hat{N}} \end{array}\endgroup\\
& \longrightarrow \Lambda^N \partial_N V^M - V^N \partial_N \Lambda^M + Y^{MN}{}_{KL} \partial_N \Lambda^K V^L + \lambda_{\text{eff.}} \partial_N \Lambda^N V^M\,,\\
& \coloneqq\Lambda^N \partial_N V^M - V^N \partial_N \Lambda^M + Y^{MN}{}_{KL} \partial_N \Lambda^K V^L + \lambda_{\text{eff.}} \partial_N \Lambda^N V^M\,.
\end{align}
In this picture, we thus obtain a transfer of weight from the universal weight to the weight of $V$ such that the effective weight in the two theories remains the same:
\begin{align}
\hat{\lambda} - 1= \lambda - \frac{1}{2} = \lambda_{\text{eff.}}
\end{align}
Recall that the generalised gauge field that forms the starting point of the rather intricate tensor hierarchy has an effective weight of 0 in both theories. The fact that this is not disturbed in this reduction is perhaps to be expected.\par
For completeness, we note the following restrictions to the $(\mathbf{56,2})$ piece of the $E_{8(8)}$ projectors:
\begin{align}\label{eq:ProjRestrictions}
{\left( {\hat{\mathbb{P}}}_{\mathbf{1}} \right)}_{Kc Ld}{}^{Ma Nb} & = \frac{1}{248} \Omega_{KL} \Omega^{MN} \varepsilon^{ab} \varepsilon_{cd}\\
{\left( {\hat{\mathbb{P}}}_{\mathbf{248}} \right)}_{Kc Ld}{}^{Ma Nb} & = - \frac{1}{5} {(t_\alpha)}_{KL} {(t^\alpha)}^{MN} \varepsilon^{ab} \varepsilon_{cd} + \frac{1}{60} \Omega_{KL} \Omega^{MN} \delta^{(a}_d \delta^{b)}_c\\
\begin{split}
{\left( {\hat{\mathbb{P}}}_{\mathbf{3875}} \right)}_{Kc Ld}{}^{Ma Nb} & = \left( \frac{1}{14} \delta^M_K \delta^N_L - \frac{3}{7} {(t_\alpha)}_K{}^M {(t^\alpha)}_L{}^N - \frac{1}{56} \delta^N_K \delta^M_L \right) \delta^a_c \delta^b_d\\
& \qquad + \left( \frac{1}{14} \delta^N_K \delta^M_L - \frac{3}{7} {(t_\alpha)}_K{}^N {(t^\alpha)}_L{}^M - \frac{1}{56} \delta^M_K \delta^N_L \right) \delta^b_c \delta^a_d\\
& \qquad + \frac{1}{56} \left( 2 \delta^{(M}_K \delta^{N)}_L - \Omega_{KL} \Omega^{NM} \right) \varepsilon^{ab} \varepsilon_{cd}
\end{split}\\
\begin{split}
{\left( {\hat{\mathbb{P}}}_{\mathbf{27000}} \right)}_{Kc Ld}{}^{Ma Nb} & = \left( \frac{6}{14} \delta^M_K \delta^N_L + \frac{3}{7} {(t_\alpha)}_K{}^M {(t^\alpha)}_L{}^N + \frac{1}{56} \delta^N_K \delta^M_L \right) \delta^a_c \delta^b_d\\
& \qquad + \left( \frac{6}{14} \delta^N_K \delta^M_L  + \frac{3}{7} {(t_\alpha)}_K{}^N {(t^\alpha)}_L{}^M + \frac{1}{56} \delta^M_K \delta^N_L \right) \delta^b_c \delta^a_d\\
& \qquad - \frac{1}{56} \left( 2 \delta^{(M}_K \delta^{N)}_L - \frac{24}{31} \Omega_{KL} \Omega^{NM} \right) \varepsilon^{ab} \varepsilon_{cd}
\end{split}\\
{\left( {\hat{\mathbb{P}}}_{\mathbf{30380}} \right)}_{Kc Ld}{}^{Ma Nb} & = \frac{1}{5} {(t_\alpha)}_{KL} {(t^\alpha)}^{MN} \varepsilon^{ab} \varepsilon_{cd} - \frac{1}{60} \Omega_{KL} \Omega^{MN} \delta^{(a}_d \delta^{b)}_c + \delta^{[Ma}_{Kc} \delta^{Nb]}_{Ld}\,.
\end{align}
Note that the (anti-)symmetrisation of the composite indices $Ma$ is given by
\begin{align}
\delta^{(Ma}_{Kc} \delta^{Nb)}_{Ld} & = \frac{1}{2} \left( \delta^M_K \delta^N_L \delta^a_c \delta^b_d + \delta^N_K \delta^M_L \delta^b_c \delta^a_d \right)\,,\\
\delta^{[Ma}_{Kc} \delta^{Nb]}_{Ld} & = \frac{1}{2} \left( \delta^M_K \delta^N_L \delta^a_c \delta^b_d - \delta^N_K \delta^M_L \delta^b_c \delta^a_d \right)\,.
\end{align}
The sum of the relevant projectors then still satisfy
\begin{align}
{\left( {\hat{\mathbb{P}}}_{\mathbf{1}} \right)}_{Kc Ld}{}^{Ma Nb} + {\left( {\hat{\mathbb{P}}}_{\mathbf{3875}} \right)}_{Kc Ld}{}^{Ma Nb} + {\left( {\hat{\mathbb{P}}}_{\mathbf{27000}} \right)}_{Kc Ld}{}^{Ma Nb} & = \delta^{(Ma}_{Kc} \delta^{Nb)}_{Ld}\,,\\
{\left( {\hat{\mathbb{P}}}_{\mathbf{248}} \right)}_{Kc Ld}{}^{Ma Nb} + {\left( {\hat{\mathbb{P}}}_{\mathbf{30380}} \right)}_{Kc Ld}{}^{Ma Nb} & = \delta^{[Ma}_{Kc} \delta^{Nb]}_{Ld}\,.
\end{align}
We now look in more detail at the $E_{8(8)}$ section constraints (which is typically viewed as the vanishing of some other representation $R_2 \subset R_1 \otimes R_1$)
\begin{align}\label{eq:E8Section}
{\left( {\hat{\mathbb{P}}}_{\mathbf{1} \oplus \mathbf{248} \oplus \mathbf{3875}} \right)}_{\hat{K} \hat{L}}{}^{\hat{M} \hat{N}} {\hat{C}}_{\hat{M}} \otimes {{\hat{C}^\prime}}_{\hat{N}} & =  0\,,
\end{align}
where ${\hat{C}}_{\hat{M}}, {\hat{C}}^\prime_{\hat{M}} \in \{ {\hat{\partial}}_{\hat{M}}, {\hat{B}}_{\mu \hat{M}}, {\hat{\Sigma}}_{\hat{M}} , {\hat{C}}_{\hat{\mu} \hat{\nu} \hat{M}}{}^{\hat{N}}, \ldots\}$ are covariantly constrained objects. However, since the projectors do not reduce exactly (e.g.\ in \eqref{eq:ProjectorReduction}), it is much more convenient to consider the section constraint in terms of the $Y$-tensor which does reduce exactly \eqref{eq:E8ToE7YTensor}. We thus consider the combination
\begin{align}
{\hat{Y}}^{\hat{M} \hat{N}}{}_{\hat{K} \hat{L}} {\hat{C}}_{\hat{M}} \otimes {\hat{C}^\prime}_{\hat{N}} & = { \left(62 {\hat{\mathbb{P}}}_{\mathbf{1}} + 30 {\hat{\mathbb{P}}}_{\mathbf{248}} +  14 {\hat{\mathbb{P}}}_{\mathbf{3875}} \right)}_{\hat{L} \hat{K}}{}^{\hat{M} \hat{N}} {\hat{C}}_{\hat{M}} \otimes {{\hat{C}}^\prime}_{\hat{N}} = 0\,.
\end{align}
Substituting in the explicit forms of the projectors, the section constraints read
\begin{align}
\hat{\kappa}^{\hat{K} \hat{L}} {\hat{C}}_{\hat{K}} \otimes {\hat{C}}^\prime_{\hat{L}} & = 0\,,\\
\hat{f}^{\hat{P} \hat{K} \hat{L}} {\hat{C}}_{\hat{K}} \otimes {\hat{C}}^\prime_{\hat{L}} & = 0\,,\\
\left( \delta^{(\hat{K}}_{\hat{M}} \delta^{\hat{L})}_{\hat{N}} - 30 {\hat{f}}^{\hat{P} (\hat{K}}{}_{\hat{M}} {\hat{f}}_{\hat{P}}{}^{\hat{L})}{}_{\hat{N}} \right)  {\hat{C}}_{\hat{K}} \otimes {\hat{C}}^\prime_{\hat{L}} & = 0,
\end{align}
The $E_{7(7)}$ analogue of \eqref{eq:E8Section} is
\begin{align}
{\left( {\mathbb{P}}_{\mathbf{1} \oplus \mathbf{133}} \right)}_{MN}{}^{KL} \partial_K \otimes \partial_L & = 0
\end{align}
which can equivalently be obtained from the $Y$-tensor:
\begin{align}
Y^{KL}{}_{MN} \partial_K \otimes \partial_L = {\left( 12 {\mathbb{P}}_{\mathbf{133}} + 28 {\mathbb{P}}_{\mathbf{1}} \right)}_{NM}{}^{KL} \partial_K \otimes \partial_L = 0\,.
\end{align}
Expanding the projectors in terms of invariants, this becomes
\begin{align}
\Omega^{KL} \partial_K \otimes \partial_L & = 0\,, \qquad {(t^{\alpha})}^{KL} \partial_K \otimes \partial_L = 0\,.
\end{align}
We now expand each of the $E_{8(8)}$ constraints to explicitly verify that they reduce to the $E_{7(7)}$ section constraints. The only non-trivial constraint obtainable from the first of the $E_{8(8)}$ constraints, under the ansatz that only ${\hat{\partial}}_{M1} \neq 0$, is
\begin{align}
\Omega^{KL} \varepsilon^{ab} \partial_{Ka} \otimes \partial_{Lb} \biggr\vert_{a=b=1}=0 
\end{align}
which always vanishes and is thus vacuous on the $E_{7(7)}$ section. Looking at \eqref{eq:E8StructureConstants}, the only two non-vanishing structure constants give
\begin{align}
\hat{f}^{\alpha Ka Lb} {\partial}_{Ka} \otimes {\partial}_{Lb} \biggr\vert_{a=b=1} & = 0 \qquad \longrightarrow \qquad {(t^\alpha)}^{KL} \varepsilon^{ab} {\partial}_{Ka} \otimes {\partial}_{Lb} \biggr\vert_{a=b=1} = 0\,,\\
\hat{f}^{i Ka Lb} {\partial}_{Ka} \otimes {\partial}_{Lb} \biggr\vert_{a=b=1} & = 0 \qquad \longrightarrow \qquad \Omega^{KL} {(D^i)}^{ab} {\partial}_{Ka} \otimes {\partial}_{Lb} \biggr\vert_{a=b=1} = 0\,.
\end{align}
The first is, again, vacuous but the second gives one of the $E_{7(7)}$ section constraints
\begin{align}
\Omega^{KL} \partial_K \otimes \partial_L = 0\,.
\end{align}
The final constraint is symmetric under $\hat{M} \leftrightarrow \hat{N}$ and so there are 6 possible independent constraints. If [$\hat{M} = \alpha$ and $\hat{N} = \beta$] or [$\hat{M} = i$ and $\hat{N} = j$], then it is vacuous since the result is proportional to $\varepsilon^{ab} {\hat{\partial}}_{Ka} \otimes {\hat{\partial}}_{Lb}$. If [$\hat{M} = M$ and $\hat{N} = N$], then we obtain
\begin{align}
\left( \frac{3}{4} \delta^{(K}_M \delta^{L)}_N - 6 {(t_\alpha)}_M{}^{(K} {(t^\alpha)}_N{}^{L)} \right) \partial_{K} \otimes \partial_{L} = 0\,,
\end{align}
which can be brought to the form
\begin{align}
-6 {(t_\alpha)}_{MN} {(t^\alpha)}^{KL} \partial_K \otimes \partial_L = 0
\end{align}
upon substituting in the relation \eqref{eq:E7Gen}, thereby recovering the second $E_{7(7)}$ section constraint. In fact, choosing $[\hat{M} = \alpha \text{ and }  \hat{N} = j]$ gives
\begin{align}
-\frac{\sqrt{6}}{2} {(t_\alpha)}^{KL} {(D_j)}^{(cd)} {\hat{\partial}}_{Kc} \otimes {\hat{\partial}}_{Ld} = 0
\end{align}
and so also recovers the second $E_{7(7)}$ section condition. The remaining two choices, [$\hat{M} = \alpha$ and $\hat{N} = Nc$] or [$\hat{M} = j$ and $\hat{N} = Nc$], both vanish trivially as there are no structure constants of the required index structures. Thus the set of all $E_{8(8)}$ constraints, subject to \eqref{eq:E8E7Ansatz}, recovers both of the $E_{7(7)}$ constraints and nothing else.\par
Of particular interest here is that the maximal subgroup of $E_{8(8)}$ has an additional $\operatorname{SL}(2)$ symmetry over the $E_{7(7)}$ symmetry that is made manifest in the usual EFT. In particular, we have an $\operatorname{SL}(2)$ doublet worth of $E_{7(7)}$ coordinate representations inside the $\mathbf{248}$ of $E_{8(8)}$. In reducing the section constraints above, we explicitly broke the $\operatorname{SL}(2)$ covariance and chose ${\hat{Y}}^{M1} \equiv Y^M$ to be the $E_{7(7)}$ extended coordinates. However, this choice was entirely arbitrary; we could equally have chosen ${\hat{Y}}^{M2} \equiv Y^{\overbar{M}}$ (in the notation of Section~\ref{sec:ReductionLargerGenMetric}) to be the $E_{7(7)}$ extended coordinates instead. More generally, we could have chosen $a=1$ to be the extended coordinates on one local patch and $a=2$ to be the extended coordinates on another local patch such that we require an $\operatorname{SL}(2)$ transformation to patch the two together in a manner reminiscent of the T- and U-folds of Hull. However, the interpretation of such configurations from the $E_{7(7)}$ perspective is not clear. It is not a non-geometric configuration of the sort that has been previously studied in the literature since the patching $\operatorname{SL}(2)$ transformation is not even a $G$-transformation and is thus not generated by the local symmetries of the supergravity fields. These may be considered as examples of `truly' non-geometric backgrounds, of the sort hypothesised in \cite{Berman:2018okd,Otsuki:2019owg} and in much earlier works such as \cite{Dabholkar:2005ve}, in the sense that they cannot be related to any geometric configurations by duality transformations.
\subsection{\texorpdfstring{$\operatorname{SL}(5)$}{SL(5)} EFT to \texorpdfstring{$\operatorname{SL}(3) \times \operatorname{SL}(2)$}{SL(3)xSL(2)} EFT and \texorpdfstring{$\operatorname{O}(3,3)$}{O(3,3)} DFT}
We begin with the $\operatorname{SL(5)}$ section constraint
\begin{align}\label{eq:SL5Section}
{\hat{Y}}^{\hat{M}\hat{N}}{}_{\hat{K}\hat{L}} \partial_{\hat{M}} \otimes \partial_{\hat{N}} & = 3! \delta^{{\hat{\underbar{m}}}_1 {\hat{\underbar{m}}}_2 {\hat{\underbar{n}}}_1 {\hat{\underbar{n}}}_2}_{{\hat{\underbar{k}}}_1 {\hat{\underbar{k}}}_2 {\hat{\underbar{l}}}_1 {\hat{\underbar{l}}}_2}\partial_{{\hat{\underbar{m}}}_1 {\hat{\underbar{m}}}_2} \otimes \partial_{{\hat{\underbar{n}}}_1 {\hat{\underbar{n}}}_2} = 0\,,
\end{align}
where we have used the fact that the $R_1 = \mathbf{10}$ indices can be written in terms of the 5-dimensional indices ${{\hat{\underbar{m}}}}_1 = 1, \ldots, 5$ as an antisymmetric pair $\hat{M} = [{\hat{\underbar{m}}}_1 {\hat{\underbar{m}}}_2]$. It is well-known that the section condition has only two inequivalent solutions; the so-called M-theory section and the Type IIB section. These are solutions in the sense that they give rise to theories with no further constraints. Here, we consider the $\operatorname{SL}(3) \times \operatorname{SL}(2)$ and $\operatorname{O}(3,3)$ ExFTs as arising from \emph{partial} solutions to the section conditions; choices of dropped coordinate dependences that give rise to theories with residual constraints.\par
These come in two classes. The first are obtained from imposing only a subset of the constraints that solve the section condition in the usual manner and these are expected to give the section conditions of the lower-dimensional EFTs. The $\operatorname{SL}(5) \rightarrow \operatorname{SL}(3) \times \operatorname{SL}(2)$ reduction falls under this category. Further imposing the M-theory or Type IIB sections of the child ExFT should yield the M-theory or the Type IIB solutions of the child theory; this partial solution should be extensible to a full solution of the parent ExFT's section constraint. The second is a partial solution that cannot be extended to a full M-theory or Type IIB solution in the manner described above. Whilst it may sound like this in conflict with the usual narrative that there are only two inequivalent solutions, we emphasise that this is not the case since it is not a full solution of the section condition. The $\operatorname{SL}(5) \rightarrow \operatorname{O}(3,3)$ reduction is an example of such a reduction.\par
We begin with a brief description of the usual story. For the M-theory section, we decompose $\operatorname{SL}(5)$ under $\operatorname{GL}(4)$ according to
\begin{align}
\mathbf{10} \rightarrow \mathbf{4}_{-3} \oplus \mathbf{6}_2
\end{align}
which corresponds to the decomposition of the generalised coordinates
\begin{align}
{\hat{Y}}^{[\hat{\underbar{m}}_1 \hat{\underbar{m}}_2]} = ({\hat{Y}}^{\hat{m}_1 5}, {\hat{Y}}^{\hat{m}_1 \hat{m}_2}) = \left( Y^{\hat{m}_1 \hat{m}_2}, \frac{1}{2} \varepsilon^{\hat{m}_1 \hat{m}_2 \hat{n}_1 \hat{n}_2} Y_{\hat{n}_1 \hat{n}_2} \right)\,.
\end{align}
The section condition decomposes under this branching as (here, we have rewritten the generalised Kronecker delta in terms of $\varepsilon$-symbols)
\begin{align}\label{eq:SL5GL4Section}
\begin{aligned}
\varepsilon^{\hat{k} \hat{m}_1 \hat{m}_2 \hat{n}_1 5} \left( \partial_{\hat{m}_1 \hat{m}_2} \otimes \partial_{\hat{n}_1 5} + \partial_{\hat{n}_1 5} \otimes \partial_{\hat{m}_1 \hat{m}_2} \right) & = 0\,,\\
\varepsilon^{5 \hat{m}_1 \hat{m}_2 \hat{n}_1 \hat{n}_2} \partial_{\hat{m}_1 \hat{m}_2} \otimes \partial_{\hat{n}_1 \hat{n}_2} & = 0\,.
\end{aligned}
\end{align}
It is then easy to see that these are solved by dropping all coordinate dependences on the membrane wrapping coordinates
\begin{align}\label{eq:MThSoln}
\partial_{\hat{m}_1 \hat{m}_2} \sim \partial^{\hat{n}_1 \hat{n}_2} = 0 \rightarrow \text{M-theory}\,.
\end{align}
For the Type IIB section, we instead decompose $\operatorname{SL}(5)$ under $\operatorname{SL}(3) \times \operatorname{SL}(2)$ as
\begin{align}
\mathbf{10} \rightarrow {(\overbar{\mathbf{3}}, \mathbf{1})}_{4} \oplus {(\mathbf{3,2})}_{-1} \oplus {(\mathbf{1,1})}_{-6}\,.
\end{align}
The index splitting $\hat{\underbar{m}} = (m, \alpha)$ induces the decomposition of the generalised coordinates
\begin{align}
{\hat{Y}}^{[{\hat{\underbar{m}}}_1 {\hat{\underbar{m}}}_2]} \rightarrow \left( Y^{[m_1 m_2]}, Y^{m_1 \alpha}, Y^{[\alpha \beta]} \right)\,,
\end{align}
where $m =1,2,3$ and $\alpha = 4,5$, and the decomposition of the section condition 
\begin{align}\label{eq:IIBSplitting}
\begin{aligned}
\delta^{m_1 m_2 n_1}_{k_1 l_1 l_2} \delta^\alpha_\gamma \left( \partial_{m_1 m_2} \otimes \partial_{n_1 \alpha} + \partial_{n_1 \alpha} \otimes \partial_{m_1 m_2} \right) & = 0\,,\\
\delta^{m_1 n_1}_{k_1 l_1} \delta^{\alpha \beta}_{\gamma \delta} \left( \partial_{m_1 \alpha} \otimes \partial_{n_1 \beta} - \partial_{m_1 n_1} \otimes \partial_{\alpha \beta} - \partial_{\alpha \beta} \otimes \partial_{m_1 n_1} \right) & = 0\,.
\end{aligned}
\end{align}
The IIB section is then given by the choice 
\begin{align}\label{eq:IIBSoln}
\partial_{m \alpha} = \partial_{\alpha \beta} = 0 \rightarrow \text{IIB}\,.
\end{align}
We now turn to how we can recover the $\operatorname{SL}(3) \times \operatorname{SL}(2)$ EFT and $\operatorname{O}(3,3)$ DFT section conditions from the above. Rather than follow the M-theory reduction, we consider what happens if we choose to set $\partial_{\hat{m}_1 5} = 0$ instead of the M-theory solution \eqref{eq:MThSoln}. This does \emph{not} fully solve the section condition and \eqref{eq:SL5GL4Section} instead reduces to
\begin{align}
\varepsilon^{\hat{m}_1 \hat{m}_2 \hat{n}_1 \hat{n}_2}  \partial_{\hat{m}_1 \hat{m}_2} \otimes \partial_{\hat{n}_1 \hat{n}_2}= 0\,.
\end{align}
To find solutions for this we further split $\hat{m} = ( m, 4)$, similar to the Type IIB solution, for which 
\begin{align}
\varepsilon^{m_1 m_2 n_1 4} \left( \partial_{m_1 m_2} \otimes \partial_{n_1 4} + \partial_{n_1 4} \otimes \partial_{m_1 m_2} \right) = 0\,.
\end{align}
Owing to the notation $\otimes$, this can be compressed down to a single term
\begin{align}
\varepsilon^{m_1 m_2 n_1} \partial_{m_1 m_2} \otimes \partial_{n_1 4} = 0\,.
\end{align}
 If we identify $\varepsilon^{m_1 m_2 n_1} \partial_{m_1 m_2} \sim \partial^{n_1}$ and $\partial_{n_1 4} \sim \partial_{n_1}$ we obtain the $\operatorname{O}(3,3)$ DFT section condition
\begin{align}
\partial^m \otimes \partial_m = 0\,.
\end{align}
We stress that, despite yielding a 3-dimensional section, this is \emph{not} obtainable from the Type IIB solution since the usual coordinates of the DFT descend wholly from the usual coordinates of the M-theory section (recall that a Type IIB section shares two coordinates with the M-theory section but takes a third coordinate from the membrane wrapping directions). However, nor should it be thought of as a Type IIA `section', since it was constructed as an independent (partial) solution to the M-theory section. It is perhaps better thought of an independent path through which one can obtain the Type IIA theory from the $\operatorname{SL}(5)$ theory.\par
The $\operatorname{SL}(3) \times \operatorname{SL}(2)$ section condition can be recovered from a similar analysis applied to the alternate splitting \eqref{eq:IIBSplitting}. Rather than taking the solution that gives the Type IIB section \eqref{eq:IIBSoln}, if we instead choose to set
\begin{align}\label{eq:SL3SL2SectionSolution}
\partial_{m_1 m_2} = 0
\end{align}
then the only non-trivial remaining constraint is
\begin{align}
\delta^{m_1 n_1}_{k_1 l_1} \delta^{\alpha \beta}_{\gamma \delta} \partial_{m_1 \alpha} \otimes \partial_{n_1 \beta} = 0
\end{align}
which is the $\operatorname{SL}(3) \times \operatorname{SL}(2)$ section condition. We have summarised the ways that one obtains the various theories that we discussed above in Table~\ref{tab:MTheorySolutions}. We mention a couple of things to note. Firstly, since we obtained the $\operatorname{SL}(3) \times \operatorname{SL}(2)$ generalised metric from a KK reduction, we should further impose the KK isometry $\partial_{\alpha \beta} = 0$. However, since this is independent of the section constraint, we have not listed this in the table. Additionally, the table is not exhaustive; a circle reduction of the generalised metric in the Type IIB parametrisation is also likely to give further partial solutions that we have not discussed here but we expect that their respective section conditions can be recovered in an analogous fashion.\par
\begin{table}
\centering
\begin{tabulary}{0.7\textwidth}{LCL}
\toprule
Constraint & Residual Section Constraint & Resulting Theory\\
\midrule
$\partial_{\hat{m}\hat{n}} = 0$ & $-$ & M-theory\\
$\partial_{m \alpha} = \partial_{\alpha \beta} = 0$ & $-$ & IIB\\
$\partial_{\hat{m}5} = 0$ & $\partial^m \otimes \partial_m = 0$ & $\operatorname{O}(3,3) \text{ DFT}$\\
$\partial_{mn} = 0$ & $\delta^{mn}_{pq} \delta^{\alpha \beta}_{\gamma \delta} \partial_{m\alpha} \otimes \partial_{n \beta} = 0$ & $\operatorname{SL}(3) \times \operatorname{SL}(2) \text{ EFT}$\\
\bottomrule
\end{tabulary}
\caption{The possible solutions and partial solutions that can be obtained from the $\operatorname{SL}(5)$ section constraint.}
\label{tab:MTheorySolutions}
\end{table}
However, in the absence of such explicit calculations, it is not clear which choices of partial solutions admit such interpretations. For example, rather than taking the full IIB section, one might be tempted to consider setting only $\partial_{m \alpha} = 0$ to give rise to a constrained theory with some residual section condition. However, there is no reason to expect that such arbitrary choices will lead to recognisable ExFTs. A full classification would likely be based on considering subgroups of $\hat{G}$ which have representations that are consistent with the branching of the parent ExFT's representations under that reduction (whether it be a circle/torus reduction or more general ansatzes). However, it is hopefully clear that the much more involved section conditions of larger EFTs may house more derivative EFTs than one might initially expect.
\section{Rewriting the \texorpdfstring{$\mathcal{B}$-$\mathcal{F}$}{BF} Term in \texorpdfstring{$E_{8(8)}$}{E8(8)} EFT}\label{sec:BF}
Despite the similarity in the way that the EFTs are constructed, even a cursory look reveals that the details of the theories differ wildly by dimension. Focusing on the $d=3$ and $d=4$ EFTs, we have already outlined the novel features of the $E_{8(8)}$ coordinates, generalised Lie derivative and generalised metric but its differences from $E_{7(7)}$ EFT extends to the action as well. The former possesses the signature vector-scalar duality of $d=3$ theories and a full Lagrangian description whilst the vector-vector duality of the latter is demoted to an extra condition on a pseudo-action as a twisted self-duality constraint on the generalised fieldstrength $\mathcal{F}_{\mu \nu}{}^M$. Additionally, the Chern-Simons term of $E_{8(8)}$ EFT must somehow source part of the topological and Yang-Mills terms of $E_{7(7)}$ EFT (since these are where the gauge sectors are encoded in the respective theories), but the identification is far from obvious.\par
In this section, we study how a subset of the terms in the $E_{8(8)}$ EFT Lagrangian density rearrange themselves into the terms found in the $E_{7(7)}$ EFT Lagrangian density. The full identification of the reduction will also involve a careful study of the topological terms in the two theories but this is complicated by the fact that one can add and remove terms that vanish under section condition (including terms that involve covariantly constrained objects) to both theories. Such topological terms may also involve boundary terms of the type described in \cite{Berman:2011kg} and are well beyond the scope of the present work.\par
Before we proceed, we first make a list of the pieces of the $E_{7(7)}$ fields that need to be identified from amongst the $E_{8(8)}$ fields. The external coordinates of $E_{7(7)}$ EFT are given by $\mu = (\hat{\mu}, 4)$ where $\hat{\mu} = 1,2,3$ are the external coordinates of the $E_{8(8)}$ EFT and the direction $4$ is to be identified from the internal coordinates in $E_{8(8)}$. The $E_{7(7)}$ EFT gauge fields are $\{\mathcal{A}_\mu{}^M, \mathcal{B}_{\mu \nu M}, \mathcal{B}_{\mu \nu\alpha}\}$ where $M$ and $\alpha$ index the $\mathbf{56}$ coordinate- and $\mathbf{133}$ adjoint-representations of $E_{7(7)}$ EFT respectively. They decompose under this $3+1$ splitting to components of the form
\begin{align}
{\mathcal{A}}_{\mu}{}^M = \begin{pmatrix} {\mathcal{A}}_{\hat{\mu}}{}^M\\ {\mathcal{A}}_4{}^M \end{pmatrix}\,, \qquad {\mathcal{B}}_{\mu \nu M} = \begin{pmatrix} {\mathcal{B}}_{\hat{\mu} \hat{\nu} M}\\ {\mathcal{B}}_{\hat{\mu} 4M} \end{pmatrix}\,, \qquad {\mathcal{B}}_{\mu \nu \alpha} = \begin{pmatrix} {\mathcal{B}}_{\hat{\mu} \hat{\nu} \alpha}\\ {\mathcal{B}}_{\hat{\mu} 4 \alpha} \end{pmatrix}\,.
\end{align}
Some of these have an obvious $E_{8(8)}$ origin as follows:
\begin{align}\label{eq:E7E8GaugeFields}
{\mathcal{A}}_{\hat{\mu}}{}^M = {\hat{\mathcal{A}}}_{{\hat{\mu}}}{}^M\,, \qquad {\mathcal{B}}_{\hat{\mu} 4M} \sim {\hat{\mathcal{B}}}_{\hat{\mu} M} + \ldots \,, \qquad {\mathcal{B}}_{\hat{\mu} 4\alpha} \sim {\hat{\mathcal{B}}}_{\hat{\mu} \alpha} + \ldots\,,
\end{align}
which are sourced from the $E_{8(8)}$ generalised gauge fields, decomposed under $E_{7(7)}$, as ${\hat{\mathcal{A}}}_{{\hat{\mu}}}{}^{\hat{M}} = ( {\hat{\mathcal{A}}}_{{\hat{\mu}}}{}^{M}, \ldots )$ and ${\hat{\mathcal{B}}}_{\hat{\mu} \hat{M}} = ({\hat{\mathcal{B}}}_{\hat{\mu} M}, {\hat{\mathcal{B}}}_{\hat{\mu} \alpha} , \ldots )$.  We are thus left with three components which we have yet to determine the $E_{8(8)}$ origins of and we denote them as $\hat{\Phi}$ to differentiate them from the other fields:
\begin{align}
{\mathcal{A}}_{\mu}{}^M = \begin{pmatrix} {\hat{\mathcal{A}}}_{\hat{\mu}}{}^M\\ {\hat{\Phi}}^M \end{pmatrix}\,, \qquad {\mathcal{B}}_{\mu \nu M} = \begin{pmatrix} {\hat{\Phi}}_{\hat{\mu} \hat{\nu} M}\\ {\hat{\mathcal{B}}}_{\hat{\mu}M} + \ldots \end{pmatrix}\,, \qquad {\mathcal{B}}_{\mu \nu \alpha} = \begin{pmatrix} {\hat{\Phi}}_{\hat{\mu} \hat{\nu} \alpha}\\ {\hat{\mathcal{B}}}_{\hat{\mu} \alpha} + \ldots \end{pmatrix}\,.
\end{align}
We have used ellipses to denote corrections to the na\"{i}ve identification. Similarly, the $E_{7(7)}$ generalised field strength of $\mathcal{A}$ decomposes under the KK splitting into the components
\begin{align}
\mathcal{F}_{\mu \nu}{}^M = \begin{pmatrix} \mathcal{F}_{\hat{\mu} \hat{\nu}}{}^M\\ \mathcal{F}_{\hat{\mu}4}{}^M \end{pmatrix}\,.
\end{align}
The upper components have an obvious origin in the $E_{8(8)}$ analogue ${\hat{\mathcal{F}}}_{\hat{\mu} \hat{\nu}}{}^{\hat{M}} = ({\hat{\mathcal{F}}}_{\hat{\mu} \hat{\nu}}{}^M , \ldots )$ such that (we have already shown the generalised Lie derivative already reduces correctly)
\begin{align}
\mathcal{F}_{\hat{\mu} \hat{\nu}}{}^M = {\hat{\mathcal{F}}}_{\hat{\mu} \hat{\nu}}{}^M\,.
\end{align}
The unidentified piece $\mathcal{F}_{\hat{\mu} 4}{}^M$ is given in terms of $E_{7(7)}$ variables as
\begin{align}
F_{\hat{\mu} 4}{}^M & = \partial_{\hat{\mu}} {\mathcal{A}}_4{}^M - \cancel{\partial_4 {\mathcal{A}}_{\hat{\mu}}{}^M} - \frac{1}{2} \left( \mathbb{L}_{\mathcal{A}_{\hat{\mu}}} {\mathcal{A}}_4{}^M - \mathbb{L}_{{\mathcal{A}}_4}  {\mathcal{A}}_{\hat{\mu}}{}^M \right)\\
	& = \mathcal{D}_{\hat{\mu}} {\mathcal{A}}_4{}^M + {(\! ( \mathcal{A}_{\hat{\mu}},  {\mathcal{A}}_4{})\!)}^M\,,\\
{\mathcal{F}}_{\hat{\mu} 4}{}^M & = F_{\hat{\mu} 4}{}^M  - 12 {(t^\alpha)}^{MN} \partial_N {\mathcal{B}}_{\hat{\mu} 4 \alpha} - \frac{1}{2} \Omega^{MN} {\mathcal{B}}_{\hat{\mu} 4 N}\,,
\end{align}
where $(\!( \cdot, \cdot )\!)$ is the symmetric part of the generalised Lie derivative and $\mathcal{D}_\mu \coloneqq \partial_\mu - \mathbb{L}_{\mathcal{A}_\mu}$ is the Lie-covariantised derivative. We rewrite this in terms of the $E_{8(8)}$ variables that we identified above, giving
\begin{align}
{F}_{\hat{\mu} 4}{}^M & = {\hat{\mathcal{D}}}_{\hat{\mu}} {\hat{\Phi}}^M + {( \! ( \hat{\mathcal{A}}_{\hat{\mu}}, \hat{\Phi} )\!)}^M\\
	& =\begingroup \renewcommand{\arraystretch}{1.5} \begin{array}[t]{l} {\hat{\mathcal{D}}}_{\hat{\mu}} {\hat{\Phi}}^M - 12 {(t^\alpha)}^{MN} \partial_N \left( {(t_{\alpha})}_{PQ} {\hat{\mathcal{A}}}_{\hat{\mu}}{}^P {\hat{\Phi}}^Q \right)\\ + \frac{1}{2} \Omega^{MN} (\partial_N {\hat{\mathcal{A}}}_{\hat{\mu}}{}^P {\hat{\Phi}}_P + \partial_N {\hat{\Phi}}^P {\hat{\mathcal{A}}}_{\hat{\mu} P})\end{array} \endgroup\\
{\mathcal{F}}_{\hat{\mu} 4}{}^M	& = \begingroup \renewcommand{\arraystretch}{1.5} \begin{array}[t]{l} {\hat{\mathcal{D}}}_{\hat{\mu}} {\hat{\Phi}}^M - 12 {(t^\alpha)}^{MN} \partial_N \left( {\mathcal{B}}_{\hat{\mu} 4 \alpha} + {(t_{\alpha})}_{PQ} {\hat{\mathcal{A}}}_{\hat{\mu}}{}^P {\hat{\Phi}}^Q \right)\\ - \frac{1}{2} \Omega^{MN} ({\mathcal{B}}_{\hat{\mu} 4 N} - \partial_N {\hat{\mathcal{A}}}_{\hat{\mu}}{}^P {\hat{\Phi}}_P - \partial_N {\hat{\Phi}}^P {\hat{\mathcal{A}}}_{\hat{\mu} P}) \end{array}\endgroup
\end{align}
and so the components of the $E_{7(7)}$ field strength are given in terms of $E_{8(8)}$ fields as
\begin{align}
\mathcal{F}_{\mu \nu}{}^M & = \begin{pmatrix} {\hat{\mathcal{F}}}_{\hat{\mu} \hat{\nu}}{}^M\\ {\hat{\mathcal{D}}}_{\hat{\mu}} {\hat{\Phi}}^M - 12 {(t^\alpha)}^{MN} \partial_N {\hat{b}}_{\hat{\mu} \alpha} - \frac{1}{2} \Omega^{MN} {\hat{b}}_{\hat{\mu} N}
\end{pmatrix}\,,\\
{\hat{b}}_{\hat{\mu} \alpha} & = {\mathcal{B}}_{\hat{\mu} 4 \alpha} + {(t_{\alpha})}_{PQ} {\hat{\mathcal{A}}}_{\hat{\mu}}{}^P {\hat{\Phi}}^Q\,,\\
{\hat{b}}_{\hat{\mu} M} & = {\mathcal{B}}_{\hat{\mu} 4 N} - \partial_N {\hat{\mathcal{A}}}_{\hat{\mu}}{}^P {\hat{\Phi}}_P - \partial_N {\hat{\Phi}}^P {\hat{\mathcal{A}}}_{\hat{\mu} P}\,.
\end{align}
Note, in particular, that $\hat{b}_{\hat{\mu} \bullet}$ is still given in terms of a mixture of $E_{8(8)}$ objects $(\hat{\mathcal{A}}_{\hat{\mu}}{}^{\hat{M}}, \hat{\Phi}^{\hat{M}})$ and $E_{7(7)}$ objects ($\mathcal{B}_{\hat{\mu}\hat{4} \bullet}$). This is due to the fact that are not identifying the $E_{8(8)}$ $\hat{\mathcal{B}}_{\hat{\mu}}$ field as the $\hat{\mu}$-$4$ component of the $E_{7(7)}$ ${\mathcal{B}}_{\mu \nu}$ directly, as we indicated by ellipses previously. We shall clear up the relation between all these fields below.\par
Our starting point is the sum of the kinetic term for the scalar sector and the $\hat{\mathcal{B}}$-$\hat{F}$ contribution to the Chern-Simons term of $E_{8(8)}$ EFT (as before, we adorn all objects and indices in $d=3$ with hats). The kinetic term can be written in terms of the scalar current,
\begin{align}{\hat{\jmath}}_{\hat{\mu}}{}^{\hat{M}} \coloneqq \frac{1}{60} {\hat{f}}^{\hat{M}}{}_{\hat{K}}{}^{\hat{L}} {\hat{\mathcal{M}}}^{\hat{K} \hat{Q}} {\hat{\mathcal{D}}}_{\hat{\mu}}{\hat{\mathcal{M}}}_{\hat{Q} \hat{L}}\,,
\end{align}
and the uncovariantised field strength $\hat{F}$ may be improved to the fully covariantised field strength $\hat{\mathcal{F}}$ since the two differ only by terms that vanish under the section condition. Our starting point is the Lagrangian
\begin{align}\label{eq:E8ActionPart}
{\hat{\mathcal{L}}} \coloneqq {\hat{\mathcal{L}}}_{\text{kin.}} + {\hat{\mathcal{L}}}_{\text{CS}} & = \frac{1}{240} {\hat{\mathcal{D}}}_{\hat{\mu}} {\hat{\mathcal{M}}}_{\hat{M} \hat{N}} {\hat{\mathcal{D}}}^{\hat{\mu}} {\hat{\mathcal{M}}}^{\hat{M} \hat{N}} + \frac{1}{2} \varepsilon^{\hat{\mu} \hat{\nu} \hat{\rho}} {\hat{F}}_{\hat{\mu} \hat{\nu}}{}^{\hat{M}} {\hat{\mathcal{B}}}_{\hat{\rho}\hat{M}} + \ldots\\
	& = - \frac{\hat{e}}{4} {\hat{g}}^{\hat{\mu} \hat{\nu}} {\hat{\jmath}}_{\hat{\mu}}{}^{\hat{M}} {\hat{\jmath}}_{\hat{\nu} \hat{M}} + \frac{1}{2} \varepsilon^{\hat{\mu} \hat{\nu} \hat{\rho}} {\hat{\mathcal{F}}}_{\hat{\mu} \hat{\nu}}{}^{\hat{M}} {\hat{\mathcal{B}}}_{\hat{\rho}\hat{M}} + \ldots\label{eq:KinCS}\,,
\end{align}
where the ellipses represent terms that are independent of $\hat{\mathcal{B}}$-field and so it is sufficient to vary \eqref{eq:KinCS} to obtain the $\hat{\mathcal{B}}$-field equations of motion. In doing so, we use the fact that the scalar current is given explicitly by
\begin{align}
{\hat{\jmath}}_{\hat{\mu}}{}^{\hat{M}} & = \frac{1}{60} {\hat{f}}^{\hat{M}}{}_{\hat{K}}{}^{\hat{L}} {\hat{\mathcal{M}}}^{\hat{K}\hat{Q}} (\partial_{\hat{\mu}} - {\hat{\mathcal{A}}}_{\hat{\mu}}{}^{\hat{R}} \partial_{\hat{R}} ) {\hat{\mathcal{M}}}_{\hat{Q}\hat{L}} + ({\hat{\mathcal{M}}}^{\hat{M}\hat{N}} + {\hat{\kappa}}^{\hat{M}\hat{N}})({\hat{f}}_{\hat{N}}{}^{\hat{S}}{}_{\hat{T}} \partial_{\hat{S}} {\hat{\mathcal{A}}}_{\hat{\mu}}{}^{\hat{T}} + {\hat{\mathcal{B}}}_{\hat{\mu} \hat{N}})\label{eq:JBRelation}\,.
\end{align}
The resulting equation of motion for $\hat{\mathcal{B}}$ is then
\begin{align}\label{eq:VSDuality}
{\hat{\epsilon}}^{\hat{\mu} \hat{\rho} \hat{\sigma}} {\hat{\mathcal{F}}}_{\hat{\rho} \hat{\sigma}}{}^{\hat{M}} = 2 {\hat{g}}^{\hat{\mu} \hat{\nu}} {\hat{\jmath}}_{\hat{\nu}}{}^{\hat{M}}\,,
\end{align}
which relates the fields in the scalar coset to the vector fields just as in a vector-scalar duality relation. Multiplying both sides of \eqref{eq:JBRelation} by ${\hat{\jmath}}_{\hat{\nu} \hat{M}}$ gives an expression of the form ${\hat{\mathcal{B}}}_{\hat{\mu} \hat{M}} {\hat{\jmath}}_{\hat{\nu}}{}^{\hat{M}} = {\hat{\jmath}}_{\hat{\mu} \hat{M}} {\hat{\jmath}}_{\hat{\nu}}{}^{\hat{M}} + \ldots$, from which we may rewrite both terms in \eqref{eq:E8ActionPart} in terms of ${\hat{\jmath}}^2$ terms as
\begin{align}\label{eq:KinCS2}
{\hat{\mathcal{L}}} & = \frac{3}{4} \hat{e} {\hat{g}}^{\hat{\mu} \hat{\nu}} {\hat{\jmath}}_{\hat{\mu}}{}^{\hat{M}} {\hat{\jmath}}_{\hat{\nu} \hat{M}} + \ldots\,.
\end{align}
The final ingredient that we need to proceed is to note that, in analogy with the left-invariant currents in $\sigma$-models, we may write\footnote{One way to see this is to start with the Bianchi identity for $\mathcal{F}$:
\begin{align}
0 & = {\hat{\mathcal{D}}}_{\hat{\mu}} \left(  \varepsilon^{\hat{\mu} \hat{\rho} \hat{\sigma}} \hat{\mathcal{F}}_{\hat{\rho} \hat{\sigma}}{}^{\hat{M}} \right) \otimes {\hat{C}}_{\hat{N}}\qquad \Rightarrow \qquad 0 = {\hat{\mathcal{D}}}_{\hat{\mu}} \left(  2 {\hat{e}} \hat{\jmath}^{\hat{\mu} \hat{M}}\right) \otimes {\hat{C}}_{\hat{N}}\,,
\end{align}
where ${\hat{C}}_{\hat{M}}$ is a covariantly constrained object. We implement this by a Lagrange multiplier $+2 {\hat{\mathcal{D}}}_{\hat{\mu}} \left( \hat{e} {\hat{\jmath}}^{\hat{\mu} \hat{M}} \right) {\hat{\chi}}_{\hat{N}}$, where ${\hat{\chi}}_{\hat{N}}$ is covariantly constrained. Varying \eqref{eq:KinCS2} with respect to ${\hat{\jmath}}^{\hat{\mu} \hat{M}}$ yields that it can be written as the total derivative of ${\hat{\chi}}_{\hat{N}}$ (actually this requires more care since the variation of $\hat{\jmath}$ is not unconstrained but rather given in terms of the variations of ${\hat{\mathcal{A}}}$ and $\hat{\mathcal{B}}$ and so, strictly speaking, we may need some projectors to act on an unconstrained $\hat{\chi}$).}
\begin{align}\label{eq:JDChi}
{\hat{\jmath}}_{\hat{\mu} \hat{M}} = {\hat{\mathcal{D}}}_{\hat{\mu}} {\hat{\chi}}_{\hat{M}}\,,
\end{align}
for some covariantly constrained ${\hat{\chi}}_{\hat{M}}$ (a scalar on the external space).\par
We now return to \eqref{eq:KinCS2} and split it into 3 terms so that we can proceed to identify the necessary pieces needed for the $E_{7(7)}$ theory:
\begin{align}
{\hat{\mathcal{L}}} & = a \hat{e} {\hat{g}}^{\hat{\mu} \hat{\nu}} {\hat{\jmath}}_{\hat{\mu}}{}^{\hat{M}} {\hat{\jmath}}_{\hat{\nu} \hat{M}} + b \hat{e} {\hat{g}}^{\hat{\mu} \hat{\nu}} {\hat{\jmath}}_{\hat{\mu}}{}^{\hat{M}} {\hat{\jmath}}_{\hat{\nu} \hat{M}} + c \hat{e} {\hat{g}}^{\hat{\mu} \hat{\nu}} {\hat{\jmath}}_{\hat{\mu}}{}^{\hat{M}} {\hat{\jmath}}_{\hat{\nu} \hat{M}} + \ldots\,,
\end{align}
subject to $a + b + c = \frac{3}{4}$. The first term, we rewrite back in terms of the scalars$\mathcal{D} \mathcal{\hat{M}} \mathcal{D} {\hat{\mathcal{M}}}^{-1}$ as
\begin{align}
a \hat{e} {\hat{g}}^{\hat{\mu} \hat{\nu}} {\hat{\jmath}}_{\hat{\mu}}{}^{\hat{M}} {\hat{\jmath}}_{\hat{\nu} \hat{M}} & = - \frac{a}{60} \hat{e} {\hat{g}}^{\hat{\mu} \hat{\nu}} {\hat{\mathcal{D}}}_{\hat{\mu}} {\hat{\mathcal{M}}}_{\hat{M} \hat{N}} {\hat{\mathcal{D}}}_{\hat{\nu}} {\hat{\mathcal{M}}}^{\hat{M} \hat{N}} = - \frac{a}{60} \hat{e} {\hat{g}}^{\hat{\mu} \hat{\nu}} {\hat{\mathcal{D}}}_{\hat{\mu}} \mathcal{M}_{MN} {\hat{\mathcal{D}}}_{\hat{\nu}} \mathcal{M}^{MN} + \ldots\,,\nonumber
\end{align}
where we have expanded the $E_{8(8)}$ coordinates under the $E_{7(7)}$ decomposition \eqref{eq:E8E7Decomp} and restricted to the $E_{7(7)}$ coordinates (the $\operatorname{SL}(2)$ index fixed to $a=1$). We have also used the fact that the principal contribution to the $M1$-$N1$ component of the generalised metric should be the $E_{7(7)}$ generalised metric $\hat{\mathcal{M}}_{M1 N1} = \mathcal{M}_{MN} + \ldots$.
 If we assume that the fourth direction is a (generalised) isometry of the fields, which we can describe by 
\begin{align}
\mathbb{L}_{\mathcal{A}_4} \bullet = 0 \qquad \Rightarrow \qquad \partial_4 \bullet = 0 \, , \label{isoconst}
\end{align}
 we can simply replace the covariant derivatives with the 4-dimensional completions as ${\hat{\mathcal{D}}}_{\hat{\mu}} \rightarrow \mathcal{D}_\mu$. Finally, using the KK ansatz of the inverse external metric \eqref{eq:KKInverse}, we have $g^{\hat{\mu} \hat{\nu}} = e^{- 2 \alpha \phi} {\hat{g}}^{\hat{\mu}\hat{\nu}}$ for which $e = e^{3 \alpha \phi} \hat{e}$. Thus, we end up with
\begin{align}
a \hat{e} {\hat{g}}^{\hat{\mu} \hat{\nu}} {\hat{\jmath}}_{\hat{\mu}}{}^{\hat{M}} {\hat{\jmath}}_{\hat{\nu} \hat{M}} & = - \frac{a}{60} e e^{\alpha \phi} g^{\mu \nu} \mathcal{D}_\mu \mathcal{M}_{MN} \mathcal{D}_{\nu} \mathcal{M}^{MN} + \ldots\,.\nonumber
\end{align}
For the second term, we write one of the $\hat{\jmath}$ in terms of $\hat{\mathcal{F}}$ using vector-scalar relation \eqref{eq:VSDuality} and the other using \eqref{eq:JDChi} to give
\begin{align}
b \hat{e} {\hat{g}}^{\hat{\mu} \hat{\nu}} {\hat{\jmath}}_{\hat{\mu}}{}^{\hat{M}} {\hat{\jmath}}_{\hat{\nu} \hat{M}} & = \frac{be}{2} e^{-3 \alpha \phi} \epsilon^{\hat{\nu} \hat{\rho} \hat{\sigma}} {\hat{\mathcal{F}}}_{\hat{\rho} \hat{\sigma}}{}^{\hat{M}} {\hat{\mathcal{D}}}_{\hat{\nu}} {\hat{\chi}}_{\hat{M}} = \frac{be}{2} e^{-3 \alpha \phi} \epsilon^{\hat{\nu} \hat{\rho} \hat{\sigma}} {\hat{\mathcal{F}}}_{\hat{\rho} \hat{\sigma}}{}^{M} {\hat{\mathcal{D}}}_{\hat{\nu}} {\hat{\chi}}^{N} \Omega_{NM}+ \ldots\,.\nonumber
\end{align}
Finally, for the third term, we use the relations \eqref{eq:JBRelation} and \eqref{eq:VSDuality}, to rewrite it as
\begin{align}
c \hat{e} {\hat{g}}^{\hat{\mu} \hat{\nu}} {\hat{\jmath}}_{\hat{\mu}}{}^{\hat{M}} {\hat{\jmath}}_{\hat{\nu} \hat{M}} & =  \frac{c\hat{e}}{2} \epsilon^{\hat{\mu} \hat{\rho} \hat{\sigma}} {\hat{\mathcal{F}}}_{\hat{\rho} \hat{\sigma}}{}^{\hat{M}} ( {\hat{\mathcal{B}}}_{\hat{\mu} \hat{M}} + {\hat{f}}_{\hat{M}}{}^{\hat{K}}{}_{\hat{L}} \partial_{\hat{K}} {\hat{\mathcal{A}}}_{\hat{\mu}}{}^{\hat{L}} + \ldots)\nonumber\\
	& = \frac{c \hat{e}}{2} \epsilon^{\hat{\mu} \hat{\rho} \hat{\sigma}} {\hat{\mathcal{F}}}_{\hat{\rho} \hat{\sigma}}{}^{Ma} \Omega_{MN} \left( - \Omega^{NK} {\hat{\mathcal{B}}}_{\hat{\mu}K a} + \sqrt{12} {(t_\alpha)}^{NK} \partial_{Ka} {\hat{\mathcal{A}}}_{\hat{\mu}}{}^{\alpha} + \ldots \right)\,,\nonumber
\end{align}
where we have expanded the contracted $E_{8(8)}$ indices in terms of $E_{7(7)} \times \operatorname{SL}(2)$ indices and focused on the $\hat{M} = Ma$ pieces (the $(\mathbf{56,2})$ representations) of the contracted indices (the decomposition of the Killing form and structure constants are given in \eqref{eq:KillingDecomp} and \eqref{eq:E8StructureConstants} respectively\footnote{Strictly speaking, we should rescale the structure constants by $\sqrt{60}$, relative to \eqref{eq:E8StructureConstants}, since we have reverted to the conventions of \cite{Hohm:2014fxa} for this section. However, we shall demonstrate later that we do not need to worry about the precise scaling (at least at the classical level).}). Since we only want non-trivial derivatives in the $a=1$ direction---the directions that form the $E_{7(7)}$ generalised coordinates---we pick out only the $M1 \equiv M$ indices:
\begin{align*}
c \hat{e} {\hat{g}}^{\hat{\mu} \hat{\nu}} {\hat{\jmath}}_{\hat{\mu}}{}^{\hat{M}} {\hat{\jmath}}_{\hat{\nu} \hat{M}} & = \frac{c e}{2} e^{-3\alpha \phi} \epsilon^{\hat{\mu} \hat{\rho} \hat{\sigma}} {\hat{\mathcal{F}}}_{\hat{\rho} \hat{\sigma}}{}^{M} \Omega_{MN} \left( - \Omega^{NK} {\hat{\mathcal{B}}}_{\hat{\mu}K } + \sqrt{12} {(t_\alpha)}^{NK} \partial_{K} {\hat{\mathcal{A}}}_{\hat{\mu}}{}^{\alpha} + \ldots \right)\,.
\end{align*}
Then, the sum of the three terms can be written as
\begingroup
\renewcommand{\arraystretch}{1.5}
\begin{align}
\begin{array}{ll}
{\hat{\mathcal{L}}} & = - \frac{a}{60} e e^{\alpha \phi} g^{\mu \nu} \mathcal{D}_\mu \mathcal{M}_{MN} \mathcal{D}_{\nu} \mathcal{M}^{MN}\\
& \quad + e^{-3 \alpha \phi}  \varepsilon^{\hat{\mu} \hat{\rho} \hat{\sigma}} {\mathcal{F}}_{\hat{\rho} \hat{\sigma}}{}^{M} \Omega_{MN} \left( - \frac{b}{2} {\hat{\mathcal{D}}}_{\hat{\mu}} {\hat{\chi}}^{N} -  \frac{c}{2} \Omega^{NK} {\hat{\mathcal{B}}}_{\hat{\mu} K} + c\sqrt{3} {(t^\alpha)}^{NK} \partial_K {\hat{\mathcal{A}}}_{\hat{\mu} \alpha} \right) + \ldots \,,
\end{array}\nonumber
\end{align}
\endgroup
Then, for judicious choices of $b$ and $c$, the term in parentheses is of the form of ${\mathcal{F}}_{\hat{\mu} 4}{}^{N}$ once we identify
\begin{align}
- \frac{b}{2} {\hat{\chi}}^{\hat{N}} & = {\hat{\Phi}}^{\hat{N}}\,,\label{eq:ChiScaling}\\
c {\hat{\mathcal{B}}}_{\hat{\mu} K} & = {\hat{b}}_{\hat{\mu} K} = {\mathcal{B}}_{\hat{\mu} 4 K} - \partial_K {\hat{\mathcal{A}}}_{\hat{\mu}}{}^P {\hat{\Phi}}_P - \partial_K {\hat{\Phi}}^P {\hat{\mathcal{A}}}_{\hat{\mu} P}\,,\\
c \sqrt{3} {\hat{\mathcal{A}}}_{\hat{\mu} \alpha} & = -12 {\hat{b}}_{\hat{\mu} \alpha} = -12 ({\mathcal{B}}_{\hat{\mu} 4 \alpha} + {(t_{\alpha})}_{PQ} {\hat{\mathcal{A}}}_{\hat{\mu}}{}^P {\hat{\Phi}}^Q )\,.
\end{align}
The last two equations then give the precise form of the $E_{7(7)}$ components in terms of $E_{8(8)}$ fields that we previously denoted by ellipses in \eqref{eq:E7E8GaugeFields}. After the above manipulations, we are then left with
\begin{align}
{\hat{\mathcal{L}}} & = - \frac{a}{60} e e^{\alpha \phi} g^{\mu \nu} \mathcal{D}_\mu \mathcal{M}_{MN} \mathcal{D}_{\nu} \mathcal{M}^{MN} + e^{-3 \alpha \phi}  \varepsilon^{\hat{\mu} \hat{\rho} \hat{\sigma}} {\mathcal{F}}_{\hat{\rho} \hat{\sigma}}{}^M \Omega_{MN} {\mathcal{F}}_{\hat{\mu}4}{}^N + \ldots\\
	&  = - \frac{a}{60} e e^{\alpha \phi} g^{\mu \nu} \mathcal{D}_\mu \mathcal{M}_{MN} \mathcal{D}_{\nu} \mathcal{M}^{MN} + \frac{1}{4} e^{-3 \alpha \phi} \varepsilon^{\mu \nu \rho \sigma} {\mathcal{F}}_{\mu \nu}{}^M \Omega_{MN} {\mathcal{F}}_{\rho \sigma}{}^N + \ldots\,.
\end{align}
We are now very close to the desired result. The final piece of the puzzle is to use the $E_{7(7)}$ twisted self-duality constraint for $\mathcal{F}_{\mu \nu}{}^M$:
\begin{align}
\mathcal{F}_{\mu \nu}{}^M= \frac{1}{2} {\mathcal{M}}^M{}_N \sqrt{-g} \varepsilon_{\mu \nu}{}^{\rho \sigma} \mathcal{F}_{\rho \sigma}{}^N \, , \label{twisted}
\end{align}
where $\mathcal{M}^M{}_N = \mathcal{M}^{MP} \Omega_{PN}$.
With this final ingredient we may rewrite the second term as the Yang-Mills term in the $E_{7(7)}$ pseudo-action\footnote{The careful reader should rightly be worried about the status of this pseudo-action and the insertion of the self duality constraint. The pragmatic way to think about is that it is like the pseudo-action for chiral boson where the chirality constraint is imposed on the resulting equations of motion. For a more rigorous approach it is be possible to construct a PST type action \cite{Pasti:2012wv} or one could follow the recent approach of Sen \cite{Sen:2019qit}}:
\begin{align}
{\mathcal{L}}_{\text{YM}}=\frac{1}{4}\mathcal{M}_{MN}\mathcal{F}_{\mu \nu}{}^M\mathcal{F}^{\mu \nu}{}^N \, .
\end{align}
Equivalently, the equations of motion of $E_{8(8)}$ EFT reduces (for appropriate choices of the coefficients $a,b$ and $c$) to the on-shell equations of motion of $E_{7(7)}$ EFT. Note that, since we started with (part of) the topological term in $E_{8(8)}$, the ellipsis will include terms that enter into the topological term in $E_{7(7)}$.\par
We make one final remark regarding the relative scaling coefficients $a,b$ and $c$ which were constrained to satisfy $a + b + c = \frac{3}{4}$. We actually have an additional freedom in rescaling ${\hat{\chi}}_{\hat{M}}$, which affects $b$ in \eqref{eq:ChiScaling}, and so we can always find $a, b$ and $c$ such that the reduction of the terms to those in the $E_{7(7)}$ Lagrangian is exact. \par
We stress that the recombination of terms to produce the required scalar kinetic terms and Yang-Mills term for the $E_{7(7)}$ theory is highly non-trivial. It rests on:
\begin{itemize}
\item
the equations of motion of the $\hat{\mathcal{B}}$ field in the $E_{8(8)}$ EFT giving the $d=3$ vector-scalar relation \eqref{eq:VSDuality}; 
\item
the twisted self-duality relation of the $E_{7(7)}$ vector fields \eqref{twisted};
\item
and the isometry condition \eqref{isoconst}.
\end{itemize}
Finally let us comment that getting from the $\hat{\mathcal{B}}$-$\hat{\mathcal{F}}$ term to a Yang-Mills term by spontaneously breaking the symmetry and integrating out a field is very similar to the Papageorgakis-Mukhi mechanism in Bagger-Lambert theory \cite{Mukhi:2008ux}. There the Chern-Simons scalar theory was turned into a Yang-Mills theory using the equations of motion from integrating out one of the vector fields after one of the scalars is given a constant vacuum expectation value. The situation is similar here where the $\hat{\mathcal{B}}$-field is integrated out after the $E_{8(8)}$ symmetry is broken to $E_{7(7)}$.
\section{Discussion}
In this chapter, we have considered some of the aspects of reductions between ExFTs, with a particular focus on EFT-to-EFT reductions. We began with explicit examples of the dimensional reduction of the generalised metric in ExFTs. For the $\operatorname{SL}(5)$ generalised metric, we described how both the $\operatorname{SL}(3) \times \operatorname{SL}(2)$ and $\operatorname{O}(3,3)$ generalised metrics could be obtained by a Kaluza-Klein reduction by different identifications of the section. In doing so, we suggested a generalised Kaluza-Klein ansatz that both the DFT and $\operatorname{SL}(5)$ EFT generalised metrics respected and argued that both the conventional Kaluza-Klein ansatz and the ansatz of \cite{Thompson:2011uw} could be understood as a special case of this generalised Kaluza-Klein reduction for which certain components of the $Y$-tensor vanish.\par
We outlined some of the difficulties faced when trying to reduce the generalised metrics of larger EFTs. These include the appearance of more on-shell degrees of freedom (the $C_{(6)}$ entering in $E_{6(6)}$ EFT and above and the dual graviton appearing in $E_{8(8)}$ EFT) as well as the increase in the number of blocks that appear in the generalised metric. These suggest that the generalised KK ansatz may not be the whole story. \par
We then considered the reduction of the section condition for $E_{8(8)}$ and $\operatorname{SL}(5)$ EFTs. The former emphasised the fact that we require a consistent set of conventions between the parent and child theories. We showed explicitly that, amongst other things, the projector onto the adjoint does not reduce exactly (rather acquiring a $\delta \delta$ term) but that the $Y$-tensor does and gave the interpretation that the \emph{effective} weight remains unchanged between theories, even if though universal weight of the two theories differ. The reduction of the $\operatorname{SL}(5)$ section condition showed how the section conditions of smaller ExFTs can be obtained from the parent EFT as \emph{partial} solutions to the section condition. In both instances, we raised the prospect of some rather intriguing phenomena. For the $E_{8(8)} \rightarrow E_{7(7)}$ reduction, we suggested that the $(\mathbf{56},\mathbf{2})$ within the $\mathbf{248}$ could allow for the section conditions to be solved independently on different local patches in such a way that they require the residual $\operatorname{SL}(2)$ symmetry to patch together correctly. For the $\operatorname{SL}(5)$ case, we instead suggested that taking partial solutions into consideration may allow for more ExFT reductions than one may have initially expected.\par
In the final section, we described how one of the topological terms of $E_{8(8)}$ EFT is better understood as a $B$-$F$ term that, taken together with the kinetic term for the scalar sector, reproduces the kinetic and Yang-Mills term of $E_{7(7)}$ EFT (more precisely, the equations of motion that we obtain agree with the on-shell equations of motion of $E_{7(7)}$ EFT) upon employing vector-scalar duality and the twisted self-duality condition.\par
There is still plenty left to explore in terms of reductions between ExFTs. The most conspicuous omission in this work is the reduction of the full tensor hierarchy; it is a non-trivial problem to determine how the on-shell degrees of freedom need to be reshuffled into the tensor hierarchy of the lower-dimensional EFT. On a related note, the reduction of the topological terms in each theory remains to be studied. Obvious extensions to the ideas presented here would be to consider reductions of EFTs on more general spaces \cite{Malek:2016vsh} or even the Scherk-Schwarz reductions of EFTs \cite{Berman:2012uy}. It would also be interesting to construct explicit solutions using the KK type gauge fields in the reduction along the lines of \cite{Berman:2014hna}.

\part{Conclusion}
	\chapter{Conclusion}\label{ch:Conclusion}
T-duality arose out of string theory from the possibility that closed strings could wind around compact directions. Its emergence was quite unexpected; with no analogous phenomenon for particles, there was no reason to expect the compactified string spectrum to possess such an invariance and even less reason for it to be as counter-intuitive as one that requires a compact direction in spacetime. One might be forgiven for dismissing it as nothing more than a quirk of strings and yet its existence has turned out to be of profound importance within string theory. Its action on the open string finally gave credence to the view that D-branes---previously only viewed as boundary conditions to the string---should be considered as dynamical objects in their own right and the realisation that both S- and T-duality descended from a more fundamental duality of branes, going by the name U-duality, in one dimension higher acted as the main driver of the developments in the mid- to late- '90s that saw the unification of the five distinct superstring theories into a single framework called M-theory.\par
Extended field theories (ExFTs) were constructed as a formalism in which such remarkable dualities (or rather the continuous solution-generating groups) are promoted to a manifest symmetry of a theory in higher dimensions. Dual solutions are encoded as ambiguities in how one identifies the physical section in the presence of isometries. The idea itself is not new; the unification of the local symmetries of the supergravity fields into the solution-generating group acting on a higher-dimensional space followed in the footsteps of the old ideas of Theodor Kaluza and Oskar Klein. However the construction itself is tricky since Kaluza-Klein theory alone turned out to be insufficient since the unified transformations do not close on an algebra. We thus began this thesis with an introduction of how this is resolved, giving a description of the construction of ExFTs that tried to emphasise the underlying themes common to both DFT (for T-duality) and EFT for (U-duality).\par
Despite a construction that might suggest that ExFTs are a mere rewriting of supergravities, they have proven to be far more powerful than one might initially expect. We thus dedicated the bulk of this thesis to studying some of the aspects that reach beyond conventional supergravities by presenting three studies of various exotica arising in ExFTs. We began with exotic branes, of which there is now a significant body of work studying them even within ExFTs. These include the globally non-geometric T-folds and U-folds of Hull that require patching of local descriptions by duality transformations and thus do not admit a global geometric description. These led to a natural extension to objects called `locally non-geometric objects' which do not even admit local descriptions in terms of supergravity fields since they generically depend on winding or wrapping coordinates in ExFTs. A more careful analysis of such objects using the GLSM model suggests that these coordinate dependences are to be interpreted in terms of worldsheet instanton corrections \cite{Tong:2002rq,Harvey:2005ab,Kimura:2013fda,Lust:2017jox,Kimura:2018hph}.\par
In Chapter~\ref{ch:NonGeometricE7}, we constructed an explicit solution in $E_{7(7)}$ EFT that unified many of the codimension-2 exotic branes that were argued to exist within string theory in \cite{Obers:1998fb,deBoer:2010ud,deBoer:2012ma,Plauschinn:2018wbo}. We showed explicitly how the M-theory and Type IIB sections of EFT reproduced those duality-related codimension-2 objects, placing a particular emphasis on how they are related by simple rotations in the extended space. The fact that any consistent rotation of that solution led to some known background in string theory is suggestive of the fact that \emph{every} rotation from a valid solution produces another solution that, in the string theory picture, can be obtained by a judicious choice of duality transformations.\par
We thus began an algorithmic enumeration of such backgrounds, producing the complex webs outlined in Appendices~\ref{app:ExoticWeb} and \ref{app:MOrigin}. We found agreement with the literature on mixed-symmetry potentials (to which these exotic branes couple to) that were obtained from separate representation-theoretic arguments \cite{Kleinschmidt:2011vu,Bergshoeff:2016ncb,Bergshoeff:2017gpw,Lombardo:2016swq,Fernandez-Melgarejo:2018yxq,Fernandez-Melgarejo:2019mgd}. We ended our discussion of exotic branes, emboldened by multiple examples of ExFTs allowing the unification of multiple solutions \cite{Berkeley:2014nza,Berman:2014jsa,Berman:2014hna,Bakhmatov:2016kfn,Bakhmatov:2017les,Kimura:2018hph}, with the intriguing possibility that all branes (including exotic branes) could be unified into a single solution under an $E_{11}$ symmetry.\par
In the second of our studies, we considered a completely different exotic aspect of ExFTs. The study of non-Riemannian geometries---backgrounds that do not admit even a local description in terms of a Riemannian metric---was started for DFT in \cite{Lee:2013hma,Cho:2018alk,Morand:2017fnv,Cho:2019ofr} and culminated in a full classification of admissible backgrounds that satisfied the $\operatorname{O}(D,D)$ constraints. These included various atypical geometries, such as the non-relativistic Gomis-Ooguri limit of string theory or the contrasting ultra-relativistic Carroll geometry. In particular, we demonstrated in Chapter~\ref{ch:DFTNonRiemannian} that the former could be obtained from a T-duality transformation of the fundamental string along a time-like direction to yield the poorly understood negative-tension string. The power of ExFTs is that they can still be used to describe such singular geometries since, even if the spacetime metric is singular, the full generalised metric remains non-singular and well-behaved everywhere.\par
The focus of Chapter~\ref{ch:E8NonRiemannian} was an EFT analogue of the \emph{maximally non-Riemannian} solution of DFT which admits a moduli-free reduction of supergravities. In both DFT and $E_{8(8)}$ EFT, the generalised metric becomes non-dynamical and all fluctuations are projected out. We argued in detail that these particular solutions should be interpreted as a $G/G$ coset structure. Perhaps unexpectedly, the $E_{8(8)}$ maximally non-Riemannian solution landed us on a particular 3-dimensional topological theory \cite{Hohm:2018ybo} which had only previously been obtained by a truncation of the full EFT. Along with the previously mentioned backgrounds, this is exemplary of the fact that non-Riemannian backgrounds seem to have an affinity for appearing in rather unexpected places.\par
The final aspect of ExFTs that we considered was working towards filling in a hole in the literature, namely the reduction of EFTs to EFTs. In particular, we focused on the reduction of the generalised metric, coordinates, section condition and pieces of the $E_{8(8)}$ EFT action to $E_{7(7)}$ EFT. In doing so, we mostly confirmed what one might expect from such reductions. However, we still found some indications that even the simplest reductions to one EFT smaller could harbour some rather unexpected possibilities. For example, the section in $E_{8(8)}$ EFT is actually large enough to contain two copies of $E_{7(7)}$ EFT section and the $\operatorname{SL}(2)$ symmetry relating them must be broken by hand to land on $E_{7(7)}$ EFT. This raises the intriguing possibility of partially solving the $E_{8(8)}$ section conditions differently on two different patches in such a way that the two are not even related by a solution-generating $E_{7(7)}$ transformation in the reduced theory. However, whilst the solution-generating transformations (and hence duality transformations) have an origin in the local symmetries of supergravities, the supergravity interpretation of the requisite $\operatorname{SL}(2)$ transformation relating these two sections is far from clear and remains to be understood.\par
As we have hopefully demonstrated, our understanding of ExFTs is far from complete. Despite their constructions being intimately linked with the conventional supergravity lore they have demonstrated rather unexpected ties with much more recent developments in string and M-theory. With the underlying theme of ExFTs being unification, it remains to be seen what other strands they can pull together. We hope that studying such exotica (both inside and outside of ExFTs) will stimulate more developments in these areas. Indeed, having demonstrated the considerable power of ExFTs, perhaps it is time to turn the tables and start searching for predictions of ExFTs within string- and M-theory.

\part*{Appendices}
\appendix
\addcontentsline{toc}{part}{Appendices}
	\chapter{The Generalised Field strength in the Non-geometric Solution}\label{app:ExoticFieldstrength}
Here we show, by explicit computation, that the field strength of the generalised gauge field, defined by the ansatz \eqref{eq:AAnsatz} together with $\mathcal{B}_{\mu \nu, \bullet} = 0$, in the non-geometric solution of $E_{7(7)}\times \mathbb{R}^+$ EFT satisfies the twisted self-duality constraint \eqref{eq:TwistedSelfDuality} and the generalised Bianchi identity \eqref{eq:Bianchi}.
\section{Twisted Self-Duality}
Applying the simplifications $\mathcal{B}_{\mu \nu, \bullet}= 0$ and demanding that the generalised vectors do not depend on the internal coordinates ($\partial_N \mathcal{A}_\mu{}^M = 0$) the covariantised generalised field strength reduces to the Abelian field strength
\begin{align}
\mathcal{F}_{\mu \nu}{}^M \rightarrow F_{\mu \nu}{}^M = 2 \partial_{[\mu} \mathcal{A}_{\nu]}{}^M\,.
\end{align}
Recall that the external coordinates are taken to be $x^\mu = (t,r, \theta, z)$. For the ansatz given above, the only non-vanishing components of the field strength in the $5^3$ frame are
\begin{align}
\mathcal{F}_{\mu \nu}{}^M & = ( \mathcal{F}_{tr}{}^{\xi \chi}, \mathcal{F}_{t \theta}{}^{\xi \chi}, \mathcal{F}_{rz, \xi \chi}, \mathcal{F}_{\theta z, \xi \chi})\\
	& = \left( \frac{\sigma(\sigma^2 \theta^2 - H^2)}{rH^2} , \frac{2 \sigma^2 \theta}{H}, - \frac{2 \sigma^2 \theta H}{r K^2}, \frac{\sigma ( \sigma^2 \theta^2 - H^2)}{K^2}\right)\,.
\end{align}
We begin by considering $\mathcal{F}_{tr}{}^{\xi \chi}$ component:
\begin{align}
\mathcal{F}_{tr}{}^{\xi \chi} & = - {\left| g_{(4)}\right|}^{\frac{1}{2}} \varepsilon_{tr \theta z} g^{\theta \theta} g^{zz} \Omega^{\xi \chi}{}_{\xi \chi} \mathcal{M}^{\xi \chi, \xi \chi} \mathcal{F}_{\theta z}{}^{\xi \chi}\,.
\end{align}
Substituting $\Omega^{\xi \chi}{}_{\xi \chi} = - e^{\Delta} \delta^{\xi \chi}_{\xi \chi} = - {|g_{(4)}|}^{\frac{1}{4}} \delta^{\xi \chi}_{\xi \chi}$ and $g^{\theta \theta},g^{zz}$ (read off from \eqref{eq:Ext}), we obtain
\begin{align}
\mathcal{F}_{tr}{}^{\xi \chi} & = {\left| g_{(4)}\right|}^{\frac{3}{4}} \varepsilon_{tr \theta z} \frac{\sigma ( \sigma^2 \theta^2 - H^2)}{r^2HK^2} \mathcal{M}^{\xi \chi, \xi \chi} \,.
\end{align}
Finally, using $\mathcal{M}^{\xi \chi, \xi \chi} = {|g_{(4)}|}^{-\frac{1}{4}} {(HK^{-1})}^{\frac{3}{2}}$ (read off from \eqref{eq:53GenMetric}) and $|g_{(4)}| = r^2 HK$, we obtain
\begin{align}
\mathcal{F}_{tr}{}^{\xi \chi} & = {(r^2 HK)}^{\frac{1}{2}} \frac{\sigma ( \sigma^2 \theta^2 - H^2)}{r^2HK^2} {(HK^{-1})}^{-\frac{3}{2}}\\
	& = \frac{\sigma(\sigma^2 \theta^2 - H^2)}{rH^2}
\end{align}
which thus satisfies the self-duality relation. Likewise, the remaining relations all follow upon using the fact that the scaling factors $e^{-\Delta} = {|g_{(4)}|}^{-\frac{1}{4}}$ cancel such that $\Omega^{MN} \mathcal{M}_{NK} = {\tilde{\Omega}}^{MN} {\tilde{\mathcal{M}}}_{NK}$. Explicitly, we have
\begin{align}
\mathcal{F}_{\theta z, \xi \chi} & = - \frac{{|g_{(4)}|}^{\frac{1}{2}}}{2} \times 2 \epsilon_{\theta z t r} g^{tt} g^{rr} {\tilde{\Omega}}_{\xi \chi}{}^{\xi \chi} {\tilde{\mathcal{M}}}^{\xi \chi, \xi \chi} \mathcal{F}_{tr}{}^{\xi \chi}\\
	& = \frac{\sigma ( \sigma^2 \theta^2 - H^2)}{K^2}\,,\\
\mathcal{F}_{t \theta}{}^{\xi \chi} & = - \frac{{|g_{(4)}|}^{\frac{1}{2}}}{2} \times 2 \epsilon_{t \theta r z} g^{rr} g^{zz} {\tilde{\Omega}}^{\xi \chi}{}_{\xi \chi} {\tilde{\mathcal{M}}}_{\xi \chi, \xi \chi} \mathcal{F}_{tr,\xi \chi}\\
	& = \frac{2 \sigma^2 \theta}{H}\,,\\
\mathcal{F}_{rz, \xi \chi} & = - \frac{{|g_{(4)}|}^{\frac{1}{2}}}{2} \times 2 \epsilon_{r z t \theta} g^{tt} g^{\theta \theta} {\tilde{\Omega}}_{\xi \chi}{}^{\xi \chi} {\tilde{\mathcal{M}}}_{\xi \chi, \xi \chi} \mathcal{F}_{t\theta}{}^{\xi \chi}\\
	& = - \frac{2 \sigma^2 \theta H}{K^2 r}\,.
\end{align}
\section{Bianchi Identity}\label{sec:Bianchi}
For the solution discussed here, the Bianchi identity \eqref{eq:Bianchi} reduces to
\begin{align}
\mathbb{L}_{\mathcal{A}_{[\mu}} \mathcal{F}_{\nu \rho]}{}^\mu & = 0
\end{align}
but each term vanishes independently since $\partial_M \mathcal{A}_\nu{}^N = 0$ and so the Bianchi identity is satisfied trivially.

	\chapter{Exotic Backgrounds}\label{app:ExoticBackgrounds}
In this appendix, we have compiled a list of the backgrounds of the exotic branes that appeared in the non-geometric solution of Section \ref{sec:NonGeomSoln}. The familiar solutions (fundamental, Dirichlet and solitonic branes) can be found in most standard textbooks (see, for example \cite{Ortin:2015hya}) and also, closer to the context described in this thesis, in \cite{Rudolph:2016sxe}. For the Type II branes we have adopted the Einstein frame, which is related to the string frame via $\textrm{d} s^2_{\textrm{s}} = e^{\frac{\phi}{2}} \textrm{d} s^2_{\text{E}}$ where $\phi$ is the dilaton, and indicated this with a subscript $E$. These were obtained through a sequence of T- and S-duality transformations, starting from a smeared NS5-brane. As such, all harmonic functions here $H$ are harmonic in $(r,\theta)$ only, with
\begin{align}
H(r) = h_0 + \sigma \ln \frac{\mu}{r}\,.
\end{align}
Here $h_0$ is a diverging bare quantity, $\mu$ a renormalisation scale and $\sigma$ a dimensionless constant which is irrelevant for the discussion here. We also define $K \coloneqq H^2 + \sigma^2 \theta^2$.
\section{Codimension-2 Exotic M-Theory Branes}
\subsection{\texorpdfstring{$5^3$}{53}-brane}
\begin{gather}
\begin{gathered}
\textrm{d} s^2 = {(HK^{-1})}^{-\frac{1}{3}} ( - \textrm{d} t^2 + \textrm{d} {\vec{x}}^2_{(5)} ) + {(HK^{-1})}^{\frac{2}{3}} \textrm{d} {\vec{y}}^2_{(3)} +  H^{\frac{2}{3}} K^{\frac{1}{3}} (\textrm{d} r^2 + r^2 \textrm{d} \theta^2)\\
A_{(3)} = - K^{-1} \sigma \theta \textrm{d} y^1 \wedge \textrm{d} y^2 \wedge \textrm{d} y^3, \qquad A_{(6)} = - H^{-1} K \textrm{d} t \wedge \textrm{d} x^1 \wedge \ldots \wedge \textrm{d} x^5
\end{gathered}
\end{gather}
\subsection{\texorpdfstring{$2^6$}{26}-brane}
\begin{gather}
\begin{gathered}
\textrm{d} s^2 = {(HK^{-1})}^{-\frac{2}{3}} ( - \textrm{d} t^2 + \textrm{d} {\vec{x}}^2_{(2)} ) + {(HK^{-1})}^{\frac{1}{3}} \textrm{d} {\vec{y}}^2_{(6)} +  H^{\frac{1}{3}} K^{\frac{2}{3}} (\textrm{d} r^2 + r^2 \textrm{d} \theta^2)\\
A_{(3)} = - H^{-1} K \textrm{d} t \wedge \textrm{d} x^1 \wedge \textrm{d} x^2, \qquad A_{(6)} = - K^{-1} \sigma \theta \textrm{d} y^1 \wedge \ldots \wedge \textrm{d} y^6
\end{gathered}
\end{gather}
\subsection{\texorpdfstring{$0^{(1,7)}$}{0(1,7)}-brane}
\begin{gather}
\begin{gathered}
\textrm{d} s^2 = - H^{-1} K \textrm{d} t^2 + \textrm{d} {\vec{x}}^2_{(7)} + HK^{-1} {( \textrm{d} z - H^{-1}K \textrm{d} t)}^2 + K (\textrm{d} r^2 + r^2 \textrm{d} \theta^2)
\end{gathered}
\end{gather}
\section{Codimension-2 Exotic Type II Branes}
Both Type IIA and IIB possess $5_2^2$-, $1_4^6$- and $0_4^{(1,6)}$-branes. The Type IIA theory further admits $p_3^{7-p}$-branes for $p = 1, 3, 5,7$ whilst Type IIB instead admits $p_3^{7-p}$-branes for $p = 0,2,4,6$. Note that the $7_3$-brane in Type IIB is better described, together with the D7-brane, as part of the $(p,q)$ 7-brane of F-theory.
\subsection{\texorpdfstring{$5^2_2$}{522}-brane}
\begin{gather}
\begin{gathered}
\textrm{d} s^2_{\text{E}} = {(HK^{-1})}^{-\frac{1}{4}} \left( - \textrm{d} t^2 + \textrm{d} {\vec{x}}^2_{(5)} \right) + {(HK^{-1})}^{\frac{3}{4}} \textrm{d} {\vec{y}}^2_{(2)} + H^{\frac{3}{4}} K^{\frac{1}{4}} (\textrm{d} r^2 + r^2 \textrm{d} \theta^2)\\
B_{(2)} = - K^{-1} \sigma \theta \textrm{d} y^1 \wedge \textrm{d} y^2, \qquad B_{(6)} = - H^{-1} K \textrm{d} t \wedge \textrm{d} x^1 \wedge \ldots \wedge \textrm{d} x^5\\
e^{2(\phi - \phi_0)} = HK^{-1}
\end{gathered}
\end{gather}
\subsection{\texorpdfstring{$p_3^{7-p}$}{p37-p}-brane}
\begin{gather}
\textrm{d} s^2_{\text{E}} = {(HK^{-1})}^{\frac{p-7}{8}} \left( - \textrm{d} t^2 + \textrm{d} {\vec{x}}^2_{(p)} \right) + {(HK^{-1})}^{\frac{p+1}{8}} \textrm{d} {\vec{y}}^2_{(7-p)} + H^{\frac{p+1}{8}} K^{\frac{7-p}{8}} \left( \textrm{d} r^2 + r^2 \textrm{d} \theta^2\right)\nonumber\\
C_{(7-p)} = - K^{-1} \sigma \theta \textrm{d} y^1 \wedge \ldots \wedge \textrm{d} y^{7-p},\qquad  C_{(p+1)} = - H^{-1} K \textrm{d} t \wedge \textrm{d} x^1 \wedge \ldots \wedge \textrm{d} x^p\nonumber\\
e^{2(\phi- \phi_0)} = {(HK^{-1})}^{-\frac{p-3}{2}}
\end{gather}
\subsection{\texorpdfstring{$1_4^6$}{146}-brane}
\begin{gather}
\begin{gathered}
\textrm{d} s^2_{\text{E}} = {(HK^{-1})}^{-\frac{3}{4}} \left( - \textrm{d} t^2 + \textrm{d} x^2 \right) + {(HK^{-1})}^{\frac{1}{4}} \textrm{d} {\vec{y}}^2_{(6)} + H^{\frac{1}{4}} K^{\frac{3}{4}} \left( \textrm{d} r^2 + r^2 \textrm{d} \theta^2\right)\\
B_{(6)} = - K^{-1} \sigma \theta \textrm{d} y^1 \wedge \ldots \wedge \textrm{d} y^{6},\qquad  B_{(2)} = - H^{-1} K \textrm{d} t \wedge \textrm{d} x^1 \wedge\textrm{d} x^2\\ 
e^{2(\phi- \phi_0)} = H^{-1} K
\end{gathered}
\end{gather}
\subsection{\texorpdfstring{$0_4^{(1,6)}$}{04(1,6)}-brane}
\begin{gather}
\begin{gathered}
\textrm{d} s^2_{\text{E}} = - {(HK^{-1})}^{-1} \textrm{d} t^2 + HK^{-1} {(\textrm{d} z - H^{-1} K \textrm{d} t)}^2 + \textrm{d} {\vec{x}}^2_{(6)} + K (\textrm{d} r^2 + r^2 \textrm{d} \theta^2)\\
e^{2(\phi - \phi_0)} = 1
\end{gathered}
\end{gather}
Note that the $(1_4^6, 1_3^6)$ and $(5_3^2, 5_2^2)$ each form S-duality doublets and thus share the same metric in the Einstein frame, exchanging only $B_{(p)} \xleftrightarrow{\text{S}} C_{(p)}$ and inverting the dilaton.

	\begingroup
\setlength{\parindent}{0pt}
\newcommand{\forceindent}{\leavevmode{\parindent=1em\indent}}
\chapter{Map of Exotic Branes}\label{app:ExoticWeb}
In this Appendix, we list the duality web of branes whose tension scale down to $g_s^{-7}$. The first three diagrams consist mostly of the standard branes but the codimension-1 and -0 extensions of the 5-brane chain at $g_s^{-2}$ may be less familiar to the reader. These diagrams are believed to be complete down to $g_s^{-7}$ in the sense that they were generated by an exhaustive search, subject to the constraints outlined in the main text.
\section{\texorpdfstring{$g_s^0$}{gs0} Duality Orbits}
\begin{figure}[H]
\centering
\begin{tikzpicture}
\matrix(M)[matrix of math nodes, row sep=2em, column sep=2 em, minimum width=2em]{
\clap{\text{A}} & \text{P}/{}^10_0 & \text{F1}/1_0\\
\clap{\text{B}} & \text{P}/{}^10_0 & \text{F1}/1_0\\
};
\draw[latex-latex] (M-1-2) -- (M-2-2);
\draw[latex-latex] (M-1-3) -- (M-2-3);
\draw[latex-latex] (M-1-2) -- (M-2-3);
\draw[latex-latex] (M-1-3) -- (M-2-2);
\end{tikzpicture}
\caption{The T-duality orbit of the $\text{F1}=1_0$.}
\label{fig:10Orbit}
\end{figure}
\begin{multicols}{2}
\forceindent S-dualities:
\begin{itemize}
	\item $\text{P}={}^10_0 \leftrightarrow \text{P}={}^10_0$
	\item $\text{F1}=1_0 \leftrightarrow \text{D1}=1_1$ See Fig. \ref{fig:11Orbit}
\end{itemize}
\columnbreak
\forceindent M-theory origins:
\begin{itemize}
	\item $\text{P}={}^10_0 \rightarrow \text{WM} = 0$
	\item $\text{F1}=1_0 \rightarrow \text{M2} = 2$
\end{itemize}
\end{multicols}
\vspace{-0.5cm}
Note that the massless WM must be treated separately from the remaining branes; one instead uses $P^2=0$ such that the masses of the PA and D0 are obtained from the radius of the ${11}^{\text{th}}$ direction. 
\begin{landscape}
\section{\texorpdfstring{$g_s^{-1}$}{gs-1} Duality Orbits}
\begin{figure}[H]
\centering
\begin{tikzpicture}
\matrix(M)[matrix of math nodes, row sep=3em, column sep=2em, minimum width=2em]{
\clap{\text{A}} & \text{D0}/0_1 & & \text{D2}/2_1 & & \text{D4}/4_1 && \text{D6}/6_1 && \text{D8}/8_1\\
\clap{\text{B}} & & \text{D1}/1_1 & & \text{D3}/3_1 & & \text{D5}/5_1 && \text{D7}/7_1 && \text{D9}/9_1\\
};
\draw[latex-latex] (M-1-2) -- (M-2-3);
\draw[latex-latex] (M-2-3) -- (M-1-4);
\draw[latex-latex] (M-1-4) -- (M-2-5);
\draw[latex-latex] (M-2-5) -- (M-1-6);
\draw[latex-latex] (M-1-6) -- (M-2-7);
\draw[latex-latex] (M-2-7) -- (M-1-8);
\draw[latex-latex] (M-1-8) -- (M-2-9);
\draw[latex-latex] (M-2-9) -- (M-1-10);
\draw[latex-latex] (M-1-10) -- (M-2-11);
\end{tikzpicture}
\caption{The T-duality orbit of the $\text{D1}=1_1$.}
\label{fig:11Orbit}
\end{figure}
\begin{multicols}{2}
\forceindent S-dualities:
\begin{itemize}
	\item $\text{D1}=1_1 \leftrightarrow \text{F1}=1_0$ See Fig. \ref{fig:10Orbit}
	\item $\text{D3}=3_1 \leftrightarrow \text{D3}=3_1$ Self-dual
	\item $\text{D5}=5_1 \leftrightarrow \text{NS5}=5_2$ See Fig. \ref{fig:52Orbit}
	\item $\text{D7}=7_1 \leftrightarrow \text{NS7}=7_3$ See Fig. \ref{fig:532Orbit}
	\item $\text{D9}=9_1 \leftrightarrow 9_4$ See Fig. \ref{fig:7420Orbit}
\end{itemize}
\columnbreak
\forceindent M-theory origins:
\begin{itemize}
	\item $\text{D0}=0_1 \rightarrow \text{WM}=0$
	\item $\text{D2}=2_1 \rightarrow \text{M2}=2$
	\item $\text{D4}=4_1 \rightarrow \text{M5}=5$
	\item $\text{D6}=6_1 \rightarrow \text{KK6M}=6^1$
	\item $\text{D8}=8_1 \rightarrow \text{KK8M}=8^{(1,0)}$
\end{itemize}
\end{multicols}
\clearpage
\end{landscape}
\section{\texorpdfstring{$g_s^{-2}$}{gs-2} Duality Orbits}
\begin{figure}[H]
\centering
\begin{tikzpicture}
\matrix(M)[matrix of math nodes, row sep=3em, column sep=1 em, minimum width=4em]{
\clap{\text{A}} & \text{NS5}/5_2 & \text{KK5A}/5_2^1 & 5_2^2 & 5_2^3 & 5_2^4\\
\clap{\text{B}} & \text{NS5}/5_2 & \text{KK5A}/5_2^1 & 5_2^2 & 5_2^3 & 5_2^4\\
};
\draw[latex-latex] (M-1-2) -- (M-2-3);
\draw[latex-latex] (M-1-2) -- (M-2-2);
\draw[latex-latex] (M-2-3) -- (M-1-4);
\draw[latex-latex] (M-2-3) -- (M-1-3);
\draw[latex-latex] (M-1-4) -- (M-2-5);
\draw[latex-latex] (M-1-4) -- (M-2-4);
\draw[latex-latex] (M-2-5) -- (M-1-6);
\draw[latex-latex] (M-2-5) -- (M-1-5);
\draw[latex-latex] (M-2-6) -- (M-1-6);
\draw[latex-latex] (M-2-2) -- (M-1-3);
\draw[latex-latex] (M-1-3) -- (M-2-4);
\draw[latex-latex] (M-2-4) -- (M-1-5);
\draw[latex-latex] (M-1-5) -- (M-2-6);
\end{tikzpicture}
\caption{The T-duality orbit of the $5_2^3$.}
\label{fig:52Orbit}
\end{figure}
\begin{multicols}{2}
\forceindent S-dualities:
\begin{itemize}
	\item $\text{NS5}/5_2 \leftrightarrow \text{D5}/5_1$ See Fig. \ref{fig:11Orbit}
	\item $\text{KK5A}/5_2^1 \leftrightarrow \text{KK5A}/5_2^1$ Self-dual
	\item $5_2^2 \leftrightarrow 5_3^2$ See Fig. \ref{fig:532Orbit}
	\item $5_2^3 \leftrightarrow 5_4^3$ See Fig. \ref{fig:4413Orbit}
	\item $5_2^4 \leftrightarrow 5_5^4$ See Fig. \ref{fig:5522Orbit}
\end{itemize}
\columnbreak
\forceindent M-theory origins:
\begin{itemize}
	\item $\text{NS5}/5_2 \rightarrow \text{M5}/5$
	\item $\text{KK5A}/5_2^1 \rightarrow \text{KK6M}/6^1$
	\item $5_2^2 \rightarrow 5^3$
	\item $5_2^3 \rightarrow 5^{(1,3)}$
	\item $5_2^4 \rightarrow 5^{(1,0,4)}$
\end{itemize}
\end{multicols}
\clearpage
\section{\texorpdfstring{$g_s^{-3}$}{gs-3} Duality Orbits}
\begin{figure}[H]
\centering
\begin{tikzpicture}
\matrix(M)[matrix of math nodes, row sep=2.1em, column sep=1.8em, minimum width=2em]{
& & \mathclap{0} & \mathclap{1} & \mathclap{2} & \mathclap{3} & \mathclap{4} & \mathclap{5} & \mathclap{6} & \mathclap{7}\\
\clap{\text{codim-2}} & \clap{\text{A}} & \mathclap{0_3^7} & & \mathclap{2_3^5} & & \mathclap{4_3^3} & & \mathclap{6_3^1} &\\
\clap{\text{codim-2}} & \clap{\text{B}} & & \mathclap{1_3^6} & & \mathclap{3_3^4} & & \mathclap{5_3^2} & & \mathclap{7_3}\\
\clap{\text{codim-1}} & \clap{\text{A}} & & \mathclap{1_3^{(1,6)}} & & \mathclap{3_3^{(1,4)}} & & \mathclap{5_3^{(1,2)}} & & \mathclap{7_3^{(1,0)}}\\
\clap{\text{codim-1}} & \clap{\text{B}} & \mathclap{0_3^{(1,7)}} & & \mathclap{2_3^{(1,5)}} & & \mathclap{4_3^{(1,3)}} & & \mathclap{6_3^{(1,1)}} &\\
\clap{\text{codim-0}} & \clap{\text{A}} & \mathclap{0_3^{(2,7)}} & & \mathclap{2_3^{(2,5)}} & & \mathclap{4_3^{(2,3)}} & & \mathclap{6_3^{(2,1)}} &\\
\clap{\text{codim-0}} & \clap{\text{B}} & & \mathclap{1_3^{(2,6)}} & & \mathclap{3_3^{(2,4)}} & & \mathclap{5_3^{(2,2)}} & & \mathclap{7_3^{(2,0)}}\\
};
\draw[latex-latex] (M-2-3) -- (M-3-4);
\draw[latex-latex] (M-3-4) -- (M-2-5);
\draw[latex-latex] (M-2-5) -- (M-3-6);
\draw[latex-latex] (M-3-6) -- (M-2-7);
\draw[latex-latex] (M-2-7) -- (M-3-8);
\draw[latex-latex] (M-3-8) -- (M-2-9);
\draw[latex-latex] (M-2-9) -- (M-3-10);
\draw[latex-latex] (M-5-3) -- (M-4-4);
\draw[latex-latex] (M-4-4) -- (M-5-5);
\draw[latex-latex] (M-5-5) -- (M-4-6);
\draw[latex-latex] (M-4-6) -- (M-5-7);
\draw[latex-latex] (M-5-7) -- (M-4-8);
\draw[latex-latex] (M-4-8) -- (M-5-9);
\draw[latex-latex] (M-5-9) -- (M-4-10);
\draw[latex-latex] (M-6-3) -- (M-7-4);
\draw[latex-latex] (M-7-4) -- (M-6-5);
\draw[latex-latex] (M-6-5) -- (M-7-6);
\draw[latex-latex] (M-7-6) -- (M-6-7);
\draw[latex-latex] (M-6-7) -- (M-7-8);
\draw[latex-latex] (M-7-8) -- (M-6-9);
\draw[latex-latex] (M-6-9) -- (M-7-10);
\draw[latex-latex] (M-2-3) -- (M-5-3);
\draw[latex-latex] (M-5-3) -- (M-6-3);
\draw[latex-latex] (M-2-5) -- (M-5-5);
\draw[latex-latex] (M-5-5) -- (M-6-5);
\draw[latex-latex] (M-2-7) -- (M-5-7);
\draw[latex-latex] (M-5-7) -- (M-6-7);
\draw[latex-latex] (M-2-9) -- (M-5-9);
\draw[latex-latex] (M-5-9) -- (M-6-9);
\draw[latex-latex] (M-3-4) -- (M-4-4);
\draw[latex-latex] (M-4-4) -- (M-7-4);
\draw[latex-latex] (M-3-6) -- (M-4-6);
\draw[latex-latex] (M-4-6) -- (M-7-6);
\draw[latex-latex] (M-3-8) -- (M-4-8);
\draw[latex-latex] (M-4-8) -- (M-7-8);
\draw[latex-latex] (M-3-10) -- (M-4-10);
\draw[latex-latex] (M-4-10) -- (M-7-10);
\end{tikzpicture}
\caption{The T-duality orbit of the $5_3^2$.}
\label{fig:532Orbit}
\end{figure}
\begin{multicols}{2}
\forceindent S-dualities:
\begin{itemize}
 	\item $1_3^6 \leftrightarrow 1_4^6$ See Fig. \ref{fig:146Orbit}
	\item $3_3^4 \leftrightarrow 3_3^4$ Self-dual
	\item $5_3^2 \leftrightarrow 5_2^2$ See Fig. \ref{fig:52Orbit}
	\item $7_3 \leftrightarrow 7_1$ See Fig. \ref{fig:11Orbit}
	\item $0_3^{(1,7)} \leftrightarrow 0_6^{(1,7)}$ See Fig. \ref{fig:0617Orbit}
	\item $2_3^{(1,5)} \leftrightarrow 2_5^{(1,5)}$ See Fig. \ref{fig:2515Orbit}
	\item $4_3^{(1,3)} \leftrightarrow 4_4^{(1,3)}$ See Fig. \ref{fig:4413Orbit}
	\item $6_3^{(1,1)} \leftrightarrow 6_3^{(1,1)}$ Self-dual
	\item $1_3^{(2,6)} \leftrightarrow 1_7^{(2,6)}$ See Fig. \ref{fig:1726Orbit}
	\item $3_3^{(2,4)} \leftrightarrow 3_6^{(2,4)}$ See Fig. \ref{fig:3624Orbit}
	\item $5_3^{(2,2)} \leftrightarrow 5_5^{(2,2)}$ See Fig. \ref{fig:5522Orbit}
	\item $7_3^{(2,0)} \leftrightarrow 7_4^{(2,0)}$ See Fig. \ref{fig:7420Orbit}
\end{itemize}
\columnbreak
\forceindent M-theory origins:
\begin{itemize}
	\item $0_3^7 \rightarrow 0^{(1,7)}$
	\item $2_3^5 \rightarrow 2^6$
	\item $4_3^3 \rightarrow 5^3$
	\item $6_3^1 \rightarrow 6^1=\text{KK6M}$
	\item $1_3^{(1,6)} \rightarrow 1^{(1,1,6)}$
	\item $3_3^{(1,4)} \rightarrow 3^{(2,4)}$
	\item $5_3^{(1,2)} \rightarrow 5^{(1,3)}$
	\item $7_3^{(1,0)}=\text{KK7A} \rightarrow 8^{(1,0)}=\text{KK8M}$
	\item $0_3^{(2,7)} \rightarrow 0^{(1,0,0,2,7)}$
	\item $2_3^{(2,5)} \rightarrow 2^{(1,0,2,5)}$
	\item $4_3^{(2,3)} \rightarrow 4^{(1,2,3)}$
	\item $6_3^{(2,1)} \rightarrow 6^{(3,1)}$
\end{itemize}
\end{multicols}
\clearpage
\section{\texorpdfstring{$g_s^{-4}$}{gs-4} Duality Orbits}
\begin{figure}[H]
\centering
\begin{tikzpicture}
\matrix(M)[matrix of math nodes, row sep=3em, column sep=4 em, minimum width=2em]{
& & \mathclap{0} & \mathclap{1}\\
\clap{\text{codim-2}} & \clap{\text{A}} & \mathclap{0_4^{(1,6)}} & \mathclap{1_4^6}\\
\clap{\text{codim-2}} & \clap{\text{B}} & \mathclap{0_4^{(1,6)}} & \mathclap{1_4^6}\\
\clap{\text{codim-1}} & \clap{\text{A}} & \mathclap{0_4^{(1,1,6)}} & \mathclap{1_4^{(1,0,6)}}\\
\clap{\text{codim-1}} & \clap{\text{B}} & \mathclap{0_4^{(1,1,6)}} & \mathclap{1_4^{(1,0,6)}}\\
\clap{\text{codim-0}} & \clap{\text{A}} & \mathclap{0_4^{(2,1,6)}} & \mathclap{1_4^{(2,0,6)}}\\
\clap{\text{codim-0}} & \clap{\text{B}} & \mathclap{0_4^{(2,1,6)}} & \mathclap{1_4^{(2,0,6)}}\\
};
\draw[latex-latex] (M-2-3) -- (M-3-3);
\draw[latex-latex] (M-3-3) -- (M-4-3);
\draw[latex-latex] (M-4-3) -- (M-5-3);
\draw[latex-latex] (M-5-3) -- (M-6-3);
\draw[latex-latex] (M-6-3) -- (M-7-3);
\draw[latex-latex] (M-2-4) -- (M-3-4);
\draw[latex-latex] (M-3-4) -- (M-4-4);
\draw[latex-latex] (M-4-4) -- (M-5-4);
\draw[latex-latex] (M-5-4) -- (M-6-4);
\draw[latex-latex] (M-6-4) -- (M-7-4);
\draw[latex-latex] (M-2-3) to[out=225, in=135] (M-5-3);
\draw[latex-latex] (M-4-3) to[out=225, in=135] (M-7-3);
\draw[latex-latex] (M-2-4) to[out=315, in=45] (M-5-4);
\draw[latex-latex] (M-4-4) to[out=315, in=45] (M-7-4);
\draw[latex-latex] (M-2-3) -- (M-3-4);
\draw[latex-latex] (M-3-3) -- (M-2-4);
\draw[latex-latex] (M-4-3) -- (M-5-4);
\draw[latex-latex] (M-5-3) -- (M-4-4);
\draw[latex-latex] (M-6-3) -- (M-7-4);
\draw[latex-latex] (M-7-3) -- (M-6-4);
\end{tikzpicture}
\caption{The T-duality orbit of the $1_4^6$.}
\label{fig:146Orbit}
\end{figure}
\begin{multicols}{2}
\forceindent S-dualties:
\begin{itemize}
	\item $0_4^{(1,6)} \leftrightarrow 0_4^{(1,6)}$ Self-dual
	\item $0_4^{(1,1,6)} \leftrightarrow 0_6^{(1,1,6)}$ See Fig. \ref{fig:0617Orbit}
	\item $0_4^{(2,1,6)} \leftrightarrow 0_8^{(2,1,6)}$
	\item $1_4^6 \leftrightarrow 1_3^6$ See Fig. \ref{fig:532Orbit}
	\item $1_4^{(1,0,6)} \leftrightarrow 1_5^{(1,0,6)}$ See Fig. \ref{fig:2515Orbit}
	\item $1_4^{(2,0,6)} \leftrightarrow 1_7^{(2,0,6)}$ See Fig. \ref{fig:1726Orbit}
\end{itemize}
\columnbreak
\forceindent M-theory origins:
\begin{itemize}
	\item $0_4^{(1,6)} \rightarrow 0^{(1,7)}$
	\item $0_4^{(1,1,6)} \rightarrow 0^{(2,1,6)}$
	\item $0_4^{(2,1,6)} \rightarrow 0^{(1,0,2,1,6)}$
	\item $1_4^6 \rightarrow 2^6$
	\item $1_4^{(1,0,6)} \rightarrow 1^{(1,1,6)}$
	\item $1_4^{(2,0,6)} \rightarrow 1^{(1,2,0,6)}$
\end{itemize}
\end{multicols}
\clearpage
\begin{landscape}

\begin{figure}[H]
\centering
\begin{tikzpicture}
\matrix(M)[matrix of math nodes, row sep=2em, column sep=4 em, minimum width=2em]{
& & \mathclap{0} & \mathclap{1} & \mathclap{2} & \mathclap{3} & \mathclap{4} & \mathclap{5}\\
\clap{\text{codim-1}} & \clap{\text{A}} & \mathclap{0_4^{(5,3)}} & \mathclap{1_4^{(4,3)}} & \mathclap{2_4^{(3,3)}} & \mathclap{3_4^{(2,3)}} & \mathclap{4_4^{(1,3)}} & \mathclap{5_4^3}\\
\clap{\text{codim-1}} & \clap{\text{B}} & \mathclap{0_4^{(5,3)}} & \mathclap{1_4^{(4,3)}} & \mathclap{2_4^{(3,3)}} & \mathclap{3_4^{(2,3)}} & \mathclap{4_4^{(1,3)}} & \mathclap{5_4^3}\\
\clap{\text{codim-0}} & \clap{\text{A}} & \mathclap{0_4^{(1,5,3)}} & \mathclap{1_4^{(1,4,3)}} & \mathclap{2_4^{(1,3,3)}} & \mathclap{3_4^{(1,2,3)}} & \mathclap{4_4^{(1,1,3)}} & \mathclap{5_4^{(1,0,3)}}\\
\clap{\text{codim-0}} & \clap{\text{B}} & \mathclap{0_4^{(1,5,3)}} & \mathclap{1_4^{(1,4,3)}} & \mathclap{2_4^{(1,3,3)}} & \mathclap{3_4^{(1,2,3)}} & \mathclap{4_4^{(1,1,3)}} & \mathclap{5_4^{(1,0,3)}}\\
};
\draw[latex-latex] (M-2-3) -- (M-3-3);
\draw[latex-latex] (M-3-3) -- (M-4-3);
\draw[latex-latex] (M-4-3) -- (M-5-3);
\draw[latex-latex] (M-2-3) to[out=225, in=135] (M-5-3);
\draw[latex-latex] (M-2-4) -- (M-3-4);
\draw[latex-latex] (M-3-4) -- (M-4-4);
\draw[latex-latex] (M-4-4) -- (M-5-4);
\draw[latex-latex] (M-2-4) to[out=225, in=135] (M-5-4);
\draw[latex-latex] (M-2-5) -- (M-3-5);
\draw[latex-latex] (M-3-5) -- (M-4-5);
\draw[latex-latex] (M-4-5) -- (M-5-5);
\draw[latex-latex] (M-2-5) to[out=225, in=135] (M-5-5);
\draw[latex-latex] (M-2-6) -- (M-3-6);
\draw[latex-latex] (M-3-6) -- (M-4-6);
\draw[latex-latex] (M-4-6) -- (M-5-6);
\draw[latex-latex] (M-2-6) to[out=315, in=45] (M-5-6);
\draw[latex-latex] (M-2-7) -- (M-3-7);
\draw[latex-latex] (M-3-7) -- (M-4-7);
\draw[latex-latex] (M-4-7) -- (M-5-7);
\draw[latex-latex] (M-2-7) to[out=315, in=45] (M-5-7);
\draw[latex-latex] (M-2-8) -- (M-3-8);
\draw[latex-latex] (M-3-8) -- (M-4-8);
\draw[latex-latex] (M-4-8) -- (M-5-8);
\draw[latex-latex] (M-2-8) to[out=315, in=45] (M-5-8);
\draw[latex-latex] (M-2-3) -- (M-3-4);
\draw[latex-latex] (M-3-3) -- (M-2-4);
\draw[latex-latex] (M-2-4) -- (M-3-5);
\draw[latex-latex] (M-3-4) -- (M-2-5);
\draw[latex-latex] (M-2-5) -- (M-3-6);
\draw[latex-latex] (M-3-5) -- (M-2-6);
\draw[latex-latex] (M-2-6) -- (M-3-7);
\draw[latex-latex] (M-3-6) -- (M-2-7);
\draw[latex-latex] (M-2-7) -- (M-3-8);
\draw[latex-latex] (M-3-7) -- (M-2-8);
\draw[latex-latex] (M-4-3) -- (M-5-4);
\draw[latex-latex] (M-5-3) -- (M-4-4);
\draw[latex-latex] (M-4-4) -- (M-5-5);
\draw[latex-latex] (M-5-4) -- (M-4-5);
\draw[latex-latex] (M-4-5) -- (M-5-6);
\draw[latex-latex] (M-5-5) -- (M-4-6);
\draw[latex-latex] (M-4-6) -- (M-5-7);
\draw[latex-latex] (M-5-6) -- (M-4-7);
\draw[latex-latex] (M-4-7) -- (M-5-8);
\draw[latex-latex] (M-5-7) -- (M-4-8);
\end{tikzpicture}
\caption{The T-duality orbit of the $4_4^{(1,3)}$.}
\label{fig:4413Orbit}
\end{figure}
\vskip8pt
\begin{minipage}{0.595\linewidth}
\forceindent S-dualities:
\begin{multicols}{2}
\begin{itemize}
	\item $0_4^{(5,3)} \leftrightarrow 0_7^{(5,3)}$ See Fig. \ref{fig:0753Orbit}
	\item $1_4^{(4,3)} \leftrightarrow 1_6^{(4,3)}$ See Fig. \ref{fig:1643Orbit}
	\item $2_4^{(3,3)} \leftrightarrow 2_5^{(3,3)}$ See Fig. \ref{fig:2515Orbit}
	\item $3_4^{(2,3)} \leftrightarrow 3_4^{(2,3)}$ Self-dual
	\item $4_4^{(1,3)} \leftrightarrow 4_3^{(1,3)}$ See Fig. \ref{fig:532Orbit}
	\item $5_4^3 \leftrightarrow 5_2^3$ See Fig. \ref{fig:52Orbit}
	\item $0_4^{(1,5,3)} \leftrightarrow 0_9^{(1,5,3)}$
	\item $1_4^{(1,4,3)} \leftrightarrow 1_8^{(1,4,3)}$
	\item $2_4^{(1,3,3)} \leftrightarrow 2_7^{(1,3,3)}$ See Fig. \ref{fig:27133Orbit}
	\item $3_4^{(1,2,3)} \leftrightarrow 3_6^{(1,2,3)}$ See Fig. \ref{fig:3624Orbit}
	\item $4_4^{(1,1,3)} \leftrightarrow 4_5^{(1,1,3)}$ See Fig. \ref{fig:5522Orbit}
	\item $5_4^{(1,0,3)} \leftrightarrow 5_4^{(1,0,3)}$ Self-dual
\end{itemize}
\end{multicols}
\end{minipage}
\begin{minipage}{0.4\linewidth}
\forceindent M-theory origins:
\begin{multicols}{2}
\begin{itemize}
	\item $0_4^{(5,3)} \rightarrow 0^{(1,0,5,3)}$
	\item $1_4^{(4,3)} \rightarrow 1^{(1,4,3)}$
	\item $2_4^{(3,3)} \rightarrow 2^{(4,3)}$
	\item $3_4^{(2,3)} \rightarrow 3^{(2,4)}$
	\item $4_4^{(1,3)} \rightarrow 5^{(1,3)}$
	\item $5_4^3 \rightarrow 5^3$
	\item $0_4^{(1,5,3)} \rightarrow 0^{(1,0,0,1,5,3)}$
	\item $1_4^{(1,4,3)} \rightarrow 1^{(1,0,1,4,3)}$
	\item $2_4^{(1,3,3)} \rightarrow 2^{(1,1,3,3)}$
	\item $3_4^{(1,2,3)} \rightarrow 3^{(2,2,3)}$
	\item $4_4^{(1,1,3)} \rightarrow 4^{(1,2,3)}$
	\item $5_4^{(1,0,3)} \rightarrow 5^{(1,0,4)}$
\end{itemize}
\end{multicols}
\end{minipage}
\clearpage

\begin{figure}[H]
\centering
\begin{tikzpicture}
\matrix(M)[matrix of math nodes, row sep=3em, column sep=1 em, minimum width=4em]{
\clap{\text{A}} & \mathclap{0_4^{(9,0)}} & & \mathclap{2_4^{(7,0)}} & & \mathclap{4_4^{(5,0)}} & & \mathclap{6_4^{(3,0)}} & & \mathclap{8_4^{(1,0)}} &\\
\clap{\text{B}} & & \mathclap{1_4^{(8,0)}} & & \mathclap{3_4^{(6,0)}} & & \mathclap{5_4^{(4,0)}} & & \mathclap{7_4^{(2,0)}} & & \mathclap{\text{NS9B}/9_4}\\
};
\draw[latex-latex] (M-1-2) -- (M-2-3);
\draw[latex-latex] (M-2-3) -- (M-1-4);
\draw[latex-latex] (M-1-4) -- (M-2-5);
\draw[latex-latex] (M-2-5) -- (M-1-6);
\draw[latex-latex] (M-1-6) -- (M-2-7);
\draw[latex-latex] (M-2-7) -- (M-1-8);
\draw[latex-latex] (M-1-8) -- (M-2-9);
\draw[latex-latex] (M-2-9) -- (M-1-10);
\draw[latex-latex] (M-1-10) -- (M-2-11);
\end{tikzpicture}
\caption[The T-duality orbit of the $7_4^{(2,0)}$.]{The T-duality orbit of the $7_4^{(2,0)}$.}
\label{fig:7420Orbit}
\end{figure}
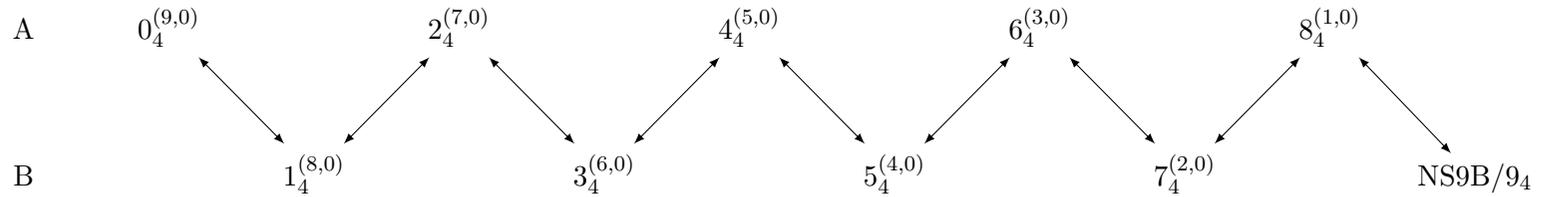
\begin{multicols}{2}
\forceindent S-dualities:
\begin{itemize}
	\item $0_4^{(9,0)} \leftrightarrow 0_{10}^{(9,0)}$
	\item $1_4^{(8,0)} \leftrightarrow 1_9^{(8,0)}$
	\item $3_4^{(6,0)} \leftrightarrow 3_7^{(6,0)}$ See Fig. \ref{fig:3760Orbit}
	\item $5_4^{(4,0)} \leftrightarrow 5_5^{(4,0)}$ See Fig. \ref{fig:5522Orbit}
	\item $7_4^{(2,0)} \leftrightarrow 7_3^{(2,0)}$ See Fig. \ref{fig:532Orbit}
	\item $\text{NS9B}/9_4 \leftrightarrow \text{D9}/9_1$ See Fig. \ref{fig:11Orbit}
\end{itemize}
\columnbreak
\forceindent M-theory origins:
\begin{itemize}
	\item $0_4^{(9,0)} \rightarrow 0^{(1,0,0,0,0,9,0)}$
	\item $2_4^{(7,0)} \rightarrow 2^{(1,0,0,7,0)}$
	\item $4_4^{(5,0)} \rightarrow 4^{(1,5,0)}$
	\item $6_4^{(3,0)} \rightarrow 6^{(3,1)}$
	\item $8_4^{(1,0)}=\text{KK8A} \rightarrow 8^{(1,0)}=\text{KK8M}$
\end{itemize}
\end{multicols}
\clearpage
\end{landscape}
\section{\texorpdfstring{$g_s^{-5}$}{gs-5} Duality Orbits}
\begin{figure}[H]
\centering
\begin{tikzpicture}
\matrix(M)[matrix of math nodes, row sep=2.75em, column sep=2.8em, minimum width=1.8em]{
\clap{\text{A}} & \mathclap{0_5^{(2,6,0)}} & & \mathclap{0_5^{(2,4,2)}} & & \mathclap{0_5^{(2,2,4)}} & & \mathclap{0_5^{(2,0,6)}}\\
\clap{\text{B}} & & \mathclap{0_5^{(2,5,1)}} & & \mathclap{0_5^{(2,3,3)}} & & \mathclap{0_5^{(2,1,5)}} &\\
\clap{\text{A}} & & \mathclap{1_5^{(1,5,1)}} & & \mathclap{1_5^{(1,3,3)}} & & \mathclap{1_5^{(1,1,5)}} &\\
\clap{\text{B}} & \mathclap{1_5^{(1,6,0)}} & & \mathclap{1_5^{(1,4,2)}} & & \mathclap{1_5^{(1,2,4)}} & & \mathclap{1_5^{(1,0,6)}}\\
\clap{\text{A}} & \mathclap{2_5^{(6,0)}} & & \mathclap{2_5^{(4,2)}} & & \mathclap{2_5^{(2,4)}} & & \mathclap{2_5^{6}}\\
\clap{\text{B}} & & \mathclap{2_5^{(5,1)}} & & \mathclap{2_5^{(3,3)}} & & \mathclap{2_5^{(1,5)}} &\\
\clap{\text{A}} & & \mathclap{0_5^{(1,2,5,1)}} & & \mathclap{0_5^{(1,2,3,3)}} & & \mathclap{0_5^{(1,2,1,5)}} &\\
\clap{\text{B}} & \mathclap{0_5^{(1,2,6,0)}} & & \mathclap{0_5^{(1,2,4,2)}} & & \mathclap{0_5^{(1,2,2,4)}} & & \mathclap{0_5^{(1,2,0,6)}}\\
\clap{\text{A}} & \mathclap{1_5^{(1,1,6,0)}} & & \mathclap{1_5^{(1,1,4,2)}} & & \mathclap{1_5^{(1,1,2,4)}} & & \mathclap{1_5^{(1,1,0,6)}}\\
\clap{\text{B}} & & \mathclap{1_5^{(1,1,5,1)}} & & \mathclap{1_5^{(1,1,3,3)}} & & \mathclap{1_5^{(1,1,1,5)}} &\\
\clap{\text{A}} & & \mathclap{2_5^{(1,0,5,1)}} & & \mathclap{2_5^{(1,0,3,3)}} & & \mathclap{2_5^{(1,0,1,5)}} &\\
\clap{\text{B}} & \mathclap{2_5^{(1,0,6,0)}} & & \mathclap{2_5^{(1,0,4,2)}} & & \mathclap{2_5^{(1,0,2,4)}} & & \mathclap{2_5^{(1,0,0,6)}}\\
};
\draw[latex-latex] (M-1-2) -- (M-4-2);
\draw[latex-latex] (M-4-2) -- (M-5-2);
\draw[latex-latex] (M-8-2) -- (M-9-2);
\draw[latex-latex] (M-9-2) -- (M-12-2);
\draw[latex-latex] (M-1-4) -- (M-4-4);
\draw[latex-latex] (M-4-4) -- (M-5-4);
\draw[latex-latex] (M-8-4) -- (M-9-4);
\draw[latex-latex] (M-9-4) -- (M-12-4);
\draw[latex-latex] (M-1-6) -- (M-4-6);
\draw[latex-latex] (M-4-6) -- (M-5-6);
\draw[latex-latex] (M-8-6) -- (M-9-6);
\draw[latex-latex] (M-9-6) -- (M-12-6);
\draw[latex-latex] (M-1-8) -- (M-4-8);
\draw[latex-latex] (M-4-8) -- (M-5-8);
\draw[latex-latex] (M-5-8) -- (M-8-8);
\draw[latex-latex] (M-8-8) -- (M-9-8);
\draw[latex-latex] (M-9-8) -- (M-12-8);
\draw[latex-latex] (M-2-3) -- (M-3-3);
\draw[latex-latex] (M-3-3) -- (M-6-3);
\draw[latex-latex] (M-7-3) -- (M-10-3);
\draw[latex-latex] (M-10-3) -- (M-11-3);
\draw[latex-latex] (M-2-5) -- (M-3-5);
\draw[latex-latex] (M-3-5) -- (M-6-5);
\draw[latex-latex] (M-7-5) -- (M-10-5);
\draw[latex-latex] (M-10-5) -- (M-11-5);
\draw[latex-latex] (M-2-7) -- (M-3-7);
\draw[latex-latex] (M-3-7) -- (M-6-7);
\draw[latex-latex] (M-7-7) -- (M-10-7);
\draw[latex-latex] (M-10-7) -- (M-11-7);
\draw[latex-latex] (M-1-2) -- (M-2-3);
\draw[latex-latex] (M-2-3) -- (M-1-4);
\draw[latex-latex] (M-1-4) -- (M-2-5);
\draw[latex-latex] (M-2-5) -- (M-1-6);
\draw[latex-latex] (M-1-6) -- (M-2-7);
\draw[latex-latex] (M-2-7) -- (M-1-8);
\draw[latex-latex] (M-5-2) -- (M-6-3);
\draw[latex-latex] (M-6-3) -- (M-5-4);
\draw[latex-latex] (M-5-4) -- (M-6-5);
\draw[latex-latex] (M-6-5) -- (M-5-6);
\draw[latex-latex] (M-5-6) -- (M-6-7);
\draw[latex-latex] (M-6-7) -- (M-5-8);
\draw[latex-latex] (M-9-2) -- (M-10-3);
\draw[latex-latex] (M-10-3) -- (M-9-4);
\draw[latex-latex] (M-9-4) -- (M-10-5);
\draw[latex-latex] (M-10-5) -- (M-9-6);
\draw[latex-latex] (M-9-6) -- (M-10-7);
\draw[latex-latex] (M-10-7) -- (M-9-8);
\draw[latex-latex] (M-4-2) -- (M-3-3);
\draw[latex-latex] (M-3-3) -- (M-4-4);
\draw[latex-latex] (M-4-4) -- (M-3-5);
\draw[latex-latex] (M-3-5) -- (M-4-6);
\draw[latex-latex] (M-4-6) -- (M-3-7);
\draw[latex-latex] (M-3-7) -- (M-4-8);
\draw[latex-latex] (M-8-2) -- (M-7-3);
\draw[latex-latex] (M-7-3) -- (M-8-4);
\draw[latex-latex] (M-8-4) -- (M-7-5);
\draw[latex-latex] (M-7-5) -- (M-8-6);
\draw[latex-latex] (M-8-6) -- (M-7-7);
\draw[latex-latex] (M-7-7) -- (M-8-8);
\draw[latex-latex] (M-12-2) -- (M-11-3);
\draw[latex-latex] (M-11-3) -- (M-12-4);
\draw[latex-latex] (M-12-4) -- (M-11-5);
\draw[latex-latex] (M-11-5) -- (M-12-6);
\draw[latex-latex] (M-12-6) -- (M-11-7);
\draw[latex-latex] (M-11-7) -- (M-12-8);
\draw[latex-latex] (M-1-8) to[out=280, in=80] (M-8-8);
\draw[latex-latex] (M-4-8) to[out=292, in=67] (M-9-8);
\draw[latex-latex] (M-5-8) to[out=280, in=80] (M-12-8);
\draw[latex-latex] (M-1-6) to[out=280, in=80] (M-8-6);
\draw[latex-latex] (M-4-6) to[out=292, in=67] (M-9-6);
\draw[latex-latex] (M-5-6) to[out=280, in=80] (M-12-6);
\draw[latex-latex] (M-1-4) to[out=260, in=100] (M-8-4);
\draw[latex-latex] (M-4-4) to[out=247, in=112] (M-9-4);
\draw[latex-latex] (M-5-4) to[out=260, in=100] (M-12-4);
\draw[latex-latex] (M-1-2) to[out=260, in=100] (M-8-2);
\draw[latex-latex] (M-4-2) to[out=247, in=112] (M-9-2);
\draw[latex-latex] (M-5-2) to[out=260, in=100] (M-12-2);
\draw[latex-latex] (M-2-7) to[out=280, in=80] (M-7-7);
\draw[latex-latex] (M-3-7) to[out=292, in=67] (M-10-7);
\draw[latex-latex] (M-6-7) to[out=280, in=80] (M-11-7);
\draw[latex-latex] (M-2-3) to[out=260, in=100] (M-7-3);
\draw[latex-latex] (M-3-3) to[out=247, in=112] (M-10-3);
\draw[latex-latex] (M-6-3) to[out=260, in=100] (M-11-3);
\draw[latex-latex] (M-2-5) to[out=247, in=112] (M-7-5);
\draw[latex-latex] (M-3-5) to[out=292, in=67] (M-10-5);
\draw[latex-latex] (M-6-5) to[out=247, in=112] (M-11-5);
\end{tikzpicture}
\caption{The T-duality orbit of the $2_5^{(1,5)}$.}
\label{fig:2515Orbit}
\end{figure}
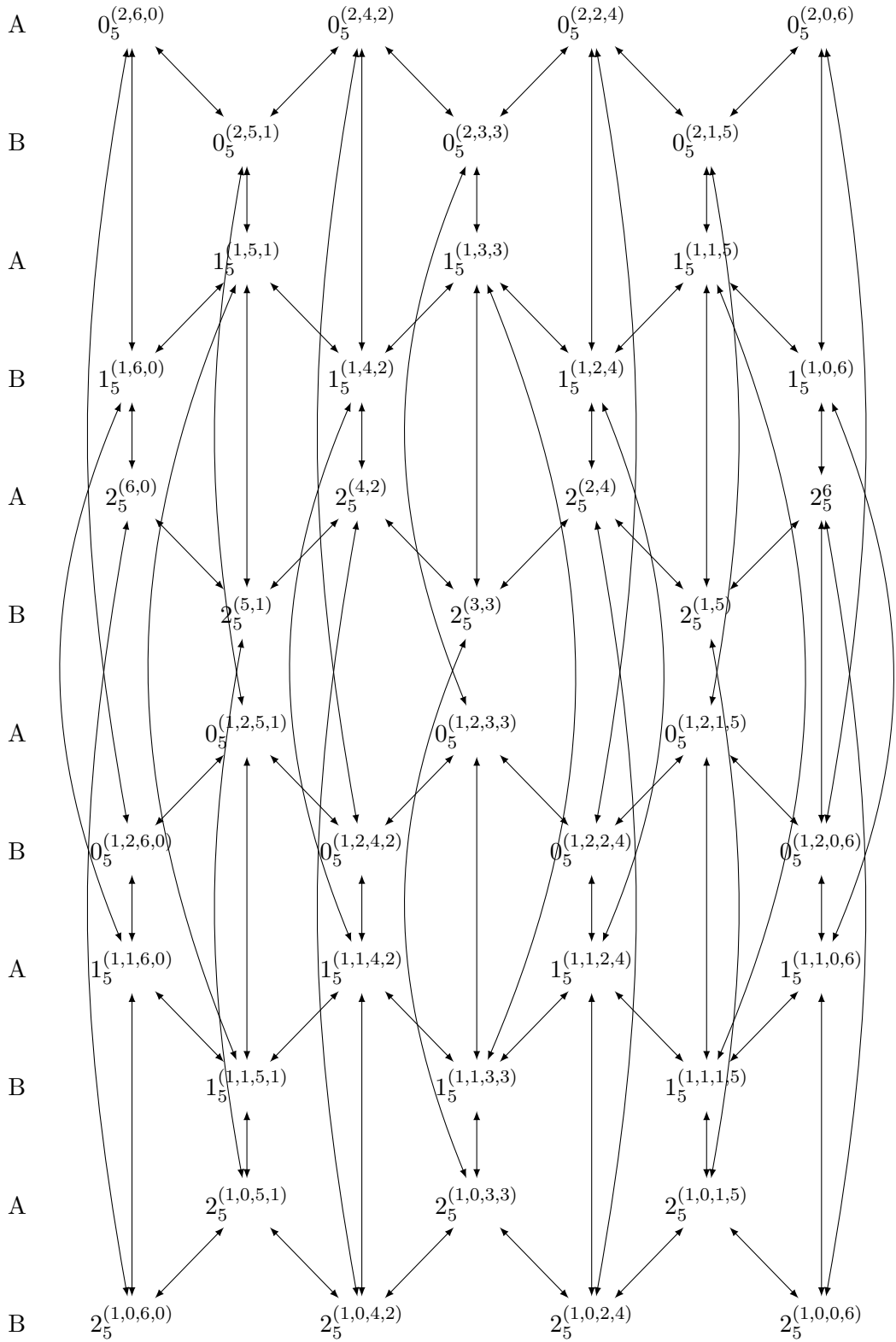
\begin{multicols}{2}
\forceindent S-dualities:
\begin{itemize}
	\item $0_5^{(2,5,1)} \leftrightarrow 0_8^{(2,5,1)}$
	\item $0_5^{(2,3,3)} \leftrightarrow 0_7^{(2,3,3)}$ See Fig. \ref{fig:0753Orbit}
	\item $0_5^{(2,1,5)} \leftrightarrow 0_6^{(2,1,5)}$ See Fig. \ref{fig:0617Orbit}
	\item $1_5^{(1,6,0)} \leftrightarrow 1_7^{(1,6,0)}$ See Fig. \ref{fig:17160Orbit}
	\item $1_5^{(1,4,2)} \leftrightarrow 1_6^{(1,4,2)}$ See Fig. \ref{fig:1643Orbit}
	\item $1_5^{(1,2,4)} \leftrightarrow 1_5^{(1,2,4)}$ Self-dual
	\item $1_5^{(1,0,6)} \leftrightarrow 1_4^{(1,0,6)}$ See Fig. \ref{fig:146Orbit}
	\item $2_5^{(5,1)} \leftrightarrow 2_5^{(5,1)}$ Self-dual
	\item $2_5^{(3,3)} \leftrightarrow 2_4^{(3,3)}$ See Fig. \ref{fig:4413Orbit}
	\item $2_5^{(1,5)} \leftrightarrow 2_3^{(1,5)}$ See Fig. \ref{fig:532Orbit}
	\item $0_5^{(1,2,6,0)} \leftrightarrow 0_{11}^{(1,2,6,0)}$
	\item $0_5^{(1,2,4,2)} \leftrightarrow 0_{10}^{(1,2,4,2)}$
	\item $0_5^{(1,2,2,4)} \leftrightarrow 0_9^{(1,2,2,4)}$
	\item $0_5^{(1,2,0,6)} \leftrightarrow 0_8^{(1,2,0,6)}$
	\item $1_5^{(1,1,5,1)} \leftrightarrow 1_9^{(1,1,5,1)}$
	\item $1_5^{(1,1,3,3)} \leftrightarrow 1_8^{(1,1,3,3)}$
	\item $1_5^{(1,1,1,5)} \leftrightarrow 1_7^{(1,1,1,5)}$ See Fig. \ref{fig:1726Orbit}
	\item $2_5^{(1,0,6,0)} \leftrightarrow 2_8^{(1,0,6,0)}$
	\item $2_5^{(1,0,4,2)} \leftrightarrow 2_7^{(1,0,4,2)}$ See Fig. \ref{fig:27133Orbit}
	\item $2_5^{(1,0,2,4)} \leftrightarrow 2_6^{(1,0,2,4)}$ See Fig. \ref{fig:3624Orbit}
	\item $2_5^{(1,0,0,6)} \leftrightarrow 2_5^{(1,0,0,6)}$ Self-dual
\end{itemize}
\columnbreak
\forceindent M-theory origins:
\begin{itemize}
	\item $0_5^{(2,6,0)} \rightarrow 0^{(1,0,2,6,0)}$
	\item $0_5^{(2,4,2)} \rightarrow 0^{(1,2,4,2)}$
	\item $0_5^{(2,2,4)} \rightarrow 0^{(3,2,4)}$
	\item $0_5^{(2,0,6)} \rightarrow 0^{(2,1,6)}$
	\item $1_5^{(1,5,1)} \rightarrow 1^{(2,5,1)}$
	\item $1_5^{(1,3,3)} \rightarrow 1^{(1,4,3)}$
	\item $1_5^{(1,1,5)} \rightarrow 1^{(1,1,6)}$
	\item $2_5^{(6,0)} \rightarrow 2^{(1,0,0,6,0)}$
	\item $2_5^{(4,2)} \rightarrow 2^{(4,3)}$
	\item $2_5^{(2,4)} \rightarrow 3^{(2,4)}$
	\item $2_5^6 \rightarrow 2^6$
	\item $0_5^{(1,2,5,1)} \rightarrow 0^{(1,0,1,2,5,1)}$
	\item $0_5^{(1,2,3,3)} \rightarrow 0^{(1,0,1,2,3,3)}$
	\item $0_5^{(1,2,1,5)} \rightarrow 0^{(1,1,2,1,5)}$
	\item $1_5^{(1,1,6,0)} \rightarrow 1^{(1,0,1,1,6,0)}$
	\item $1_5^{(1,1,4,2)} \rightarrow 1^{(1,1,1,4,2)} $
	\item $1_5^{(1,1,2,4)} \rightarrow 1^{(2,1,2,4)}$
	\item $1_5^{(1,1,0,6)} \rightarrow 1^{(1,2,0,6)}$
	\item $2_5^{(1,0,5,1)} \rightarrow 2^{(2,0,5,1)}$
	\item $2_5^{(1,0,3,3)} \rightarrow 2^{(1,1,3,3)}$
	\item $2_5^{(1,0,1,5)} \rightarrow 2^{(1,0,2,5)}$
\end{itemize}
\end{multicols}
\clearpage
\vspace*{\fill}
\begin{figure}[H]
\centering
\begin{tikzpicture}
\matrix(M)[matrix of math nodes, row sep=2.9em, column sep=3em, minimum width=3em]{
\clap{\text{A}} & \mathclap{0_5^{(5,0,4)}} & & \mathclap{0_5^{(5,2,2)}} & & \mathclap{0_5^{(5,4,0)}}\\
\clap{\text{B}} & & \mathclap{0_5^{(5,1,3)}} & & \mathclap{0_5^{(5,3,1)}} &\\
\clap{\text{A}} & & \mathclap{1_5^{(4,1,3)}} & & \mathclap{1_5^{(4,3,1)}} &\\
\clap{\text{B}} & \mathclap{1_5^{(4,0,4)}} & & \mathclap{1_5^{(4,2,2)}} & & \mathclap{1_5^{(4,4,0)}}\\
\clap{\text{A}} & \mathclap{2_5^{(3,0,4)}} & & \mathclap{2_5^{(3,2,2)}} & & \mathclap{2_5^{(3,4,0)}} &\\
\clap{\text{B}} & & \mathclap{2_5^{(3,1,3)}} & & \mathclap{2_5^{(3,3,1)}} &\\
\clap{\text{A}} & & \mathclap{3_5^{(2,1,3)}} & & \mathclap{3_5^{(2,3,1)}} &\\
\clap{\text{B}} & \mathclap{3_5^{(2,0,4)}} & & \mathclap{3_5^{(2,2,2)}} & & \mathclap{3_5^{(2,4,0)}}\\
\clap{\text{A}} & \mathclap{4_5^{(1,0,4)}} & & \mathclap{4_5^{(1,2,2)}} & & \mathclap{4_5^{(1,4,0)}}\\
\clap{\text{B}} & & \mathclap{4_5^{(1,1,3)}} & & \mathclap{4_5^{(1,3,1)}} &\\
\clap{\text{A}} & & \mathclap{5_5^{(1,3)}} & & \mathclap{5_5^{(3,1)}} &\\
\clap{\text{B}} & \mathclap{5_5^4} & & \mathclap{5_5^{(2,2)}} & & \mathclap{5_5^{(4,0)}} & \\
};
\draw[latex-latex] (M-1-2) -- (M-4-2);
\draw[latex-latex] (M-4-2) -- (M-5-2);
\draw[latex-latex] (M-5-2) -- (M-8-2);
\draw[latex-latex] (M-8-2) -- (M-9-2);
\draw[latex-latex] (M-9-2) -- (M-12-2);
\draw[latex-latex] (M-1-4) -- (M-4-4);
\draw[latex-latex] (M-4-4) -- (M-5-4);
\draw[latex-latex] (M-5-4) -- (M-8-4);
\draw[latex-latex] (M-8-4) -- (M-9-4);
\draw[latex-latex] (M-9-4) -- (M-12-4);
\draw[latex-latex] (M-1-6) -- (M-4-6);
\draw[latex-latex] (M-4-6) -- (M-5-6);
\draw[latex-latex] (M-5-6) -- (M-8-6);
\draw[latex-latex] (M-8-6) -- (M-9-6);
\draw[latex-latex] (M-9-6) -- (M-12-6);
\draw[latex-latex] (M-2-3) -- (M-3-3);
\draw[latex-latex] (M-3-3) -- (M-6-3);
\draw[latex-latex] (M-6-3) -- (M-7-3);
\draw[latex-latex] (M-7-3) -- (M-10-3);
\draw[latex-latex] (M-10-3) -- (M-11-3);
\draw[latex-latex] (M-2-5) -- (M-3-5);
\draw[latex-latex] (M-3-5) -- (M-6-5);
\draw[latex-latex] (M-6-5) -- (M-7-5);
\draw[latex-latex] (M-7-5) -- (M-10-5);
\draw[latex-latex] (M-10-5) -- (M-11-5);
\draw[latex-latex] (M-1-2) -- (M-2-3);
\draw[latex-latex] (M-2-3) -- (M-1-4);
\draw[latex-latex] (M-1-4) -- (M-2-5);
\draw[latex-latex] (M-2-5) -- (M-1-6);
\draw[latex-latex] (M-5-2) -- (M-6-3);
\draw[latex-latex] (M-6-3) -- (M-5-4);
\draw[latex-latex] (M-5-4) -- (M-6-5);
\draw[latex-latex] (M-6-5) -- (M-5-6);
\draw[latex-latex] (M-9-2) -- (M-10-3);
\draw[latex-latex] (M-10-3) -- (M-9-4);
\draw[latex-latex] (M-9-4) -- (M-10-5);
\draw[latex-latex] (M-10-5) -- (M-9-6);
\draw[latex-latex] (M-4-2) -- (M-3-3);
\draw[latex-latex] (M-3-3) -- (M-4-4);
\draw[latex-latex] (M-4-4) -- (M-3-5);
\draw[latex-latex] (M-3-5) -- (M-4-6);
\draw[latex-latex] (M-8-2) -- (M-7-3);
\draw[latex-latex] (M-7-3) -- (M-8-4);
\draw[latex-latex] (M-8-4) -- (M-7-5);
\draw[latex-latex] (M-7-5) -- (M-8-6);
\draw[latex-latex] (M-12-2) -- (M-11-3);
\draw[latex-latex] (M-11-3) -- (M-12-4);
\draw[latex-latex] (M-12-4) -- (M-11-5);
\draw[latex-latex] (M-11-5) -- (M-12-6);
\end{tikzpicture}
\caption{The T-duality orbit of the $5_5^{(2,2)}$.}
\label{fig:5522Orbit}
\end{figure}
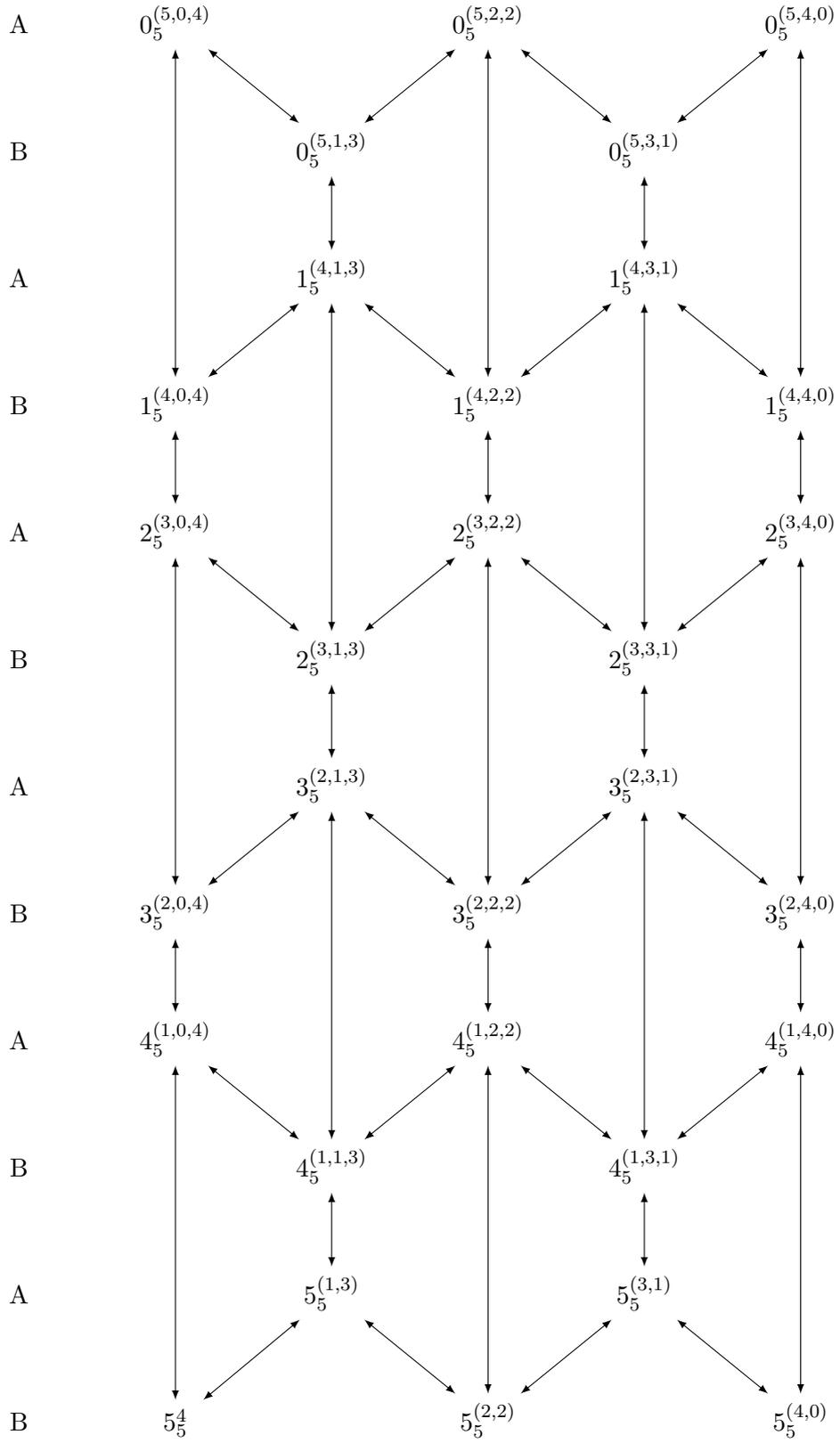
\begin{multicols}{2}
\forceindent S-dualities:
\begin{itemize}
	\item $0_5^{(5,1,3)} \leftrightarrow 0_{10}^{(5,1,3)}$
	\item $0_5^{(5,3,1)} \leftrightarrow 0_{11}^{(5,3,1)}$
	\item $1_5^{(4,0,4)} \leftrightarrow 1_8^{(4,0,4)}$
	\item $1_5^{(4,2,2)} \leftrightarrow 1_9^{(4,2,2)}$
	\item $1_5^{(4,4,0)} \leftrightarrow 1_{10}^{(4,4,0)}$
	\item $2_5^{(3,1,3)} \leftrightarrow 2_7^{(3,1,3)}$ See Fig. \ref{fig:27133Orbit}
	\item $2_5^{(3,3,1)} \leftrightarrow 2_8^{(3,3,1)}$
	\item $3_5^{(2,0,4)} \leftrightarrow 3_5^{(2,0,4)}$ Self-dual
	\item $3_5^{(2,2,2)} \leftrightarrow 3_6^{(2,2,2)}$ See Fig. \ref{fig:3624Orbit}
	\item $3_5^{(2,4,0)} \leftrightarrow 3_7^{(2,4,0)}$ See Fig. \ref{fig:3760Orbit}
	\item $4_5^{(1,1,3)} \leftrightarrow 4_4^{(1,1,3)}$ See Fig. \ref{fig:4413Orbit}
	\item $4_5^{(1,3,1)} \leftrightarrow 4_5^{(1,3,1)}$ Self-dual
	\item $5_5^4 \leftrightarrow 5_2^4$ See Fig. \ref{fig:52Orbit}
	\item $5_5^{(2,2)} \leftrightarrow 5_3^{(2,2)}$ See Fig. \ref{fig:532Orbit}
	\item $5_5^{(4,0)} \leftrightarrow 5_4^{(4,0)}$ See Fig. \ref{fig:7420Orbit}
\end{itemize}
\columnbreak
\forceindent M-theory origins:
\begin{itemize}
	\item $0_5^{(5,0,4)} \rightarrow 0^{(1,0,0,5,0,4)}$
	\item $0_5^{(5,2,2)} \rightarrow 0^{(1,0,0,0,5,2,2)}$
	\item $0_5^{(5,4,0)} \rightarrow 0^{(1,0,0,0,0,5,4,0)}$
	\item $1_5^{(4,1,3)} \rightarrow 1^{(1,0,4,1,3)}$
	\item $1_5^{(4,3,1)} \rightarrow 1^{(1,0,0,4,3,1)}$
	\item $2_5^{(3,0,4)} \rightarrow 2^{(4,0,4)}$
	\item $2_5^{(3,2,2)} \rightarrow 2^{(1,3,2,2)}$
	\item $2_5^{(3,4,0)} \rightarrow 2^{(1,0,3,4,0)}$
	\item $3_5^{(2,1,3)} \rightarrow 3^{(2,2,3)}$
	\item $3_5^{(2,3,1)} \rightarrow 3^{(3,3,1)}$
	\item $4_5^{(1,0,4)} \rightarrow 5^{(1,0,4)}$
	\item $4_5^{(1,2,2)} \rightarrow 4^{(1,2,3)}$
	\item $4_5^{(1,4,0)} \rightarrow 4^{(1,5,0)}$
	\item $5_5^{(1,3)} \rightarrow 5^{(1,3)}$
	\item $5_5^{(3,1)} \rightarrow 6^{(3,1)}$
\end{itemize}
\end{multicols}
\section{\texorpdfstring{$g_s^{-6}$}{gs-6} Duality Orbits}
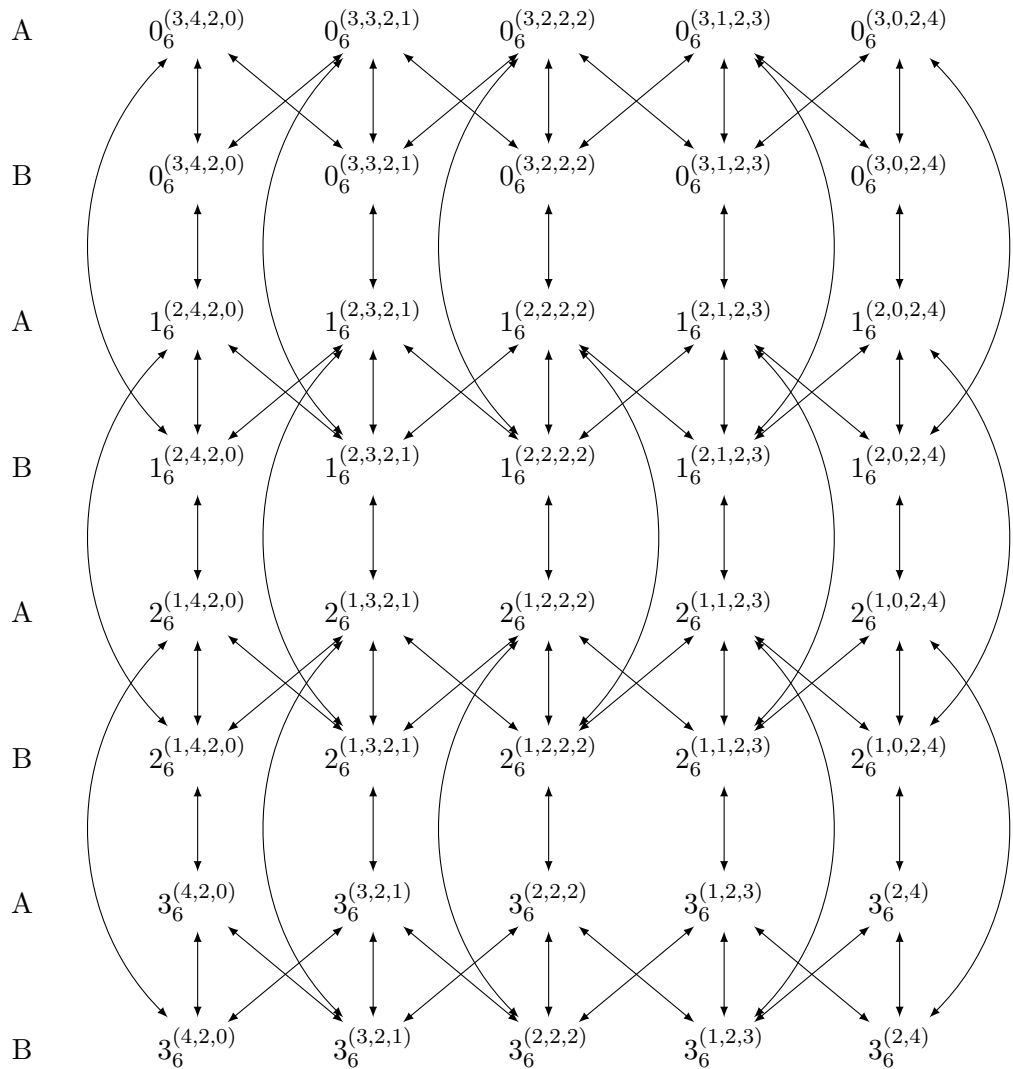
\begin{figure}[H]
\centering
\begin{tikzpicture}
\matrix(M)[matrix of math nodes, row sep=3em, column sep=4 em, minimum width=2em]{
\clap{\text{A}} & \mathclap{0_6^{(3,4,2,0)}} & \mathclap{0_6^{(3,3,2,1)}} & \mathclap{0_6^{(3,2,2,2)}} & \mathclap{0_6^{(3,1,2,3)}} & \mathclap{0_6^{(3,0,2,4)}}\\
\clap{\text{B}} & \mathclap{0_6^{(3,4,2,0)}} & \mathclap{0_6^{(3,3,2,1)}} & \mathclap{0_6^{(3,2,2,2)}} & \mathclap{0_6^{(3,1,2,3)}} & \mathclap{0_6^{(3,0,2,4)}}\\
\clap{\text{A}} & \mathclap{1_6^{(2,4,2,0)}} & \mathclap{1_6^{(2,3,2,1)}} & \mathclap{1_6^{(2,2,2,2)}} & \mathclap{1_6^{(2,1,2,3)}} & \mathclap{1_6^{(2,0,2,4)}}\\
\clap{\text{B}} & \mathclap{1_6^{(2,4,2,0)}} & \mathclap{1_6^{(2,3,2,1)}} & \mathclap{1_6^{(2,2,2,2)}} & \mathclap{1_6^{(2,1,2,3)}} & \mathclap{1_6^{(2,0,2,4)}}\\
\clap{\text{A}} & \mathclap{2_6^{(1,4,2,0)}} & \mathclap{2_6^{(1,3,2,1)}} & \mathclap{2_6^{(1,2,2,2)}} & \mathclap{2_6^{(1,1,2,3)}} & \mathclap{2_6^{(1,0,2,4)}}\\
\clap{\text{B}} & \mathclap{2_6^{(1,4,2,0)}} & \mathclap{2_6^{(1,3,2,1)}} & \mathclap{2_6^{(1,2,2,2)}} & \mathclap{2_6^{(1,1,2,3)}} & \mathclap{2_6^{(1,0,2,4)}}\\
\clap{\text{A}} & \mathclap{3_6^{(4,2,0)}} & \mathclap{3_6^{(3,2,1)}} & \mathclap{3_6^{(2,2,2)}} & \mathclap{3_6^{(1,2,3)}} & \mathclap{3_6^{(2,4)}}\\
\clap{\text{B}} & \mathclap{3_6^{(4,2,0)}} & \mathclap{3_6^{(3,2,1)}} & \mathclap{3_6^{(2,2,2)}} & \mathclap{3_6^{(1,2,3)}} & \mathclap{3_6^{(2,4)}}\\
};
\draw[latex-latex] (M-1-2) -- (M-2-2);
\draw[latex-latex] (M-2-2) -- (M-3-2);
\draw[latex-latex] (M-3-2) -- (M-4-2);
\draw[latex-latex] (M-4-2) -- (M-5-2);
\draw[latex-latex] (M-5-2) -- (M-6-2);
\draw[latex-latex] (M-6-2) -- (M-7-2);
\draw[latex-latex] (M-7-2) -- (M-8-2);
\draw[latex-latex] (M-1-3) -- (M-2-3);
\draw[latex-latex] (M-2-3) -- (M-3-3);
\draw[latex-latex] (M-3-3) -- (M-4-3);
\draw[latex-latex] (M-4-3) -- (M-5-3);
\draw[latex-latex] (M-5-3) -- (M-6-3);
\draw[latex-latex] (M-6-3) -- (M-7-3);
\draw[latex-latex] (M-7-3) -- (M-8-3);
\draw[latex-latex] (M-1-4) -- (M-2-4);
\draw[latex-latex] (M-2-4) -- (M-3-4);
\draw[latex-latex] (M-3-4) -- (M-4-4);
\draw[latex-latex] (M-4-4) -- (M-5-4);
\draw[latex-latex] (M-5-4) -- (M-6-4);
\draw[latex-latex] (M-6-4) -- (M-7-4);
\draw[latex-latex] (M-7-4) -- (M-8-4);
\draw[latex-latex] (M-1-5) -- (M-2-5);
\draw[latex-latex] (M-2-5) -- (M-3-5);
\draw[latex-latex] (M-3-5) -- (M-4-5);
\draw[latex-latex] (M-4-5) -- (M-5-5);
\draw[latex-latex] (M-5-5) -- (M-6-5);
\draw[latex-latex] (M-6-5) -- (M-7-5);
\draw[latex-latex] (M-7-5) -- (M-8-5);
\draw[latex-latex] (M-1-6) -- (M-2-6);
\draw[latex-latex] (M-2-6) -- (M-3-6);
\draw[latex-latex] (M-3-6) -- (M-4-6);
\draw[latex-latex] (M-4-6) -- (M-5-6);
\draw[latex-latex] (M-5-6) -- (M-6-6);
\draw[latex-latex] (M-6-6) -- (M-7-6);
\draw[latex-latex] (M-7-6) -- (M-8-6);
\draw[latex-latex] (M-1-2) -- (M-2-3);
\draw[latex-latex] (M-2-2) -- (M-1-3);
\draw[latex-latex] (M-1-3) -- (M-2-4);
\draw[latex-latex] (M-2-3) -- (M-1-4);
\draw[latex-latex] (M-1-4) -- (M-2-5);
\draw[latex-latex] (M-2-4) -- (M-1-5);
\draw[latex-latex] (M-1-5) -- (M-2-6);
\draw[latex-latex] (M-2-5) -- (M-1-6);
\draw[latex-latex] (M-3-2) -- (M-4-3);
\draw[latex-latex] (M-4-2) -- (M-3-3);
\draw[latex-latex] (M-3-3) -- (M-4-4);
\draw[latex-latex] (M-4-3) -- (M-3-4);
\draw[latex-latex] (M-3-4) -- (M-4-5);
\draw[latex-latex] (M-4-4) -- (M-3-5);
\draw[latex-latex] (M-3-5) -- (M-4-6);
\draw[latex-latex] (M-4-5) -- (M-3-6);
\draw[latex-latex] (M-5-2) -- (M-6-3);
\draw[latex-latex] (M-6-2) -- (M-5-3);
\draw[latex-latex] (M-5-3) -- (M-6-4);
\draw[latex-latex] (M-6-3) -- (M-5-4);
\draw[latex-latex] (M-5-4) -- (M-6-5);
\draw[latex-latex] (M-6-4) -- (M-5-5);
\draw[latex-latex] (M-5-5) -- (M-6-6);
\draw[latex-latex] (M-6-5) -- (M-5-6);
\draw[latex-latex] (M-7-2) -- (M-8-3);
\draw[latex-latex] (M-8-2) -- (M-7-3);
\draw[latex-latex] (M-7-3) -- (M-8-4);
\draw[latex-latex] (M-8-3) -- (M-7-4);
\draw[latex-latex] (M-7-4) -- (M-8-5);
\draw[latex-latex] (M-8-4) -- (M-7-5);
\draw[latex-latex] (M-7-5) -- (M-8-6);
\draw[latex-latex] (M-8-5) -- (M-7-6);
\draw[latex-latex] (M-1-2) to[out=225, in=135] (M-4-2);
\draw[latex-latex] (M-3-2) to[out=225, in=135] (M-6-2);
\draw[latex-latex] (M-5-2) to[out=225, in=135] (M-8-2);
\draw[latex-latex] (M-1-3) to[out=225, in=135] (M-4-3);
\draw[latex-latex] (M-3-3) to[out=225, in=135] (M-6-3);
\draw[latex-latex] (M-5-3) to[out=225, in=135] (M-8-3);
\draw[latex-latex] (M-1-4) to[out=225, in=135] (M-4-4);
\draw[latex-latex] (M-3-4) to[out=315, in=45] (M-6-4);
\draw[latex-latex] (M-5-4) to[out=225, in=135] (M-8-4);
\draw[latex-latex] (M-1-5) to[out=315, in=45] (M-4-5);
\draw[latex-latex] (M-3-5) to[out=315, in=45] (M-6-5);
\draw[latex-latex] (M-5-5) to[out=315, in=45] (M-8-5);
\draw[latex-latex] (M-1-6) to[out=315, in=45] (M-4-6);
\draw[latex-latex] (M-3-6) to[out=315, in=45] (M-6-6);
\draw[latex-latex] (M-5-6) to[out=315, in=45] (M-8-6);
\end{tikzpicture}
\caption{The T-duality orbit of the $3_6^{(2,4)}$.}
\label{fig:3624Orbit}
\end{figure}
\clearpage
\begin{multicols}{2}
\forceindent S-dualities:
\begin{itemize}
	\item $0_6^{(3,4,2,0)} \leftrightarrow 0_{13}^{(3,4,2,0)}$
	\item $0_6^{(3,3,2,1)} \leftrightarrow 0_{12}^{(3,3,2,1)}$
	\item $0_6^{(3,2,2,2)} \leftrightarrow 0_{11}^{(3,2,2,2)}$
	\item $0_6^{(3,1,2,3)} \leftrightarrow 0_{10}^{(3,1,2,3)}$
	\item $0_6^{(3,0,2,4)} \leftrightarrow 0_9^{(3,0,2,4)}$
	\item $1_6^{(2,4,2,0)} \leftrightarrow 1_{11}^{(2,4,2,0)}$
	\item $1_6^{(2,3,2,1)} \leftrightarrow 1_{10}^{(2,3,2,1)}$
	\item $1_6^{(2,2,2,2)} \leftrightarrow 1_9^{(2,2,2,2)}$
	\item $1_6^{(2,1,2,3)} \leftrightarrow 1_8^{(2,1,2,3)}$
	\item $1_6^{(2,0,2,4)} \leftrightarrow 1_7^{(2,0,2,4)}$ See Fig. \ref{fig:1726Orbit}
	\item $2_6^{(1,4,2,0)} \leftrightarrow 2_9^{(1,4,2,0)}$
	\item $2_6^{(1,3,2,1)} \leftrightarrow 2_8^{(1,3,2,1)}$
	\item $2_6^{(1,2,2,2)} \leftrightarrow 2_7^{(1,2,2,2)}$ See Fig. \ref{fig:27133Orbit}
	\item $2_6^{(1,1,2,3)} \leftrightarrow 2_6^{(1,1,2,3)}$ Self-dual
	\item $2_6^{(1,0,2,4)} \leftrightarrow 2_5^{(1,0,2,4)}$ See Fig. \ref{fig:2515Orbit}
	\item $3_6^{(4,2,0)} \leftrightarrow 3_7^{(4,2,0)}$ See Fig. \ref{fig:3760Orbit}
	\item $3_6^{(3,2,1)} \leftrightarrow 3_6^{(3,2,1)}$ Self-dual
	\item $3_6^{(2,2,2)} \leftrightarrow 3_5^{(2,2,2)}$ See Fig. \ref{fig:5522Orbit}
	\item $3_6^{(1,2,3)} \leftrightarrow 3_4^{(1,2,3)}$ See Fig. \ref{fig:4413Orbit}
	\item $3_6^{(2,4)} \leftrightarrow 3_3^{(2,4)}$ See Fig. \ref{fig:532Orbit}
\end{itemize}
\columnbreak
\forceindent M-theory origins:
\begin{itemize}
	\item $0_6^{(3,4,2,0)} \rightarrow 0^{(1,0,0,0,0,3,4,2,0)}$
	\item $0_6^{(3,3,2,1)} \rightarrow 0^{(1,0,0,0,3,3,2,1)}$
	\item $0_6^{(3,2,2,2)} \rightarrow 0^{(1,0,0,3,2,2,2)}$
	\item $0_6^{(3,1,2,3)} \rightarrow 0^{(1,0,3,1,2,3)}$
	\item $0_6^{(3,0,2,4)} \rightarrow 0^{(1,3,0,2,4)}$
	\item $1_6^{(2,4,2,0)} \rightarrow 1^{(1,0,0,2,4,2,0)}$
	\item $1_6^{(2,3,2,1)} \rightarrow 1^{(1,0,2,3,2,1)}$
	\item $1_6^{(2,2,2,2)} \rightarrow 1^{(1,2,2,2,2)}$
	\item $1_6^{(2,1,2,3)} \rightarrow 1^{(3,1,2,3)}$
	\item $1_6^{(2,0,2,4)} \rightarrow 1^{(2,1,2,4)}$
	\item $2_6^{(1,4,2,0)} \rightarrow 2^{(1,1,4,2,0)}$
	\item $2_6^{(1,3,2,1)} \rightarrow 2^{(2,3,2,1)}$
	\item $2_6^{(1,2,2,2)} \rightarrow 2^{(1,3,2,2)}$
	\item $2_6^{(1,1,2,3)} \rightarrow 2^{(1,1,3,3)}$
	\item $2_6^{(1,0,2,4)} \rightarrow 2^{(1,0,2,5)}$
	\item $3_6^{(4,2,0)} \rightarrow 3^{(5,2,0)}$
	\item $3_6^{(3,2,1)} \rightarrow 3^{(3,3,1)}$
	\item $3_6^{(2,2,2)} \rightarrow 3^{(2,2,3)}$
	\item $3_6^{(1,2,3)} \rightarrow 4^{(1,2,3)}$
	\item $3_6^{(2,4)} \rightarrow 3^{(2,4)}$
\end{itemize}
\end{multicols}
\begin{landscape}
\thispagestyle{empty}
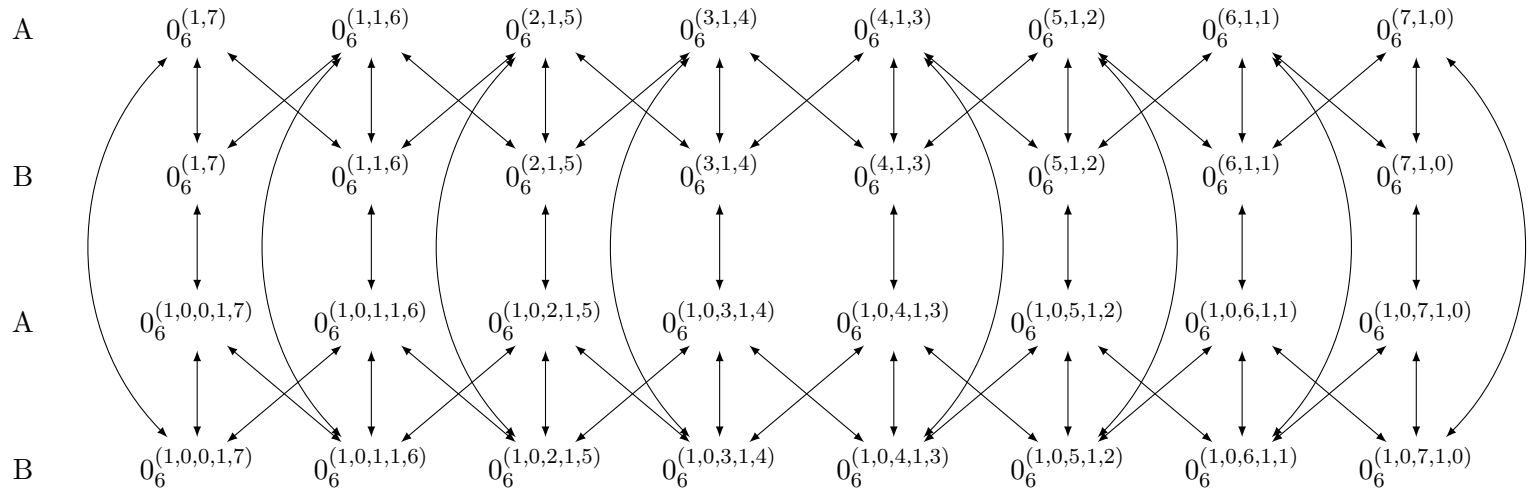
\begin{figure}[H]
\centering
\begin{tikzpicture}
\matrix(M)[matrix of math nodes, row sep=3em, column sep=4em, minimum width=2em]{
\clap{\text{A}} & \mathclap{0_6^{(1,7)}} & \mathclap{0_6^{(1,1,6)}} & \mathclap{0_6^{(2,1,5)}} & \mathclap{0_6^{(3,1,4)}} & \mathclap{0_6^{(4,1,3)}} & \mathclap{0_6^{(5,1,2)}} & \mathclap{0_6^{(6,1,1)}} & \mathclap{0_6^{(7,1,0)}}\\
\clap{\text{B}} & \mathclap{0_6^{(1,7)}} & \mathclap{0_6^{(1,1,6)}} & \mathclap{0_6^{(2,1,5)}} & \mathclap{0_6^{(3,1,4)}} & \mathclap{0_6^{(4,1,3)}} & \mathclap{0_6^{(5,1,2)}} & \mathclap{0_6^{(6,1,1)}} & \mathclap{0_6^{(7,1,0)}}\\
\clap{\text{A}} & \mathclap{0_6^{(1,0,0,1,7)}} & \mathclap{0_6^{(1,0,1,1,6)}} & \mathclap{0_6^{(1,0,2,1,5)}} & \mathclap{0_6^{(1,0,3,1,4)}} & \mathclap{0_6^{(1,0,4,1,3)}} & \mathclap{0_6^{(1,0,5,1,2)}} & \mathclap{0_6^{(1,0,6,1,1)}} & \mathclap{0_6^{(1,0,7,1,0)}}\\
\clap{\text{B}} & \mathclap{0_6^{(1,0,0,1,7)}} & \mathclap{0_6^{(1,0,1,1,6)}} & \mathclap{0_6^{(1,0,2,1,5)}} & \mathclap{0_6^{(1,0,3,1,4)}} & \mathclap{0_6^{(1,0,4,1,3)}} & \mathclap{0_6^{(1,0,5,1,2)}} & \mathclap{0_6^{(1,0,6,1,1)}} & \mathclap{0_6^{(1,0,7,1,0)}}\\
};
\draw[latex-latex] (M-1-2) -- (M-2-2);
\draw[latex-latex] (M-2-2) -- (M-3-2);
\draw[latex-latex] (M-3-2) -- (M-4-2);
\draw[latex-latex] (M-1-3) -- (M-2-3);
\draw[latex-latex] (M-2-3) -- (M-3-3);
\draw[latex-latex] (M-3-3) -- (M-4-3);
\draw[latex-latex] (M-1-4) -- (M-2-4);
\draw[latex-latex] (M-2-4) -- (M-3-4);
\draw[latex-latex] (M-3-4) -- (M-4-4);
\draw[latex-latex] (M-1-5) -- (M-2-5);
\draw[latex-latex] (M-2-5) -- (M-3-5);
\draw[latex-latex] (M-3-5) -- (M-4-5);
\draw[latex-latex] (M-1-6) -- (M-2-6);
\draw[latex-latex] (M-2-6) -- (M-3-6);
\draw[latex-latex] (M-3-6) -- (M-4-6);
\draw[latex-latex] (M-1-7) -- (M-2-7);
\draw[latex-latex] (M-2-7) -- (M-3-7);
\draw[latex-latex] (M-3-7) -- (M-4-7);
\draw[latex-latex] (M-1-8) -- (M-2-8);
\draw[latex-latex] (M-2-8) -- (M-3-8);
\draw[latex-latex] (M-3-8) -- (M-4-8);
\draw[latex-latex] (M-1-9) -- (M-2-9);
\draw[latex-latex] (M-2-9) -- (M-3-9);
\draw[latex-latex] (M-3-9) -- (M-4-9);
\draw[latex-latex] (M-1-2) -- (M-2-3);
\draw[latex-latex] (M-2-2) -- (M-1-3);
\draw[latex-latex] (M-1-3) -- (M-2-4);
\draw[latex-latex] (M-2-3) -- (M-1-4);
\draw[latex-latex] (M-1-4) -- (M-2-5);
\draw[latex-latex] (M-2-4) -- (M-1-5);
\draw[latex-latex] (M-1-5) -- (M-2-6);
\draw[latex-latex] (M-2-5) -- (M-1-6);
\draw[latex-latex] (M-1-6) -- (M-2-7);
\draw[latex-latex] (M-2-6) -- (M-1-7);
\draw[latex-latex] (M-1-7) -- (M-2-8);
\draw[latex-latex] (M-2-7) -- (M-1-8);
\draw[latex-latex] (M-1-8) -- (M-2-9);
\draw[latex-latex] (M-2-8) -- (M-1-9);
\draw[latex-latex] (M-3-2) -- (M-4-3);
\draw[latex-latex] (M-4-2) -- (M-3-3);
\draw[latex-latex] (M-3-3) -- (M-4-4);
\draw[latex-latex] (M-4-3) -- (M-3-4);
\draw[latex-latex] (M-3-4) -- (M-4-5);
\draw[latex-latex] (M-4-4) -- (M-3-5);
\draw[latex-latex] (M-3-5) -- (M-4-6);
\draw[latex-latex] (M-4-5) -- (M-3-6);
\draw[latex-latex] (M-3-6) -- (M-4-7);
\draw[latex-latex] (M-4-6) -- (M-3-7);
\draw[latex-latex] (M-3-7) -- (M-4-8);
\draw[latex-latex] (M-4-7) -- (M-3-8);
\draw[latex-latex] (M-3-8) -- (M-4-9);
\draw[latex-latex] (M-4-8) -- (M-3-9);
\draw[latex-latex] (M-1-2) to[out=225, in=135] (M-4-2);
\draw[latex-latex] (M-1-3) to[out=225, in=135] (M-4-3);
\draw[latex-latex] (M-1-4) to[out=225, in=135] (M-4-4);
\draw[latex-latex] (M-1-5) to[out=225, in=135] (M-4-5);
\draw[latex-latex] (M-1-6) to[out=315, in=45] (M-4-6);
\draw[latex-latex] (M-1-7) to[out=315, in=45] (M-4-7);
\draw[latex-latex] (M-1-8) to[out=315, in=45] (M-4-8);
\draw[latex-latex] (M-1-9) to[out=315, in=45] (M-4-9);
\end{tikzpicture}
\caption{The T-duality orbit of the $0_6^{(1,7)}$.}
\label{fig:0617Orbit}
\end{figure}
\clearpage
\end{landscape}
\begingroup
\forceindent S-dualities:
\begin{multicols}{2}
\begin{itemize}
	\item $0_6^{(1,7)} \leftrightarrow 0_3^{(1,7)}$ See Fig. \ref{fig:532Orbit}
	\item $0_6^{(1,1,6)} \leftrightarrow 0_4^{(1,1,6)}$ See Fig. \ref{fig:146Orbit}
	\item $0_6^{(2,1,5)} \leftrightarrow 0_5^{(2,1,5)}$ See Fig. \ref{fig:2515Orbit}
	\item $0_6^{(3,1,4)} \leftrightarrow 0_6^{(3,1,4)}$ Self-dual
	\item $0_6^{(4,1,3)} \leftrightarrow 0_7^{(4,1,3)}$ See Fig. \ref{fig:0753Orbit}
	\item $0_6^{(5,1,2)} \leftrightarrow 0_8^{(5,1,2)}$
	\item $0_6^{(6,1,1)} \leftrightarrow 0_9^{(6,1,1)}$
	\item $0_6^{(7,1,0)} \leftrightarrow 0_{10}^{(7,1,0)}$
	\item $0_6^{(1,0,0,1,7)} \leftrightarrow 0_6^{(1,0,0,1,7)}$ Self-dual
	\item $0_6^{(1,0,1,1,6)} \leftrightarrow 0_7^{(1,0,1,1,6)}$ See Fig. \ref{fig:1726Orbit}
	\item $0_6^{(1,0,2,1,5)} \leftrightarrow 0_8^{(1,0,2,1,5)}$
	\item $0_6^{(1,0,3,1,4)} \leftrightarrow 0_9^{(1,0,3,1,4)}$
	\item $0_6^{(1,0,4,1,3)} \leftrightarrow 0_{10}^{(1,0,4,1,3)}$
	\item $0_6^{(1,0,5,1,2)} \leftrightarrow 0_{11}^{(1,0,5,1,2)}$
	\item $0_6^{(1,0,6,1,1)} \leftrightarrow 0_{12}^{(1,0,6,1,1)}$
	\item $0_6^{(1,0,7,1,0)} \leftrightarrow 0_{13}^{(1,0,7,1,0)}$
\end{itemize}
\columnbreak
\forceindent M-theory origins:
\begin{itemize}
	\item $0_6^{(1,7)} \rightarrow 0^{(1,7)}$
	\item $0_6^{(1,1,6)} \rightarrow 1^{(1,1,6)} $
	\item $0_6^{(2,1,5)} \rightarrow 0^{(2,1,6)}$
	\item $0_6^{(3,1,4)} \rightarrow 0^{(3,2,4)}$
	\item $0_6^{(4,1,3)} \rightarrow 0^{(5,1,3)}$
	\item $0_6^{(5,1,2)} \rightarrow 0^{(1,5,1,2)}$
	\item $0_6^{(6,1,1)} \rightarrow 0^{(1,0,6,1,1)}$
	\item $0_6^{(7,1,0)} \rightarrow 0^{(1,0,0,7,1,0)}$
	\item $0_6^{(1,0,0,1,7)} \rightarrow 0^{(1,0,0,2,7)}$
	\item $0_6^{(1,0,1,1,6)} \rightarrow 0^{(1,0,2,1,6)}$
	\item $0_6^{(1,0,2,1,5)} \rightarrow 0^{(1,1,2,1,5)}$
	\item $0_6^{(1,0,3,1,4)} \rightarrow 0^{(2,0,3,1,4)}$
	\item $0_6^{(1,0,4,1,3)} \rightarrow 0^{(1,1,0,4,1,3)}$
	\item $0_6^{(1,0,5,1,2)} \rightarrow 0^{(1,0,1,0,5,1,2)}$
	\item $0_6^{(1,0,6,1,1)} \rightarrow 0^{(1,0,0,1,0,6,1,1)}$
	\item $0_6^{(1,0,7,1,0)} \rightarrow 0^{(1,0,0,0,1,0,7,1,0)}$
\end{itemize}
\end{multicols}
\endgroup
\clearpage
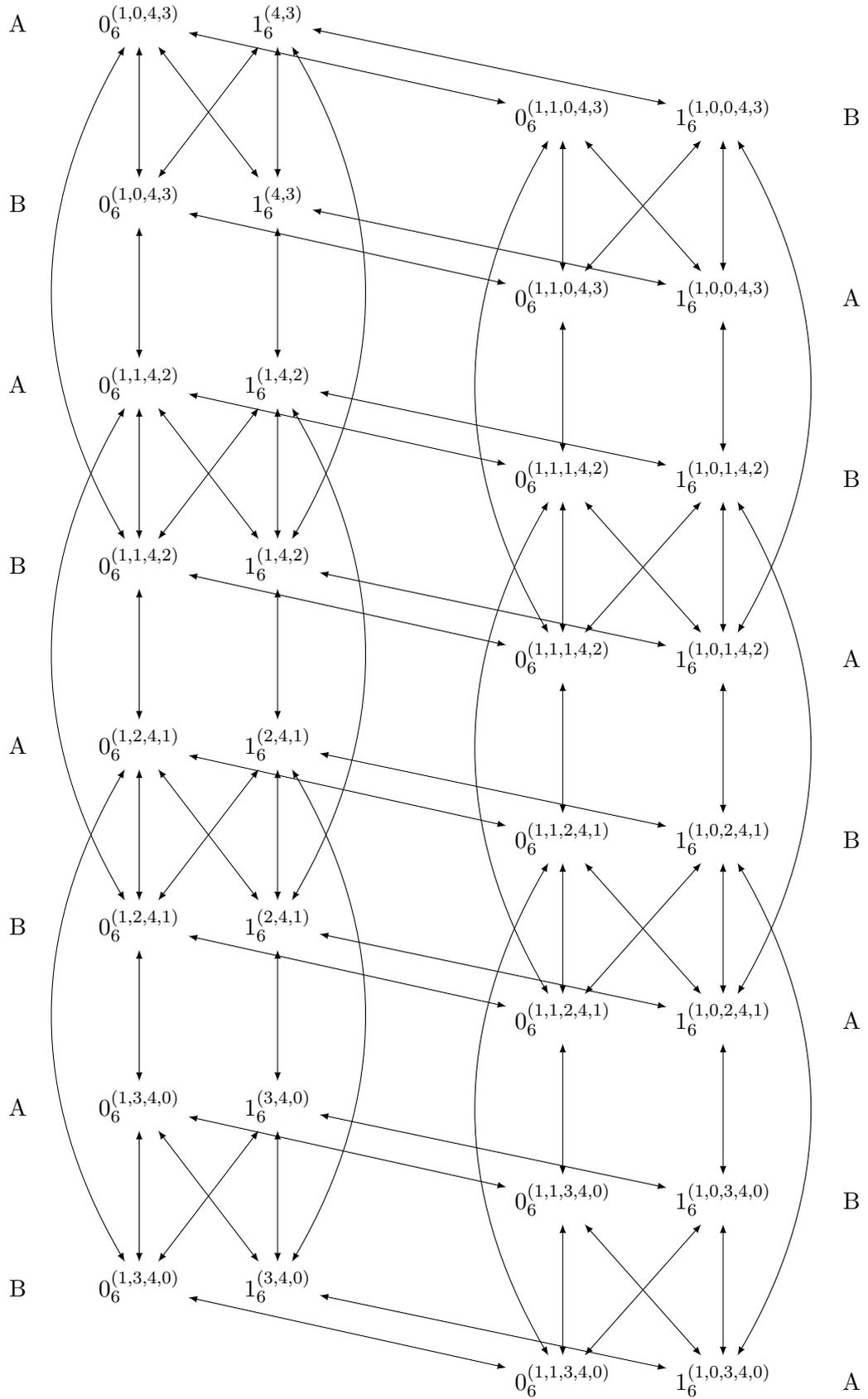
\begin{figure}[H]
\centering
\begin{tikzpicture}
\begin{scope}
\matrix(M)[matrix of math nodes, row sep=5.5em, column sep=2em, minimum width=2em]{
\clap{\text{A}} & 0_6^{(1,0,4,3)} & 1_6^{(4,3)}\\
\clap{\text{B}} & 0_6^{(1,0,4,3)} & 1_6^{(4,3)}\\
\clap{\text{A}} & 0_6^{(1,1,4,2)} & 1_6^{(1,4,2)}\\
\clap{\text{B}} & 0_6^{(1,1,4,2)} & 1_6^{(1,4,2)}\\
\clap{\text{A}} & 0_6^{(1,2,4,1)} & 1_6^{(2,4,1)}\\
\clap{\text{B}} & 0_6^{(1,2,4,1)} & 1_6^{(2,4,1)}\\
\clap{\text{A}} & 0_6^{(1,3,4,0)} & 1_6^{(3,4,0)}\\
\clap{\text{B}} & 0_6^{(1,3,4,0)} & 1_6^{(3,4,0)}\\
};
\end{scope}
\begin{scope}[xshift=8.5cm, yshift=-1.5cm]
\matrix(N)[matrix of math nodes, row sep=5.5em, column sep=2em, minimum width=2em]{
0_6^{(1,1,0,4,3)} & 1_6^{(1,0,0,4,3)} & \clap{\text{B}}\\
0_6^{(1,1,0,4,3)} & 1_6^{(1,0,0,4,3)} & \clap{\text{A}}\\
0_6^{(1,1,1,4,2)} & 1_6^{(1,0,1,4,2)} & \clap{\text{B}}\\
0_6^{(1,1,1,4,2)} & 1_6^{(1,0,1,4,2)} & \clap{\text{A}}\\
0_6^{(1,1,2,4,1)} & 1_6^{(1,0,2,4,1)} & \clap{\text{B}}\\
0_6^{(1,1,2,4,1)} & 1_6^{(1,0,2,4,1)} & \clap{\text{A}}\\
0_6^{(1,1,3,4,0)} & 1_6^{(1,0,3,4,0)} & \clap{\text{B}}\\
0_6^{(1,1,3,4,0)} & 1_6^{(1,0,3,4,0)} & \clap{\text{A}}\\
};
\end{scope}
\draw[latex-latex] (M-1-2) -- (M-2-2);
\draw[latex-latex] (M-1-3) -- (M-2-3);
\draw[latex-latex] (M-2-2) -- (M-3-2);
\draw[latex-latex] (M-2-3) -- (M-3-3);
\draw[latex-latex] (M-3-2) -- (M-4-2);
\draw[latex-latex] (M-3-3) -- (M-4-3);
\draw[latex-latex] (M-4-2) -- (M-5-2);
\draw[latex-latex] (M-4-3) -- (M-5-3);
\draw[latex-latex] (M-5-2) -- (M-6-2);
\draw[latex-latex] (M-5-3) -- (M-6-3);
\draw[latex-latex] (M-6-2) -- (M-7-2);
\draw[latex-latex] (M-6-3) -- (M-7-3);
\draw[latex-latex] (M-7-2) -- (M-8-2);
\draw[latex-latex] (M-7-3) -- (M-8-3);
\draw[latex-latex] (N-1-1) -- (N-2-1);
\draw[latex-latex] (N-1-2) -- (N-2-2);
\draw[latex-latex] (N-2-1) -- (N-3-1);
\draw[latex-latex] (N-2-2) -- (N-3-2);
\draw[latex-latex] (N-3-1) -- (N-4-1);
\draw[latex-latex] (N-3-2) -- (N-4-2);
\draw[latex-latex] (N-4-1) -- (N-5-1);
\draw[latex-latex] (N-4-2) -- (N-5-2);
\draw[latex-latex] (N-5-1) -- (N-6-1);
\draw[latex-latex] (N-5-2) -- (N-6-2);
\draw[latex-latex] (N-6-1) -- (N-7-1);
\draw[latex-latex] (N-6-2) -- (N-7-2);
\draw[latex-latex] (N-7-1) -- (N-8-1);
\draw[latex-latex] (N-7-2) -- (N-8-2);
\draw[latex-latex] (M-1-2) to[out=240, in=120] (M-4-2);
\draw[latex-latex] (M-1-3) to[out=300, in=60] (M-4-3);
\draw[latex-latex] (M-3-2) to[out=240, in=120] (M-6-2);
\draw[latex-latex] (M-3-3) to[out=300, in=60] (M-6-3);
\draw[latex-latex] (M-5-2) to[out=240, in=120] (M-8-2);
\draw[latex-latex] (M-5-3) to[out=300, in=60] (M-8-3);
\draw[latex-latex] (N-1-1) to[out=240, in=120] (N-4-1);
\draw[latex-latex] (N-1-2) to[out=300, in=60] (N-4-2);
\draw[latex-latex] (N-3-1) to[out=240, in=120] (N-6-1);
\draw[latex-latex] (N-3-2) to[out=300, in=60] (N-6-2);
\draw[latex-latex] (N-5-1) to[out=240, in=120] (N-8-1);
\draw[latex-latex] (N-5-2) to[out=300, in=60] (N-8-2);
\draw[latex-latex] (M-1-2) -- (M-2-3);
\draw[latex-latex] (M-2-2) -- (M-1-3);
\draw[latex-latex] (M-3-2) -- (M-4-3);
\draw[latex-latex] (M-4-2) -- (M-3-3);
\draw[latex-latex] (M-5-2) -- (M-6-3);
\draw[latex-latex] (M-6-2) -- (M-5-3);
\draw[latex-latex] (M-7-2) -- (M-8-3);
\draw[latex-latex] (M-8-2) -- (M-7-3);
\draw[latex-latex] (N-1-1) -- (N-2-2);
\draw[latex-latex] (N-2-1) -- (N-1-2);
\draw[latex-latex] (N-3-1) -- (N-4-2);
\draw[latex-latex] (N-4-1) -- (N-3-2);
\draw[latex-latex] (N-5-1) -- (N-6-2);
\draw[latex-latex] (N-6-1) -- (N-5-2);
\draw[latex-latex] (N-7-1) -- (N-8-2);
\draw[latex-latex] (N-8-1) -- (N-7-2);
\draw[latex-latex] (M-1-2) --(N-1-1);
\draw[latex-latex] (M-1-3) --(N-1-2);
\draw[latex-latex] (M-2-2) --(N-2-1);
\draw[latex-latex] (M-2-3) --(N-2-2);
\draw[latex-latex] (M-3-2) --(N-3-1);
\draw[latex-latex] (M-3-3) --(N-3-2);
\draw[latex-latex] (M-4-2) --(N-4-1);
\draw[latex-latex] (M-4-3) --(N-4-2);
\draw[latex-latex] (M-5-2) --(N-5-1);
\draw[latex-latex] (M-5-3) --(N-5-2);
\draw[latex-latex] (M-6-2) --(N-6-1);
\draw[latex-latex] (M-6-3) --(N-6-2);
\draw[latex-latex] (M-7-2) --(N-7-1);
\draw[latex-latex] (M-7-3) --(N-7-2);
\draw[latex-latex] (M-8-2) --(N-8-1);
\draw[latex-latex] (M-8-3) --(N-8-2);
\end{tikzpicture}
\caption{The T-duality orbit of the $1_6^{(4,3)}$.}
\label{fig:1643Orbit}
\end{figure}
\begin{multicols}{2}
\forceindent S-dualities:
\begin{itemize}
	\item $0_6^{(1,0,4,3)} \leftrightarrow 0_6^{(1,0,4,3)}$ Self-dual
	\item $0_6^{(1,1,4,2)} \leftrightarrow 0_7^{(1,1,4,2)}$ See Fig. \ref{fig:0753Orbit}
	\item $0_6^{(1,2,4,1)} \leftrightarrow 0_8^{(1,2,4,1)}$
	\item $0_6^{(1,3,4,0)} \leftrightarrow 0_9^{(1,3,4,0)}$
	\item $0_6^{(1,1,0,4,3)} \leftrightarrow 0_9^{(1,1,0,4,3)}$
	\item $0_6^{(1,1,1,4,2)} \leftrightarrow 0_{10}^{(1,1,1,4,2)}$
	\item $0_6^{(1,1,2,4,1)} \leftrightarrow 0_{11}^{(1,1,2,4,1)}$
	\item $0_6^{(1,1,3,4,0)} \leftrightarrow 0_{12}^{(1,1,3,4,0)}$
	\item $1_6^{(4,3)} \leftrightarrow 1_4^{(4,3)}$ See Fig. \ref{fig:4413Orbit}
	\item $1_6^{(1,4,2)} \leftrightarrow 1_5^{(1,4,2)}$ See Fig. \ref{fig:2515Orbit}
	\item $1_6^{(2,4,1)} \leftrightarrow 1_6^{(2,4,1)}$ Self-dual
	\item $1_6^{(3,4,0)} \leftrightarrow 1_7^{(3,4,0)}$ See Fig. \ref{fig:17160Orbit}
	\item $1_6^{(1,0,0,4,3)} \leftrightarrow 1_7^{(1,0,0,4,3)}$ See Fig. \ref{fig:27133Orbit}
	\item $1_6^{(1,0,1,4,2)} \leftrightarrow 1_8^{(1,0,1,4,2)}$
	\item $1_6^{(1,0,2,4,1)} \leftrightarrow 1_9^{(1,0,2,4,1)}$
	\item $1_6^{(1,0,3,4,0)} \leftrightarrow 1_{10}^{(1,0,3,4,0)}$
\end{itemize}
\columnbreak
\forceindent M-theory origins:
\begin{itemize}
	\item $0_6^{(1,0,4,3)} \rightarrow 0^{(1,0,5,3)}$
	\item $0_6^{(1,1,4,2)} \rightarrow 0^{(1,2,4,2)}$
	\item $0_6^{(1,2,4,1)} \rightarrow 0^{(2,2,4,1)}$
	\item $0_6^{(1,3,4,0)} \rightarrow 0^{(1,1,3,4,0)}$
	\item $0_6^{(1,1,0,4,3)} \rightarrow 0^{(2,1,0,4,3)}$
	\item $0_6^{(1,1,1,4,2)} \rightarrow 0^{(1,1,1,1,4,2)}$
	\item $0_6^{(1,1,2,4,1)} \rightarrow 0^{(1,0,1,1,2,4,1)}$
	\item $0_6^{(1,1,3,4,0)} \rightarrow 0^{(1,0,0,1,1,3,4,0)}$
	\item $1_6^{(4,3)} \rightarrow 2^{(4,3)}$
	\item $1_6^{(1,4,2)} \rightarrow 1^{(1,4,3)} $
	\item $1_6^{(2,4,1)} \rightarrow 1^{(2,5,1)}$
	\item $1_6^{(3,4,0)} \rightarrow 1^{(4,4,0)}$
	\item $1_6^{(1,0,0,4,3)} \rightarrow 1^{(1,0,1,4,3)}$
	\item $1_6^{(1,0,1,4,2)} \rightarrow 1^{(1,1,1,4,2)}$
	\item $1_6^{(1,0,2,4,1)} \rightarrow 1^{(2,0,2,4,1)}$
	\item $1_6^{(1,0,3,4,0)} \rightarrow 1^{(1,1,0,3,4,0)}$
\end{itemize}
\end{multicols}
\section{\texorpdfstring{$g_s^{-7}$}{gs-7} Duality Orbits}
\thispagestyle{empty}
\begin{figure}[H]
\centering
\begin{tikzpicture}
\begin{scope}
\matrix(M)[matrix of math nodes, row sep=1.75em, column sep=1em, minimum width=1em]{
\clap{\text{A}} & 0_7^{(1,0,0,2,6)} & & 0_7^{(1,0,2,0,6)}\\
\clap{\text{B}} & & 0_7^{(1,0,1,1,6)} &\\
\clap{\text{A}} & & 0_7^{(1,1,1,1,5)} &\\
\clap{\text{B}} & 0_7^{(1,1,0,2,5)} & & 0_7^{(1,1,2,0,5)}\\
\clap{\text{A}} & 0_7^{(1,2,0,2,4)} & & 0_7^{(1,2,2,0,4)}\\
\clap{\text{B}} & & 0_7^{(1,2,1,1,4)} &\\
\clap{\text{A}} & & 0_7^{(1,3,1,1,3)} &\\
\clap{\text{B}} & 0_7^{(1,3,0,2,3)} & & 0_7^{(1,3,2,0,3)}\\
\clap{\text{A}} & 0_7^{(1,4,0,2,2)} & & 0_7^{(1,4,2,0,2)}\\
\clap{\text{B}} & & 0_7^{(1,4,1,1,2)} &\\
\clap{\text{A}} & & 0_7^{(1,5,1,1,1)} & \\
\clap{\text{B}} & 0_7^{(1,5,0,2,1)} & & 0_7^{(1,5,2,0,1)}\\
\clap{\text{A}} & 0_7^{(1,6,0,2,0)} & & 0_7^{(1,6,2,0,0)}\\
\clap{\text{B}} & & 0_7^{(1,6,1,1,0)} & \\
};
\end{scope}
\begin{scope}[xshift=7.6cm, yshift=-1.5cm]
\matrix(N)[matrix of math nodes, row sep=1.75em, column sep=1em, minimum width=1em]{
1_7^{(2,6)} & & 1_7^{(2,0,6)} & \clap{\text{B}}\\
& 1_7^{(1,1,6)} & & \clap{\text{A}}\\
& 1_7^{(1,1,1,5)} & & \clap{\text{B}}\\
1_7^{(1,0,2,5)} & & 1_7^{(1,2,0,5)} & \clap{\text{A}}\\
1_7^{(2,0,2,4)} & & 1_7^{(2,2,0,4)} & \clap{\text{B}}\\
& 1_7^{(2,1,1,4)} & & \clap{\text{A}}\\
& 1_7^{(3,1,1,3)} & & \clap{\text{B}}\\
1_7^{(3,0,2,3)} & & 1_7^{(3,2,0,3)} & \clap{\text{A}}\\
1_7^{(4,0,2,2)} & & 1_7^{(4,2,0,2)} & \clap{\text{B}}\\
& 1_7^{(4,1,1,2)} & & \clap{\text{A}}\\
& 1_7^{(5,1,1,1)} & & \clap{\text{B}}\\
1_7^{(5,0,2,1)} & & 1_7^{(5,2,0,1)} & \clap{\text{A}}\\
1_7^{(6,0,2,0)} & & 1_7^{(6,2,0,0)} & \clap{\text{B}}\\
& 1_7^{(6,1,1,0)} & & \clap{\text{A}}\\
};
\end{scope}
\draw[latex-latex] (M-1-2) -- (M-4-2);
\draw[latex-latex] (M-4-2) -- (M-5-2);
\draw[latex-latex] (M-5-2) -- (M-8-2);
\draw[latex-latex] (M-8-2) -- (M-9-2);
\draw[latex-latex] (M-9-2) -- (M-12-2);
\draw[latex-latex] (M-12-2) -- (M-13-2);
\draw[latex-latex] (M-2-3) -- (M-3-3);
\draw[latex-latex] (M-3-3) -- (M-6-3);
\draw[latex-latex] (M-6-3) -- (M-7-3);
\draw[latex-latex] (M-7-3) -- (M-10-3);
\draw[latex-latex] (M-10-3) -- (M-11-3);
\draw[latex-latex] (M-11-3) -- (M-14-3);
\draw[latex-latex] (M-1-4) -- (M-4-4);
\draw[latex-latex] (M-4-4) -- (M-5-4);
\draw[latex-latex] (M-5-4) -- (M-8-4);
\draw[latex-latex] (M-8-4) -- (M-9-4);
\draw[latex-latex] (M-9-4) -- (M-12-4);
\draw[latex-latex] (M-12-4) -- (M-13-4);
\draw[latex-latex] (N-1-1) -- (N-4-1);
\draw[latex-latex] (N-4-1) -- (N-5-1);
\draw[latex-latex] (N-5-1) -- (N-8-1);
\draw[latex-latex] (N-8-1) -- (N-9-1);
\draw[latex-latex] (N-9-1) -- (N-12-1);
\draw[latex-latex] (N-12-1) -- (N-13-1);
\draw[latex-latex] (N-2-2) -- (N-3-2);
\draw[latex-latex] (N-3-2) -- (N-6-2);
\draw[latex-latex] (N-6-2) -- (N-7-2);
\draw[latex-latex] (N-7-2) -- (N-10-2);
\draw[latex-latex] (N-10-2) -- (N-11-2);
\draw[latex-latex] (N-11-2) -- (N-14-2);
\draw[latex-latex] (N-1-3) -- (N-4-3);
\draw[latex-latex] (N-4-3) -- (N-5-3);
\draw[latex-latex] (N-5-3) -- (N-8-3);
\draw[latex-latex] (N-8-3) -- (N-9-3);
\draw[latex-latex] (N-9-3) -- (N-12-3);
\draw[latex-latex] (N-12-3) -- (N-13-3);
\draw[latex-latex] (M-1-2) -- (M-2-3);
\draw[latex-latex] (M-2-3) -- (M-1-4);
\draw[latex-latex] (M-4-2) -- (M-3-3);
\draw[latex-latex] (M-3-3) -- (M-4-4);
\draw[latex-latex] (M-5-2) -- (M-6-3);
\draw[latex-latex] (M-6-3) -- (M-5-4);
\draw[latex-latex] (M-8-2) -- (M-7-3);
\draw[latex-latex] (M-7-3) -- (M-8-4);
\draw[latex-latex] (M-9-2) -- (M-10-3);
\draw[latex-latex] (M-10-3) -- (M-9-4);
\draw[latex-latex] (M-12-2) -- (M-11-3);
\draw[latex-latex] (M-11-3) -- (M-12-4);
\draw[latex-latex] (M-13-2) -- (M-14-3);
\draw[latex-latex] (M-14-3) -- (M-13-4);
\draw[latex-latex] (N-1-1) -- (N-2-2);
\draw[latex-latex] (N-2-2) -- (N-1-3);
\draw[latex-latex] (N-4-1) -- (N-3-2);
\draw[latex-latex] (N-3-2) -- (N-4-3);
\draw[latex-latex] (N-5-1) -- (N-6-2);
\draw[latex-latex] (N-6-2) -- (N-5-3);
\draw[latex-latex] (N-8-1) -- (N-7-2);
\draw[latex-latex] (N-7-2) -- (N-8-3);
\draw[latex-latex] (N-9-1) -- (N-10-2);
\draw[latex-latex] (N-10-2) -- (N-9-3);
\draw[latex-latex] (N-12-1) -- (N-11-2);
\draw[latex-latex] (N-11-2) -- (N-12-3);
\draw[latex-latex] (N-13-1) -- (N-14-2);
\draw[latex-latex] (N-14-2) -- (N-13-3);
\draw[latex-latex] (M-1-2) --(N-1-1);
\draw[latex-latex] (M-2-3) --(N-2-2);
\draw[latex-latex] (M-1-4) --(N-1-3);
\draw[latex-latex] (M-4-2) --(N-4-1);
\draw[latex-latex] (M-3-3) --(N-3-2);
\draw[latex-latex] (M-4-4) --(N-4-3);
\draw[latex-latex] (M-5-2) --(N-5-1);
\draw[latex-latex] (M-6-3) --(N-6-2);
\draw[latex-latex] (M-5-4) --(N-5-3);
\draw[latex-latex] (M-8-2) --(N-8-1);
\draw[latex-latex] (M-7-3) --(N-7-2);
\draw[latex-latex] (M-8-4) --(N-8-3);
\draw[latex-latex] (M-9-2) --(N-9-1);
\draw[latex-latex] (M-10-3) --(N-10-2);
\draw[latex-latex] (M-9-4) --(N-9-3);
\draw[latex-latex] (M-12-2) --(N-12-1);
\draw[latex-latex] (M-11-3) --(N-11-2);
\draw[latex-latex] (M-12-4) --(N-12-3);
\draw[latex-latex] (M-13-2) --(N-13-1);
\draw[latex-latex] (M-14-3) --(N-14-2);
\draw[latex-latex] (M-13-4) --(N-13-3);
\end{tikzpicture}
\caption{The T-duality orbit of the $1_7^{(2,6)}$.}
\label{fig:1726Orbit}
\end{figure}
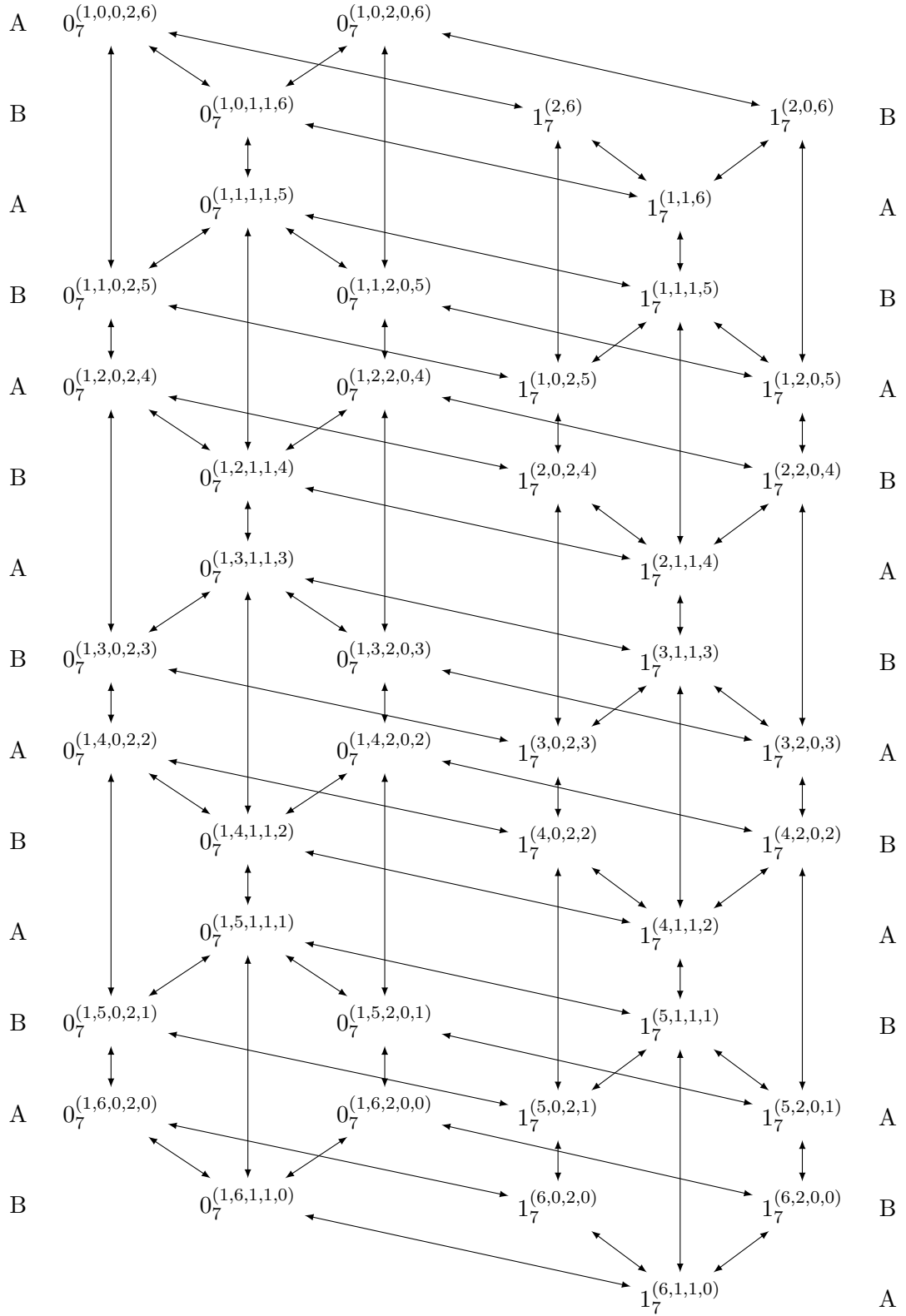
\begin{multicols}{2}
\forceindent S-dualities: 
\begin{itemize}
	\item $0_7^{(1,0,1,1,6)} \leftrightarrow 0_6^{(1,0,1,1,6)}$ See Fig. \ref{fig:0617Orbit}
	\item $0_7^{(1,1,0,2,5)} \leftrightarrow 0_7^{(1,1,0,2,5)}$ Self-dual
	\item $0_7^{(1,1,2,0,5)} \leftrightarrow 0_8^{(1,1,2,0,5)}$
	\item $0_7^{(1,2,1,1,4)} \leftrightarrow 0_9^{(1,2,1,1,4)}$
	\item $0_7^{(1,3,0,2,3)} \leftrightarrow 0_{10}^{(1,3,0,2,3)}$
	\item $0_7^{(1,3,2,0,3)} \leftrightarrow 0_{11}^{(1,3,2,0,3)}$
	\item $0_7^{(1,4,1,1,2)} \leftrightarrow 0_{12}^{(1,4,1,1,2)}$
	\item $0_7^{(1,5,0,2,1)} \leftrightarrow 0_{13}^{(1,5,0,2,1)}$
	\item $0_7^{(1,5,2,0,1)} \leftrightarrow 0_{14}^{(1,5,2,0,1)}$
	\item $0_7^{(1,6,1,1,0)} \leftrightarrow 0_{15}^{(1,6,1,1,0)}$
	\item $1_7^{(2,6)} \leftrightarrow 1_3^{(2,6)}$ See Fig. \ref{fig:532Orbit}
	\item $1_7^{(2,0,6)} \leftrightarrow 1_4^{(2,0,6)}$ See Fig. \ref{fig:146Orbit}
	\item $1_7^{(1,1,1,5)} \leftrightarrow 1_5^{(1,1,1,5)}$ See Fig. \ref{fig:2515Orbit}
	\item $1_7^{(2,0,2,4)} \leftrightarrow 1_6^{(2,0,2,4)}$ See Fig. \ref{fig:3624Orbit}
	\item $1_7^{(2,2,0,4)} \leftrightarrow 1_7^{(2,2,0,4)}$ Self-dual
	\item $1_7^{(3,1,1,3)} \leftrightarrow 1_8^{(3,1,1,3)}$
	\item $1_7^{(4,0,2,2)} \leftrightarrow 1_9^{(4,0,2,2)}$
	\item $1_7^{(4,2,0,2)} \leftrightarrow 1_{10}^{(4,2,0,2)}$
	\item $1_7^{(5,1,1,1)} \leftrightarrow 1_{11}^{(5,1,1,1)}$
	\item $1_7^{(6,0,2,0)} \leftrightarrow 1_{12}^{(6,0,2,0)}$
	\item $1_7^{(6,2,0,0)} \leftrightarrow 1_{13}^{(6,2,0,0)}$
\end{itemize}
\columnbreak
\forceindent M-theory origins:
\begin{itemize}
	\item $0_7^{(1,0,0,2,6)} \rightarrow 0^{(1,0,0,2,7)}$
	\item $0_7^{(1,0,2,0,6)} \rightarrow 0^{(1,0,2,1,6)}$
	\item $0_7^{(1,1,1,1,5)} \rightarrow 0^{(1,1,2,1,5)}$
	\item $0_7^{(1,2,0,2,4)} \rightarrow 0^{(1,3,0,2,4)}$
	\item $0_7^{(1,2,2,0,4)} \rightarrow 0^{(2,2,2,0,4)}$
	\item $0_7^{(1,3,1,1,3)} \rightarrow 0^{(1,1,3,1,1,3)}$
	\item $0_7^{(1,4,0,2,2)} \rightarrow 0^{(1,0,1,4,0,2,2)}$
	\item $0_7^{(1,4,2,0,2)} \rightarrow 0^{(1,0,0,1,4,2,0,2)}$
	\item $0_7^{(1,5,1,1,1)} \rightarrow 0^{(1,0,0,0,1,5,1,1,1)}$
	\item $0_7^{(1,6,0,2,0)} \rightarrow 0^{(1,0,0,0,0,1,6,0,2,0)}$
	\item $0_7^{(1,6,2,0,0)} \rightarrow 0^{(1,0,0,0,0,0,1,6,2,0,0)}$
	\item $1_7^{(1,1,6)} \rightarrow 1^{(1,1,6)}$
	\item $1_7^{(1,0,2,5)} \rightarrow 2^{(1,0,2,5)}$
	\item $1_7^{(1,2,0,5)} \rightarrow 1^{(1,2,0,6)}$
	\item $1_7^{(2,1,1,4)} \rightarrow 1^{(2,1,2,4)}$
	\item $1_7^{(3,0,2,3)} \rightarrow 1^{(3,1,2,3)}$
	\item $1_7^{(3,2,0,3)} \rightarrow 1^{(4,2,0,3)}$
	\item $1_7^{(4,1,1,2)} \rightarrow 1^{(1,4,1,1,2)}$
	\item $1_7^{(5,0,2,1)} \rightarrow 1^{(1,0,5,0,2,1)}$
	\item $1_7^{(5,2,0,1)} \rightarrow 1^{(1,0,0,5,2,0,1)}$
	\item $1_7^{(6,1,1,0)} \rightarrow 1^{(1,0,0,0,6,1,1,0)}$
\end{itemize}
\end{multicols}
\clearpage
\thispagestyle{empty}
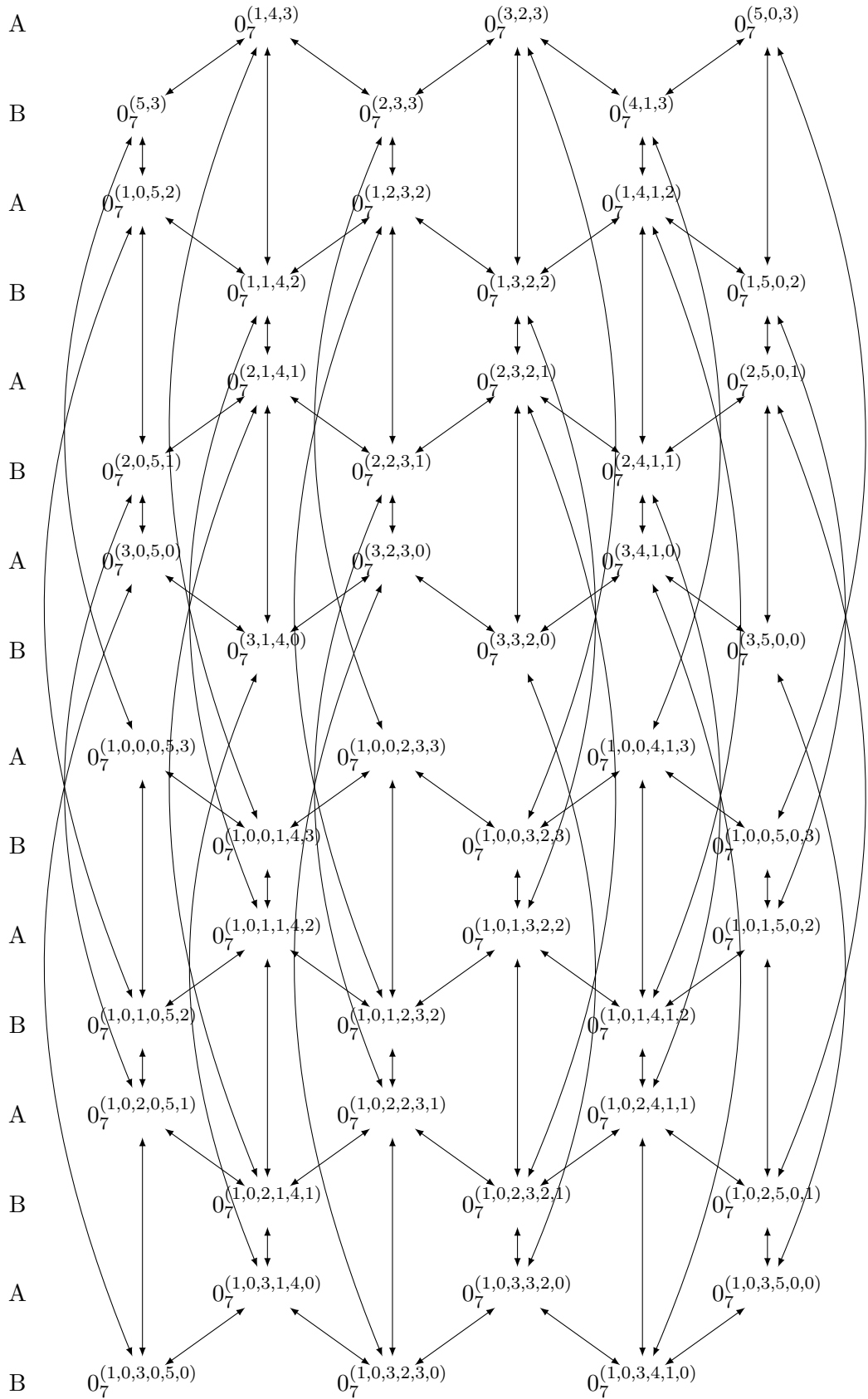
\begin{figure}[H]
\centering
\begin{tikzpicture}
\begin{scope}
\matrix(M)[matrix of math nodes, row sep=1.7em, column sep=3.35em, minimum width=1.8em]{
\clap{\text{A}} & & \mathclap{0_7^{(1,4,3)}} & & \mathclap{0_7^{(3,2,3)}} & & \mathclap{0_7^{(5,0,3)}}\\
\clap{\text{B}} & \mathclap{0_7^{(5,3)}} & & \mathclap{0_7^{(2,3,3)}} & & \mathclap{0_7^{(4,1,3)}}\\
\clap{\text{A}} & \mathclap{0_7^{(1,0,5,2)}} & & \mathclap{0_7^{(1,2,3,2)}} & & \mathclap{0_7^{(1,4,1,2)}}\\
\clap{\text{B}} & & \mathclap{0_7^{(1,1,4,2)}} & & \mathclap{0_7^{(1,3,2,2)}} & & \mathclap{0_7^{(1,5,0,2)}}\\
\clap{\text{A}} & &\mathclap{0_7^{(2,1,4,1)}} & & \mathclap{0_7^{(2,3,2,1)}} & & \mathclap{0_7^{(2,5,0,1)}}\\
\clap{\text{B}} & \mathclap{0_7^{(2,0,5,1)}} & & \mathclap{0_7^{(2,2,3,1)}} & & \mathclap{0_7^{(2,4,1,1)}}\\
\clap{\text{A}} & \mathclap{0_7^{(3,0,5,0)}} & & \mathclap{0_7^{(3,2,3,0)}} & & \mathclap{0_7^{(3,4,1,0)}}\\
\clap{\text{B}} & & \mathclap{0_7^{(3,1,4,0)}} & & \mathclap{0_7^{(3,3,2,0)}} & & \mathclap{0_7^{(3,5,0,0)}}\\
};
\end{scope}
\begin{scope}[yshift=-11.7cm]
\matrix(N)[matrix of math nodes, row sep=1.7em, column sep=3.35em, minimum width=1.8em]{
\clap{\text{A}} & \mathclap{0_7^{(1,0,0,0,5,3)}} & & \mathclap{0_7^{(1,0,0,2,3,3)}} & & \mathclap{0_7^{(1,0,0,4,1,3)}}\\
\clap{\text{B}} & & \mathclap{0_7^{(1,0,0,1,4,3)}} & & \mathclap{0_7^{(1,0,0,3,2,3)}} & & \mathclap{0_7^{(1,0,0,5,0,3)}}\\
\clap{\text{A}} & & \mathclap{0_7^{(1,0,1,1,4,2)}} & & \mathclap{0_7^{(1,0,1,3,2,2)}} & & \mathclap{0_7^{(1,0,1,5,0,2)}}\\
\clap{\text{B}} &  \mathclap{0_7^{(1,0,1,0,5,2)}} & & \mathclap{0_7^{(1,0,1,2,3,2)}} & & \mathclap{0_7^{(1,0,1,4,1,2)}}\\
\clap{\text{A}} & \mathclap{0_7^{(1,0,2,0,5,1)}} & & \mathclap{0_7^{(1,0,2,2,3,1)}} & & \mathclap{0_7^{(1,0,2,4,1,1)}}\\
\clap{\text{B}} & &\mathclap{0_7^{(1,0,2,1,4,1)}} & & \mathclap{0_7^{(1,0,2,3,2,1)}} & & \mathclap{0_7^{(1,0,2,5,0,1)}}\\
\clap{\text{A}} & & \mathclap{0_7^{(1,0,3,1,4,0)}} & & \mathclap{0_7^{(1,0,3,3,2,0)}} & & \mathclap{0_7^{(1,0,3,5,0,0)}}\\
\clap{\text{B}} &  \mathclap{0_7^{(1,0,3,0,5,0)}} & & \mathclap{0_7^{(1,0,3,2,3,0)}} & & \mathclap{0_7^{(1,0,3,4,1,0)}}\\
};
\draw[latex-latex] (M-2-2) -- (M-1-3);
\draw[latex-latex] (M-1-3) -- (M-2-4);
\draw[latex-latex] (M-2-4) -- (M-1-5);
\draw[latex-latex] (M-1-5) -- (M-2-6);
\draw[latex-latex] (M-2-6) -- (M-1-7);
\draw[latex-latex] (M-3-2) -- (M-4-3);
\draw[latex-latex] (M-4-3) -- (M-3-4);
\draw[latex-latex] (M-3-4) -- (M-4-5);
\draw[latex-latex] (M-4-5) -- (M-3-6);
\draw[latex-latex] (M-3-6) -- (M-4-7);
\draw[latex-latex] (M-6-2) -- (M-5-3);
\draw[latex-latex] (M-5-3) -- (M-6-4);
\draw[latex-latex] (M-6-4) -- (M-5-5);
\draw[latex-latex] (M-5-5) -- (M-6-6);
\draw[latex-latex] (M-6-6) -- (M-5-7);
\draw[latex-latex] (M-7-2) -- (M-8-3);
\draw[latex-latex] (M-8-3) -- (M-7-4);
\draw[latex-latex] (M-7-4) -- (M-8-5);
\draw[latex-latex] (M-8-5) -- (M-7-6);
\draw[latex-latex] (M-7-6) -- (M-8-7);
\draw[latex-latex] (M-2-2) -- (M-3-2);
\draw[latex-latex] (M-3-2) -- (M-6-2);
\draw[latex-latex] (M-6-2) -- (M-7-2);
\draw[latex-latex] (M-1-3) -- (M-4-3);
\draw[latex-latex] (M-4-3) -- (M-5-3);
\draw[latex-latex] (M-5-3) -- (M-8-3);
\draw[latex-latex] (M-2-4) -- (M-3-4);
\draw[latex-latex] (M-3-4) -- (M-6-4);
\draw[latex-latex] (M-6-4) -- (M-7-4);
\draw[latex-latex] (M-1-5) -- (M-4-5);
\draw[latex-latex] (M-4-5) -- (M-5-5);
\draw[latex-latex] (M-5-5) -- (M-8-5);
\draw[latex-latex] (M-2-6) -- (M-3-6);
\draw[latex-latex] (M-3-6) -- (M-6-6);
\draw[latex-latex] (M-6-6) -- (M-7-6);
\draw[latex-latex] (M-1-7) -- (M-4-7);
\draw[latex-latex] (M-4-7) -- (M-5-7);
\draw[latex-latex] (M-5-7) -- (M-8-7);
\draw[latex-latex] (N-1-2) -- (N-2-3);
\draw[latex-latex] (N-2-3) -- (N-1-4);
\draw[latex-latex] (N-1-4) -- (N-2-5);
\draw[latex-latex] (N-2-5) -- (N-1-6);
\draw[latex-latex] (N-1-6) -- (N-2-7);
\draw[latex-latex] (N-4-2) -- (N-3-3);
\draw[latex-latex] (N-3-3) -- (N-4-4);
\draw[latex-latex] (N-4-4) -- (N-3-5);
\draw[latex-latex] (N-3-5) -- (N-4-6);
\draw[latex-latex] (N-4-6) -- (N-3-7);
\draw[latex-latex] (N-5-2) -- (N-6-3);
\draw[latex-latex] (N-6-3) -- (N-5-4);
\draw[latex-latex] (N-5-4) -- (N-6-5);
\draw[latex-latex] (N-6-5) -- (N-5-6);
\draw[latex-latex] (N-5-6) -- (N-6-7);
\draw[latex-latex] (N-8-2) -- (N-7-3);
\draw[latex-latex] (N-7-3) -- (N-8-4);
\draw[latex-latex] (N-8-4) -- (N-7-5);
\draw[latex-latex] (N-7-5) -- (N-8-6);
\draw[latex-latex] (N-8-6) -- (N-7-7);
\draw[latex-latex] (N-1-2) -- (N-4-2);
\draw[latex-latex] (N-4-2) -- (N-5-2);
\draw[latex-latex] (N-5-2) -- (N-8-2);
\draw[latex-latex] (N-2-3) -- (N-3-3);
\draw[latex-latex] (N-3-3) -- (N-6-3);
\draw[latex-latex] (N-6-3) -- (N-7-3);
\draw[latex-latex] (N-1-4) -- (N-4-4);
\draw[latex-latex] (N-4-4) -- (N-5-4);
\draw[latex-latex] (N-5-4) -- (N-8-4);
\draw[latex-latex] (N-2-5) -- (N-3-5);
\draw[latex-latex] (N-3-5) -- (N-6-5);
\draw[latex-latex] (N-6-5) -- (N-7-5);
\draw[latex-latex] (N-1-6) -- (N-4-6);
\draw[latex-latex] (N-4-6) -- (N-5-6);
\draw[latex-latex] (N-5-6) -- (N-8-6);
\draw[latex-latex] (N-2-7) -- (N-3-7);
\draw[latex-latex] (N-3-7) -- (N-6-7);
\draw[latex-latex] (N-6-7) -- (N-7-7);
\draw[latex-latex] (M-2-2) to[out=247, in=112] (N-1-2);
\draw[latex-latex] (M-3-2) to[out=247, in=112] (N-4-2);
\draw[latex-latex] (M-6-2) to[out=247, in=112] (N-5-2);
\draw[latex-latex] (M-7-2) to[out=247, in=112] (N-8-2);
\draw[latex-latex] (M-1-3) to[out=247, in=112] (N-2-3);
\draw[latex-latex] (M-4-3) to[out=247, in=112] (N-3-3);
\draw[latex-latex] (M-5-3) to[out=247, in=112] (N-6-3);
\draw[latex-latex] (M-8-3) to[out=247, in=112] (N-7-3);
\draw[latex-latex] (M-2-4) to[out=247, in=112] (N-1-4);
\draw[latex-latex] (M-3-4) to[out=247, in=112] (N-4-4);
\draw[latex-latex] (M-6-4) to[out=247, in=112] (N-5-4);
\draw[latex-latex] (M-7-4) to[out=247, in=112] (N-8-4);
\draw[latex-latex] (M-1-5) to[out=292, in=67] (N-2-5);
\draw[latex-latex] (M-4-5) to[out=292, in=67] (N-3-5);
\draw[latex-latex] (M-5-5) to[out=292, in=67] (N-6-5);
\draw[latex-latex] (M-8-5) to[out=292, in=67] (N-7-5);
\draw[latex-latex] (M-2-6) to[out=292, in=67] (N-1-6);
\draw[latex-latex] (M-3-6) to[out=292, in=67] (N-4-6);
\draw[latex-latex] (M-6-6) to[out=292, in=67] (N-5-6);
\draw[latex-latex] (M-7-6) to[out=292, in=67] (N-8-6);
\draw[latex-latex] (M-1-7) to[out=292, in=67] (N-2-7);
\draw[latex-latex] (M-4-7) to[out=292, in=67] (N-3-7);
\draw[latex-latex] (M-5-7) to[out=292, in=67] (N-6-7);
\draw[latex-latex] (M-8-7) to[out=292, in=67] (N-7-7);
\end{scope}
\end{tikzpicture}
\caption{The T-duality orbit of the $0_7^{(5,3)}$.}
\label{fig:0753Orbit}
\end{figure}
\begin{multicols}{2}
\forceindent S-dualities:
\begin{itemize}
	\item $0_7^{(5,3)} \leftrightarrow 0_4^{(5,3)}$ See Fig. \ref{fig:4413Orbit}
	\item $0_7^{(2,3,3)} \leftrightarrow 0_5^{(2,3,3)}$ See Fig. \ref{fig:2515Orbit}
	\item $0_7^{(4,1,3)} \leftrightarrow 0_6^{(4,1,3)}$ See Fig. \ref{fig:0617Orbit}
	\item $0_7^{(1,1,4,2)} \leftrightarrow 0_6^{(1,1,4,2)}$ See Fig. \ref{fig:1643Orbit}
	\item $0_7^{(1,3,2,2)} \leftrightarrow 0_7^{(1,3,2,2)}$ Self-dual
	\item $0_7^{(1,5,0,2)} \leftrightarrow 0_8^{(1,5,0,2)}$
	\item $0_7^{(2,0,5,1)} \leftrightarrow 0_7^{(2,0,5,1)}$ Self-dual
	\item $0_7^{(2,2,3,1)} \leftrightarrow 0_8^{(2,2,3,1)}$
	\item $0_7^{(2,4,1,1)} \leftrightarrow 0_9^{(2,4,1,1)}$
	\item $0_7^{(3,1,4,0)} \leftrightarrow 0_9^{(3,1,4,0)}$ 
	\item $0_7^{(3,3,2,0)} \leftrightarrow 0_{10}^{(3,3,2,0)}$ 
	\item $0_7^{(3,5,0,0)} \leftrightarrow 0_{11}^{(3,5,0,0)}$ 
	\item $0_7^{(1,0,0,1,4,3)} \leftrightarrow 0_8^{(1,0,0,1,4,3)}$ 
	\item $0_7^{(1,0,0,3,2,3)} \leftrightarrow 0_9^{(1,0,0,3,2,3)}$ 
	\item $0_7^{(1,0,0,5,0,3)} \leftrightarrow 0_{10}^{(1,0,0,5,0,3)}$ 
	\item $0_7^{(1,0,1,0,5,2)} \leftrightarrow 0_9^{(1,0,1,0,5,2)}$ 
	\item $0_7^{(1,0,1,2,3,2)} \leftrightarrow 0_{10}^{(1,0,1,2,3,2)}$ 
	\item $0_7^{(1,0,1,4,1,2)} \leftrightarrow 0_{11}^{(1,0,1,4,1,2)}$ 
	\item $0_7^{(1,0,2,1,4,1)} \leftrightarrow 0_{11}^{(1,0,2,1,4,1)}$ 
	\item $0_7^{(1,0,2,3,2,1)} \leftrightarrow 0_{12}^{(1,0,2,3,2,1)}$ 
	\item $0_7^{(1,0,2,5,0,1)} \leftrightarrow 0_{13}^{(1,0,2,5,0,1)}$ 
	\item $0_7^{(1,0,3,0,5,0)} \leftrightarrow 0_{12}^{(1,0,3,0,5,0)}$ 
	\item $0_7^{(1,0,3,2,3,0)} \leftrightarrow 0_{13}^{(1,0,3,2,3,0)}$ 
	\item $0_7^{(1,0,3,4,1,0)} \leftrightarrow 0_{14}^{(1,0,3,4,1,0)}$ 
\end{itemize}
\columnbreak
\forceindent M-theory origins:
\begin{itemize}
	\item $0_7^{(1,4,3)} \rightarrow 1^{(1,4,3)}$
	\item $0_7^{(3,2,3)} \rightarrow 0^{(3,2,4)}$
	\item $0_7^{(5,0,3)} \rightarrow 0^{(5,1,3)}$
	\item $0_7^{(1,0,5,2)} \rightarrow 0^{(1,0,5,3)}$
	\item $0_7^{(1,2,3,2)} \rightarrow 0^{(1,2,4,2)}$
	\item $0_7^{(1,4,1,2)} \rightarrow 0^{(1,5,1,2)}$
	\item $0_7^{(2,1,4,1)} \rightarrow 0^{(2,2,4,1)}$
	\item $0_7^{(2,3,2,1)} \rightarrow 0^{(3,3,2,1)}$
	\item $0_7^{(2,5,0,1)} \rightarrow 0^{(1,2,5,0,1)}$
	\item $0_7^{(3,0,5,0)} \rightarrow 0^{(4,0,5,0)}$
	\item $0_7^{(3,2,3,0)} \rightarrow 0^{(1,3,2,3,0)}$
	\item $0_7^{(3,4,1,0)} \rightarrow 0^{(1,0,3,4,1,0)}$
	\item $0_7^{(1,0,0,0,5,3)} \rightarrow 0^{(1,0,0,1,5,3)}$
	\item $0_7^{(1,0,0,2,3,3)} \rightarrow 0^{(1,0,1,2,3,3)}$
	\item $0_7^{(1,0,0,4,1,3)} \rightarrow 0^{(1,1,0,4,1,3)}$
	\item $0_7^{(1,0,1,1,4,2)} \rightarrow 0^{(1,1,1,1,4,2)}$
	\item $0_7^{(1,0,1,3,2,2)} \rightarrow 0^{(2,0,1,3,2,2)}$
	\item $0_7^{(1,0,1,5,0,2)} \rightarrow 0^{(1,1,0,1,5,0,2)}$
	\item $0_7^{(1,0,2,0,5,1)} \rightarrow 0^{(2,0,2,0,5,1)}$
	\item $0_7^{(1,0,2,2,3,1)} \rightarrow 0^{(1,1,0,2,2,3,1)}$
	\item $0_7^{(1,0,2,4,1,1)} \rightarrow 0^{(1,0,1,0,2,4,1,1)}$
	\item $0_7^{(1,0,3,1,4,0)} \rightarrow 0^{(1,0,1,0,3,1,4,0)}$
	\item $0_7^{(1,0,3,3,2,0)} \rightarrow 0^{(1,0,0,1,0,3,3,2,0)}$
	\item $0_7^{(1,0,3,5,0,0)} \rightarrow 0^{(1,0,0,0,1,0,3,5,0,0)}$
\end{itemize}
\end{multicols}
\clearpage
\thispagestyle{empty}
\begin{landscape}
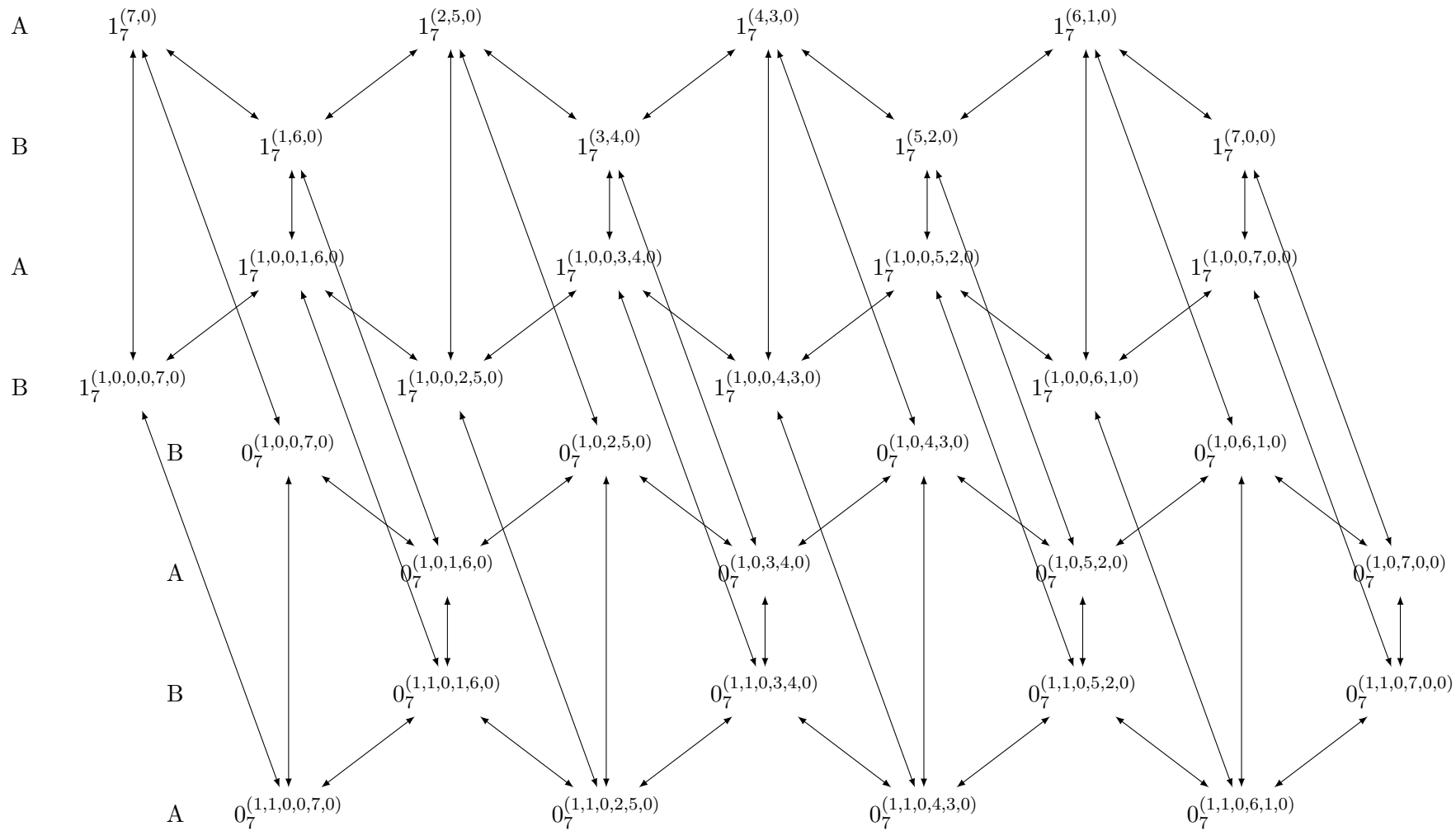
\begin{figure}[H]
\centering
\begin{tikzpicture}
\begin{scope}
\matrix(M)[matrix of math nodes, row sep=2.98em, column sep=1.41em, minimum width=0em]{
\text{A} & 1_7^{(7,0)} & & 1_7^{(2,5,0)} & & 1_7^{(4,3,0)} & & 1_7^{(6,1,0)} &\\
\text{B} & & 1_7^{(1,6,0)} & & 1_7^{(3,4,0)} & & 1_7^{(5,2,0)} & & 1_7^{(7,0,0)} \\
\text{A} & & 1_7^{(1,0,0,1,6,0)} & & 1_7^{(1,0,0,3,4,0)} & & 1_7^{(1,0,0,5,2,0)} & & 1_7^{(1,0,0,7,0,0)} \\
\text{B} & 1_7^{(1,0,0,0,7,0)} & & 1_7^{(1,0,0,2,5,0)} & & 1_7^{(1,0,0,4,3,0)} & & 1_7^{(1,0,0,6,1,0)} &\\
};
\end{scope}
\begin{scope}[yshift=-6.8cm,xshift=2.5cm]
\matrix(N)[matrix of math nodes, row sep=2.98em, column sep=1.41em, minimum width=0em]{
\text{B} & 0_7^{(1,0,0,7,0)} & & 0_7^{(1,0,2,5,0)} & & 0_7^{(1,0,4,3,0)} & & 0_7^{(1,0,6,1,0)} &\\
\text{A} & & 0_7^{(1,0,1,6,0)} & & 0_7^{(1,0,3,4,0)} & & 0_7^{(1,0,5,2,0)} & & 0_7^{(1,0,7,0,0)} \\
\text{B} & & 0_7^{(1,1,0,1,6,0)} & & 0_7^{(1,1,0,3,4,0)} & & 0_7^{(1,1,0,5,2,0)} & & 0_7^{(1,1,0,7,0,0)} \\
\text{A} & 0_7^{(1,1,0,0,7,0)} & & 0_7^{(1,1,0,2,5,0)} & & 0_7^{(1,1,0,4,3,0)} & & 0_7^{(1,1,0,6,1,0)} &\\
};
\end{scope}
\draw[latex-latex] (M-1-2) -- (M-2-3);
\draw[latex-latex] (M-2-3) -- (M-1-4);
\draw[latex-latex] (M-1-4) -- (M-2-5);
\draw[latex-latex] (M-2-5) -- (M-1-6);
\draw[latex-latex] (M-1-6) -- (M-2-7);
\draw[latex-latex] (M-2-7) -- (M-1-8);
\draw[latex-latex] (M-1-8) -- (M-2-9);
\draw[latex-latex] (M-4-2) -- (M-3-3);
\draw[latex-latex] (M-3-3) -- (M-4-4);
\draw[latex-latex] (M-4-4) -- (M-3-5);
\draw[latex-latex] (M-3-5) -- (M-4-6);
\draw[latex-latex] (M-4-6) -- (M-3-7);
\draw[latex-latex] (M-3-7) -- (M-4-8);
\draw[latex-latex] (M-4-8) -- (M-3-9);
\draw[latex-latex] (M-1-2) -- (M-4-2);
\draw[latex-latex] (M-2-3) -- (M-3-3);
\draw[latex-latex] (M-1-4) -- (M-4-4);
\draw[latex-latex] (M-2-5) -- (M-3-5);
\draw[latex-latex] (M-1-6) -- (M-4-6);
\draw[latex-latex] (M-2-7) -- (M-3-7);
\draw[latex-latex] (M-1-8) -- (M-4-8);
\draw[latex-latex] (M-2-9) -- (M-3-9);
\draw[latex-latex] (N-1-2) -- (N-2-3);
\draw[latex-latex] (N-2-3) -- (N-1-4);
\draw[latex-latex] (N-1-4) -- (N-2-5);
\draw[latex-latex] (N-2-5) -- (N-1-6);
\draw[latex-latex] (N-1-6) -- (N-2-7);
\draw[latex-latex] (N-2-7) -- (N-1-8);
\draw[latex-latex] (N-1-8) -- (N-2-9);
\draw[latex-latex] (N-4-2) -- (N-3-3);
\draw[latex-latex] (N-3-3) -- (N-4-4);
\draw[latex-latex] (N-4-4) -- (N-3-5);
\draw[latex-latex] (N-3-5) -- (N-4-6);
\draw[latex-latex] (N-4-6) -- (N-3-7);
\draw[latex-latex] (N-3-7) -- (N-4-8);
\draw[latex-latex] (N-4-8) -- (N-3-9);
\draw[latex-latex] (N-1-2) -- (N-4-2);
\draw[latex-latex] (N-2-3) -- (N-3-3);
\draw[latex-latex] (N-1-4) -- (N-4-4);
\draw[latex-latex] (N-2-5) -- (N-3-5);
\draw[latex-latex] (N-1-6) -- (N-4-6);
\draw[latex-latex] (N-2-7) -- (N-3-7);
\draw[latex-latex] (N-1-8) -- (N-4-8);
\draw[latex-latex] (N-2-9) -- (N-3-9);
\draw[latex-latex] (M-1-2) -- (N-1-2);
\draw[latex-latex] (M-4-2) -- (N-4-2);
\draw[latex-latex] (M-2-3) -- (N-2-3);
\draw[latex-latex] (M-3-3) -- (N-3-3);
\draw[latex-latex] (M-1-4) -- (N-1-4);
\draw[latex-latex] (M-4-4) -- (N-4-4);
\draw[latex-latex] (M-2-5) -- (N-2-5);
\draw[latex-latex] (M-3-5) -- (N-3-5);
\draw[latex-latex] (M-1-6) -- (N-1-6);
\draw[latex-latex] (M-4-6) -- (N-4-6);
\draw[latex-latex] (M-2-7) -- (N-2-7);
\draw[latex-latex] (M-3-7) -- (N-3-7);
\draw[latex-latex] (M-1-8) -- (N-1-8);
\draw[latex-latex] (M-4-8) -- (N-4-8);
\draw[latex-latex] (M-2-9) -- (N-2-9);
\draw[latex-latex] (M-3-9) -- (N-3-9);
\end{tikzpicture}
\caption{The T-duality orbit of the $1_7^{(1,6,0)}$.}
\label{fig:17160Orbit}
\end{figure}
\end{landscape}
\clearpage
\begin{multicols}{2}
\forceindent S-dualities:
\begin{itemize}
	\item $1_7^{(1,6,0)} \leftrightarrow 1_5^{(1,6,0)}$ See Fig. \ref{fig:2515Orbit}
	\item $1_7^{(3,4,0)} \leftrightarrow 1_6^{(3,4,0)}$ See Fig. \ref{fig:1643Orbit}
	\item $1_7^{(5,2,0)} \leftrightarrow 1_7^{(5,2,0)}$ Self-dual
	\item $1_7^{(7,0,0)} \leftrightarrow 1_8^{(7,0,0)}$
	\item $1_7^{(1,0,0,0,7,0)} \leftrightarrow 1_8^{(1,0,0,0,7,0)}$
	\item $1_7^{(1,0,0,2,5,0)} \leftrightarrow 1_9^{(1,0,0,2,5,0)}$
	\item $1_7^{(1,0,0,4,3,0)} \leftrightarrow 1_{10}^{(1,0,0,4,3,0)}$
	\item $1_7^{(1,0,0,6,1,0)} \leftrightarrow 1_{11}^{(1,0,0,6,1,0)}$
	\item $0_7^{(1,0,0,7,0)} \leftrightarrow 0_7^{(1,0,0,7,0)}$ Self-dual
	\item $0_7^{(1,0,2,5,0)} \leftrightarrow 0_8^{(1,0,2,5,0)}$
	\item $0_7^{(1,0,4,3,0)} \leftrightarrow 0_9^{(1,0,4,3,0)}$
	\item $0_7^{(1,0,6,1,0)} \leftrightarrow 0_{10}^{(1,0,6,1,0)}$
	\item $0_7^{(1,1,0,1,6,0)} \leftrightarrow 0_{11}^{(1,1,0,1,6,0)}$
	\item $0_7^{(1,1,0,3,4,0)} \leftrightarrow 0_{12}^{(1,1,0,3,4,0)}$
	\item $0_7^{(1,1,0,5,2,0)} \leftrightarrow 0_{13}^{(1,1,0,5,2,0)}$
	\item $0_7^{(1,1,0,7,0,0)} \leftrightarrow 0_{14}^{(1,1,0,7,0,0)}$
\end{itemize}
\columnbreak
\forceindent M-theory origins:
\begin{itemize}
	\item $1_7^{(7,0)} \rightarrow 2^{(7,0)}$
	\item $1_7^{(2,5,0)} \rightarrow 1^{(2,5,1)}$
	\item $1_7^{(4,3,0)} \rightarrow 1^{(4,4,0)}$
	\item $1_7^{(6,1,0)} \rightarrow 1^{(7,1,0)}$
	\item $1_7^{(1,0,0,1,6,0)} \rightarrow 1^{(1,0,1,1,6,0)}$
	\item $1_7^{(1,0,0,3,4,0)} \rightarrow 1^{(1,1,0,3,4,0)}$
	\item $1_7^{(1,0,0,5,2,0)} \rightarrow 1^{(2,0,0,5,2,0)}$
	\item $1_7^{(1,0,0,7,0,0)} \rightarrow 1^{(1,1,0,0,7,0,0)}$
	\item $0_7^{(1,0,1,6,0)} \rightarrow 0^{(1,0,2,6,0)}$
	\item $0_7^{(1,0,3,4,0)} \rightarrow 0^{(1,1,3,4,0)}$
	\item $0_7^{(1,0,5,2,0)} \rightarrow 0^{(2,0,5,2,0)}$
	\item $0_7^{(1,0,7,0,0)} \rightarrow 0^{(1,1,0,7,0,0)}$
	\item $0_7^{(1,1,0,0,7,0)} \rightarrow 0^{(2,1,0,0,7,0)}$
	\item $0_7^{(1,1,0,2,5,0)} \rightarrow 0^{(1,1,1,0,2,5,0)}$
	\item $0_7^{(1,1,0,4,3,0)} \rightarrow 0^{(1,0,1,1,0,4,3,0)}$
	\item $0_7^{(1,1,0,6,1,0)} \rightarrow 0^{(1,0,0,1,1,0,6,1,0)}$
\end{itemize}
\end{multicols}
\clearpage
\thispagestyle{empty}
\begin{landscape}
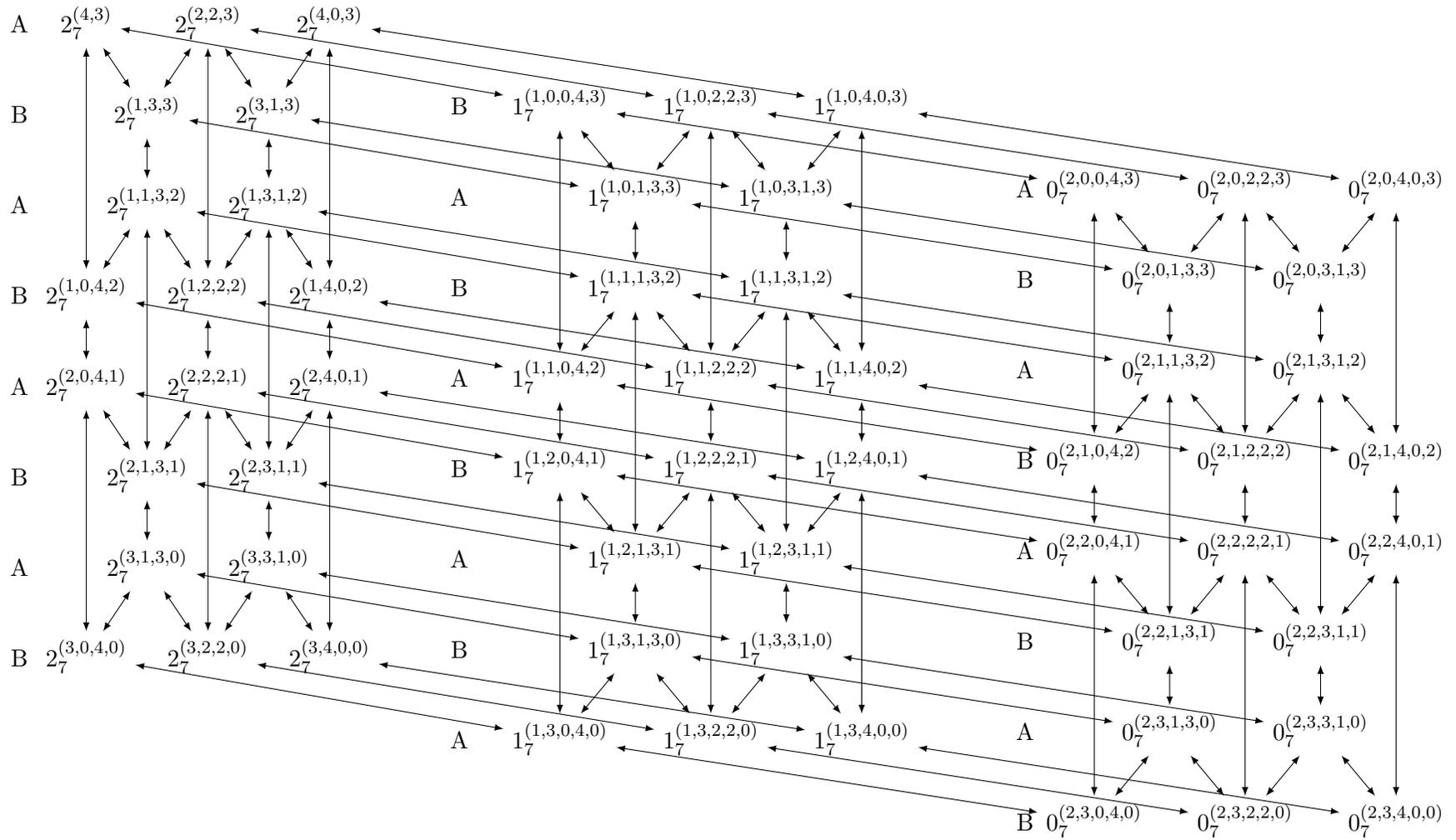
\begin{figure}[H]
\centering
\begin{tikzpicture}
\begin{scope}[xshift=-2.5cm]
\matrix(m)[matrix of math nodes, row sep=1.7em, column sep=-1em]{
\text{A}\phantom{\mathclap{2_7^{(4,3)}}}\\
\text{B}\phantom{\mathclap{2_7^{(1,3,3)}}}\\
\text{A}\phantom{\mathclap{2_7^{(1,1,3,2)}}}\\
\text{B}\phantom{\mathclap{2_7^{(1,0,4,2)}}}\\
\text{A}\phantom{\mathclap{2_7^{(2,0,4,1)}}}\\
\text{B}\phantom{\mathclap{2_7^{(2,1,3,1)}}}\\
\text{A}\phantom{\mathclap{2_7^{(3,1,3,0)}}}\\
\text{B}\phantom{\mathclap{2_7^{(3,0,4,0)}}}\\
};
\end{scope}
\begin{scope}[xshift=0.5cm]
\matrix(M)[matrix of math nodes, row sep=1.7em, column sep=-1.5em, minimum width=2em]{
2_7^{(4,3)} & & 2_7^{(2,2,3)} & & 2_7^{(4,0,3)}\\
& 2_7^{(1,3,3)} & & 2_7^{(3,1,3)} & \\
& 2_7^{(1,1,3,2)} & & 2_7^{(1,3,1,2)} & \\
2_7^{(1,0,4,2)} & & 2_7^{(1,2,2,2)} & & 2_7^{(1,4,0,2)}\\
2_7^{(2,0,4,1)} & & 2_7^{(2,2,2,1)} & & 2_7^{(2,4,0,1)}\\
& 2_7^{(2,1,3,1)} & & 2_7^{(2,3,1,1)} & \\
& 2_7^{(3,1,3,0)} & & 2_7^{(3,3,1,0)} & \\
2_7^{(3,0,4,0)} & & 2_7^{(3,2,2,0)} & & 2_7^{(3,4,0,0)}\\
};
\end{scope}
\begin{scope}[xshift=4.5cm, yshift=-1.3cm]
\matrix(n)[matrix of math nodes, row sep=1.7em]{
\text{B}\phantom{\mathclap{1_7^{(1,0,0,4,3)}}}\\
\text{A}\phantom{\mathclap{1_7^{(1,0,1,3,3)}}}\\
\text{B}\phantom{\mathclap{1_7^{(1,1,1,3,2)}}}\\
\text{A}\phantom{\mathclap{1_7^{(1,1,0,4,2)}}}\\
\text{B}\phantom{\mathclap{1_7^{(1,2,0,4,1)}}}\\
\text{A}\phantom{\mathclap{1_7^{(1,2,1,3,1)}}}\\
\text{B}\phantom{\mathclap{1_7^{(1,3,1,3,0)}}}\\
\text{A}\phantom{\mathclap{1_7^{(1,3,0,4,0)}}}\\
};
\end{scope}
\begin{scope}[xshift=8.5cm,yshift=-1.3cm]
\matrix(N)[matrix of math nodes, row sep=1.7em, column sep=-1.5em, minimum width=2em]{
1_7^{(1,0,0,4,3)} & & 1_7^{(1,0,2,2,3)} & & 1_7^{(1,0,4,0,3)}\\
& 1_7^{(1,0,1,3,3)} & & 1_7^{(1,0,3,1,3)} & \\
& 1_7^{(1,1,1,3,2)} & & 1_7^{(1,1,3,1,2)} & \\
1_7^{(1,1,0,4,2)} & & 1_7^{(1,1,2,2,2)} & & 1_7^{(1,1,4,0,2)}\\
1_7^{(1,2,0,4,1)} & & 1_7^{(1,2,2,2,1)} & & 1_7^{(1,2,4,0,1)}\\
& 1_7^{(1,2,1,3,1)} & & 1_7^{(1,2,3,1,1)} & \\
& 1_7^{(1,3,1,3,0)} & & 1_7^{(1,3,3,1,0)} & \\
1_7^{(1,3,0,4,0)} & & 1_7^{(1,3,2,2,0)} & & 1_7^{(1,3,4,0,0)}\\
};
\end{scope}
\begin{scope}[xshift=13.5cm,yshift=-2.6cm]
\matrix(p)[matrix of math nodes, row sep=1.7em]{
\text{A}\phantom{\mathclap{0_7^{(2,0,0,4,3)}}}\\
\text{B}\phantom{\mathclap{0_7^{(2,0,1,3,3)}}}\\
\text{A}\phantom{\mathclap{0_7^{(2,1,1,3,2)}}}\\
\text{B}\phantom{\mathclap{0_7^{(2,1,0,4,2)}}}\\
\text{A}\phantom{\mathclap{0_7^{(2,2,0,4,1)}}}\\
\text{B}\phantom{\mathclap{0_7^{(2,2,1,3,1)}}}\\
\text{A}\phantom{\mathclap{0_7^{(2,3,1,3,0)}}}\\
\text{B}\phantom{\mathclap{0_7^{(2,3,0,4,0)}}}\\
};
\end{scope}
\begin{scope}[xshift=17cm,yshift=-2.6cm]
\matrix(P)[matrix of math nodes, row sep=1.7em, column sep=-1.5em, minimum width=2em]{
0_7^{(2,0,0,4,3)} & & 0_7^{(2,0,2,2,3)} & & 0_7^{(2,0,4,0,3)}\\
& 0_7^{(2,0,1,3,3)} & & 0_7^{(2,0,3,1,3)} & \\
& 0_7^{(2,1,1,3,2)} & & 0_7^{(2,1,3,1,2)} & \\
0_7^{(2,1,0,4,2)} & & 0_7^{(2,1,2,2,2)} & & 0_7^{(2,1,4,0,2)}\\
0_7^{(2,2,0,4,1)} & & 0_7^{(2,2,2,2,1)} & & 0_7^{(2,2,4,0,1)}\\
& 0_7^{(2,2,1,3,1)} & & 0_7^{(2,2,3,1,1)} & \\
& 0_7^{(2,3,1,3,0)} & & 0_7^{(2,3,3,1,0)} & \\
0_7^{(2,3,0,4,0)} & & 0_7^{(2,3,2,2,0)} & & 0_7^{(2,3,4,0,0)}\\
};
\end{scope}
\draw[latex-latex] (M-1-1) -- (M-2-2);
\draw[latex-latex] (M-2-2) -- (M-1-3);
\draw[latex-latex] (M-1-3) -- (M-2-4);
\draw[latex-latex] (M-2-4) -- (M-1-5);
\draw[latex-latex] (M-4-1) -- (M-3-2);
\draw[latex-latex] (M-3-2) -- (M-4-3);
\draw[latex-latex] (M-4-3) -- (M-3-4);
\draw[latex-latex] (M-3-4) -- (M-4-5);
\draw[latex-latex] (M-5-1) -- (M-6-2);
\draw[latex-latex] (M-6-2) -- (M-5-3);
\draw[latex-latex] (M-5-3) -- (M-6-4);
\draw[latex-latex] (M-6-4) -- (M-5-5);
\draw[latex-latex] (M-8-1) -- (M-7-2);
\draw[latex-latex] (M-7-2) -- (M-8-3);
\draw[latex-latex] (M-8-3) -- (M-7-4);
\draw[latex-latex] (M-7-4) -- (M-8-5);
\draw[latex-latex] (M-1-1) -- (M-4-1);
\draw[latex-latex] (M-4-1) -- (M-5-1);
\draw[latex-latex] (M-5-1) -- (M-8-1);
\draw[latex-latex] (M-2-2) -- (M-3-2);
\draw[latex-latex] (M-3-2) -- (M-6-2);
\draw[latex-latex] (M-6-2) -- (M-7-2);
\draw[latex-latex] (M-1-3) -- (M-4-3);
\draw[latex-latex] (M-4-3) -- (M-5-3);
\draw[latex-latex] (M-5-3) -- (M-8-3);
\draw[latex-latex] (M-2-4) -- (M-3-4);
\draw[latex-latex] (M-3-4) -- (M-6-4);
\draw[latex-latex] (M-6-4) -- (M-7-4);
\draw[latex-latex] (M-1-5) -- (M-4-5);
\draw[latex-latex] (M-4-5) -- (M-5-5);
\draw[latex-latex] (M-5-5) -- (M-8-5);
\draw[latex-latex] (N-1-1) -- (N-2-2);
\draw[latex-latex] (N-2-2) -- (N-1-3);
\draw[latex-latex] (N-1-3) -- (N-2-4);
\draw[latex-latex] (N-2-4) -- (N-1-5);
\draw[latex-latex] (N-4-1) -- (N-3-2);
\draw[latex-latex] (N-3-2) -- (N-4-3);
\draw[latex-latex] (N-4-3) -- (N-3-4);
\draw[latex-latex] (N-3-4) -- (N-4-5);
\draw[latex-latex] (N-5-1) -- (N-6-2);
\draw[latex-latex] (N-6-2) -- (N-5-3);
\draw[latex-latex] (N-5-3) -- (N-6-4);
\draw[latex-latex] (N-6-4) -- (N-5-5);
\draw[latex-latex] (N-8-1) -- (N-7-2);
\draw[latex-latex] (N-7-2) -- (N-8-3);
\draw[latex-latex] (N-8-3) -- (N-7-4);
\draw[latex-latex] (N-7-4) -- (N-8-5);
\draw[latex-latex] (N-1-1) -- (N-4-1);
\draw[latex-latex] (N-4-1) -- (N-5-1);
\draw[latex-latex] (N-5-1) -- (N-8-1);
\draw[latex-latex] (N-2-2) -- (N-3-2);
\draw[latex-latex] (N-3-2) -- (N-6-2);
\draw[latex-latex] (N-6-2) -- (N-7-2);
\draw[latex-latex] (N-1-3) -- (N-4-3);
\draw[latex-latex] (N-4-3) -- (N-5-3);
\draw[latex-latex] (N-5-3) -- (N-8-3);
\draw[latex-latex] (N-2-4) -- (N-3-4);
\draw[latex-latex] (N-3-4) -- (N-6-4);
\draw[latex-latex] (N-6-4) -- (N-7-4);
\draw[latex-latex] (N-1-5) -- (N-4-5);
\draw[latex-latex] (N-4-5) -- (N-5-5);
\draw[latex-latex] (N-5-5) -- (N-8-5);
\draw[latex-latex] (P-1-1) -- (P-2-2);
\draw[latex-latex] (P-2-2) -- (P-1-3);
\draw[latex-latex] (P-1-3) -- (P-2-4);
\draw[latex-latex] (P-2-4) -- (P-1-5);
\draw[latex-latex] (P-4-1) -- (P-3-2);
\draw[latex-latex] (P-3-2) -- (P-4-3);
\draw[latex-latex] (P-4-3) -- (P-3-4);
\draw[latex-latex] (P-3-4) -- (P-4-5);
\draw[latex-latex] (P-5-1) -- (P-6-2);
\draw[latex-latex] (P-6-2) -- (P-5-3);
\draw[latex-latex] (P-5-3) -- (P-6-4);
\draw[latex-latex] (P-6-4) -- (P-5-5);
\draw[latex-latex] (P-8-1) -- (P-7-2);
\draw[latex-latex] (P-7-2) -- (P-8-3);
\draw[latex-latex] (P-8-3) -- (P-7-4);
\draw[latex-latex] (P-7-4) -- (P-8-5);
\draw[latex-latex] (P-1-1) -- (P-4-1);
\draw[latex-latex] (P-4-1) -- (P-5-1);
\draw[latex-latex] (P-5-1) -- (P-8-1);
\draw[latex-latex] (P-2-2) -- (P-3-2);
\draw[latex-latex] (P-3-2) -- (P-6-2);
\draw[latex-latex] (P-6-2) -- (P-7-2);
\draw[latex-latex] (P-1-3) -- (P-4-3);
\draw[latex-latex] (P-4-3) -- (P-5-3);
\draw[latex-latex] (P-5-3) -- (P-8-3);
\draw[latex-latex] (P-2-4) -- (P-3-4);
\draw[latex-latex] (P-3-4) -- (P-6-4);
\draw[latex-latex] (P-6-4) -- (P-7-4);
\draw[latex-latex] (P-1-5) -- (P-4-5);
\draw[latex-latex] (P-4-5) -- (P-5-5);
\draw[latex-latex] (P-5-5) -- (P-8-5);
\draw[latex-latex] (M-1-1) -- (N-1-1);
\draw[latex-latex] (M-1-3) -- (N-1-3);
\draw[latex-latex] (M-1-5) -- (N-1-5);
\draw[latex-latex] (M-2-2) -- (N-2-2);
\draw[latex-latex] (M-2-4) -- (N-2-4);
\draw[latex-latex] (M-3-2) -- (N-3-2);
\draw[latex-latex] (M-3-4) -- (N-3-4);
\draw[latex-latex] (M-4-1) -- (N-4-1);
\draw[latex-latex] (M-4-3) -- (N-4-3);
\draw[latex-latex] (M-4-5) -- (N-4-5);
\draw[latex-latex] (M-5-1) -- (N-5-1);
\draw[latex-latex] (M-5-3) -- (N-5-3);
\draw[latex-latex] (M-5-5) -- (N-5-5);
\draw[latex-latex] (M-6-2) -- (N-6-2);
\draw[latex-latex] (M-6-4) -- (N-6-4);
\draw[latex-latex] (M-7-2) -- (N-7-2);
\draw[latex-latex] (M-7-4) -- (N-7-4);
\draw[latex-latex] (M-8-1) -- (N-8-1);
\draw[latex-latex] (M-8-3) -- (N-8-3);
\draw[latex-latex] (M-8-5) -- (N-8-5);
\draw[latex-latex] (N-1-1) -- (P-1-1);
\draw[latex-latex] (N-1-3) -- (P-1-3);
\draw[latex-latex] (N-1-5) -- (P-1-5);
\draw[latex-latex] (N-2-2) -- (P-2-2);
\draw[latex-latex] (N-2-4) -- (P-2-4);
\draw[latex-latex] (N-3-2) -- (P-3-2);
\draw[latex-latex] (N-3-4) -- (P-3-4);
\draw[latex-latex] (N-4-1) -- (P-4-1);
\draw[latex-latex] (N-4-3) -- (P-4-3);
\draw[latex-latex] (N-4-5) -- (P-4-5);
\draw[latex-latex] (N-5-1) -- (P-5-1);
\draw[latex-latex] (N-5-3) -- (P-5-3);
\draw[latex-latex] (N-5-5) -- (P-5-5);
\draw[latex-latex] (N-6-2) -- (P-6-2);
\draw[latex-latex] (N-6-4) -- (P-6-4);
\draw[latex-latex] (N-7-2) -- (P-7-2);
\draw[latex-latex] (N-7-4) -- (P-7-4);
\draw[latex-latex] (N-8-1) -- (P-8-1);
\draw[latex-latex] (N-8-3) -- (P-8-3);
\draw[latex-latex] (N-8-5) -- (P-8-5);
\end{tikzpicture}
\caption{The T-duality orbit of the $2_7^{(1,3,3)}$.}
\label{fig:27133Orbit}
\end{figure}
\end{landscape}
\clearpage
\forceindent S-dualities:
\begin{multicols}{2}
\begin{itemize}
	\item $2_7^{(1,3,3)} \leftrightarrow 2_4^{(1,3,3)}$ See Fig. \ref{fig:4413Orbit}
	\item $2_7^{(3,1,3)} \leftrightarrow 2_5^{(3,1,3)}$ See Fig. \ref{fig:5522Orbit}
	\item $2_7^{(1,0,4,2)} \leftrightarrow 2_5^{(1,0,4,2)}$ See Fig. \ref{fig:2515Orbit}
	\item $2_7^{(1,2,2,2)} \leftrightarrow 2_6^{(1,2,2,2)}$ See Fig. \ref{fig:3624Orbit}
	\item $2_7^{(1,4,0,2)} \leftrightarrow 2_7^{(1,4,0,2)}$ Self-dual
	\item $2_7^{(2,1,3,1)} \leftrightarrow 2_7^{(2,1,3,1)}$ Self-dual
	\item $2_7^{(2,3,1,1)} \leftrightarrow 2_8^{(2,3,1,1)}$
	\item $2_7^{(3,0,4,0)} \leftrightarrow 2_8^{(3,0,4,0)}$
	\item $2_7^{(3,2,2,0)} \leftrightarrow 2_9^{(3,2,2,0)}$
	\item $2_7^{(3,4,0,0)} \leftrightarrow 2_{10}^{(3,4,0,0)}$
	\item $1_7^{(1,0,0,4,3)} \leftrightarrow 1_6^{(1,0,0,4,3)}$ See Fig. \ref{fig:1643Orbit}
	\item $1_7^{(1,0,2,2,3)} \leftrightarrow 1_7^{(1,0,2,2,3)}$ Self-dual
	\item $1_7^{(1,0,4,0,3)} \leftrightarrow 1_8^{(1,0,4,0,3)}$
	\item $1_7^{(1,1,1,3,2)} \leftrightarrow 1_9^{(1,1,1,3,2)}$
	\item $1_7^{(1,1,3,1,2)} \leftrightarrow 1_{10}^{(1,1,3,1,2)}$
	\item $1_7^{(1,2,0,4,1)} \leftrightarrow 1_9^{(1,2,0,4,1)}$
	\item $1_7^{(1,2,2,2,1)} \leftrightarrow 1_{10}^{(1,2,2,2,1)}$
	\item $1_7^{(1,2,4,0,1)} \leftrightarrow 1_{11}^{(1,2,4,0,1)}$
	\item $1_7^{(1,3,1,3,0)} \leftrightarrow 1_{11}^{(1,3,1,3,0)}$
	\item $1_7^{(1,3,3,1,0)} \leftrightarrow 1_{12}^{(1,3,3,1,0)}$
	\item $0_7^{(2,0,1,3,3)} \leftrightarrow 0_9^{(2,0,1,3,3)}$
	\item $0_7^{(2,0,3,1,3)} \leftrightarrow 0_{10}^{(2,0,3,1,3)}$
	\item $0_7^{(2,1,0,4,2)} \leftrightarrow 0_{10}^{(2,1,0,4,2)}$
	\item $0_7^{(2,1,2,2,2)} \leftrightarrow 0_{11}^{(2,1,2,2,2)}$
	\item $0_7^{(2,1,4,0,2)} \leftrightarrow 0_{12}^{(2,1,4,0,2)}$
	\item $0_7^{(2,2,1,3,1)} \leftrightarrow 0_{12}^{(2,2,1,3,1)}$
	\item $0_7^{(2,2,3,1,1)} \leftrightarrow 0_{13}^{(2,2,3,1,1)}$
	\item $0_7^{(2,3,0,4,0)} \leftrightarrow 0_{13}^{(2,3,0,4,0)}$
	\item $0_7^{(2,3,2,2,0)} \leftrightarrow 0_{14}^{(2,3,2,2,0)}$
	\item $0_7^{(2,3,4,0,0)} \leftrightarrow 0_{15}^{(2,3,4,0,0)}$
\end{itemize}
\end{multicols}
\clearpage
\forceindent M-theory origins:
\begin{multicols}{2}
\begin{itemize}
	\item $2_7^{(4,3)} \rightarrow 2^{(4,3)}$
	\item $2_7^{(2,2,3)} \rightarrow 3^{(2,2,3)}$
	\item $2_7^{(4,0,3)} \rightarrow 2^{(4,0,4)}$
	\item $2_7^{(1,1,3,2)} \rightarrow 2^{(1,1,3,3)}$
	\item $2_7^{(1,3,1,2)} \rightarrow 2^{(1,3,2,2)}$
	\item $2_7^{(2,0,4,1)} \rightarrow 2^{(2,0,5,1)}$
	\item $2_7^{(2,2,2,1)} \rightarrow 2^{(2,3,2,1)}$
	\item $2_7^{(2,4,0,1)} \rightarrow 2^{(3,4,0,1)}$
	\item $2_7^{(3,1,3,0)} \rightarrow 2^{(4,1,3,0)}$
	\item $2_7^{(3,3,1,0)} \rightarrow 2^{1,3,3,1,0)}$
	\item $1_7^{(1,0,1,3,3)} \rightarrow 1^{(1,0,1,4,3)}$
	\item $1_7^{(1,0,3,1,3)} \rightarrow 1^{(1,0,4,1,3)}$
	\item $1_7^{(1,1,0,4,2)} \rightarrow 1^{(1,1,1,4,2)}$
	\item $1_7^{(1,1,2,2,2)} \rightarrow 1^{(1,2,2,2,2)}$
	\item $1_7^{(1,1,4,0,2)} \rightarrow 1^{(1,1,1,4,0,2)}$
	\item $1_7^{(1,2,1,3,1)} \rightarrow 1^{(2,2,1,3,1)}$
	\item $1_7^{(1,2,3,1,1)} \rightarrow 1^{(1,1,2,3,1,1)}$
	\item $1_7^{(1,3,0,4,0)} \rightarrow 1^{(1,1,3,0,4,0)}$
	\item $1_7^{(1,3,2,2,0)} \rightarrow 1^{(1,0,1,3,2,2,0)}$
	\item $1_7^{(1,3,4,0,0)} \rightarrow 1^{(1,0,0,1,3,4,0,0)}$
	\item $0_7^{(2,0,0,4,3)} \rightarrow 0^{(2,1,0,4,3)}$
	\item $0_7^{(2,0,2,2,3)} \rightarrow 0^{(3,0,2,2,3)}$
	\item $0_7^{(2,0,4,0,3)} \rightarrow 0^{(1,2,0,4,0,3)}$
	\item $0_7^{(2,1,1,3,2)} \rightarrow 0^{(1,2,1,1,3,2)}$
	\item $0_7^{(2,1,3,1,2)} \rightarrow 0^{(1,0,2,1,3,1,2)}$
	\item $0_7^{(2,2,0,4,1)} \rightarrow 0^{(1,0,2,2,0,4,1)}$
	\item $0_7^{(2,2,2,2,1)} \rightarrow 0^{(1,0,0,2,2,2,2,1)}$
	\item $0_7^{(2,2,4,0,1)} \rightarrow 0^{(1,0,0,0,2,2,4,0,1)}$
	\item $0_7^{(2,3,1,3,0)} \rightarrow 0^{(1,0,0,0,2,3,1,3,0)}$
	\item $0_7^{(2,3,3,1,0)} \rightarrow 0^{(1,0,0,0,0,2,3,3,1,0)}$
\end{itemize}
\end{multicols}
\clearpage
\begin{landscape}
\begin{figure}[H]
\centering
\begin{tikzpicture}
\matrix(M)[matrix of math nodes, row sep=2.5em, column sep=3em, minimum width=2em]{
\text{A} & & 3_7^{(1,5,0)} & & 3_7^{(3,3,0)} & & 3_7^{(5,1,0)} &\\
\text{B} & 3_7^{(6,0)} & & 3_7^{(2,4,0)} & & 3_7^{(4,2,0)} & & 3_7^{(6,0,0)}\\
\text{A} & 2_7^{(1,0,0,6,0)} & & 2_7^{(1,0,2,4,0)} & & 2_7^{(1,0,4,2,0)} & & 2_7^{(1,0,6,0,0)}\\
\text{B} & & 2_7^{(1,0,1,5,0)} & & 2_7^{(1,0,3,3,0)} & & 2_7^{(1,0,5,1,0)} &\\
\text{A} & & 1_7^{(2,0,1,5,0)} & & 1_7^{(2,0,3,3,0)} & & 1_7^{(2,0,5,1,0)} &\\
\text{B} & 1_7^{(2,0,0,6,0)} & & 1_7^{(2,0,2,4,0)} & & 1_7^{(2,0,4,2,0)} & & 1_7^{(2,0,6,0,0)}\\
\text{A} & 0_7^{(3,0,0,6,0)} & & 0_7^{(3,0,2,4,0)} & & 0_7^{(3,0,4,2,0)} & & 0_7^{(3,0,6,0,0)}\\
\text{B} & & 0_7^{(3,0,1,5,0)} & & 0_7^{(3,0,3,3,0)} & & 0_7^{(3,0,5,1,0)} &\\
};
\draw[latex-latex] (M-2-2) -- (M-1-3);
\draw[latex-latex] (M-1-3) -- (M-2-4);
\draw[latex-latex] (M-2-4) -- (M-1-5);
\draw[latex-latex] (M-1-5) -- (M-2-6);
\draw[latex-latex] (M-2-6) -- (M-1-7);
\draw[latex-latex] (M-1-7) -- (M-2-8);
\draw[latex-latex] (M-3-2) -- (M-4-3);
\draw[latex-latex] (M-4-3) -- (M-3-4);
\draw[latex-latex] (M-3-4) -- (M-4-5);
\draw[latex-latex] (M-4-5) -- (M-3-6);
\draw[latex-latex] (M-3-6) -- (M-4-7);
\draw[latex-latex] (M-4-7) -- (M-3-8);
\draw[latex-latex] (M-6-2) -- (M-5-3);
\draw[latex-latex] (M-5-3) -- (M-6-4);
\draw[latex-latex] (M-6-4) -- (M-5-5);
\draw[latex-latex] (M-5-5) -- (M-6-6);
\draw[latex-latex] (M-6-6) -- (M-5-7);
\draw[latex-latex] (M-5-7) -- (M-6-8);
\draw[latex-latex] (M-7-2) -- (M-8-3);
\draw[latex-latex] (M-8-3) -- (M-7-4);
\draw[latex-latex] (M-7-4) -- (M-8-5);
\draw[latex-latex] (M-8-5) -- (M-7-6);
\draw[latex-latex] (M-7-6) -- (M-8-7);
\draw[latex-latex] (M-8-7) -- (M-7-8);
\draw[latex-latex] (M-2-2) -- (M-3-2);
\draw[latex-latex] (M-3-2) -- (M-6-2);
\draw[latex-latex] (M-6-2) -- (M-7-2);
\draw[latex-latex] (M-2-4) -- (M-3-4);
\draw[latex-latex] (M-3-4) -- (M-6-4);
\draw[latex-latex] (M-6-4) -- (M-7-4);
\draw[latex-latex] (M-2-6) -- (M-3-6);
\draw[latex-latex] (M-3-6) -- (M-6-6);
\draw[latex-latex] (M-6-6) -- (M-7-6);
\draw[latex-latex] (M-2-8) -- (M-3-8);
\draw[latex-latex] (M-3-8) -- (M-6-8);
\draw[latex-latex] (M-6-8) -- (M-7-8);
\draw[latex-latex] (M-1-3) -- (M-4-3);
\draw[latex-latex] (M-4-3) -- (M-5-3);
\draw[latex-latex] (M-5-3) -- (M-8-3);
\draw[latex-latex] (M-1-5) -- (M-4-5);
\draw[latex-latex] (M-4-5) -- (M-5-5);
\draw[latex-latex] (M-5-5) -- (M-8-5);
\draw[latex-latex] (M-1-7) -- (M-4-7);
\draw[latex-latex] (M-4-7) -- (M-5-7);
\draw[latex-latex] (M-5-7) -- (M-8-7);
\end{tikzpicture}
\caption{The T-duality orbit of the $3_7^{(6,0)}$.}
\label{fig:3760Orbit}
\end{figure}
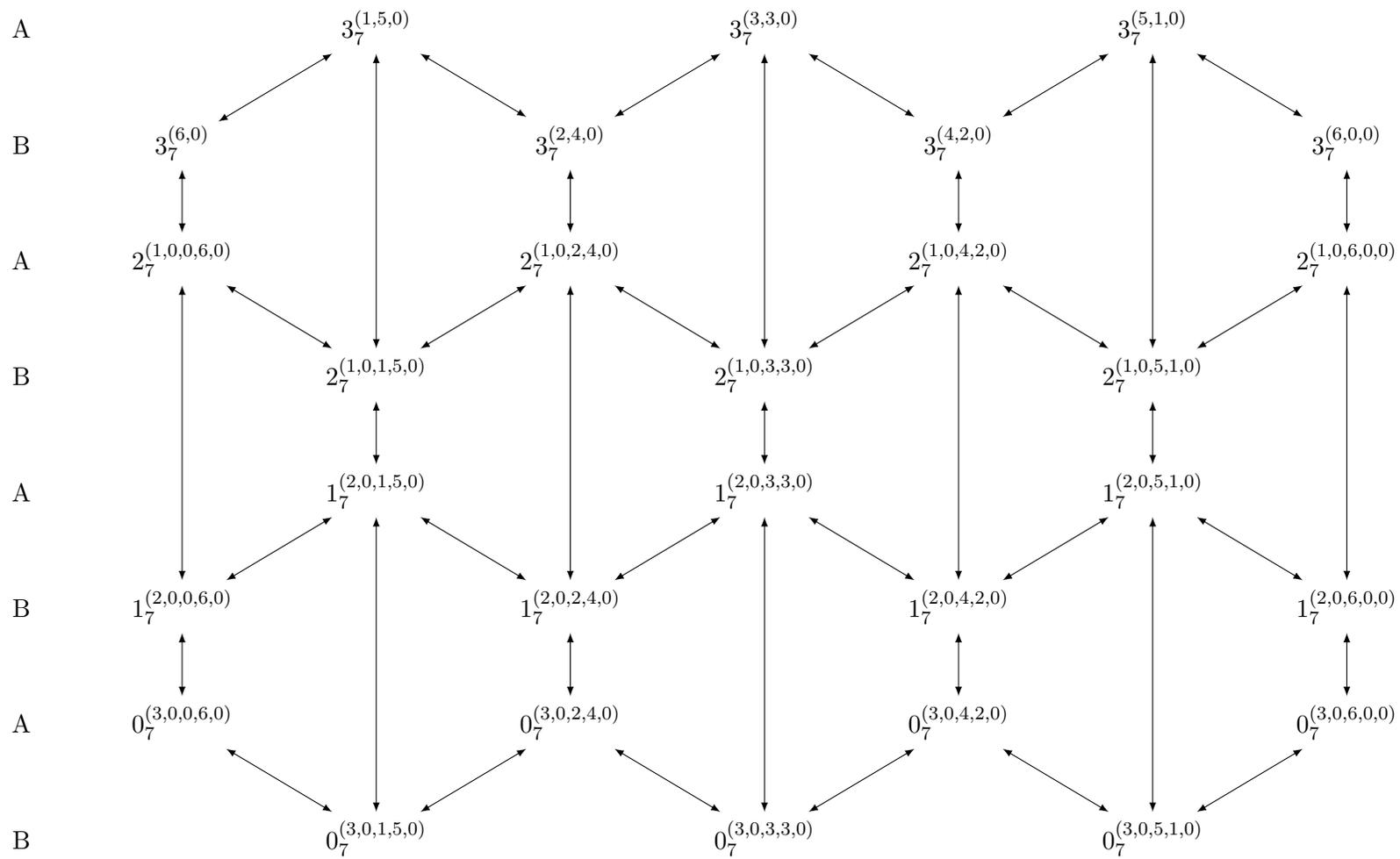
\end{landscape}
\begin{multicols}{2}
\forceindent S-dualities:
\begin{itemize}
	\item $3_7^{(6,0)} \leftrightarrow 3_4^{(6,0)}$ See Fig. \ref{fig:7420Orbit}
	\item $3_7^{(2,4,0)} \leftrightarrow 3_5^{(2,4,0)}$ See Fig. \ref{fig:5522Orbit}
	\item $3_7^{(4,2,0)} \leftrightarrow 3_6^{(4,2,0)}$ See Fig. \ref{fig:3624Orbit}
	\item $3_7^{(6,0,0)} \leftrightarrow 3_7^{(6,0,0)}$ Self-dual
	\item $2_7^{(1,0,1,5,0)} \leftrightarrow 2_7^{(1,0,1,5,0)}$ Self-dual
	\item $2_7^{(1,0,3,3,0)} \leftrightarrow 2_8^{(1,0,3,3,0)}$ 
	\item $2_7^{(1,0,5,1,0)} \leftrightarrow 2_9^{(1,0,5,1,0)}$
	\item $1_7^{(2,0,0,6,0)} \leftrightarrow 1_9^{(2,0,0,6,0)}$
	\item $1_7^{(2,0,2,4,0)} \leftrightarrow 1_{10}^{(2,0,2,4,0)}$
	\item $1_7^{(2,0,4,2,0)} \leftrightarrow 1_{11}^{(2,0,4,2,0)}$
	\item $1_7^{(2,0,6,0,0)} \leftrightarrow 1_{12}^{(2,0,6,0,0)}$
	\item $0_7^{(3,0,1,5,0)} \leftrightarrow 0_{12}^{(3,0,1,5,0)}$
	\item $0_7^{(3,0,3,3,0)} \leftrightarrow 0_{13}^{(3,0,3,3,0)}$
	\item $0_7^{(3,0,5,1,0)} \leftrightarrow 0_{14}^{(3,0,5,1,0)}$
\end{itemize}
\columnbreak
\forceindent M-theory origins:
\begin{itemize}
	\item $3_7^{(1,5,0)} \rightarrow 4^{(1,5,0)}$
	\item $3_7^{(3,3,0)} \rightarrow 3^{(3,3,1)}$
	\item $3_7^{(5,1,0)} \rightarrow 3^{(5,2,0)}$
	\item $2_7^{(1,0,0,6,0)} \rightarrow 2^{(1,0,0,7,0)}$
	\item $2_7^{(1,0,2,4,0)} \rightarrow 2^{(1,0,3,4,0)}$
	\item $2_7^{(1,0,4,2,0)} \rightarrow 2^{(1,1,4,2,0)}$
	\item $2_7^{(1,0,6,0,0)} \rightarrow 2^{(2,0,6,0,0)}$
	\item $1_7^{(2,0,1,5,0)} \rightarrow 1^{(3,0,1,5,0)}$
	\item $1_7^{(2,0,3,3,0)} \rightarrow 1^{(1,2,0,3,3,0)}$
	\item $1_7^{(2,0,5,1,0)} \rightarrow 1^{(1,0,2,0,5,1,0)}$
	\item $0_7^{(3,0,0,6,0)} \rightarrow 0^{(1,0,3,0,0,6,0)}$
	\item $0_7^{(3,0,2,4,0)} \rightarrow 0^{(1,0,0,3,0,2,4,0)}$
	\item $0_7^{(3,0,4,2,0)} \rightarrow 0^{(1,0,0,0,3,0,4,2,0)}$
	\item $0_7^{(3,0,6,0,0)} \rightarrow 0^{(1,0,0,0,0,3,0,6,0,0)}$
\end{itemize}
\end{multicols}
\endgroup

	\newgeometry{vmargin=2cm, hmargin=2cm}
\chapter{The M-Theory Origin of Exotic Branes}\label{app:MOrigin}
Here, we group the branes by their M-theory origins. Unlike the M2 or M5 brane, exotic branes can have more than two branes that they can descend to in Type IIA owing to their more intricate structure. Just as the circle fibration of the KK6M allows for an additional possible reduction for a total of three possibilities (KK5A, D6 $6_3^1$), these exotic branes can be reduced transverse or longitudinal to their distinguished directions, leading to more possible branes. We have compiled this list by taking all of the M-theory branes that were computed in Appendix~\ref{app:ExoticWeb} and then computing all their possible reductions without truncating the list to $0 \leq n \leq 7$.
\begingroup
\setlength\columnsep{-28pt}
\begin{multicols}{2}
\begin{itemize}
	\item $0=\text{WM}
		\begin{cases}
			{}^10_0\text{=P; See Fig.\ \ref{fig:10Orbit}}\\
			0_1\text{=D0; See Fig.\ \ref{fig:11Orbit}}
		\end{cases}$
	\item $0^{(1,7)}
		\begin{cases}
			0_3^7\text{; See Fig.\ \ref{fig:532Orbit}}\\
			0_4^{(1,6)}\text{; See Fig.\ \ref{fig:146Orbit}}\\
			0_6^{(1,7)}\text{; See Fig.\ \ref{fig:0617Orbit}}
		\end{cases}$
	\item $0^{(2,1,6)}
		\begin{cases}
			0_4^{(1,1,6)}\text{; See Fig.\ \ref{fig:146Orbit}}\\
			0_5^{(2,0,6)}\text{; See Fig.\ \ref{fig:2515Orbit}}\\
			0_6^{(2,1,5)}\text{; See Fig.\ \ref{fig:0617Orbit}}\\
			0_8^{(2,1,6)}
		\end{cases}$
	\item $0^{(3,2,4)}
		\begin{cases}
			0_5^{(2,2,4)}\text{; See Fig.\ \ref{fig:2515Orbit}}\\
			0_6^{(3,1,4)}\text{; See Fig.\ \ref{fig:0617Orbit}}\\
			0_7^{(3,2,3)}\text{; See Fig.\ \ref{fig:0753Orbit}}\\
			0_9^{(3,2,4)}
		\end{cases}$
	\item $0^{(5,1,3)}
		\begin{cases}
			0_6^{(4,1,3)}\text{; See Fig.\ \ref{fig:0617Orbit}}\\
			0_7^{(5,0,3)}\text{; See Fig.\ \ref{fig:0753Orbit}}\\
			0_8^{(5,1,2)}\\
			0_{10}^{(5,1,3)}
		\end{cases}$
	\item $0^{(1,0,5,3)}
		\begin{cases}
			0_4^{(5,3)}\text{; See Fig.\ \ref{fig:4413Orbit}}\\
			0_6^{(1,0,4,3)}\text{; See Fig.\ \ref{fig:1643Orbit}}\\
			0_7^{(1,0,5,2)}\text{; See Fig.\ \ref{fig:0753Orbit}}\\
			0_9^{(1,0,5,3)}
		\end{cases}$
	\item $0^{(1,2,4,2)}
		\begin{cases}
			0_5^{(2,4,2)}\text{; See Fig.\ \ref{fig:2515Orbit}}\\
			0_6^{(1,1,4,2)}\text{; See Fig.\ \ref{fig:1643Orbit}}\\
			0_7^{(1,2,3,2)}\text{; See Fig.\ \ref{fig:0753Orbit}}\\
			0_8^{(1,2,4,1)}\\
			0_{10}^{(1,2,4,2)}
		\end{cases}$
	\item $0^{(1,5,1,2)}
		\begin{cases}
			0_6^{(5,1,2)}\text{; See Fig.\ \ref{fig:0617Orbit}}\\
			0_7^{(1,4,1,2)}\text{; See Fig.\ \ref{fig:0753Orbit}}\\
			0_8^{(1,5,0,2)}\\
			0_9^{(1,5,1,1)}\\
			0_{11}^{(1,5,1,2)}
		\end{cases}$
	\item $0^{(2,2,4,1)}
		\begin{cases}
			0_6^{(1,2,4,1)}\text{; See Fig.\ \ref{fig:1643Orbit}}\\
			0_7^{(2,1,4,1)}\text{; See Fig.\ \ref{fig:0753Orbit}}\\
			0_8^{(2,2,3,1)}\\
			0_9^{(2,2,4,0)}\\
			0_{11}^{(2,2,4,1)}
		\end{cases}$
	\item $0^{(3,3,2,1)}
		\begin{cases}
			0_7^{(2,3,2,1)}\text{; See Fig.\ \ref{fig:0753Orbit}}\\
			0_8^{(3,2,2,1)}\\
			0_9^{(3,3,1,1)}\\
			0_{10}^{(3,3,2,0)}\\
			0_{12}^{(3,3,2,1)}
		\end{cases}$
	\item $0^{(4,0,5,0)}
		\begin{cases}
			0_7^{(3,0,5,0)}\text{; See Fig.\ \ref{fig:0753Orbit}}\\
			0_9^{(4,0,4,0)}\\
			0_{12}^{(4,0,5,0)}
		\end{cases}$
	\item $0^{(1,0,0,2,7)}
		\begin{cases}
			0_3^{(2,7)}\text{; See Fig.\ \ref{fig:532Orbit}}\\
			0_6^{(1,0,0,1,7)}\text{; See Fig.\ \ref{fig:0617Orbit}}\\
			0_7^{(1,0,0,2,6)}\text{; See Fig.\ \ref{fig:1726Orbit}}
		\end{cases}$
	\item $0^{(1,0,2,1,6)}
		\begin{cases}
			0_4^{(2,1,6)}\text{; See Fig.\ \ref{fig:146Orbit}}\\
			0_6^{(1,0,1,1,6)}\text{; See Fig.\ \ref{fig:0617Orbit}}\\
			0_7^{(1,0,2,0,6)}\text{; See Fig.\ \ref{fig:1726Orbit}}\\
			0_8^{(1,0,2,1,5)}
		\end{cases}$
	\item $0^{(1,0,2,6,0)}
		\begin{cases}
			0_5^{(2,6,0)}\text{; See Fig.\ \ref{fig:2515Orbit}}\\
			0_7^{(1,0,1,6,0)}\text{; See Fig.\ \ref{fig:17160Orbit}}\\
			0_8^{(1,0,2,5,0)}\\
			0_{11}^{(1,0,2,6,0)}
		\end{cases}$
	\item $0^{(1,0,6,1,1)}
		\begin{cases}
			0_6^{(6,1,1)}\text{; See Fig.\ \ref{fig:0617Orbit}}\\
			0_8^{(1,0,5,1,1)}\\
			0_9^{(1,0,5,0,1)}\\
			0_{10}^{(1,0,5,1,0)}\\
			0_{12}^{(1,0,6,1,1)}
		\end{cases}$
	\item $0^{(1,1,2,1,5)}
		\begin{cases}
			0_5^{(1,2,1,5)}\text{; See Fig.\ \ref{fig:2515Orbit}}\\
			0_6^{(1,0,2,1,5)}\text{; See Fig.\ \ref{fig:0617Orbit}}\\
			0_7^{(1,1,1,1,5)}\text{; See Fig.\ \ref{fig:1726Orbit}}\\
			0_8^{(1,1,2,0,5)}\\
			0_9^{(1,1,2,1,4)}
		\end{cases}$
	\item $0^{(1,1,3,4,0)}
		\begin{cases}
			0_6^{(1,3,4,0)}\text{; See Fig.\ \ref{fig:1643Orbit}}\\
			0_7^{(1,0,3,4,0)}\text{; See Fig.\ \ref{fig:17160Orbit}}\\
			0_8^{(1,1,2,4,0)}\\
			0_9^{(1,1,3,3,0)}\\
			0_{12}^{(1,1,3,4,0)}
		\end{cases}$
	\item $0^{(1,2,5,0,1)}
		\begin{cases}
			0_7^{(2,5,0,1)}\text{; See Fig.\ \ref{fig:0753Orbit}}\\
			0_8^{(1,1,5,0,1)}\\
			0_9^{(1,2,4,0,1)}\\
			0_{11}^{(1,2,5,0,0)}\\
			0_{13}^{(1,2,5,0,1)}
		\end{cases}$
	\item $0^{(1,3,0,2,4)}
		\begin{cases}
			0_6^{(3,0,2,4)}\text{; See Fig.\ \ref{fig:3624Orbit}}\\
			0_7^{(1,2,0,2,4)}\text{; See Fig.\ \ref{fig:1726Orbit}}\\
			0_9^{(1,3,0,1,4)}\\
			0_{10}^{(1,3,0,2,3)}
		\end{cases}$
	\item $0^{(1,3,2,3,0)}
		\begin{cases}
			0_7^{(3,2,3,0)}\text{; See Fig.\ \ref{fig:0753Orbit}}\\
			0_8^{(1,2,2,3,0)}\\
			0_9^{(1,3,1,3,0)}\\
			0_{10}^{(1,3,2,2,0)}\\
			0_{13}^{(1,3,2,3,0)}
		\end{cases}$
	\item $0^{(2,0,3,1,4)}
		\begin{cases}
			0_6^{(1,0,3,1,4)}\text{; See Fig.\ \ref{fig:0617Orbit}}\\
			0_8^{(2,0,2,1,4)}\\
			0_9^{(2,0,3,0,4)}\\
			0_{10}^{(2,0,3,1,1)}
		\end{cases}$
	\item $0^{(2,0,5,2,0)}
		\begin{cases}
			0_7^{(1,0,5,2,0)}\text{; See Fig.\ \ref{fig:17160Orbit}}\\
			0_9^{(2,0,4,2,0)}\\
			0_{10}^{(2,0,5,1,0)}\\
			0_{13}^{(2,0,5,2,0)}
		\end{cases}$
	\item $0^{(2,1,0,4,3)}
		\begin{cases}
			0_6^{(1,1,0,4,3)}\text{; See Fig.\ \ref{fig:1643Orbit}}\\
			0_7^{(2,0,0,4,3)}\text{; See Fig.\ \ref{fig:27133Orbit}}\\
			0_9^{(2,1,0,3,3)}\\
			0_{10}^{(2,1,0,4,2)}
		\end{cases}$
	\item $0^{(2,2,2,0,4)}
		\begin{cases}
			0_7^{(1,2,2,0,4)}\text{; See Fig.\ \ref{fig:1726Orbit}}\\
			0_8^{(2,1,2,0,4)}\\
			0_9^{(2,2,1,0,4)}\\
			0_{11}^{(2,2,2,0,3)}
		\end{cases}$
	\item $0^{(3,0,2,2,3)}
		\begin{cases}
			0_7^{(2,0,2,2,3)}\text{; See Fig.\ \ref{fig:27133Orbit}}\\
			0_9^{(3,0,1,2,3)}\\
			0_{10}^{(3,0,2,1,3)}\\
			0_{11}^{(3,0,2,2,2)}
		\end{cases}$
	\item $0^{(1,0,0,1,5,3)}
		\begin{cases}
			0_4^{(1,5,3)}\text{; See Fig.\ \ref{fig:4413Orbit}}\\
			0_7^{(1,0,0,0,5,3)}\text{; See Fig.\ \ref{fig:0753Orbit}}\\
			0_8^{(1,0,0,1,4,3)}\\
			0_9^{(1,0,0,1,5,2)}
		\end{cases}$
	\item $0^{(1,0,0,5,0,4)}
		\begin{cases}
			0_5^{(5,0,4)}\text{; See Fig.\ \ref{fig:5522Orbit}}\\
			0_8^{(1,0,0,4,0,4)}\\
			0_{10}^{(1,0,0,5,0,3)}
		\end{cases}$
	\item $0^{(1,0,0,7,1,0)}
		\begin{cases}
			0_6^{(7,1,0)}\text{; See Fig.\ \ref{fig:0617Orbit}}\\
			0_9^{(1,0,0,6,1,0)}\\
			0_{10}^{(1,0,0,7,0,0)}\\
			0_{13}^{(1,0,0,7,1,0)}
		\end{cases}$
	\item $0^{(1,0,1,2,3,3)}
		\begin{cases}
			0_5^{(1,2,3,3)}\text{; See Fig.\ \ref{fig:2515Orbit}}\\
			0_7^{(1,0,0,2,3,3)}\text{; See Fig.\ \ref{fig:0753Orbit}}\\
			0_8^{(1,0,1,1,3,3)}\\
			0_9^{(1,0,1,2,2,3)}\\
			0_{10}^{(1,0,1,2,3,2)}
		\end{cases}$
	\item $0^{(1,0,3,1,2,3)}
		\begin{cases}
			0_6^{(3,1,2,3)}\text{; See Fig.\ \ref{fig:3624Orbit}}\\
			0_8^{(1,0,2,1,2,3)}\\
			0_9^{(1,0,3,0,2,3)}\\
			0_{10}^{(1,0,3,1,1,3)}\\
			0_{11}^{(1,0,3,1,2,2)}
		\end{cases}$
	\item $0^{(1,0,3,4,1,0)}
		\begin{cases}
			0_7^{(3,4,1,0)}\text{; See Fig.\ \ref{fig:0753Orbit}}\\
			0_9^{(1,0,2,4,1,0)}\\
			0_{10}^{(1,0,3,3,1,0)}\\
			0_{11}^{(1,0,3,4,0,0)}\\
			0_{14}^{(1,0,3,4,1,0)}
		\end{cases}$
	\item $0^{(1,1,0,4,1,3)}
		\begin{cases}
			0_6^{(1,0,4,1,3)}\text{; See Fig.\ \ref{fig:0617Orbit}}\\
			0_7^{(1,0,0,4,1,3)}\text{; See Fig.\ \ref{fig:0753Orbit}}\\
			0_9^{(1,1,0,3,1,3)}\\
			0_{10}^{(1,1,0,4,0,3)}\\
			0_{11}^{(1,1,0,4,1,2)}
		\end{cases}$
	\item $0^{(1,1,0,7,0,0)}
		\begin{cases}
			0_7^{(1,0,7,0,0)}\text{; See Fig.\ \ref{fig:17160Orbit}}\\
			0_8^{(1,0,0,7,0,0)}\\
			0_{10}^{(1,1,0,6,0,0)}\\
			0_{14}^{(1,1,0,7,0,0)}
		\end{cases}$
	\item $0^{(1,1,1,1,4,2)}
		\begin{cases}
			0_6^{(1,1,1,4,2)}\text{; See Fig.\ \ref{fig:1643Orbit}}\\
			0_7^{(1,0,1,1,4,2)}\text{; See Fig.\ \ref{fig:0753Orbit}}\\
			0_8^{(1,1,0,1,4,2)}\\
			0_9^{(1,1,1,0,4,2)}\\
			0_{10}^{(1,1,1,1,3,2)}\\
			0_{11}^{(1,1,1,1,4,1)}
		\end{cases}$
	\item $0^{(1,1,3,1,1,3)}
		\begin{cases}
			0_7^{(1,3,1,1,3)}\text{; See Fig.\ \ref{fig:1726Orbit}}\\
			0_8^{(1,0,3,1,1,3)}\\
			0_9^{(1,1,2,1,1,3)}\\
			0_{10}^{(1,1,3,0,1,3)}\\
			0_{11}^{(1,1,3,1,0,3)}\\
			0_{12}^{(1,1,3,1,1,2)}
		\end{cases}$
	\item $0^{(1,2,0,4,0,3)}
		\begin{cases}
			0_7^{(2,0,4,0,3)}\text{; See Fig.\ \ref{fig:27133Orbit}}\\
			0_8^{(1,1,0,4,0,3)}\\
			0_{10}^{(1,2,0,3,0,3)}\\
			0_{12}^{(1,2,0,4,0,2)}
		\end{cases}$
	\item $0^{(1,2,1,1,3,2)}
		\begin{cases}
			0_7^{(2,1,1,3,2)}\text{; See Fig.\ \ref{fig:27133Orbit}}\\
			0_8^{(1,1,1,1,3,2)}\\
			0_9^{(1,2,0,1,3,2)}\\
			0_{10}^{(1,2,1,0,3,2)}\\
			0_{11}^{(1,2,1,1,2,2)}\\
			0_{12}^{(1,2,1,1,3,1)}
		\end{cases}$
	\item $0^{(2,0,1,3,2,2)}
		\begin{cases}
			0_7^{(1,0,1,3,2,2)}\text{; See Fig.\ \ref{fig:0753Orbit}}\\
			0_9^{(2,0,0,3,2,2)}\\
			0_{10}^{(2,0,1,2,2,2)}\\
			0_{11}^{(2,0,1,3,1,2)}\\
			0_{12}^{(2,0,1,3,2,1)}
		\end{cases}$
	\item $0^{(2,0,2,0,5,1)}
		\begin{cases}
			0_7^{(1,0,2,0,5,1)}\text{; See Fig.\ \ref{fig:0753Orbit}}\\
			0_9^{(2,0,1,5,0,1)}\\
			0_{11}^{(2,0,2,0,4,1)}\\
			0_{12}^{(2,0,2,0,5,0)}
		\end{cases}$
	\item $0^{(2,1,0,0,7,0)}
		\begin{cases}
			0_7^{(1,1,0,0,7,0)}\text{; See Fig.\ \ref{fig:17160Orbit}}\\
			0_8^{(2,0,0,0,7,0)}\\
			0_{11}^{(2,1,0,0,6,0)}
		\end{cases}$
	\item $0^{(1,0,0,0,0,9,0)}
		\begin{cases}
			0_4^{(9,0)}\text{; See Fig.\ \ref{fig:7420Orbit}}\\
			0_9^{(1,0,0,0,0,8,0)}
		\end{cases}$
	\item $0^{(1,0,0,0,5,2,2)}
		\begin{cases}
			0_5^{(5,2,2)}\text{; See Fig.\ \ref{fig:5522Orbit}}\\
			0_9^{(1,0,0,0,4,2,2)}\\
			0_{10}^{(1,0,0,0,5,1,2)}\\
			0_{11}^{(1,0,0,0,5,2,1)}
		\end{cases}$
	\item $0^{(1,0,0,1,2,5,1)}
		\begin{cases}
			0_5^{(1,2,5,1)}\text{; See Fig.\ \ref{fig:2515Orbit}}\\
			0_8^{(1,0,0,0,2,5,1)}\\
			0_9^{(1,0,0,1,1,5,1)}\\
			0_{10}^{(1,0,0,1,2,4,1)}\\
			0_{11}^{(1,0,0,1,2,5,0)}
		\end{cases}$
	\item $0^{(1,0,0,3,2,2,2)}
		\begin{cases}
			0_6^{(3,2,2,2)}\text{; See Fig.\ \ref{fig:3624Orbit}}\\
			0_9^{(1,0,0,2,2,2,2)}\\
			0_{10}^{(1,0,0,3,1,2,2)}\\
			0_{11}^{(1,0,0,3,2,1,2)}\\
			0_{12}^{(1,0,0,3,2,2,1)}
		\end{cases}$
	\item $0^{(1,0,1,0,5,1,2)}
		\begin{cases}
			0_6^{(1,0,5,1,2)}\text{; See Fig.\ \ref{fig:0617Orbit}}\\
			0_8^{(1,0,0,0,5,1,2)}\\
			0_{10}^{(1,0,1,0,4,1,2)}\\
			0_{11}^{(1,0,1,0,5,0,2)}\\
			0_{12}^{(1,0,1,0,5,1,1)}
		\end{cases}$
	\item $0^{(1,0,1,1,2,4,1)}
		\begin{cases}
			0_6^{(1,1,2,4,1)}\text{; See Fig.\ \ref{fig:1643Orbit}}\\
			0_8^{(1,0,0,1,2,4,1)}\\
			0_9^{(1,0,1,0,2,4,1)}\\
			0_{10}^{(1,0,1,1,1,4,1)}\\
			0_{11}^{(1,0,1,1,2,3,1)}\\
			0_{12}^{(1,0,1,1,2,4,0)}
		\end{cases}$
	\item $0^{(1,0,1,4,0,2,2)}
		\begin{cases}
			0_7^{(1,4,0,2,2)}\text{; See Fig.\ \ref{fig:1726Orbit}}\\
			0_9^{(1,0,0,4,0,2,2)}\\
			0_{10}^{(1,0,1,3,0,2,2)}\\
			0_{12}^{(1,0,1,4,0,1,2)}\\
			0_{13}^{(1,0,1,4,0,2,1)}
		\end{cases}$
	\item $0^{(1,0,2,1,3,1,2)}
		\begin{cases}
			0_7^{(2,1,3,1,2)}\text{; See Fig.\ \ref{fig:27133Orbit}}\\
			0_9^{(1,0,1,1,3,1,2)}\\
			0_{10}^{(1,0,2,0,3,1,2)}\\
			0_{11}^{(1,0,2,1,2,1,2)}\\
			0_{12}^{(1,0,2,1,3,0,2)}\\
			0_{13}^{(1,0,2,1,3,1,1)}
		\end{cases}$
	\item $0^{(1,0,2,2,0,4,1)}
		\begin{cases}
			0_7^{(2,2,0,4,1)}\text{; See Fig.\ \ref{fig:27133Orbit}}\\
			0_9^{(1,0,1,2,0,4,1)}\\
			0_{10}^{(1,0,2,1,0,4,1)}\\
			0_{12}^{(1,0,2,2,0,3,1)}\\
			0_{13}^{(1,0,2,2,0,4,0)}
		\end{cases}$
	\item $0^{(1,0,3,0,0,6,0)}
		\begin{cases}
			0_7^{(3,0,0,6,0)}\text{; See Fig.\ \ref{fig:3760Orbit}}\\
			0_9^{(1,0,2,0,0,6,0)}\\
			0_{12}^{(1,0,3,0,0,5,0)}
		\end{cases}$
	\item $0^{(1,1,0,1,5,0,2)}
		\begin{cases}
			0_7^{(1,0,1,5,0,2)}\text{; See Fig.\ \ref{fig:0753Orbit}}\\
			0_8^{(1,0,0,1,5,0,2)}\\
			0_{10}^{(1,1,0,0,5,0,2)}\\
			0_{11}^{(1,1,0,1,4,0,2)}\\
			0_{13}^{(1,1,0,1,5,0,1)}
		\end{cases}$
	\item $0^{(1,1,0,2,2,3,1)}
		\begin{cases}
			0_7^{(1,0,2,2,3,1)}\text{; See Fig.\ \ref{fig:0753Orbit}}\\
			0_8^{(1,0,0,2,2,3,1)}\\
			0_{10}^{(1,1,0,1,2,3,1)}\\
			0_{11}^{(1,1,0,2,1,3,1)}\\
			0_{12}^{(1,1,0,2,2,2,1)}\\
			0_{13}^{(1,1,0,2,2,3,0)}
		\end{cases}$
	\item $0^{(1,1,1,0,2,5,0)}
		\begin{cases}
			0_7^{(1,1,0,2,5,0)}\text{; See Fig.\ \ref{fig:17160Orbit}}\\
			0_8^{(1,0,1,0,2,5,0)}\\
			0_9^{(1,1,0,0,2,5,0)}\\
			0_{11}^{(1,1,1,0,1,5,0)}\\
			0_{12}^{(1,1,1,0,2,4,0)}
		\end{cases}$
	\item $0^{(1,0,0,0,0,5,4,0)}
		\begin{cases}
			0_5^{(5,4,0)}\text{; See Fig.\ \ref{fig:5522Orbit}}\\
			0_{10}^{(1,0,0,0,0,4,4,0)}\\
			0_{11}^{(1,0,0,0,0,5,3,0)}
		\end{cases}$
	\item $0^{(1,0,0,0,3,3,2,1)}
		\begin{cases}
			0_6^{(3,3,2,1)}\text{; See Fig.\ \ref{fig:3624Orbit}}\\
			0_{10}^{(1,0,0,0,2,3,2,1)}\\
			0_{11}^{(1,0,0,0,3,2,2,1)}\\
			0_{12}^{(1,0,0,0,3,3,1,1)}\\
			0_{13}^{(1,0,0,0,3,3,2,0)}
		\end{cases}$
	\item $0^{(1,0,0,1,0,6,1,1)}
		\begin{cases}
			0_6^{(1,0,6,1,1)}\text{; See Fig.\ \ref{fig:0617Orbit}}\\
			0_9^{(1,0,0,0,0,6,1,1)}\\
			0_{11}^{(1,0,0,1,0,5,1,1)}\\
			0_{12}^{(1,0,0,1,0,6,0,1)}\\
			0_{13}^{(1,0,0,1,0,6,1,0)}
		\end{cases}$
	\item $0^{(1,0,0,1,1,3,4,0)}
		\begin{cases}
			0_6^{(1,1,3,4,0)}\text{; See Fig.\ \ref{fig:1643Orbit}}\\
			0_9^{(1,0,0,0,1,3,4,0)}\\
			0_{10}^{(1,0,0,1,0,3,4,0)}\\
			0_{11}^{(1,0,0,1,1,2,4,0)}\\
			0_{12}^{(1,0,0,1,1,3,3,0)}
		\end{cases}$
	\item $0^{(1,0,0,1,4,2,0,2)}
		\begin{cases}
			0_7^{(1,4,2,0,2)}\text{; See Fig.\ \ref{fig:1726Orbit}}\\
			0_{10}^{(1,0,0,0,4,2,0,2)}\\
			0_{11}^{(1,0,0,1,3,2,0,2)}\\
			0_{12}^{(1,0,0,1,4,1,0,2)}\\
			0_{14}^{(1,0,0,1,4,2,0,1)}
		\end{cases}$
	\item $0^{(1,0,0,2,2,2,2,1)}
		\begin{cases}
			0_7^{(2,2,2,2,1)}\text{; See Fig.\ \ref{fig:27133Orbit}}\\
			0_{10}^{(1,0,0,1,2,2,2,1)}\\
			0_{11}^{(1,0,0,2,1,2,2,1)}\\
			0_{12}^{(1,0,0,2,2,1,2,1)}\\
			0_{13}^{(1,0,0,2,2,2,1,1)}\\
			0_{14}^{(1,0,0,2,2,2,2,0)}
		\end{cases}$
	\item $0^{(1,0,0,3,0,2,4,0)}
		\begin{cases}
			0_7^{(3,0,2,4,0)}\text{; See Fig.\ \ref{fig:3760Orbit}}\\
			0_{10}^{(1,0,0,2,0,2,4,0)}\\
			0_{12}^{(1,0,0,3,0,1,4,0)}\\
			0_{13}^{(1,0,0,3,0,2,3,0}
		\end{cases}$
	\item $0^{(1,0,1,0,2,4,1,1)}
		\begin{cases}
			0_7^{(1,0,2,4,1,1)}\text{; See Fig.\ \ref{fig:0753Orbit}}\\
			0_9^{(1,0,0,0,2,4,1,1)}\\
			0_{11}^{(1,0,1,0,1,4,1,1)}\\
			0_{12}^{(1,0,1,0,2,3,1,1)}\\
			0_{13}^{(1,0,1,0,2,4,0,1)}\\
			0_{14}^{(1,0,1,0,2,4,1,0)}
		\end{cases}$
	\item $0^{(1,0,1,0,3,1,4,0)}
		\begin{cases}
			0_7^{(1,0,3,1,4,0)}\text{; See Fig.\ \ref{fig:0753Orbit}}\\
			0_9^{(1,0,0,0,0,3,1,4,0)}\\
			0_{11}^{(1,0,1,0,2,1,4,0)}\\
			0_{12}^{(1,0,1,0,3,0,4,0)}\\
			0_{13}^{(1,0,1,0,3,1,3,0)}
		\end{cases}$
	\item $0^{(1,0,1,1,0,4,3,0)}
		\begin{cases}
			0_7^{(1,1,0,4,3,0)}\text{; See Fig.\ \ref{fig:17160Orbit}}\\
			0_9^{(1,0,0,1,0,4,3,0)}\\
			0_{10}^{(1,0,1,0,0,4,3,0)}\\
			0_{12}^{(1,0,1,1,0,3,3,0)}\\
			0_{13}^{(1,0,1,1,0,4,2,0)}
		\end{cases}$
	\item $0^{(1,0,0,0,0,3,4,2,0)}
		\begin{cases}
			0_6^{(3,4,2,0)}\text{; See Fig.\ \ref{fig:3624Orbit}}\\
			0_{10}^{(1,0,0,0,0,2,4,2,0)}\\
			0_{11}^{(1,0,0,0,0,3,3,2,0)}\\
			0_{12}^{(1,0,0,0,0,3,4,1,0)}
		\end{cases}$
	\item $0^{(1,0,0,0,1,0,7,1,0)}
		\begin{cases}
			0_6^{(1,0,7,1,0)}\text{; See Fig.\ \ref{fig:0617Orbit}}\\
			0_{10}^{(1,0,0,0,0,0,7,1,0)}\\
			0_{12}^{(1,0,0,0,1,0,6,1,0)}\\
			0_{13}^{(1,0,0,0,1,0,7,0,0)}
		\end{cases}$
	\item $0^{(1,0,0,0,1,5,1,1,1)}
		\begin{cases}
			0_7^{(1,5,1,1,1)}\text{; See Fig.\ \ref{fig:1726Orbit}}\\
			0_{11}^{(1,0,0,0,0,5,1,1,1)}\\
			0_{12}^{(1,0,0,0,1,4,1,1,1)}\\
			0_{13}^{(1,0,0,0,1,5,0,1,1)}\\
			0_{14}^{(1,0,0,0,1,5,1,0,1)}\\
			0_{15}^{(1,0,0,0,1,5,1,1,0)}
		\end{cases}$
	\item $0^{(1,0,0,0,2,2,4,0,1)}
		\begin{cases}
			0_7^{(2,2,4,0,1)}\text{; See Fig.\ \ref{fig:27133Orbit}}\\
			0_{11}^{(1,0,0,0,1,2,4,0,1)}\\
			0_{12}^{(1,0,0,0,2,1,4,0,1)}\\
			0_{13}^{(1,0,0,0,2,2,3,0,1)}\\
			0_{15}^{(1,0,0,0,2,2,4,0,0)}
		\end{cases}$
	\item $0^{(1,0,0,0,3,0,4,2,0)}
		\begin{cases}
			0_7^{(3,0,4,2,0)}\text{; See Fig.\ \ref{fig:3760Orbit}}\\
			0_{11}^{(1,0,0,0,2,0,4,2,0)}\\
			0_{13}^{(1,0,0,0,3,0,3,2,0)}\\
			0_{14}^{(1,0,0,0,3,0,4,1,0)}
		\end{cases}$
	\item $0^{(1,0,0,1,0,3,3,2,0)}
		\begin{cases}
			0_7^{(1,0,3,3,2,0)}\text{; See Fig.\ \ref{fig:0753Orbit}}\\
			0_{10}^{(1,0,0,0,0,3,3,2,0)}\\
			0_{12}^{(1,0,0,1,0,2,3,2,0)}\\
			0_{13}^{(1,0,0,1,0,3,2,2,0)}\\
			0_{14}^{(1,0,0,1,0,3,3,1,0)}
		\end{cases}$
	\item $0^{(1,0,0,1,1,0,6,1,0)}
		\begin{cases}
			0_7^{(1,1,0,6,1,0)}\text{; See Fig.\ \ref{fig:17160Orbit}}\\
			0_{10}^{(1,0,0,0,1,0,6,1,0)}\\
			0_{11}^{(1,0,0,1,0,0,6,1,0)}\\
			0_{13}^{(1,0,0,1,1,0,5,1,0)}\\
			0_{14}^{(1,0,0,1,1,0,6,0,0)}
		\end{cases}$
	\item $0^{(1,0,0,0,2,3,1,3,0)}
		\begin{cases}
			0_7^{(2,3,1,3,0)}\text{; See Fig.\ \ref{fig:27133Orbit}}\\
			0_{11}^{(1,0,0,0,1,3,1,3,0)}\\
			0_{12}^{(1,0,0,0,2,2,1,3,0)}\\
			0_{13}^{(1,0,0,0,2,3,0,3,0)}\\
			0_{14}^{(1,0,0,0,2,3,1,2,0)}
		\end{cases}$
	\item $0^{(1,0,0,0,1,0,3,5,0,0)}
		\begin{cases}
			0_7^{(1,0,3,5,0,0)}\text{; See Fig.\ \ref{fig:0753Orbit}}\\
			0_{11}^{(1,0,0,0,0,0,3,5,0,0)}\\
			0_{13}^{(1,0,0,0,1,0,2,5,0,0)}\\
			0_{14}^{(1,0,0,0,1,0,3,4,0,0)}
		\end{cases}$
	\item $0^{(1,0,0,0,0,1,6,0,2,0)}
		\begin{cases}
			0_7^{(1,6,0,2,0)}\text{; See Fig.\ \ref{fig:1726Orbit}}\\
			0_{12}^{(1,0,0,0,0,0,6,0,2,0)}\\
			0_{13}^{(1,0,0,0,0,1,5,0,2,0)}\\
			0_{15}^{(1,0,0,0,0,1,6,0,1,0)}
		\end{cases}$
	\item $0^{(1,0,0,0,0,2,3,3,1,0)}
		\begin{cases}
			0_7^{(2,3,3,1,0)}\text{; See Fig.\ \ref{fig:27133Orbit}}\\
			0_{12}^{(1,0,0,0,0,1,3,3,1,0)}\\
			0_{13}^{(1,0,0,0,0,2,2,3,1,0)}\\
			0_{14}^{(1,0,0,0,0,2,3,2,1,0)}\\
			0_{15}^{(1,0,0,0,0,2,3,3,0,0)}
		\end{cases}$
	\item $0^{(1,0,0,0,0,3,0,6,0,0)}
		\begin{cases}
			0_7^{(3,0,6,0,0)}\text{; See Fig.\ \ref{fig:3760Orbit}}\\
			0_{12}^{(1,0,0,0,0,2,0,6,0,0)}\\
			0_{14}^{(1,0,0,0,0,3,0,5,0,0)}
		\end{cases}$
	\item $0^{(1,0,0,0,0,0,1,6,2,0,0)}
		\begin{cases}
			0_7^{(1,6,2,0,0)}\text{; See Fig.\ \ref{fig:1726Orbit}}\\
			0_{13}^{(1,0,0,0,0,0,0,6,2,0,0)}\\
			0_{14}^{(1,0,0,0,0,0,1,5,2,0,0)}\\
			0_{15}^{(1,0,0,0,0,0,1,6,1,0,0)}
		\end{cases}$
	\item $1^{(1,1,6)}
		\begin{cases}
			1_3^{(1,6)}\text{; See Fig.\ \ref{fig:532Orbit}}\\
			1_4^{(1,0,6)}\text{; See Fig.\ \ref{fig:146Orbit}}\\
			1_5^{(1,1,5)}\text{; See Fig.\ \ref{fig:2515Orbit}}\\
			0_6^{(1,1,6)}\text{; See Fig.\ \ref{fig:0617Orbit}}\\
			1_7^{(1,1,6)}\text{; See Fig.\ \ref{fig:1726Orbit}}
		\end{cases}$
	\item $1^{(1,4,3)}
		\begin{cases}
			1_4^{(4,3)}\text{; See Fig.\ \ref{fig:4413Orbit}}\\
			1_5^{(1,3,3)}\text{; See Fig.\ \ref{fig:2515Orbit}}\\
			1_6^{(1,4,2)}\text{; See Fig.\ \ref{fig:1643Orbit}}\\
			0_7^{(1,4,3)}\text{; See Fig.\ \ref{fig:0753Orbit}}\\
			1_8^{(1,4,3)}
		\end{cases}$
	\item $1^{(2,5,1)}
		\begin{cases}
			1_5^{(1,5,1)}\text{; See Fig.\ \ref{fig:2515Orbit}}\\
			1_6^{(2,4,1)}\text{; See Fig.\ \ref{fig:1643Orbit}}\\
			1_7^{(2,5,0)}\text{; See Fig.\ \ref{fig:17160Orbit}}\\
			0_8^{(2,5,1)}\\
			1_9^{(2,5,1)}
		\end{cases}$
	\item $1^{(4,4,0)}
		\begin{cases}
			1_6^{(3,4,0)}\text{; See Fig.\ \ref{fig:1643Orbit}}\\
			1_7^{(4,3,0)}\text{; See Fig.\ \ref{fig:17160Orbit}}\\
			0_9^{(4,4,0)}\\
			1_{10}^{(4,4,0)}
		\end{cases}$
	\item $1^{(7,1,0)}
		\begin{cases}
			1_7^{(6,1,0)}\text{; See Fig.\ \ref{fig:17160Orbit}}\\
			1_8^{(7,0,0)}\\
			0_{10}^{(7,1,0)}\\
			1_{11}^{(7,1,0)}
		\end{cases}$
	\item $1^{(1,2,0,6)}
		\begin{cases}
			1_4^{(2,0,6)}\text{; See Fig.\ \ref{fig:146Orbit}}\\
			1_5^{(1,1,0,6)}\text{; See Fig.\ \ref{fig:2515Orbit}}\\
			1_7^{(1,2,0,5)}\text{; See Fig.\ \ref{fig:1726Orbit}}\\
			0_8^{(1,2,0,6)}
		\end{cases}$
	\item $1^{(2,1,2,4)}
		\begin{cases}
			1_5^{(1,1,2,4)}\text{; See Fig.\ \ref{fig:2515Orbit}}\\
			1_6^{(2,0,2,4)}\text{; See Fig.\ \ref{fig:3624Orbit}}\\
			1_7^{(2,1,1,4)}\text{; See Fig.\ \ref{fig:1726Orbit}}\\
			1_8^{(2,1,2,3)}\\
			0_9^{(2,1,2,4)}
		\end{cases}$
	\item $1^{(3,1,2,3)}
		\begin{cases}
			1_6^{(2,1,2,3)}\text{; See Fig.\ \ref{fig:3624Orbit}}\\
			1_7^{(3,0,2,3)}\text{; See Fig.\ \ref{fig:1726Orbit}}\\
			1_8^{(3,1,1,3)}\\
			1_9^{(3,1,2,2)}\\
			0_{10}^{(3,1,2,3)}
		\end{cases}$
	\item $1^{(4,2,0,3)}
		\begin{cases}
			1_7^{(3,2,0,3)}\text{; See Fig.\ \ref{fig:1726Orbit}}\\
			1_8^{(4,1,0,3)}\\
			1_{10}^{(4,2,0,2)}\\
			0_{11}^{(4,2,0,3)}
		\end{cases}$
	\item $1^{(1,0,1,4,3)}
		\begin{cases}
			1_4^{(1,4,3)}\text{; See Fig.\ \ref{fig:4413Orbit}}\\
			1_6^{(1,0,0,4,3)}\text{; See Fig.\ \ref{fig:1643Orbit}}\\
			1_7^{(1,0,1,3,3)}\text{; See Fig.\ \ref{fig:27133Orbit}}\\
			1_8^{(1,0,1,4,2)}\\
			1_9^{(1,0,1,4,3)}
		\end{cases}$
	\item $1^{(2,0,2,4,1)}
		\begin{cases}
			1_6^{(1,0,2,4,1)}\text{; See Fig.\ \ref{fig:1643Orbit}}\\
			1_8^{(2,0,1,4,1)}\\
			1_9^{(2,0,2,3,1)}\\
			1_{10}^{(2,0,2,4,0)}\\
			0_{12}^{(2,0,2,4,1)}
		\end{cases}$
	\item $1^{(1,0,4,1,3)}
		\begin{cases}
			1_5^{(4,1,3)}\text{; See Fig.\ \ref{fig:5522Orbit}}\\
			1_7^{(1,0,3,1,3)}\text{; See Fig.\ \ref{fig:27133Orbit}}\\
			1_8^{(1,0,4,0,3)}\\
			1_9^{(1,0,4,1,2)}\\
			0_{10}^{(1,0,4,1,3)}
		\end{cases}$
	\item $1^{(1,1,1,4,2)}
		\begin{cases}
			1_5^{(1,1,4,2)}\text{; See Fig.\ \ref{fig:2515Orbit}}\\
			1_6^{(1,0,1,4,2)}\text{; See Fig.\ \ref{fig:1643Orbit}}\\
			1_7^{(1,1,0,4,2)}\text{; See Fig.\ \ref{fig:27133Orbit}}\\
			1_8^{(1,1,1,3,2)}\\
			1_9^{(1,1,1,4,1)}\\
			0_{10}^{(1,1,1,4,2)}
		\end{cases}$
	\item $1^{(1,2,2,2,2)}
		\begin{cases}
			1_6^{(2,2,2,2)}\text{; See Fig.\ \ref{fig:3624Orbit}}\\
			1_7^{(1,1,2,2,2)}\text{; See Fig.\ \ref{fig:27133Orbit}}\\
			1_8^{(1,2,1,2,2)}\\
			1_9^{(1,2,2,1,2)}\\
			1_{10}^{(1,2,2,2,1)}\\
			0_{11}^{(1,2,2,2,2)}
		\end{cases}$
	\item $1^{(1,4,1,1,2)}
		\begin{cases}
			1_7^{(4,1,1,2)}\text{; See Fig.\ \ref{fig:1726Orbit}}\\
			1_8^{(1,3,1,1,2)}\\
			1_9^{(1,4,0,1,2)}\\
			1_{10}^{(1,4,1,0,2)}\\
			1_{11}^{(1,4,1,1,1)}\\
			0_{12}^{(1,4,1,1,2)}
		\end{cases}$
	\item $1^{(2,1,4,0,2)}
		\begin{cases}
			1_7^{(1,1,4,0,2)}\text{; See Fig.\ \ref{fig:27133Orbit}}\\
			1_8^{(2,0,4,0,2)}\\
			1_9^{(2,1,3,0,2)}\\
			1_{11}^{(2,1,4,0,1)}\\
			0_{12}^{(2,1,4,0,2)}
		\end{cases}$
	\item $1^{(2,2,1,3,1)}
		\begin{cases}
			1_7^{(1,2,1,3,1)}\text{; See Fig.\ \ref{fig:27133Orbit}}\\
			1_8^{(2,1,1,3,1)}\\
			1_9^{(2,2,0,3,1)}\\
			1_{10}^{(2,2,1,2,1)}\\
			1_{11}^{(2,2,1,3,0)}\\
			0_{12}^{(2,2,1,3,1)}
		\end{cases}$
	\item $1^{(3,0,1,5,0)}
		\begin{cases}
			1_7^{(2,0,1,5,0)}\text{; See Fig.\ \ref{fig:3760Orbit}}\\
			1_9^{(3,0,0,5,0)};\\
			1_{10}^{(3,0,1,4,0)};\\
			0_{12}^{(3,0,1,5,0)}
		\end{cases}$
	\item $1^{(1,0,0,4,3,1)}
		\begin{cases}
			1_5^{(4,1,3)}\text{; See Fig.\ \ref{fig:5522Orbit}}\\
			1_8^{(1,0,0,3,3,1)}\\
			1_9^{(1,0,0,4,2,1)}\\
			1_{10}^{(1,0,0,4,3,0)}\\
			0_{11}^{(1,0,0,4,3,1)}
		\end{cases}$
	\item $1^{(1,0,1,1,6,0)}
		\begin{cases}
			1_5^{(1,1,6,0)}\text{; See Fig.\ \ref{fig:2515Orbit}}\\
			1_7^{(1,0,0,1,6,0)}\text{; See Fig.\ \ref{fig:17160Orbit}}\\
			1_8^{(1,0,1,0,6,0)}\\
			1_9^{(1,0,1,1,5,0)}\\
			1_{11}^{(1,0,1,1,6,0)}
		\end{cases}$
	\item $1^{(1,0,2,3,2,1)}
		\begin{cases}
			1_6^{(2,3,2,1)}\text{; See Fig.\ \ref{fig:3624Orbit}}\\
			1_8^{(1,0,1,3,2,1)}\\
			1_9^{(1,0,2,2,2,1)}\\
			1_{10}^{(1,0,2,3,1,1)}\\
			1_{11}^{(1,0,2,3,2,0)}\\
			0_{12}^{(1,0,2,3,2,1)}
		\end{cases}$
	\item $1^{(1,0,5,0,2,1)}
		\begin{cases}
			1_7^{(5,0,2,1)}\text{: See Fig.\ \ref{fig:1726Orbit}}\\
			1_9^{(1,0,4,0,2,1)}\\
			1_{11}^{(1,0,5,0,1,1)}\\
			1_{12}^{(1,0,5,0,2,0)}\\
			0_{13}^{(1,0,5,0,2,1)}
		\end{cases}$
	\item $1^{(1,1,0,3,4,0)}
		\begin{cases}
			1_6^{(1,0,3,4,0)}\text{; See Fig.\ \ref{fig:1643Orbit}}\\
			1_7^{(1,0,0,3,4,0)}\text{; See Fig.\ \ref{fig:17160Orbit}}\\
			1_9^{(1,1,0,2,4,0)}\\
			1_{10}^{(1,1,0,3,3,0)}\\
			0_{12}^{(1,1,0,3,4,0)}
		\end{cases}$
	\item $1^{(1,1,2,3,1,1)}
		\begin{cases}
			1_7^{(1,2,3,1,1)}\text{; See Fig.\ \ref{fig:27133Orbit}}\\
			1_8^{(1,0,2,3,1,1)}\\
			1_9^{(1,1,1,3,1,1)}\\
			1_{10}^{(1,1,2,2,1,1)}\\
			1_{11}^{(1,1,2,3,0,1)}\\
			1_{12}^{(1,1,2,3,1,0)}\\
			0_{13}^{(1,1,2,3,1,1)}
		\end{cases}$
	\item $1^{(1,1,3,0,4,0)}
		\begin{cases}
			1_7^{(1,3,0,4,0)}\text{; See Fig.\ \ref{fig:27133Orbit}}\\
			1_8^{(1,0,3,0,4,0)}\\
			1_9^{(1,1,2,0,4,0)}\\
			1_{11}^{(1,1,3,0,3,0)}\\
			0_{13}^{(1,1,3,0,4,0)}
		\end{cases}$
	\item $1^{(1,2,0,3,3,0)}
		\begin{cases}
			1_7^{(2,0,3,3,0)}\text{; See Fig.\ \ref{fig:3760Orbit}}\\
			1_8^{(1,1,0,3,3,0)}\\
			1_{10}^{(1,2,0,2,3,0)}\\
			1_{11}^{(1,2,0,3,2,0)}\\
			0_{13}^{(1,2,0,3,3,0)}
		\end{cases}$
	\item $1^{(2,0,0,5,2,0)}
		\begin{cases}
			1_7^{(1,0,0,5,2,0)}\text{; See Fig.\ \ref{fig:17160Orbit}}\\
			1_{10}^{(2,0,0,4,2,0)}\\
			1_{11}^{(2,0,0,5,1,0)}\\
			0_{13}^{(2,0,0,5,2,0)}
		\end{cases}$
	\item $1^{(1,0,0,2,4,2,0)}
		\begin{cases}
			1_6^{(2,4,2,0)}\text{; See Fig.\ \ref{fig:3624Orbit}}\\
			1_9^{(1,0,0,1,4,2,0)}\\
			1_{10}^{(1,0,0,2,3,2,0)}\\
			1_{11}^{(1,0,0,2,4,1,0)}\\
			0_{13}^{(1,0,0,2,4,2,0)}
		\end{cases}$
	\item $1^{(1,0,0,5,2,0,1)}
		\begin{cases}
			1_7^{(5,2,0,1)}\text{; See Fig.\ \ref{fig:1726Orbit}}\\
			1_{10}^{(1,0,0,4,2,0,1)}\\
			1_{11}^{(1,0,0,5,1,0,1)}\\
			1_{13}^{(1,0,0,5,2,0,0)}\\
			0_{14}^{(1,0,0,5,2,0,1)}
		\end{cases}$
	\item $1^{(1,0,1,3,2,2,0)}
		\begin{cases}
			1_7^{(1,3,2,2,0)}\text{; See Fig.\ \ref{fig:27133Orbit}}\\
			1_9^{(1,0,0,3,2,2,0)}\\
			1_{10}^{(1,0,1,2,2,2,0)}\\
			1_{11}^{(1,0,1,3,1,2,0)}\\
			1_{12}^{(1,0,1,3,2,1,0)}\\
			0_{14}^{(1,0,1,3,2,2,0)}
		\end{cases}$
	\item $1^{(1,0,2,0,5,1,0)}
		\begin{cases}
			1_7^{(2,0,5,1,0)}\text{; See Fig.\ \ref{fig:3760Orbit}}\\
			1_9^{(1,0,1,0,5,1,0)}\\
			1_{11}^{(1,0,2,0,4,1,0)}\\
			1_{12}^{(1,0,2,0,5,0,0)}\\
			0_{14}^{(1,0,2,0,5,1,0)}
		\end{cases}$
	\item $1^{(1,0,0,0,6,1,1,0)}
		\begin{cases}
			1_7^{(6,1,1,0)}\text{; See Fig.\ \ref{fig:1726Orbit}}\\
			1_{11}^{(1,0,0,0,5,1,1,0)}\\
			1_{12}^{(1,0,0,0,6,0,1,0)}\\
			1_{13}^{(1,0,0,0,6,1,0,0)}\\
			0_{15}^{(1,0,0,0,6,1,1,0}
		\end{cases}$
	\item $1^{(1,1,0,0,7,0,0)}
		\begin{cases}
			1_7^{(1,0,0,7,0,0)}\text{; See Fig.\ \ref{fig:17160Orbit}}\\
			1_8^{(1,0,0,0,7,0,0)}\\
			1_{11}^{(1,1,0,0,6,0,0)}\\
			0_{14}^{(1,1,0,0,7,0,0)}
		\end{cases}$
	\item $1^{(1,0,0,1,3,4,0,0)}
		\begin{cases}
			1_7^{(1,3,4,0,0)}\text{; See Fig.\ \ref{fig:27133Orbit}}\\
			1_{10}^{(1,0,0,0,3,4,0,0)}\\
			1_{11}^{(1,0,0,1,2,4,0,0)}\\
			1_{12}^{(1,0,0,1,3,3,0,0)}\\
			1_{15}^{(1,0,0,1,3,4,0,0)}
		\end{cases}$
	\item $2=\text{M2}
		\begin{cases}
			1_0=\text{F1; See Fig.\ \ref{fig:10Orbit}}\\
			2_1=\text{D2; See Fig.\ \ref{fig:11Orbit}}
		\end{cases}$
	\item $2^6
		\begin{cases}
			2_3^5\text{; See Fig.\ \ref{fig:532Orbit}}\\
			1_4^6\text{; See Fig.\ \ref{fig:146Orbit}}\\
			2_5^6\text{; See Fig.\ \ref{fig:2515Orbit}}
		\end{cases}$
	\item $2^{(4,3)}
		\begin{cases}
			2_4^{(3,3)}\text{; See Fig.\ \ref{fig:4413Orbit}}\\
			2_5^{(4,2)}\text{; See Fig.\ \ref{fig:2515Orbit}}\\
			1_6^{(4,3)}\text{; See Fig.\ \ref{fig:1643Orbit}}\\
			2_7^{(4,3)}\text{; See Fig.\ \ref{fig:27133Orbit}}
		\end{cases}$
	\item $2^{(7,0)}
		\begin{cases}
			2_5^{(6,0)}\text{; See Fig.\ \ref{fig:2515Orbit}}\\
			1_7^{(7,0)}\text{; See Fig.\ \ref{fig:17160Orbit}}\\
			2_8^{(7,0)}
		\end{cases}$
	\item $2^{(4,0,4)}
		\begin{cases}
			2_5^{(3,0,4)}\text{; See Fig.\ \ref{fig:5522Orbit}}\\
			2_7^{(4,0,3)}\text{; See Fig.\ \ref{fig:27133Orbit}}\\
			1_8^{(4,0,4)}
		\end{cases}$
	\item $2^{(1,0,2,5)}
		\begin{cases}
			2_3^{(2,5)}\text{; See Fig.\ \ref{fig:532Orbit}}\\
			2_5^{(1,0,1,5)}\text{; See Fig.\ \ref{fig:2515Orbit}}\\
			2_6^{(1,0,2,4)}\text{; See Fig.\ \ref{fig:3624Orbit}}\\
			1_7^{(1,0,2,5)}\text{; See Fig.\ \ref{fig:1726Orbit}}\\
		\end{cases}$
	\item $2^{(1,1,3,3)}
		\begin{cases}
			2_4^{(1,3,3)}\text{; See Fig.\ \ref{fig:4413Orbit}}\\
			2_5^{(1,0,3,3)}\text{; See Fig.\ \ref{fig:2515Orbit}}\\
			2_6^{(1,1,2,3)}\text{; See Fig.\ \ref{fig:3624Orbit}}\\
			2_7^{(1,1,3,2)}\text{; See Fig.\ \ref{fig:27133Orbit}}\\
			1_8^{(1,1,3,3)}
		\end{cases}$
	\item $2^{(1,3,2,2)}
		\begin{cases}
			2_5^{(3,2,2)}\text{; See Fig.\ \ref{fig:5522Orbit}}\\
			2_6^{(1,2,2,2)}\text{; See Fig.\ \ref{fig:3624Orbit}}\\
			2_7^{(1,3,1,2)}\text{; See Fig.\ \ref{fig:27133Orbit}}\\
			2_8^{(1,3,2,1)}\\
			1_9^{(1,3,2,2)}
		\end{cases}$
	\item $2^{(2,0,5,1)}
		\begin{cases}
			2_5^{(1,0,5,1)}\text{; See Fig.\ \ref{fig:2515Orbit}}\\
			2_7^{(2,0,4,1)}\text{; See Fig.\ \ref{fig:27133Orbit}}\\
			2_8^{(2,0,5,0)}\\
			1_9^{(2,0,5,1)}
		\end{cases}$
	\item $2^{(2,3,2,1)}
		\begin{cases}
			2_6^{(1,3,2,1)}\text{; See Fig.\ \ref{fig:3624Orbit}}\\
			2_7^{(2,2,2,1)}\text{; See Fig.\ \ref{fig:27133Orbit}}\\
			2_8^{(2,3,1,1)}\\
			2_9^{(2,3,2,0)}\\
			1_{10}^{(2,3,2,1)}
		\end{cases}$
	\item $2^{(3,4,0,1)}
		\begin{cases}
			2_7^{(2,4,0,1)}\text{; See Fig.\ \ref{fig:27133Orbit}}\\
			2_8^{(3,3,0,1)}\\
			2_{10}^{(3,4,0,0)}\\
			1_{11}^{(3,4,0,1)}
		\end{cases}$
	\item $2^{(4,1,3,0)}
		\begin{cases}
			2_7^{(3,1,3,0)}\text{; See Fig.\ \ref{fig:27133Orbit}}\\
			2_8^{(4,0,3,0)}\\
			2_9^{(4,1,2,0)}\\
			1_{11}^{(4,1,3,0)}
		\end{cases}$
	\item $2^{(1,0,0,7,0)}
		\begin{cases}
			2_4^{(7,0)}\text{; See Fig.\ \ref{fig:7420Orbit}}\\
			2_7^{(1,0,0,6,0)}\text{; See Fig.\ \ref{fig:3760Orbit}}\\
			1_9^{(1,0,0,7,0)}
		\end{cases}$
	\item $2^{(1,0,3,4,0)}
		\begin{cases}
			2_5^{(3,4,0)}\text{; See Fig.\ \ref{fig:5522Orbit}}\\
			2_7^{(1,0,2,4,0)}\text{; See Fig.\ \ref{fig:3760Orbit}}\\
			2_8^{(1,0,3,3,0)}\\
			1_{10}^{(1,0,3,4,0)}
		\end{cases}$
	\item $2^{(1,1,4,2,0)}
		\begin{cases}
			2_6^{(1,4,2,0)}\text{; See Fig.\ \ref{fig:3624Orbit}}\\
			2_7^{(1,0,4,2,0)}\text{; See Fig.\ \ref{fig:3760Orbit}}\\
			2_9^{(1,1,4,1,0)}\\
			1_{11}^{(1,1,4,2,0)}
		\end{cases}$
	\item $2^{(1,3,3,1,0)}
		\begin{cases}
			2_7^{(3,3,1,0)}\text{; See Fig.\ \ref{fig:27133Orbit}}\\
			2_8^{(1,2,3,1,0)}\\
			2_9^{(1,3,2,1,0)}\\
			2_{10}^{(1,3,3,0,0)}\\
			1_{12}^{(1,3,3,1,0)}
		\end{cases}$
	\item $2^{(2,0,6,0,0)}
		\begin{cases}
			2_7^{(1,0,6,0,0)}\text{; See Fig.\ \ref{fig:3760Orbit}}\\
			2_9^{(2,0,5,0,0)};\\
			1_{12}^{(2,0,6,0,0)}
		\end{cases}$
	\item $3^{(2,4)}
		\begin{cases}
			3_3^{(1,4)}\text{; See Fig.\ \ref{fig:532Orbit}}\\
			3_4^{(2,3)}\text{; See Fig.\ \ref{fig:4413Orbit}}\\
			2_5^{(2,4)}\text{; See Fig.\ \ref{fig:2515Orbit}}\\
			3_6^{(2,4)}\text{; See Fig.\ \ref{fig:3624Orbit}}
		\end{cases}$
	\item $3^{(2,2,3)}
		\begin{cases}
			3_4^{(1,2,3)}\text{; See Fig.\ \ref{fig:4413Orbit}}\\
			3_5^{(2,1,3)}\text{; See Fig.\ \ref{fig:5522Orbit}}\\
			3_6^{(2,2,2)}\text{; See Fig.\ \ref{fig:3624Orbit}}\\
			2_7^{(2,2,3)}\text{; See Fig.\ \ref{fig:27133Orbit}}
		\end{cases}$
	\item $3^{(3,3,1)}
		\begin{cases}
			3_5^{(2,3,1)}\text{; See Fig.\ \ref{fig:5522Orbit}}\\
			3_6^{(3,2,1)}\text{; See Fig.\ \ref{fig:3624Orbit}}\\
			3_7^{(3,3,0)}\text{; See Fig.\ \ref{fig:3760Orbit}}\\
			2_8^{(3,3,1)}
		\end{cases}$
	\item $3^{(5,2,0)}
		\begin{cases}
			3_6^{(4,2,0)}\text{; See Fig.\ \ref{fig:3624Orbit}}\\
			3_7^{(5,1,0)}\text{; See Fig.\ \ref{fig:3760Orbit}}\\
			2_9^{(5,2,0)}
		\end{cases}$
	\item $4^{(1,2,3)}
		\begin{cases}
			4_3^{(2,3)}\text{; See Fig.\ \ref{fig:532Orbit}}\\
			4_4^{(1,1,3)}\text{; See Fig.\ \ref{fig:4413Orbit}}\\
			4_5^{(1,2,2)}\text{; See Fig.\ \ref{fig:5522Orbit}}\\
			3_6^{(1,2,3)}\text{; See Fig.\ \ref{fig:3624Orbit}}
		\end{cases}$
	\item $4^{(1,5,0)}
		\begin{cases}
			4_4^{(5,0)}\text{; See Fig.\ \ref{fig:7420Orbit}}\\
			4_5^{(1,4,0)}\text{; See Fig.\ \ref{fig:5522Orbit}}\\
			3_7^{(1,5,0)}\text{; See Fig.\ \ref{fig:3760Orbit}}
		\end{cases}$
	\item $5=\text{M5}
		\begin{cases}
			4_1=\text{D4; See Fig.\ \ref{fig:11Orbit}}\\
			5_2=\text{NS5; See Fig.\ \ref{fig:52Orbit}}
		\end{cases}$
	\item $5^3
		\begin{cases}
			5_2^2\text{; See Fig.\ \ref{fig:52Orbit}}\\
			4_3^3\text{; See Fig.\ \ref{fig:532Orbit}}\\
			5_4^3\text{; See Fig.\ \ref{fig:4413Orbit}}
		\end{cases}$
	\item $5^{(1,3)}
		\begin{cases}
			5_2^3=\text{R-monopole; See Fig.\ \ref{fig:52Orbit}}\\
			5_3^{(1,2)}\text{; See Fig.\ \ref{fig:532Orbit}}\\
			4_4^{(1,3)}\text{; See Fig.\ \ref{fig:4413Orbit}}\\
			5_5^{(1,3)}\text{; See Fig.\ \ref{fig:5522Orbit}}
		\end{cases}$
	\item $5^{(1,0,4)}
		\begin{cases}
			5_2^4\text{; See Fig.\ \ref{fig:52Orbit}}\\
			5_4^{(1,0,3)}\text{; See Fig.\ \ref{fig:4413Orbit}}\\
			4_5^{(1,0,4)}\text{; See Fig.\ \ref{fig:5522Orbit}}
		\end{cases}$
	\item $6^1=\text{KK6M}
		\begin{cases}
			6_1=\text{D6; See Fig.\ \ref{fig:11Orbit}}\\
			5_2^1=\text{KK5A; See Fig.\ \ref{fig:52Orbit}}\\
			6_3^1 \text{; See Fig.\ \ref{fig:532Orbit}}
		\end{cases}$
	\item $6^{(3,1)}
		\begin{cases}
			6_3^{(2,1)}\text{; See Fig.\ \ref{fig:532Orbit}}\\
			6_4^{(3,0)}\text{; See Fig.\ \ref{fig:7420Orbit}}\\
			5_5^{(3,1)}\text{; See Fig.\ \ref{fig:5522Orbit}}
		\end{cases}$
	\item $8^{(1,0)}=\text{KK8M}
		\begin{cases}
			8_1=\text{D8; See Fig.\ \ref{fig:11Orbit}}\\
			7_3^{(1,0)}=\text{KK7A; See Fig.\ \ref{fig:532Orbit}}\\
			8_4^{(1,0)}=\text{KK8A; See Fig.\ \ref{fig:7420Orbit}}
		\end{cases}$
\end{itemize}
\end{multicols}
\endgroup
\restoregeometry

	\chapter{Computing the traces of the Coset Projectors}\label{app:CosetProjectorTraces}
In Section~\ref{sec:CosetProjectorsForAllExFTs}, we claimed that the coset projector 
\begin{align}
\mathcal{P}_{MN}{}^{KL} = \frac{1}{\alpha} \left( \delta^{(K}_M \delta^{L)}_N - \omega \mathcal{M}_{MN} \mathcal{M}^{KL} - \mathcal{M}_{MQ} Y^{Q(K}{}_{RN} \mathcal{M}^{L)R} \right)\,.
\end{align}
projected onto a space of dimension $G/H$. Here, we demonstrate that this is indeed the case. The DFT, and $n=4,5$ EFT projectors were already treated in \cite{Berkeley:2014nza,Rudolph:2016sxe} and so we shall consider only $n=3,6,7,8$ EFT as well as GR, for completeness. The trace of $\mathcal{P}$ is given by
\begin{align}
\mathcal{P}_{MN}{}^{MN} & = \frac{1}{2\alpha} \left( \operatorname{dim} R_1 (1 - \omega) + \alpha \operatorname{dim} \operatorname{adj.} - \mathcal{M}_{MN} Y^{MN}{}_{QP}\mathcal{M}^{PQ} \right)\,.
\end{align}
For convenience, we shall also denote the trace of the $Y$-tensor as
\begin{align}\label{eq:r2}
r \coloneqq \frac{1}{2\alpha} \mathcal{M}_{MN} Y^{MN}{}_{QP}\mathcal{M}^{PQ}\,.
\end{align}
\section{General Relativity}
We begin with the simplest case of GR, reinterpreted as an ExFT, for which we have $\alpha =1$, $\omega = 0$ and $Y^{MN}{}_{KL} = 0$ (see Table~\ref{tab:Summary}). The coordinate representation is $R_1 = \mathbf{D}$ whilst the adjoint representation is $\mathbf{D^2}$. Substituting in these numbers we obtain
\begin{align}
\mathcal{P}_{MN}{}^{MN} = \frac{D(D+1)}{2}\,,
\end{align}
which is just the dimension of the coset $\operatorname{GL}(D) / \operatorname{SO}(D)$ that is parametrised by the metric.
\section{\texorpdfstring{$\operatorname{SL}(3) \times \operatorname{SL}(2)$}{SL(3)xSL(2)} EFT}
For $n=3$, we recall that the coset projector is given by the sum of projectors onto the adjoint representations of the two factors in $G$:
\begin{align}
\begin{aligned}
\delta_\Lambda \mathcal{M}_{MN} = \Lambda^P\partial_P \mathcal{M}_{MN} & + 2 \times 2 \mathcal{M}_{MQ} {\left( \mathbb{P}_{(\mathbf{8,1})} \right)}^Q{}_N{}^{(K}{}_R \mathcal{M}^{L)R} \partial_K \Lambda^P \mathcal{M}_{LP}\\
	& + 2 \times 3 \mathcal{M}_{MQ} {\left( \mathbb{P}_{(\mathbf{1,3})} \right)}^Q{}_N{}^{(K}{}_R \mathcal{M}^{L)R} \partial_K \Lambda^P \mathcal{M}_{LP}\,.
\end{aligned}
\end{align}
The traces of the two coset projectors are given by
\begin{align}
\mathcal{P}^{(\mathbf{8,1})}_{MN}{}^{MN} & = \frac{1}{2} \mathcal{M}_{MQ} \left( \mathbb{P}_{(\mathbf{8,1})}{}^Q{}_N{}^{M}{}_R  \mathcal{M}^{NR} + \mathbb{P}_{(\mathbf{8,1})}{}^Q{}_N{}^{N}{}_R  \mathcal{M}^{MR} \right)\\
	& = \frac{1}{2} \mathcal{M}_{MQ} \mathbb{P}_{(\mathbf{8,1})}{}^Q{}_N{}^{M}{}_R  \mathcal{M}^{NR} + 4\,,\\
\mathcal{P}^{(\mathbf{1,3})}_{MN}{}^{MN} & = \frac{1}{2} \mathcal{M}_{MQ} \left( \mathbb{P}_{(\mathbf{1,3})}{}^Q{}_N{}^{M}{}_R  \mathcal{M}^{NR} + \mathbb{P}_{(\mathbf{1,3})}{}^Q{}_N{}^{N}{}_R  \mathcal{M}^{MR} \right)\\
	& = \frac{1}{2} \mathcal{M}_{MQ} \mathbb{P}_{(\mathbf{1,3})}{}^Q{}_N{}^{M}{}_R  \mathcal{M}^{NR} + \frac{3}{2}\,.
\end{align}
We may use the explicit form of the adjoint projectors given in \cite{Hohm:2015xna},
\begin{align}
{\left( {\mathbb{P}}_{\mathbf{(8,1)}} \right)}^M{}_N{}^K{}_L = {\left( {\mathbb{P}}_{\mathbf{(8,1)}} \right)}^{m \alpha}{}_{l \delta}{}^{j \beta}{}_{k \gamma} = \frac{1}{2} \delta^i_k \delta^j_l \delta^\alpha_\delta \delta^\beta_\gamma - \frac{1}{6} \delta^j_k \delta^i_l \delta^\alpha_\delta \delta^\beta_\gamma\,,\\
{\left( {\mathbb{P}}_{\mathbf{(1,3)}} \right)}^M{}_N{}^K{}_L = {\left( {\mathbb{P}}_{\mathbf{(1,3)}} \right)}^{m \alpha}{}_{l \delta}{}^{j \beta}{}_{k \gamma} = \frac{1}{3} \delta^i_l \delta^j_k \delta^\alpha_\gamma \delta^\beta_\delta - \frac{1}{6} \delta^j_k \delta^i_l \delta^\alpha_\delta \delta^\beta_\gamma\,,
\end{align}
to compute the trace of the total coset projector:
\begin{align}
\mathcal{P}_{MN}{}^{MN} & = \mathcal{P}^{(\mathbf{8,1})}_{MN}{}^{MN} + \mathcal{P}^{(\mathbf{1,3})}_{MN}{}^{MN}\\
	& = \frac{1}{2} \mathcal{M}_{MQ} \left(  {\left(\mathbb{P}_{(\mathbf{8,1})} \right)}^Q{}_N{}^{M}{}_R  + {\left( \mathbb{P}_{(\mathbf{1,3})} \right)}^Q{}_N{}^{M}{}_R  \right) \mathcal{M}^{NR} + \frac{11}{2}\,.
\end{align}
Using the fact that $\mathcal{M}_{i \alpha, j \beta} = g_{ij} \otimes g_{\alpha \beta}$ in the usual parametrisation, we obtain
\begin{align}
\mathcal{P}_{MN}{}^{MN} = \frac{3}{2} + \frac{11}{2} = 7
\end{align}
which is the dimension of the coset $(\operatorname{SL}(3) \times \operatorname{SL}(2))/(\operatorname{SO}(3) \times \operatorname{SO}(2))$ as required.
\section{\texorpdfstring{$E_{6(6)}$}{E6(6)} EFT}
We focus on the value of $r$ defined in \eqref{eq:r2}. For $G=E_{6(6)}$ EFT, the coordinate representation is the fundamental representation $R_1 = \mathbf{27}$ and so we take the generalised indices to run from $M, N = 1, \ldots, 27$. The Y-tensor is given by
\begin{align}
Y^{MN}{}_{PQ} = 10 d^{MNK} d_{PQK}\,,
\end{align}
with $d^{MNK}$ and $d_{MNK}$ the symmetric cubic invariants of $E_{6(6)}$. We use the $\operatorname{USp}(8)$ construction of \cite{Musaev:2014lna} in which the generalised metric is defined in terms of the generalised vielbein ${\mathcal{E}}_M{}^{ij}$, carrying an antisymmetric pair of indices $i,j=1, \ldots, 8$ that transforms in the fundamental representation of $H = \operatorname{USp}(8)$. We shall follow their conventions and raise and lower indices with the symplectic form $\Omega_{ij}$ with the conventions
\begin{align}\label{eq:VOmega}
{\mathcal{E}}_{Mij} = {\mathcal{E}}_M{}^{kl} \Omega_{ki} \Omega_{lj}\,, \qquad {\mathcal{E}}_M{}^{ij} \Omega_{ij} = 0\,, \qquad \Omega_{ik} \Omega^{jk} = \delta_i^j\,.
\end{align}
In terms of this generalised vielbein, we have $\mathcal{M}_{MN} = {\mathcal{E}}_M{}^{ij} {\mathcal{E}}_{Nij}$. The orthogonality relations with the inverse vielbein are given by
\begin{align}
{\mathcal{E}}_M{}^{ij} {\mathcal{E}}_{ij}{}^N = \delta_M^N, \qquad {\mathcal{E}}_M{}^{kl} {\mathcal{E}}_{ij}{}^M = \delta^{kl}_{ij} - \frac{1}{8} \Omega_{ij} \Omega^{kl}.
\end{align}
Finally, the totally symmetric invariant $d^{MNK}$ is given in terms of the symplectic form as
\begin{align}
d^{MNK}  &= \frac{2}{\sqrt{5}} {\mathcal{E}}_{ij}{}^M {\mathcal{E}}_{kl}{}^N {\mathcal{E}}_{mn}{}^P \Omega^{jk} \Omega^{lm} \Omega^{ni}\,,\\
d_{MNK} & = \frac{2}{\sqrt{5}} {\mathcal{E}}_M{}^{ij} {\mathcal{E}}_N{}^{kl} {\mathcal{E}}_P{}^{mn} \Omega_{jk} \Omega_{lm} \Omega_{ni}\,.
\end{align}
We can then compute
\begin{align}
\mathcal{M}_{MN} d^{MNK} & =\frac{2}{\sqrt{5}} {\mathcal{E}}_M{}^{pq} {\mathcal{E}}_N{}^{rs}\Omega_{rp} \Omega_{sq} {\mathcal{E}}_{ij}{}^M {\mathcal{E}}_{kl}{}^N {\mathcal{E}}_{mn}{}^K \Omega^{jk} \Omega^{lm} \Omega^{ni}\\ 
& = \frac{2}{\sqrt{5}} \left(\delta^{pq}_{ij} - \frac{1}{8} \Omega_{ij} \Omega^{pq} \right) \left(\delta^{rs}_{kl} - \frac{1}{8} \Omega_{kl} \Omega^{rs} \right) {\mathcal{E}}_{mn}{}^P \Omega_{rp} \Omega_{sq} \Omega^{jk} \Omega^{lm} \Omega^{ni}\\
& = \frac{2}{\sqrt{5}} {\mathcal{E}}_{mn}{}^K \left( \frac{1}{2} ( \Omega_{ki} \Omega_{lj} - \Omega_{kj} \Omega_{li} ) - \frac{1}{8} \Omega_{ij} \Omega_{kl} \right) \Omega^{jk} \Omega^{lm} \Omega^{ni}\\
& \propto {\mathcal{E}}_{mn}{}^K \Omega^{mn}
\end{align}
which vanishes by \eqref{eq:VOmega}. With $r=0$, it is then simple to use the relevant values from Table~\ref{tab:Summary} to compute
\begin{align}
\mathcal{P}_{MN}{}^{MN} = \frac{1}{2 \times 6} \left( 27 \left( 1 + \frac{1}{3} \right) + 6 \times 27 - 0 \right) = 42
\end{align}
which is the appropriate dimension of the coset $E_{6(6)}/\operatorname{USp}(8)$.
\section{\texorpdfstring{$E_{7(7)}$}{E7(7)} EFT}
For $E_{7(7)}$, we use the conventions outlined in Section~\ref{sec:E7EFT} but recount some basic facts for convenience. We index the coordinate representation $R_1 = \mathbf{56}$ by $M, N =1, \ldots, 56$ (which are raised and lowered by the symplectic form $\Omega$) and the adjoint representation as $\alpha = 1, \ldots, 133$. The generators, valued in the fundamental representation are denoted ${\left(t_\alpha\right)}_M{}^N$, in terms of which the $Y$-tensor is given by
\begin{equation}
Y^{MN}{}_{KL} = - 12 {\left( t_\alpha \right)}^{MN} {\left( t^\alpha \right)}_{KL} - \frac{1}{2} \Omega^{MN} \Omega_{KL} \,,
\end{equation}
where ${\left( t_\alpha \right)}_{MN} =  {\left( t_\alpha \right)}_M{}^K \Omega_{KN}$, ${\left( t_\alpha \right)}^{MN} = \Omega^{MK} {\left( t_\alpha \right)}_K{}^N $ are both symmetric in $MN$. We now introduce a generalised vielbein carrying antisymmetrised $\mathrm{SU}(8)$ indices, ${\mathcal{E}}_M{}^{A} = ( {\mathcal{E}}_M{}^{ij}, {\mathcal{E}}_{M ij} )$, such that \cite{Godazgar:2014nqa}
\begin{align}
\mathcal{M}_{MN} = {\mathcal{E}}_M{}^A {\mathcal{E}}_N{}^B {\overbar{\mathcal{M}}}_{AB} =  {\mathcal{E}}_M{}^{ij} {\mathcal{E}}_{N ij} +  {\mathcal{E}}_{M ij} {\mathcal{E}}_{N}{}^{ij} \,.
\end{align}
Since $\Omega^{MN} \mathcal{M}_{MN} = 0$, we can show that $\mathcal{M}_{MN} Y^{MN}{}_{KL} \mathcal{M}^{KL}$ vanishes provided that ${\left( t_\alpha \right)}^{MN} \mathcal{M}_{MN} = 0$. We begin with
\begin{align}
{\left( t_\alpha \right)}^{MN} \mathcal{M}_{MN} = {\left( t_{\alpha} \right)}^{MN}  {\mathcal{E}}_M{}^{A} {\mathcal{E}}_{N}^{B} {\overbar{\mathcal{M}}}_{AB} = {\mathcal{E}}_\alpha{}^{\overbar{\alpha}} {\left( t_{\overbar{\alpha}} \right)}^{AB} {\overbar{\mathcal{M}}}_{AB}\,,
\end{align}
where ${\mathcal{E}}_\alpha{}^{\overbar{\alpha}}$ is the adjoint representation of the vielbein (which we do not need) and ${ \left( t_{\overbar{\alpha}} \right)}^{AB}$ corresponds to the $E_{7(7)}$ generator in the $\operatorname{SU}(8)$ basis. In this basis ${\mathcal{E}}^{A} = ( {\mathcal{E}}^{ij}, {\mathcal{E}}_{ij} )$ and ${\mathcal{E}}_{\overbar{\alpha}} = ( {\mathcal{E}}_i{}^j, {\mathcal{E}}_{ijkl} )$, $\Omega_{ij}{}^{kl} = \delta_{ij}^{kl}$ and (see e.g. appendix of \cite{LeDiffon:2011wt}) the components of ${ \left( t_{\overbar{\alpha}} \right)}^{AB}$ are given by
\begin{equation}
\begin{split} 
{\left( t_i{}^j \right)}_{kl}{}^{mn}& = - \delta^j_{[k} \delta^{mn}_{l]i} - \frac{1}{8} \delta_i^j \delta^{mn}_{kl} = + {\left( t_i{}^j \right)}^{mn}{}_{kl}{}\,,\\
{\left( t_{ijkl} \right)}_{mnpq} &= \frac{1}{4!} \eta_{ijklmnpa} \,,\quad {\left( t_{ijkl} \right)}^{mnpq} = - \delta_{ijkl}^{mnpq}\,.
\end{split}
\end{equation}
We then want to compute
\begin{equation}
{ \left( t_{\overbar{\alpha}} \right)}^{AB} {\overbar{\mathcal{M}}}_{AB} = 2 { \left( t_{\overbar{\alpha}} \right)}^{ij}{}{}_{kl} {\overbar{\mathcal{M}}}_{ij}{}^{kl}\,,
\end{equation}
which is automatically zero for $\overbar{\alpha} = {}_{ijkl}$ and for $\overbar{\alpha} = {}_i{}^j$ turns out to vanish on evaluating the contractions.
We conclude that ${\left( t_{\alpha} \right)}^{MN} \mathcal{M}_{MN} = 0$, and so $Y^{PQ}{}_{MN} \mathcal{M}_{PQ} = 0$.
\section{\texorpdfstring{$E_{8(8)}$}{E8(8)} EFT}
For $E_{8(8)}$ EFT, the $Y$-tensor is given by
\begin{align}
Y^{MN}{}_{KL} & = - f^M{}_{LR} f^{RN}{}_K + 2 \delta^{(M}_K \delta^{N)}_L
\end{align}
and so we compute $r$ as
\begin{align}
r & = \frac{1}{120} \mathcal{M}_{MN} \mathcal{M}^{KL} \left( - f^{M}{}_{LP} f^{PN}{}_K +2 \delta^{(M}_K \delta^{N)}_L \right)\\
	& = \frac{1}{120} \mathcal{M}_{MN} f^M{}_{LP} \mathcal{M}^{PQ} \mathcal{M}^{NR} f_{QR}{}^{L} + \frac{1}{60} \mathcal{M}_{MN} \mathcal{M}^{MN}\\
	& = - \frac{60}{120} \kappa^{PQ} \mathcal{M}_{PQ} + \frac{248}{60} =  \frac{2}{15}\,,
\end{align}
where we have used the following identities:
\begin{align}
\mathcal{M}^{PM} \mathcal{M}^{QN} f_{PQ}{}^K = - f^{MN}{}_L \mathcal{M}^{LK}\,, \qquad \kappa^{MN} \mathcal{M}_{MN} & = 8\,.
\end{align}
The latter follows from defining a generalised vielbein ${\mathcal{E}}_M{}^{\mathcal{A}}$ for the usual coset. In particular, we split the indices into $\operatorname{SO}(16)$ spinor indices $A,B = 1, \ldots, 128$ and adjoint indices (equivalently, antisymmetrised vector indices) $[ij] = 1, \ldots, 120$ such that the vielbein decomposes to ${\mathcal{E}}_M{}^{\mathcal{A}} = ( {\mathcal{E}}_M{}^A, {\mathcal{E}}_M{}^{ij})$. These satisfy (see e.g.\ \cite{Baguet:2016jph})
\begin{align}
\kappa^{MN} {\mathcal{E}}_M{}^A {\mathcal{E}}_N{}^B = \delta^{AB} \,,\qquad  \kappa^{MN} {\mathcal{E}}_M{}^{ij} {\mathcal{E}}_N{}^{kl} = - 2 \delta^{i[k} \delta^{l]j} \,.
\end{align}
The generalised metric is then given by $\mathcal{M}_{MN} = {\mathcal{E}}_M{}^A {\mathcal{E}}_N{}^B \delta_{AB} + \frac{1}{2} {\mathcal{E}}_M{}^{ij} {\mathcal{E}}_N{}^{kl} \delta_{ik} \delta_{jl}$ and it follows from the defining properties of the vielbein that $\kappa^{MN} \mathcal{M}_{MN} = 128 - 120 =8$. Using the values given in Table~\ref{tab:Summary}, we obtain $\mathcal{P}_{MN}{}^{MN} = 128$ as expected.

\backmatter
	\linespread{0.95}\selectfont
	\addcontentsline{toc}{part}{Bibliography}
	\bibliographystyle{JHEP}
	\bibliography{\jobname}

\providecommand{\href}[2]{#2}\begingroup\raggedright\begin{thebibliography}{100}

\bibitem{Berman:2018okd}
D.~S. Berman, E.~T. Musaev and R.~Otsuki, \emph{{Exotic Branes in Exceptional
  Field Theory: $E_{7(7)}$ and Beyond}},
  \href{https://doi.org/10.1007/JHEP12(2018)053}{\emph{JHEP} {\bfseries 12}
  (2018) 053} [\href{https://arxiv.org/abs/1806.00430}{{\ttfamily
  1806.00430}}].

\bibitem{Berman:2019izh}
D.~S. Berman, C.~D.~A. Blair and R.~Otsuki, \emph{{Non-Riemannian geometry of
  M-theory}}, \href{https://doi.org/10.1007/JHEP07(2019)175}{\emph{JHEP}
  {\bfseries 07} (2019) 175}
  [\href{https://arxiv.org/abs/1902.01867}{{\ttfamily 1902.01867}}].

\bibitem{Otsuki:2019owg}
D.~S. Berman, E.~T. Musaev and R.~Otsuki, \emph{{Exotic Branes in M-Theory}},
  \href{https://doi.org/10.22323/1.347.0138}{\emph{PoS} {\bfseries CORFU2018}
  (2019) 138} [\href{https://arxiv.org/abs/1903.10247}{{\ttfamily
  1903.10247}}].

\bibitem{Berman:2019efr}
D.~S. Berman and R.~Otsuki, \emph{{Reductions of Exceptional Field Theories}},
  \href{https://doi.org/10.1007/JHEP03(2020)066}{\emph{JHEP} {\bfseries 03}
  (2020) 066} [\href{https://arxiv.org/abs/1911.06150}{{\ttfamily
  1911.06150}}].

\bibitem{Bakhmatov:2017les}
I.~Bakhmatov, D.~Berman, A.~Kleinschmidt, E.~Musaev and R.~Otsuki,
  \emph{{Exotic branes in Exceptional Field Theory: the SL(5) duality group}},
  \href{https://doi.org/10.1007/JHEP08(2018)021}{\emph{JHEP} {\bfseries 08}
  (2018) 021} [\href{https://arxiv.org/abs/1710.09740}{{\ttfamily
  1710.09740}}].

\bibitem{BUSCHER198759}
T.~Buscher, \emph{A symmetry of the string background field equations},
  \href{https://doi.org/https://doi.org/10.1016/0370-2693(87)90769-6}{\emph{Physics
  Letters B} {\bfseries 194} (1987) 59 }.

\bibitem{Alvarez:1994dn}
E.~Alvarez, L.~Alvarez-Gaume and Y.~Lozano, \emph{{An Introduction to T duality
  in string theory}},
  \href{https://doi.org/10.1016/0920-5632(95)00429-D}{\emph{Nucl. Phys. Proc.
  Suppl.} {\bfseries 41} (1995) 1}
  [\href{https://arxiv.org/abs/hep-th/9410237}{{\ttfamily hep-th/9410237}}].

\bibitem{Hull:2004in}
C.~M. Hull, \emph{{A Geometry for non-geometric string backgrounds}},
  \href{https://doi.org/10.1088/1126-6708/2005/10/065}{\emph{JHEP} {\bfseries
  10} (2005) 065} [\href{https://arxiv.org/abs/hep-th/0406102}{{\ttfamily
  hep-th/0406102}}].

\bibitem{Hull:2006va}
C.~M. Hull, \emph{{Doubled Geometry and T-Folds}},
  \href{https://doi.org/10.1088/1126-6708/2007/07/080}{\emph{JHEP} {\bfseries
  07} (2007) 080} [\href{https://arxiv.org/abs/hep-th/0605149}{{\ttfamily
  hep-th/0605149}}].

\bibitem{Siegel:1993th}
W.~Siegel, \emph{{Superspace duality in low-energy superstrings}},
  \href{https://doi.org/10.1103/PhysRevD.48.2826}{\emph{Phys. Rev.} {\bfseries
  D48} (1993) 2826} [\href{https://arxiv.org/abs/hep-th/9305073}{{\ttfamily
  hep-th/9305073}}].

\bibitem{Siegel:1993xq}
W.~Siegel, \emph{{Two vierbein formalism for string inspired axionic gravity}},
  \href{https://doi.org/10.1103/PhysRevD.47.5453}{\emph{Phys. Rev.} {\bfseries
  D47} (1993) 5453} [\href{https://arxiv.org/abs/hep-th/9302036}{{\ttfamily
  hep-th/9302036}}].

\bibitem{Hull:2009mi}
C.~Hull and B.~Zwiebach, \emph{{Double Field Theory}},
  \href{https://doi.org/10.1088/1126-6708/2009/09/099}{\emph{JHEP} {\bfseries
  09} (2009) 099} [\href{https://arxiv.org/abs/0904.4664}{{\ttfamily
  0904.4664}}].

\bibitem{Hohm:2010xe}
O.~Hohm and S.~K. Kwak, \emph{{Frame-like Geometry of Double Field Theory}},
  \href{https://doi.org/10.1088/1751-8113/44/8/085404}{\emph{J. Phys.}
  {\bfseries A44} (2011) 085404}
  [\href{https://arxiv.org/abs/1011.4101}{{\ttfamily 1011.4101}}].

\bibitem{Hohm:2010jy}
O.~Hohm, C.~Hull and B.~Zwiebach, \emph{{Background independent action for
  double field theory}},
  \href{https://doi.org/10.1007/JHEP07(2010)016}{\emph{JHEP} {\bfseries 07}
  (2010) 016} [\href{https://arxiv.org/abs/1003.5027}{{\ttfamily 1003.5027}}].

\bibitem{Hohm:2010pp}
O.~Hohm, C.~Hull and B.~Zwiebach, \emph{{Generalized metric formulation of
  double field theory}},
  \href{https://doi.org/10.1007/JHEP08(2010)008}{\emph{JHEP} {\bfseries 08}
  (2010) 008} [\href{https://arxiv.org/abs/1006.4823}{{\ttfamily 1006.4823}}].

\bibitem{Hohm:2011dv}
O.~Hohm, S.~K. Kwak and B.~Zwiebach, \emph{{Double Field Theory of Type II
  Strings}}, \href{https://doi.org/10.1007/JHEP09(2011)013}{\emph{JHEP}
  {\bfseries 09} (2011) 013} [\href{https://arxiv.org/abs/1107.0008}{{\ttfamily
  1107.0008}}].

\bibitem{Hohm:2017wtr}
O.~Hohm, E.~T. Musaev and H.~Samtleben, \emph{{O($d+1, d+1$) enhanced double
  field theory}}, \href{https://doi.org/10.1007/JHEP10(2017)086}{\emph{JHEP}
  {\bfseries 10} (2017) 086}
  [\href{https://arxiv.org/abs/1707.06693}{{\ttfamily 1707.06693}}].

\bibitem{Jeon:2011vx}
I.~Jeon, K.~Lee and J.-H. Park, \emph{{Incorporation of fermions into double
  field theory}}, \href{https://doi.org/10.1007/JHEP11(2011)025}{\emph{JHEP}
  {\bfseries 11} (2011) 025} [\href{https://arxiv.org/abs/1109.2035}{{\ttfamily
  1109.2035}}].

\bibitem{Jeon:2012kd}
I.~Jeon, K.~Lee and J.-H. Park, \emph{{Ramond-Ramond Cohomology and O(D,D)
  T-duality}}, \href{https://doi.org/10.1007/JHEP09(2012)079}{\emph{JHEP}
  {\bfseries 09} (2012) 079} [\href{https://arxiv.org/abs/1206.3478}{{\ttfamily
  1206.3478}}].

\bibitem{Lee:2013hma}
K.~Lee and J.-H. Park, \emph{{Covariant action for a string in "doubled yet
  gauged" spacetime}},
  \href{https://doi.org/10.1016/j.nuclphysb.2014.01.003}{\emph{Nucl. Phys.}
  {\bfseries B880} (2014) 134}
  [\href{https://arxiv.org/abs/1307.8377}{{\ttfamily 1307.8377}}].

\bibitem{Geissbuhler:2013uka}
D.~Geissbuhler, D.~Marques, C.~Nunez and V.~Penas, \emph{{Exploring Double
  Field Theory}}, \href{https://doi.org/10.1007/JHEP06(2013)101}{\emph{JHEP}
  {\bfseries 06} (2013) 101} [\href{https://arxiv.org/abs/1304.1472}{{\ttfamily
  1304.1472}}].

\bibitem{Berman:2013eva}
D.~S. Berman and D.~C. Thompson, \emph{{Duality Symmetric String and
  M-Theory}}, \href{https://doi.org/10.1016/j.physrep.2014.11.007}{\emph{Phys.
  Rept.} {\bfseries 566} (2014) 1}
  [\href{https://arxiv.org/abs/1306.2643}{{\ttfamily 1306.2643}}].

\bibitem{Aldazabal:2013sca}
G.~Aldazabal, D.~Marques and C.~Nunez, \emph{{Double Field Theory: A
  Pedagogical Review}},
  \href{https://doi.org/10.1088/0264-9381/30/16/163001}{\emph{Class. Quant.
  Grav.} {\bfseries 30} (2013) 163001}
  [\href{https://arxiv.org/abs/1305.1907}{{\ttfamily 1305.1907}}].

\bibitem{Hohm:2011cp}
O.~Hohm and S.~K. Kwak, \emph{{Massive Type II in Double Field Theory}},
  \href{https://doi.org/10.1007/JHEP11(2011)086}{\emph{JHEP} {\bfseries 11}
  (2011) 086} [\href{https://arxiv.org/abs/1108.4937}{{\ttfamily 1108.4937}}].

\bibitem{Hohm:2011zr}
O.~Hohm, S.~K. Kwak and B.~Zwiebach, \emph{{Unification of Type II Strings and
  T-duality}},
  \href{https://doi.org/10.1103/PhysRevLett.107.171603}{\emph{Phys. Rev. Lett.}
  {\bfseries 107} (2011) 171603}
  [\href{https://arxiv.org/abs/1106.5452}{{\ttfamily 1106.5452}}].

\bibitem{Hohm:2011ex}
O.~Hohm and S.~K. Kwak, \emph{{Double Field Theory Formulation of Heterotic
  Strings}}, \href{https://doi.org/10.1007/JHEP06(2011)096}{\emph{JHEP}
  {\bfseries 06} (2011) 096} [\href{https://arxiv.org/abs/1103.2136}{{\ttfamily
  1103.2136}}].

\bibitem{Malek:2016vsh}
E.~Malek, \emph{{From Exceptional Field Theory to Heterotic Double Field Theory
  via K3}}, \href{https://doi.org/10.1007/JHEP03(2017)057}{\emph{JHEP}
  {\bfseries 03} (2017) 057}
  [\href{https://arxiv.org/abs/1612.01990}{{\ttfamily 1612.01990}}].

\bibitem{Hohm:2013nja}
O.~Hohm and H.~Samtleben, \emph{{Gauge theory of Kaluza-Klein and winding
  modes}}, \href{https://doi.org/10.1103/PhysRevD.88.085005}{\emph{Phys. Rev.}
  {\bfseries D88} (2013) 085005}
  [\href{https://arxiv.org/abs/1307.0039}{{\ttfamily 1307.0039}}].

\bibitem{Coimbra:2011nw}
A.~Coimbra, C.~Strickland-Constable and D.~Waldram, \emph{{Supergravity as
  Generalised Geometry I: Type II Theories}},
  \href{https://doi.org/10.1007/JHEP11(2011)091}{\emph{JHEP} {\bfseries 11}
  (2011) 091} [\href{https://arxiv.org/abs/1107.1733}{{\ttfamily 1107.1733}}].

\bibitem{Jeon:2011sq}
I.~Jeon, K.~Lee and J.-H. Park, \emph{{Supersymmetric Double Field Theory:
  Stringy Reformulation of Supergravity}},
  \href{https://doi.org/10.1103/PhysRevD.86.089903, 10.1103/PhysRevD.85.081501,
  10.1103/PhysRevD.85.089908}{\emph{Phys. Rev.} {\bfseries D85} (2012) 081501}
  [\href{https://arxiv.org/abs/1112.0069}{{\ttfamily 1112.0069}}].

\bibitem{Jeon:2012hp}
I.~Jeon, K.~Lee, J.-H. Park and Y.~Suh, \emph{{Stringy Unification of Type IIA
  and IIB Supergravities under N=2 D=10 Supersymmetric Double Field Theory}},
  \href{https://doi.org/10.1016/j.physletb.2013.05.016}{\emph{Phys. Lett.}
  {\bfseries B723} (2013) 245}
  [\href{https://arxiv.org/abs/1210.5078}{{\ttfamily 1210.5078}}].

\bibitem{Godazgar:2013bja}
H.~Godazgar and M.~Godazgar, \emph{{Duality completion of higher derivative
  corrections}}, \href{https://doi.org/10.1007/JHEP09(2013)140}{\emph{JHEP}
  {\bfseries 09} (2013) 140} [\href{https://arxiv.org/abs/1306.4918}{{\ttfamily
  1306.4918}}].

\bibitem{Hohm:2013jaa}
O.~Hohm, W.~Siegel and B.~Zwiebach, \emph{{Doubled $\alpha^\prime$-geometry}},
  \href{https://doi.org/10.1007/JHEP02(2014)065}{\emph{JHEP} {\bfseries 02}
  (2014) 065} [\href{https://arxiv.org/abs/1306.2970}{{\ttfamily 1306.2970}}].

\bibitem{Hohm:2014xsa}
O.~Hohm and B.~Zwiebach, \emph{{Double field theory at order $\alpha^\prime$}},
  \href{https://doi.org/10.1007/JHEP11(2014)075}{\emph{JHEP} {\bfseries 11}
  (2014) 075} [\href{https://arxiv.org/abs/1407.3803}{{\ttfamily 1407.3803}}].

\bibitem{Hohm:2015mka}
O.~Hohm and B.~Zwiebach, \emph{{Double metric, generalized metric, and
  $\alpha^\prime$-deformed double field theory}},
  \href{https://doi.org/10.1103/PhysRevD.93.064035}{\emph{Phys. Rev.}
  {\bfseries D93} (2016) 064035}
  [\href{https://arxiv.org/abs/1509.02930}{{\ttfamily 1509.02930}}].

\bibitem{Bedoya:2014pma}
O.~A. Bedoya, D.~Marques and C.~Nunez, \emph{{Heterotic
  $\alpha^\prime$-corrections in Double Field Theory}},
  \href{https://doi.org/10.1007/JHEP12(2014)074}{\emph{JHEP} {\bfseries 12}
  (2014) 074} [\href{https://arxiv.org/abs/1407.0365}{{\ttfamily 1407.0365}}].

\bibitem{Coimbra:2014qaa}
A.~Coimbra, R.~Minasian, H.~Triendl and D.~Waldram, \emph{{Generalised geometry
  for string corrections}},
  \href{https://doi.org/10.1007/JHEP11(2014)160}{\emph{JHEP} {\bfseries 11}
  (2014) 160} [\href{https://arxiv.org/abs/1407.7542}{{\ttfamily 1407.7542}}].

\bibitem{Marques:2015vua}
D.~Marques and C.~A. Nunez, \emph{{T-duality and $\alpha^\prime$-corrections}},
  \href{https://doi.org/10.1007/JHEP10(2015)084}{\emph{JHEP} {\bfseries 10}
  (2015) 084} [\href{https://arxiv.org/abs/1507.00652}{{\ttfamily
  1507.00652}}].

\bibitem{Lee:2015kba}
K.~Lee, \emph{{Quadratic $\alpha^\prime$-corrections to heterotic double field
  theory}}, \href{https://doi.org/10.1016/j.nuclphysb.2015.08.013}{\emph{Nucl.
  Phys.} {\bfseries B899} (2015) 594}
  [\href{https://arxiv.org/abs/1504.00149}{{\ttfamily 1504.00149}}].

\bibitem{Baron:2017dvb}
W.~H. Baron, J.~J. Fernandez-Melgarejo, D.~Marques and C.~Nunez, \emph{{The Odd
  story of $\alpha^\prime$-corrections}},
  \href{https://doi.org/10.1007/JHEP04(2017)078}{\emph{JHEP} {\bfseries 04}
  (2017) 078} [\href{https://arxiv.org/abs/1702.05489}{{\ttfamily
  1702.05489}}].

\bibitem{Duff:1996aw}
M.~J. Duff, \emph{{M theory (The Theory formerly known as strings)}},
  \href{https://doi.org/10.1142/S0217751X96002583}{\emph{Int. J. Mod. Phys.}
  {\bfseries A11} (1996) 5623}
  [\href{https://arxiv.org/abs/hep-th/9608117}{{\ttfamily hep-th/9608117}}].

\bibitem{Witten:1995zh}
E.~Witten, \emph{{Some comments on string dynamics}},  in \emph{{Future
  perspectives in string theory. Proceedings, Conference, Strings'95, Los
  Angeles, USA, March 13-18, 1995}}, pp.~501--523, 1995,
  \href{https://arxiv.org/abs/hep-th/9507121}{{\ttfamily hep-th/9507121}}.

\bibitem{Townsend:1995kk}
P.~K. Townsend, \emph{{The eleven-dimensional supermembrane revisited}},
  \href{https://doi.org/10.1016/0370-2693(95)00397-4}{\emph{Phys. Lett.}
  {\bfseries B350} (1995) 184}
  [\href{https://arxiv.org/abs/hep-th/9501068}{{\ttfamily hep-th/9501068}}].

\bibitem{Obers:1998fb}
N.~Obers and B.~Pioline, \emph{{U duality and M theory}},
  \href{https://doi.org/10.1016/S0370-1573(99)00004-6}{\emph{Phys.Rept.}
  {\bfseries 318} (1999) 113}
  [\href{https://arxiv.org/abs/hep-th/9809039}{{\ttfamily hep-th/9809039}}].

\bibitem{Berman:2020tqn}
D.~S. Berman and C.~D. Blair, \emph{{The Geometry, Branes and Applications of
  Exceptional Field Theory }},
  \href{https://arxiv.org/abs/2006.09777}{{\ttfamily 2006.09777}}.

\bibitem{Hohm:2014fxa}
O.~Hohm and H.~Samtleben, \emph{{Exceptional field theory. III. E$_{8(8)}$}},
  \href{https://doi.org/10.1103/PhysRevD.90.066002}{\emph{Phys. Rev.}
  {\bfseries D90} (2014) 066002}
  [\href{https://arxiv.org/abs/1406.3348}{{\ttfamily 1406.3348}}].

\bibitem{Hohm:2013uia}
O.~Hohm and H.~Samtleben, \emph{{Exceptional field theory. II. E$_{7(7)}$}},
  \href{https://doi.org/10.1103/PhysRevD.89.066017}{\emph{Phys. Rev.}
  {\bfseries D89} (2014) 066017}
  [\href{https://arxiv.org/abs/1312.4542}{{\ttfamily 1312.4542}}].

\bibitem{Hohm:2013vpa}
O.~Hohm and H.~Samtleben, \emph{{Exceptional Field Theory I: $E_{6(6)}$
  covariant Form of M-Theory and Type IIB}},
  \href{https://doi.org/10.1103/PhysRevD.89.066016}{\emph{Phys. Rev.}
  {\bfseries D89} (2014) 066016}
  [\href{https://arxiv.org/abs/1312.0614}{{\ttfamily 1312.0614}}].

\bibitem{Abzalov:2015ega}
A.~Abzalov, I.~Bakhmatov and E.~T. Musaev, \emph{{Exceptional field theory:
  $SO(5,5)$}}, \href{https://doi.org/10.1007/JHEP06(2015)088}{\emph{JHEP}
  {\bfseries 06} (2015) 088}
  [\href{https://arxiv.org/abs/1504.01523}{{\ttfamily 1504.01523}}].

\bibitem{Musaev:2015ces}
E.~T. Musaev, \emph{{Exceptional field theory: $SL(5)$}},
  \href{https://doi.org/10.1007/JHEP02(2016)012}{\emph{JHEP} {\bfseries 02}
  (2016) 012} [\href{https://arxiv.org/abs/1512.02163}{{\ttfamily
  1512.02163}}].

\bibitem{Hohm:2015xna}
O.~Hohm and Y.-N. Wang, \emph{{Tensor hierarchy and generalized Cartan calculus
  in $\operatorname{SL}(3) \times \operatorname{SL}(2)$ exceptional field
  theory}}, \href{https://doi.org/10.1007/JHEP04(2015)050}{\emph{JHEP}
  {\bfseries 04} (2015) 050}
  [\href{https://arxiv.org/abs/1501.01600}{{\ttfamily 1501.01600}}].

\bibitem{Berman:2015rcc}
D.~S. Berman, C.~D.~A. Blair, E.~Malek and F.~J. Rudolph, \emph{{An action for
  F-theory: $\mathrm{SL}(2){{\mathbb{R}}}^{+}$ exceptional field theory}},
  \href{https://doi.org/10.1088/0264-9381/33/19/195009}{\emph{Class. Quant.
  Grav.} {\bfseries 33} (2016) 195009}
  [\href{https://arxiv.org/abs/1512.06115}{{\ttfamily 1512.06115}}].

\bibitem{Berman:2010is}
D.~S. Berman and M.~J. Perry, \emph{{Generalized Geometry and M theory}},
  \href{https://doi.org/10.1007/JHEP06(2011)074}{\emph{JHEP} {\bfseries 06}
  (2011) 074} [\href{https://arxiv.org/abs/1008.1763}{{\ttfamily 1008.1763}}].

\bibitem{Berman:2011jh}
D.~S. Berman, H.~Godazgar, M.~J. Perry and P.~West, \emph{{Duality Invariant
  Actions and Generalised Geometry}},
  \href{https://doi.org/10.1007/JHEP02(2012)108}{\emph{JHEP} {\bfseries 02}
  (2012) 108} [\href{https://arxiv.org/abs/1111.0459}{{\ttfamily 1111.0459}}].

\bibitem{Berman:2011cg}
D.~S. Berman, H.~Godazgar, M.~Godazgar and M.~J. Perry, \emph{{The Local
  symmetries of M-theory and their formulation in generalised geometry}},
  \href{https://doi.org/10.1007/JHEP01(2012)012}{\emph{JHEP} {\bfseries 01}
  (2012) 012} [\href{https://arxiv.org/abs/1110.3930}{{\ttfamily 1110.3930}}].

\bibitem{Malek:2012pw}
E.~Malek, \emph{{U-duality in three and four dimensions}},
  \href{https://doi.org/10.1142/S0217751X1750169X}{\emph{Int. J. Mod. Phys.}
  {\bfseries A32} (2017) 1750169}
  [\href{https://arxiv.org/abs/1205.6403}{{\ttfamily 1205.6403}}].

\bibitem{Berman:2012vc}
D.~S. Berman, M.~Cederwall, A.~Kleinschmidt and D.~C. Thompson, \emph{{The
  gauge structure of generalised diffeomorphisms}},
  \href{https://doi.org/10.1007/JHEP01(2013)064}{\emph{JHEP} {\bfseries 01}
  (2013) 064} [\href{https://arxiv.org/abs/1208.5884}{{\ttfamily 1208.5884}}].

\bibitem{Hillmann:2009ci}
C.~Hillmann, \emph{{Generalized E(7(7)) coset dynamics and D=11 supergravity}},
  \href{https://doi.org/10.1088/1126-6708/2009/03/135}{\emph{JHEP} {\bfseries
  03} (2009) 135} [\href{https://arxiv.org/abs/0901.1581}{{\ttfamily
  0901.1581}}].

\bibitem{Hohm:2013jma}
O.~Hohm and H.~Samtleben, \emph{{U-duality covariant gravity}},
  \href{https://doi.org/10.1007/JHEP09(2013)080}{\emph{JHEP} {\bfseries 09}
  (2013) 080} [\href{https://arxiv.org/abs/1307.0509}{{\ttfamily 1307.0509}}].

\bibitem{Bossard:2017aae}
G.~Bossard, M.~Cederwall, A.~Kleinschmidt, J.~Palmkvist and H.~Samtleben,
  \emph{{Generalized diffeomorphisms for $E_9$}},
  \href{https://doi.org/10.1103/PhysRevD.96.106022}{\emph{Phys. Rev.}
  {\bfseries D96} (2017) 106022}
  [\href{https://arxiv.org/abs/1708.08936}{{\ttfamily 1708.08936}}].

\bibitem{Bossard:2018utw}
G.~Bossard, F.~Ciceri, G.~Inverso, A.~Kleinschmidt and H.~Samtleben,
  \emph{{E$_{9}$ exceptional field theory. Part I. The potential}},
  \href{https://doi.org/10.1007/JHEP03(2019)089}{\emph{JHEP} {\bfseries 03}
  (2019) 089} [\href{https://arxiv.org/abs/1811.04088}{{\ttfamily
  1811.04088}}].

\bibitem{Baguet:2016jph}
A.~Baguet and H.~Samtleben, \emph{{E$_{8(8)}$ Exceptional Field Theory:
  Geometry, Fermions and Supersymmetry}},
  \href{https://doi.org/10.1007/JHEP09(2016)168}{\emph{JHEP} {\bfseries 09}
  (2016) 168} [\href{https://arxiv.org/abs/1607.03119}{{\ttfamily
  1607.03119}}].

\bibitem{Godazgar:2014nqa}
H.~Godazgar, M.~Godazgar, O.~Hohm, H.~Nicolai and H.~Samtleben,
  \emph{{Supersymmetric E$_{7(7)}$ Exceptional Field Theory}},
  \href{https://doi.org/10.1007/JHEP09(2014)044}{\emph{JHEP} {\bfseries 09}
  (2014) 044} [\href{https://arxiv.org/abs/1406.3235}{{\ttfamily 1406.3235}}].

\bibitem{Musaev:2015pla}
E.~T. Musaev, \emph{{Exceptional Field Theory for $E_{6(6)}$ supergravity}},
  {\emph{TSPU Bulletin} {\bfseries 12} (2014) 198}
  [\href{https://arxiv.org/abs/1503.08397}{{\ttfamily 1503.08397}}].

\bibitem{Aldazabal:2011nj}
G.~Aldazabal, W.~Baron, D.~Marques and C.~Nunez, \emph{{The effective action of
  Double Field Theory}}, \href{https://doi.org/10.1007/JHEP11(2011)052,
  10.1007/JHEP11(2011)109}{\emph{JHEP} {\bfseries 11} (2011) 052}
  [\href{https://arxiv.org/abs/1109.0290}{{\ttfamily 1109.0290}}].

\bibitem{Dibitetto:2012rk}
G.~Dibitetto, J.~J. Fernandez-Melgarejo, D.~Marques and D.~Roest,
  \emph{{Duality orbits of non-geometric fluxes}},
  \href{https://doi.org/10.1002/prop.201200078}{\emph{Fortsch. Phys.}
  {\bfseries 60} (2012) 1123}
  [\href{https://arxiv.org/abs/1203.6562}{{\ttfamily 1203.6562}}].

\bibitem{Grana:2012rr}
M.~Grana and D.~Marques, \emph{{Gauged Double Field Theory}},
  \href{https://doi.org/10.1007/JHEP04(2012)020}{\emph{JHEP} {\bfseries 04}
  (2012) 020} [\href{https://arxiv.org/abs/1201.2924}{{\ttfamily 1201.2924}}].

\bibitem{Berman:2012uy}
D.~S. Berman, E.~T. Musaev, D.~C. Thompson and D.~C. Thompson, \emph{{Duality
  Invariant M-theory: Gauged supergravities and Scherk-Schwarz reductions}},
  \href{https://doi.org/10.1007/JHEP10(2012)174}{\emph{JHEP} {\bfseries 10}
  (2012) 174} [\href{https://arxiv.org/abs/1208.0020}{{\ttfamily 1208.0020}}].

\bibitem{Aldazabal:2013mya}
G.~Aldazabal, M.~Graña, D.~Marqués and J.~A. Rosabal, \emph{{Extended
  geometry and gauged maximal supergravity}},
  \href{https://doi.org/10.1007/JHEP06(2013)046}{\emph{JHEP} {\bfseries 06}
  (2013) 046} [\href{https://arxiv.org/abs/1302.5419}{{\ttfamily 1302.5419}}].

\bibitem{Aldazabal:2013via}
G.~Aldazabal, M.~Graña, D.~Marqués and J.~A. Rosabal, \emph{{The gauge
  structure of Exceptional Field Theories and the tensor hierarchy}},
  \href{https://doi.org/10.1007/JHEP04(2014)049}{\emph{JHEP} {\bfseries 04}
  (2014) 049} [\href{https://arxiv.org/abs/1312.4549}{{\ttfamily 1312.4549}}].

\bibitem{1986PhLB..169..374R}
L.~J. {Romans}, \emph{{Massive N = 2a supergravity in ten dimensions}},
  \href{https://doi.org/10.1016/0370-2693(86)90375-8}{\emph{Physics Letters B}
  {\bfseries 169} (1986) 374}.

\bibitem{Ciceri:2016dmd}
F.~Ciceri, A.~Guarino and G.~Inverso, \emph{{The exceptional story of massive
  IIA supergravity}},
  \href{https://doi.org/10.1007/JHEP08(2016)154}{\emph{JHEP} {\bfseries 08}
  (2016) 154} [\href{https://arxiv.org/abs/1604.08602}{{\ttfamily
  1604.08602}}].

\bibitem{deBoer:2010ud}
J.~de~Boer and M.~Shigemori, \emph{{Exotic branes and non-geometric
  backgrounds}},
  \href{https://doi.org/10.1103/PhysRevLett.104.251603}{\emph{Phys. Rev. Lett.}
  {\bfseries 104} (2010) 251603}
  [\href{https://arxiv.org/abs/1004.2521}{{\ttfamily 1004.2521}}].

\bibitem{deBoer:2012ma}
J.~de~Boer and M.~Shigemori, \emph{{Exotic Branes in String Theory}},
  \href{https://doi.org/10.1016/j.physrep.2013.07.003}{\emph{Phys. Rept.}
  {\bfseries 532} (2013) 65} [\href{https://arxiv.org/abs/1209.6056}{{\ttfamily
  1209.6056}}].

\bibitem{Fernandez-Melgarejo:2018yxq}
J.~J. Fernández-Melgarejo, T.~Kimura and Y.~Sakatani, \emph{{Weaving the
  Exotic Web}}, \href{https://doi.org/10.1007/JHEP09(2018)072}{\emph{JHEP}
  {\bfseries 09} (2018) 072}
  [\href{https://arxiv.org/abs/1805.12117}{{\ttfamily 1805.12117}}].

\bibitem{Berkeley:2014nza}
J.~Berkeley, D.~S. Berman and F.~J. Rudolph, \emph{{Strings and Branes are
  Waves}}, \href{https://doi.org/10.1007/JHEP06(2014)006}{\emph{JHEP}
  {\bfseries 06} (2014) 006} [\href{https://arxiv.org/abs/1403.7198}{{\ttfamily
  1403.7198}}].

\bibitem{Berman:2014jsa}
D.~S. Berman and F.~J. Rudolph, \emph{{Branes are Waves and Monopoles}},
  \href{https://doi.org/10.1007/JHEP05(2015)015}{\emph{JHEP} {\bfseries 05}
  (2015) 015} [\href{https://arxiv.org/abs/1409.6314}{{\ttfamily 1409.6314}}].

\bibitem{Bakhmatov:2016kfn}
I.~Bakhmatov, A.~Kleinschmidt and E.~T. Musaev, \emph{{Non-geometric branes are
  DFT monopoles}}, \href{https://doi.org/10.1007/JHEP10(2016)076}{\emph{JHEP}
  {\bfseries 10} (2016) 076}
  [\href{https://arxiv.org/abs/1607.05450}{{\ttfamily 1607.05450}}].

\bibitem{Kimura:2018hph}
T.~Kimura, S.~Sasaki and K.~Shiozawa, \emph{{Worldsheet Instanton Corrections
  to Five-branes and Waves in Double Field Theory}},
  \href{https://doi.org/10.1007/JHEP07(2018)001}{\emph{JHEP} {\bfseries 07}
  (2018) 001} [\href{https://arxiv.org/abs/1803.11087}{{\ttfamily
  1803.11087}}].

\bibitem{Lee:2018gxc}
K.~Lee, \emph{{Kerr-Schild Double Field Theory and Classical Double Copy}},
  \href{https://doi.org/10.1007/JHEP10(2018)027}{\emph{JHEP} {\bfseries 10}
  (2018) 027} [\href{https://arxiv.org/abs/1807.08443}{{\ttfamily
  1807.08443}}].

\bibitem{Rudolph:2016sxe}
F.~J. Rudolph, \emph{{Duality Covariant Solutions in Extended Field Theories}},
  Ph.D. thesis, Queen Mary, U. of London, 2016.
\newblock \href{https://arxiv.org/abs/1610.03440}{{\ttfamily 1610.03440}}.

\bibitem{Blair:2013noa}
C.~D.~A. Blair, E.~Malek and A.~J. Routh, \emph{{An $O(D, D)$ invariant
  Hamiltonian action for the superstring}},
  \href{https://doi.org/10.1088/0264-9381/31/20/205011}{\emph{Class. Quant.
  Grav.} {\bfseries 31} (2014) 205011}
  [\href{https://arxiv.org/abs/1308.4829}{{\ttfamily 1308.4829}}].

\bibitem{Blair:2016xnn}
C.~D.~A. Blair, \emph{{Doubled strings, negative strings and null waves}},
  \href{https://doi.org/10.1007/JHEP11(2016)042}{\emph{JHEP} {\bfseries 11}
  (2016) 042} [\href{https://arxiv.org/abs/1608.06818}{{\ttfamily
  1608.06818}}].

\bibitem{Blair:2017hhy}
C.~D.~A. Blair and E.~T. Musaev, \emph{{Five-brane actions in double field
  theory}}, \href{https://doi.org/10.1007/JHEP03(2018)111}{\emph{JHEP}
  {\bfseries 03} (2018) 111}
  [\href{https://arxiv.org/abs/1712.01739}{{\ttfamily 1712.01739}}].

\bibitem{Ko:2015rha}
S.~M. Ko, C.~Melby-Thompson, R.~Meyer and J.-H. Park, \emph{{Dynamics of
  Perturbations in Double Field Theory \& Non-Relativistic String Theory}},
  \href{https://doi.org/10.1007/JHEP12(2015)144}{\emph{JHEP} {\bfseries 12}
  (2015) 144} [\href{https://arxiv.org/abs/1508.01121}{{\ttfamily
  1508.01121}}].

\bibitem{Gomis:2000bd}
J.~Gomis and H.~Ooguri, \emph{{Nonrelativistic closed string theory}},
  \href{https://doi.org/10.1063/1.1372697}{\emph{J. Math. Phys.} {\bfseries 42}
  (2001) 3127} [\href{https://arxiv.org/abs/hep-th/0009181}{{\ttfamily
  hep-th/0009181}}].

\bibitem{Morand:2017fnv}
K.~Morand and J.-H. Park, \emph{{Classification of non-Riemannian
  doubled-yet-gauged spacetime}},
  \href{https://doi.org/10.1140/epjc/s10052-017-5257-z,
  10.1140/epjc/s10052-018-6394-8}{\emph{Eur. Phys. J.} {\bfseries C77} (2017)
  685} [\href{https://arxiv.org/abs/1707.03713}{{\ttfamily 1707.03713}}].

\bibitem{Cho:2018alk}
K.~Cho, K.~Morand and J.-H. Park, \emph{{Kaluza–Klein reduction on a
  maximally non-Riemannian space is moduli-free}},
  \href{https://doi.org/10.1016/j.physletb.2019.04.042}{\emph{Phys. Lett.}
  {\bfseries B793} (2019) 65}
  [\href{https://arxiv.org/abs/1808.10605}{{\ttfamily 1808.10605}}].

\bibitem{Cho:2019ofr}
K.~Cho and J.-H. Park, \emph{{Remarks on the non-Riemannian sector in Double
  Field Theory}},
  \href{https://doi.org/10.1140/epjc/s10052-020-7648-9}{\emph{Eur. Phys. J.}
  {\bfseries C80} (2020) 101}
  [\href{https://arxiv.org/abs/1909.10711}{{\ttfamily 1909.10711}}].

\bibitem{Hull:2007zu}
C.~M. Hull, \emph{{Generalised Geometry for M-Theory}},
  \href{https://doi.org/10.1088/1126-6708/2007/07/079}{\emph{JHEP} {\bfseries
  07} (2007) 079} [\href{https://arxiv.org/abs/hep-th/0701203}{{\ttfamily
  hep-th/0701203}}].

\bibitem{Pacheco:2008ps}
P.~Pires~Pacheco and D.~Waldram, \emph{{M-theory, exceptional generalised
  geometry and superpotentials}},
  \href{https://doi.org/10.1088/1126-6708/2008/09/123}{\emph{JHEP} {\bfseries
  09} (2008) 123} [\href{https://arxiv.org/abs/0804.1362}{{\ttfamily
  0804.1362}}].

\bibitem{Coimbra:2011ky}
A.~Coimbra, C.~Strickland-Constable and D.~Waldram, \emph{{$E_{d(d)} \times
  \mathbb{R}^+$ generalised geometry, connections and M theory}},
  \href{https://doi.org/10.1007/JHEP02(2014)054}{\emph{JHEP} {\bfseries 02}
  (2014) 054} [\href{https://arxiv.org/abs/1112.3989}{{\ttfamily 1112.3989}}].

\bibitem{Coimbra:2012af}
A.~Coimbra, C.~Strickland-Constable and D.~Waldram, \emph{{Supergravity as
  Generalised Geometry II: $E_{d(d)} \times \mathbb{R}^+$ and M theory}},
  \href{https://doi.org/10.1007/JHEP03(2014)019}{\emph{JHEP} {\bfseries 03}
  (2014) 019} [\href{https://arxiv.org/abs/1212.1586}{{\ttfamily 1212.1586}}].

\bibitem{Coimbra:2012yy}
A.~Coimbra, C.~Strickland-Constable and D.~Waldram, \emph{{Generalised Geometry
  and type II Supergravity}},
  \href{https://doi.org/10.1002/prop.201100096}{\emph{Fortsch. Phys.}
  {\bfseries 60} (2012) 982} [\href{https://arxiv.org/abs/1202.3170}{{\ttfamily
  1202.3170}}].

\bibitem{Hitchin:2004ut}
N.~Hitchin, \emph{{Generalized Calabi-Yau manifolds}},
  \href{https://doi.org/10.1093/qjmath/54.3.281}{\emph{Quart. J. Math.}
  {\bfseries 54} (2003) 281}
  [\href{https://arxiv.org/abs/math/0209099}{{\ttfamily math/0209099}}].

\bibitem{Gualtieri:2003dx}
M.~Gualtieri, \emph{{Generalized complex geometry}}, Ph.D. thesis, Oxford U.,
  2003.
\newblock \href{https://arxiv.org/abs/math/0401221}{{\ttfamily math/0401221}}.

\bibitem{Tumanov:2016abm}
A.~G. Tumanov and P.~West, \emph{{E11 in 11D}},
  \href{https://doi.org/10.1016/j.physletb.2016.04.058}{\emph{Phys. Lett.}
  {\bfseries B758} (2016) 278}
  [\href{https://arxiv.org/abs/1601.03974}{{\ttfamily 1601.03974}}].

\bibitem{Kleinschmidt:2003jf}
A.~Kleinschmidt and P.~C. West, \emph{{Representations of G+++ and the role of
  space-time}},
  \href{https://doi.org/10.1088/1126-6708/2004/02/033}{\emph{JHEP} {\bfseries
  02} (2004) 033} [\href{https://arxiv.org/abs/hep-th/0312247}{{\ttfamily
  hep-th/0312247}}].

\bibitem{West:2004kb}
P.~C. West, \emph{{E(11) origin of brane charges and U-duality multiplets}},
  \href{https://doi.org/10.1088/1126-6708/2004/08/052}{\emph{JHEP} {\bfseries
  08} (2004) 052} [\href{https://arxiv.org/abs/hep-th/0406150}{{\ttfamily
  hep-th/0406150}}].

\bibitem{West:2011mm}
P.~West, \emph{{Generalised geometry, eleven dimensions and E11}},
  \href{https://doi.org/10.1007/JHEP02(2012)018}{\emph{JHEP} {\bfseries 02}
  (2012) 018} [\href{https://arxiv.org/abs/1111.1642}{{\ttfamily 1111.1642}}].

\bibitem{Riccioni:2009xr}
F.~Riccioni, D.~Steele and P.~West, \emph{{The E(11) origin of all maximal
  supergravities: The Hierarchy of field-strengths}},
  \href{https://doi.org/10.1088/1126-6708/2009/09/095}{\emph{JHEP} {\bfseries
  09} (2009) 095} [\href{https://arxiv.org/abs/0906.1177}{{\ttfamily
  0906.1177}}].

\bibitem{West:2003fc}
P.~C. West, \emph{{E(11), SL(32) and central charges}},
  \href{https://doi.org/10.1016/j.physletb.2003.09.059}{\emph{Phys. Lett.}
  {\bfseries B575} (2003) 333}
  [\href{https://arxiv.org/abs/hep-th/0307098}{{\ttfamily hep-th/0307098}}].

\bibitem{West:2004iz}
P.~C. West, \emph{{Brane dynamics, central charges and E(11)}},
  \href{https://doi.org/10.1088/1126-6708/2005/03/077}{\emph{JHEP} {\bfseries
  03} (2005) 077} [\href{https://arxiv.org/abs/hep-th/0412336}{{\ttfamily
  hep-th/0412336}}].

\bibitem{Cook:2008bi}
P.~P. Cook and P.~C. West, \emph{{Charge multiplets and masses for E(11)}},
  \href{https://doi.org/10.1088/1126-6708/2008/11/091}{\emph{JHEP} {\bfseries
  11} (2008) 091} [\href{https://arxiv.org/abs/0805.4451}{{\ttfamily
  0805.4451}}].

\bibitem{Tumanov:2015yjd}
A.~G. Tumanov and P.~West, \emph{{E$_{11}$ must be a symmetry of strings and
  branes}}, \href{https://doi.org/10.1016/j.physletb.2016.06.011}{\emph{Phys.
  Lett.} {\bfseries B759} (2016) 663}
  [\href{https://arxiv.org/abs/1512.01644}{{\ttfamily 1512.01644}}].

\bibitem{Lee:2016qwn}
K.~Lee, S.-J. Rey and Y.~Sakatani, \emph{{Effective action for non-geometric
  fluxes from duality covariant actions}},
  \href{https://doi.org/10.1007/JHEP07(2017)075}{\emph{JHEP} {\bfseries 07}
  (2017) 075} [\href{https://arxiv.org/abs/1612.08738}{{\ttfamily
  1612.08738}}].

\bibitem{Godazgar:2013rja}
H.~Godazgar, M.~fGodazgar and M.~J. Perry, \emph{{E8 duality and dual
  gravity}}, \href{https://doi.org/10.1007/JHEP06(2013)044}{\emph{JHEP}
  {\bfseries 06} (2013) 044} [\href{https://arxiv.org/abs/1303.2035}{{\ttfamily
  1303.2035}}].

\bibitem{Mori:2019slw}
H.~Mori, S.~Sasaki and K.~Shiozawa, \emph{{Doubled Aspects of Vaisman Algebroid
  and Gauge Symmetry in Double Field Theory}},
  \href{https://doi.org/10.1063/1.5108783}{\emph{J. Math. Phys.} {\bfseries 61}
  (2020) 013505} [\href{https://arxiv.org/abs/1901.04777}{{\ttfamily
  1901.04777}}].

\bibitem{Chatzistavrakidis:2019huz}
A.~Chatzistavrakidis, L.~Jonke, F.~S. Khoo and R.~J. Szabo, \emph{{The
  Algebroid Structure of Double Field Theory}},
  \href{https://doi.org/10.22323/1.347.0132}{\emph{PoS} {\bfseries CORFU2018}
  (2019) 132} [\href{https://arxiv.org/abs/1903.01765}{{\ttfamily
  1903.01765}}].

\bibitem{Vaisman:2012ke}
I.~Vaisman, \emph{{On the geometry of double field theory}},
  \href{https://doi.org/10.1063/1.3694739}{\emph{J. Math. Phys.} {\bfseries 53}
  (2012) 033509} [\href{https://arxiv.org/abs/1203.0836}{{\ttfamily
  1203.0836}}].

\bibitem{Cederwall:2015ica}
M.~Cederwall and J.~A. Rosabal, \emph{{E$_{8}$ geometry}},
  \href{https://doi.org/10.1007/JHEP07(2015)007}{\emph{JHEP} {\bfseries 07}
  (2015) 007} [\href{https://arxiv.org/abs/1504.04843}{{\ttfamily
  1504.04843}}].

\bibitem{Fernandez-Melgarejo:2019pvx}
J.~J. Fernández-Melgarejo, Y.~Sakatani and S.~Uehara, \emph{{Exotic branes and
  mixed-symmetry potentials II: duality rules and exceptional $p$-form gauge
  fields}},  \href{https://arxiv.org/abs/1909.01335}{{\ttfamily 1909.01335}}.

\bibitem{Blair:2014zba}
C.~D.~A. Blair and E.~Malek, \emph{{Geometry and fluxes of SL(5) exceptional
  field theory}}, \href{https://doi.org/10.1007/JHEP03(2015)144}{\emph{JHEP}
  {\bfseries 03} (2015) 144} [\href{https://arxiv.org/abs/1412.0635}{{\ttfamily
  1412.0635}}].

\bibitem{Berman:2014hna}
D.~S. Berman and F.~J. Rudolph, \emph{{Strings, Branes and the Self-dual
  Solutions of Exceptional Field Theory}},
  \href{https://doi.org/10.1007/JHEP05(2015)130}{\emph{JHEP} {\bfseries 05}
  (2015) 130} [\href{https://arxiv.org/abs/1412.2768}{{\ttfamily 1412.2768}}].

\bibitem{Berman:2019biz}
D.~S. Berman, \emph{{A Kaluza-Klein Approach to Double and Exceptional Field
  Theory}},  in \emph{{Durham Symposium, Higher Structures in M-Theory Durham,
  UK, August 12-18, 2018}}, 2019,
  \href{https://arxiv.org/abs/1903.02860}{{\ttfamily 1903.02860}}.

\bibitem{Bergshoeff:2011se}
E.~A. Bergshoeff, T.~Ortin and F.~Riccioni, \emph{{Defect Branes}},
  \href{https://doi.org/10.1016/j.nuclphysb.2011.10.037}{\emph{Nucl. Phys.}
  {\bfseries B856} (2012) 210}
  [\href{https://arxiv.org/abs/1109.4484}{{\ttfamily 1109.4484}}].

\bibitem{Bergshoeff:2012ex}
E.~A. Bergshoeff, A.~Marrani and F.~Riccioni, \emph{{Brane orbits}},
  \href{https://doi.org/10.1016/j.nuclphysb.2012.03.014}{\emph{Nucl. Phys.}
  {\bfseries B861} (2012) 104}
  [\href{https://arxiv.org/abs/1201.5819}{{\ttfamily 1201.5819}}].

\bibitem{Bergshoeff:2012pm}
E.~A. Bergshoeff, A.~Kleinschmidt and F.~Riccioni, \emph{{Supersymmetric Domain
  Walls}}, \href{https://doi.org/10.1103/PhysRevD.86.085043}{\emph{Phys. Rev.}
  {\bfseries D86} (2012) 085043}
  [\href{https://arxiv.org/abs/1206.5697}{{\ttfamily 1206.5697}}].

\bibitem{Corley:2001zk}
S.~Corley, A.~Jevicki and S.~Ramgoolam, \emph{{Exact correlators of giant
  gravitons from dual N=4 SYM theory}},
  \href{https://doi.org/10.4310/ATMP.2001.v5.n4.a6}{\emph{Adv. Theor. Math.
  Phys.} {\bfseries 5} (2002) 809}
  [\href{https://arxiv.org/abs/hep-th/0111222}{{\ttfamily hep-th/0111222}}].

\bibitem{Sorkin:1983ns}
R.~d. Sorkin, \emph{{Kaluza-Klein Monopole}},
  \href{https://doi.org/10.1103/PhysRevLett.51.87}{\emph{Phys. Rev. Lett.}
  {\bfseries 51} (1983) 87}.

\bibitem{Gross:1983hb}
D.~J. Gross and M.~J. Perry, \emph{{Magnetic Monopoles in Kaluza-Klein
  Theories}}, \href{https://doi.org/10.1016/0550-3213(83)90462-5}{\emph{Nucl.
  Phys.} {\bfseries B226} (1983) 29}.

\bibitem{Sakatani:2014hba}
Y.~Sakatani, \emph{{Exotic branes and non-geometric fluxes}},
  \href{https://doi.org/10.1007/JHEP03(2015)135}{\emph{JHEP} {\bfseries 03}
  (2015) 135} [\href{https://arxiv.org/abs/1412.8769}{{\ttfamily 1412.8769}}].

\bibitem{Bergshoeff:2015cba}
E.~A. Bergshoeff, V.~A. Penas, F.~Riccioni and S.~Risoli, \emph{{Non-geometric
  fluxes and mixed-symmetry potentials}},
  \href{https://doi.org/10.1007/JHEP11(2015)020}{\emph{JHEP} {\bfseries 11}
  (2015) 020} [\href{https://arxiv.org/abs/1508.00780}{{\ttfamily
  1508.00780}}].

\bibitem{Bergshoeff:2016ncb}
E.~A. Bergshoeff, O.~Hohm, V.~A. Penas and F.~Riccioni, \emph{{Dual Double
  Field Theory}}, \href{https://doi.org/10.1007/JHEP06(2016)026}{\emph{JHEP}
  {\bfseries 06} (2016) 026}
  [\href{https://arxiv.org/abs/1603.07380}{{\ttfamily 1603.07380}}].

\bibitem{Lombardo:2016swq}
D.~M. Lombardo, F.~Riccioni and S.~Risoli, \emph{{$P$ fluxes and exotic
  branes}}, \href{https://doi.org/10.1007/JHEP12(2016)114}{\emph{JHEP}
  {\bfseries 12} (2016) 114}
  [\href{https://arxiv.org/abs/1610.07975}{{\ttfamily 1610.07975}}].

\bibitem{West:2001as}
P.~C. West, \emph{{E(11) and M theory}},
  \href{https://doi.org/10.1088/0264-9381/18/21/305}{\emph{Class. Quant. Grav.}
  {\bfseries 18} (2001) 4443}
  [\href{https://arxiv.org/abs/hep-th/0104081}{{\ttfamily hep-th/0104081}}].

\bibitem{Cook:2009ri}
P.~P. Cook, \emph{{Exotic E(11) branes as composite gravitational solutions}},
  \href{https://doi.org/10.1088/0264-9381/26/23/235023}{\emph{Class. Quant.
  Grav.} {\bfseries 26} (2009) 235023}
  [\href{https://arxiv.org/abs/0908.0485}{{\ttfamily 0908.0485}}].

\bibitem{Kimura:2013fda}
T.~Kimura and S.~Sasaki, \emph{{Gauged Linear Sigma Model for Exotic
  Five-brane}},
  \href{https://doi.org/10.1016/j.nuclphysb.2013.08.017}{\emph{Nucl. Phys.}
  {\bfseries B876} (2013) 493}
  [\href{https://arxiv.org/abs/1304.4061}{{\ttfamily 1304.4061}}].

\bibitem{Kimura:2013zva}
T.~Kimura and S.~Sasaki, \emph{{Worldsheet instanton corrections to
  $5^2_2$-brane geometry}},
  \href{https://doi.org/10.1007/JHEP08(2013)126}{\emph{JHEP} {\bfseries 08}
  (2013) 126} [\href{https://arxiv.org/abs/1305.4439}{{\ttfamily 1305.4439}}].

\bibitem{Chatzistavrakidis:2013jqa}
A.~Chatzistavrakidis, F.~F. Gautason, G.~Moutsopoulos and M.~Zagermann,
  \emph{{Effective actions of nongeometric five-branes}},
  \href{https://doi.org/10.1103/PhysRevD.89.066004}{\emph{Phys. Rev.}
  {\bfseries D89} (2014) 066004}
  [\href{https://arxiv.org/abs/1309.2653}{{\ttfamily 1309.2653}}].

\bibitem{Kimura:2014upa}
T.~Kimura, S.~Sasaki and M.~Yata, \emph{{World-volume Effective Actions of
  Exotic Five-branes}},
  \href{https://doi.org/10.1007/JHEP07(2014)127}{\emph{JHEP} {\bfseries 07}
  (2014) 127} [\href{https://arxiv.org/abs/1404.5442}{{\ttfamily 1404.5442}}].

\bibitem{Kimura:2016anf}
T.~Kimura, S.~Sasaki and M.~Yata, \emph{{World-volume Effective Action of
  Exotic Five-brane in M-theory}},
  \href{https://doi.org/10.1007/JHEP02(2016)168}{\emph{JHEP} {\bfseries 02}
  (2016) 168} [\href{https://arxiv.org/abs/1601.05589}{{\ttfamily
  1601.05589}}].

\bibitem{Plauschinn:2018wbo}
E.~Plauschinn, \emph{{Non-geometric backgrounds in string theory}},
  \href{https://doi.org/10.1016/j.physrep.2018.12.002}{\emph{Phys. Rept.}
  {\bfseries 798} (2019) 1} [\href{https://arxiv.org/abs/1811.11203}{{\ttfamily
  1811.11203}}].

\bibitem{Kimura:2016xzd}
T.~Kimura, \emph{{Supersymmetry projection rules on exotic branes}},
  \href{https://doi.org/10.1093/ptep/ptw052}{\emph{PTEP} {\bfseries 2016}
  (2016) 053B05} [\href{https://arxiv.org/abs/1601.02175}{{\ttfamily
  1601.02175}}].

\bibitem{Lombardo:2017yme}
D.~M. Lombardo, F.~Riccioni and S.~Risoli, \emph{{Non-geometric fluxes \&
  tadpole conditions for exotic branes}},
  \href{https://doi.org/10.1007/JHEP10(2017)134}{\emph{JHEP} {\bfseries 10}
  (2017) 134} [\href{https://arxiv.org/abs/1704.08566}{{\ttfamily
  1704.08566}}].

\bibitem{Bergshoeff:2017gpw}
E.~A. Bergshoeff and F.~Riccioni, \emph{{Wrapping rules (in) string theory}},
  \href{https://doi.org/10.1007/JHEP01(2018)046}{\emph{JHEP} {\bfseries 01}
  (2018) 046} [\href{https://arxiv.org/abs/1710.00642}{{\ttfamily
  1710.00642}}].

\bibitem{Kleinschmidt:2011vu}
A.~Kleinschmidt, \emph{{Counting supersymmetric branes}},
  \href{https://doi.org/10.1007/JHEP10(2011)144}{\emph{JHEP} {\bfseries 10}
  (2011) 144} [\href{https://arxiv.org/abs/1109.2025}{{\ttfamily 1109.2025}}].

\bibitem{Sakatani:2017nfr}
Y.~Sakatani and S.~Uehara, \emph{{Connecting M-theory and type IIB
  parameterizations in Exceptional Field Theory}},
  \href{https://doi.org/10.1093/ptep/ptx038}{\emph{PTEP} {\bfseries 2017}
  (2017) 043B05} [\href{https://arxiv.org/abs/1701.07819}{{\ttfamily
  1701.07819}}].

\bibitem{Andriot:2014uda}
D.~Andriot and A.~Betz, \emph{{NS-branes, source corrected Bianchi identities,
  and more on backgrounds with non-geometric fluxes}},
  \href{https://doi.org/10.1007/JHEP07(2014)059}{\emph{JHEP} {\bfseries 07}
  (2014) 059} [\href{https://arxiv.org/abs/1402.5972}{{\ttfamily 1402.5972}}].

\bibitem{Hassler:2013wsa}
F.~Hassler and D.~Lust, \emph{{Non-commutative/non-associative IIA (IIB) Q- and
  R-branes and their intersections}},
  \href{https://doi.org/10.1007/JHEP07(2013)048}{\emph{JHEP} {\bfseries 07}
  (2013) 048} [\href{https://arxiv.org/abs/1303.1413}{{\ttfamily 1303.1413}}].

\bibitem{Blair:2013gqa}
C.~D.~A. Blair, E.~Malek and J.-H. Park, \emph{{M-theory and Type IIB from a
  Duality Manifest Action}},
  \href{https://doi.org/10.1007/JHEP01(2014)172}{\emph{JHEP} {\bfseries 01}
  (2014) 172} [\href{https://arxiv.org/abs/1311.5109}{{\ttfamily 1311.5109}}].

\bibitem{Park:2013gaj}
J.-H. Park and Y.~Suh, \emph{{U-geometry: SL(5)}},
  \href{https://doi.org/10.1007/JHEP11(2013)210,
  10.1007/JHEP04(2013)147}{\emph{JHEP} {\bfseries 04} (2013) 147}
  [\href{https://arxiv.org/abs/1302.1652}{{\ttfamily 1302.1652}}].

\bibitem{LeDiffon:2008sh}
A.~Le~Diffon and H.~Samtleben, \emph{{Supergravities without an Action: Gauging
  the Trombone}},
  \href{https://doi.org/10.1016/j.nuclphysb.2008.11.010}{\emph{Nucl. Phys.}
  {\bfseries B811} (2009) 1} [\href{https://arxiv.org/abs/0809.5180}{{\ttfamily
  0809.5180}}].

\bibitem{LeDiffon:2011wt}
A.~Le~Diffon, H.~Samtleben and M.~Trigiante, \emph{{N=8 Supergravity with Local
  Scaling Symmetry}},
  \href{https://doi.org/10.1007/JHEP04(2011)079}{\emph{JHEP} {\bfseries 04}
  (2011) 079} [\href{https://arxiv.org/abs/1103.2785}{{\ttfamily 1103.2785}}].

\bibitem{Tong:2002rq}
D.~Tong, \emph{{NS5-branes, T duality and world sheet instantons}},
  \href{https://doi.org/10.1088/1126-6708/2002/07/013}{\emph{JHEP} {\bfseries
  07} (2002) 013} [\href{https://arxiv.org/abs/hep-th/0204186}{{\ttfamily
  hep-th/0204186}}].

\bibitem{Harvey:2005ab}
J.~A. Harvey and S.~Jensen, \emph{{Worldsheet instanton corrections to the
  Kaluza-Klein monopole}},
  \href{https://doi.org/10.1088/1126-6708/2005/10/028}{\emph{JHEP} {\bfseries
  10} (2005) 028} [\href{https://arxiv.org/abs/hep-th/0507204}{{\ttfamily
  hep-th/0507204}}].

\bibitem{Lust:2017jox}
D.~Lüst, E.~Plauschinn and V.~Vall~Camell, \emph{{Unwinding strings in
  semi-flatland}}, \href{https://doi.org/10.1007/JHEP07(2017)027}{\emph{JHEP}
  {\bfseries 07} (2017) 027}
  [\href{https://arxiv.org/abs/1706.00835}{{\ttfamily 1706.00835}}].

\bibitem{Kimura:2018ain}
T.~Kimura, S.~Sasaki and K.~Shiozawa, \emph{{Semi-doubled Gauged Linear Sigma
  Model for Five-branes of Codimension Two}},
  \href{https://doi.org/10.1007/JHEP12(2018)095}{\emph{JHEP} {\bfseries 12}
  (2018) 095} [\href{https://arxiv.org/abs/1810.02169}{{\ttfamily
  1810.02169}}].

\bibitem{Ortin:2015hya}
T.~Ortin, \emph{{Gravity and Strings}}, Cambridge Monographs on Mathematical
  Physics. Cambridge University Press, 2015,
  \href{https://doi.org/10.1017/CBO9781139019750}{10.1017/CBO9781139019750}.

\bibitem{Gibbons:1995vg}
G.~W. Gibbons, M.~B. Green and M.~J. Perry, \emph{{Instantons and seven-branes
  in type IIB superstring theory}},
  \href{https://doi.org/10.1016/0370-2693(95)01565-5}{\emph{Phys. Lett.}
  {\bfseries B370} (1996) 37}
  [\href{https://arxiv.org/abs/hep-th/9511080}{{\ttfamily hep-th/9511080}}].

\bibitem{Tseytlin:1996ne}
A.~A. Tseytlin, \emph{{Type IIB instanton as a wave in twelve-dimensions}},
  \href{https://doi.org/10.1103/PhysRevLett.78.1864}{\emph{Phys. Rev. Lett.}
  {\bfseries 78} (1997) 1864}
  [\href{https://arxiv.org/abs/hep-th/9612164}{{\ttfamily hep-th/9612164}}].

\bibitem{LozanoTellechea:2000mc}
E.~Lozano-Tellechea and T.~Ortin, \emph{{7-branes and higher Kaluza-Klein
  branes}}, \href{https://doi.org/10.1016/S0550-3213(01)00177-8}{\emph{Nucl.
  Phys.} {\bfseries B607} (2001) 213}
  [\href{https://arxiv.org/abs/hep-th/0012051}{{\ttfamily hep-th/0012051}}].

\bibitem{Kimura:2016yqa}
T.~Kimura, \emph{{Exotic Brane Junctions from F-theory}},
  \href{https://doi.org/10.1007/JHEP05(2016)060}{\emph{JHEP} {\bfseries 05}
  (2016) 060} [\href{https://arxiv.org/abs/1602.08606}{{\ttfamily
  1602.08606}}].

\bibitem{Hull:1998mh}
C.~M. Hull, \emph{{Gravitational duality, branes and charges}},
  \href{https://doi.org/10.1016/S0920-5632(97)00682-8}{\emph{Nucl. Phys. Proc.
  Suppl.} {\bfseries 62} (1998) 412}.

\bibitem{Bergshoeff:1998bs}
E.~Bergshoeff and J.~P. van~der Schaar, \emph{{On M nine-branes}},
  \href{https://doi.org/10.1088/0264-9381/16/1/002}{\emph{Class. Quant. Grav.}
  {\bfseries 16} (1999) 23}
  [\href{https://arxiv.org/abs/hep-th/9806069}{{\ttfamily hep-th/9806069}}].

\bibitem{Fernandez-Melgarejo:2019mgd}
J.~J. Fernández-Melgarejo, Y.~Sakatani and S.~Uehara, \emph{{Exotic branes and
  mixed-symmetry potentials I: predictions from $E_{11}$ symmetry}},
  \href{https://arxiv.org/abs/1907.07177}{{\ttfamily 1907.07177}}.

\bibitem{Cook:2011ir}
P.~P. Cook, \emph{{Bound States of String Theory and Beyond}},
  \href{https://doi.org/10.1007/JHEP03(2012)028}{\emph{JHEP} {\bfseries 03}
  (2012) 028} [\href{https://arxiv.org/abs/1109.6595}{{\ttfamily 1109.6595}}].

\bibitem{Hohm:2018ybo}
O.~Hohm and H.~Samtleben, \emph{{Leibniz–Chern–Simons Theory and Phases of
  Exceptional Field Theory}},
  \href{https://doi.org/10.1007/s00220-019-03347-1}{\emph{Commun. Math. Phys.}
  {\bfseries 369} (2019) 1055}
  [\href{https://arxiv.org/abs/1805.03220}{{\ttfamily 1805.03220}}].

\bibitem{Jeon:2011cn}
I.~Jeon, K.~Lee and J.-H. Park, \emph{{Stringy differential geometry, beyond
  Riemann}}, \href{https://doi.org/10.1103/PhysRevD.84.044022}{\emph{Phys.
  Rev.} {\bfseries D84} (2011) 044022}
  [\href{https://arxiv.org/abs/1105.6294}{{\ttfamily 1105.6294}}].

\bibitem{Siegel:2015axg}
W.~Siegel, \emph{{Amplitudes for left-handed strings}},
  \href{https://arxiv.org/abs/1512.02569}{{\ttfamily 1512.02569}}.

\bibitem{Casali:2016atr}
E.~Casali and P.~Tourkine, \emph{{On the null origin of the ambitwistor
  string}}, \href{https://doi.org/10.1007/JHEP11(2016)036}{\emph{JHEP}
  {\bfseries 11} (2016) 036}
  [\href{https://arxiv.org/abs/1606.05636}{{\ttfamily 1606.05636}}].

\bibitem{Casali:2017mss}
E.~Casali and P.~Tourkine, \emph{{Windings of twisted strings}},
  \href{https://doi.org/10.1103/PhysRevD.97.061902}{\emph{Phys. Rev.}
  {\bfseries D97} (2018) 061902}
  [\href{https://arxiv.org/abs/1710.01241}{{\ttfamily 1710.01241}}].

\bibitem{Lee:2017utr}
K.~Lee, S.-J. Rey and J.~A. Rosabal, \emph{{A string theory which isn’t about
  strings}}, \href{https://doi.org/10.1007/JHEP11(2017)172}{\emph{JHEP}
  {\bfseries 11} (2017) 172}
  [\href{https://arxiv.org/abs/1708.05707}{{\ttfamily 1708.05707}}].

\bibitem{Lee:2017crr}
K.~Lee and J.~A. Rosabal, \emph{{A Note on Circle Compactification of Tensile
  Ambitwistor String}},
  \href{https://doi.org/10.1016/j.nuclphysb.2018.06.016}{\emph{Nucl. Phys.}
  {\bfseries B933} (2018) 482}
  [\href{https://arxiv.org/abs/1712.05874}{{\ttfamily 1712.05874}}].

\bibitem{Dijkgraaf:2016lym}
R.~Dijkgraaf, B.~Heidenreich, P.~Jefferson and C.~Vafa, \emph{{Negative Branes,
  Supergroups and the Signature of Spacetime}},
  \href{https://doi.org/10.1007/JHEP02(2018)050}{\emph{JHEP} {\bfseries 02}
  (2018) 050} [\href{https://arxiv.org/abs/1603.05665}{{\ttfamily
  1603.05665}}].

\bibitem{Hull:1998vg}
C.~M. Hull, \emph{{Timelike T duality, de Sitter space, large N gauge theories
  and topological field theory}},
  \href{https://doi.org/10.1088/1126-6708/1998/07/021}{\emph{JHEP} {\bfseries
  07} (1998) 021} [\href{https://arxiv.org/abs/hep-th/9806146}{{\ttfamily
  hep-th/9806146}}].

\bibitem{Hull:1998ym}
C.~M. Hull, \emph{{Duality and the signature of space-time}},
  \href{https://doi.org/10.1088/1126-6708/1998/11/017}{\emph{JHEP} {\bfseries
  11} (1998) 017} [\href{https://arxiv.org/abs/hep-th/9807127}{{\ttfamily
  hep-th/9807127}}].

\bibitem{Hohm:2018qhd}
O.~Hohm and H.~Samtleben, \emph{{The dual graviton in duality covariant
  theories}}, \href{https://doi.org/10.1002/prop.201900021}{\emph{Fortsch.
  Phys.} {\bfseries 67} (2019) 1900021}
  [\href{https://arxiv.org/abs/1807.07150}{{\ttfamily 1807.07150}}].

\bibitem{Rosabal:2014rga}
J.~A. Rosabal, \emph{{On the exceptional generalised Lie derivative for
  $d\geq7$}}, \href{https://doi.org/10.1007/JHEP09(2015)153}{\emph{JHEP}
  {\bfseries 09} (2015) 153} [\href{https://arxiv.org/abs/1410.8148}{{\ttfamily
  1410.8148}}].

\bibitem{Tseytlin:1981ks}
A.~A. Tseytlin, \emph{{On the First Order Formalism in Quantum Gravity}},
  \href{https://doi.org/10.1088/0305-4470/15/3/005}{\emph{J. Phys.} {\bfseries
  A15} (1982) L105}.

\bibitem{Witten:1988xj}
E.~Witten, \emph{{Topological Sigma Models}},
  \href{https://doi.org/10.1007/BF01466725}{\emph{Commun. Math. Phys.}
  {\bfseries 118} (1988) 411}.

\bibitem{Horowitz:1989ng}
G.~T. Horowitz, \emph{{Exactly Soluble Diffeomorphism Invariant Theories}},
  \href{https://doi.org/10.1007/BF01218410}{\emph{Commun. Math. Phys.}
  {\bfseries 125} (1989) 417}.

\bibitem{Thompson:2011uw}
D.~C. Thompson, \emph{{Duality Invariance: From M-theory to Double Field
  Theory}}, \href{https://doi.org/10.1007/JHEP08(2011)125}{\emph{JHEP}
  {\bfseries 08} (2011) 125} [\href{https://arxiv.org/abs/1106.4036}{{\ttfamily
  1106.4036}}].

\bibitem{Riccioni:2007au}
F.~Riccioni and P.~C. West, \emph{{The E(11) origin of all maximal
  supergravities}},
  \href{https://doi.org/10.1088/1126-6708/2007/07/063}{\emph{JHEP} {\bfseries
  07} (2007) 063} [\href{https://arxiv.org/abs/0705.0752}{{\ttfamily
  0705.0752}}].

\bibitem{West:2012qz}
P.~West, \emph{{E11, generalised space-time and equations of motion in four
  dimensions}}, \href{https://doi.org/10.1007/JHEP12(2012)068}{\emph{JHEP}
  {\bfseries 12} (2012) 068} [\href{https://arxiv.org/abs/1206.7045}{{\ttfamily
  1206.7045}}].

\bibitem{Obers:1999um}
N.~A. Obers and B.~Pioline, \emph{{Eisenstein series and string thresholds}},
  \href{https://doi.org/10.1007/s002200050022}{\emph{Commun. Math. Phys.}
  {\bfseries 209} (2000) 275}
  [\href{https://arxiv.org/abs/hep-th/9903113}{{\ttfamily hep-th/9903113}}].

\bibitem{Bossard:2016hgy}
G.~Bossard and B.~Pioline, \emph{{Exact $\nabla^4 R^4$ couplings and helicity
  supertraces}}, \href{https://doi.org/10.1007/JHEP01(2017)050}{\emph{JHEP}
  {\bfseries 01} (2017) 050}
  [\href{https://arxiv.org/abs/1610.06693}{{\ttfamily 1610.06693}}].

\bibitem{Bossard:2015foa}
G.~Bossard and A.~Kleinschmidt, \emph{{Loops in exceptional field theory}},
  \href{https://doi.org/10.1007/JHEP01(2016)164}{\emph{JHEP} {\bfseries 01}
  (2016) 164} [\href{https://arxiv.org/abs/1510.07859}{{\ttfamily
  1510.07859}}].

\bibitem{Keurentjes:2002xc}
A.~Keurentjes, \emph{{The Group theory of oxidation}},
  \href{https://doi.org/10.1016/S0550-3213(03)00178-0}{\emph{Nucl. Phys.}
  {\bfseries B658} (2003) 303}
  [\href{https://arxiv.org/abs/hep-th/0210178}{{\ttfamily hep-th/0210178}}].

\bibitem{Bossard:2014lra}
G.~Bossard and V.~Verschinin, \emph{{Minimal unitary representations from
  supersymmetry}}, \href{https://doi.org/10.1007/JHEP10(2014)008}{\emph{JHEP}
  {\bfseries 10} (2014) 008} [\href{https://arxiv.org/abs/1406.5527}{{\ttfamily
  1406.5527}}].

\bibitem{Blair:2018lbh}
C.~D.~A. Blair, E.~Malek and D.~C. Thompson, \emph{{O-folds: Orientifolds and
  Orbifolds in Exceptional Field Theory}},
  \href{https://doi.org/10.1007/JHEP09(2018)157}{\emph{JHEP} {\bfseries 09}
  (2018) 157} [\href{https://arxiv.org/abs/1805.04524}{{\ttfamily
  1805.04524}}].

\bibitem{Marrani:2010de}
A.~Marrani, E.~Orazi and F.~Riccioni, \emph{{Exceptional Reductions}},
  \href{https://doi.org/10.1088/1751-8113/44/15/155207}{\emph{J. Phys.}
  {\bfseries A44} (2011) 155207}
  [\href{https://arxiv.org/abs/1012.5797}{{\ttfamily 1012.5797}}].

\bibitem{deWit:2002vt}
B.~de~Wit, H.~Samtleben and M.~Trigiante, \emph{{On Lagrangians and gaugings of
  maximal supergravities}},
  \href{https://doi.org/10.1016/S0550-3213(03)00059-2}{\emph{Nucl. Phys.}
  {\bfseries B655} (2003) 93}
  [\href{https://arxiv.org/abs/hep-th/0212239}{{\ttfamily hep-th/0212239}}].

\bibitem{Dabholkar:2005ve}
A.~Dabholkar and C.~Hull, \emph{{Generalised T-duality and non-geometric
  backgrounds}},
  \href{https://doi.org/10.1088/1126-6708/2006/05/009}{\emph{JHEP} {\bfseries
  05} (2006) 009} [\href{https://arxiv.org/abs/hep-th/0512005}{{\ttfamily
  hep-th/0512005}}].

\bibitem{Berman:2011kg}
D.~S. Berman, E.~T. Musaev and M.~J. Perry, \emph{{Boundary Terms in
  Generalized Geometry and doubled field theory}},
  \href{https://doi.org/10.1016/j.physletb.2011.11.019}{\emph{Phys. Lett.}
  {\bfseries B706} (2011) 228}
  [\href{https://arxiv.org/abs/1110.3097}{{\ttfamily 1110.3097}}].

\bibitem{Pasti:2012wv}
P.~Pasti, D.~Sorokin and M.~Tonin, \emph{{Covariant actions for models with
  non-linear twisted self-duality}},
  \href{https://doi.org/10.1103/PhysRevD.86.045013}{\emph{Phys. Rev.}
  {\bfseries D86} (2012) 045013}
  [\href{https://arxiv.org/abs/1205.4243}{{\ttfamily 1205.4243}}].

\bibitem{Sen:2019qit}
A.~Sen, \emph{{Self-dual forms: Action, Hamiltonian and Compactification}},
  \href{https://doi.org/10.1088/1751-8121/ab5423}{\emph{J. Phys.} {\bfseries
  A53} (2020) 084002} [\href{https://arxiv.org/abs/1903.12196}{{\ttfamily
  1903.12196}}].

\bibitem{Mukhi:2008ux}
S.~Mukhi and C.~Papageorgakis, \emph{{M2 to D2}},
  \href{https://doi.org/10.1088/1126-6708/2008/05/085}{\emph{JHEP} {\bfseries
  05} (2008) 085} [\href{https://arxiv.org/abs/0803.3218}{{\ttfamily
  0803.3218}}].

\bibitem{Musaev:2014lna}
E.~Musaev and H.~Samtleben, \emph{{Fermions and supersymmetry in E$_{6(6)}$
  exceptional field theory}},
  \href{https://doi.org/10.1007/JHEP03(2015)027}{\emph{JHEP} {\bfseries 03}
  (2015) 027} [\href{https://arxiv.org/abs/1412.7286}{{\ttfamily 1412.7286}}].

\end{thebibliography}\endgroup
\end{document}